\documentclass[useAMS,usenatbib]{mn2e}

\usepackage[ansinew]{inputenc}
\usepackage{graphicx}
\usepackage{pdflscape}
\usepackage{longtable} 
\usepackage{multicol, blindtext}
\usepackage{color}
\usepackage{amssymb}

\def\aj{AJ}
\def\araa{ARA\&A}
\def\apj{ApJ}
\def\apjl{ApJ}
\def\apjs{ApJS}

\def\aap{A\&A}

\def\mnras{MNRAS}
\def\pasp{PASP}

\title[Probing the Torus Properties in Seyfert Galaxies]{Probing the Active Galactic Nuclei Unified Model Torus Properties in Seyfert Galaxies}
\author[]{Anelise Audibert$^{1}$\thanks{E-mail:
anelise.audibert@obspm.fr (OBSPM)}, Rog{\'e}rio Riffel$^{1}$\thanks{E-mail:
riffel@ufrgs.br (UFRGS)}, Dinalva A. Sales$^{2}$, Miriani G. Pastoriza$^{1}$ \and \& Daniel Ruschel-Dutra$^{1}$ \\
$^{1}$Departamento de Astronomia, Universidade Federal do Rio Grande do Sul, 9500 Bento Gon\c{c}alves, Porto Alegre, 91501-970, Brazil\\
$^{2}$Instituto de Matem{\'a}tica, Estat{\'i}stica e F{\' i}sica, Universidade Federal do Rio Grande, Rio Grande, 96203-900, Brazil\\}
\begin{document}
\date{Accepted 2016 September 27. Received 2016 September 27; in original form 2016 January 8}

\pagerange{\pageref{firstpage}--\pageref{lastpage}} \pubyear{2016}

\maketitle

\label{firstpage}

\begin{abstract}
We studied the physical parameters of a sample comprising of {\it all} {\it Spitzer/IRS} public spectra of Seyfert galaxies in the mid-infrared (5.2-38\,$\umu$m range) under the active galactic nuclei (AGN) unified model. We compare the observed spectra with $\sim10^6$ {\sc clumpy} model spectral energy distributions, which consider a torus composed of dusty clouds. We find a slight difference in the distribution  of line-of-sight inclination angle, $i$, requiring larger angles for Seyfert 2 (Sy\,2) and a broader distribution for Seyfert 1 (Sy\,1). We found small differences in the torus angular width, $\sigma$, indicating that Sy\,1 may host a slightly narrower torus than Sy\,2. The torus thickness, together with the bolometric luminosities derived, suggest a very compact torus up to $\sim$6\,pc from the central AGN. The number of clouds along the equatorial plane, $N$, as well the index of the radial profile, $q$, are nearly the same for both types. These results imply that the torus cloud distribution is nearly the same for type 1 and type 2 objects. The torus mass is almost the same for both types of activity, with values in the range of $M_{tor}\sim$10$^{4}-$10$^{7}\rm M_{\odot}$. The main difference appears to be related to the clouds' intrinsic properties: type 2 sources present higher optical depths $\tau_V$. The results presented here reinforce the suggestion that the classification of a galaxy may depend also on the intrinsic properties of the torus clouds rather than simply on their inclination. This is in contradiction with the simple geometric idea of the unification model.

\end{abstract}


\section{Introduction}

According to the unified model, the energy from active galactic nuclei (AGN) is powered by the accretion of matter into a supermassive black hole (SMBH) \citep{lynden,beg84}. The unification scheme suggests that the different AGN types are explained by the line-of-sight (LOS) orientation of an obscuring material, which surrounds the central source and is arranged in an axisymmetric/toroidal geometry and composed primarily of gas and dust. Under edge-on views, it obscures the radiation from the accretion disk and broad line region (BLR). Such an object is classified as a type 2 AGN. When viewed face-on, the central engine can be observed directly. These galaxies are classified as type 1 AGNs \citep{ant93, urr95}. The unified model was firstly supported through spectropolarimetric observations of the Seyfert 2 (Sy\,2) galaxy NGC\,1068 \citep{am85}, revealing the polarized broad emission lines, and followed by other polarized broad line observations in type 2 AGNs \citep[e.g., by][]{tran92, tran95}. This hidden type 1 emission can be visible via light scattering in the ionizing cones, which corresponds to ionizing radiation that is collimated by the torus opening angle, providing additional indirect evidence for the unified model \citep[e.g.,][]{pogge88, thaisa91,thaisa92}.

The dusty structure of the AGN unified model is responsible for absorbing short wavelength light from the active nucleus and re-emitting it mainly in the infrared (IR) wavelengths, leading to a peculiar signature in the spectral energy distribution (SED) of a galaxy. In specific, the silicate feature at \hbox{9.7\,$\umu$m} in the mid-infrared (MIR) is frequently found in absorption in type 2 and is also expected to appear in emission in type 1. However, in most type 1 objects this feature is either mild or absent \citep{hao07,wu09}. In addition, there are some cases where silicate emission is detected in type 2 \citep[e.g., in Sy\,2 NGC\,2110 by][]{mason09,sturm06}. Consequently, the MIR spectral range hosts fundamental features necessary to study the putative torus required in the Unified Model for AGNs. The recent significant advances in observational facilities, such as ALMA and VLTI, now allow us to resolve the central parsec scales in nearby AGNs. So far, only a few VLBI observations achieve sufficient spatial resolution to isolate the emission of these obscured structures. For instance, the cases of NGC\,1068 \citep{gra15,jaffe04,lop14} and Circinus \citep{tri07,tri14}. Very recently, \citet{santi16} and \citet{ima16} presented the first resolved images of the torus of NGC\,1068, the former using the continuum and the CO(6-5) emission observed with ALMA Band\,9 and the latter using HCN(3-2) and HCO$\rm ^{+}$(3-2) emission lines. These cases are successful precedents for forthcoming ALMA observations intended to study molecular gas in the torus and its surrounding \citep{net15}. While there are no plenty of data with such detailed observations, the optimal way to probe the physical processes related to the torus is understanding the radiation reprocessing mechanisms responsible for the singular behaviour of the AGN SEDs.

In the last two decades, several models have been developed in order to understand the torus emission. For example, \citet{kro88} proposed that the torus should constitute of a large number of optically thick dusty clouds, otherwise the dust grains would be destroyed by the high AGN energy luminosity. Their presumption is reinforced by VLTI interferometric observations in the N-band range (\hbox{8--13\,$\umu$m}) performed by \citet{tri07} in the nucleus of the Circinus galaxy, providing strong evidence of a clumpy and dusty structure. Due to computational issues in modelling a clumpy medium, some studies have explored the effect of a dusty uniform distribution in a toroidal geometry \citep{pier92, gra94, efs95, dul05, fri06}. However, to explain the low resolution ($>$1'') observed SEDs and IR spectra in such homogeneous descriptions, the models force the torus size to be at scales of $\gtrsim$100\,pc. With the advent of high-spatial resolution using 8$m$-class telescopes, it was demonstrated that the surrounded dusty environment is much more compact, with sizes of a few parsecs \citep{jaffe04, tristram09, burtscher09}.

Nevertheless, in the last few years, some efforts have been made to handle a clumpy formalism and they can naturally explain the problem with the silicate issue mentioned above \citep{nkv02,honig06,nkv08a,nkv08b,scha08,sta12}. Among them, the {\sc clumpy} models presented by \citet{nkv02,nkv08a,nkv08b} are, to date, some of the most successful models for representing the re-processed MIR torus emission and allowing us to constrain some torus properties. They consist of a large database of theoretical SEDs resulting from a 1D radiative transfer code \citep[{\sc dusty},][]{ive99} taking into account the continuum emission from clumpy media with shadowed individual clouds. One of the advantages of a clumpiness formalism is that they can reproduce more realistic MIR spectra. This is because they have a wide range of dusty cloud temperatures at the same radius from the central source and even distant clouds can be directly illuminated by the AGN, contrary to the smooth density distributions. The {\sc clumpy} models have been used by several works to study the torus properties, for example, in a sample of 26 quasar by \citet{mor09}, in a analysis of 27 Sy\,2 by \citet{lir13} and in modest to small samples of Seyfert galaxies \citep[e.g.][]{ichi15} and the works from \citet{ah11,ram09,ram11}, hereafter, AH11, RA09 and RA11, respectively. One of the main differences between our work and AH11 and RA11 is the use of near infrared (NIR) data for all the galaxies in their analysis and our much larger sample.  

We investigated the torus properties of 111 Seyfert galaxies using data archive from Infrared Spectrograph \citep[IRS,][]{houck04} aboard the {\it Spitzer Space Telescope} in the \hbox{5.2--38\,$\umu$m} spectral range. We compared the sample with the {\sc clumpy} theoretical SEDs from \citet{nkv02,nkv08a,nkv08b} using two different approaches: the $\chi_{red}^2$ test as well as bayesian inference \citep[{\sc BayesClumpy},][]{ase09}. Section 2 characterizes the sample and data reduction. In section 3 we describe the {\sc clumpy} models, the PAH's decontamination from the SEDs and the different approaches utilized to fit the data. Main results and the discussion are summarized in Section 4 and the contribution of the NIR data is exploited in Section 5. Our conclusions are presented in Section 6.

\section[]{The Data}

We have performed an analysis on a sample of 111 nearby galaxies classified as Seyfert galaxies that were available in the {\it Spitzer Heritage Archive}. The sample consists of 84 galaxies that have been presented in previous works \citep{gal10,wu09}, 14 galaxies from \citet{sal10} and another 13 objects available in the Spitzer archive (presented here for the first time). The galaxies were observed with the IRS using two low spectral resolution (R$\sim$60-127) modules: Short-Low (SL) and Long-Low (LL), covering a wavelength range from 5.2 to \hbox{38\,$\umu$m} in the MIR. The SL module has an image scale of 1.8''/pixel and the LL module 5.1''/pixel. Both are sub-divided in order 1 and 2. 

\begin{figure*}
\begin{center}
\includegraphics[width=172mm]{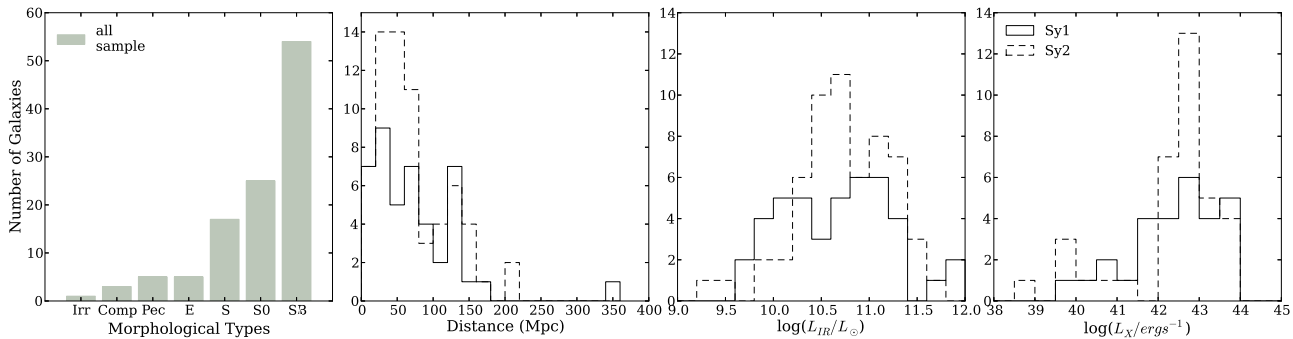}
\caption{Characteristics of the 111 AGN samples. The solid line histograms show the properties for Seyfert 1 galaxies while the dashed lines represent the Seyfert 2 sources. Morphological classification and distances were obtained from the NED - {\it NASA/IPAC Extragalactic Database} or from \citet{whi92}. In the left panel, we gathered all morphological types into 7 mean classes: irregular (Irr), compact (Comp), peculiar (Pec), elliptical (E), spiral (S), lenticular (S0) and barred spiral (SB). The references for L$_{IR}$ and L$_{X}$ can be founded in Table~\ref{obs:tab1}.}
\label{samp}
\end{center}
\end{figure*}

With the exception of 7 galaxies whose spectra are available from the SINGS Legacy program\footnote{The IRS data from the SINGS Legacy Project are available at http://irsa.ipac.caltech.edu/data/SPITZER/SINGS/. The nuclear spectra were extracted over a 50''$\times$33'' aperture.} \citep[PID 159,][]{sings}, all other data were processed using the Basic Calibration Data (BCD) pipeline\footnote{For more details, please see the IRS Instrument Handbook.} (version 18.18). The BCD pipeline manages the raw data through basic processing, such as the detection of cosmic rays, the removal of saturated pixels, dark current and flat-field subtraction and droop correction. For the sample presented in \citet{gal10}, 78 objects have the spectral mapping mode available, while the other 6 present the mapping mode only in LL and SL observations in the staring mode (NGC\,526A, NGC\,4941, NGC\,3227, IC\,5063, NGC\,7172 and NGC\,7314).

The mapping mode observations were processed by employing the CUbe Builder for IRS Spectra Maps \citep[{\sc CUBISM},][]{cubism07} to construct the data cubes. Sky subtraction were evaluated from an average spectra of the off-source orders: e.g. while the source is centred in the first order, the second order is pointed at the sky in an offset position. We used a 3.9$\times$11.1\,pixel extraction (equivalent to a 10'' circular radius aperture centred on the brightest source) to extract the spectra. In a few cases the extractions showed a mismatch between the modules or their orders that was corrected by scaling the spectra as recommended by \citet{pahfit07}.

For the remaining 27 objects the data are available in staring mode and the calibrated spectra were obtained from the Cornell Atlas of Spitzer/IRS Sources \citep[CASSIS,][]{cassis11}. CASSIS\footnote{CASSIS products are available at \newline http://cassis.astro.cornell.edu/atlas} provides optimal extractions and diagnostic tools to guarantee the most accurate background subtraction, especially for faint sources. In most cases, the optimal CASSIS extraction pointed to sky subtraction through the off-order method. However, in a few cases, CASSIS indicated the subtraction by nod positions as the best spectral extraction. Furthermore, the majority of CASSIS products were established as point-like sources\footnote{The optimal CASSIS extractions equivalent to extended sources are Mrk\,471, Mrk\,609, Mrk\,993, NGC\,5695, NGC\,5782, NGC\,7679 and NGC\,7682.}.

Our final sample is composed of 45 Sy~1 and 65 Sy~2 galaxies with redshifts between 0.002\,$\le$\,$z$\,$\le$\,0.079. The AGNs are preferentially found in host galaxies with barred spiral, lenticulars or in spiral morphological types. The mean values for the IR luminosities are L$\rm _{IR}$(Sy1)= 4.64$\times$10$^{10} \rm L_{\odot}$  for Sy\,1 and  L$\rm _{IR}$(Sy2)= 5.44$\times$10$^{10} \rm L_{\odot}$ and for the hard X-ray luminosities are L$_{\rm 2-10Kev}$(Sy1)= 1.59$\times$10$^{43}$ and L$_{\rm 2-10Kev}$(Sy2)= 1.19$\times$10$^{43}$ erg $\rm s^{-1}$. The sample properties are summarized in Figure~\ref{samp} and listed in Table~\ref{obs:tab1}.

\subsection[]{Removing the PAH Contamination}\label{pahremove}

Since the IRS {\it Spitzer} spectra were extracted in a 20'' circular diameter aperture, corresponding to $\sim$1-20\,Kpc of the galaxies (except for z=0.79, Mrk478 which represents $\sim$33\,Kpc), the host galaxy contribution is unavoidable in our sample. In order to minimize the effects from star formation and to isolate the AGN emission of the galaxy, we have adopted a similar method as that used in \citet{lir13}. They remove the stellar contribution by subtracting templates developed by \citet{pahfit07}, when the MIR is dominated by polycyclic aromatic hydrocarbons (PAHs) emission of star forming regions. 

Instead of fitting star formation templates, we chose to follow another approach in order to attenuate the PAH's contribution (and therefore, that of the host galaxy) in the spectra. We applied the {\sc pahfit} tool developed by \citet{pahfit07}. This code decomposes the emission lines of the low resolution IRS {\it Spitzer} spectra, modelling them as the sum of starlight continuum, thermal dust continuum, and emission lines (pure rotational lines of H$_2$, fine structure lines, and dust emission features). All flux intensity components are affected by dust extinction, quantified by the optical depth \citep[for more details, see][]{pahfit07}.

\begin{figure}
\begin{center}
\includegraphics[width=86mm]{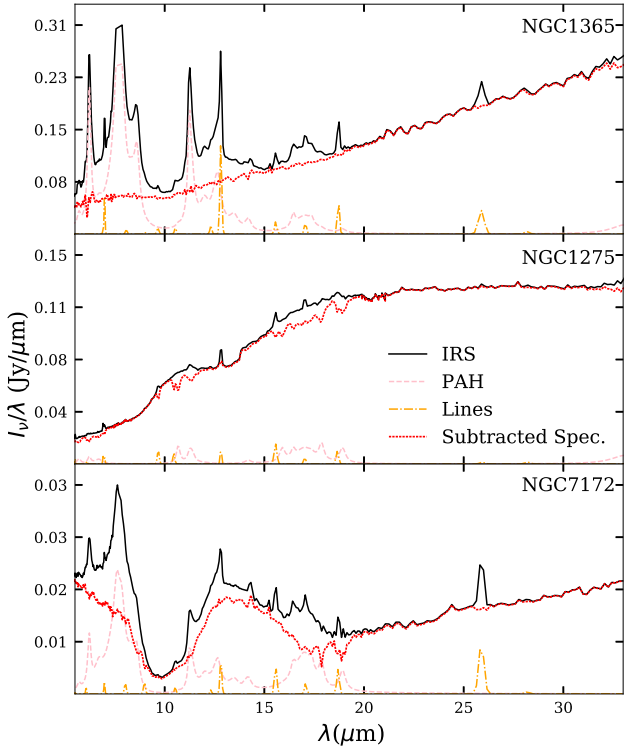}
\caption{Typical examples of the subtraction of the PAH and ionic line components from the spectra. The black lines represent the observed spectra, while the dotted/dashed orange and dashed pink lines show the resulting adjusted spectra created by fitting the PAH emission and ionic and hydrogen lines, respectively, using the {\sc pahfit} tool. In the dotted red lines are the subtracted spectra that were handled in our analysis. In the top panel the results for NGC\,1365 are shown, characterizing a galaxy with strong PAH emission. On the other hand, the middle panel shows little contribution from this feature for NGC\,1275. An example of a deep silicate absorption and PAH emission is presented in the bottom panel for the Sy\,2 NGC\,7172.}
\label{pah}
\end{center}
\end{figure}

\begin{figure}
\begin{center}
\includegraphics[width=86mm]{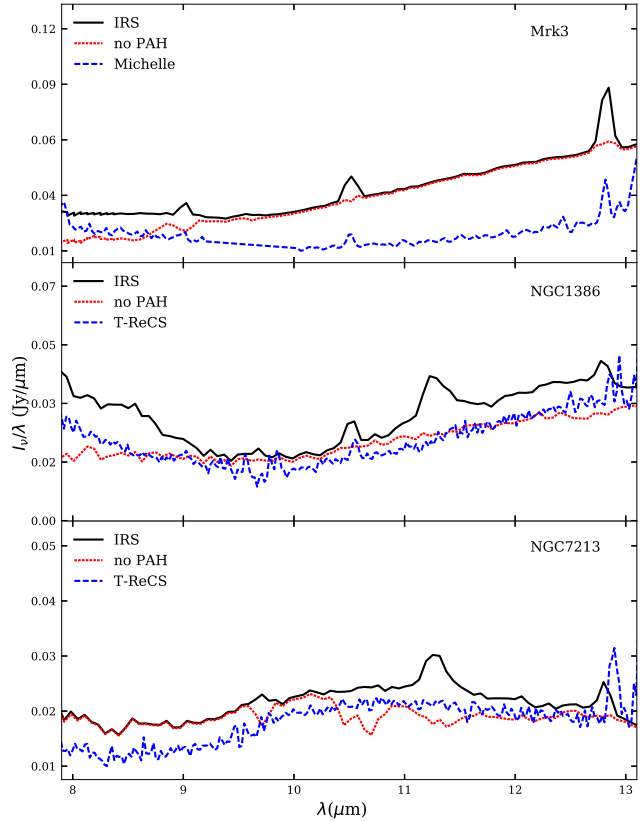}
\vspace{-0.5cm}
\caption{We show a comparison of the low resolution {\it Spitzer} IRS spectra and the ground based nuclear emission observations. The top panel shows the spectrum of Mrk\,3 from Michelle while the middle and bottom panels present the data from T-ReCs for NGC\,1386 and NGC\,7213, respectively. The black lines represent the IRS data, the dotted red lines the starburst subtracted spectra and the dashed blue lines the high resolution spectra. Except in the case of Mrk\,3, the subtracted spectra appear to well represent the emission from the active nucleus.}
\label{high}
\end{center}
\end{figure}

Since the continuum from our Seyfert sample is not only due to dust and stellar components, but also to the AGN power-law continuum, we decided to subtract only the emission lines from the H$_{2}$ and the molecular features of PAH emission from the observed spectra. It is, however, worth mentioning that most of the PAH contribution lies in 5-15\,$\mu$m, where the stellar emission is more prominent. For longer wavelengths, the host galaxy emission is very difficult to distinguish from the AGN continuum, and unfortunately, the current spectral decomposition codes are unable to separate each contribution for $\lambda\gtrsim$20\,$\mu$m \citep[e.g.,][]{pahfit07,her15}. This trend might overestimate AGN emission at longer wavelengths, which could bias the outputs of the CLUMPY models towards extended and broader tori, due to the cooler dust that peaks in the far-infrared range. Nonetheless, recently \citet{ful16} separated the AGN and PAH components using the full wavelength coverage of the {\it Spitzer}/IRS spectra of 11 Seyfert galaxies. They used the templates provided by \citet{her15} and then compared the AGN resulting spectra with the 31.5\,$\mu$m imaging photometry from the Stratospheric Observatory For Infrared Astronomy (SOFIA), finding that most of the sources are AGN dominated at 31.5\,$\mu$m.

In Figure~\ref{pah} we present some representative examples of this approach\footnote{In Appendix B we present the adjust for all the objects in the sample. The decomposed spectra files are available upon request, please contact the authors.}. Shown is a case with strong PAH emission and second with little PAH feature contribution (Sy\,1 NCG\,1365 and  Sy\,2 NGC\,1275 respectively). Also, we selected an example of deep silicate absorption in 9.7\,$\mu$m for NGC\,7172, which is surrounded by PAH emission. As can be seen, the effect of the molecular emission features is more prominent at shorter wavelengths and can alter the shape of the spectrum. The majority of galaxies (about 80\% of the sample) exhibit a substantial star forming contribution \citep[see also][]{sal10}.

Recently, some other studies have been supporting this star forming subtraction methodology. For instance, in \citet{rus14} there is no PAH emission detected using high resolution nuclear spectra from T-ReCs when compared with IRS observations of NGC\,7213 and NGC\,1386. Also, no PAH emission bands were detected in the nuclear region ($\sim$200\,pc) of Mrk\,3, a Compton-thick Sy~2, using Michelle/Gemini spectrograph \citep{sal14}. \citet{dav12} argued that the PAH molecules can not survive in a radius smaller than 50\,pc, a value corresponding to a region larger than that of the torus extension \citep[parsec scale,][]{tran92}. However, in some cases, e.g. in T-ReCs observations of NGC1808, the aromatic component was detected at 8.6 and 11.3$\umu$m in the galaxy centre ($\sim$26 pc) up to a radius of 70 pc from the nucleus \citep{sal13}.

To illustrate the effect of the starburst subtraction method, we show in Figure~\ref{high} the high resolution data from Michelle and T-ReCs of the galaxies Mrk\,3, NGC\,1386 and NGC\,7213, compared to the IRS spectra. Also shown is the final spectra resulting from the subtraction of the PAH components. The clean IRS spectra tends to better approximate the nuclear spectra from high resolution observations, except for the Mrk\,3. This is possibly due to the fact that this galaxy is a Compton-thick object and has a heavy absorbed dust/gas component \citep[$N_{H}\sim 1.1\times 10^{24} \rm cm^{-2}$,][]{sal14} obscuring the nucleus, leading to a higher continuum in the {\it Spitzer} observation. Moreover, the fact of this galaxy has a small starburst contribution may explain why we see almost no differences between the IRS spectra and the subtracted one. However, in the cases of NGC\,1386 and NGC\,7213, we believe that the PAH subtraction approach represents a good approximation of the nuclear emission. Thus, the spectral decomposition methodology was applied to all the objects used in our study.

\section[]{Modelling the SED in the MIR}

\subsection[]{{\sc clumpy} Torus Models}

A clumpy medium provides a natural explanation for the silicate absorption feature that was expected to be observed in emission in type 1 sources but is frequently mild or even flat, since it requires at least a clump obscuring the radiation at the observer's LOS. 

The most successful and up to date clumpy models are those of the Kentucky group.  \citet{nkv02,nkv08a,nkv08b} developed a formalism to handle a clumpy media, considering point-like dusty clouds distributed in a toroidal geometry around the central AGN. The {\sc clumpy} models are a large database ($\sim$10$^6$) of theoretical SEDs resulting from the radiative transfer treatment through the {\sc dusty} code \citep{ive99}. The dust grains follow the MNR size distribution \citep{mnr77} and are composed by the standard Galactic mixture of 47\% graphite with optical constants and 53\% cold silicates. While the graphite grains are the responsible for the IR emission at $\lambda\gtrsim 1\umu$m, the 9.7\,$\umu$m and 18\,$\umu$m emission and absorption features are attributed to silicate grains \citep[e.g.][]{bar87,pier92,gra94,sie04}. 

The {\sc clumpy} models assume that the torus is formed by dusty clumps constrained by the following parameters: (i) the number of clouds, $N_0$ , in the torus equatorial radius; (ii) $\tau_{V}$, the optical depth of each cloud defined at 0.55\,$\umu$m band; (iii) the radial extension of the clumpy distribution, $Y=R_{o}/R_{d}$, where $R_o$ and $R_d$ are the outer and inner radius of the torus, respectively; (iv) the radial distribution of clouds as described by a power law $r^{-q}$; (v) the torus angular width, $\sigma$, constrained by a Gaussian angular distribution width and (vi) the observers viewing angle $i$. The grid of these model parameters are listed in Table~\ref{obs:tab2}.

\begin{table}
\centering
\begin{minipage}{86mm}
\caption{Parameters values adopted in fitting}
\begin{scriptsize}
\begin{tabular}{@{}lcl@{}}
\hline
\hline
\multicolumn{1}{c}{} &  \multicolumn{1}{c}{Sampled Values} & \multicolumn{1}{c}{Description} \\
\hline
\noalign{\smallskip}
\multicolumn{3}{c}{{\sc clumpy} Models }\\
\noalign{\smallskip} 
\hline
$i$ & 0-90 steps of 10$^\circ$ & Observer's viewing angle \\ 
$N$ & 1-15 steps of 1 & Clouds along the equatorial plane \\
$q$ & 0-3 steps of 0.5 & Power law index of the radial distribution \\
$\tau_{V}$ & 5,10,20,30,40,60,80,100,150 & Optical depth of individual clouds \\
$\sigma$ & 15-70 steps of 5 & Torus angular width \\
$Y$ & 5, 10-100 steps of 10 & Torus thickness\\
\hline
\hline
\noalign{\smallskip}
\label{obs:tab2}
\end{tabular}
\end{scriptsize}
\end{minipage}
\end{table}

The model geometry also allow us to determine other parameters that are crucial to understand the obscuration effects of the central source. They are the number of clouds along the LOS, $N_{los}$, described by almost a Gaussian distribution along the equatorial plane ($N$), which depends on the inclination, $\beta=\pi/2 -i$, and angular width, $\sigma$, parameters  

\begin{equation}
N_{los}(\beta)=N exp^{- {\left ( \frac{\beta}{\sigma} \right )}^2}
\label{numlos}
\end{equation}
and the total optical depth of the torus along the LOS, product of the number of clouds and the optical depth of each cloud, or, the visual extinction:

\begin{equation}
A_{V}=1.086 N_{los} \tau_{V}
\label{averm}
\end{equation}

One of the characteristics of Nenkova et al. models is that the SEDs reproduced are not exclusively sensitive to the inclination angle, as established by the only orientation dependent unification schemes. The continuum shape and behaviour of the silicate features also have a strong dependence with the optical properties, characterized by the optical depth \hbox{$\tau_{V}$}, and the number of clouds along radial rays, specifically at the equatorial plane, $N$. In the latter, $N$ must be sufficient large, \hbox{$N\sim$5--10}, to assure the attenuation of X-rays in type 2 sources while the former one was constrained to values \hbox{$\tau_{V}\gtrsim$60} to ensure the probability of photon escape. The explanation for many problems faced by the smoothly distribution handling  are answered by the clumpiness nature of the toroidal structure and, therefore, the {\sc clumpy} models constitute a powerful tool to probe the torus physical properties proposed by the AGN's Unified Model.

\subsection[]{Fitting Procedure}

Once we applied the procedure to isolate the nuclear emission, we performed two different approaches to compare the MIR resulting spectra from IRS observations with Nenkova's theoretical models. In the following sections we describe the techniques employed.

\subsubsection[]{$\chi^2_{red}$ test}

We developed a code to compare each spectrum with all $10^6$ {\sc clumpy} models SEDs. The routine searches for the parameters which minimizes the equation:
\begin{equation}
{\chi_{red}}^2 = \frac{1}{N} \sum_{i=1}^N {\left ( \frac{F_{obs,\lambda_{i}} - F_{mod, \lambda_{i}}}{\sigma_{\lambda_i}} \right )}^2 
\label{chi2}
\end{equation}
where N is the number of data points in the spectrum, $F_{obs,\lambda_{i}}$,  and $F_{mod, \lambda_{i}}$,  are the observed and theoretical fluxes at each wavelength and $\sigma_{\lambda_i}$ are the uncertainties in $F_{obs,\lambda_{i}}$. Both $F_{obs,\lambda_{i}}$ and $F_{mod, \lambda_{i}}$ were normalized to unit at 28.0$\mu$m for all the galaxies in the sample, with the uncertainties correctly propagated. The ``decontaminated" nuclear spectrum was compared to the clumpy theoretical SEDs and we test the results for the best fit, e.g., the minimum $\chi^2_{red}$ and 5, 10, 15 and 20 per cent its deviation fractions, using a similar approach of \citet{nik09} and \citet{sal13}. In this work, we choose to represent the best fit and 10\% of deviation solutions.

\subsubsection[]{BayesCLUMPY Technique}

\begin{figure*}
\begin{center}
 \begin{minipage}{0.325\textwidth}
  \includegraphics[width=\textwidth]{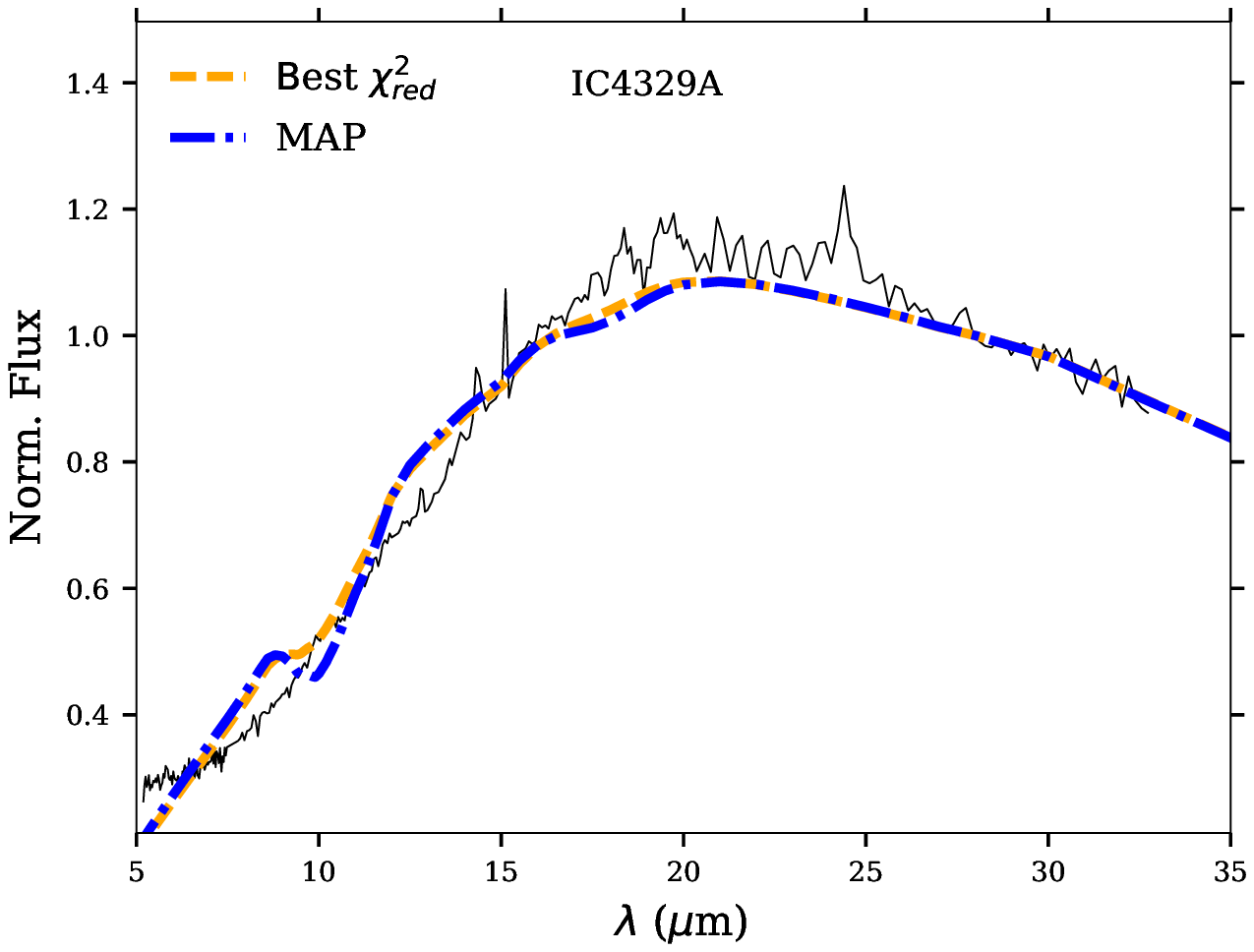}
 \end{minipage}
 \begin{minipage}{0.325\textwidth}
  \includegraphics[width=\textwidth]{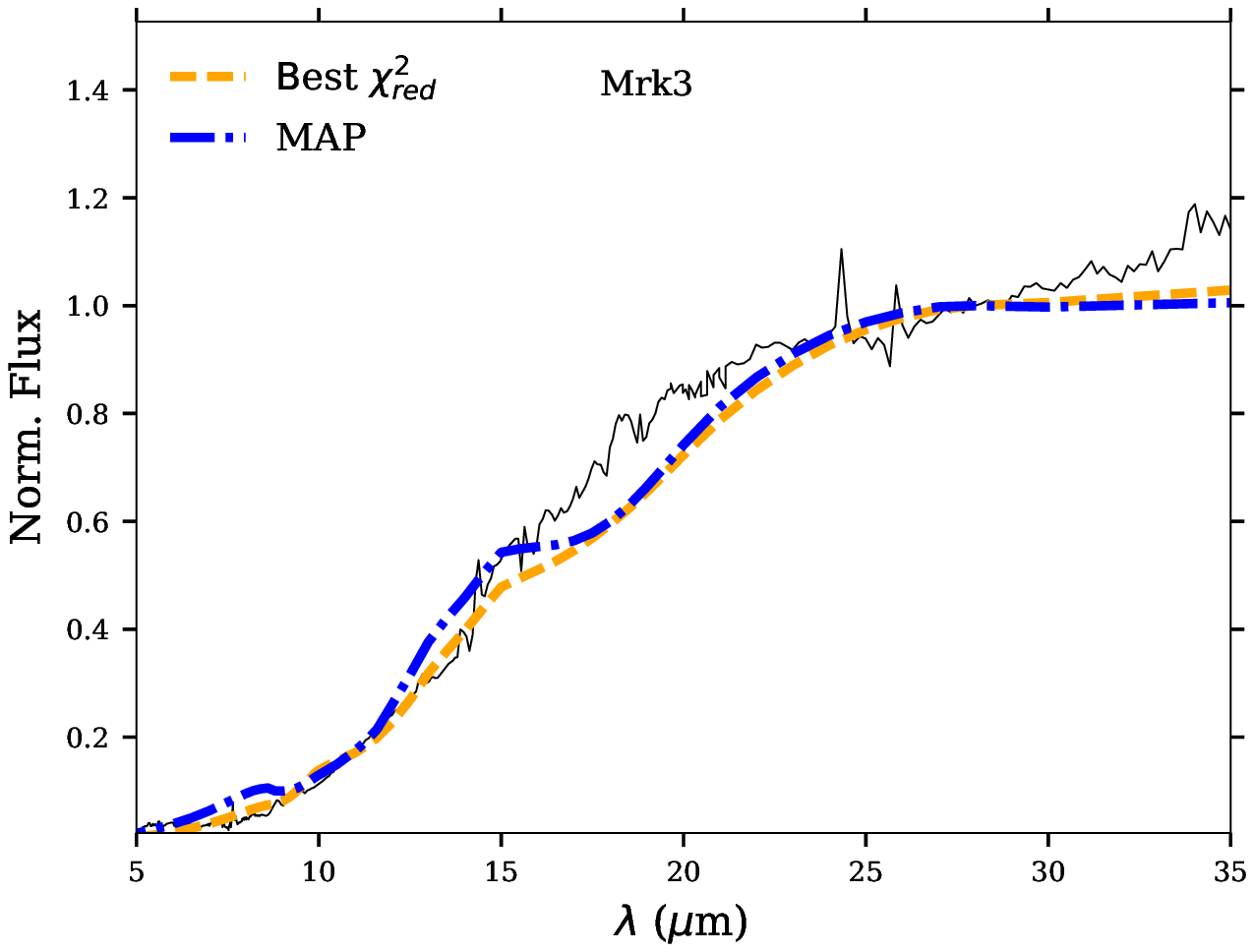}
 \end{minipage}
 \begin{minipage}{0.325\textwidth}
  \includegraphics[width=\textwidth]{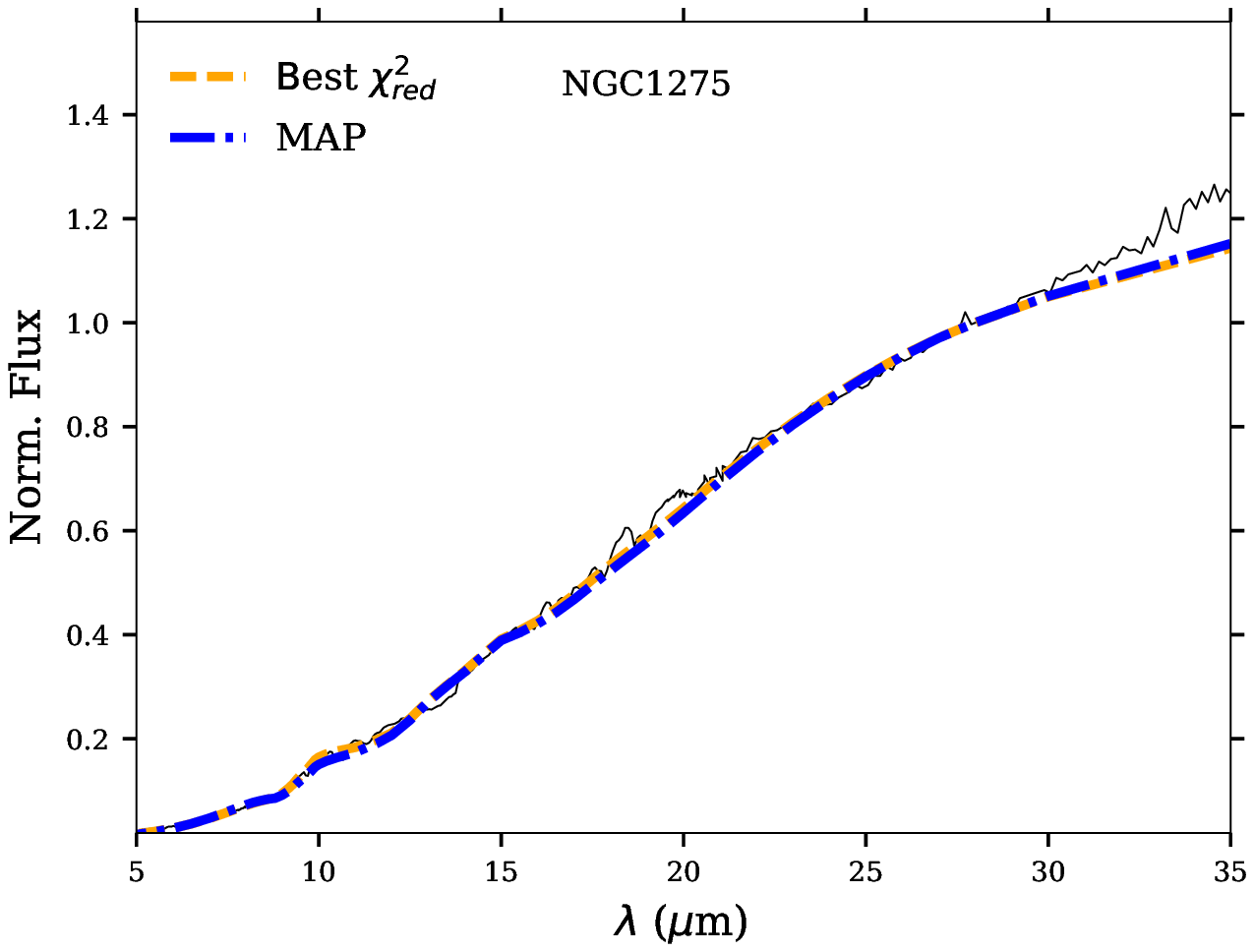}
 \end{minipage}
 \begin{minipage}{0.325\textwidth}
  \includegraphics[width=\textwidth]{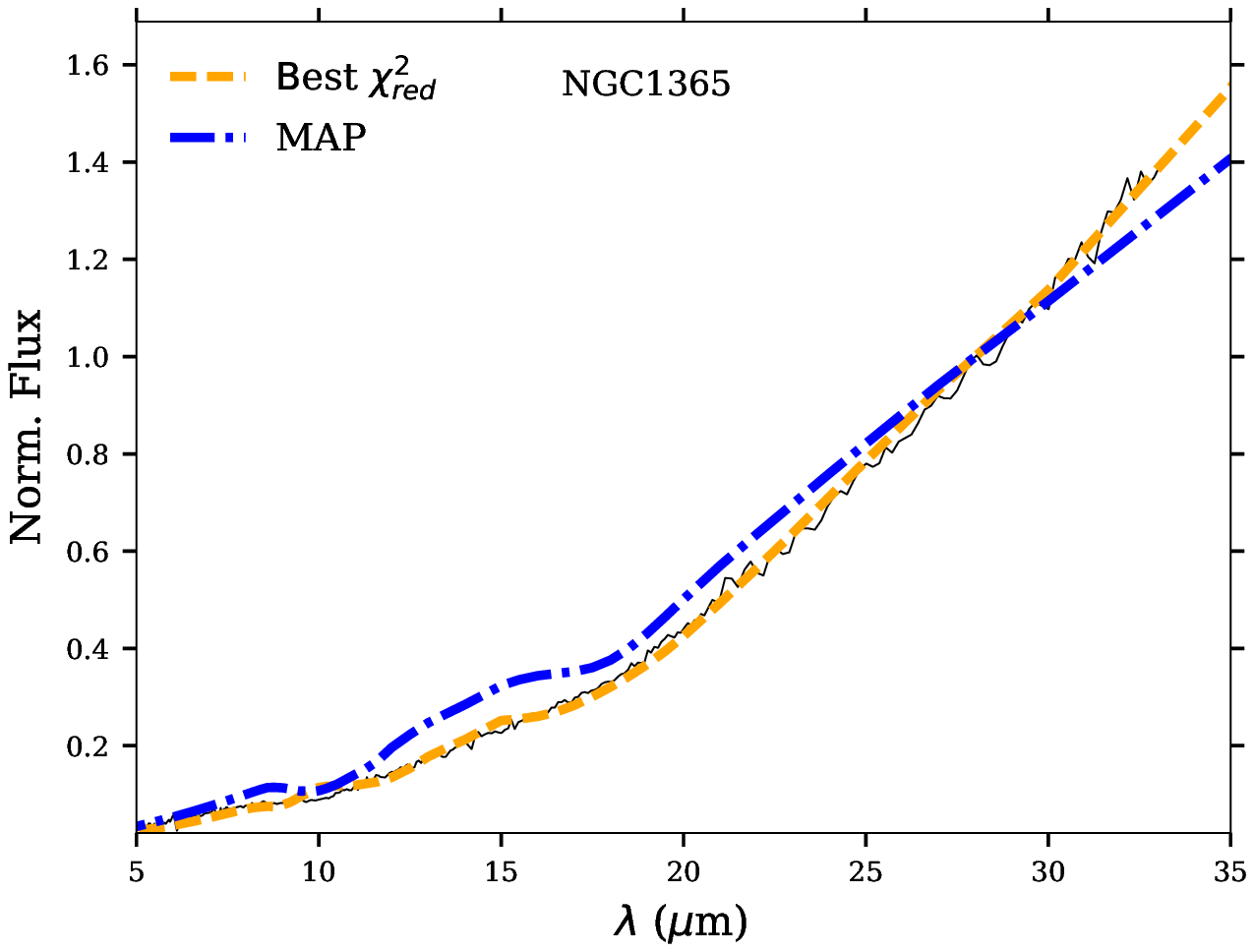}
 \end{minipage}
 \begin{minipage}{0.325\textwidth}
  \includegraphics[width=\textwidth]{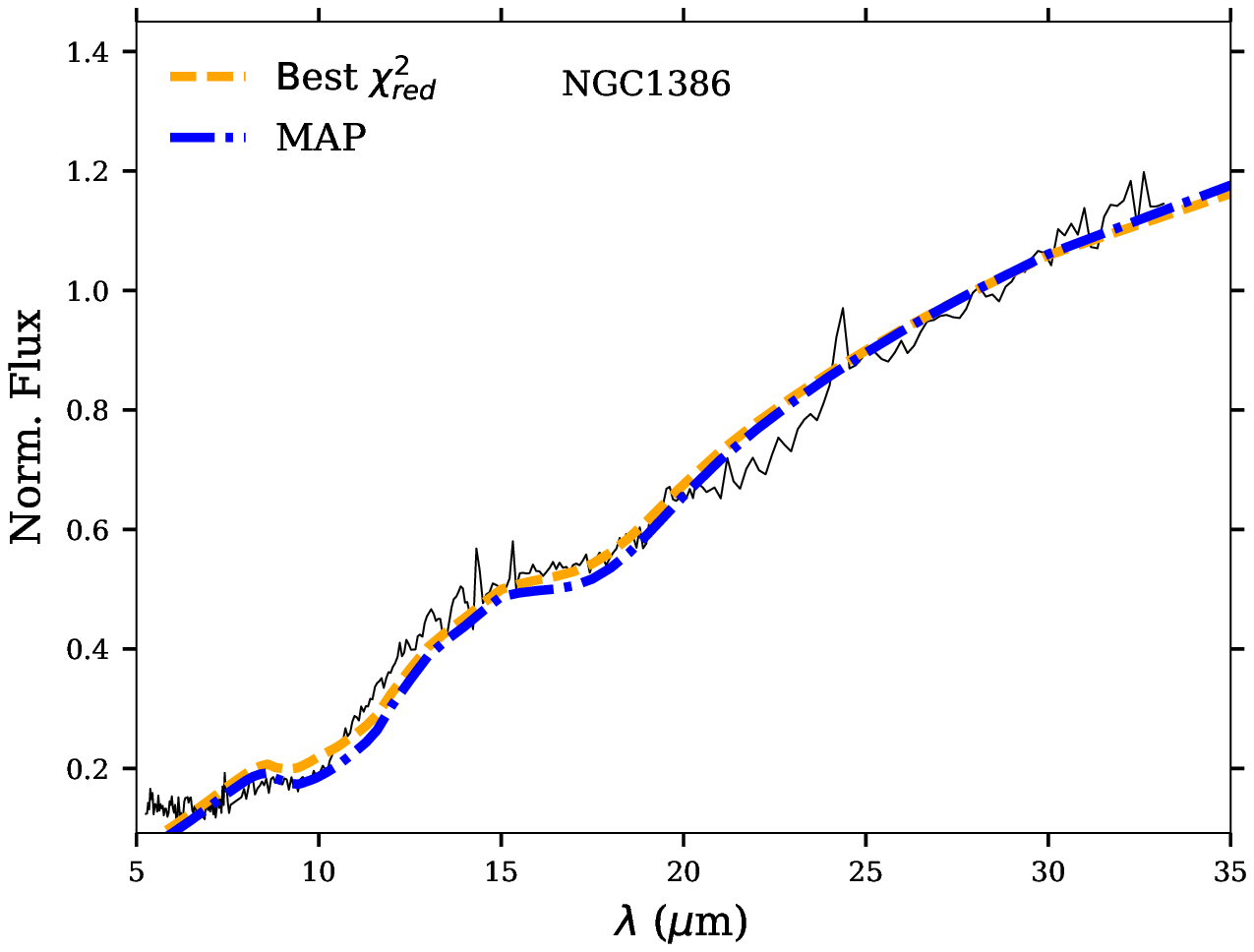}
 \end{minipage}
 \begin{minipage}{0.325\textwidth}
  \includegraphics[width=\textwidth]{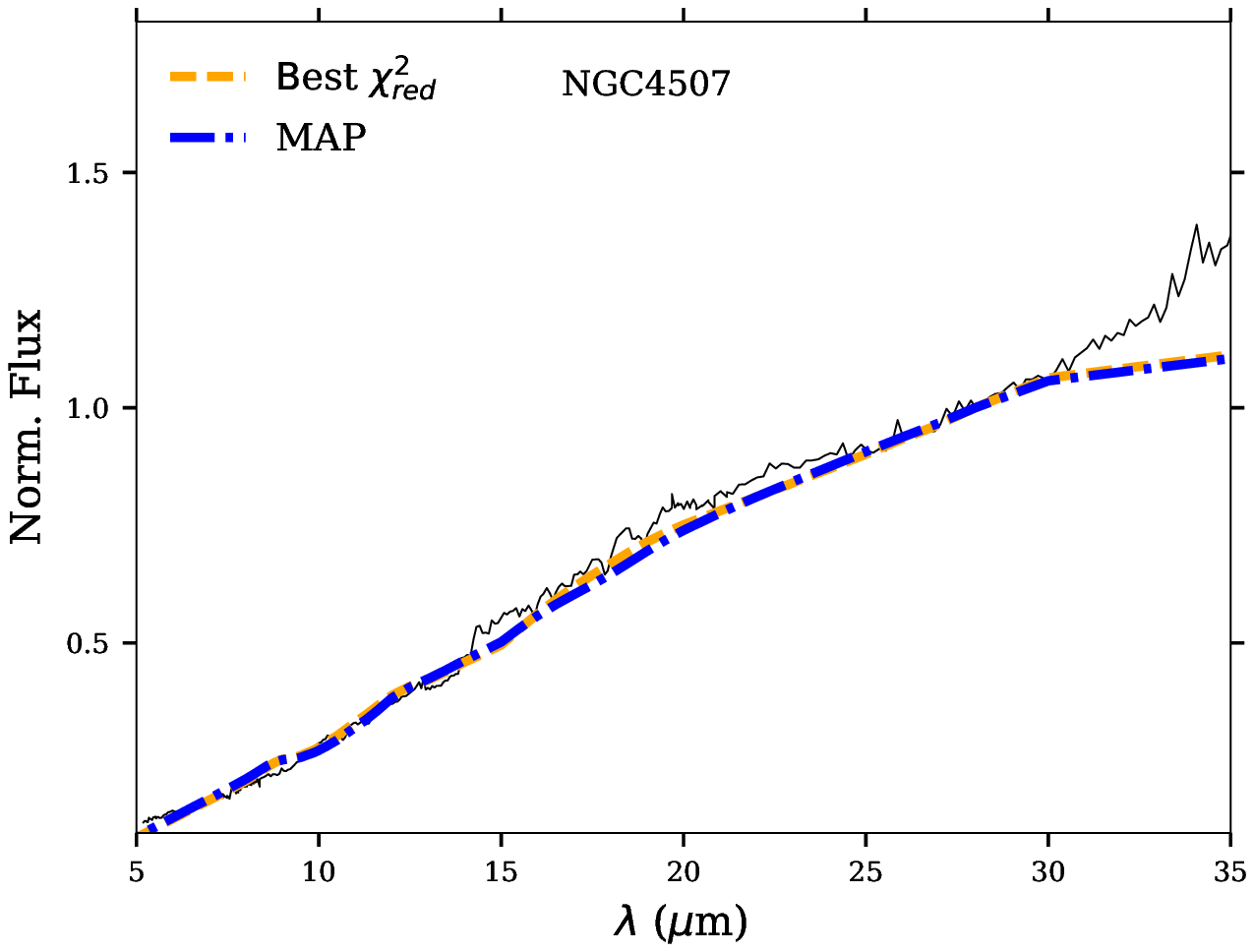}
 \end{minipage}
 \begin{minipage}{0.325\textwidth}
  \includegraphics[width=\textwidth]{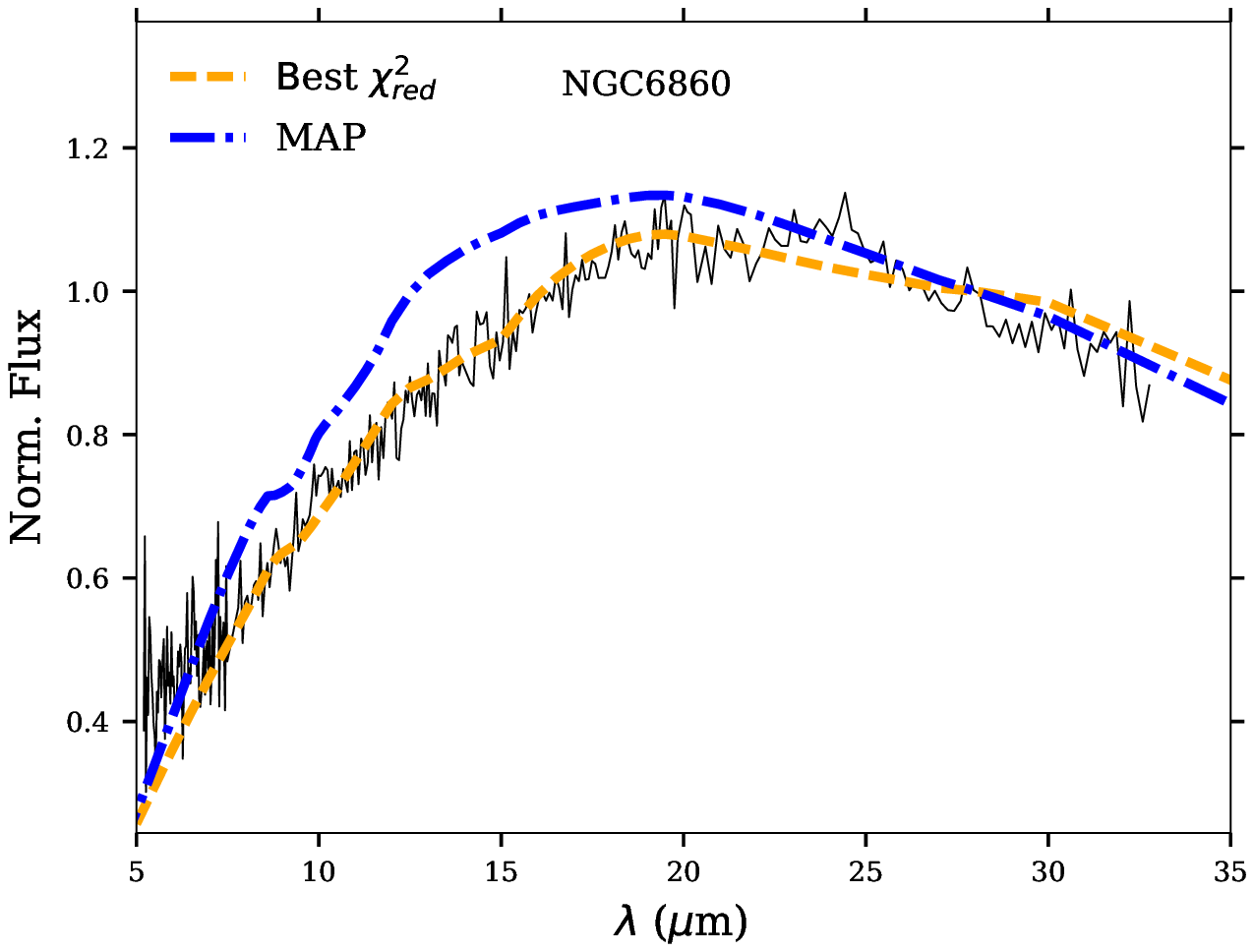}
 \end{minipage}
 \begin{minipage}{0.325\textwidth}
  \includegraphics[width=\textwidth]{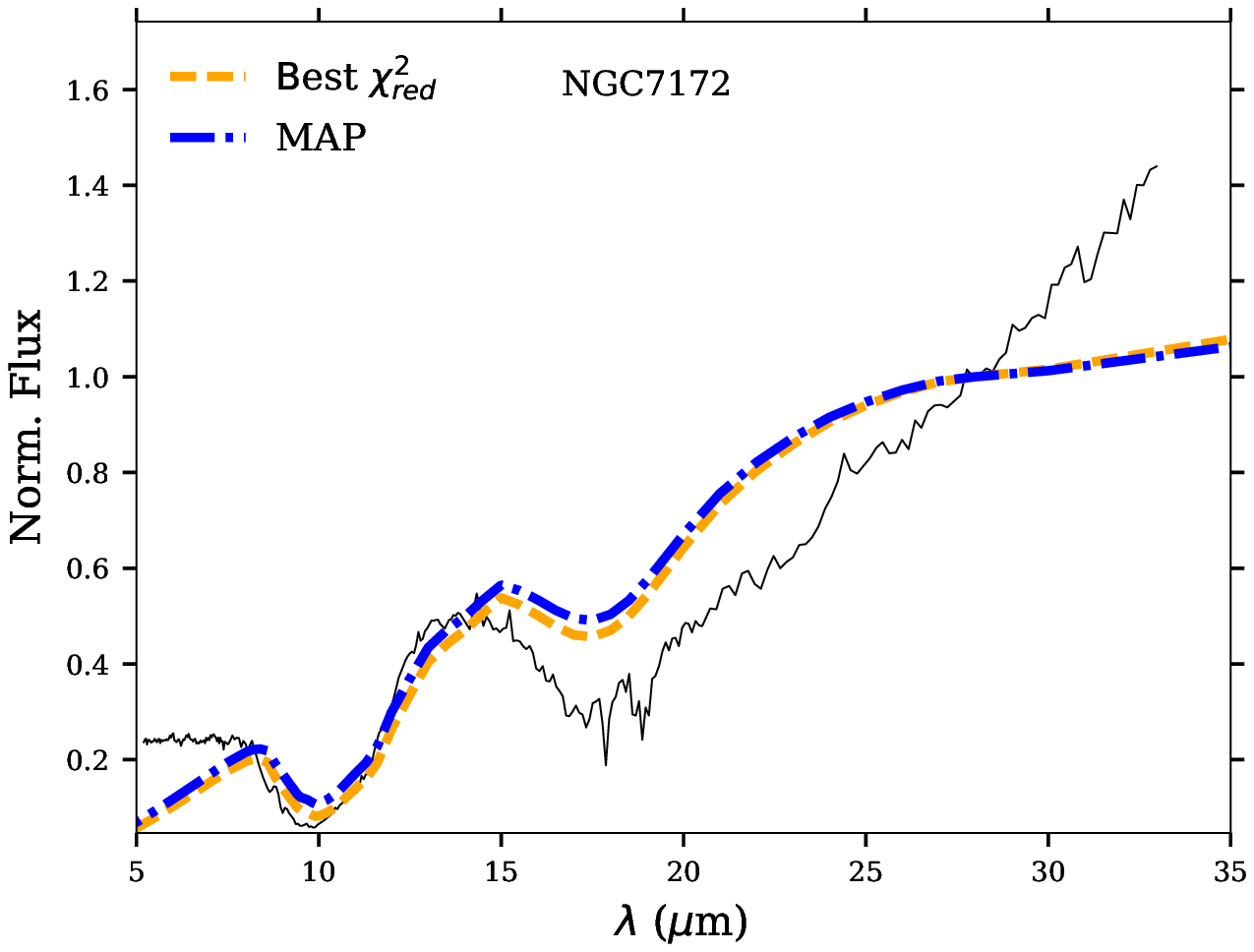}
 \end{minipage}
 \begin{minipage}{0.325\textwidth}
  \includegraphics[width=\textwidth]{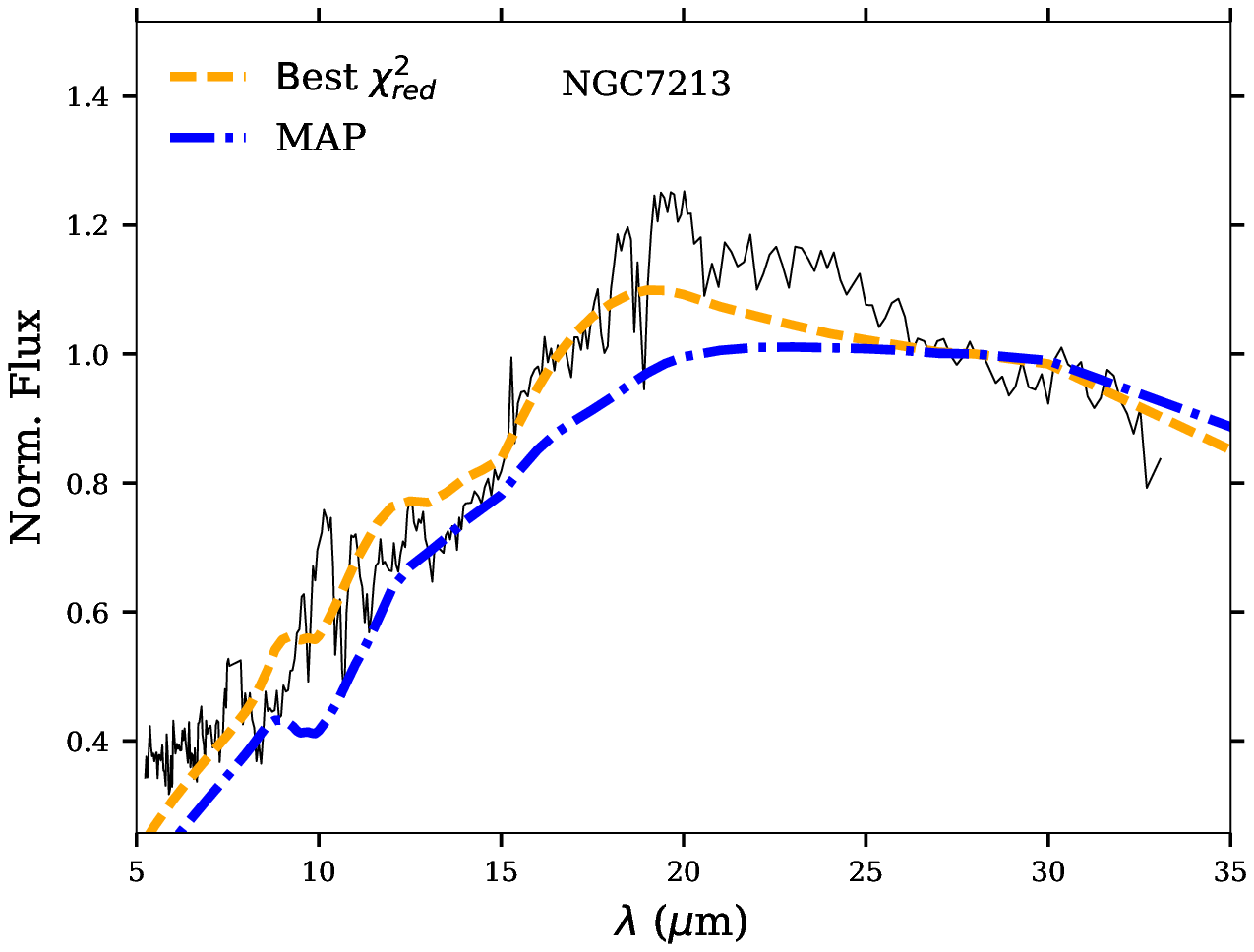}
 \end{minipage}
\caption{Examples of adjusts for the best fit using ${\chi_{red}}^2$ and the maximum-a-posteriori distribution from BayesCLUMPY. Best torus fitting to the spectrum is represented by yellow dashed line for the former and by blue dotted-dashed line in the latter case. The observed spectra and the SED models are normalized at 28$\mu$m.}
\label{adjust}
\end{center}
\end{figure*}

We apply the Bayesian inference tool BayesCLUMPY \citep{ase09} in order to achieve the best fitting parameters for the observed nuclear SEDs. The technique consists to perform a Markov Chain Monte Carlo method to investigate the parameter space defined by the first 13 eigenvectors. These values result from the combination of the principal component analysis and the artificial neural network that provides a interpolation in the database of the model grid from the theoretical {\sc clumpy} models ($\sim 10^{6}$ models). This approach allows to obtain the marginal posterior distribution for each model parameter, taking into account all a-priori constrains and the information from the observations. To assure the stability of the solution, we performed consecutive runs of the algorithm. It is important to emphasize that fitting clumpy torus models to the spectra is an intrinsically degenerate problem, as we can obtain the same observable effect for different sets of parameters. 

\subsubsection[]{Final Fitting}

An example of fitting for some galaxies is presented in Figure~\ref{adjust} and the individuals fittings are presented in the Appendix C. The yellow dashed line shows the model fitted for the minor ${\chi_{red}}^2$ value, e.g., the best fit, and blue dotted-dashed line represents the correspondent best $\chi^2$ solution for BayesCLUMPY inference, the maximum-a-posteriori (MAP) values. The derived mean parameters for both $\chi^2_{red}$ and the Bayesian method are very similar, and in general the $\chi^2_{red}$ solution is the most approximated to the observed spectrum (besides that it provides a solution within the models base). The goodness of both fitting procedures can be quantified by the values derived for the $\chi^2_{red}$. We also account an additional quality indicator, the {\it adev}, which gives the percentage mean deviation over all fitted wavelengths \citep{cid13}:

\begin{equation}
{adev} = \frac{1}{N} \sum_{i=1}^N  \frac{|F_{obs,\lambda_{i}} - F_{mod, \lambda_{i}}|}{F_{obs,\lambda_i}}
\end{equation}

\begin{figure}
\begin{center}
\includegraphics[width=86mm]{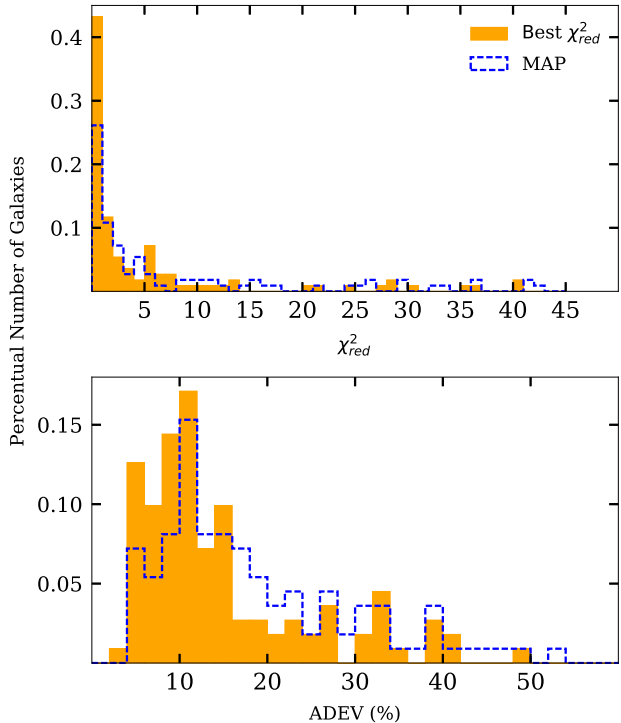}
\vspace{-0.5cm}
\caption{We show the values derived for the $\chi^2_{red}$ ({\it top panel}) and the {\it adev} ({\it bottom panel}) that quantify the goodness of the $\chi^2_{red}$ and the BayesCLUMPY fitting techniques. The orange filled histograms represent the values derived using the $\chi^2_{red}$ technique while the blue dashed lines are the distribution of the MAP values derived using the BayesCLUMPY method.}
\label{adev}
\end{center}
\end{figure}

In Figure~\ref{adev} we present the distribution of the {\it adev} and the minimum $\chi^2_{red}$ values derived for all the sample (except for the galaxies Mrk 3, NGC 1097, NGC 1566, NGC 4594, NGC 5033 and NGC 7679 that present values $\chi^2_{red}$ and MAP$>$50). For more than 50\% of the adjusted models we found $\chi^2_{red}$ values less than 5, which can be classified as satisfactory adjustments. Also, the deviation between the observation and the best model fitted is less than {\it adev}$\lesssim 20$\% for the majority of objects. In general, the Figures~\ref{adjust} and \ref{adev} indicate that the $\chi^2_{red}$ method provides more satisfactory adjusts than the MAP using the BayesCLUMPY. This is the reason we have chosen to only discuss the $\chi^2_{red}$ results in the next sections.

\section[]{Comparison between type 1 and type 2 sources}

In order to compare the two fitting methodologies, we decided to consider the best solution of the ${\chi_{red}}^2$ test and the MAP provided by the BayesCLUMPY method. However, it is important to notice that the model interpolations performed in BayesCLUMPY allows for a grid of parameters different from the ones provided by the clumpy torus models and listed in Table~\ref{obs:tab2}. 

\subsection{Direct Parameters}

\begin{figure}
\begin{center}
\includegraphics[width=86mm]{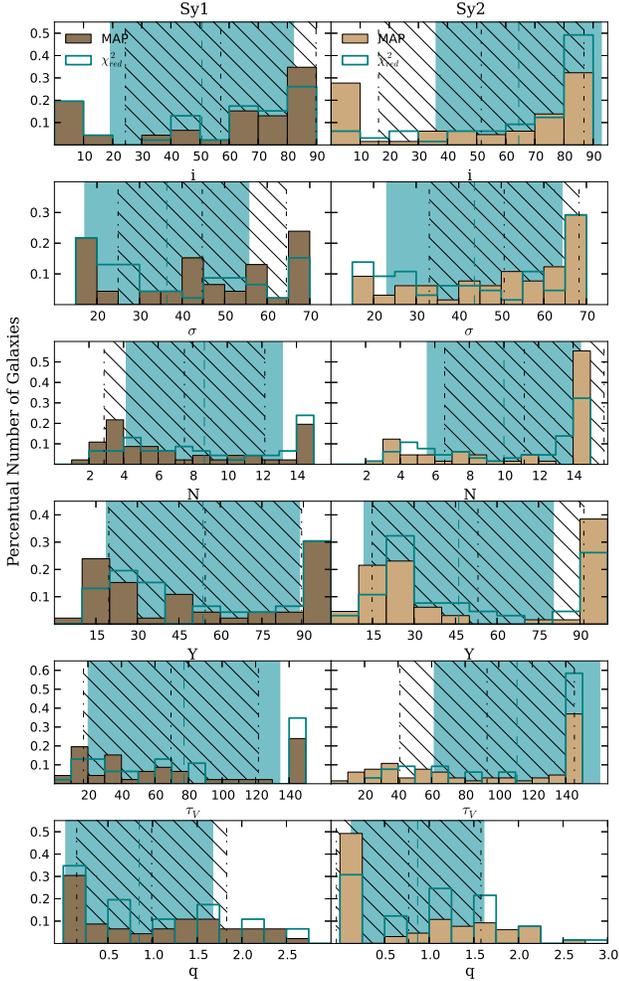}
\vspace{-0.5cm}
\caption{The frequency histograms distribution for direct parameters, $\rm i$, $\rm \sigma$, $\rm N$, $\rm Y$, $\rm \tau_V$ and  $\rm q$, derived from the {\sc clumpy} models fitting. The brown filled histograms represent the MAP distributions resulting from the employing of the BayesCLUMPY task and the stepped blue distribution shows the results of the best solution applying the $\chi^2_{red}$ test. In all panels, the distributions for the 46 Sy\,1 are plotted at the left side and for the 65 Sy\,2 at the right side. The lined-dotted lines indicate the mean value of the MAP distribution and the hatched area delineate the mean values from the $\chi^2_{red}$ and the uncertainties around the average.}
\label{direct}
\end{center}
\end{figure}

Both fitting methodologies allows for the determination of parameters within the models set, what we call {\it direct parameters}. The values obtained for those parameters are presented in frequency histograms (Figure~\ref{direct}) as well as, the  obtained mean parameters are listed in Table~\ref{obs:tabmean}. We have chosen to discuss the results obtained with the ${\chi_{red}}^2$ methodology, however, for completeness we will keep in all the histograms the results obtained with the BayesCLUMPY MAP mode. Below we discuss the results of each parameter individually.

It can be seen from Figure~\ref{direct} that the inclination angle relative to the observer's LOS, $i$, appears to be larger for Sy\,2 ($\rm \bar{i}(Sy2)=64.5^\circ \pm 28.3^\circ$) than for Sy\,1 ($\rm \bar{i}(Sy1)=50.6^\circ \pm 31.4^\circ$). This parameter was studied in previous works, with controversial results. RA11 studying a sample of 7 Sy\,1 and 14 Sy\,2 galaxies and from the 13 objects presented in AH11 sample, found no significant differences in this parameter and suggested that type 2 objects could be seen in any orientation if there is at least one cloud obscuring the observers LOS. On the other hand, \citet{mor09} studied 26 type 1 PG quasars using Spitzer data found $\bar{i}$=33$^{\circ}$ while \citet{lir13} obtained a typical value of i $\gtrsim$40 for a sample of 27 Sy\,2 with about half of their sample requiring values $i\sim$70-90$^\circ$. Our mean results for this parameter suggest that Sy\,1 do present a slightly lower value for $i$ than Sy\,2s, supporting the viewing angle orientation requirement for the AGN's Unified Model.

\begin{table}
\centering
\begin{minipage}{86mm}
\caption{Mean parameters derived from the ${\chi_{red}}^2$ and MAP of {\sc CLUMPY} models fitting}
\begin{scriptsize}
\begin{tabular}{@{}lcc@{}}
\hline
\hline
\multicolumn{1}{c}{Parameter}    &    \multicolumn{1}{c}{$\chi^2_{red}$}   &      \multicolumn{1}{c}{MAP}  \\
\noalign{\smallskip}
\multicolumn{1}{c}{}    &    \multicolumn{1}{c}{{\sc sy 1}  ---  {\sc sy 2}}   &      \multicolumn{1}{c}{{\sc sy 1}  ---  {\sc sy 2}}  \\ 
\hline
\noalign{\smallskip}
\multicolumn{3}{c}{{\sc Direct}} \\
\noalign{\smallskip} 
\hline
  i    &  50.6$\pm$31.4 --- 64.5$\pm$28.3 & 57.1$\pm$32.7 --- 51.6$\pm$35.3\\
  $\sigma$    &  36.4$\pm$19.2 --- 43.7$\pm$20.5 & 44.7$\pm$19.8 --- 50.7$\pm$17.6\\
  N    &  9.0$\pm$5.0 --- 10.0$\pm$4.0 & 8.0$\pm$5.0 --- 11.0$\pm$5.0\\
  Y    &  53.7$\pm$34.9 --- 46.1$\pm$34.1 & 54.5$\pm$34.9 --- 53.1$\pm$38.3\\
  $\tau_V$    &  77.3$\pm$57.0 --- 110.9$\pm$49.2 & 69.5$\pm$52.2 --- 93.0$\pm$52.1\\
  q    &  0.8$\pm$0.8 --- 0.9$\pm$0.7 & 1.0$\pm$0.8 --- 0.8$\pm$0.8\\
\hline
\noalign{\smallskip}
\multicolumn{3}{c}{{\sc Indirect}} \\
\noalign{\smallskip} 
\hline
  N$_{los}$    &  3.0$\pm$4.0 --- 7.0$\pm$5.0 & 4.0$\pm$4.0 --- 6.0$\pm$5.0\\
  A$_V$    &  287$\pm$595 --- 899$\pm$829 & 241$\pm$509 --- 671$\pm$799\\
  log($N_{H}/ \rm cm^{-2}$)    &  23.7$\pm$24.1 --- 24.2$\pm$24.2 & 23.7$\pm$24.0 --- 24.1$\pm$24.2\\
  P$_{esc}$    &  0.3$\pm$0.3 --- 0.1$\pm$0.2 & 0.2$\pm$0.3 --- 0.1$\pm$0.2\\
  C$_T$    &  0.7$\pm$0.2 --- 0.8$\pm$0.2 & 0.8$\pm$0.2 --- 0.9$\pm$0.2\\
  $M_{tor}(M_{\odot})$    &  2.1$\pm$3.9$\times$10$^6$ --- 2.7$\pm$5.5$\times$10$^6$ & 2.4$\pm$3.7$\times$10$^6$ --- 3.5$\pm$6.0$\times$10$^6$\\
\hline
\noalign{\smallskip}
\label{obs:tabmean}
\end{tabular}
\end{scriptsize}
\end{minipage}
\end{table}

Accordingly the {\sc clumpy} models, the dusty clouds follow a Gaussian like distribution along the equatorial ray characterized by a torus angular width ($\sigma$). Our results show that there is no significant differences for the mean $\sigma$ values in the different types of activity, being $\rm \bar{\sigma}(Sy1)=36.4^{\circ}\pm19.2^{\circ}$ and $\rm \bar{\sigma}(Sy2)=43.7^{\circ}\pm20.5^{\circ}$. Taking only the mean values into account, these results may indicate that the torus hosted by Sy\,1s is biased towards smaller values than those found in Sy\,2s. In fact, these values agree with those found by RA11 and are further supported by the findings of \citet[][$\sigma >$ 40, for 70\% of their Sy\,2]{lir13} and \citet[][$\bar{\sigma}=34$, for their type 1 sources]{mor09}. Regarding the number of clouds along the LOS, we found that both types are well represented by $\sim$ 10 clouds ($\rm \bar{N}(Sy1)=9\pm5$ and $\rm \bar{N}(Sy2)=10\pm4$). In the case of quasars this number seems to be smaller ($\sim$5)than in the case of Seyfert galaxies \citep[e.g][]{mor09,lir13}. Thus, our findings reinforce the scenario proposed by AH11 in which the number of clouds might be in an evolutionary stage of a receding torus.

Besides the above parameters, other fundamental parameter is the torus thickness, $Y$, which is calculated as the ratio of the outer $\rm R_o$ and inner radius $\rm R_d$,  $\rm Y=R_{o}/R_{d}$, where $\rm R_d$ is set as being the distance from the central source where dust sublimates and according to \citet{bar87} can be obtained by
\begin{center}
\begin{equation}
R_{d}=0.4 {\left ( \frac{L_{AGN}}{10^{45} \rm erg^{-1}} \right ) }^{0.5} {\left ( \frac{1500 \rm K}{T_{sub}} \right ) }^{2.6} ~~\rm pc
\label{rsub}
\end{equation}
\end{center}
with $\rm T_{sub}$ being the dust sublimation temperature and $L_{AGN}$ is the AGN bolometric luminosity. The values we derived for the torus thickness do not present any significant distinction on average values and also for the shape of the distribution, as it can be seeing in Figure~\ref{direct}. For both classes, we can find solutions at the edges of the distribution, indicating that the majority of objects requires or a very large value of Y or a compact torus. In this case, the mean values derived  ($\rm \bar{Y}(Sy1)=53.7\pm34.9$ and $\rm \bar{Y}(Sy2)=46.1\pm34.1$) do not represent the sample.

In fact, as pointed out by \citet{nkv08b}, when $q=2$ the IR fitting leads to a poor constrain on the torus extension, since the clouds are distributed close to $R_d$. Therefore RA11 and AH11 have chosen to restrict this parameter accordingly with observations that suggest smaller values for the torus radial extension \citep[ Y$\sim$10-20,][]{jaffe04,tri07,raban}. We also performed our MIR fitting using the same constrain of Y[5,30] adopted in AH11 and still our findings do not imply significant changes on the other parameters distribution. Hence, we decided to maintain the original $Y$ parameter space since this bi-modality found in our results can be attributed to a better constrain on $\gtrsim$20\,$\mu$m which are not included to the SEDs in the AH11 and RA11 sample. Indeed, \citet{ful16} shows that the inclusion of SOFIA photometric data in the 30-40\,$\mu$m wavelength range helps to better constrain $Y$. This is because the outer radius $R_o$ is more sensitive to the cooler dust that peaks in the far-IR, providing information about the torus size.

Due to computational limitations, the {\sc clumpy} models assume that all dust clouds have the same optical depth, $\tau_{V}$ \citep[see][for details]{nkv08b}. It is clear from Figure ~\ref{direct} that the distribution of individual clouds optical depths points is centred in high values of $\tau_{V}$. Also, approximately 60\% of the solutions for Sy\,2 galaxies require $\tau_{V} \sim 140$\,mag, presenting an average value of $\rm \bar{\tau_{V}}(Sy2) = 111 \pm 49$\,mag for type 2 sources, while for Sy\,1 we found a smaller value for $\rm \bar{\tau_{V}}(Sy1) = 77 \pm 57$\,mag. Both results are in agreement with the high optical depth condition ($\rm \tau_{V} \gtrsim 60$) of the {\sc clumpy} models, which requires such values to ensure that we do have optically thick clouds and a finite photon escape probability. However, for this parameter, our results differ from those found in the literature, which derive lower values of $\tau_V$ for Sy\,2 galaxies, for example, for the 14 Sy\,2 sample of RA11 the typical values derived are $\tau_V \sim 30$\,mag and for the 27 Sy\,2 from \citet{lir13}, the best solutions in general assume lower values ($\tau_{V} \lesssim 25$\,mag). We also tested for a possible correlation with $\tau_V$ and the galaxy inclination and no correlation was found.

In the {\sc clumpy} model the clouds distribution is described by a power law with form $ r^{-q}$. The histograms with the indexes ($q$) distribution for both types of activity show that the solutions are found to be more likely within lower values for this parameter, generally between $0<q<1$. Values of $q\sim0$ indicate a constant distribution, revealing that the number of clouds presents a weak dependence on the distance to the central AGN, while values $q\sim 1$ point to a distribution following a $1/r$ relation. The average values derived for both classes are quite similar for both types of activities, being $\rm \bar{q}(Sy1)=0.8$ and $\rm \bar{q}(Sy2)=0.9$. Our results follow the same trend as found by \citet[][$ \bar{q} = 1$]{mor09} and also by \citet[][$\rm q \sim 0$]{lir13}, since the distribution for this parameter is quite spread as can be seen in the histograms in Figure~\ref{direct}, where more than 30\% of the sample present values of $q =0$.

\begin{table}
\centering
\begin{minipage}{86mm}
\setlength{\tabcolsep}{15pt}
\caption{The K-S test for the parameters distribution of Sy1 and Sy2}
\begin{tabular}{ccc}
\hline
\hline
\noalign{\smallskip}
{\sc Parameter Distribution}  & {\sc D}  & {\sc p-value} \\
\noalign{\smallskip}
\hline
\noalign{\smallskip}
i   &   0.44   &   0.25   \\
$\sigma$   &   0.18   &   0.99   \\
N   &   0.33   &   0.31   \\
Y  &   0.20   &   0.97   \\
$\tau_V$  &   0.27   &   0.59   \\
q   &   0.17   &   0.99   \\
\hline
\noalign{\smallskip}
\label{ks_tab}
\end{tabular}
\end{minipage}
\end{table}

We also performed a two-sample Kolmogorov-Smirnov (K-S) test \citep{mises} in order to verify the results discussed above and quantify the differences between the parameters distribution for both activity types. The K-S test determine if the Sy\,1 and Sy\,2 parameters have the same distribution and the values derived for $D$, i.e. the supremum of the cumulative distribution functions (CDFs) of the Sy\,1 and Sy\,2 for each {\sc clumpy} parameter, and $p$-$value$ are shown in Table~\ref{ks_tab}. As we can see, the inclination $i$ presents a significant discrepancy between the CDFs since the $D$ value is the most considerable among the other {\sc clumpy} parameters, followed by $N$ and $\tau_V$. Since in both activity types we found $N\sim$10, the main parameters to classify a Sy\,1 or a Sy\,2 rely on a combination of $i$ and $\tau_V$: it depends on the observers' LOS orientation as well on the obscuring properties of the clouds. On the other hand, the $p$-$value$ for $\sigma$, $Y$ and $q$ suggest that Sy\,1 and Sy\,2 populations are drawn from the same distribution, e.g., we can not distinguish whether a distribution of the geometrical parameters $\sigma$, $Y$ and $q$ is from a Sy\,1 or a Sy\,2.

Some objects in our sample are common to previous works of RA09 (9 objects), AH11 (14 objects), RA11 and \citet{lir13} (20 objects). In general, our mean results are in good agreement with the literature, although the individual solutions may be quite different. We attribute these differences to the fact that each study used distinct approaches (for example, wavelength coverage, resolution, parameter constrains, methodology). For instance, that may also explain the differences in the reported parameters for the same galaxy in different papers by the same authors. In addition, our results in general tend to be more consistent with those presented by \citet{lir13}. As pointed out by the latter, there are very significant differences in the results found between RA11 and AH11 attributed to the inclusion of the 10\,$\mu$m spectroscopic observations. Since the silicate at 9.7\,$\mu$m is a important dust feature, the inclusion of detailed spectral information around this feature is crucial to properly describe the physical parameters from the SEDs.

\begin{figure}
\begin{center}
\includegraphics[width=86mm]{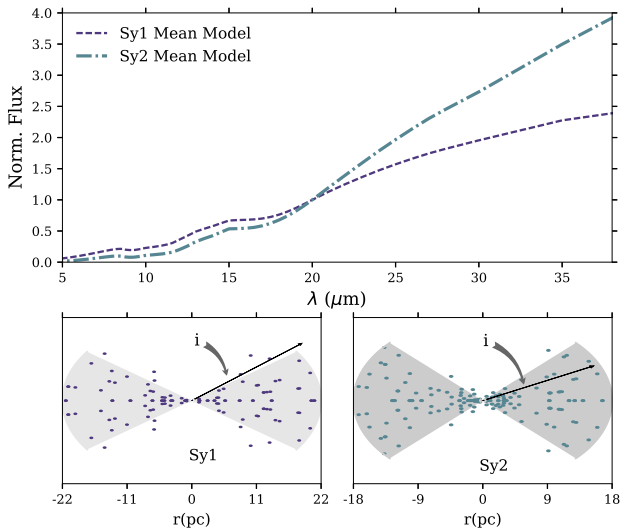}
\vspace{-0.4cm}
\caption{Combination of the mean parameters from {\sc clumpy} theoretical SEDs, represented by the dashed purple line for Sy\,1 and the dotted-dashed blue lines for Sy\,1. A schematic torus cross section is also illustrated in order to feature the main differences between the torus physical properties.}
\label{torus}
\end{center}
\end{figure}

In order to illustrate a mean SED and torus physical representation, we present a sketch in Figure~\ref{torus} that shows the mean theoretical SEDs from {\sc clumpy} to create a representative SED for each type of activity. There we combine the mean parameters derived in Table~\ref{obs:tabmean}. We also illustrate a schematic cross section view of the tori for both activity types, in order to highlight the differences in the torus physical properties, for instance, the slightly larger radial thickness for Sy\,1 and the wider angular width in Sy\,2, consequently, (given the Gaussian distribution) increasing the number of clouds in this type. These results are in agreement with the results of RA11 who found that the tori of Sy~1's are narrower and with fewer clouds than those found in Sy~2's. Furthermore, the mean SEDs do not present a turnover of the torus emission, predicted to occur between 30 - 50\,$\mu$m. This result is in agreement with those found by \citet{ful16}, where no turnover was observed below 31.5\,$\mu$m. Further nuclear far-IR observations would be essential to determine the peak of the IR emission, giving insight into the torus outskirts.

\subsection{Indirect Parameters}

\begin{figure}
\begin{center}
\includegraphics[width=86mm]{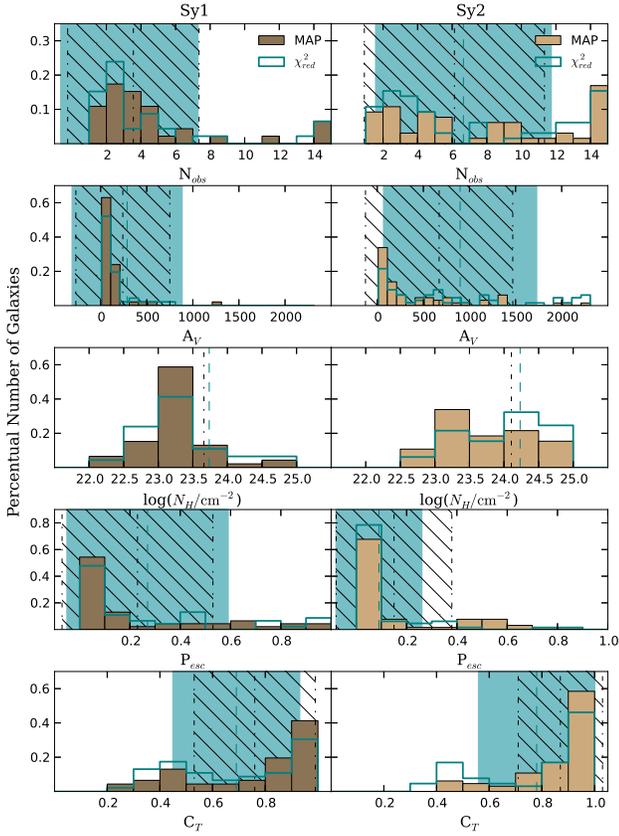}
\vspace{-0.5cm}
\caption{The frequency histograms distribution for indirect parameters, $N_{los}$, $ A_V$, $N_{H}$, $P_{esc}$ and $C_T$, derived from the {\sc clumpy} models fitting. The brown filled histograms represent the MAP distributions resulting from the employing of the BayesCLUMPY tool and the stepped blue distribution shows the results of the fit best solution applying the ${\chi_{red}}^2$ method. The panels follow the same scheme listed in Figure~\ref{direct}. } 
\label{indirect}
\end{center}
\end{figure}

As mentioned before, the model geometry enable us to estimate other important parameters that may help us to understand the physical properties of the putative torus required by the unified model. The distribution derived for the indirect parameters are described below and presented in Figure~\ref{indirect} as well as the information about mean values are summed up in Table~\ref{obs:tabmean} and described in the text. It is worth mentioning that all these indirect parameters where obtained using the results of the best direct parameters described in the previous section. 

Since we are dealing with a clumpy media one of the most important parameters in describing the torus is the number of clouds blocking our LOS ($N_{los}$). By using the model clouds  distribution (with a Gaussian like form), centred at the equatorial plane, we can compute the number of clouds along any specific direction, thus if we choose the LOS direction we can compute $N_{los}$ (Equation~\ref{numlos}). The $N_{los}$ distribution is shown in the top panel of Figure~\ref{indirect}, by inspecting this figure, it is clear that the number of clouds along the observer's LOS presents a sharp peak in its distribution for Sy~1 (centred at $\sim$3), while a spread distribution is found for Sy~2 ($\bar{N_{los}}=7$). At the same level of importance, is the extinction of the light caused by the material composing the LOS clouds. Once $N_{los}$ is know the total optical depth along the LOS is obtained with Equation~\ref{averm}. The distribution of A$_V$ is well defined for Sy~1, with small values, while in Sy~2 it is flat. In addition, the determination of $N_{los}$ is also related with the X-ray columnar hydrogen density which can be  derived using the standard Galactic ratio and the foreground extinction from \citet{boh78} via $N_{H}/A_{V}$=1.9$\times$10$^{21} \rm cm^{-2}$. In agreement with the two previous indirect parameters, the $N_{H}$ is well defined for the Sy~1 galaxies ($\bar{N_{H}}=5\times10^{23}cm^{-2}$) and with a not so centred distribution for Sy~2's ($\bar{N_{H}}=1.6\times10^{24} cm^{-2}$).

These results together with the fact that we are not finding significant difference in the observers viewing angle for the different classes, point to the fact that the most important parameter in determining if a galaxy is classified as a type 1 or 2 object is if there are clouds able to block the radiation from the BLR and central engine. This suggests that the fundamental requirements of the unified model for AGNs depends more on the intrinsic parameters of the torus than on its geometry. In fact, our results are still supported by the finding of RA11 found significant differences in the torus angular size for the different classes. However, in contradiction with our results, they do find lower optical depth in Sy2 when compared with Sy~1.

One of the fundamental requirements of the unification schemes is if a photon generated in the accretion disk is able to scape through the torus. Thus, a fundamental parameter that can be derived from \citet{nkv08b} formalism is the escape probability, $P_{esc}$. This parameter can work as a estimator whether an object is type 1 or type 2, since the putative large viewing angles in the latter is associated with the probability to have more clouds blocking the AGN radiation, leading to a finite but small probability of direct view to the AGN. When the condition $\tau_{V}\gg$1 is achieved $P_{esc}$ can be estimated as:

\begin{equation}
P_{esc} \cong e^{-N_{los}}
\label{pesc}
\end{equation}

The frequency histograms showing the $P_{esc}$ are presented in Figure~\ref{indirect}. The results agree with the predictions (i.e. lower probabilities are expected in type 2 objects). We found mean values of $\rm \bar{P}_{esc}(Sy2)=0.1$ and $\rm \bar{P}_{esc}(Sy1)=0.3$ indicating that, on average, a photon originated in the central source has a $\sim$30\% chance to escape from the torus without being absorbed. The individual clouds emission, as adopted in the {\sc clumpy} models formalism, plays a fundamental role to understand the emerging IR torus radiation. It can be originated by clouds directly illuminated by the AGN photons plus the reprocessed radiation from the shaded side of clouds which are heated by the emission of more internal clouds. Thus, what is observed is a sum of the radiation emitted by the torus and the photons generated in the accretion disc that are able to scape (i.e. $P_{esc}$), therefore knowing $P_{esc}$ is fundamental to determine the whole SED emission.

\begin{figure}
\begin{center}
\includegraphics[width=86mm]{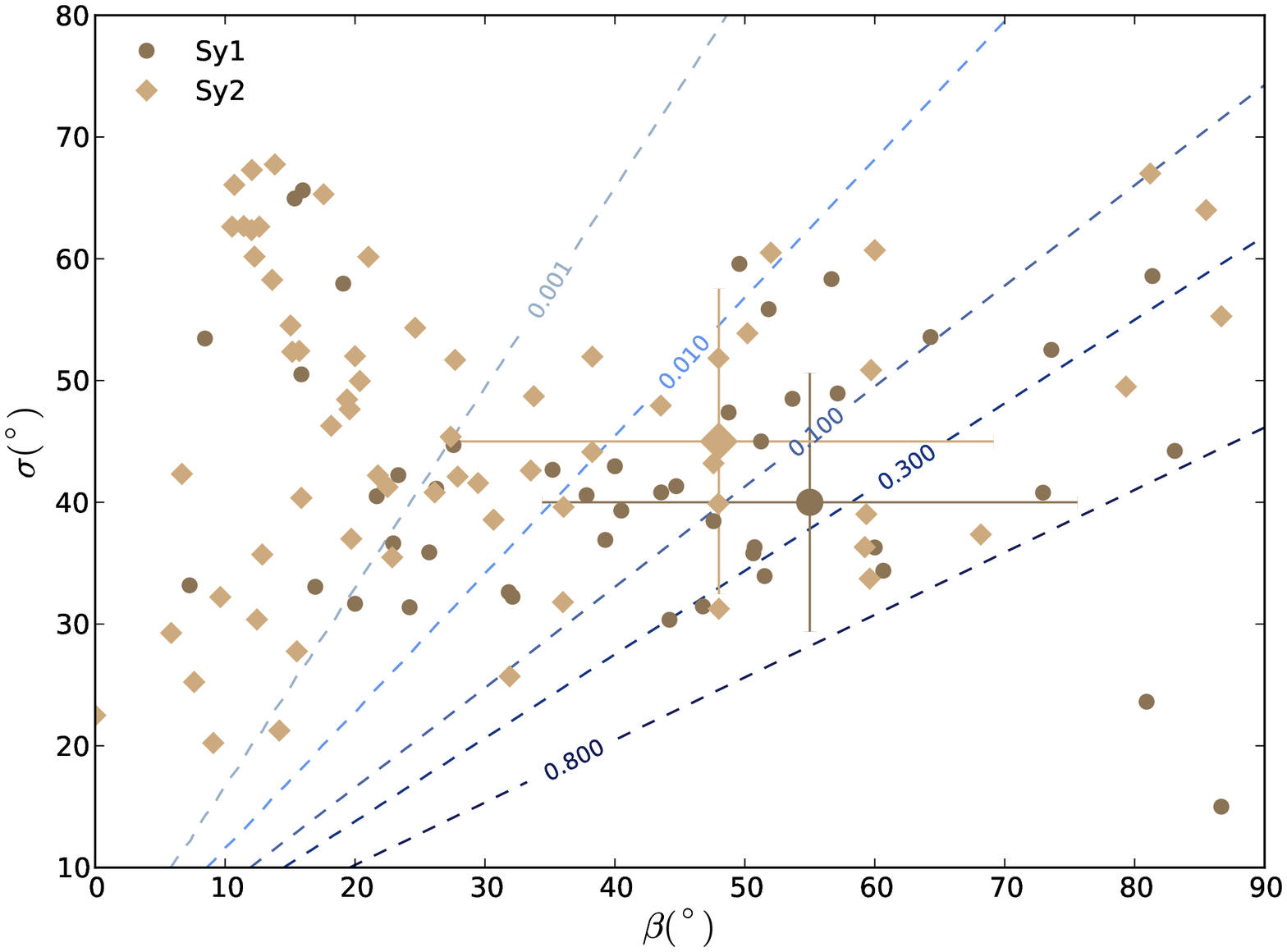}
\caption{The figure illustrates the distribution of $P_{esc}$, the photon escape probability, in function of the torus width, $\sigma$ and the complementary viewing angle, $\beta$=$\pi/2-i$, related by Equation~\ref{pesc}. In agreement with the unified model premise, which expect that Sy\,2 galaxies more likely on the left side and Sy\,1 on the center of the graph and higher  $P_{esc}$ values, we found Sy\,2 more concentrated at lower probabilities, except for some objects. As a representative value for our sample, we utilized $N$=10 to plot the $P_{esc}$ curves.}
\label{pescape}
\end{center}
\end{figure}

\begin{figure*}
\begin{center}
\includegraphics[width=172mm]{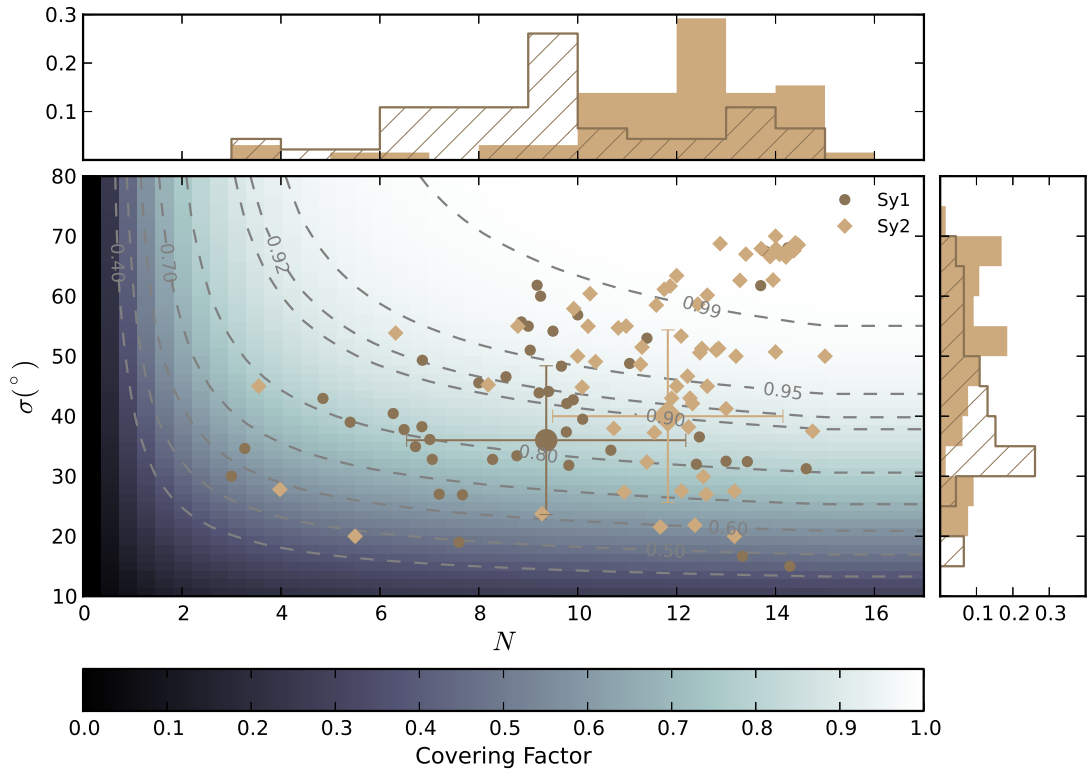}
\vspace{-1.5cm}
\caption{The graph represents the distribution of $N$ and $\sigma$ and their correlation with the covering factor, $C_{T}$. Covering factors curves are present for values from 0.2 to 0.9 according to Equation~\ref{cov}. The closed diamonds represent the Sy\,1 objects and the circles show the Sy\,2 types. The histograms for $N$ and $\sigma$ are attached at the top and right side of the main plot, respectively, for both activity types. As argued in \citet{eli12}, the Sy\,2 galaxies are more likely drawn from the distribution of higher covering factors than Sy\,1 types.}
\label{covfactor}
\end{center}
\end{figure*}

Once $P_{esc}$ is a non-linear function of $\sigma$, $\beta$ and $N$ we used the results we obtained with our Spitzer data fittings for $\sigma$ and $\beta$ and adopted a value for $N$=10 in order to determine $P_{esc}$ curves in the $\sigma \times \beta$ plane. The results are shown in Figure~\ref{pescape}, for display purposes for this figure we used a 10\% of $\chi^2_{red}$ deviations as described in \citet{sal13}. It emerges from this figure that most of the Sy\,2 galaxies present $P_{esc}\lesssim$10\%. The distribution of $P_{esc}$ is quite broad for Sy\,1, that may be a reflect of the fact that 
both $\sigma$ and $\beta$ do present a variety of values in this class (see Figure~\ref{direct}). This parameter was also studied by RA11 and AH11, our results are in good agreement with those found by these authors, in the sense that there is a significant difference between both types of activities. However, while we find almost the same fraction for Sy~1s, we find larger values for Sy~2s than the values found by these authors. We attribute this difference (in type 2 objects, $P_{esc}^{we}$=10\% and $P_{esc}^{they}$=0.1\%) to the fact they restricted the torus thickness (5$\leq$ Y $\leq$ 30) and we allow it to take all the possible values.

Another parameter provided by the model is the geometrical covering factor that can be understood as the sky fraction at the AGN centre which is being obscured by the dusty clouds. This parameter can be determined by integrating the $P_{esc}$ over all orientations, following the equation \citep{nkv08a}:
\begin{equation}
C_{T}= 1 - \int_{0}^{\pi/2}P_{esc}(\beta)\cos(\beta)d\beta  
\label{cov}
\end{equation}

The distribution of the derived values for $C_T$ are presented in Figure~\ref{indirect}. We found slightly larger $C_T$ values for type~2 sources ($\rm \bar{C}_{T}(Sy2)=0.8\pm0.2$) than for type~1 galaxies ($\rm \bar{C}_{T}(Sy1)=0.7\pm0.2$). Our values are in agreement with literature in the sense that there is a difference between both activities, however, while RA11 find typical values of 0.5 for for the Sy\,1 in their sample, these authors find almost the same values we found for the Sy~2 in their sample. \citet{mor09} also found an even smaller ($\sim$0.3) mean value in their type 1 PG quasars sample.

Accordingly to \citet{nkv08a} the definition of $C_{T}$ arises from the geometry and probabilistic nature of a {\sc clumpy} medium and can be interpreted as the fraction of randomly distributed observers whose view to the central source is blocked, or as the fraction of type 2 objects in a random sample. The covering factor can be also decisive into AGN classification, because an AGN with a larger covering factor has a higher probability to be viewed as type 2. Many questions are still opened concerning the definition of the intrinsic covering factor, if the geometrical one is related with  the ``dust'' covering factor proposed by \citet{maio07} (defined as the ration between the thermal component and the AGN contributions). Since the covering factor measures the fraction of AGN luminosity captured by the torus and converted to infrared, the AGN IR luminosity is $C_{T}L_{bol}$, where $L_{bol}$ is its bolometric luminosity \citep{eli12}. Thus, it is expected that type 2 AGNs have intrinsically higher IR luminosities than type 1. However, in disagreement with earlier expectations of a strong anisotropy at $\lambda\lesssim 8\umu$m, Spitzer observations present very similar IR fluxes of type 1 and 2 as shown by \citet{lutz,bucha}, which can be partly explained by a clumpy torus distribution.

The geometrical $C_{T}$ can be interpreted as the ``true" torus covering factor because it is independent on $i$. However, on the other hand, as can be noted from Equation~\ref{cov} $C_T$ depends on N and $\sigma$, thus to investigate this relation we show in Figure~\ref{covfactor} the results of the models fits\footnote{For display purposes we adopted the mean values of each parameter for the 10\%  deviation of  the best $\chi^2_{red}$.} in the $N$--$\sigma$ plane together with the contour plots of $C_T$. We found that type~2 objects do preferentially lie on the top right of the figure, or in other words they do have large $C_{T}$ values. However, in the case of type~1 objects (where low $C_{T}$ values are expected, since the BLR emission should be observable) we found that they are spread over the whole plane.  AH11 and RA11 suggest that type 1 and type 2 AGNs preferentially are located in different regions on the plane (with type 1 having lower values than type 2), however we do not see this trend in our work because, despite the same results for the $\sigma$, we derived larger values for the number of clouds, $N$, and therefore it is expected more clouds obscuring the central source. Since $C_T$ is very sensitive to $N$ and $\sigma$, our results always point to higher values than those found in their sample. We would also like to emphasize that our sample consist in a statistically representing sample, and this trend may be biased since the other works have a smaller sample or due to the restrictions in the parameter space adopted by the authors.

\subsection{Torus mass and size}

Beside the above torus parameters, the {\sc clumpy} formalism also allows to determine the mass of the emitting hot dust, as well as to set some constrains on the torus size. From the adjustment of the theoretical SEDs {\sc clumpy} model, the bolometric luminosity of the AGN, $ L_ {AGN} $, can be found from the scale factor ($\Theta$) of the model to the observation, by solving the equation:
\begin{equation}
\lambda_{obs} F_{obs,\lambda} = \Theta \frac{\lambda_{mod} F_{mod, \lambda}}{F_{AGN}}
\end{equation}

\begin{figure}
\begin{center}
\includegraphics[width=86mm]{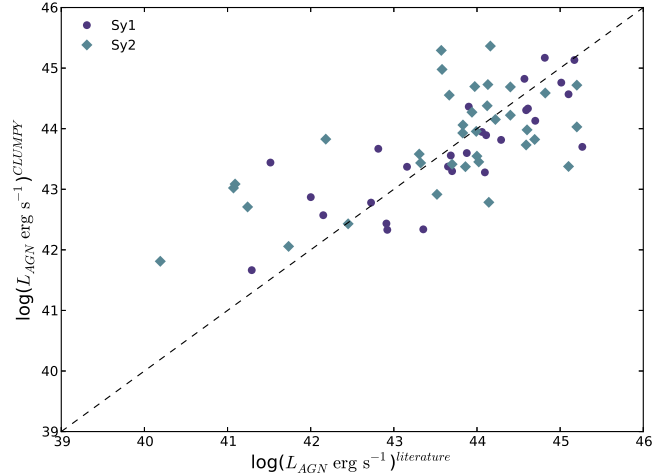}
\caption{Comparison between the bolometric luminosities, $L_{AGN}$, derived from the {\sc clumpy} models with those reported on the literature for the hard X-ray luminosities, $L_X$, applying the  bolometric corrections of \citet{elvis}. The diamonds represent the Sy\,1 objects while the circles indicate the Sy\,2 types and the dashed line indicates the identity line.}
\label{lagn}
\end{center}
\end{figure}

We can derived $L_{AGN}$ applying the distance relation $ L_{AGN} = 4 \pi D^2 \Theta $ and once it is calculated Equation~\ref{rsub} can be used to estimate the inner radius, assuming a sublimation temperature for the silicate grains (we used $ T_{sub}=1500$\,K). The torus dimension is estimate by the relation Y=$R_{o}/R_{d}$, where $Y$ is taken from the best fit. In order to test the $L_{AGN}$ values derived from the scaling of the {\sc clumpy} models, we can compare them with those reported in the literature for the X-rays by applying the bolometric correction of $\sim$20 \citep{elvis} and the comparison between the fitted and literature $L_{AGN}$ is shown in Figure~\ref{lagn}. As can be seen from the one-to-one line the observed and calculated values do agree well.

\begin{figure*}
\begin{center}
\includegraphics[width=172mm]{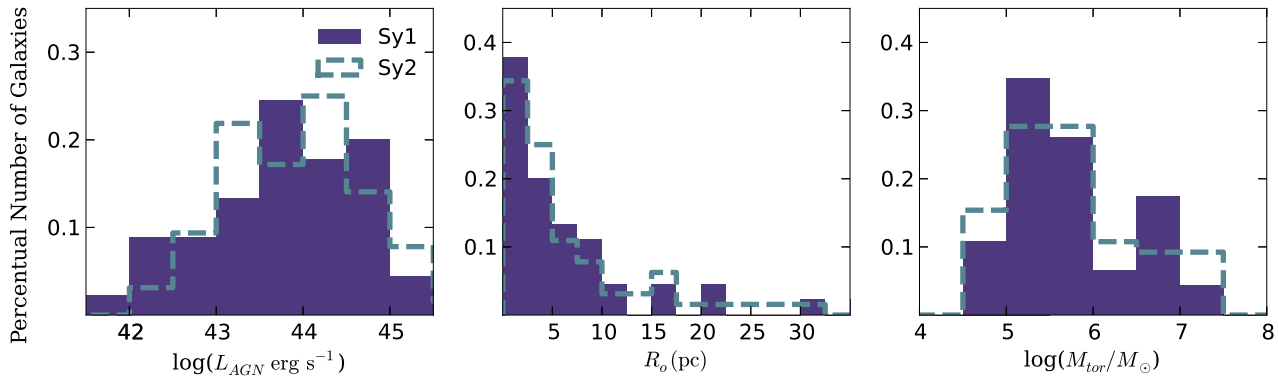}
\caption{The histograms show the distribution of the AGN bolometric luminosities (left panel), torus sizes, $R_o$, in the middle panel and the torus masses, $M_{tor}$ at the right. Histograms filled in purple represent the distributions for Sy\,1 and dashed lines in blue indicate the distributions for Sy\,2.}
\label{lumR}
\end{center}
\end{figure*}

Figure~\ref{lumR} shows histograms for the distributions of $ L_{AGN}$ and $ R_{o}$. No significant differences between type 1 and type 2 galaxies are seen. In general, the estimated values of the AGN bolometric luminosity range between $42<log(L_{AGN})<46$, and the average values in both classes are typically $L_{AGN}\sim 10^{44}$. For the sublimation radius we find averages $\rm R_d (Sy1) =$0.12\,pc for Sy\,1 galaxies and $\rm R_d(Sy2) = $ 0.16\,pc for Sy\,2, which lead to average torus sizes very close to those found in the literature \citep[RA09, AH11, RA11,][]{lir13}, with typical values of $R_o\lesssim$6\,pc. These results are further supported by observational evidences. For instance, previous works using MIR interferometric observations provide information of a relatively compact torus, with a few parsecs scale \citep[][]{jaffe04,tri07, tristram09, burtscher09}.

By adopting some approximations for the torus geometry and size, we can also estimate its total mass. Considering the mass of a single cloud as $m_{H}N_{H,c}A_{c}$, where $N_{H,c}$ is its column density and $A_c$ its cross sectional area, the total mass in clouds is given by $M_{tor}=m_{H}{N_{H}}^{1}\int \eta_{c}(r,\beta)dV$, where $\eta_{c}(r,\beta)$ indicates the clouds distribution profile. For simplicity, assuming a sharp-edge angular distribution, $M_{tor}$ can be analytically calculated \citep{nkv08b}:
\begin{equation}
M_{tor}= 4 \pi m_H sin(\sigma) {N_{H}}^{(eq)} R_{d}^2 Y I_{q}(Y)
\label{mtorus}
\end{equation}
where $Y_{q}=$1, $Y/(2lnY)$ and $Y/3$ for $q=$2,1 and 0, respectively, and ${N_{H}}^{(eq)}$ is the mean overall column density in the equatorial plane. The latter can be estimated by multiplying the number of clouds along the equatorial ray $N$ by a single cloud columnar density $N_{H,c}$ $\sim$10$^{22}$--10$^{23}\rm cm^{-2}$. Finally, since $N\sim$ 5--15,${N_{H}}^{(eq)}$ assumes typical values of $\sim$10$^{23}$--10$^{24}\rm cm^{-2}$. 

We found that $M_{tor}$ ranges from $\rm 10^4 M_\odot$ to $\rm 10^7 M_\odot$ in both activities with mean values typically with $\rm 10^6 M_\odot$ (Figure~\ref{lumR}), in agreement with the estimating by \citet{lir13} and \citet{mor09} using the {\sc clumpy} model formalism to derive $M_{tor}$. A recent work from \citet{garcia} was able to constrain $M_{tor}$ derived from {\sc clumpy} models fitting with the mass of molecular outflow observations in NGC\,1068 using ALMA observations in band 7 and 9. They estimate $M_{tor}\sim2\times10^5 M_{\odot}$, consistent to the estimated molecular gas mass detected inside the central aperture (r=20\,pc) derived from the CO(3-2) emission. 

\section{The effects of hot dust emission}

It is widely known that the inner torus radius is related with the dust grains sublimation temperature ($T\sim$800-1500\,K). Such temperature peaks at NIR wavelengths and in the case of AGNs this emission is related with the dusty torus \citep{bar87,rod06,rif09}. Therefore,  the inclusion of the NIR spectral region is crucial to probe the hotter and innermost regions of the torus. 

In fact, \citet{ram14} demonstrated the need of the inclusion of the NIR data to constrain the torus parameters, especially to constrain the torus radial extension, Y. In their work, they analysed a compilation from the literature of NIR+MIR photometry and MIR spectroscopy (8-13$\umu$m) of six Seyfert galaxies. All the objects in their study share the same characteristics: undisturbed, face-on galaxies with no prominent emission of dust lanes. In particular, they recommend a minimum combination of data in the $J$+$K$+$M$- band photometry and the $N$-band spectroscopy. 

In order to study the effects of the hotter and inner region of the torus we included in our analysis $ZJHK$-band long-slit spectroscopy data, for the 32 objects where such information was available in the literature. Data for 24 galaxies were obtained from \citet{rif06}. These data where collected using the SpeX spectrograph of the NASA 3m Infrared Telescope Facility (IRTF) in the short cross-dispersed mode (SXD, 0.8-2.4\,$\umu$m) using a slit of 0.8''$\times$15'' slit. The used spectra are those of the nuclear extraction (sampling few hundred pc). For more details about the data see \citet{rif06, rif09}. The remaining 8 sources were taken from \citep{mason15}, these spectra were observed using Gemini telescope GNIRS spectrograph in the XD mode. This instrument also covers simultaneously the 0.8-2.4\,$\umu$m wavelength range. The observations where performed using a 0.3''$\times$1.8'' slit and the spectral extractions correspond to regions of $\sim$150 to 370\,pc in the galaxies. The objects with SpeX or GNIRS spectra available are reported in Table~\ref{obs:tab1} of Appendix A. 

The same methodology described in the previous sections was used to perform the fitting of the observed spectra to the theoretical {\sc clumpy} models. Since both SpeX and GNIRS spectra have higher spectral resolution than the IRS/{\it Spitzer}, we rebined the NIR spectra in 0.16$\umu$m intervals in order to have the same sampling as the MIR . The $\chi^2_{red}$ minimization was applied once more to all 32 combination of NIR+MIR spectroscopic data. In Figure~\ref{all_fit} we present the individual results for each parameter considering all the solutions within 10\% deviation of the minimum  $\chi^2_{red}$ and compare them with the results obtained when using only the Sptizer data. 

It is shown that in most of the cases the distribution of Y is well constrain, even when only using the MIR data, despite for some differences for the solution in individual galaxies. The preference for higher values of $\tau_V$ can be found in most of the galaxies. We can also notice that the inclusion of NIR data in general better constrain better the torus width $\sigma$. In general, the distribution of the torus parameters using the NIR+MIR combination do not change dramatically compared with those found only using the IRS spectra.

\begin{figure*}
\begin{minipage}[b]{\linewidth}
\includegraphics[width=\textwidth]{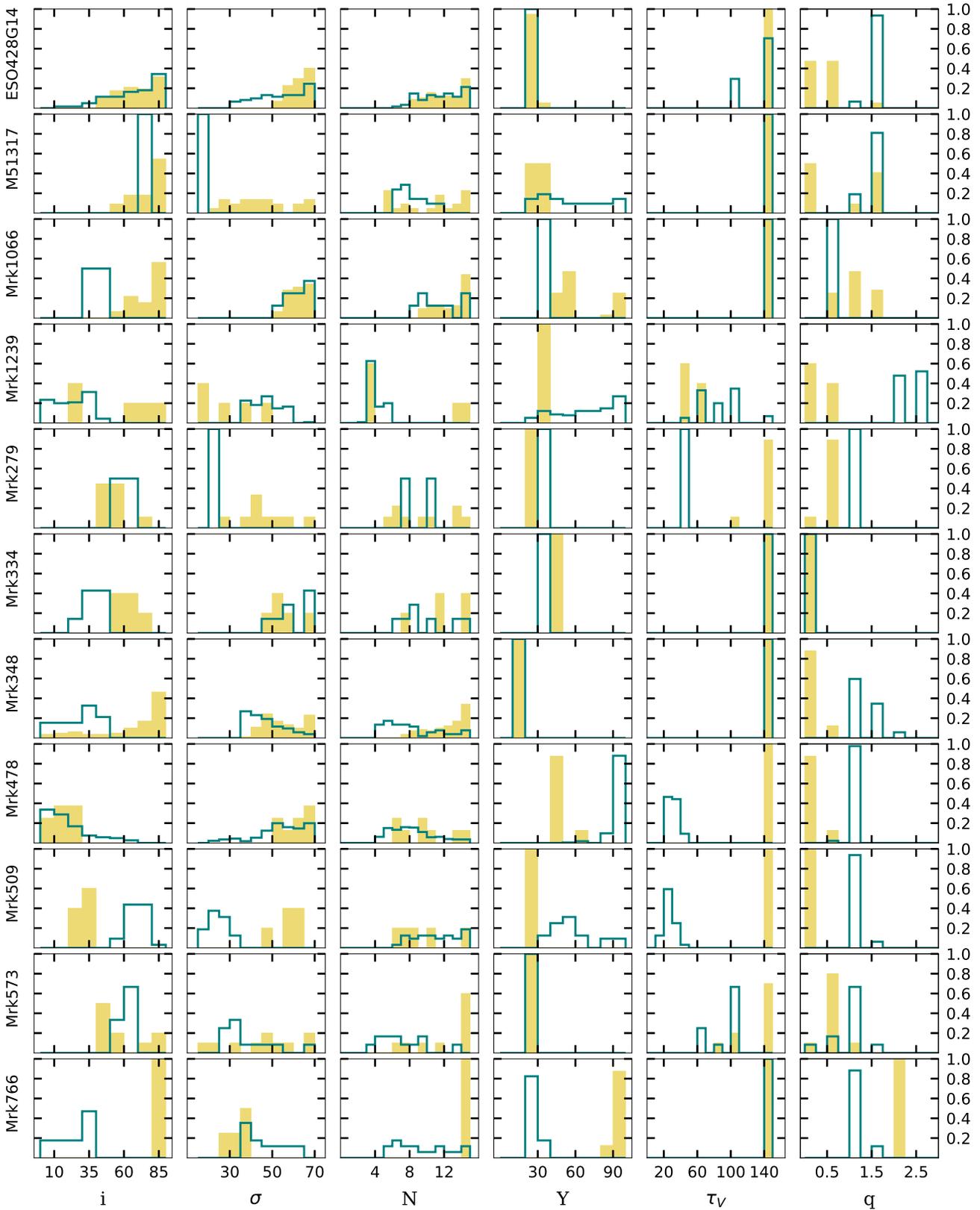}
\end{minipage}\hfill
\caption{The individual distribution of the six {\sc clumpy} parameters. The stepped histograms in blue show the parameters distribution of the JHK spectroscopic data and the IRS spectra, while the yellow filled histograms show the results only considering the MIR data.}
\label{all_fit}
\end{figure*}

\setcounter{figure}{12}
\begin{figure*}
\begin{minipage}[b]{\linewidth}
\includegraphics[width=\textwidth]{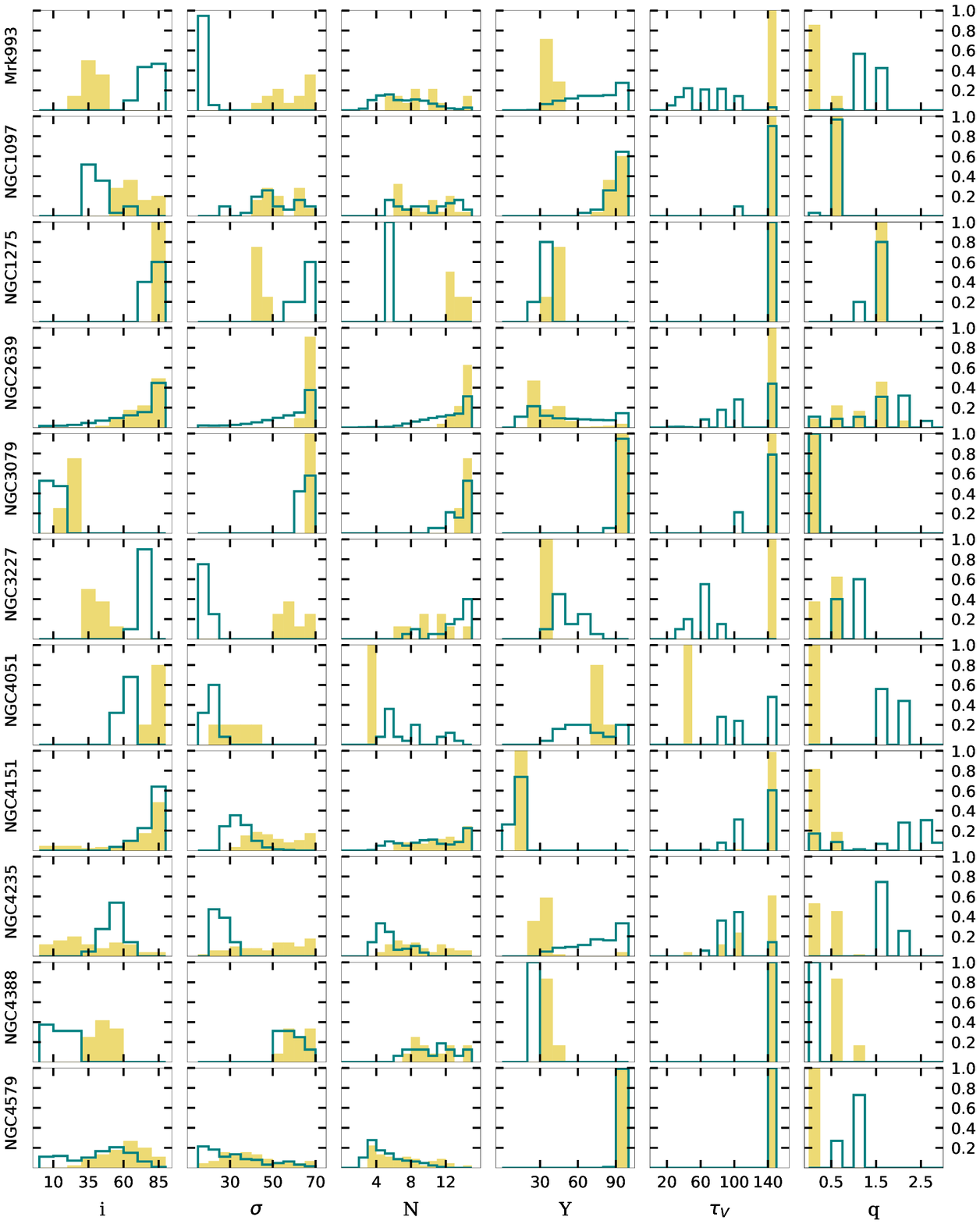}
\end{minipage}\hfill
\caption{continued from previous page.}
\end{figure*}

\setcounter{figure}{12}
\begin{figure*}
\begin{minipage}[b]{\linewidth}
\includegraphics[width=\textwidth]{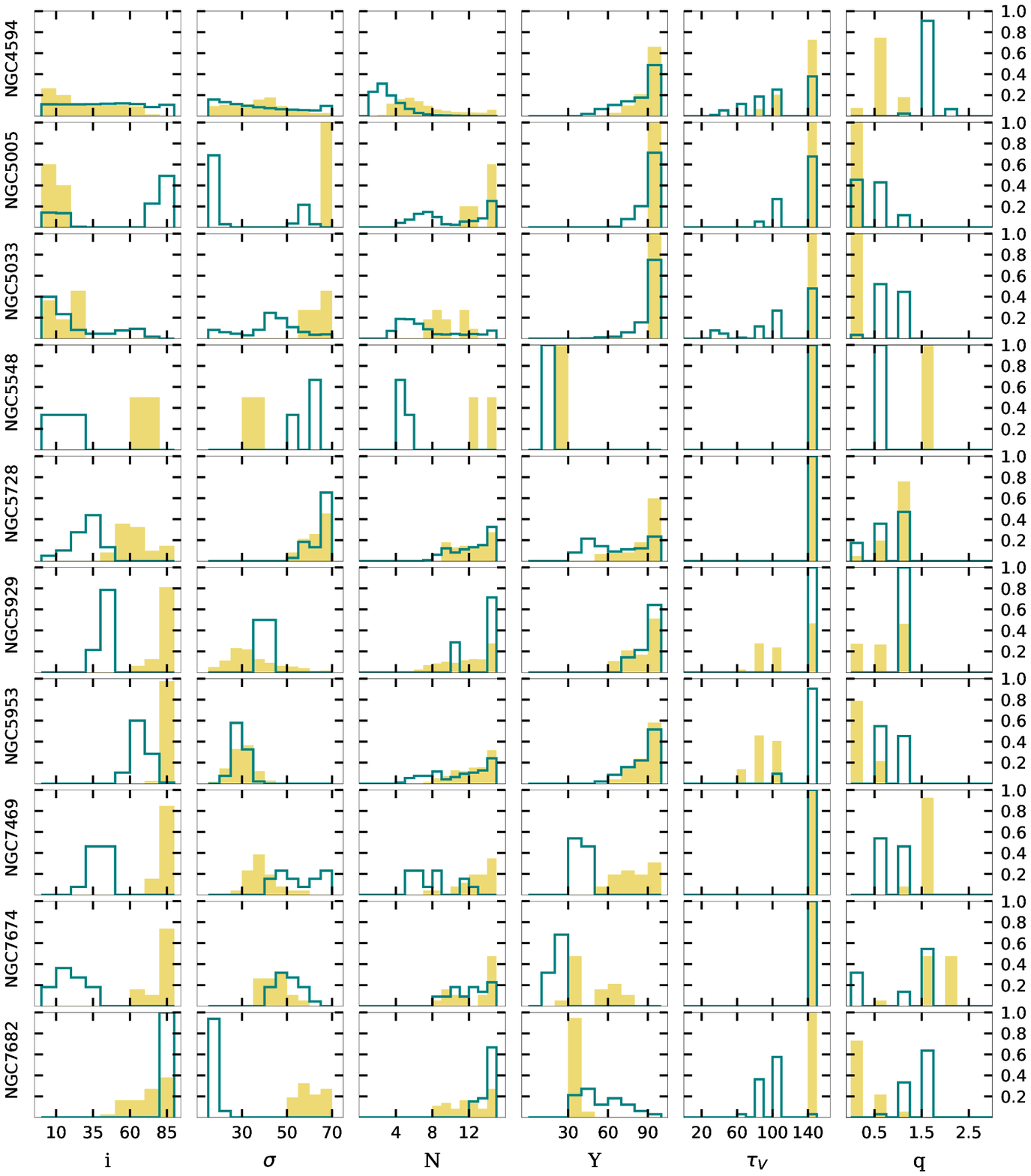}
\end{minipage}\hfill
\caption{continued from previous page.}
\end{figure*}

\begin{figure}
\begin{center}
\includegraphics[width=86mm]{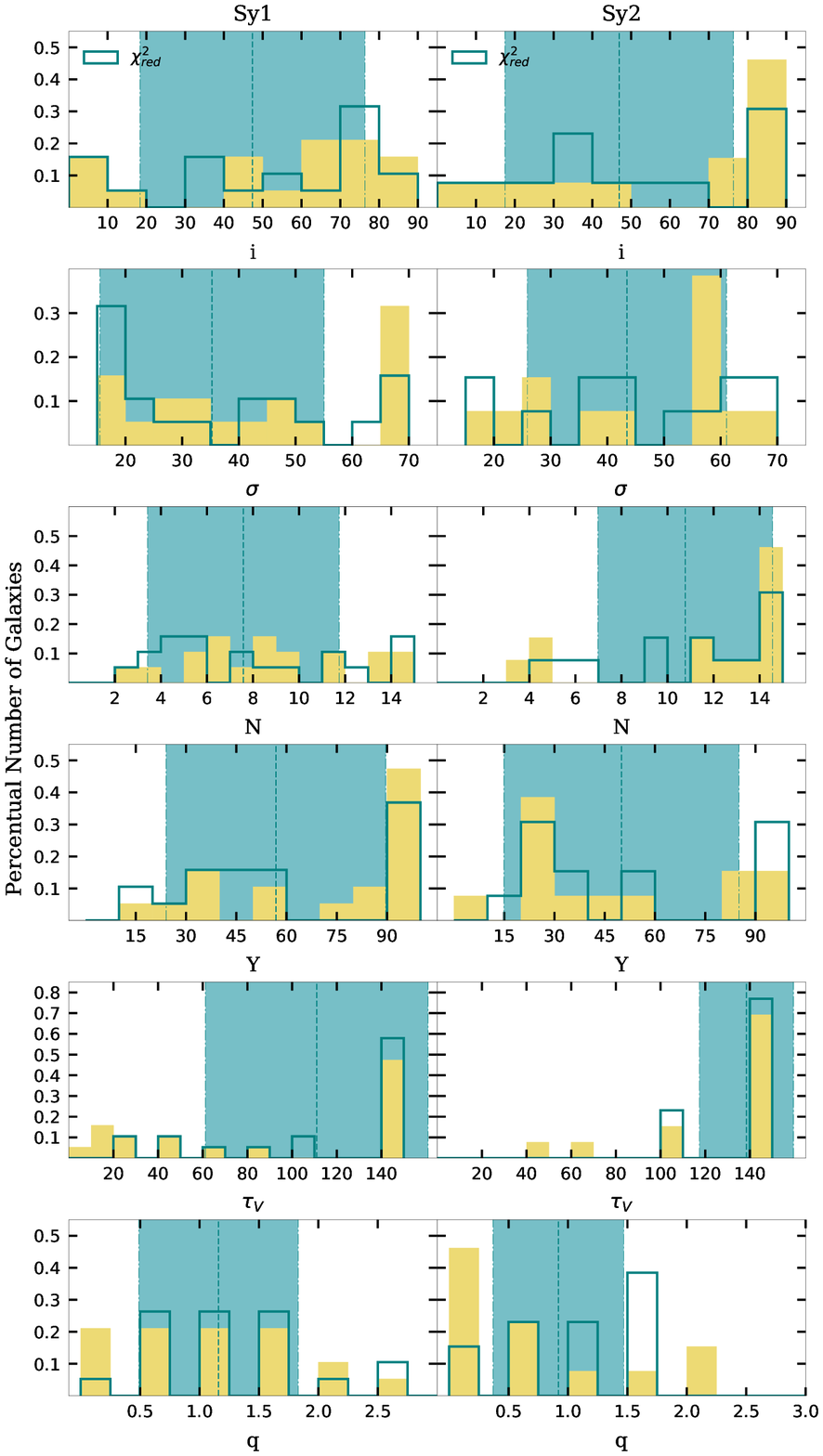}
\vspace{-0.5cm}
\caption{We present the histograms for the distribution of the direct parameters using both NIR+MIR spectra (stepped blue) and compared them with the analysis using the results for the 32 galaxies only considering the MIR data (yellow filled). The distributions for the 19 Sy\,1 are plotted at the left side and for the 13 Sy\,2 at the right side. The hatched area indicate the mean values from the $\chi^2_{red}$ and the uncertainties around the average, represented by the dashed line, for the NIR+MIR combination.}
\label{nir}
\end{center}
\end{figure}

Following the previous methodology, in order to study the main differences between type 1 and type 2 objects, we compared the mean results for the 19 Sy~1 and 13 Sy~2, as shown in Figure~\ref{nir}, obtained by including the near-IR spectral range with that obtained if we only consider the MIR data. The only noticeable change is found for the clouds' radial profile for Sy2's, where slightly higher values ($q\sim$1.5) are found when  including the NIR spectral range in the analysis.

The inclusion of the NIR to our analysis does reinforce our previous results, that the inclination angle to the observer is not the only relevant parameter to distinguish between the galaxy type. Instead, the combination of the observer's angle, the number of clouds and the physical properties of the clouds, described by the $\tau_V$,  plays a very important role in the classification. When analysed individually, in general the parameters are well constrained. However, when gathered and compared by type 1 and type 2, we found broad distributions, indicating a large dispersion of the parameters values in each type of activity. The main concern with our findings is the intrinsic degeneracy of the {\sc clumpy} models, since a combination of different parameters can reproduce almost the same SED. 

It is important to remark that the findings of \cite{ram14} are very relevant in order to constrain all the torus parameters, but they represent a small and very particular sample, without prominent dust emission and face-on galaxies. Our work aimed at exploring the generality of the {\sc clumpy} models, since they are reproducing only the torus emission and in principle they should represent any galaxy scenario. These results reinforce the fact that we are dealing with the probabilistic nature of a clumpy environment. 

We would like to highlight the importance of the use of a large and homogeneous sample in order to the determine in a reliable way the torus properties. One of the main problem when analysing SEDs and the torus properties is that usually data are not uniform. It is common to have a sample of sources in which the data are obtained with distinct filters and instruments (thus probing very different regions of the sources), and in some cases they do not have the same wavelength coverage for each individual galaxy. Consequently, this can lead to different results of the fitting \citep{ram14}.

\section[]{Conclusions}

We proposed to investigate the torus properties in a sample of 46 Sy\,1 and 65 Sy\,2 galaxies in the Unified Model scenario. The sample consists in all 5-38$\mu$m IRS/{\it Spitzer} data available in the heritage archive for Seyfert classification. To isolate the emission from the nucleus, we subtracted the host galaxy contribution in the MIR by removing the PAH bands emission via an IRS spectral decomposing code \citep[{\sc pahfit},][]{pahfit07}. 

Recently, many efforts have been made to calculate the torus emission in a clumpiness formalism. One of them are developed by \citet{nkv08a,nkv08b} ({\sc clumpy} models) and here in this work we utilize their $\sim$10$^6$ theoretical database models to derive the 6 parameters that better adjust the individual MIR spectra and analyse their distribution. 

We found differences in the derived mean values for the observers' viewing angle, $\bar{i}$=50 for Sy\,1 and $\bar{i}$=65 for Sy\,2, according to the Unified Model, which suggest that type 2 objects are observed most in edge-on angle views than type 1. Type 2 also appears to be angularly larger, with a mean Gaussian width distribution of $\sigma$=44, while the Sy\,1 class present $\bar{\sigma}$=36 and a sightly larger mean torus thickness, $\bar{Y}$=54 ($\bar{Y}$=46 for Sy\,2). Despite the fact we found almost the same number of clouds along the equatorial ray, N$\sim$10, the number of clouds obscuring the central source in the observer LOS is, on average, 7 clouds for Sy\,2 and $N_{los}\sim$4 clouds in Sy\,1, indicating most attenuated SEDs from type 2 sources. The radial index power tends to be more like a flatten radial distribution, with $q\le$1 in the majority cases, while the $\tau_V$ distribution attend the criteria of $\tau_v\gtrsim$60 for both types but do present higher values for Sy\,2, ($\rm \bar{\tau_V}(Sy1)=7$ and $\rm \bar{\tau_V}(Sy2)=111$), indicating that the cloud physical properties may be distinct.

The obscuration in type 2 objects requires higher extinction values, with an average value of $A_V\sim$900, while for Sy\,1 the average found is $\sim$290. The torus masses range from M$_{tor}\sim$10$^4$--10$^{7}M_{\odot}$ in both cases. Properties derived from de torus symmetry and random distribution nature, $C_T$ and $P_{esc}$, where the $P_{esc}$ parameter ensure non-zero AGN probabilities to edge-on inclination, we found about 30\% of probability to directly see the central source in type 1 and 10\% for Sy\,2 in the sample. Geometric covering factors, or the probabilities of absorption by the torus, are in agreement with the prediction for Sy\,2, we found that in general 80 per cent of cases the central source is obscured. However, the mean values derived for the Sy\,1 ($C_T \sim 0.7$) are larger than the values found in previous works (e.g., AH11 and RA11). 

We would like to highlight the importance of the use of a large and homogeneous sample in order to the determine in a reliable way the torus properties. In general, the inclusion of $JHK$ spectroscopic data in our MIR sample do not change the global torus properties derived for each type of activity.

Finally, the results follow the orientation dependency suggested by unification schemes, however, some properties concerning the cloud obscuration are not intrinsically the same for both types of activity. The torus geometry and cloud properties, along with orientation effects, may be crucial to characterize the differences between Sy\,1 and Sy\,2. 

On the basis of the presented results, the classification of a Seyfert galaxy may depend also on the dust intrinsic properties of the dusty torus clouds rather than only on the torus inclination angle, in contradiction with the simple geometrical requirements of the putative torus of the unification model.

\section*{Acknowledgements}

We would like to thank the anonymous referee for useful comments that helped to improve the paper.  AA and RR would like to acknowledge the financial support from CNPq during this project. D. Sales would also like to acknowledge the financial support from FAPERGS/CAPES. This research has made use of the NASA/IPAC Extragalactic Database (NED) which is operated by the Jet Propulsion Laboratory, California Institute of Technology, under contract with the National Aeronautics and Space Administration. This work is based on observations made with the Spitzer Space Telescope, which is operated by the Jet Propulsion Laboratory, California Institute of Technology under a contract with NASA.

\appendix
\onecolumn

\section[]{Sample Properties}
We present in the following Table~\ref{obs:tab1} the main properties for our 111 Seyfert sample. 

\begin{center}
\small
\setlength{\tabcolsep}{3pt}
\begin{longtable}{lrrrrccccc}
\caption[]{Sample Properties}\label{sample}  \\  

\hline \hline
\noalign{\smallskip}
\multicolumn{1}{c}{Object Name} & \multicolumn{1}{c}{RA} & \multicolumn{1}{c}{Dec} &   \multicolumn{1}{c}{z} & \multicolumn{1}{c}{Distance} & \multicolumn{1}{c}{Morphological} & \multicolumn{1}{c}{log(L$_{\rm IR}$)}  &  \multicolumn{1}{c}{L$_{\rm 2-10KeV}$}  & \multicolumn{1}{c}{Active Type} & \multicolumn{1}{c}{PID}\\
 
\multicolumn{1}{c}{} &  \multicolumn{1}{c}{J2000} & \multicolumn{1}{c}{J2000} & \multicolumn{1}{c}{} & \multicolumn{1}{c}{(Mpc)} & \multicolumn{1}{c}{Type} & \multicolumn{1}{c}{(L$_\odot$)} & \multicolumn{1}{c}{(erg $\rm s^{-1}$)} & \multicolumn{1}{c}{}  & \multicolumn{1}{c}{} \\ \noalign{\smallskip}
 \hline

\endfirsthead

\multicolumn{10}{c}%
{{\bfseries \tablename\ \thetable{} -- continued from previous page}} \\
\\
\hline \hline
\noalign{\smallskip}

\multicolumn{1}{c}{Object Name} & \multicolumn{1}{c}{RA} & \multicolumn{1}{c}{Dec} &   \multicolumn{1}{c}{z} & \multicolumn{1}{c}{Distance} & \multicolumn{1}{c}{Morphological} & \multicolumn{1}{c}{log(L$_{\rm IR}$)}  &  \multicolumn{1}{c}{L$_{\rm 2-10KeV}$}  & \multicolumn{1}{c}{Active Type} & \multicolumn{1}{c}{PID}\\
 
\multicolumn{1}{c}{} &  \multicolumn{1}{c}{J2000} & \multicolumn{1}{c}{J2000} & \multicolumn{1}{c}{} & \multicolumn{1}{c}{(Mpc)} & \multicolumn{1}{c}{Type} & \multicolumn{1}{c}{(L$_\odot$)} & \multicolumn{1}{c}{(erg $\rm s^{-1}$)} & \multicolumn{1}{c}{}  & \multicolumn{1}{c}{} \\ \noalign{\smallskip}
\hline
\endhead

\endfoot

\hline 
\caption{$^\dag$ SpeX spectra available from \citet{rif06,rif09}, $^\ddag$ GNIRS spectra from \citet{mason15}, (1) \citet{wu09}, (2) \citet{gon01}, (3) \citet{rod03}, (4) \citet{per01},(5) \citet{san03}, (6) \citet{dad07}, (7) \citet{esq14}, (8) \citet{mar12}, (9) \citet{asm11},(10) \citet{ho09}, (11) \citet{lir13}, (12) \citet{vas10}.}

\endlastfoot

\noalign{\smallskip}
Mrk334$^\dag$      &  00h03m09.6s & +21d57m37s &   0.022  &   86.7  &   Pec    & 11.02$^{(2)}$ &          &  Sy 1 & 3374 \\
Mrk335      &  00h06m19.5s & +20d12m10s &   0.026  &  110.4  & Compact  & 10.72$^{(1)}$ & 9.75e+42$^{(6)}$ &  Sy 1 & 3269 \\
Mrk938      &  00h11m06.5s & -12d06m26s &   0.020  &   84.0  &   Pec    & 11.48$^{(1)}$ &          &  Sy 2 & 3269 \\
E12-G21     &  00h40m46.1s & -79d14m24s &   0.030  &  128.6  &   E      & 11.03$^{(1)}$ &          &  Sy 1 & 3269 \\
Mrk348$^\dag$	    &  00h48m47.1s & +31d57m25s &   0.015  &   64.4  &  SA0/a   & 10.62$^{(1)}$ & 2.47e+43$^{(8)}$ &  Sy 2 & 3269 \\
NGC424      &  01h11m27.6s & -38d05m00s &   0.012  &   50.4  &  SB0/a   & 10.67$^{(1)}$ &          &  Sy 2 & 3269 \\
NGC526A     &  01h23m54.4s & -35d03m56s	&   0.019  &   81.8  & S0 pec   & 10.78$^{(1)}$ & 1.21e+43$^{(6)}$ & Sy 1 & 86, 3269 \\
NGC513      &  01h24m26.8s & +33d47m58s &   0.020  &   83.7  &  Sb/c    & 10.52$^{(1)}$ & 5.25e+42$^{(8)}$ &  Sy 2 & 3269 \\
Mrk993$^\dag$      &  01h25m31.4s & +32d08m11s &   0.016  &   60.6  &   Sa     & 10.99$^{(4)}$ &          &  Sy 1 & 40385 \\
Mrk573$^\dag$       &  01h43m57.8s & +02d21m00s &   0.017  &   67.4  &  SAB0    & 10.48$^{(2)}$ &          &  Sy 2 & 50094 \\
F01475-0740 &  01h50m02.7s & -07d25m48s &   0.018  &   75.7  &  E/S0    & 10.62$^{(1)}$ & 3.31e+43$^{(8)}$ &  Sy 2 & 3269 \\		   
NGC931      &  02h28m14.5s & +31d18m42s &   0.017  &   71.3  &  Sbc     & 10.92$^{(1)}$ &          &  Sy 1 & 3269 \\		   
NGC1056     &  02h42m48.3s & +28d34m27s &   0.005  &   22.1  &  Sa      & 9.93$^{(1)}$  &          &  Sy 2 & 3269 \\		   
NGC1097$^\dag$     &  02h46m19.0s & -30d16m30s &   0.004  &  17.5  &  SBb     & 10.78$^{(1)}$ & 7.59e+40$^{(9)}$ &  Sy 2 & 159 \\		   
NGC1125     &  02h51m40.3s & -16d39m04s &   0.011  &   46.8  &  SAB0    & 10.46$^{(1)}$ &          &  Sy 2 & 3269 \\		   
NGC1143/4   &  02h55m12.2s & -00d11m01s &   0.029  &   123.5 & S0 pec   & 10.46$^{(1)}$ &          &  Sy 2 & 159 \\
Mrk1066$^\dag$     &  02h59m58.6s & +36d49m14s &   0.012  &   47.2  &   SB0    & 10.78$^{(2)}$ & 8.32e+42$^{(8)}$ &  Sy 2 & 30572 \\
M-2-8-39    &  03h00m30.6s & -11d24m57s &   0.029  &   128   & SABa pec & 10.95$^{(1)}$ & 7.94e+43$^{(11)}$ &  Sy 2 & 3269 \\
NGC1194     &  03h03m49.1s & -01d06m13s &   0.014  &   58.2  &  SA0:    & 10.34$^{(1)}$ & 6.31e+43$^{(11)}$ &  Sy 2 & 3269 \\
NGC1241     &  03h11m14.6s & -08d55m20s &   0.014  &   57.9  &   SBb    & 10.75$^{(1)}$ &          &  Sy 2 & 3269 \\
NGC1275$^\dag$     &  03h19m48.1s & +41d30m42s &   0.017  &   70.9  &   Pec    & 11.20$^{(5)}$ & 7.24e+42$^{(10)}$ &  Sy 2 & 14   \\
NGC1320     &  03h24m48.7s & -03d02m32s &   0.009  &  37.7  &  S0/a    & 10.21$^{(1)}$ & 4.90e+42$^{(8)}$ &  Sy 2 & 159 \\
Mrk609      &  03h25m25.3s & -06d08m38s &   0.034  &   143   &  ImPec   &               & 6.63e+42$^{(6)}$ &  Sy 2 & 3374 \\ 
NGC1365     &  03h33m36.4s & -36d08m25s &   0.005  &  17.7  &   SBb    & 11.23$^{(1)}$ & 3.25e+41$^{(6)}$ &  Sy 1 & 3269 \\
NGC1386     &  03h36m46.2s & -35d59m57s &   0.003  &  16.2  &  Sa,S0   & 9.53$^{(1)}$  & 6.16e+39$^{(6)}$ &  Sy 2 & 3269 \\
F03450+0055 &  03h47m40.2s & +01d05m14s &   0.031  &  132.8  &   ?      & 11.10$^{(1)}$ &          &  Sy 1 & 3269 \\
NGC1566     &  04h20m00.4s & -54d56m16s &   0.005  &  11.8  & SABbc,Sc & 10.61$^{(1)}$ & 4.17e+41$^{(9)}$ &  Sy 1 & 159 \\
F04385-0828 &  04h40m54.9s & -08d22m22s &   0.015  &   64.7  &   S0     & 10.82$^{(1)}$ & 2.00e+43$^{(11)}$ &  Sy 2 & 3269 \\
NGC1667     &  04h48m37.1s & -06d19m12s &   0.015  &   65.0  &  SABc    & 11.02$^{(1)}$ & 3.63e+42$^{(8)}$ &  Sy 2 & 3269 \\
E33-G2      &  04h55m58.9s & -75d32m28s &   0.018  &   77.5  &  SB0     & 10.52$^{(1)}$ & 4.35e+42$^{(6)}$ &  Sy 2 & 3269 \\
M-5-13-17$^\dag$   &  05h19m35.8s & -32d39m28s &   0.012  &   54.1  &SB0/a,S0/a& 10.28$^{(1)}$ &          &  Sy 1 & 3269 \\
Mrk3       &  06h15m36.3s & +71d02m15s &   0.0135 &   55.9  &   E2 pec & 10.78$^{(2)}$ & 2.32e+42$^{(6)}$ &  Sy 2 &   14 \\ 
Mrk6        &  06h52m12.2s & +74d25m37s &   0.019  &   80.6  & SAB0:,Sa & 10.63$^{(1)}$ & 2.05e+43 &$^{(6)}$  Sy 1 & 3269 \\
ESO428-G014$^\dag$ &  07h16m31.2s & -29d19m29s &   0.006  &   26    & SAB0 pec &       &          &  Sy 2 & 30572 \\
Mrk9        &  07h36m57.0s & +58d46m13s &   0.040  &   170   &S0: pec,SB& 11.15$^{(1)}$ &          &  Sy 1 & 3269 \\
Mrk79       &  07h42m32.8s & +49d48m35s &   0.022  &   95    & SBb,SBc  & 10.90$^{(1)}$ & 2.51e+43$^{(12)}$ &  Sy 1 & 3269 \\
Mrk78       &  07h42m41.7s & +65d10m37s &   0.037  &   158   &   SB     & 11.04$^{(2)}$ &          &  Sy 2 & 50094 \\
Mrk622      &  08h07m41.0s & +39d00m15s &   0.023  &   99.6  &   S0     &       &          &  Sy 2 & 3374 \\
NGC2622     &  08h38m10.9s & +24d53m43s &   0.029  &   124   &   SBb    &       &          &  Sy 1 & 3374 \\
NGC2639$^\ddag$     &  08h43m38.1s & +50d12m20s &   0.011  &   47.7  &   SAa    & 10.34$^{(1)}$ & 7.08e+40$^{(10)}$ &  Sy 1 & 3269 \\
Mrk704      &  09h18m26.0s & +16d18m19s &   0.029  &  125.2  &   SBa    & 10.97$^{(1)}$ &          &  Sy 1 & 3269 \\
NGC2992     &  09h45m42.0s & -14d19m35s &   0.008  &  30.5  &  Sa pec  & 10.51$^{(1)}$ & 7.20e+41$^{(6)}$ &  Sy 1 & 3269 \\
Mrk1239$^\dag$     &  09h52m19.1s & -01d36m43s &   0.0199 &   85.3  &  E-S0    & 10.86$^{(1)}$ &          &  Sy 1 & 3269 \\
NGC3079$^\ddag$     &  10h01m57.8s & +55d40m47s &   0.004  &  19.7  &   SBc    & 10.62$^{(1)}$ & 1.05e+42$^{(8)}$ &  Sy 2 & 3269 \\
NGC3227$^\dag$     &  10h23m30.6s & +19d51m54s &   0.004  &  20.9  & SABa pec & 9.97$^{(1)}$  & 2.51e+42$^{(7)}$ &  Sy 1 & 668, 3269 \\
Mrk34       &  10h34m08.6s & +60d01m52s &   0.051  &   218   &   Sa     & 11.15$^{(2)}$ &          &  Sy 2 & 50094 \\
NGC3511     &  11h03m23.8s & -23d05m12s &   0.004  &  14.6  &  SAc     & 9.95$^{(1)}$  &          &  Sy 1 & 3269 \\
NGC3516     &  11h06m47.5s & +72d34m07s &   0.009  &  38.9  &  SB0     & 10.17$^{(1)}$ & 3.77e+42$^{(6)}$ &  Sy 1 & 3269 \\
M+0-29-23   &  11h21m12.2s & -02d59m03s &   0.025  &  106.6  &  SABb    & 11.36$^{(1)}$ &          &  Sy 2 & 3269 \\
NGC3660     &  11h23m32.3s & -08d39m31s &   0.012  &  52.6   &  SBbc    & 10.47$^{(1)}$ & 7.94e+42$^{(11)}$ &  Sy 2 & 3269 \\
NGC3786     &  11h39m42.5s & +31d54m33s &   0.009  &   40.9  & SABa/Pec &       &          &  Sy 2 & 3374 \\
NGC3982     &  11h56m28.1s & +55d07m31s &   0.004  &  21.8  &  SABb    &  9.81$^{(1)}$ & 1.41e+41$^{(8)}$ &  Sy 2 & 3269 \\
NGC4051$^\dag$     &  12h03m09.6s & +44d31m53s &   0.002  &  17.0  &  SABbc   &  9.66$^{(1)}$ & 1.64e+40$^{(6)}$ &  Sy 1 & 3269 \\
UGC7064     &  12h04m43.3s & +31d10m38s &   0.025  &  107.1  &  SAB     & 11.18$^{(1)}$ &          &  Sy 1 & 3269 \\
NGC4151$^\dag$     &  12h10m32.6s & +39d24m21s &   0.003  &   20.3 &  SABab   &  9.95$^{(1)}$ & 2.42e+42$^{(6)}$ &  Sy 1 & 3269 \\
NGC4235$^\ddag$     &  12h17m09.9s & +07d11m30s &   0.008  &   38    &   SAa    & 10.30$^{(4)}$ & 4.07e+41$^{(9)}$ &  Sy 1 & 40936 \\
Mrk766$^\dag$      &  12h18m26.5s & +29d48m46s &   0.013  &   55.4  &   SBa    & 10.67$^{(1)}$ & 6.44e+42$^{(6)}$ &  Sy 1 & 3269 \\
NGC4388$^\ddag$     &  12h25m46.7s & +12d39m44s &   0.008  &  18.1  &   SAb    & 10.73$^{(1)}$ & 3.37e+42$^{(6)}$ &  Sy 2 & 3269 \\
NGC4501     &  12h31m59.2s & +14d25m14s &   0.008  &  20.7  &   SAb    & 10.98$^{(1)}$ & 7.76e+38$^{(10)}$ &  Sy 2 & 3269 \\
NGC4507     &  12h35m36.6s & -39d54m33s &   0.012  &   53    &  SABab   &               & 4.70e+42$^{(6)}$ &  Sy 2 & 30572 \\
NGC4579$^\ddag$     &  12h37m43.5s & +11d49m05s &   0.005  &  16.8  &  SABc    & 10.17$^{(1)}$ & 2.66e+41$^{(6)}$ &  Sy 1 & 159 \\
NGC4593     &  12h39m39.4s & -05d20m39s &   0.009  &  44.0  &   SBb    & 10.35$^{(1)}$ & 5.74e+42$^{(6)}$ &  Sy 1 & 3269 \\
NGC4594$^\ddag$     &  12h39m59.4s & -11d37m23s &   0.003  &  10.9  &   SAa    & 9.75$^{(1)}$  & 9.77e+39$^{(9)}$ &  Sy 1 & 159 \\
NGC4602     &  12h40m36.8s & -05d07m59s &   0.008  &  34.4  &   SABbc  & 10.44$^{(1)}$ &          &  Sy 1 & 3269 \\
Tol1238-364 &  12h40m52.8s & -36d45m21s &   0.011  &   46.8  &   SBbc   & 10.87$^{(1)}$ &          &  Sy 2 & 3269 \\
M-2-33-34   &  12h52m12.4s & -13d24m53s &   0.015  &   62.7  &  Sa      & 10.49$^{(1)}$ &          &  Sy 1 & 3269 \\
NGC4941     &  13h04m13.1s & -05d33m06s &   0.004  &  13.8  &  SABab   & 9.39$^{(1)}$  & 2.71e+40$^{(6)}$ &  Sy 2 & 86, 3269 \\
NGC4968     &  13h07m06.0s & -23d40m37s &   0.010  &   42.2  &  SAB0    & 10.39$^{(1)}$ &          &  Sy 2 & 3269 \\
NGC5005$^\ddag$     &  13h10m56.2s & +37d03m33s &   0.003  &  17.5  &  SABbc   & 10.20$^{(1)}$ & 8.71e+39$^{(10)}$ &  Sy 2 & 3269 \\
NGC5033$^\ddag$     &  13h13m27.5s & +36d35m38s &   0.003  &  20.6  &  SAc     & 10.05$^{(1)}$ & 5.01e+40$^{(10)}$ &  Sy 1 & 159 \\
NGC5135     &  13h25m44.0s & -29d50m01s &   0.014  &   58.6  &  SBab    & 11.27$^{(1)}$ & 1.26e+43$^{(7)}$ &  Sy 2 & 3269 \\
M-6-30-15   &  13h35m53.8s & -34d17m44s &   0.008  &   33.2  &  S?      & 9.98$^{(1)}$  & 6.22e+42$^{(6)}$ &  Sy 1 & 3269 \\
NGC5256     &  13h38m17.5s & +48d16m37s &   0.028  &  117.3  &   Pec    & 11.51$^{(1)}$ &          &  Sy 2 & 3269 \\
I4329A      &  13h49m19.2s & -30d18m34s &   0.016  &   68.8  &  SA0     & 10.97$^{(1)}$ & 7.42e+43$^{(6)}$ &  Sy 1 & 3269 \\
Mrk279$^\dag$      &  13h53m03.4s & +69d18m30s &   0.030  &   129   &   S0     & 11.90$^{(4)}$ & 6.31e+43$^{(12)}$ &  Sy 1 & 666  \\   
NGC5347     &  13h53m17.8s & +33d29m27s &   0.008  &   36.7 & SBab     & 10.04$^{(1)}$ & 2.51e+42$^{(7)}$ &  Sy 2 & 3269 \\
Mrk463E     &  13h56m02.9s & +18d22m19s &   0.050  &   217   &  S pec   & 11.70$^{(2)}$ & 1.86e+42$^{(6)}$ &  Sy 2 & 105  \\
NGC5506     &  14h13m14.8s & -03d12m27s &   0.006  &  28.7  &  Sa pec  & 10.44$^{(1)}$ & 4.99e+42$^{(6)}$ &  Sy 2 & 3269 \\
NGC5548$^\dag$     &  14h17m59.5s & +25d08m12s &   0.017  &   73.6  &  SA0/a   & 10.66$^{(1)}$ & 1.95e+43$^{(6)}$ &  Sy 1 & 86, 3269 \\
Mrk471      &  14h22m55.4s & +32d51m03s &   0.034  &   147   &   SBa    &       &          &  Sy 2 & 3374 \\
Mrk817      &  14h36m22.1s & +58d47m39s &   0.031  &  134.7  &   SBc    & 11.35$^{(1)}$ &          &  Sy 1 & 3269 \\
NGC5695     &  14h37m22.1s & +36d34m04s &   0.014  &   60.7  &   SBb    &       &          &  Sy 2 & 30773 \\
Mrk477      &  14h40m38.1s & +53d30m16s &   0.038  &   161   &  Comp    & 11.18$^{(2)}$ &          &  Sy 2 & 30443 \\
Mrk478$^\dag$      &  14h42m07.4s & +35d26m23s &   0.079  &   347   &   S      & 11.37$^{(3)}$ & 5.16e+43$^{(6)}$ &  Sy 1 & 3187, 20142 \\
NGC5728$^\dag$     &  14h42m23.9s & -17d15m11s &   0.009  &   41.9  &   SABa   & 10.60$^{(5)}$ & 1.95e+43$^{(8)}$ &  Sy 2 & 30745 \\
Mrk841      &  15h04m01.2s & +10d26m16s &   0.036  &   157   &   E      & 11.82$^{(4)}$ & 3.27e+43$^{(6)}$ &  Sy 1 & 14 \\
NGC5929$^\dag$     &  15h26m06.1s & +41d40m14s &   0.008  &  38.5  &  Sab pec & 10.58$^{(1)}$ &          &  Sy 2 & 3269 \\
NGC5953$^\dag$     &  15h34m32.4s & +15d11m38s &   0.007  &  33.0  &  SAa pec & 10.49$^{(1)}$ &          &  Sy 2 & 3269 \\
M-2-40-4    &  15h48m24.9s & -13d45m28s &   0.025  &  107.9  &   Sc     & 11.32$^{(1)}$ &          &  Sy 2 & 3269 \\
F15480-0344 &  15h50m41.5s & -03d53m18s &   0.030  &  129.8  &   S0     & 11.14$^{(1)}$ &          &  Sy 2 & 3269 \\
Mrk883      &  16h29m52.9s & +24d26m38s &   0.037  &   159   &   Irr    &       &          &  Sy 2 & 3374 \\
NGC6810     &  19h43m34.4s & -58d39m21s &   0.007  &  29.0  &  SAab    & 10.74$^{(1)}$ &          &  Sy 2 & 3269 \\	
NGC6860     &  20h08m46.9s & -61d06m01s &   0.015  &  63.7   &  SBb     & 10.35$^{(1)}$ & 3.98e+42$^{(12)}$ &  Sy 1 & 3269 \\ 
NGC6890     &  20h18m18.1s & -44d48m25s &   0.008  &  31.8  &  SAb     & 10.27$^{(1)}$ &          &  Sy 2 & 3269 \\	
Mrk509$^\dag$      &  20h44m09.7s & -10d43m25s &   0.034  &   48.6  &  Compact & 11.21$^{(1)}$ & 9.24e+43$^{(6)}$ &  Sy 1 & 86 \\
IC5063       &  20h52m02.3s & -57d04m08s &   0.011  &  48.6   &  SA0     & 10.87$^{(1)}$ & 6.76e+42$^{(8)}$ &  Sy 2 & 86, 3269 \\
UGC11680    &  21h07m43.6s & +03d52m30s &   0.026  &  111.3  &  Scd     & 11.23$^{(1)}$ &          &  Sy 2 & 3269 \\ 
NGC7130     &  21h48m19.5s & -34d57m05s &   0.016  &  69.2   &  Sa pec  & 11.38$^{(1)}$ & 1.26e+43$^{(7)}$ &  Sy 2 & 3269 \\ 
NGC7172     &  22h02m01.9s & -31d52m11s	&   0.009  &  33.9  &  Sa pec  & 10.47$^{(1)}$ & 1.65e+42$^{(6)}$ &  Sy 2 & 86, 3269 \\
NGC7213     &  22h09m16.2s & -47d10m00s &   0.006  &  22.0  &  SAa     & 10.01$^{(1)}$ & 2.24e+42$^{(6)}$ &  Sy 1 & 3269 \\
NGC7314     &  22h35m46.2s & -26d03m01s &   0.005  &  19.0  &  SABbc   & 10.00$^{(1)}$ & 1.13e+42$^{(6)}$ &  Sy 1 & 86, 3269 \\
M-3-58-7    &  22h49m37.1s & -19d16m26s &   0.031  &  134.7  &  SAB0/a  & 11.30$^{(1)}$ & 7.94e+43$^{(11)}$ &  Sy 2 & 3269 \\ 
NGC7469$^\dag$     &  23h03m15.6s & +08d52m26s &   0.016  &  69.9   &  SABa    & 11.65$^{(1)}$ & 1.86e+43$^{(6)}$ &  Sy 1 & 3269 \\ 
NGC7496     &  23h09m47.3s & -43d25m41s &   0.006  &  20.1  &  SBb     & 10.28$^{(1)}$ &          &  Sy 2 & 3269 \\ 
NGC7582     &  23h18m23.5s & -42d22m14s &   0.005  &  18.8  &  SBab    & 10.91$^{(1)}$ & 1.01e+42$^{(6)}$ &  Sy 2 & 3269 \\ 
NGC7590     &  23h18m54.8s & -42d14m21s &   0.005  &  23.7  &  SAbc    & 10.19$^{(1)}$ & 5.89e+39$^{(9)}$ &  Sy 2 & 3269 \\ 
NGC7603     &  23h18m56.6s & +00d14m38s &   0.030  &  126.4  &  SAb pec & 11.05$^{(1)}$ &          &  Sy 1 & 3269 \\ 
NGC7674$^\dag$     &  23h27m56.7s & +08d46m45s &   0.029  &  123.9  & SAbc pec & 11.57$^{(1)}$ & 1.90e+42$^{(6)}$ &  Sy 2 & 3269 \\ 
NGC7679     &  23h28m46.7s & +03d30m41s &   0.017  &   66.2  &  SB0pec  & 11.05$^{(2)}$ & 3.39e+42$^{(6)}$ &  Sy 2 & 30323 \\
NGC7682$^\dag$     &  23h29m03.9s & +03d32m00s &   0.017  &   66.2  &  SBab    & 11.02$^{(4)}$ &          &  Sy 2 & 50588 \\
CGCG381-051 &  23h48m41.7s & +02d14m23s &   0.031  &  131.3  &   SBc    & 11.19$^{(1)}$ &          &  Sy 2 & 3269 \\
\label{obs:tab1}
\end{longtable}
\end{center}

\section[]{Individual Subtraction}

In this appendix we present the individual results of the spectral decomposition and the subtracted spectra for each galaxy in our sample.

\renewcommand{\thefigure}{B\arabic{figure}}
\setcounter{figure}{0}

\begin{figure}
\begin{minipage}[b]{0.325\linewidth}
\includegraphics[width=\textwidth]{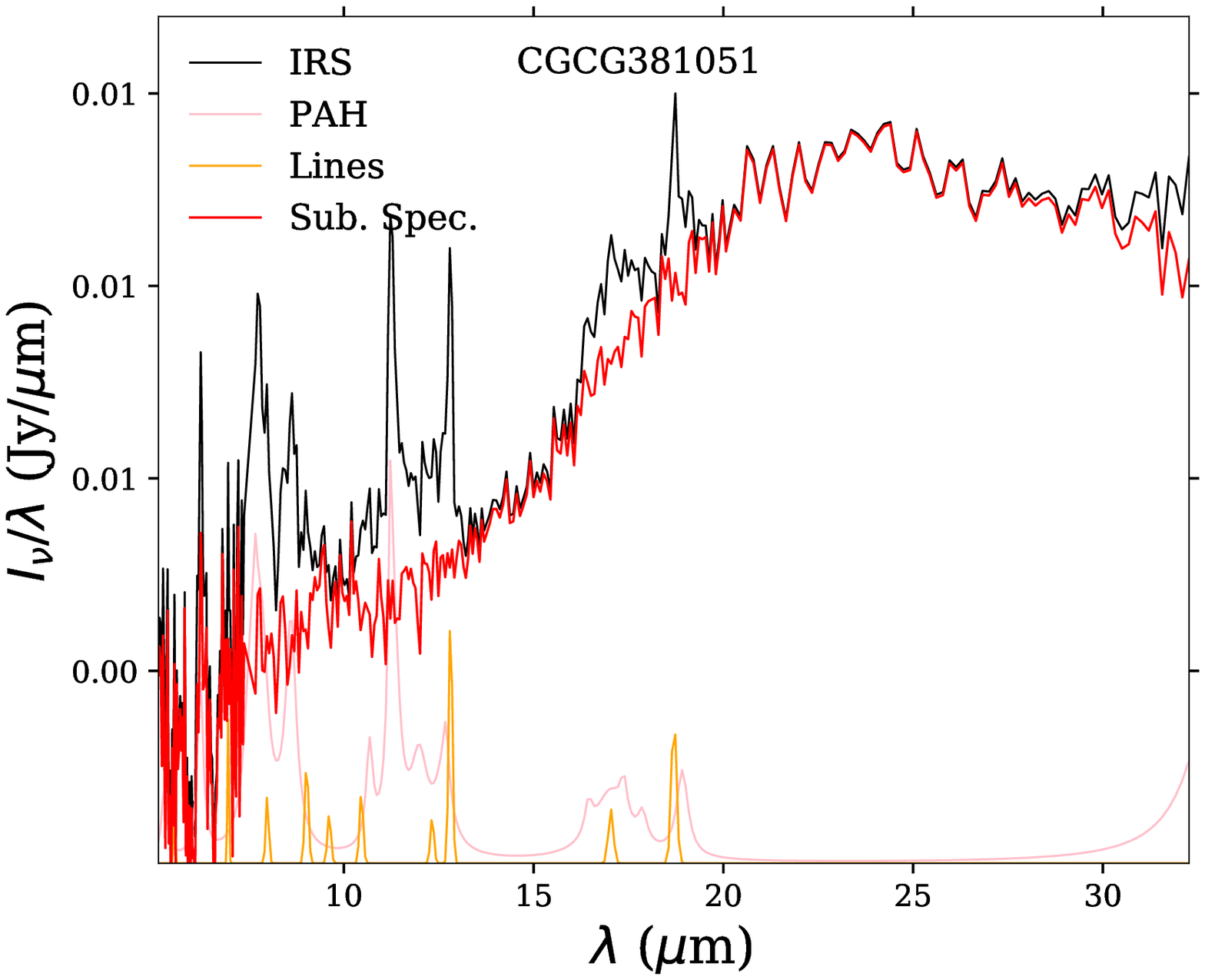}
\end{minipage} \hfill
\begin{minipage}[b]{0.325\linewidth}
\includegraphics[width=\textwidth]{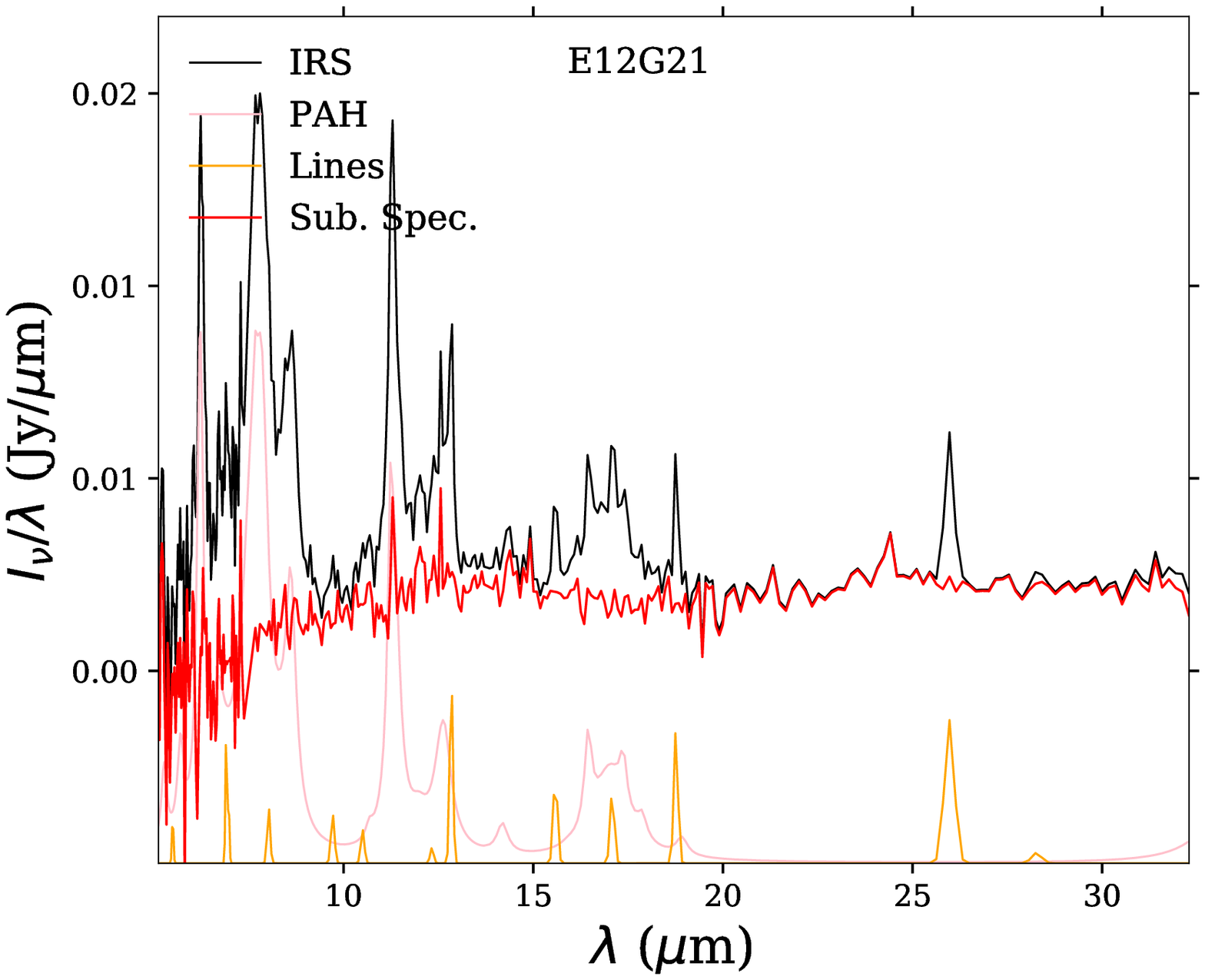}
\end{minipage} \hfill
\begin{minipage}[b]{0.325\linewidth}
\includegraphics[width=\textwidth]{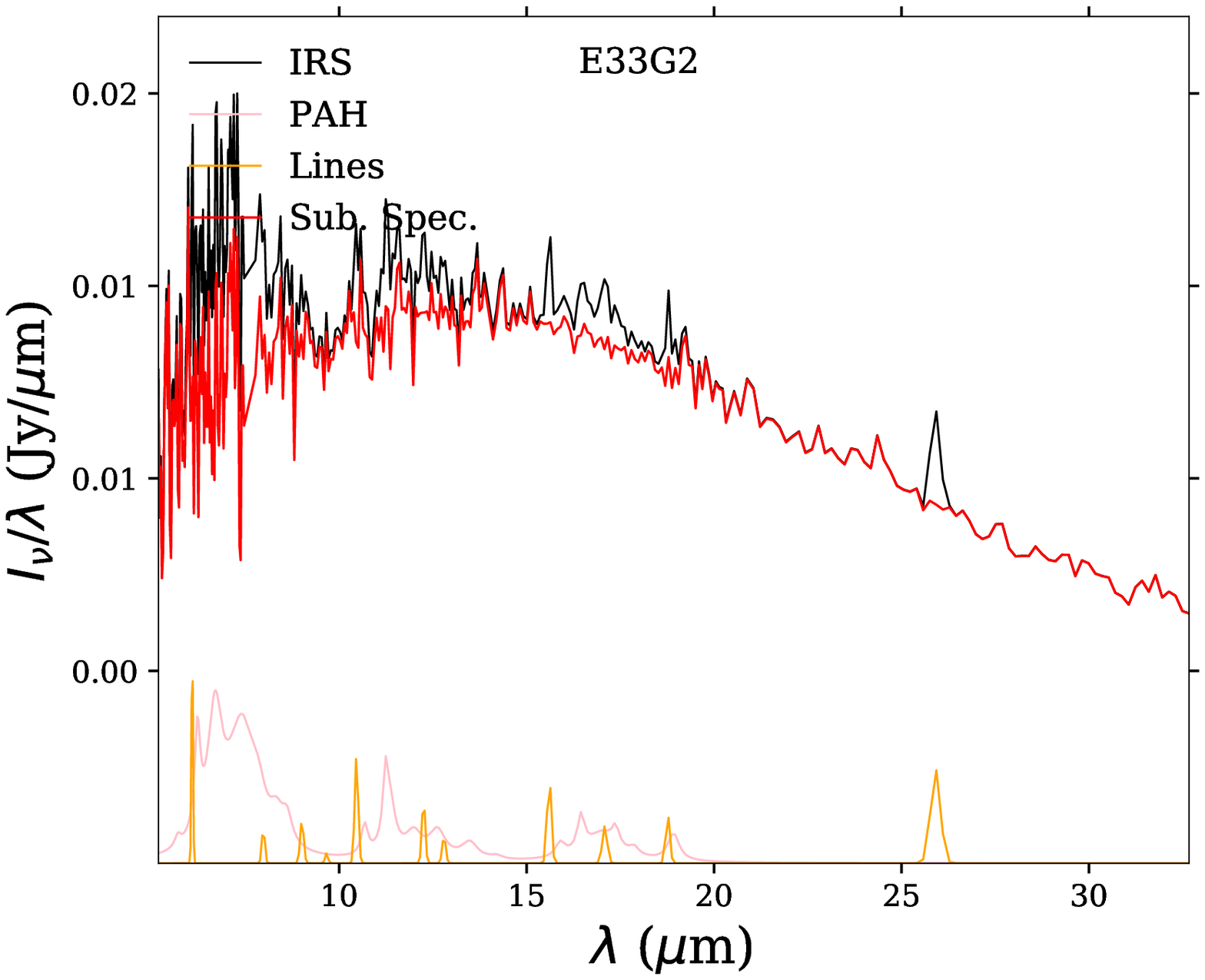}
\end{minipage} \hfill
\begin{minipage}[b]{0.325\linewidth}
\includegraphics[width=\textwidth]{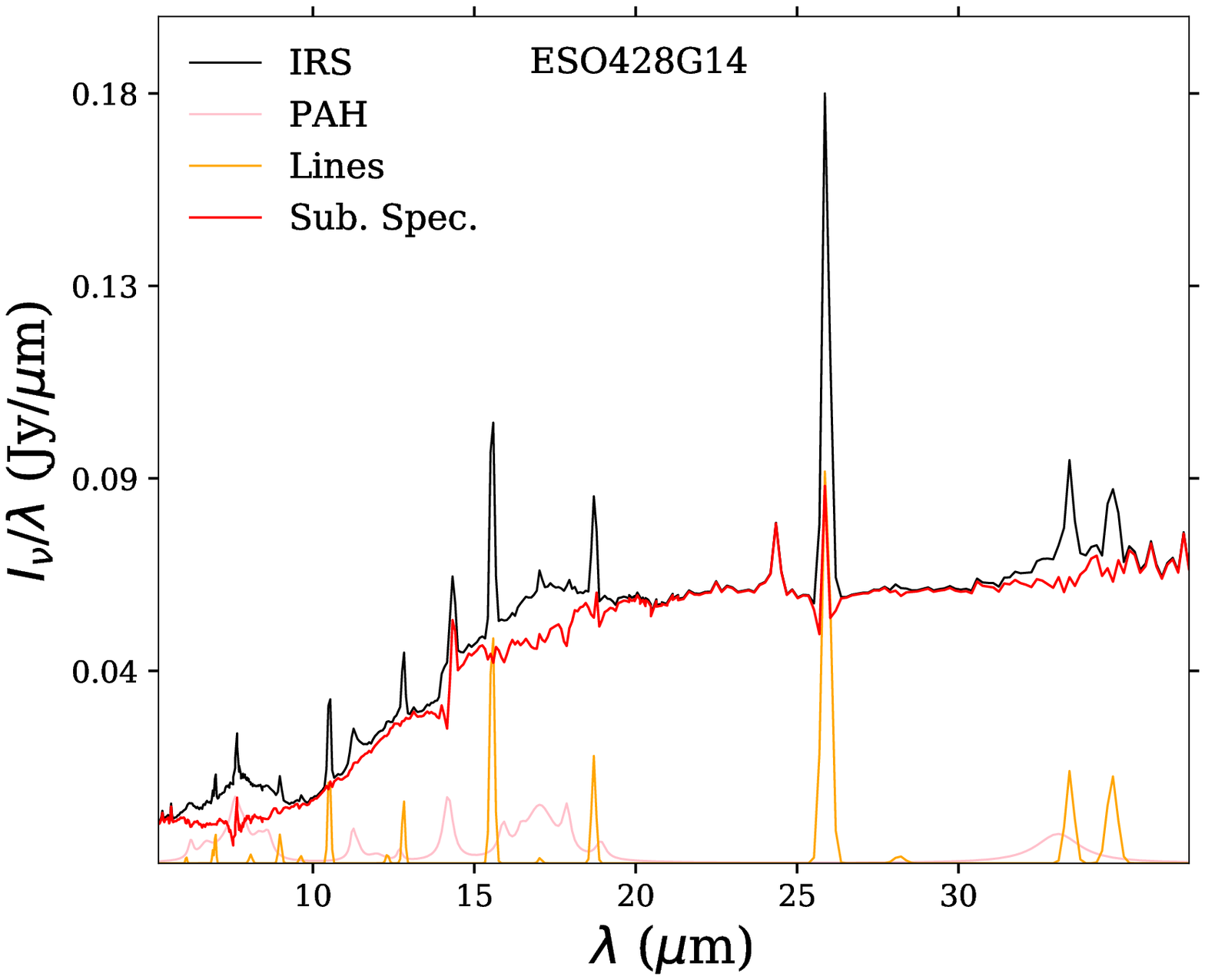}
\end{minipage} \hfill
\begin{minipage}[b]{0.325\linewidth}
\includegraphics[width=\textwidth]{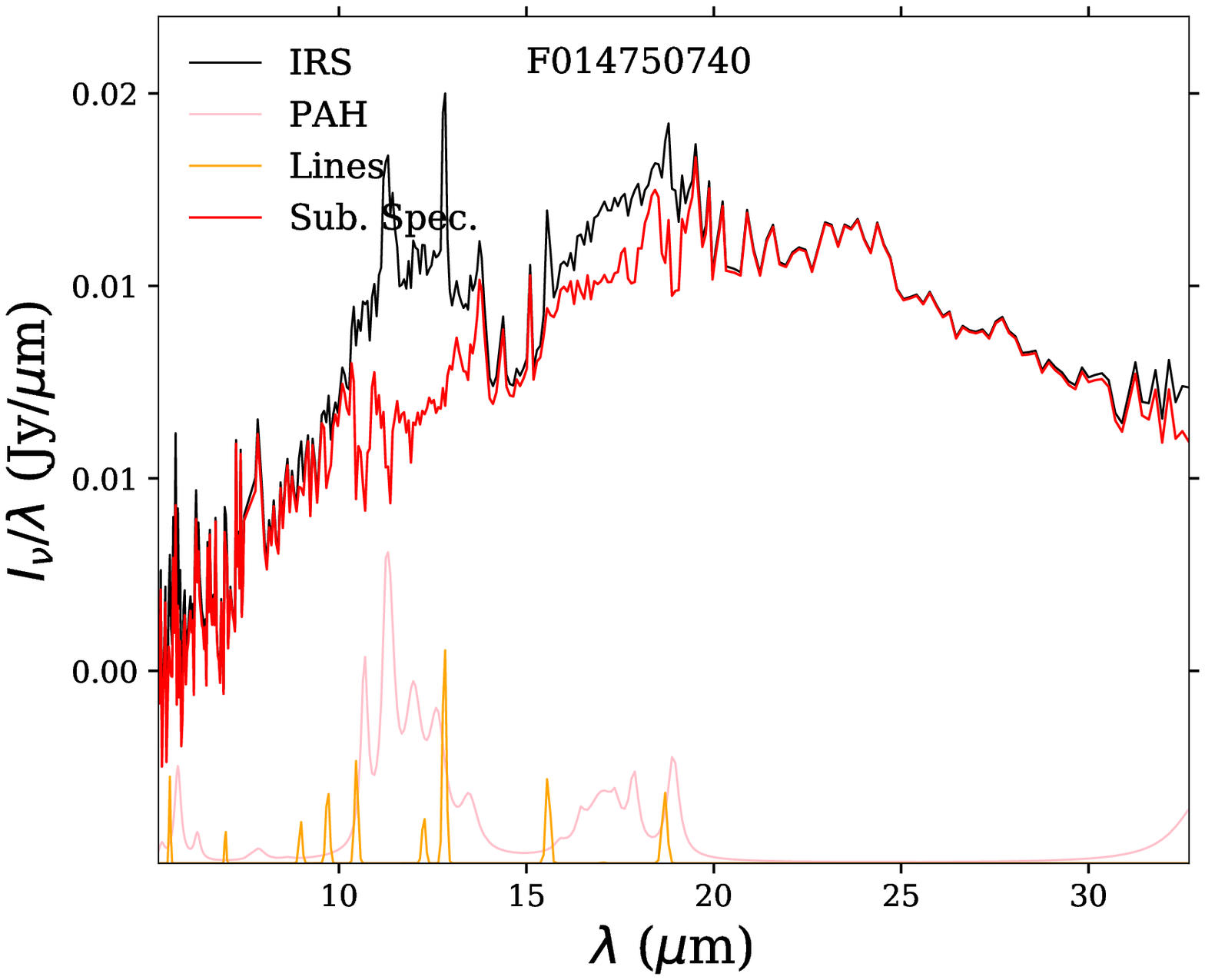}
\end{minipage} \hfill
\begin{minipage}[b]{0.325\linewidth}
\includegraphics[width=\textwidth]{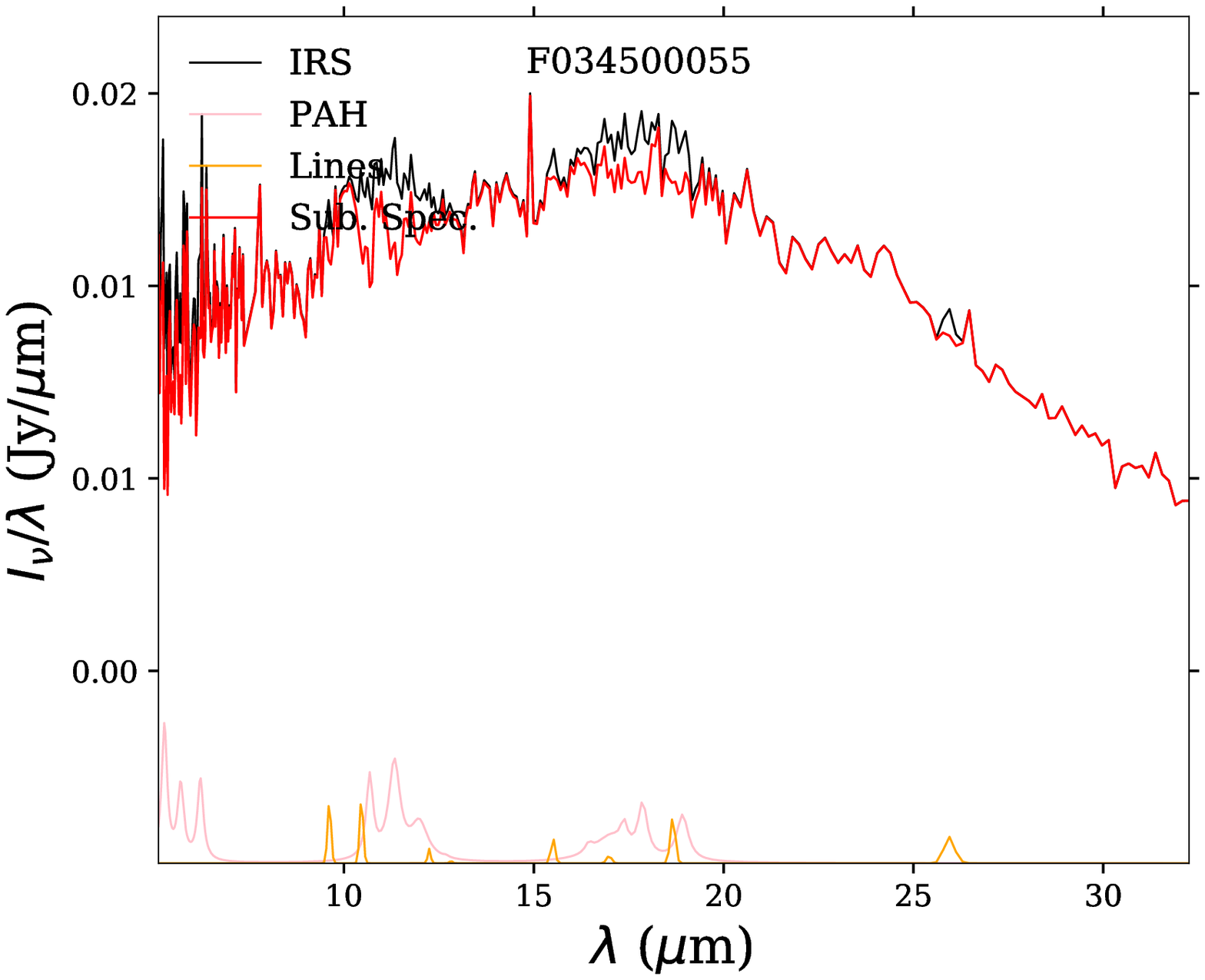}
\end{minipage} \hfill
\begin{minipage}[b]{0.325\linewidth}
\includegraphics[width=\textwidth]{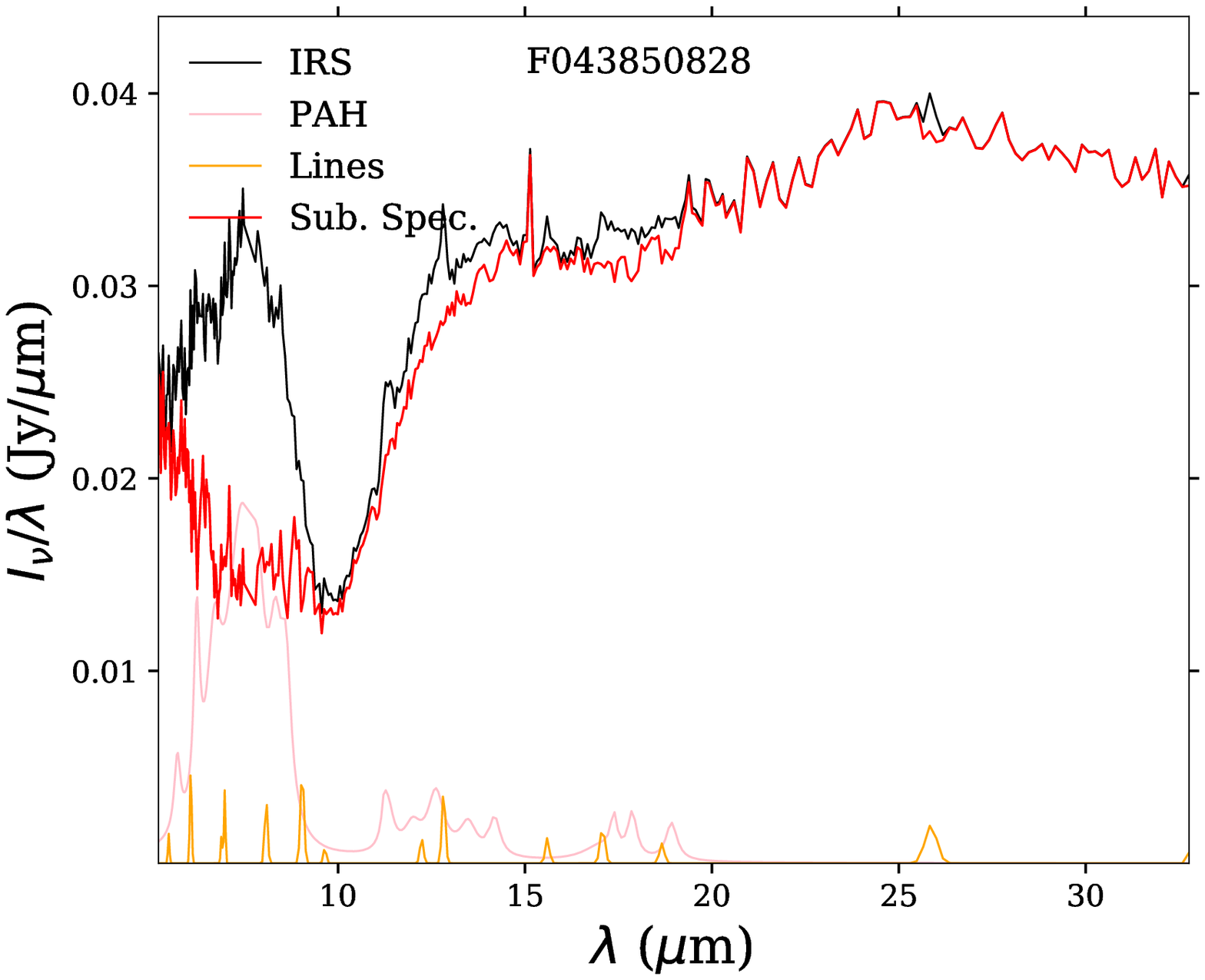}
\end{minipage} \hfill
\begin{minipage}[b]{0.325\linewidth}
\includegraphics[width=\textwidth]{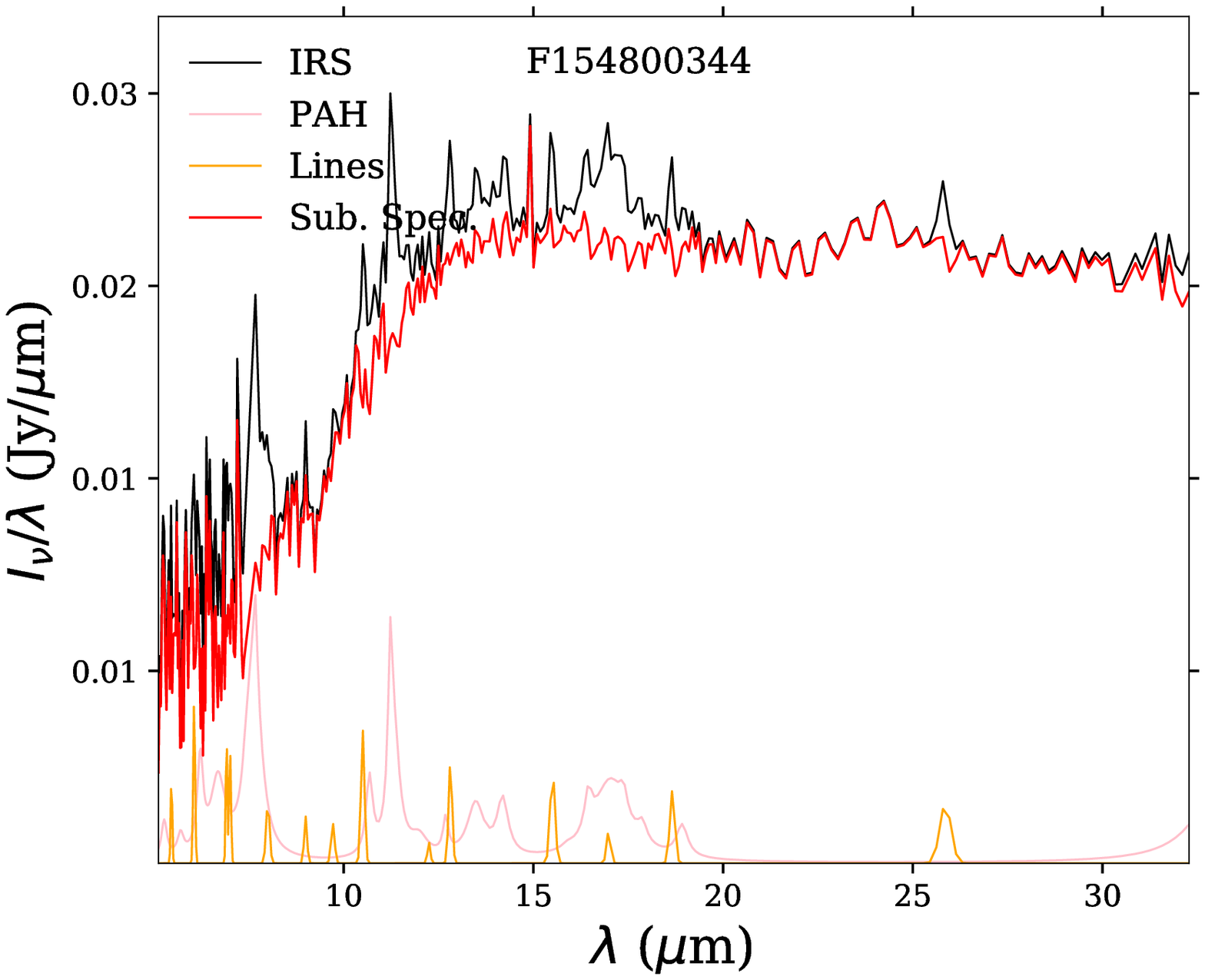}
\end{minipage} \hfill
\begin{minipage}[b]{0.325\linewidth}
\includegraphics[width=\textwidth]{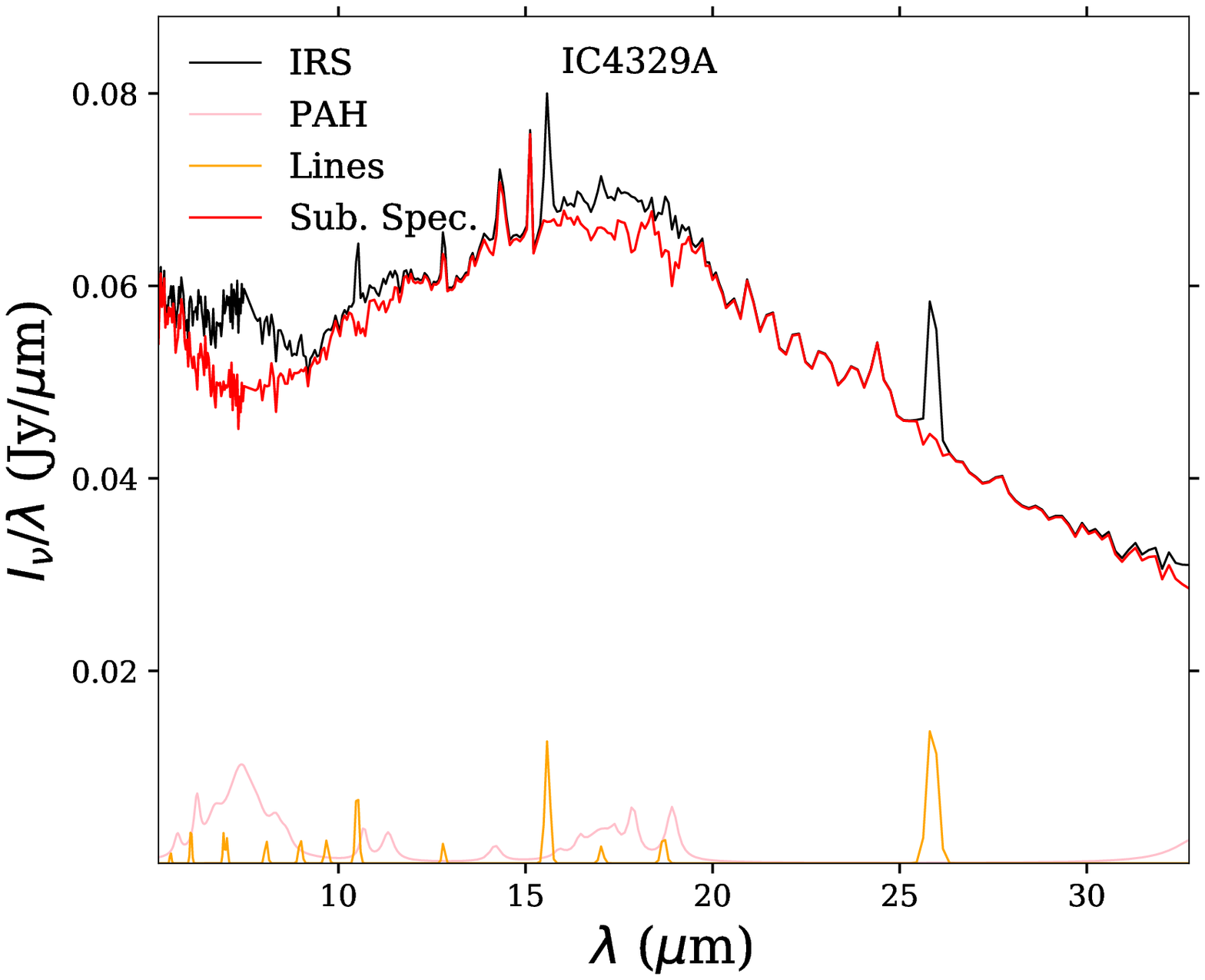}
\end{minipage} \hfill
\caption{Individual spectral decomposition. We present the subtraction of the PAH and ionic lines emission from the spectra. The black lines represent the observed IRS spectra, while the orange and pink lines show the resulting adjust by fitting the PAH emission and ionic and hydrogen lines, respectively, using the {\sc pahfit} \citep{pahfit07}. In red are presented the subtracted spectra that were adopted in our analysis.}
\setcounter{figure}{0}
\end{figure}

\begin{figure}

\begin{minipage}[b]{0.325\linewidth}
\includegraphics[width=\textwidth]{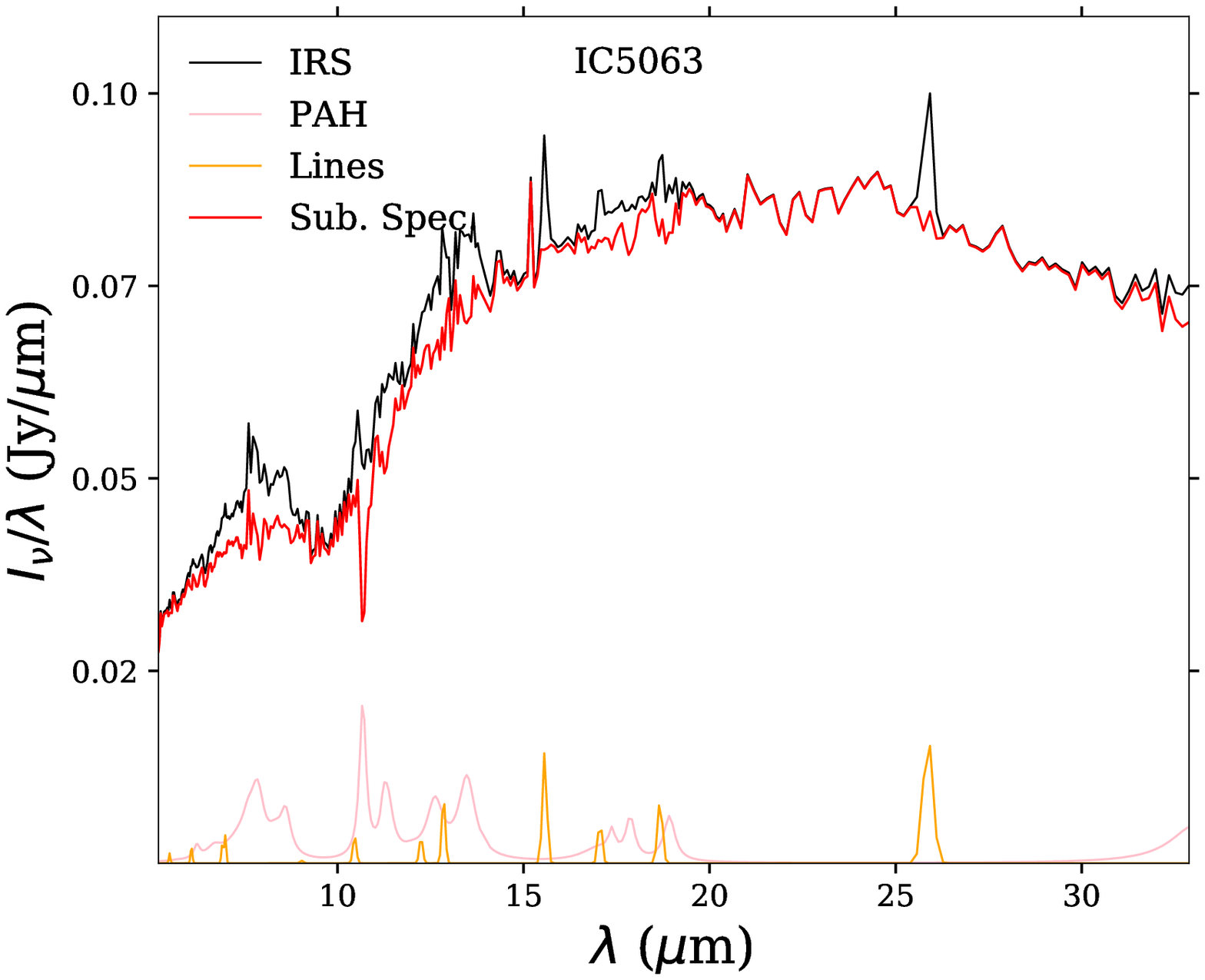}
\end{minipage} \hfill
\begin{minipage}[b]{0.325\linewidth}
\includegraphics[width=\textwidth]{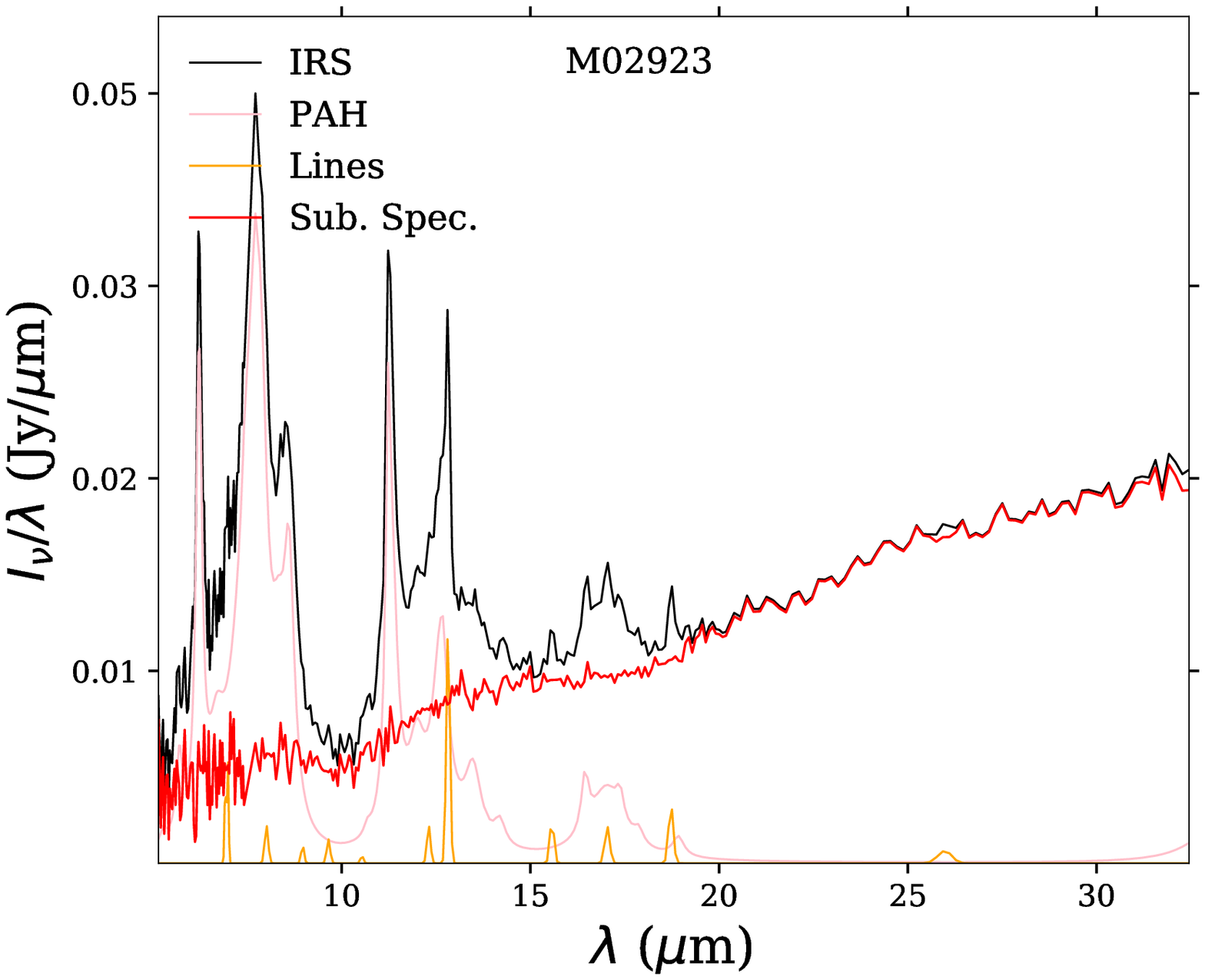}
\end{minipage} \hfill
\begin{minipage}[b]{0.325\linewidth}
\includegraphics[width=\textwidth]{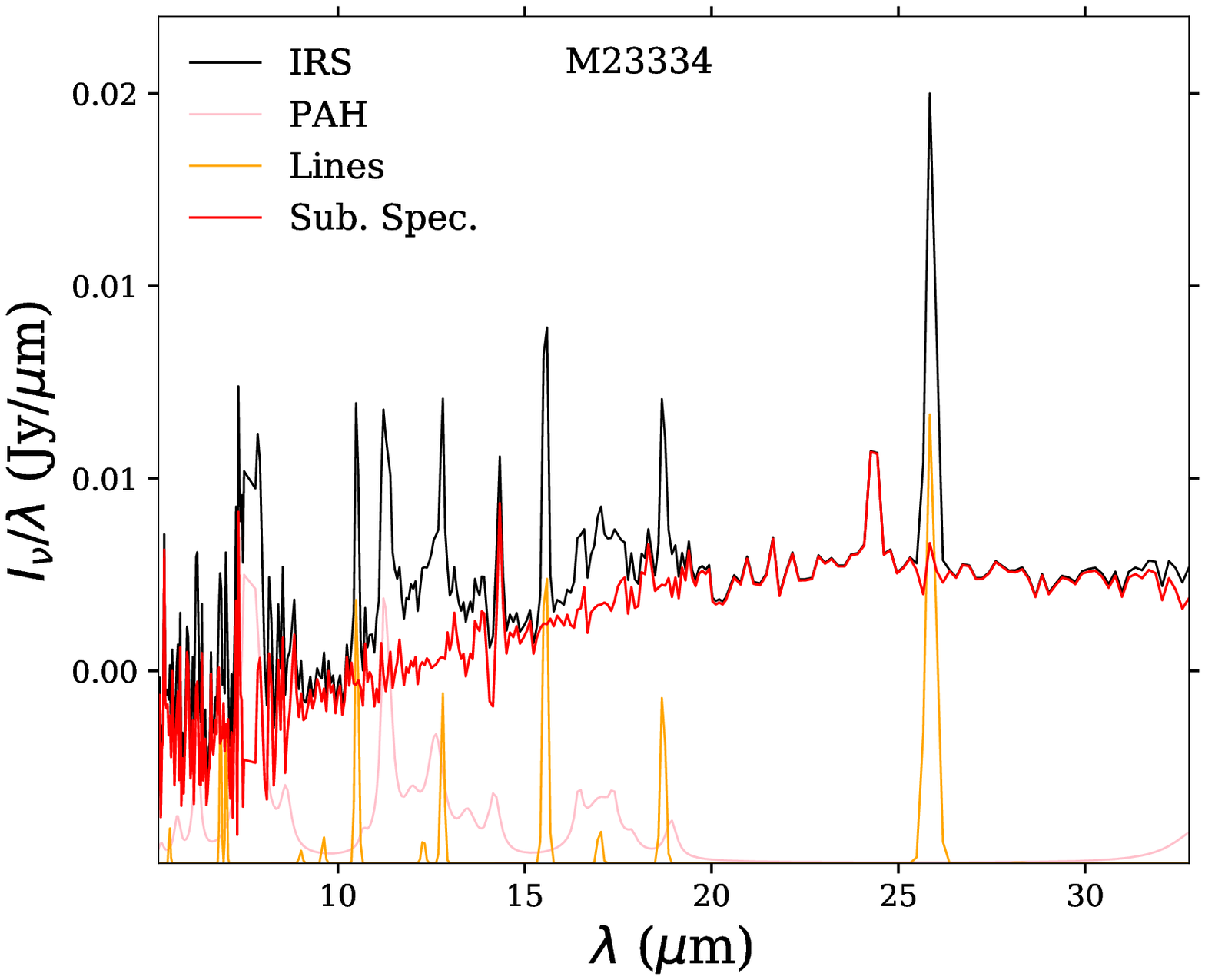}
\end{minipage} \hfill
\begin{minipage}[b]{0.325\linewidth}
\includegraphics[width=\textwidth]{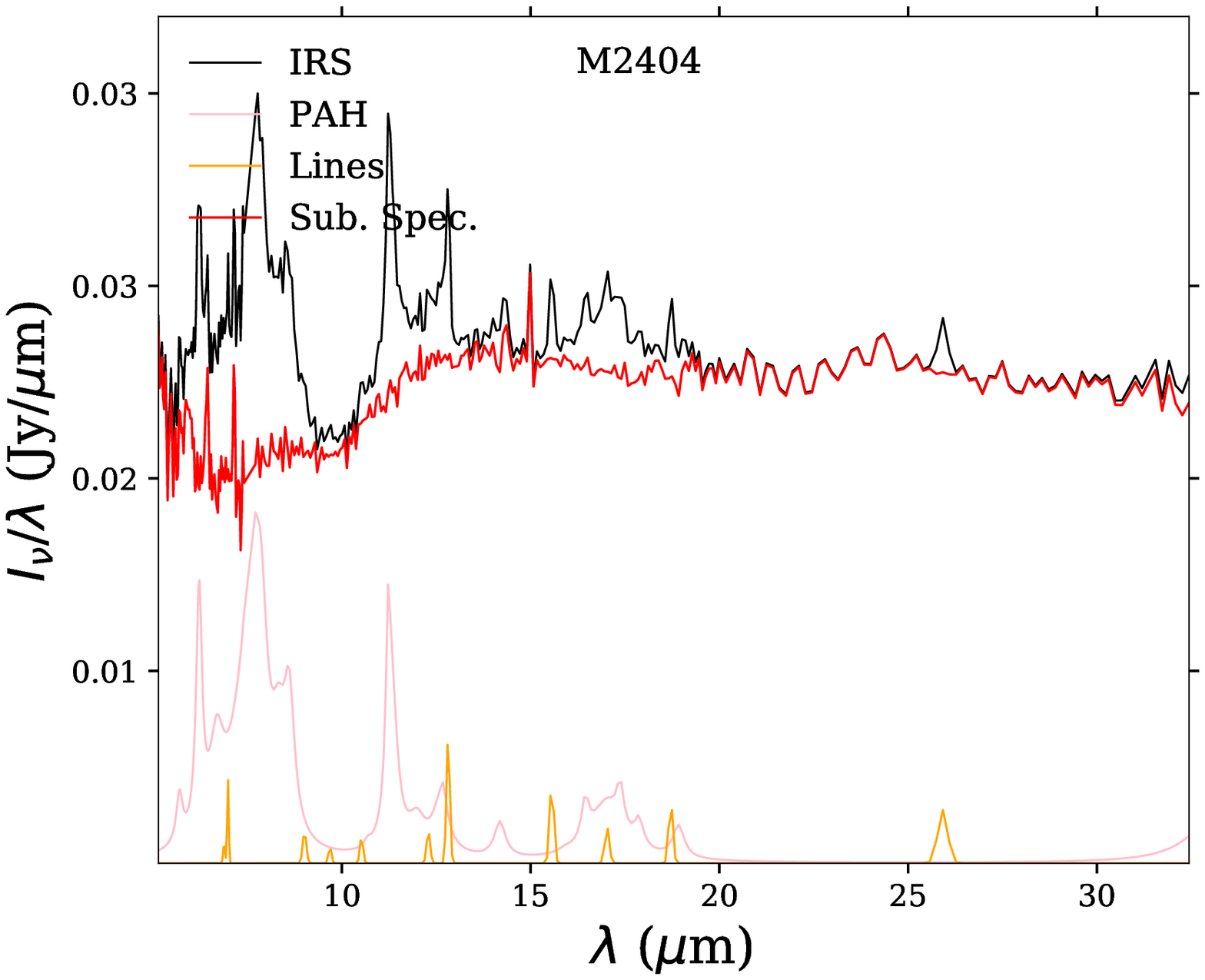}
\end{minipage} \hfill
\begin{minipage}[b]{0.325\linewidth}
\includegraphics[width=\textwidth]{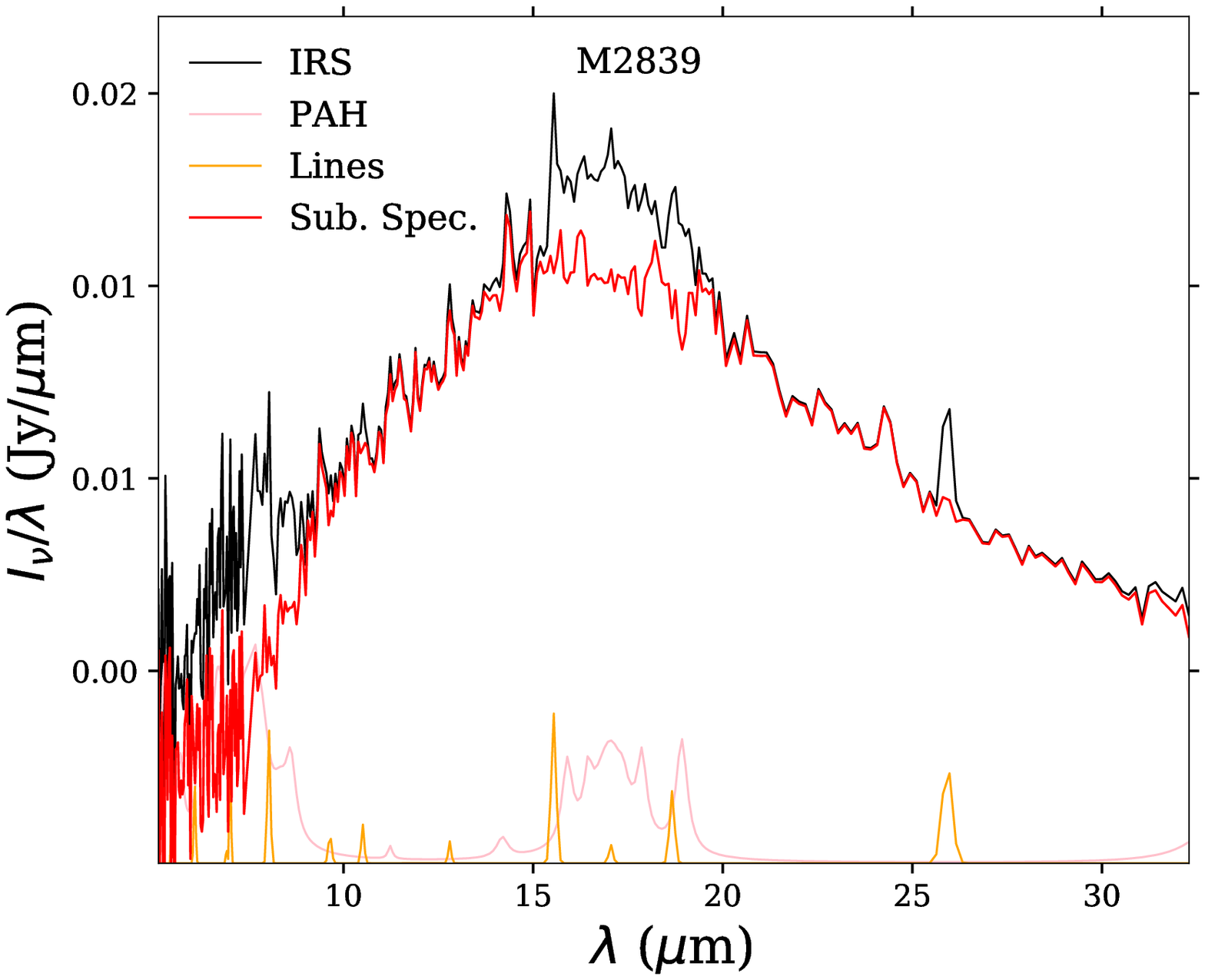}
\end{minipage} \hfill
\begin{minipage}[b]{0.325\linewidth}
\includegraphics[width=\textwidth]{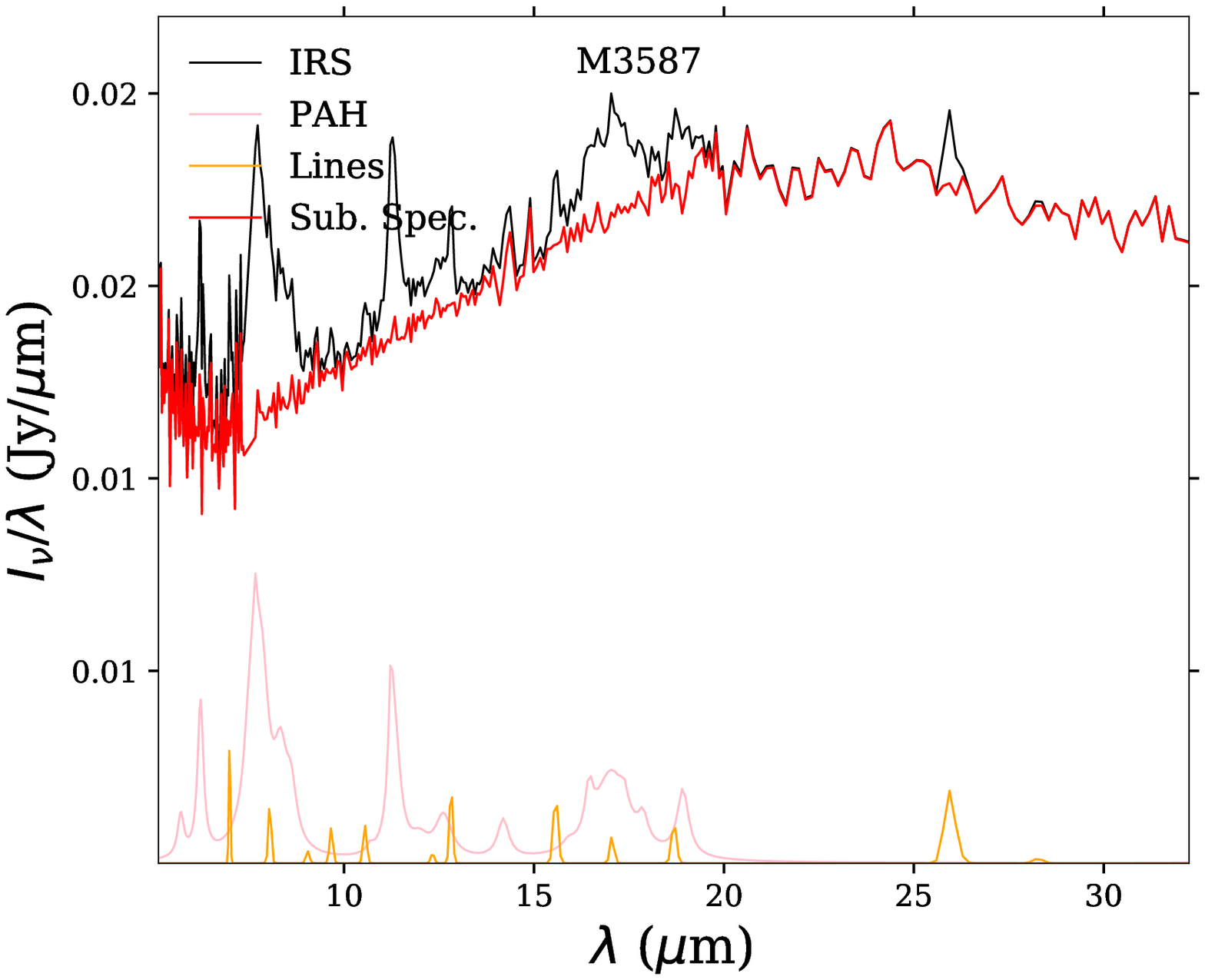}
\end{minipage} \hfill
\begin{minipage}[b]{0.325\linewidth}
\includegraphics[width=\textwidth]{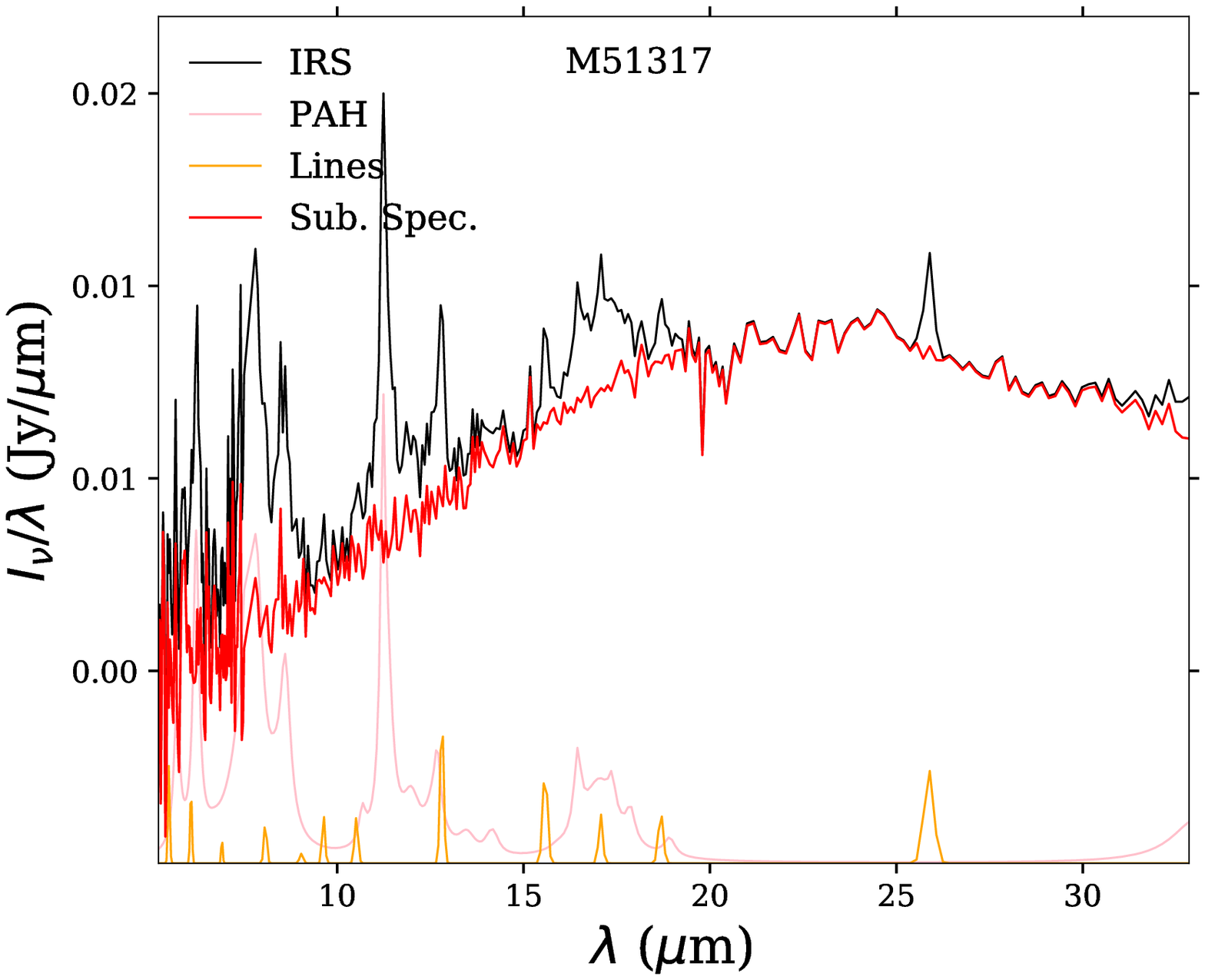}
\end{minipage} \hfill
\begin{minipage}[b]{0.325\linewidth}
\includegraphics[width=\textwidth]{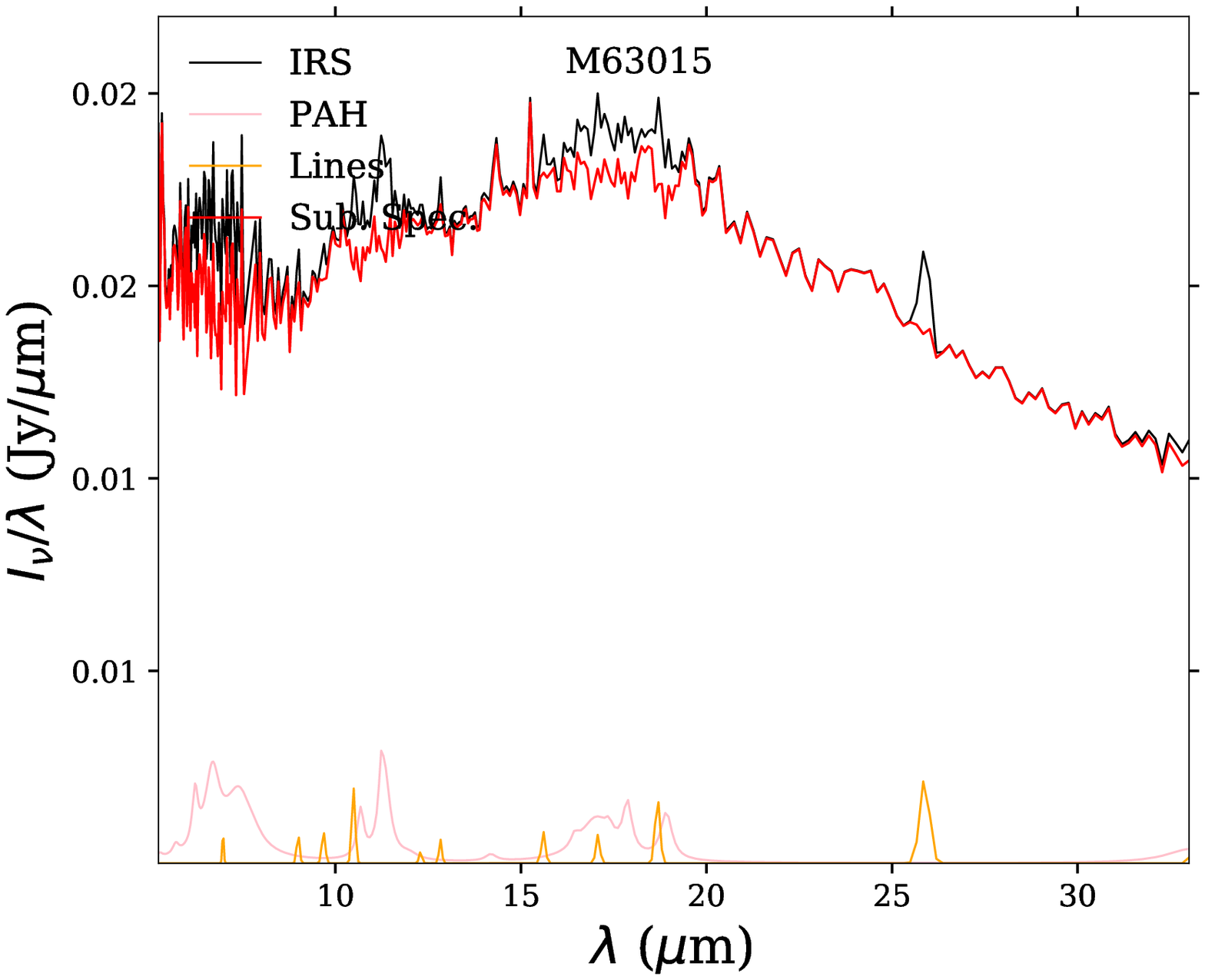}
\end{minipage} \hfill
\begin{minipage}[b]{0.325\linewidth}
\includegraphics[width=\textwidth]{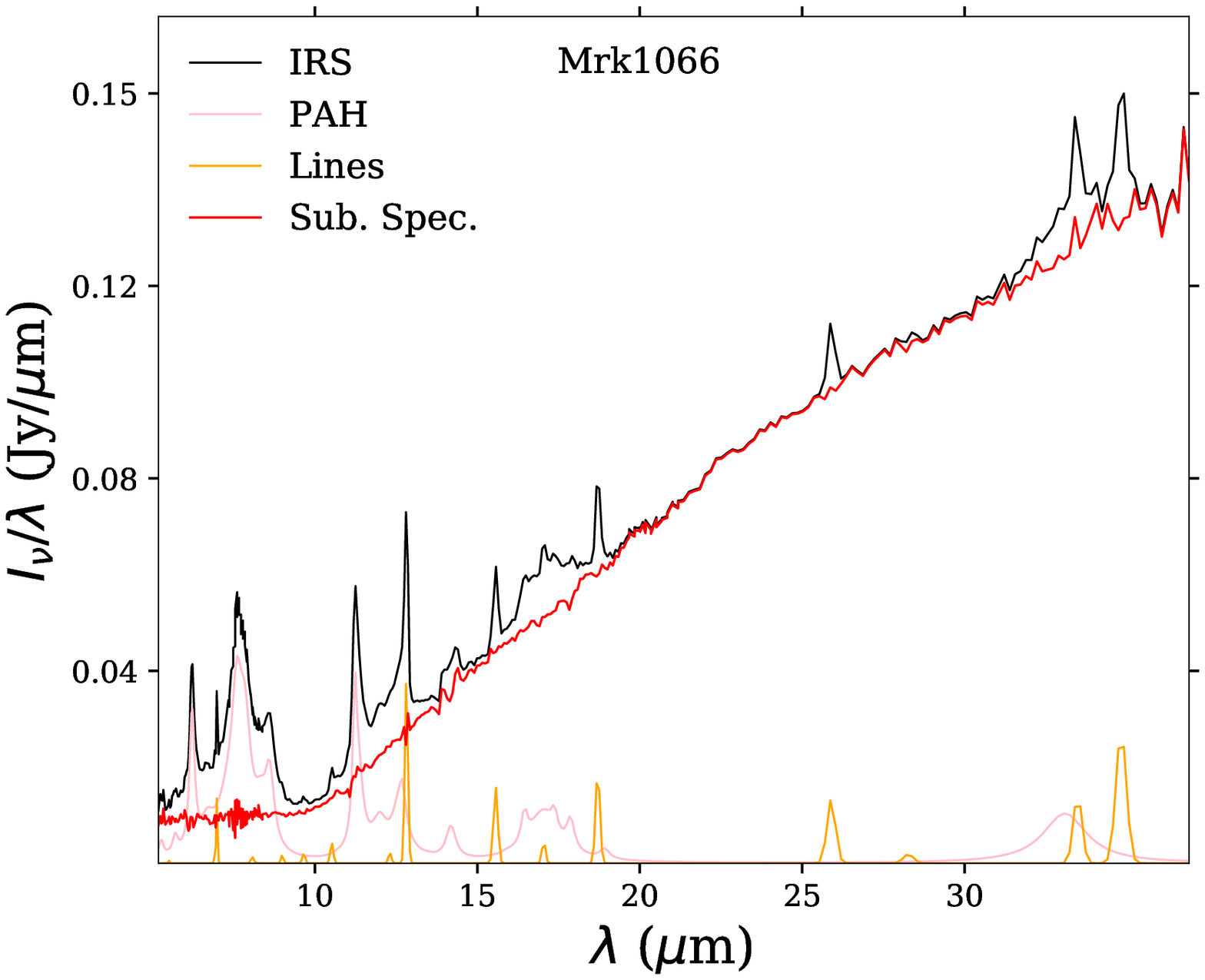}
\end{minipage} \hfill
\begin{minipage}[b]{0.325\linewidth}
\includegraphics[width=\textwidth]{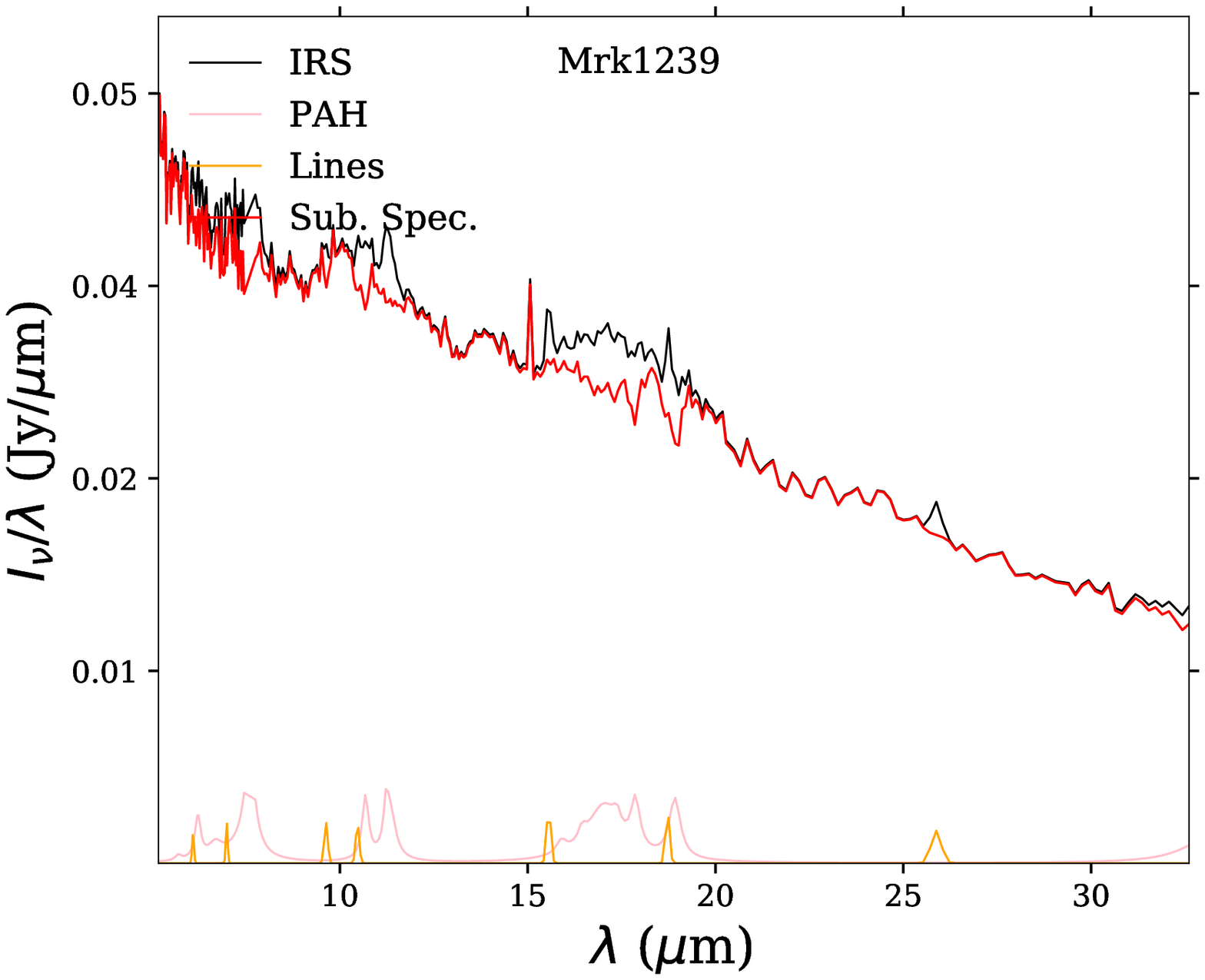}
\end{minipage} \hfill
\begin{minipage}[b]{0.325\linewidth}
\includegraphics[width=\textwidth]{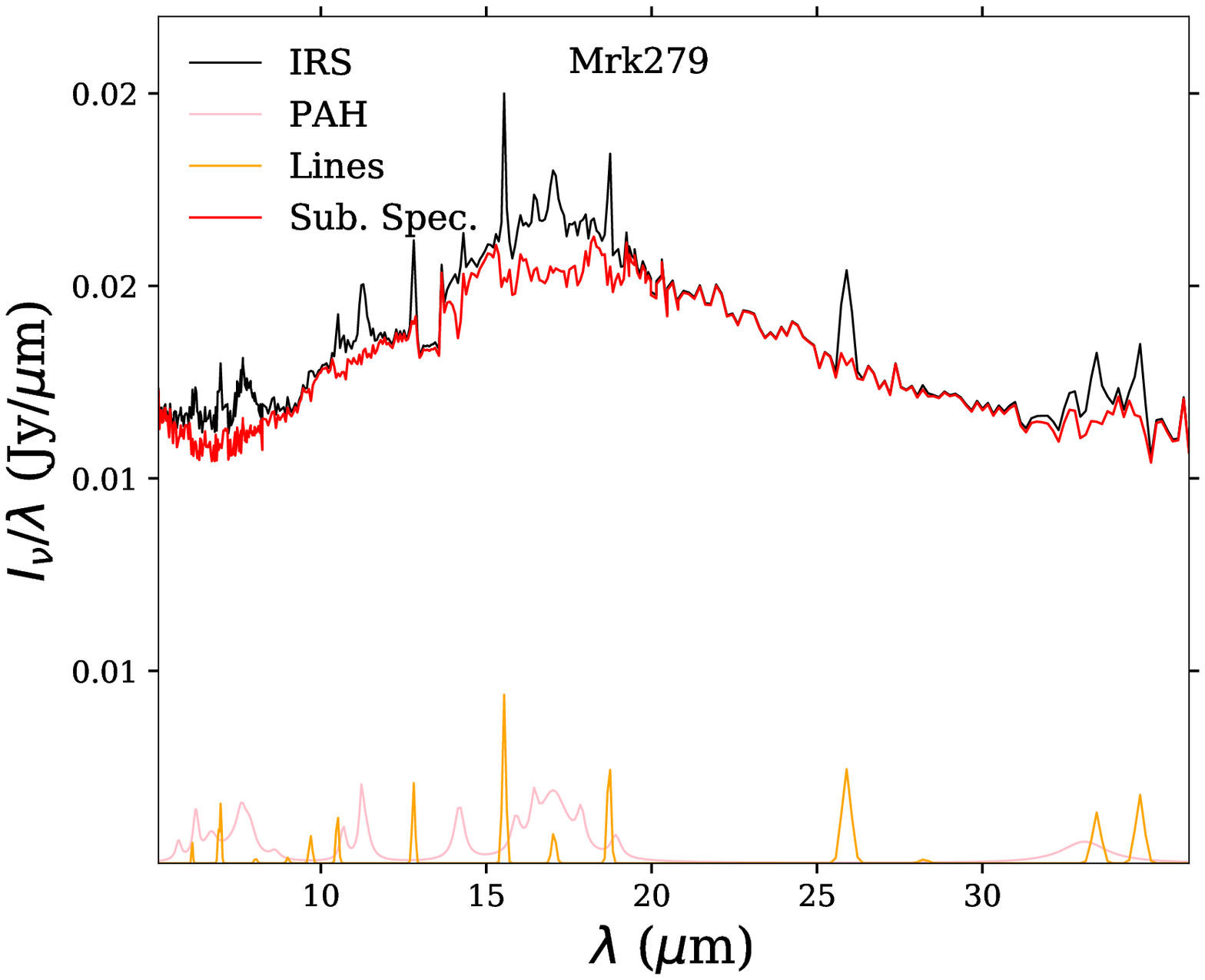}
\end{minipage} \hfill
\begin{minipage}[b]{0.325\linewidth}
\includegraphics[width=\textwidth]{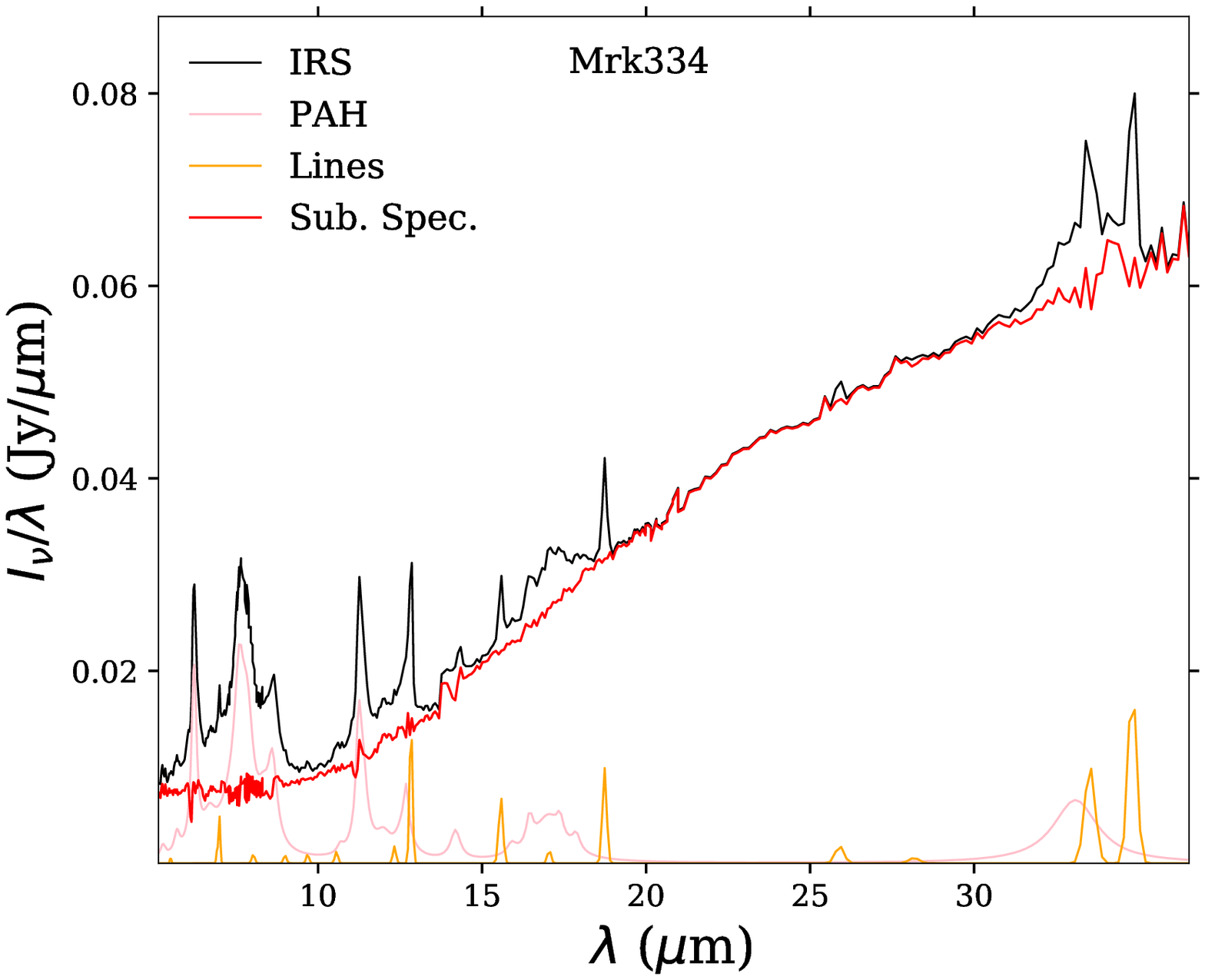}
\end{minipage} \hfill
\begin{minipage}[b]{0.325\linewidth}
\includegraphics[width=\textwidth]{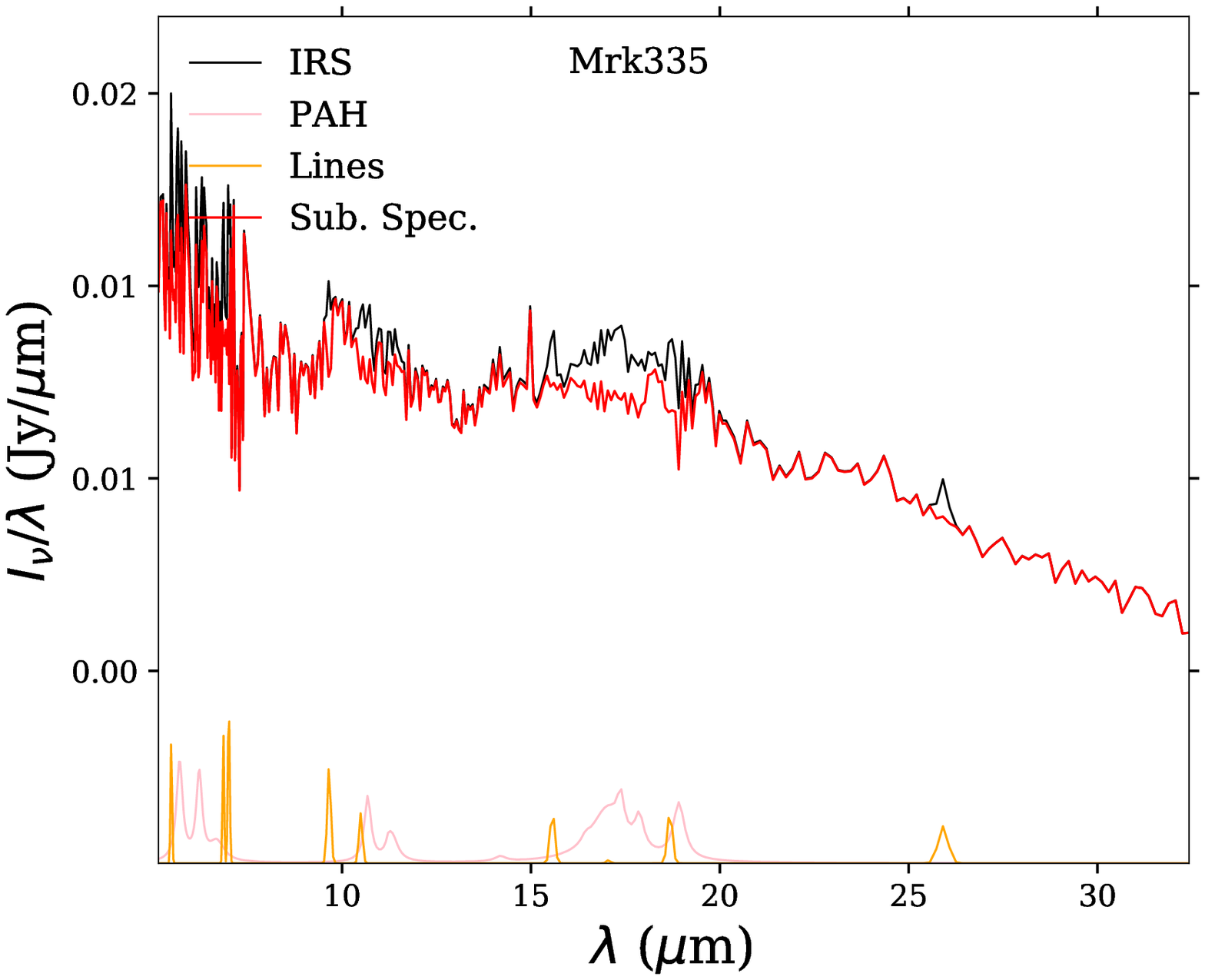}
\end{minipage} \hfill
\begin{minipage}[b]{0.325\linewidth}
\includegraphics[width=\textwidth]{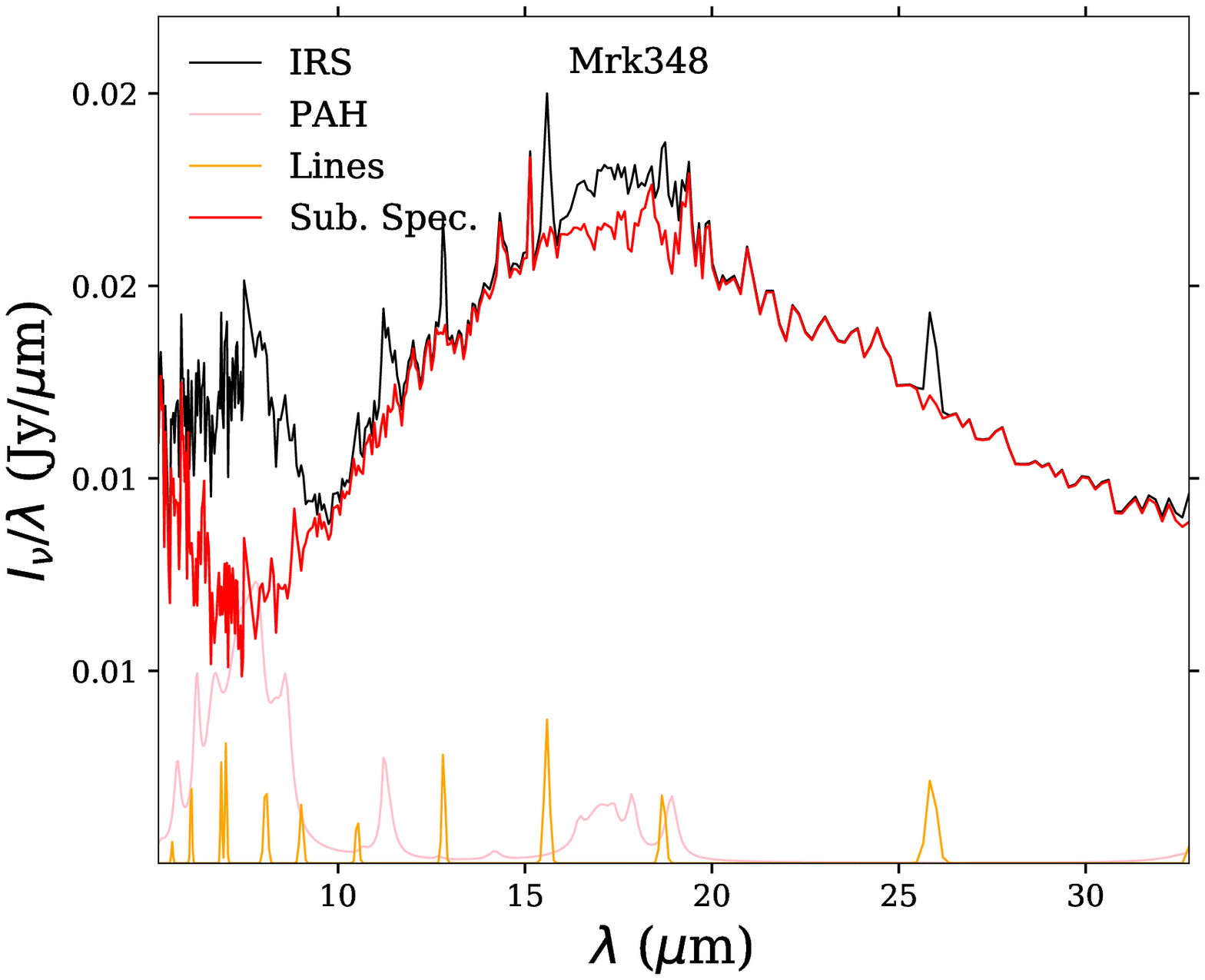}
\end{minipage} \hfill
\begin{minipage}[b]{0.325\linewidth}
\includegraphics[width=\textwidth]{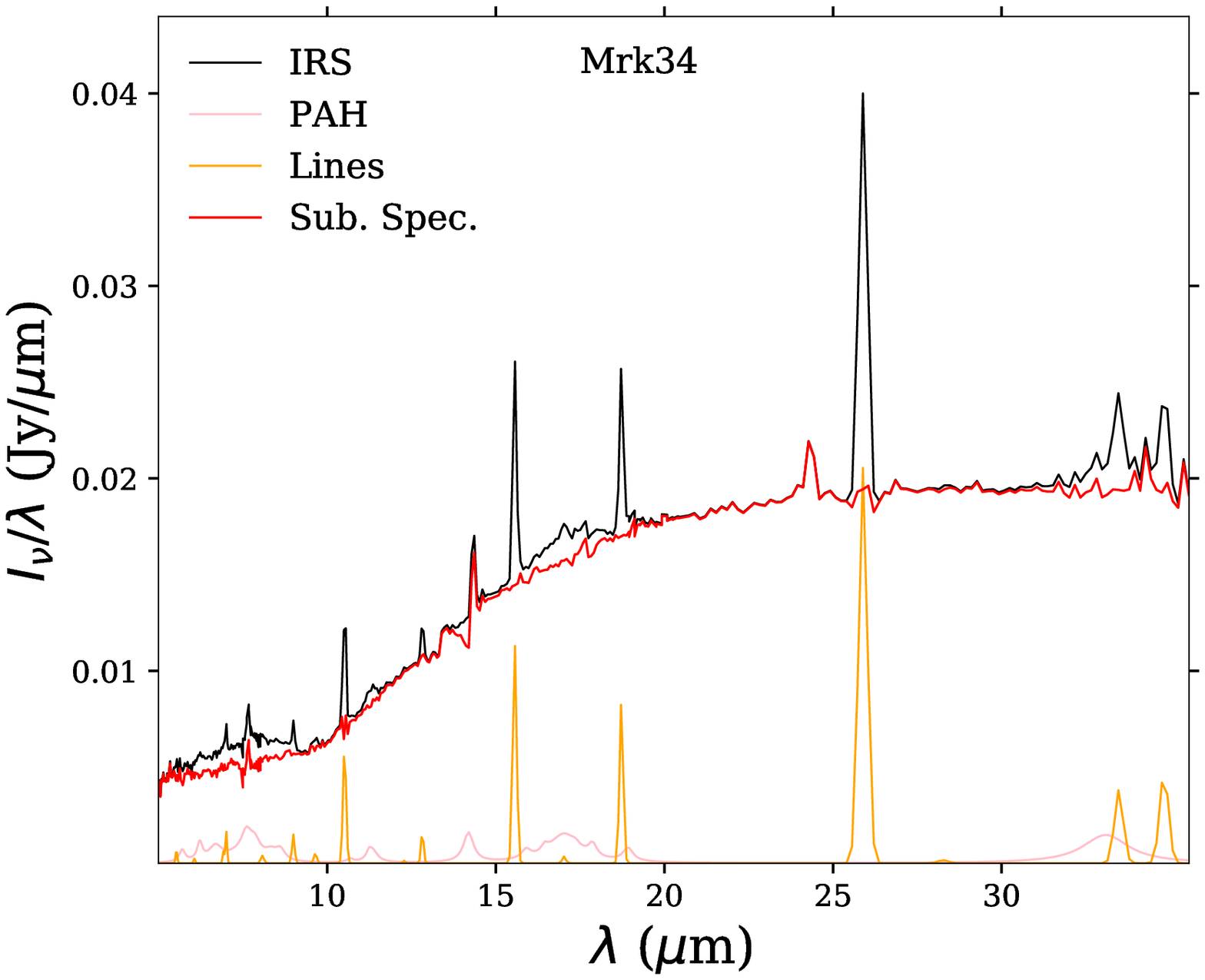}
\end{minipage} \hfill
\caption{continued from previous page.}
\setcounter{figure}{0}
\end{figure}

\begin{figure}

\begin{minipage}[b]{0.325\linewidth}
\includegraphics[width=\textwidth]{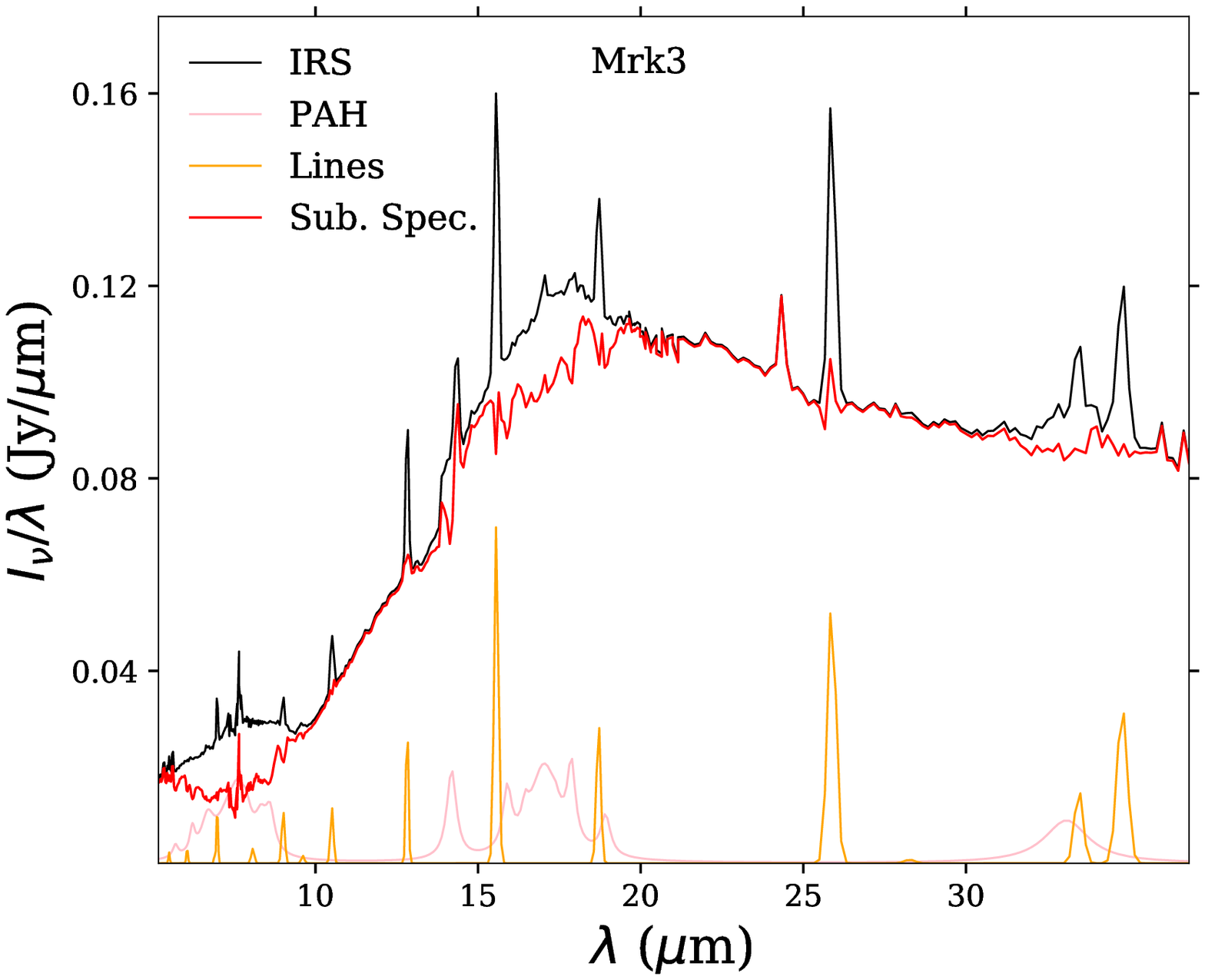}
\end{minipage} \hfill
\begin{minipage}[b]{0.325\linewidth}
\includegraphics[width=\textwidth]{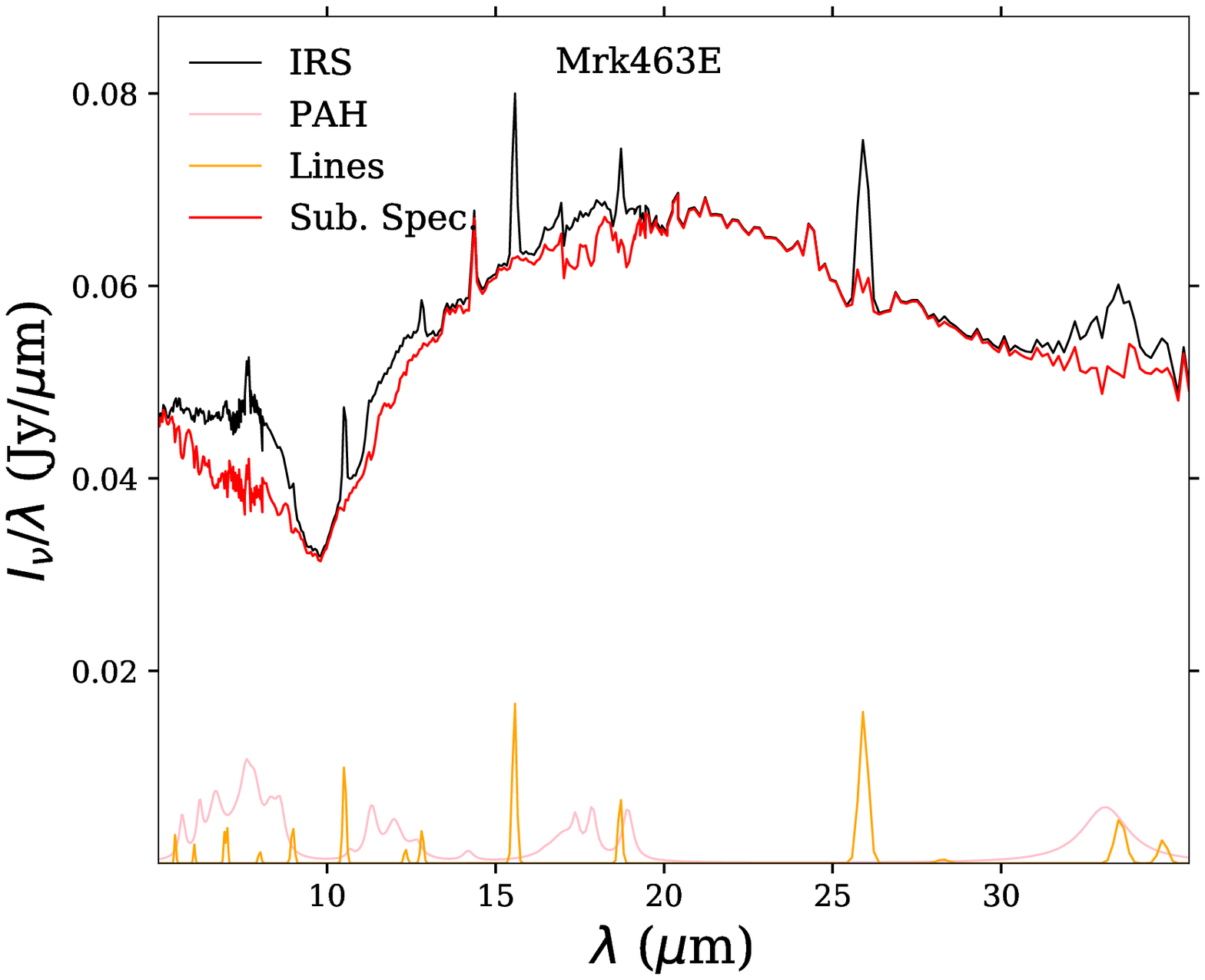}
\end{minipage} \hfill
\begin{minipage}[b]{0.325\linewidth}
\includegraphics[width=\textwidth]{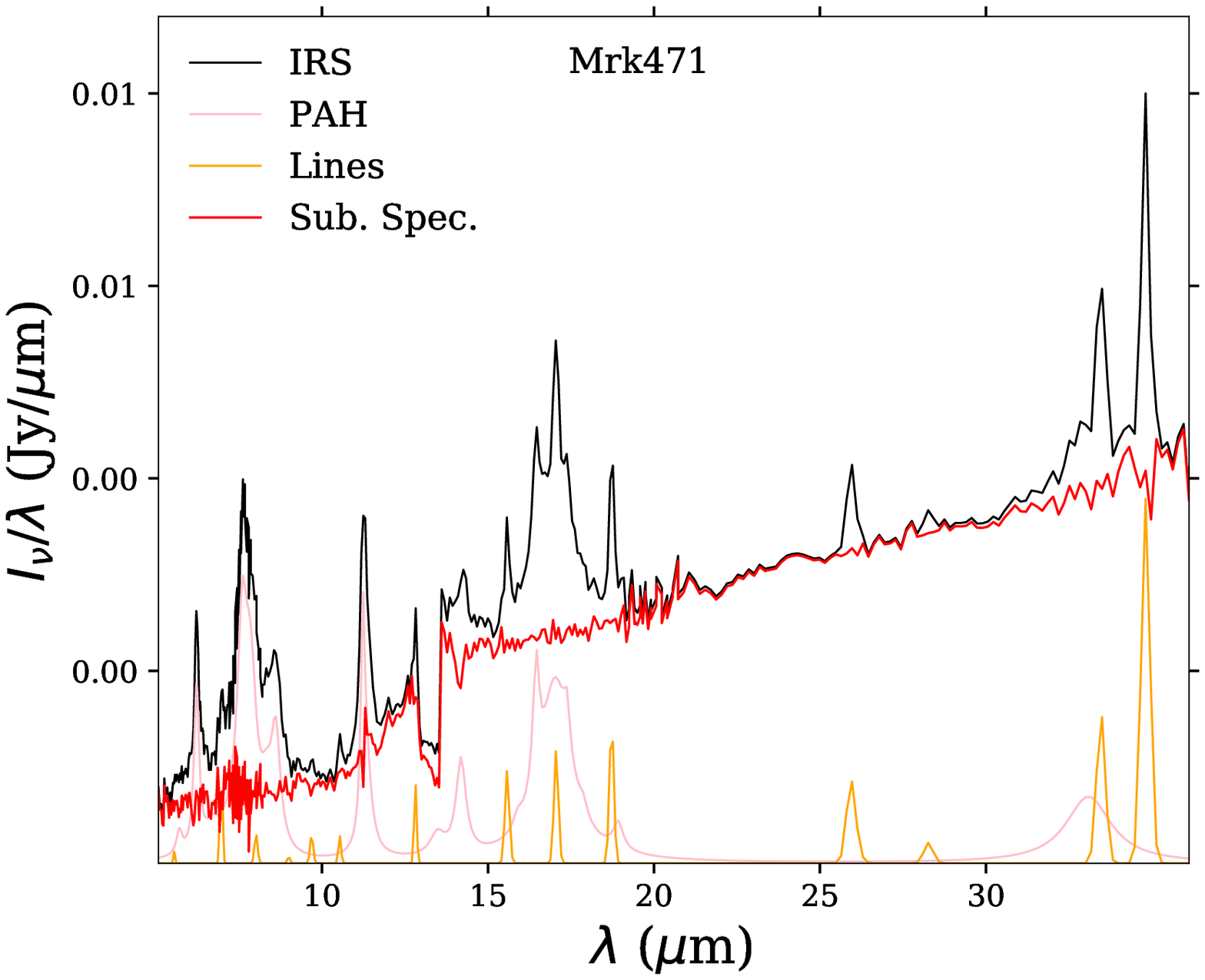}
\end{minipage} \hfill
\begin{minipage}[b]{0.325\linewidth}
\includegraphics[width=\textwidth]{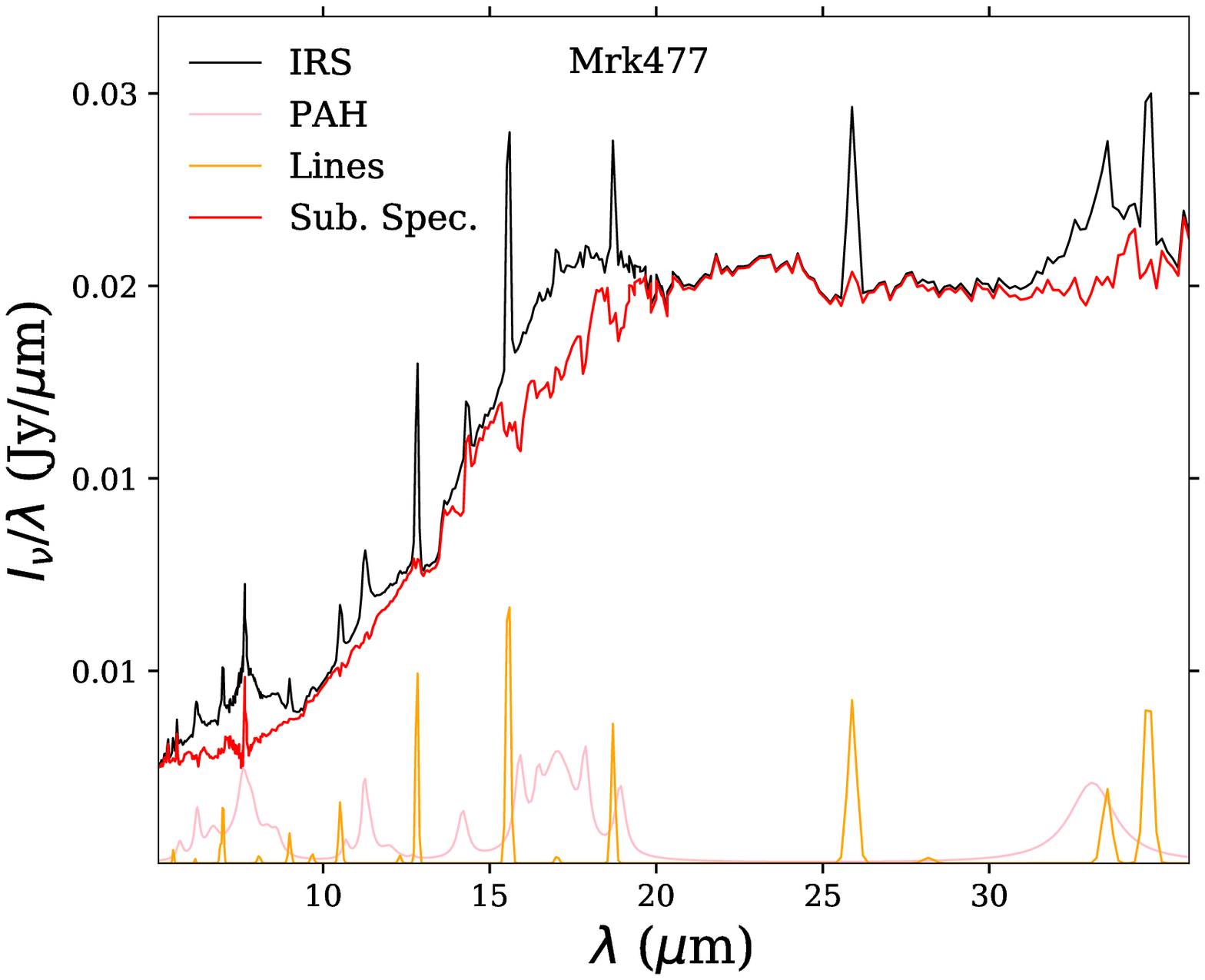}
\end{minipage} \hfill
\begin{minipage}[b]{0.325\linewidth}
\includegraphics[width=\textwidth]{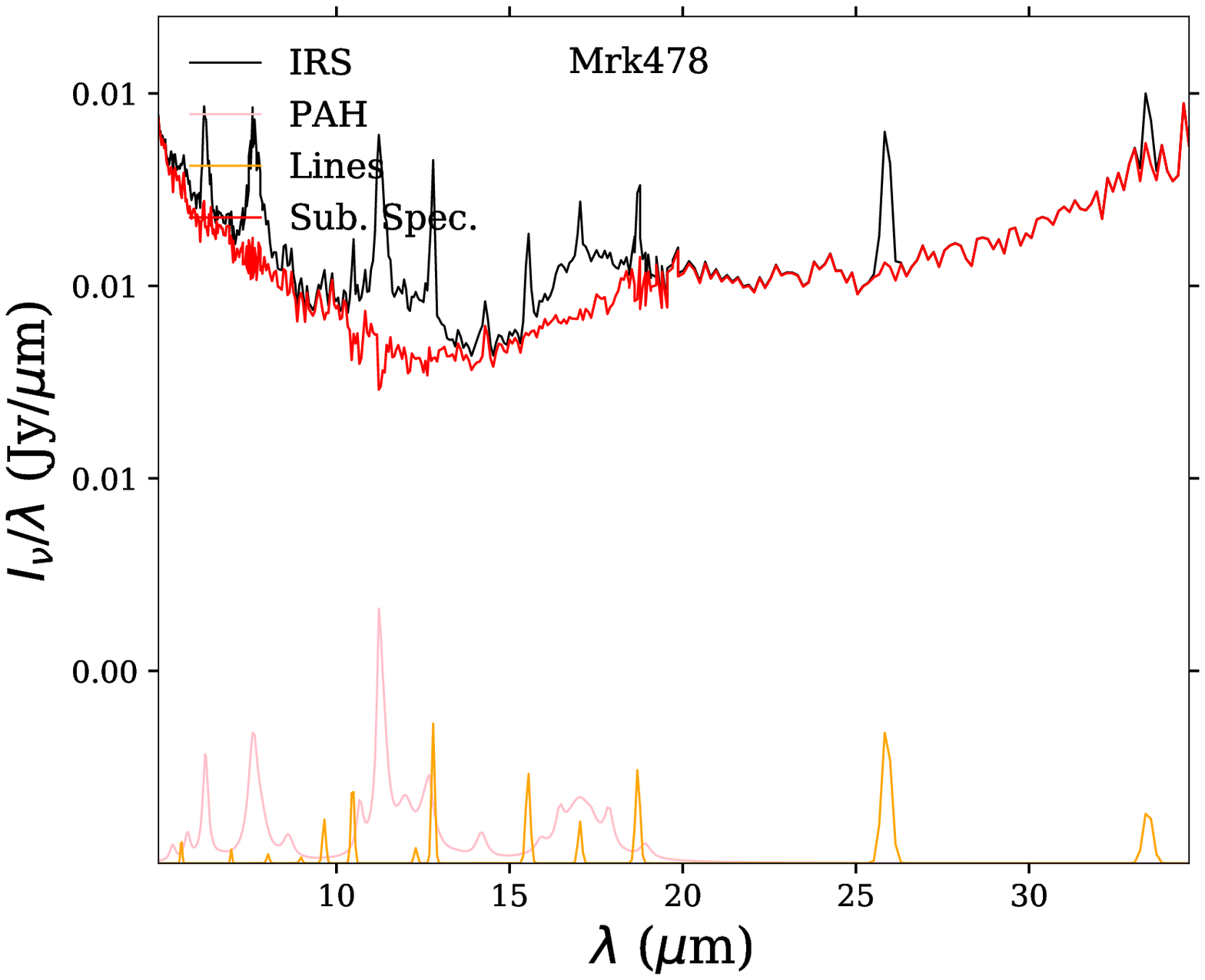}
\end{minipage} \hfill
\begin{minipage}[b]{0.325\linewidth}
\includegraphics[width=\textwidth]{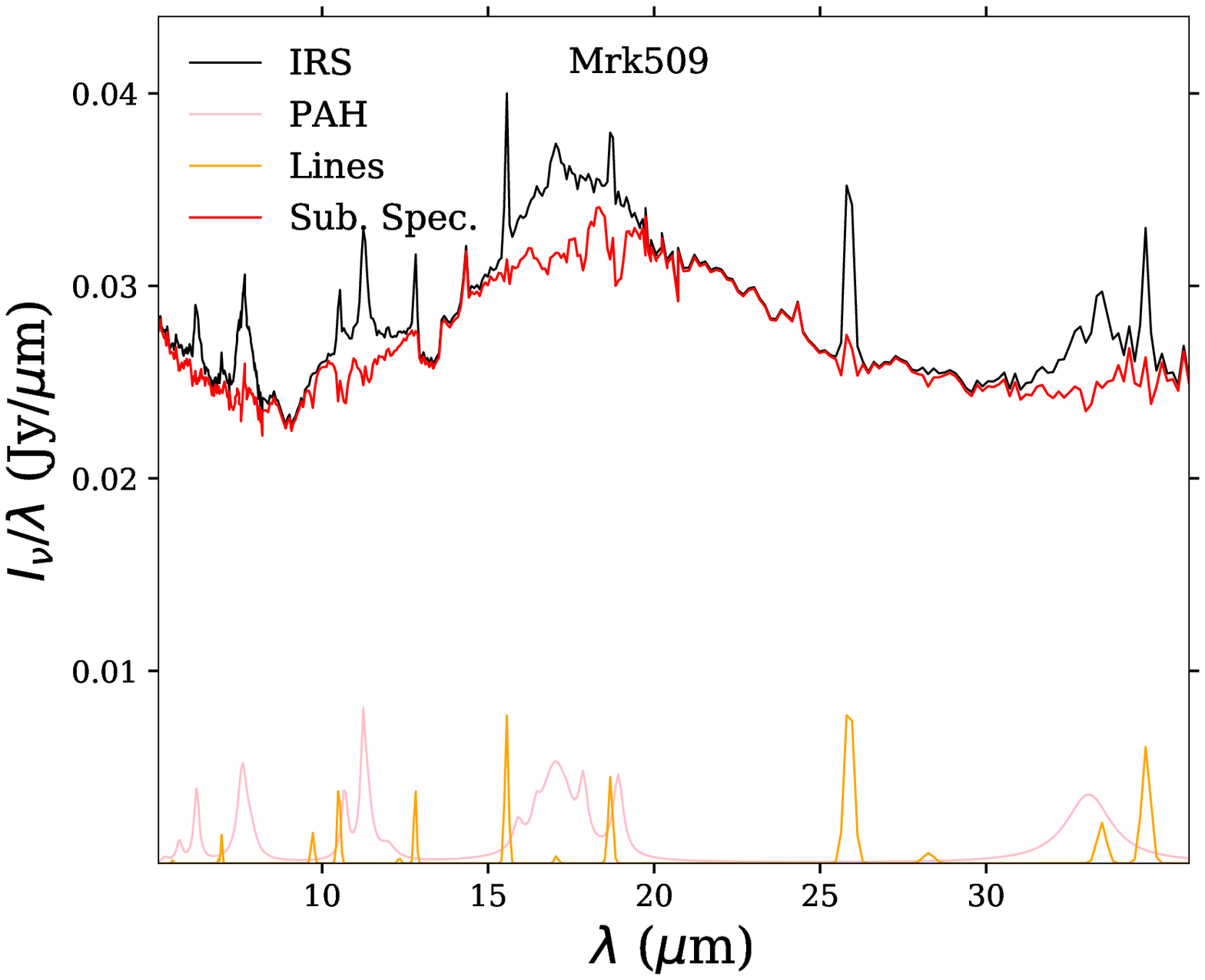}
\end{minipage} \hfill
\begin{minipage}[b]{0.325\linewidth}
\includegraphics[width=\textwidth]{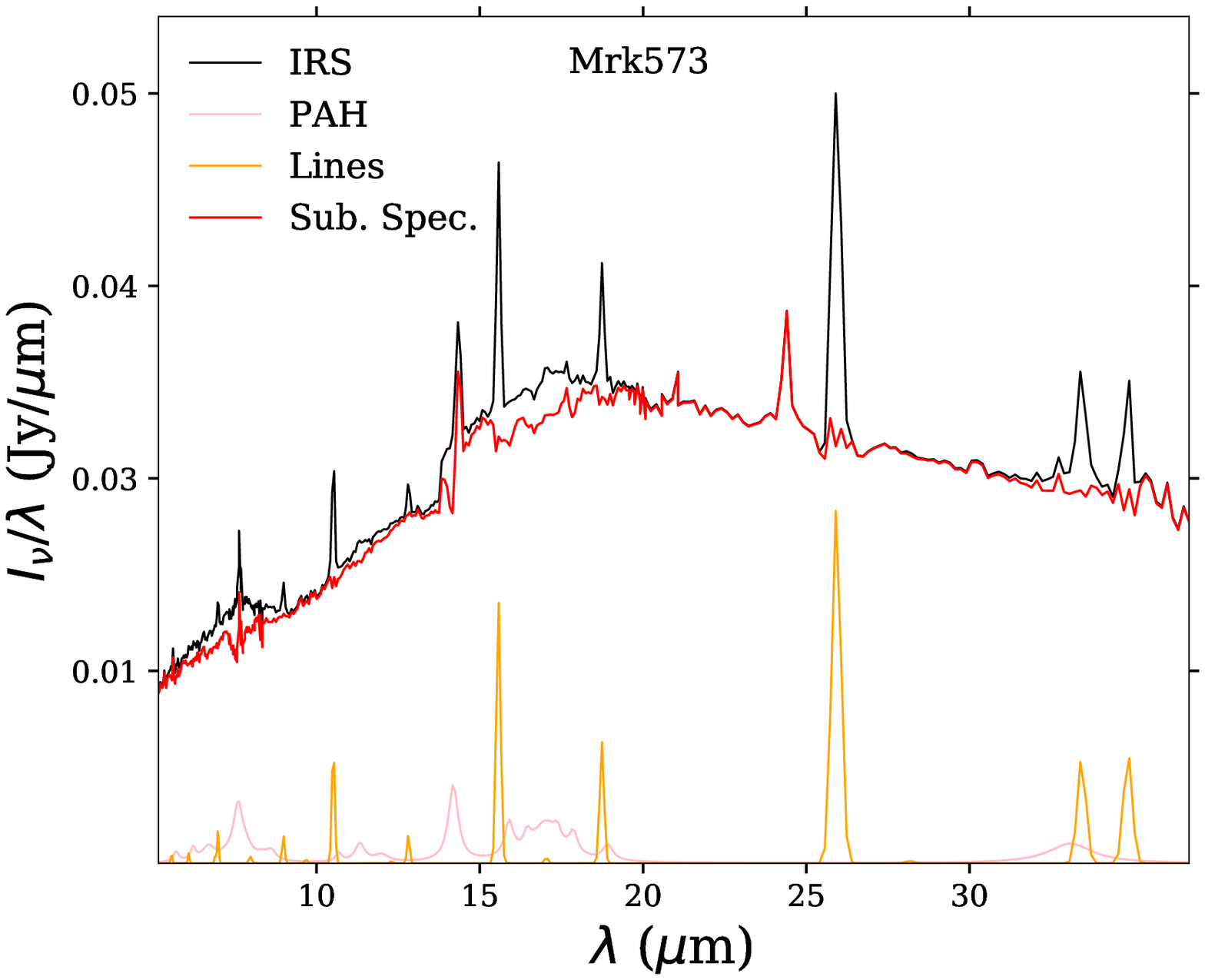}
\end{minipage} \hfill
\begin{minipage}[b]{0.325\linewidth}
\includegraphics[width=\textwidth]{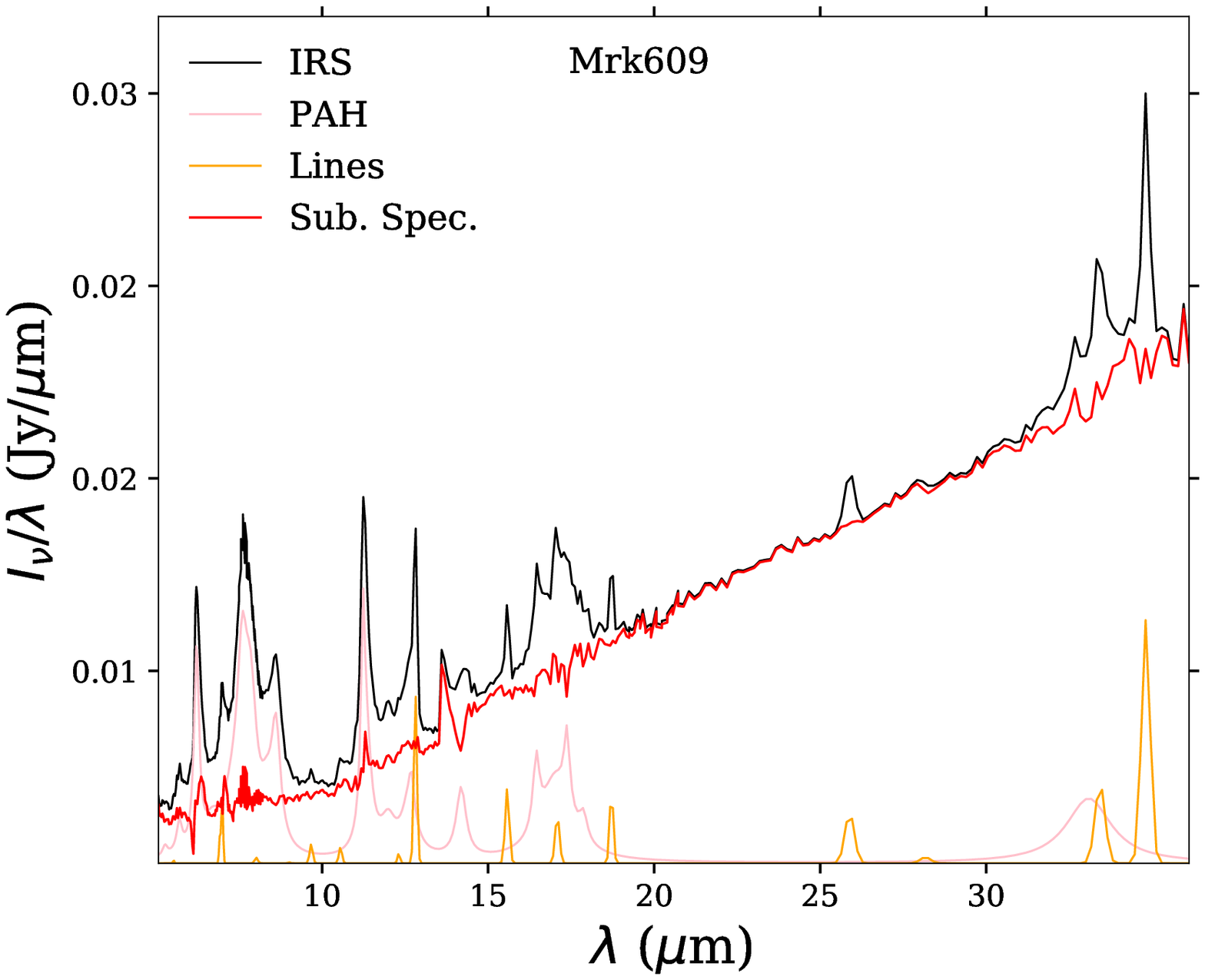}
\end{minipage} \hfill
\begin{minipage}[b]{0.325\linewidth}
\includegraphics[width=\textwidth]{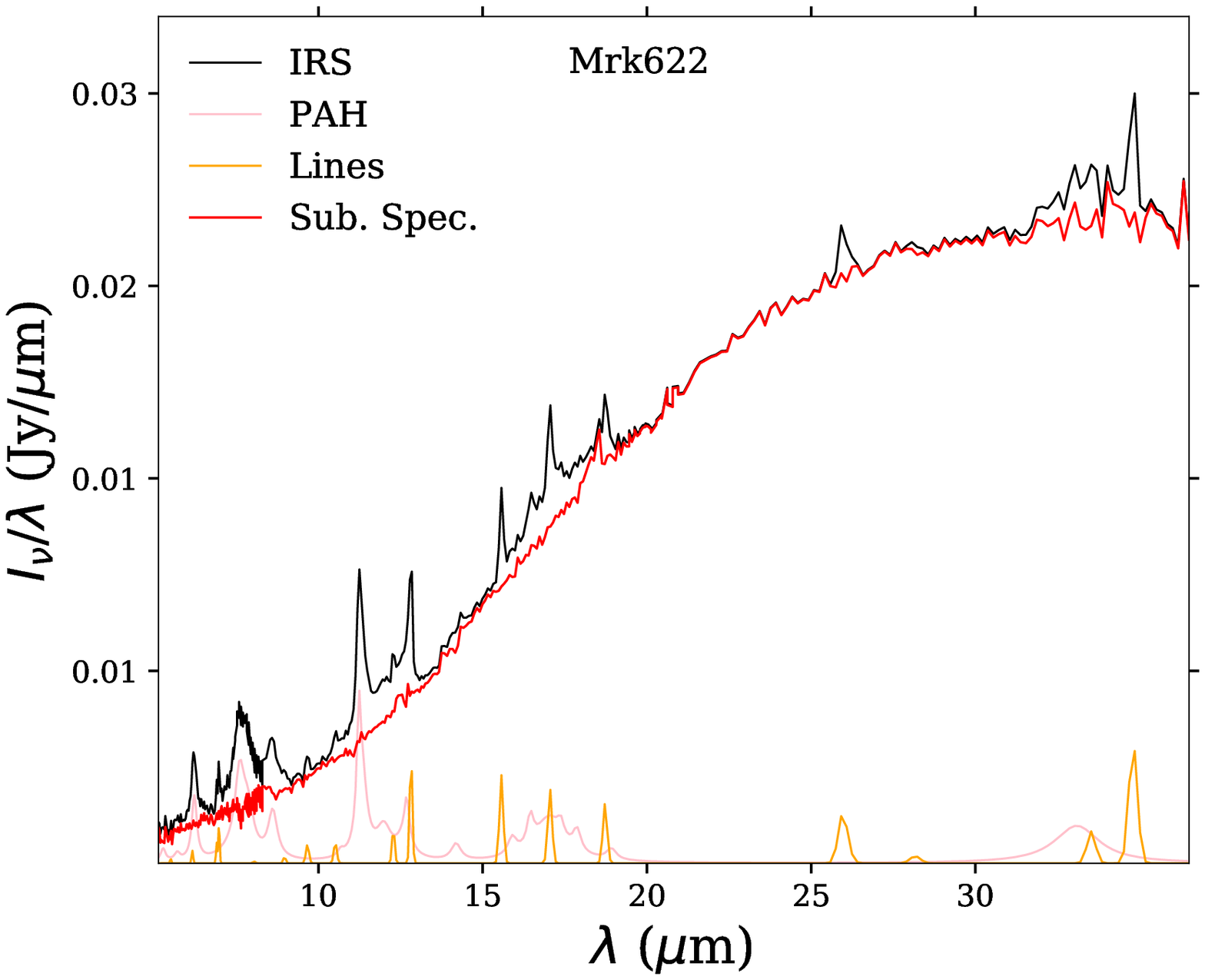}
\end{minipage} \hfill
\begin{minipage}[b]{0.325\linewidth}
\includegraphics[width=\textwidth]{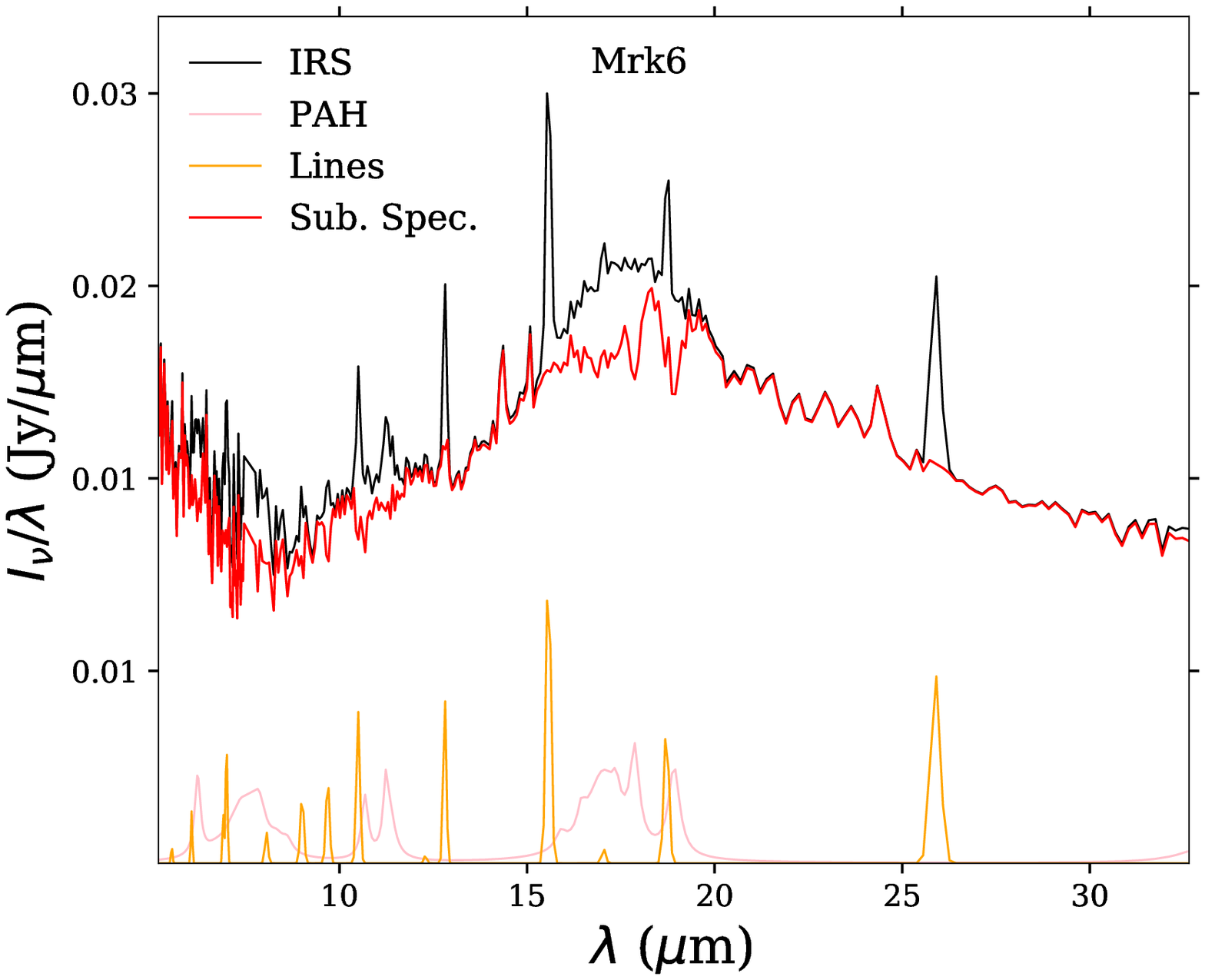}
\end{minipage} \hfill
\begin{minipage}[b]{0.325\linewidth}
\includegraphics[width=\textwidth]{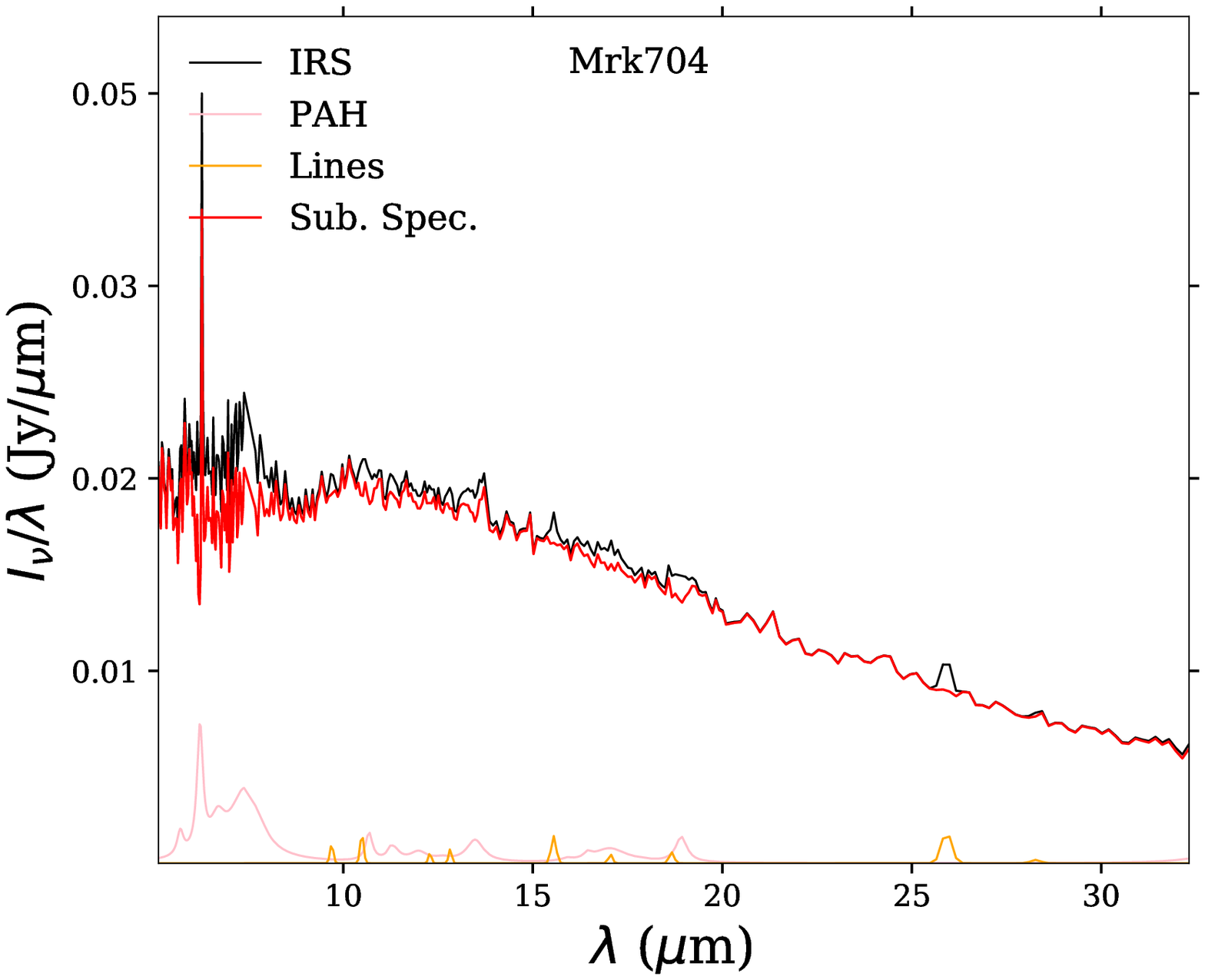}
\end{minipage} \hfill
\begin{minipage}[b]{0.325\linewidth}
\includegraphics[width=\textwidth]{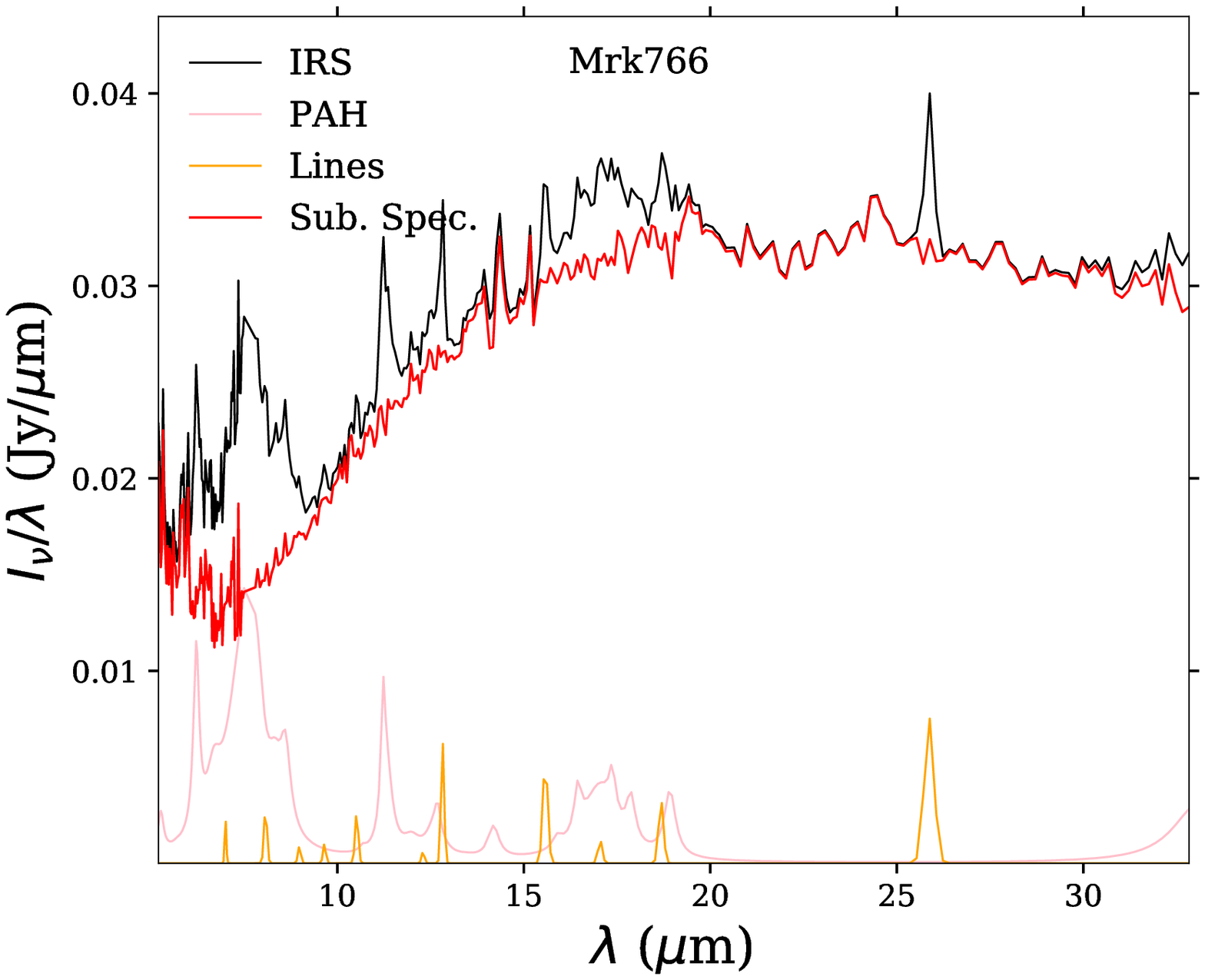}
\end{minipage} \hfill
\begin{minipage}[b]{0.325\linewidth}
\includegraphics[width=\textwidth]{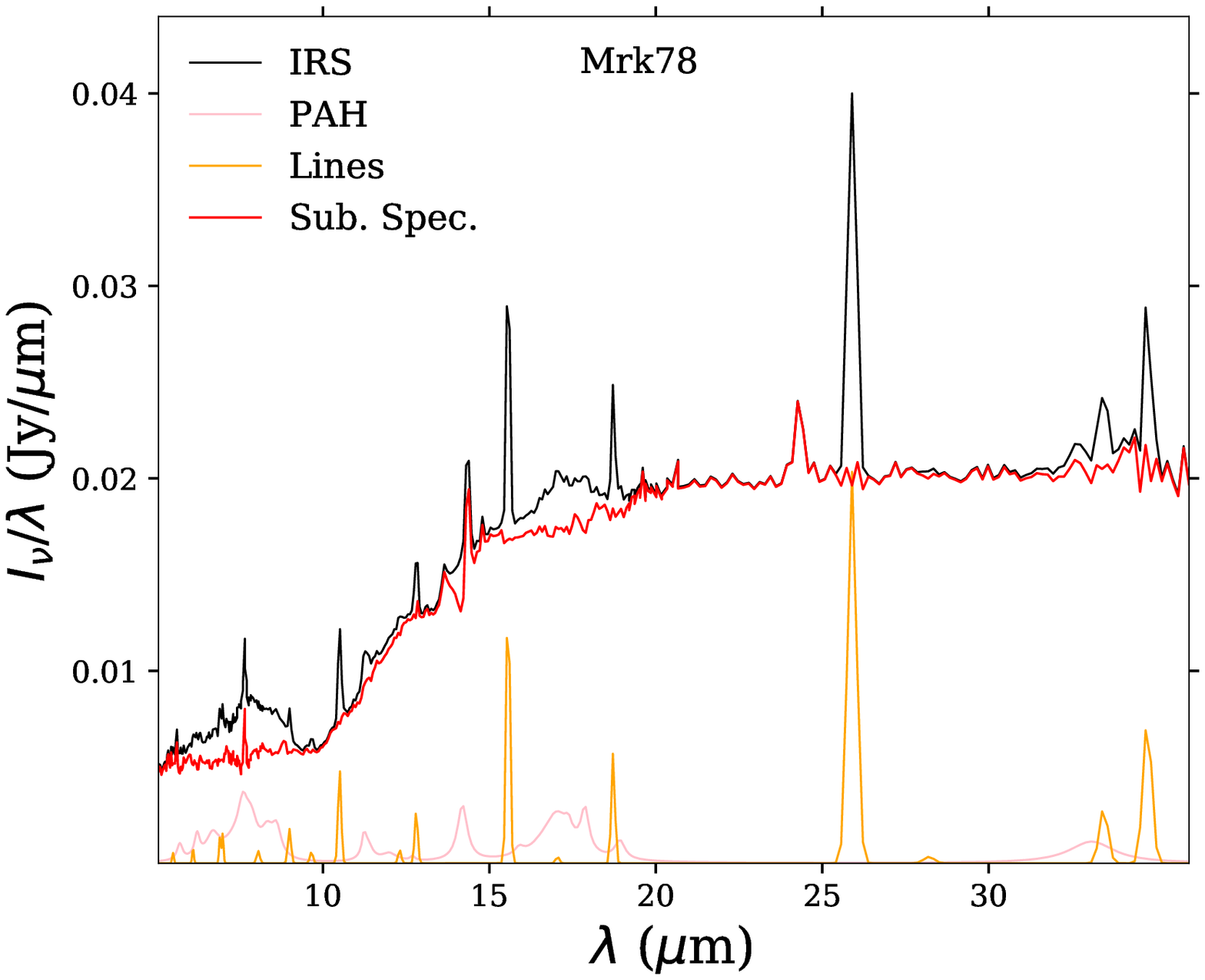}
\end{minipage} \hfill
\begin{minipage}[b]{0.325\linewidth}
\includegraphics[width=\textwidth]{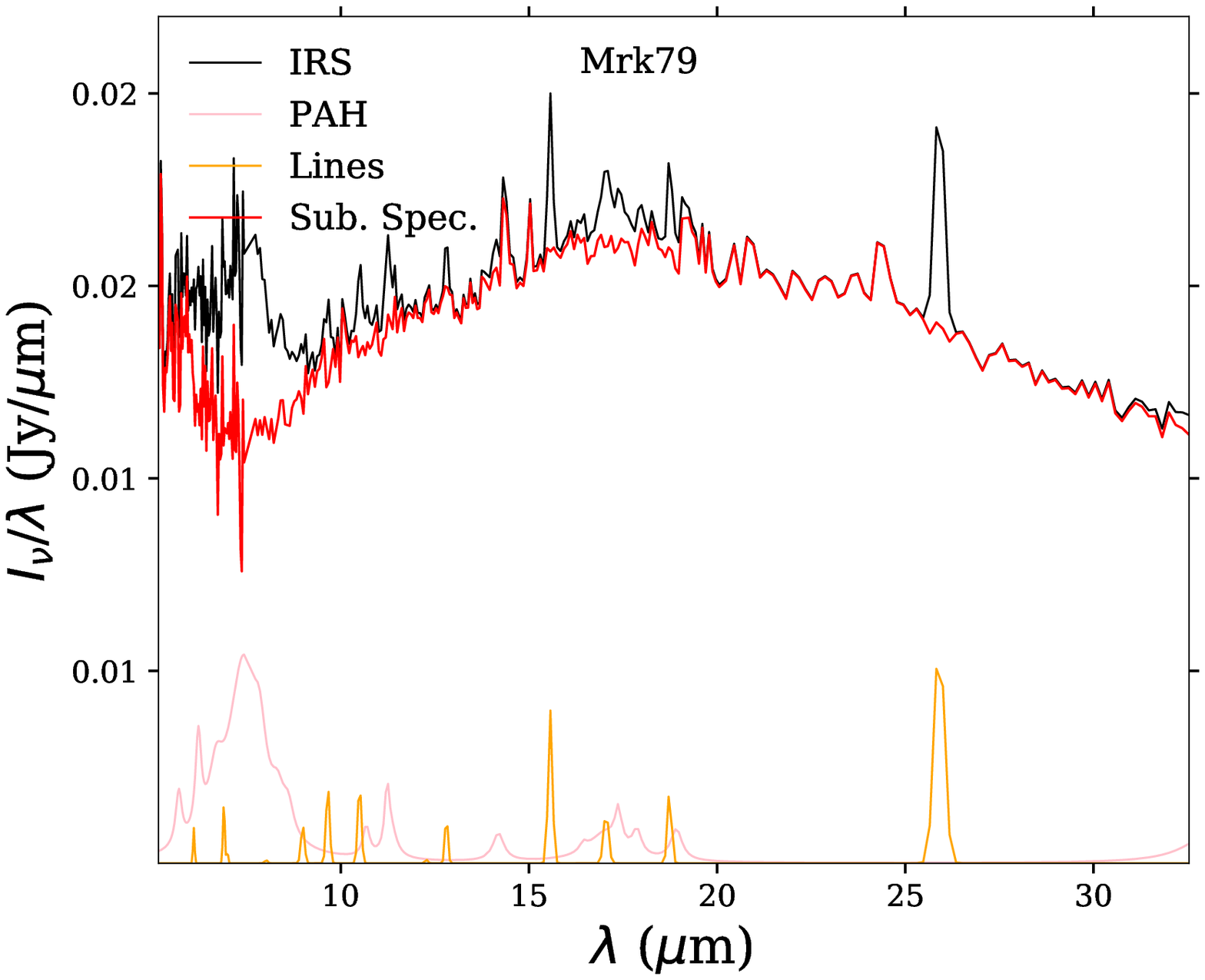}
\end{minipage} \hfill
\begin{minipage}[b]{0.325\linewidth}
\includegraphics[width=\textwidth]{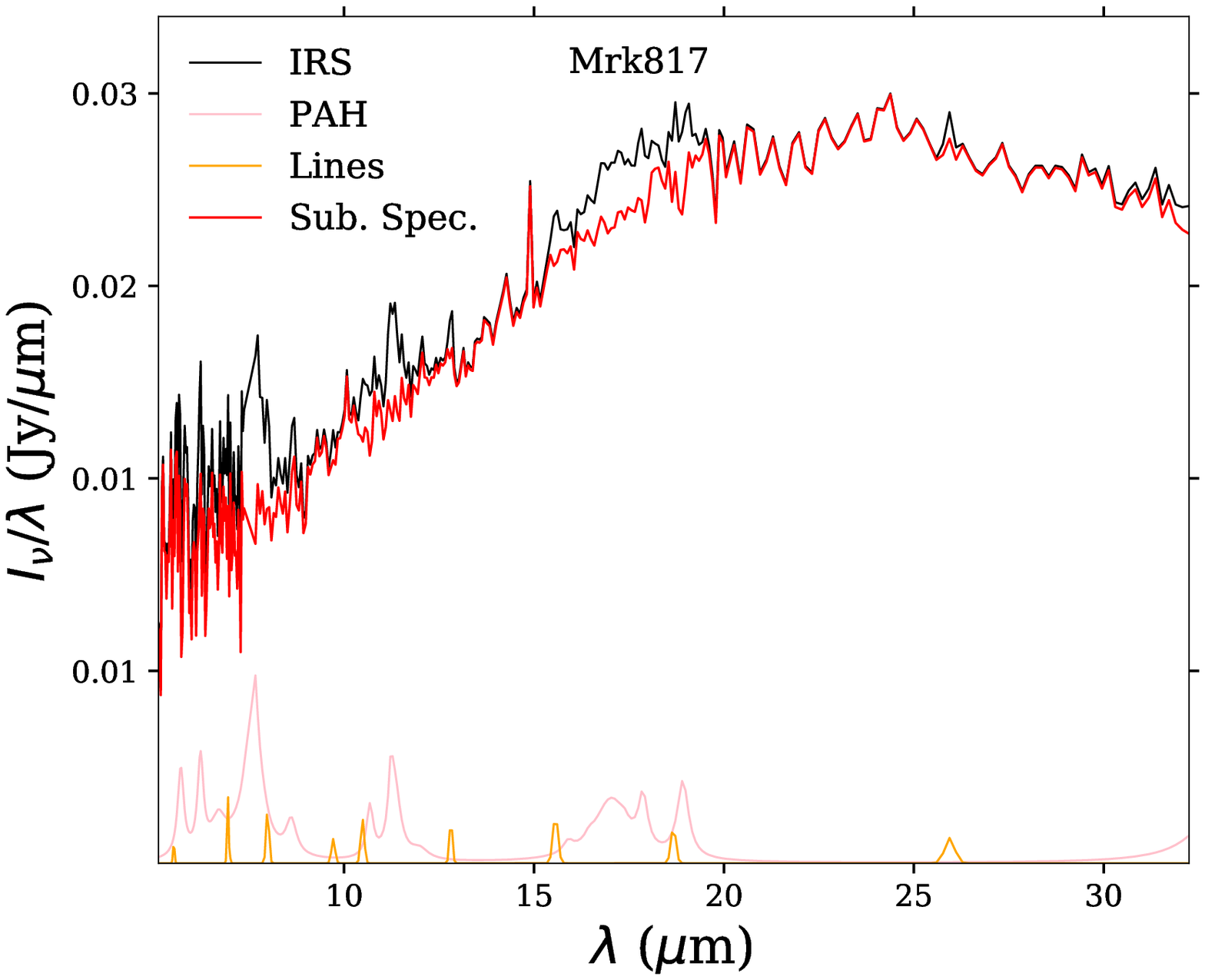}
\end{minipage} \hfill
\caption{continued from previous page.}
\setcounter{figure}{0}
\end{figure}

\begin{figure}

\begin{minipage}[b]{0.325\linewidth}
\includegraphics[width=\textwidth]{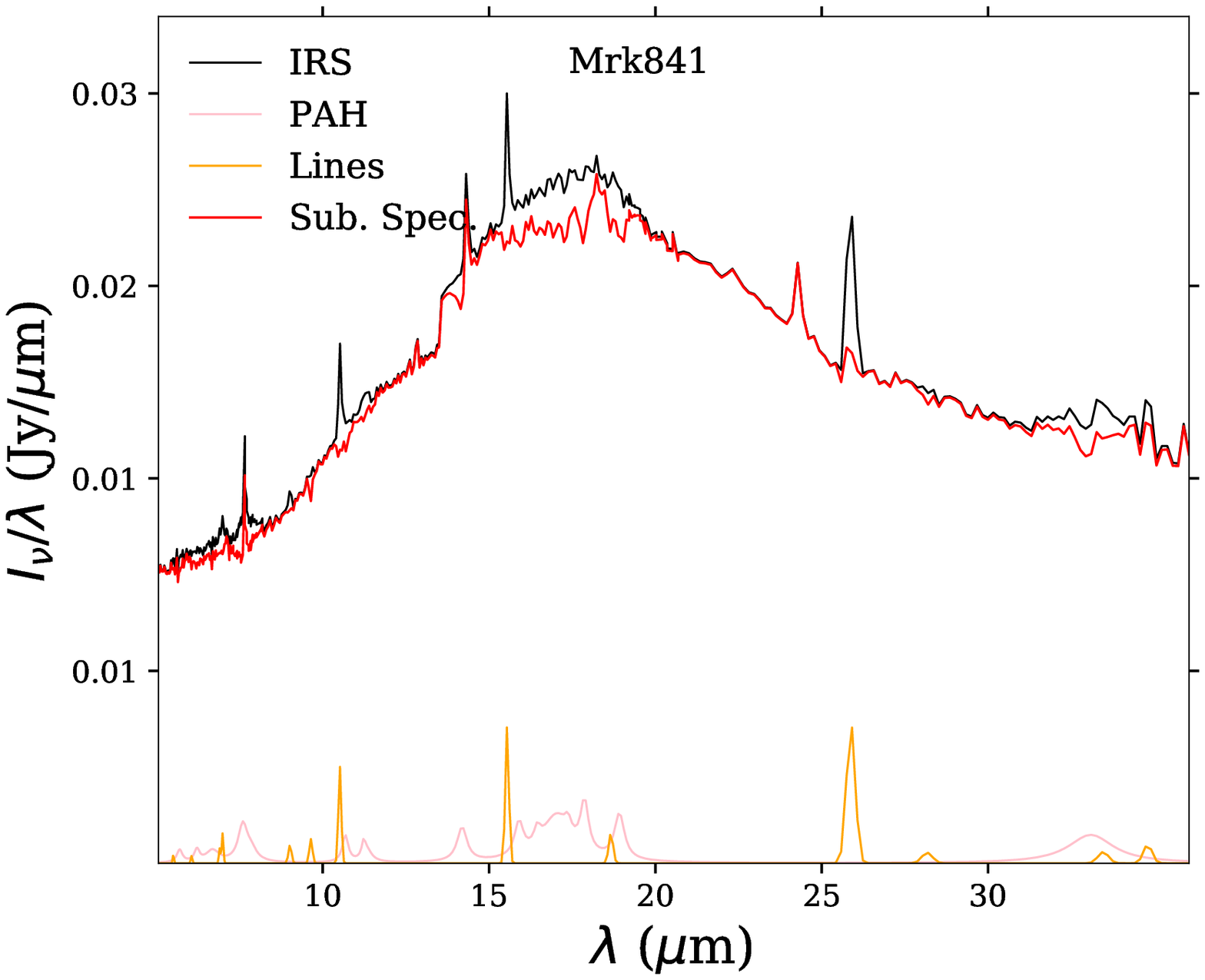}
\end{minipage} \hfill
\begin{minipage}[b]{0.325\linewidth}
\includegraphics[width=\textwidth]{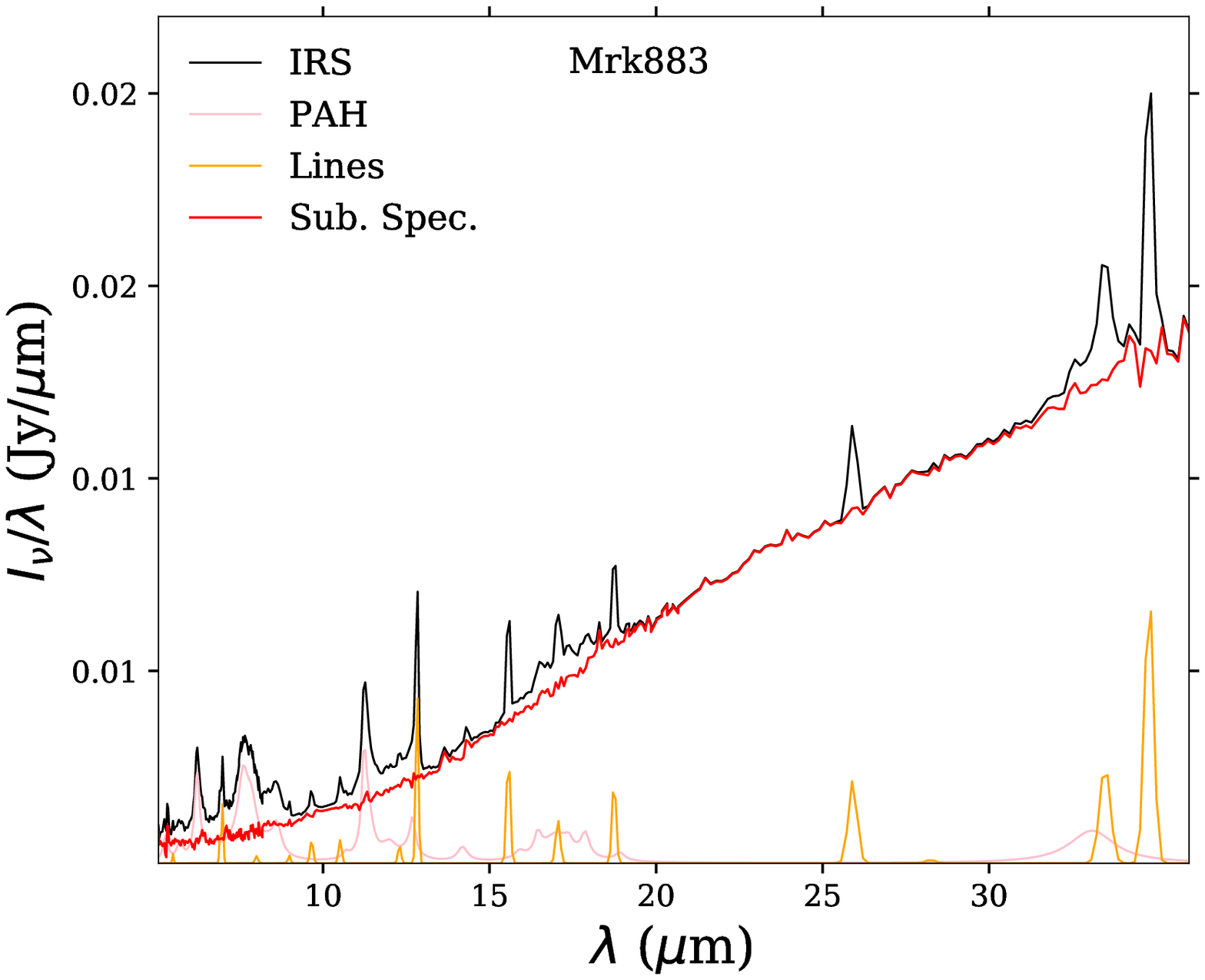}
\end{minipage} \hfill
\begin{minipage}[b]{0.325\linewidth}
\includegraphics[width=\textwidth]{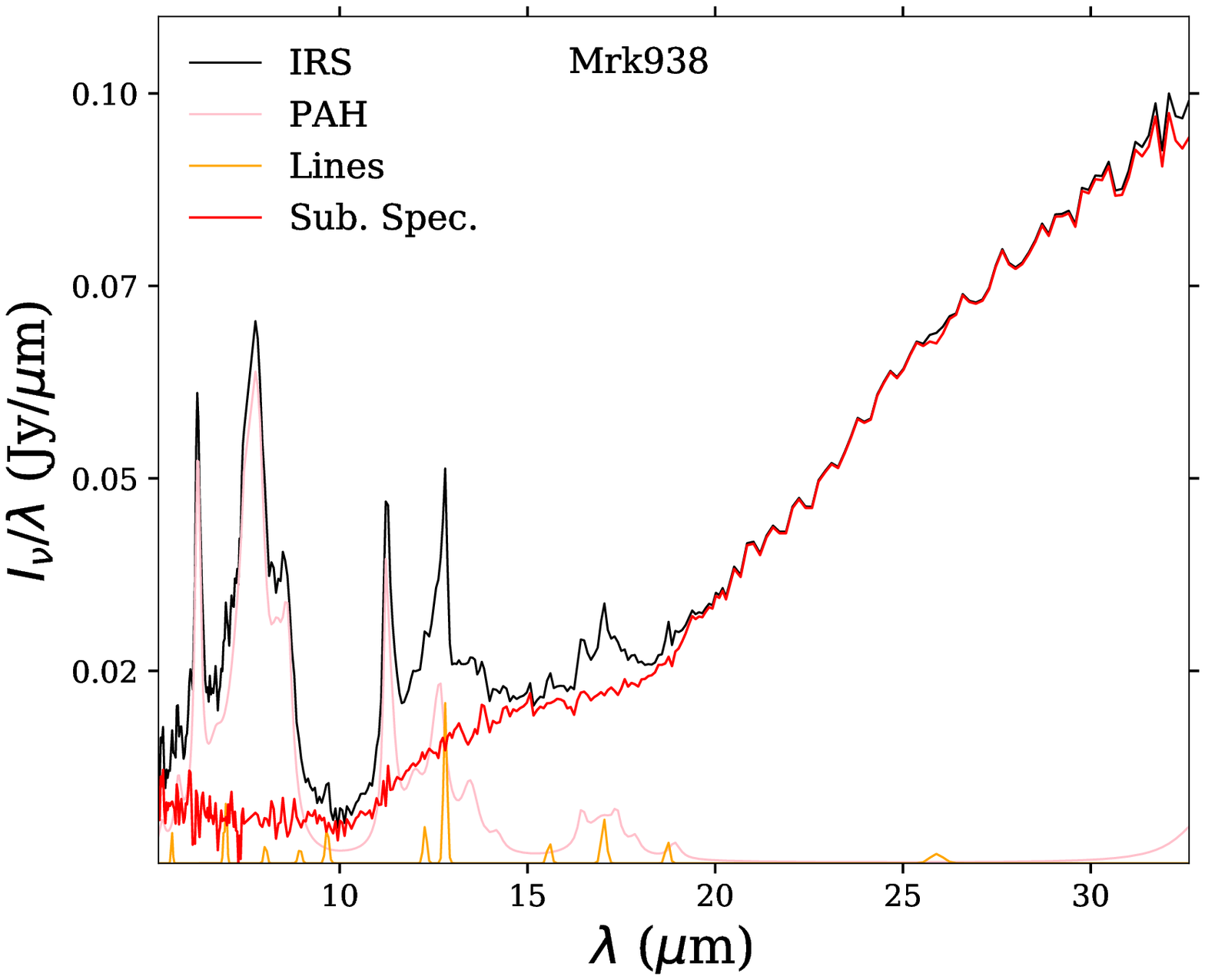}
\end{minipage} \hfill
\begin{minipage}[b]{0.325\linewidth}
\includegraphics[width=\textwidth]{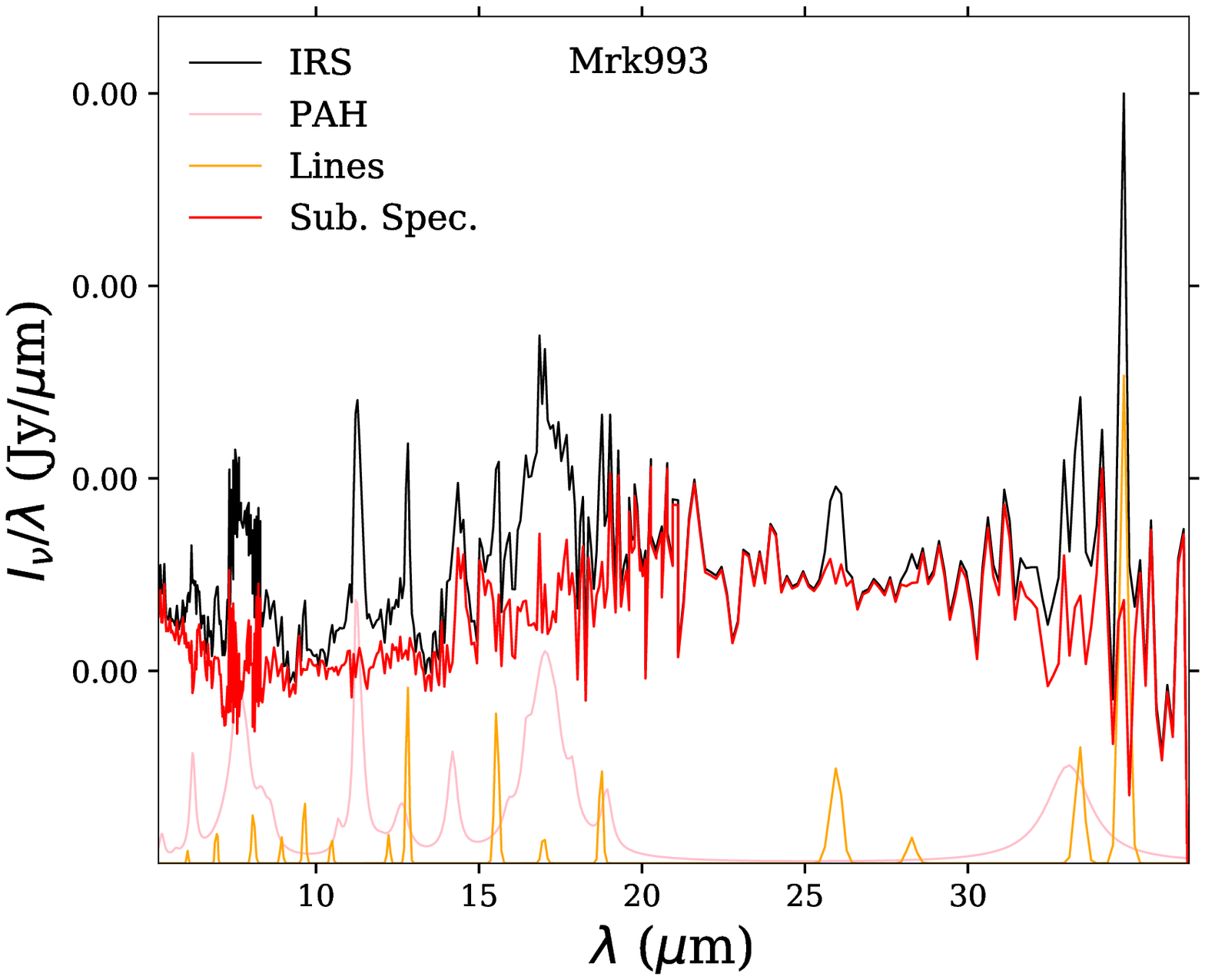}
\end{minipage} \hfill
\begin{minipage}[b]{0.325\linewidth}
\includegraphics[width=\textwidth]{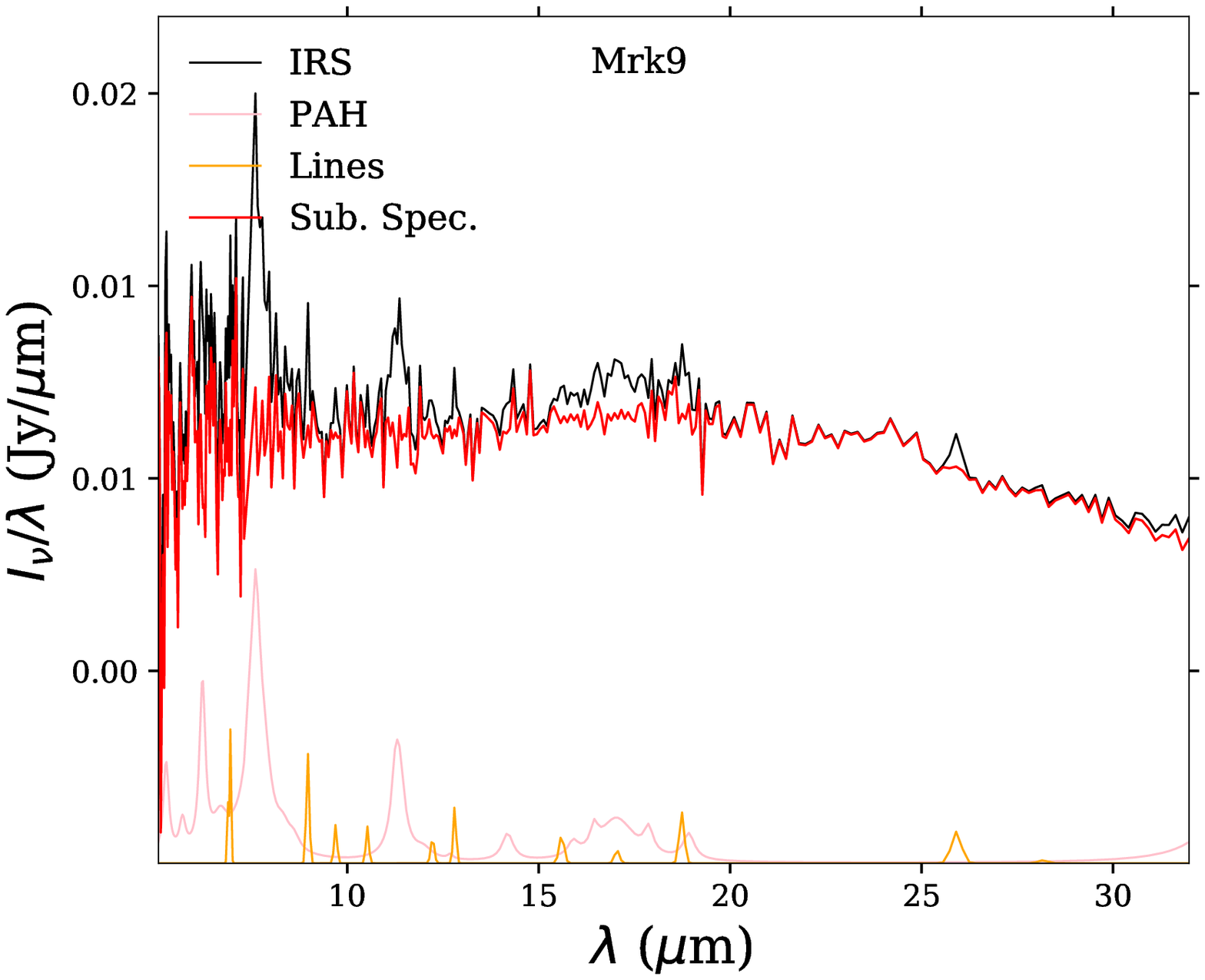}
\end{minipage} \hfill
\begin{minipage}[b]{0.325\linewidth}
\includegraphics[width=\textwidth]{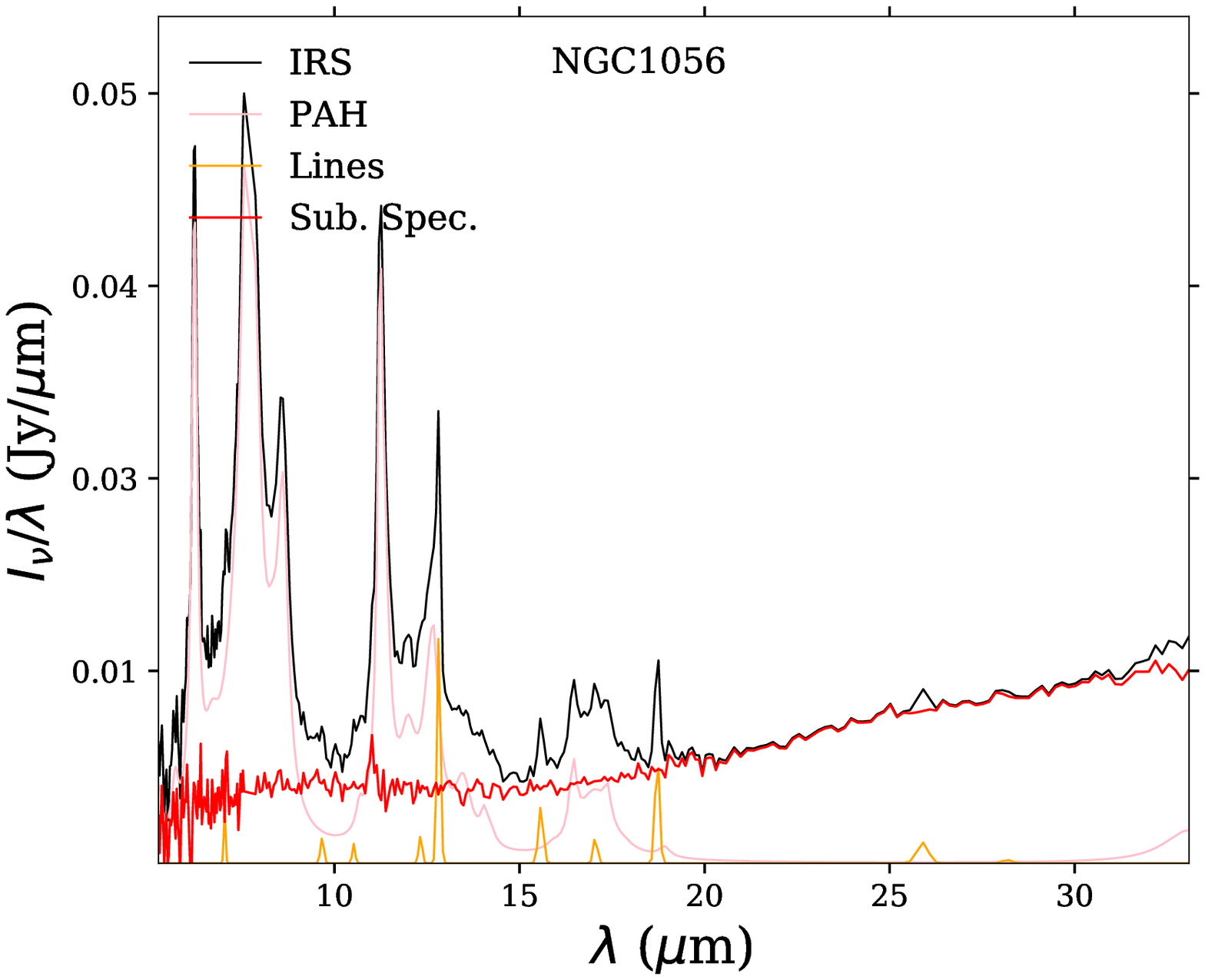}
\end{minipage} \hfill
\begin{minipage}[b]{0.325\linewidth}
\includegraphics[width=\textwidth]{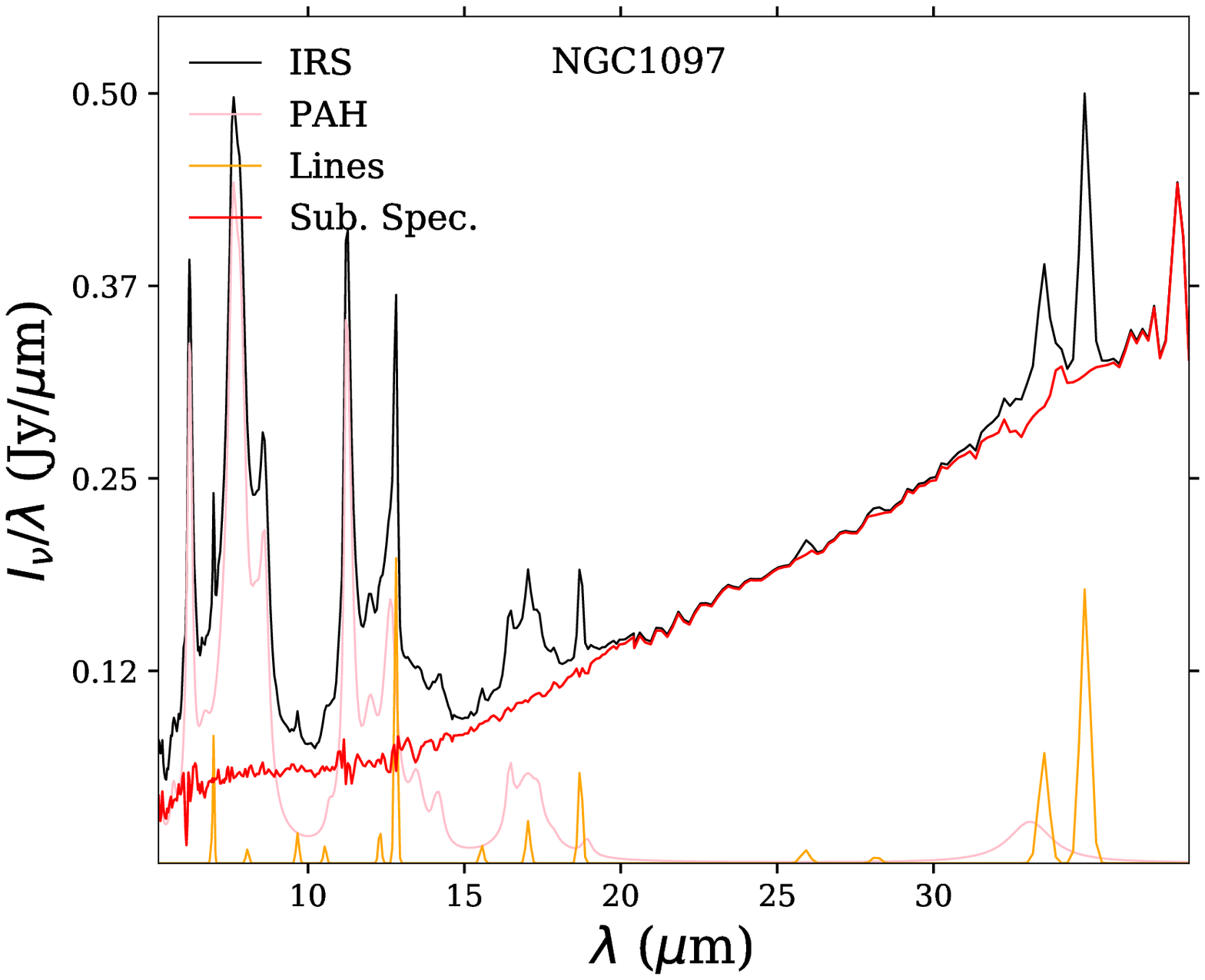}
\end{minipage} \hfill
\begin{minipage}[b]{0.325\linewidth}
\includegraphics[width=\textwidth]{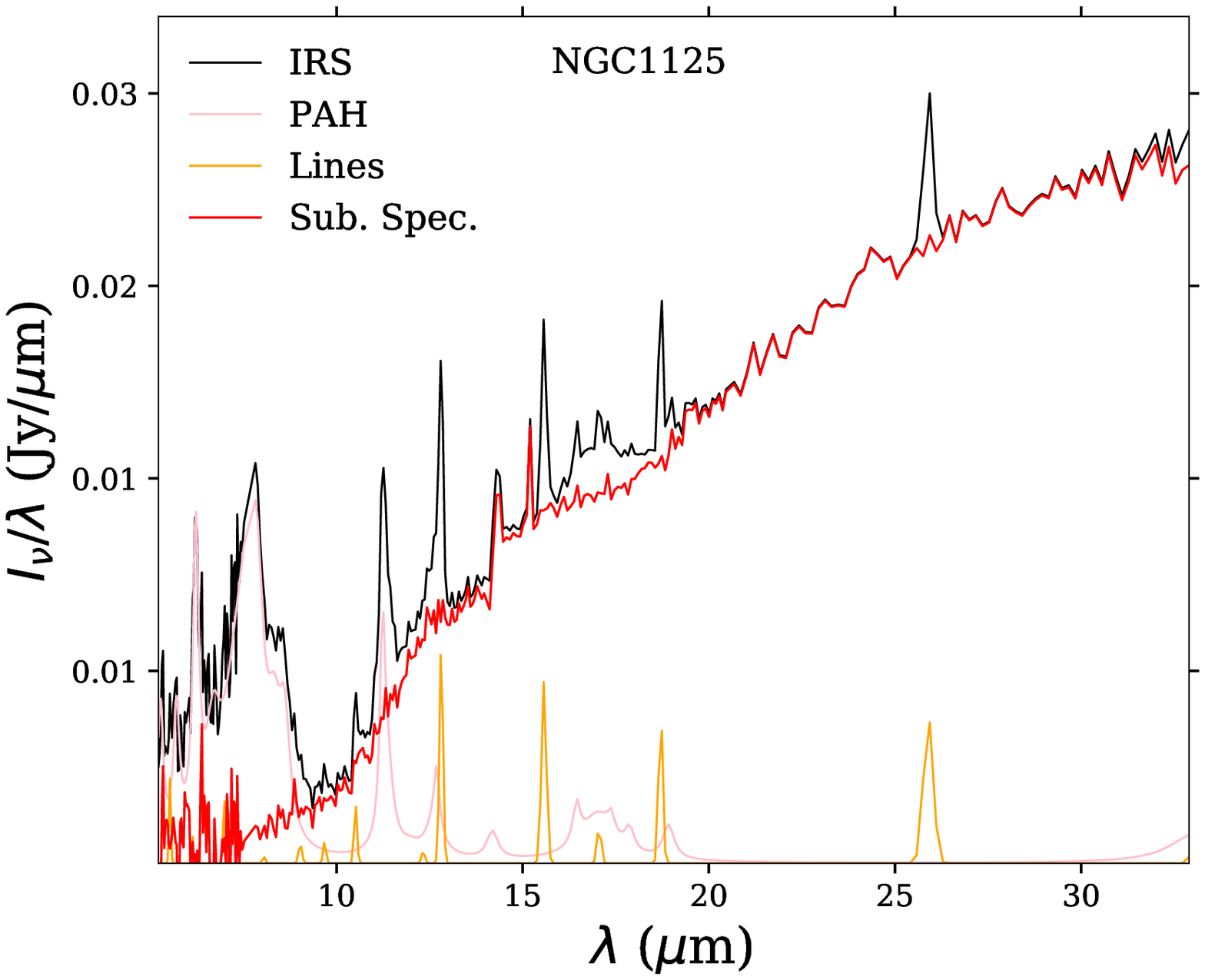}
\end{minipage} \hfill
\begin{minipage}[b]{0.325\linewidth}
\includegraphics[width=\textwidth]{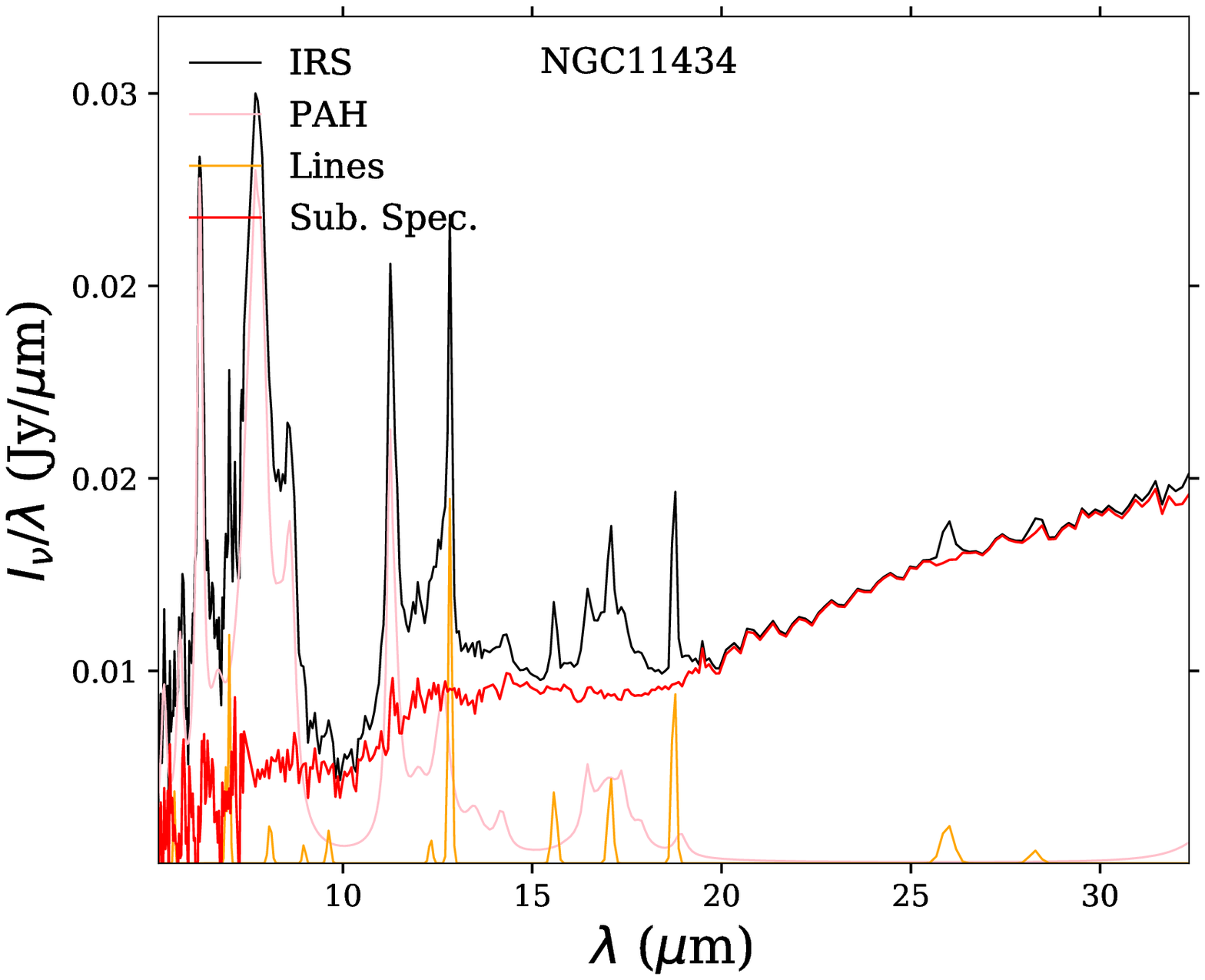}
\end{minipage} \hfill
\begin{minipage}[b]{0.325\linewidth}
\includegraphics[width=\textwidth]{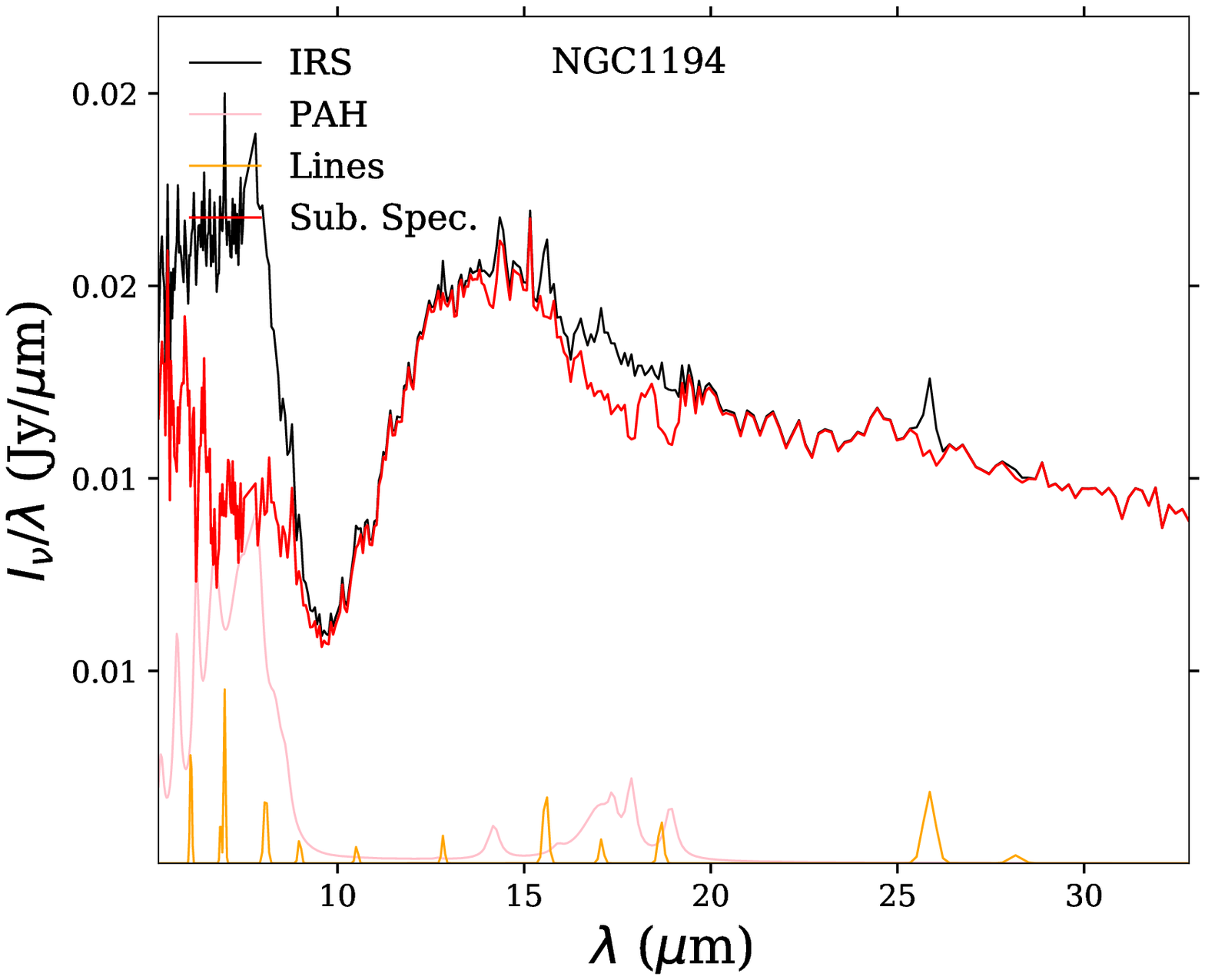}
\end{minipage} \hfill
\begin{minipage}[b]{0.325\linewidth}
\includegraphics[width=\textwidth]{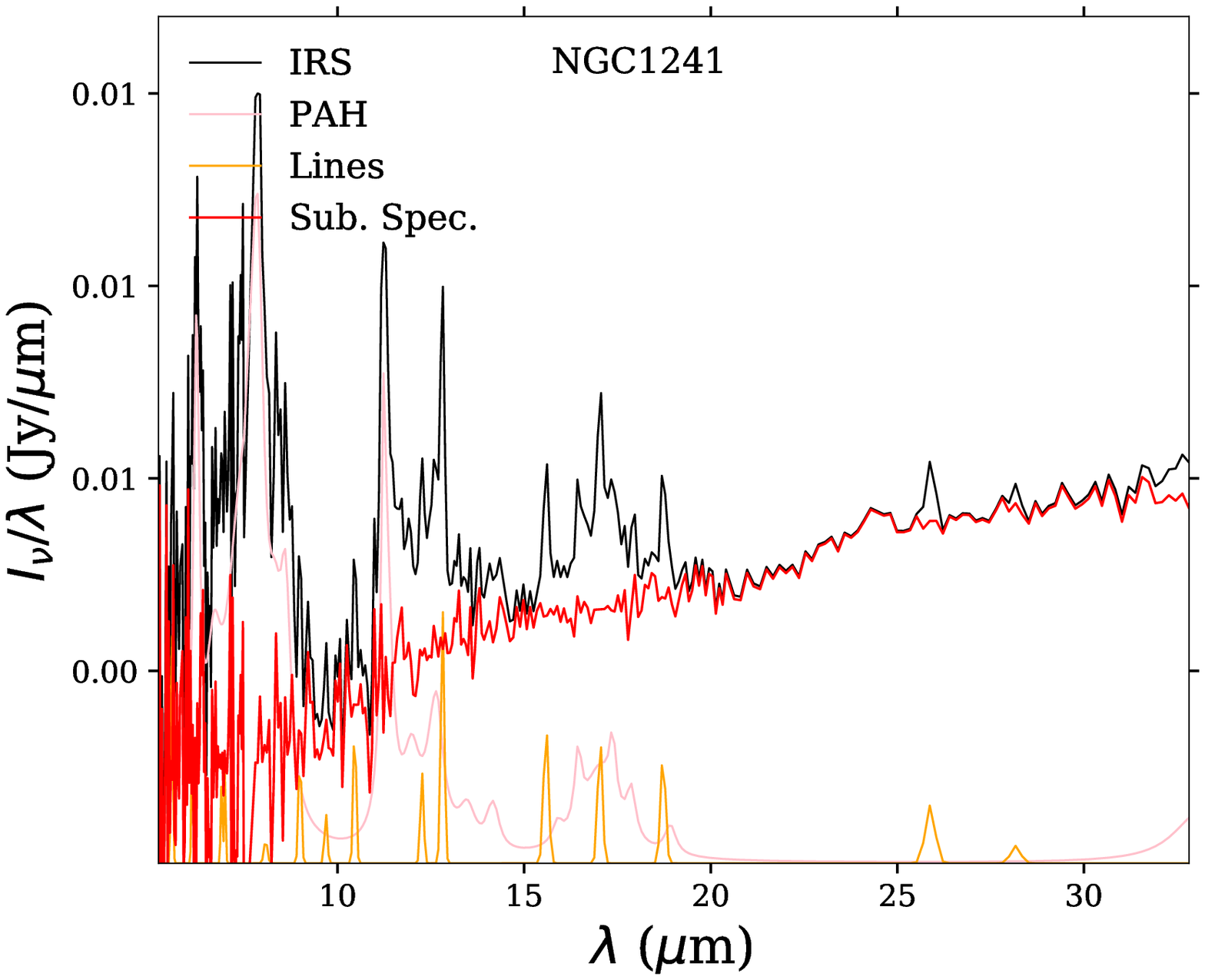}
\end{minipage} \hfill
\begin{minipage}[b]{0.325\linewidth}
\includegraphics[width=\textwidth]{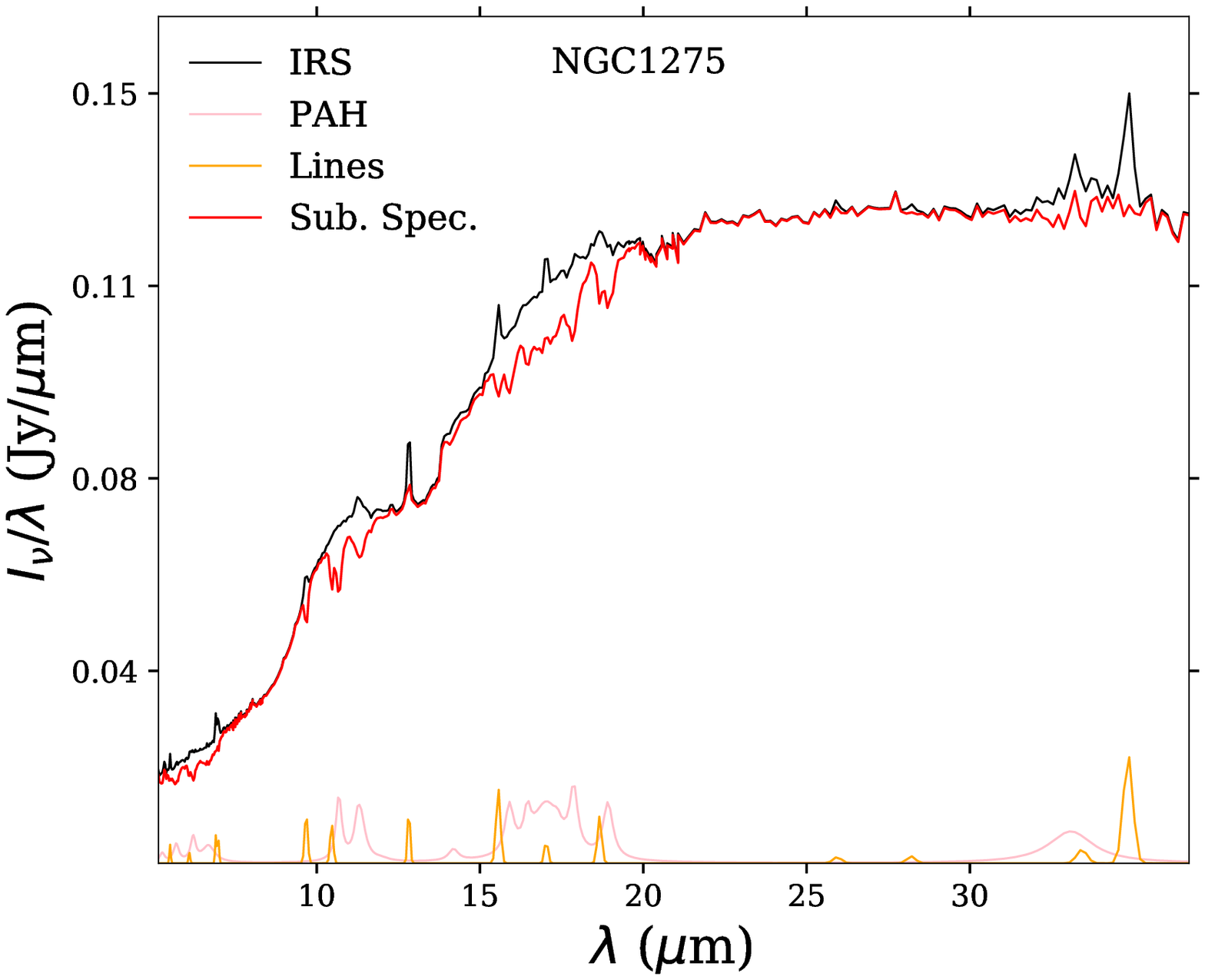}
\end{minipage} \hfill
\begin{minipage}[b]{0.325\linewidth}
\includegraphics[width=\textwidth]{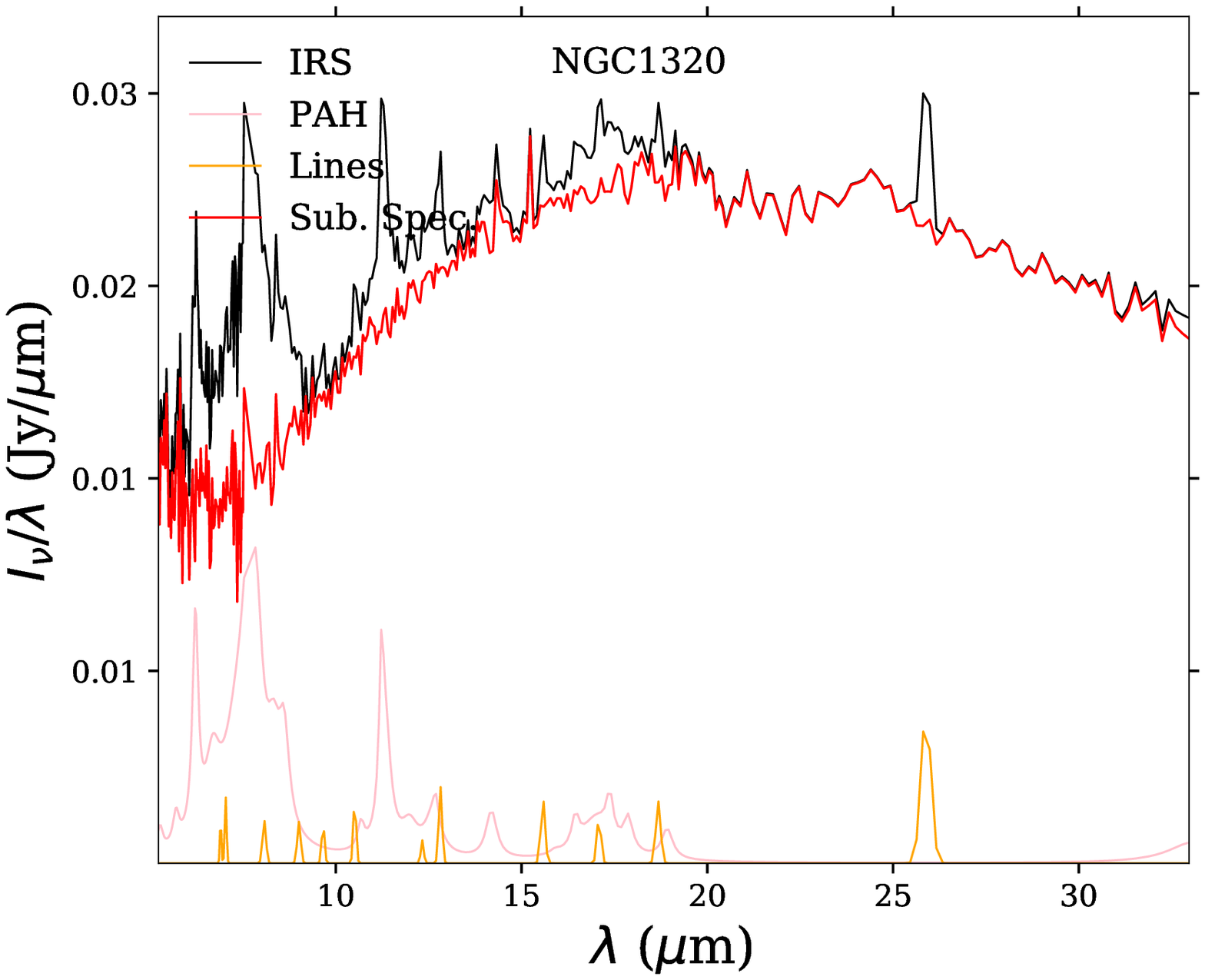}
\end{minipage} \hfill
\begin{minipage}[b]{0.325\linewidth}
\includegraphics[width=\textwidth]{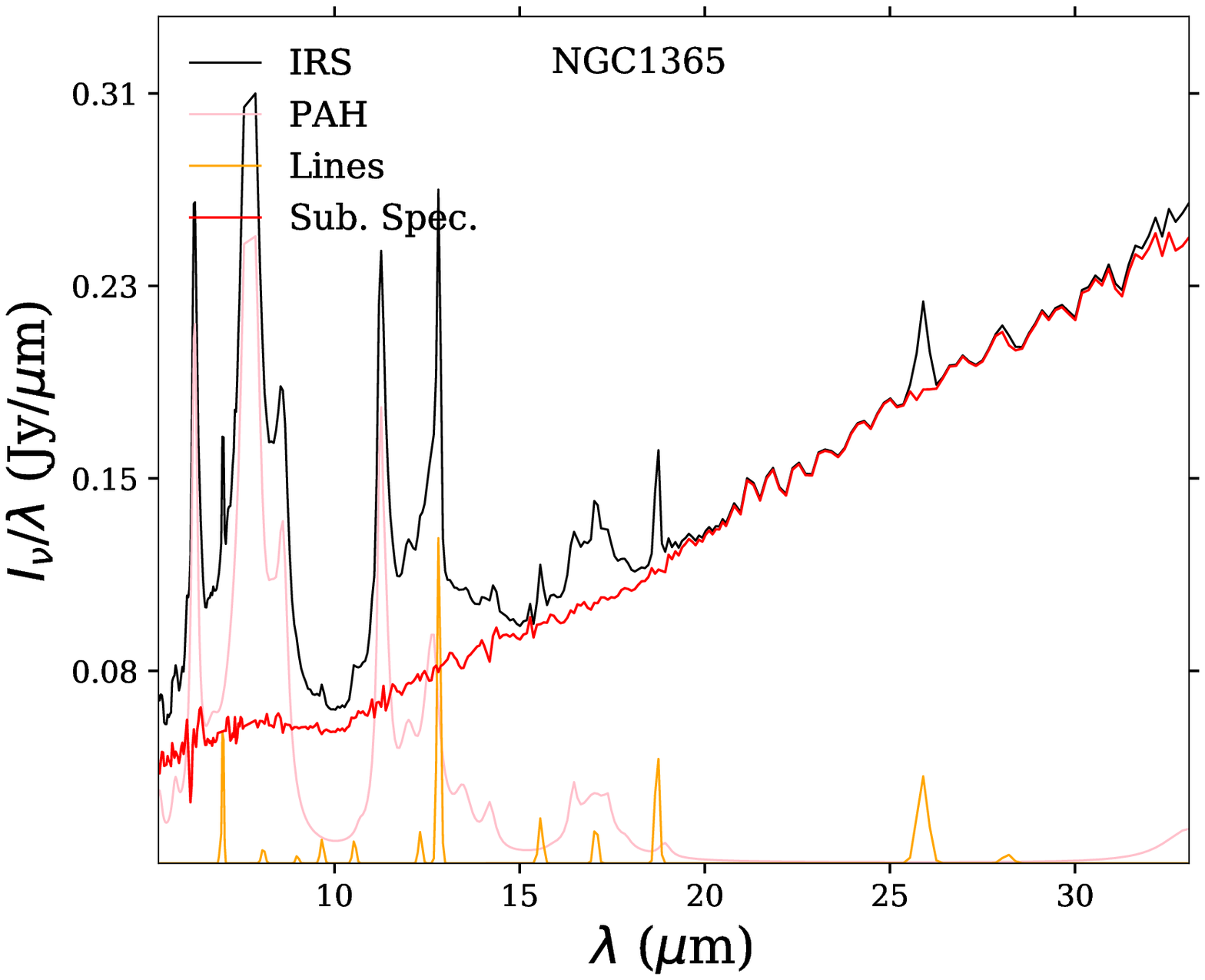}
\end{minipage} \hfill
\begin{minipage}[b]{0.325\linewidth}
\includegraphics[width=\textwidth]{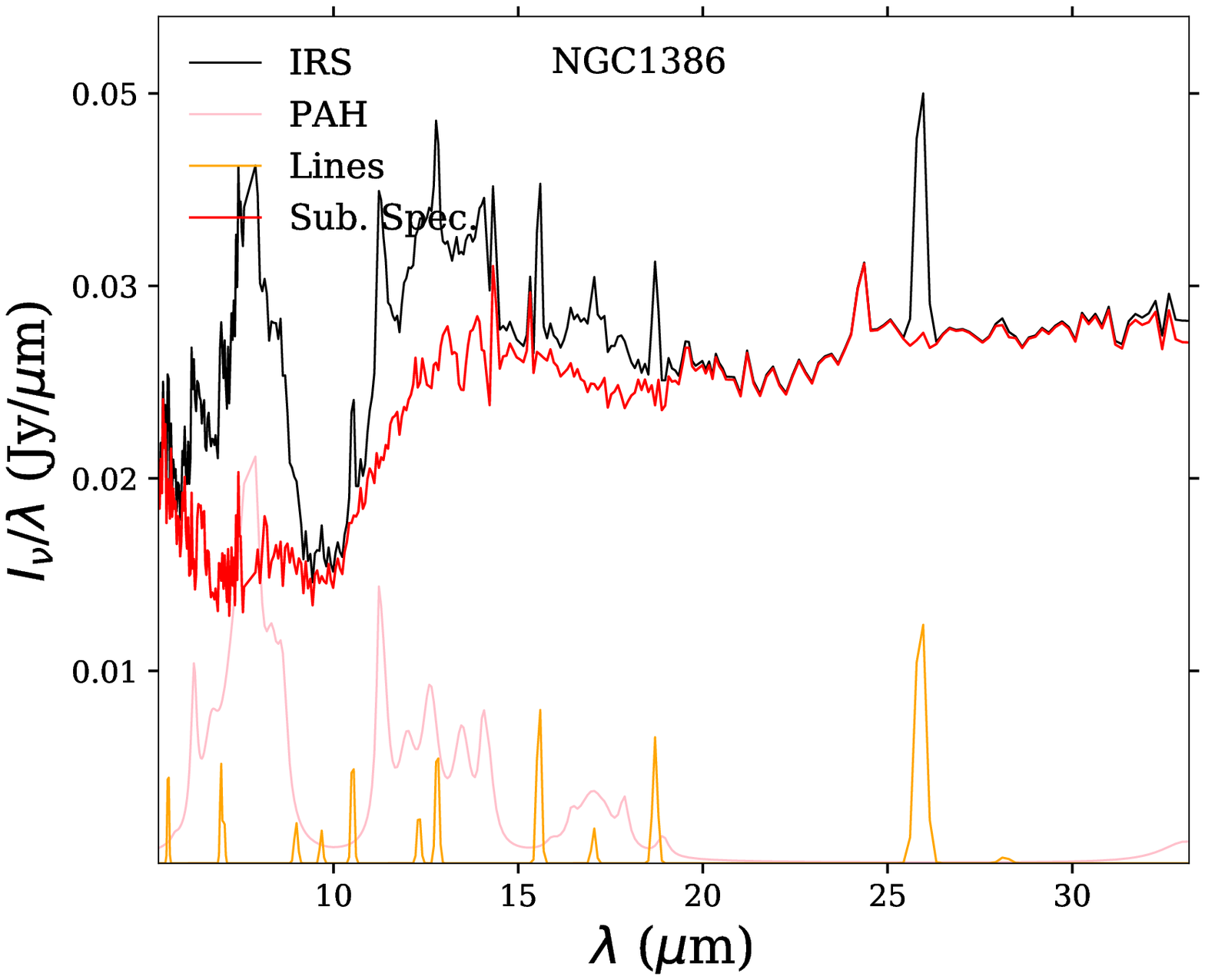}
\end{minipage} \hfill
\caption{continued from previous page.}
\setcounter{figure}{0}
\end{figure}

\begin{figure}

\begin{minipage}[b]{0.325\linewidth}
\includegraphics[width=\textwidth]{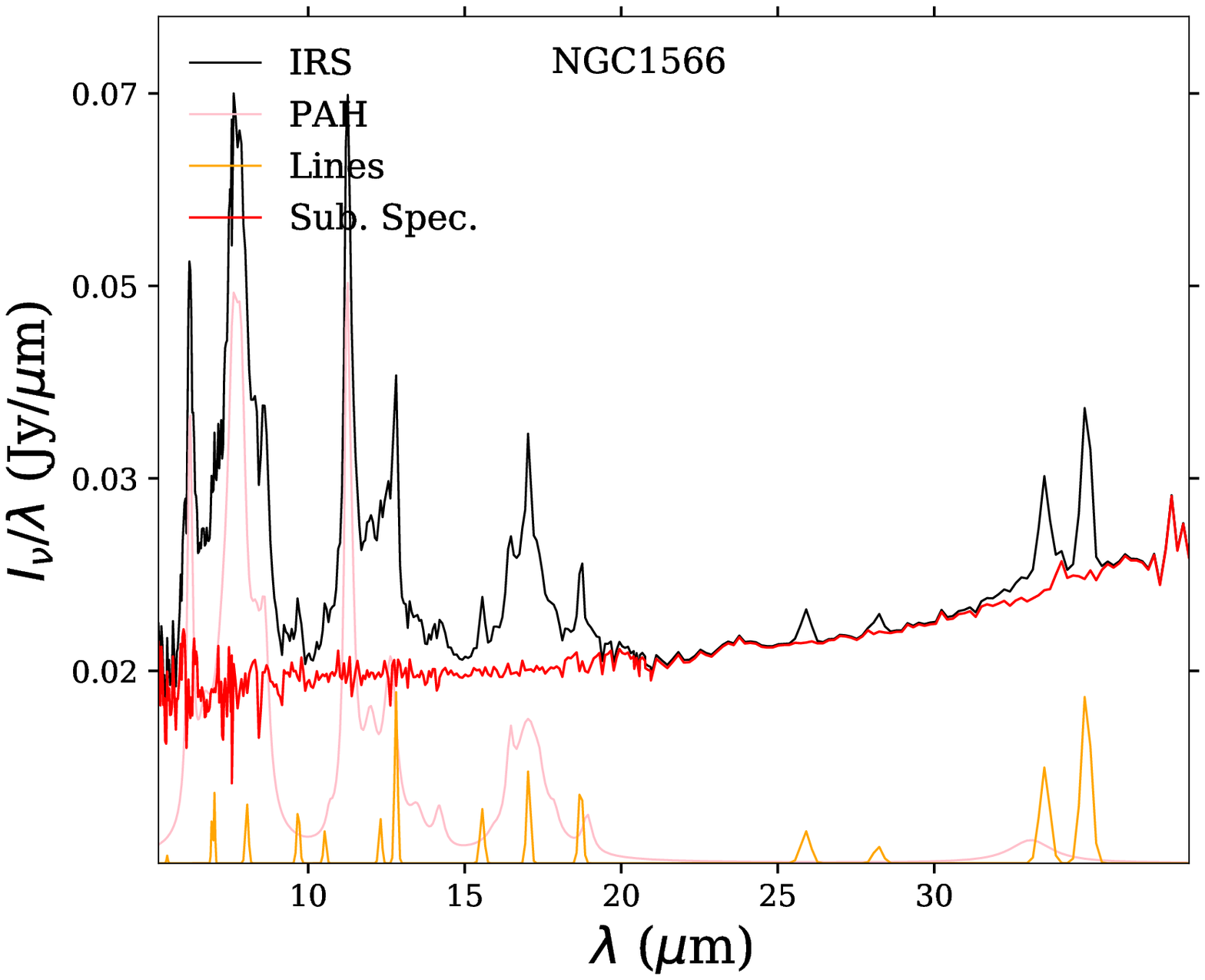}
\end{minipage} \hfill
\begin{minipage}[b]{0.325\linewidth}
\includegraphics[width=\textwidth]{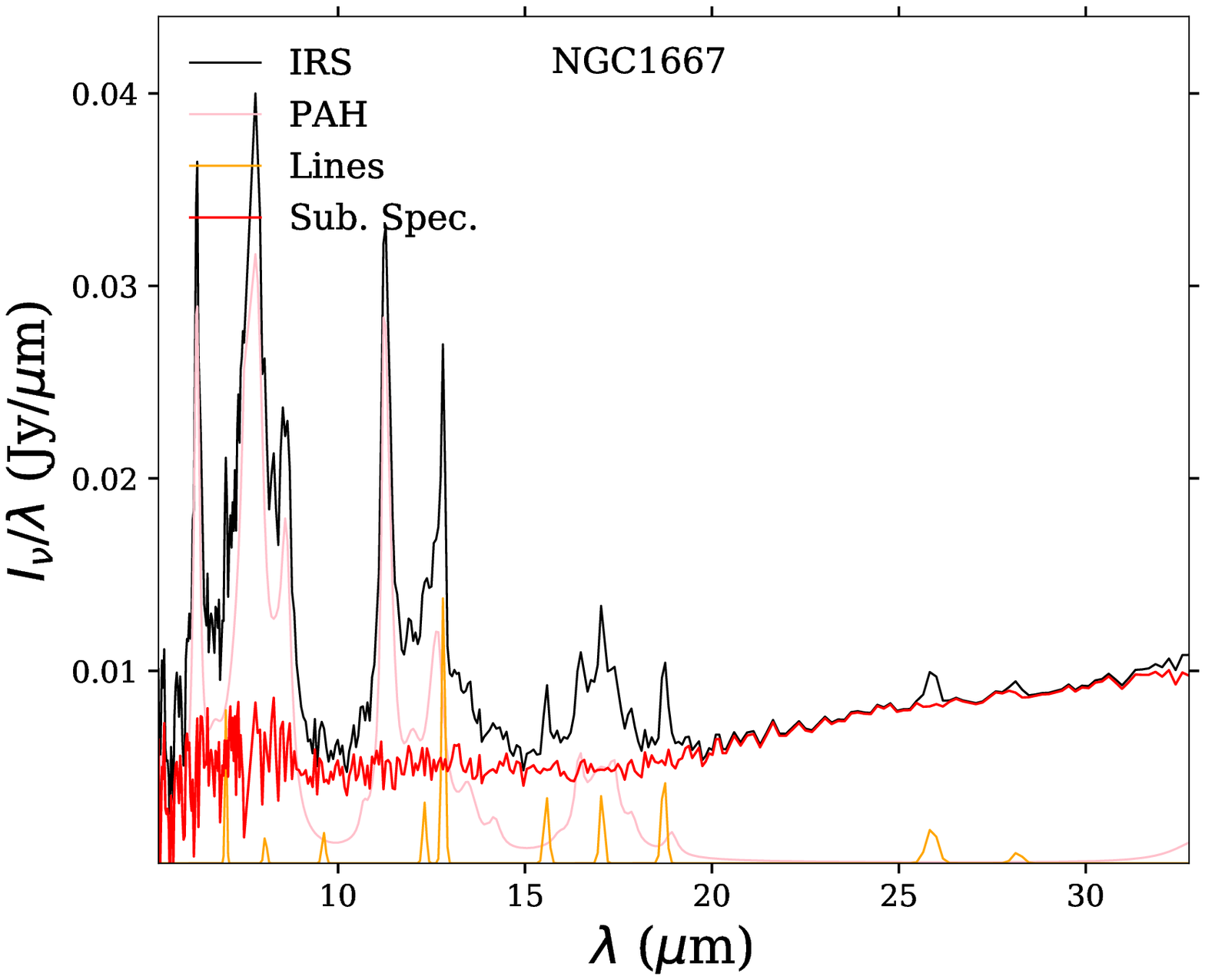}
\end{minipage} \hfill
\begin{minipage}[b]{0.325\linewidth}
\includegraphics[width=\textwidth]{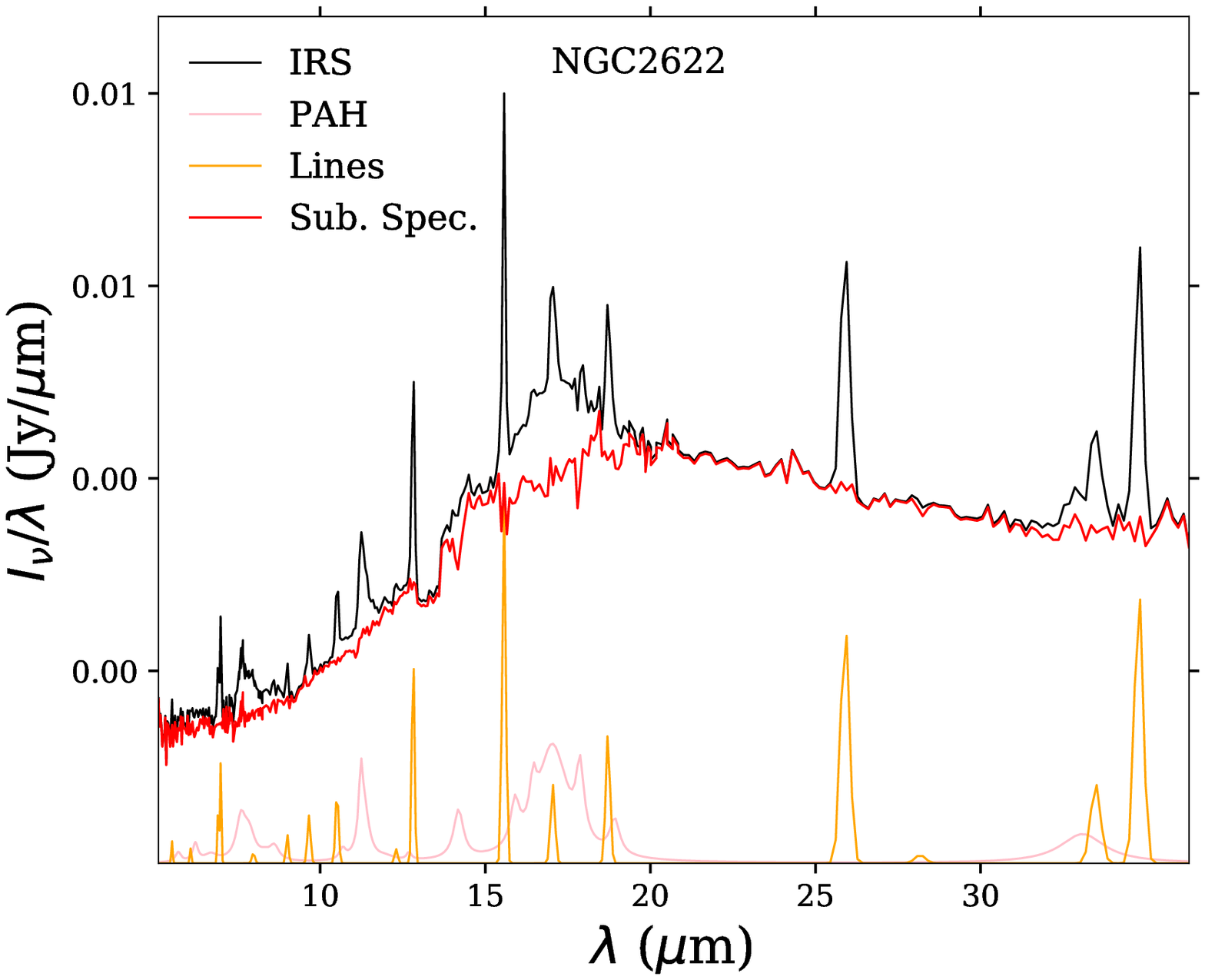}
\end{minipage} \hfill
\begin{minipage}[b]{0.325\linewidth}
\includegraphics[width=\textwidth]{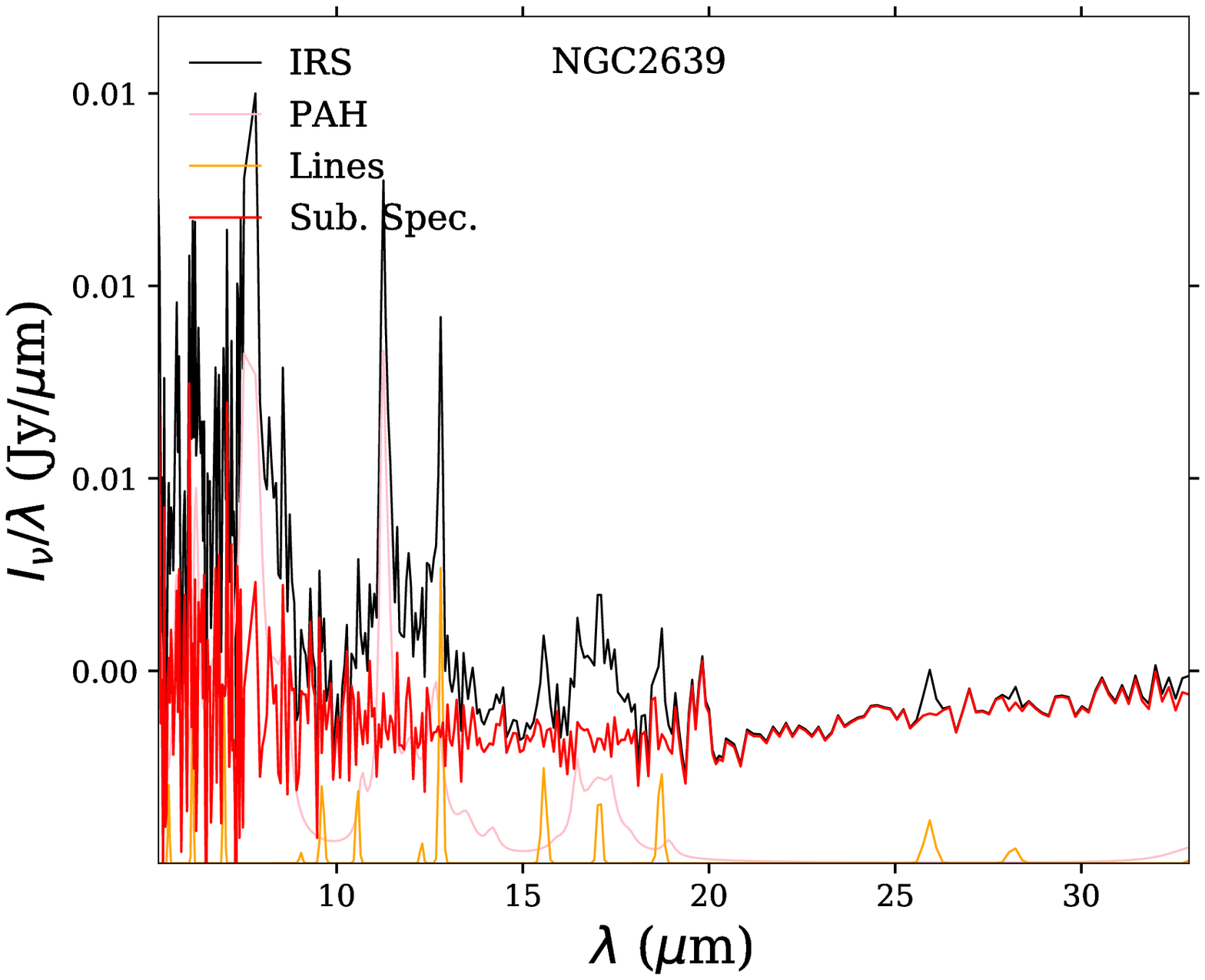}
\end{minipage} \hfill
\begin{minipage}[b]{0.325\linewidth}
\includegraphics[width=\textwidth]{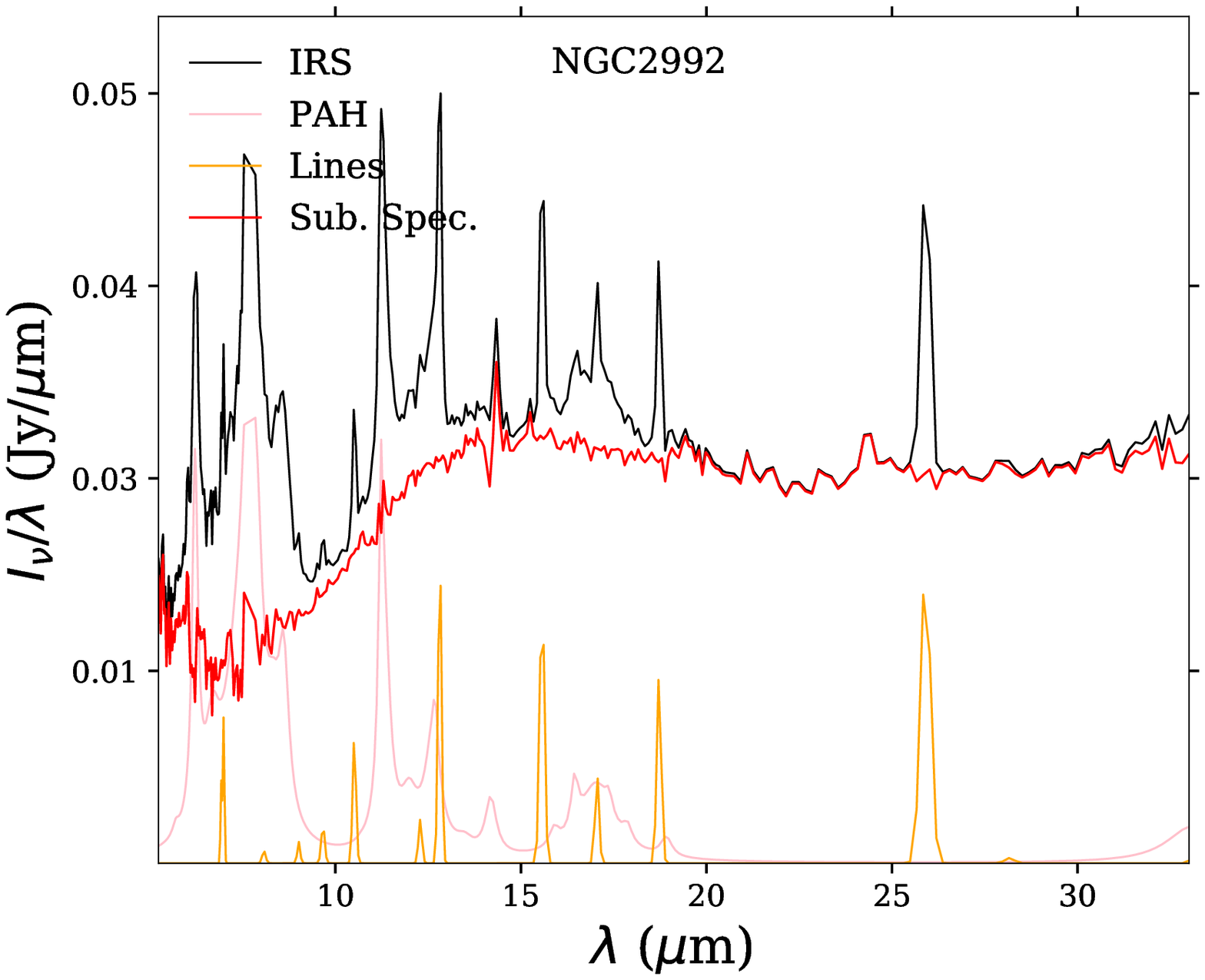}
\end{minipage} \hfill
\begin{minipage}[b]{0.325\linewidth}
\includegraphics[width=\textwidth]{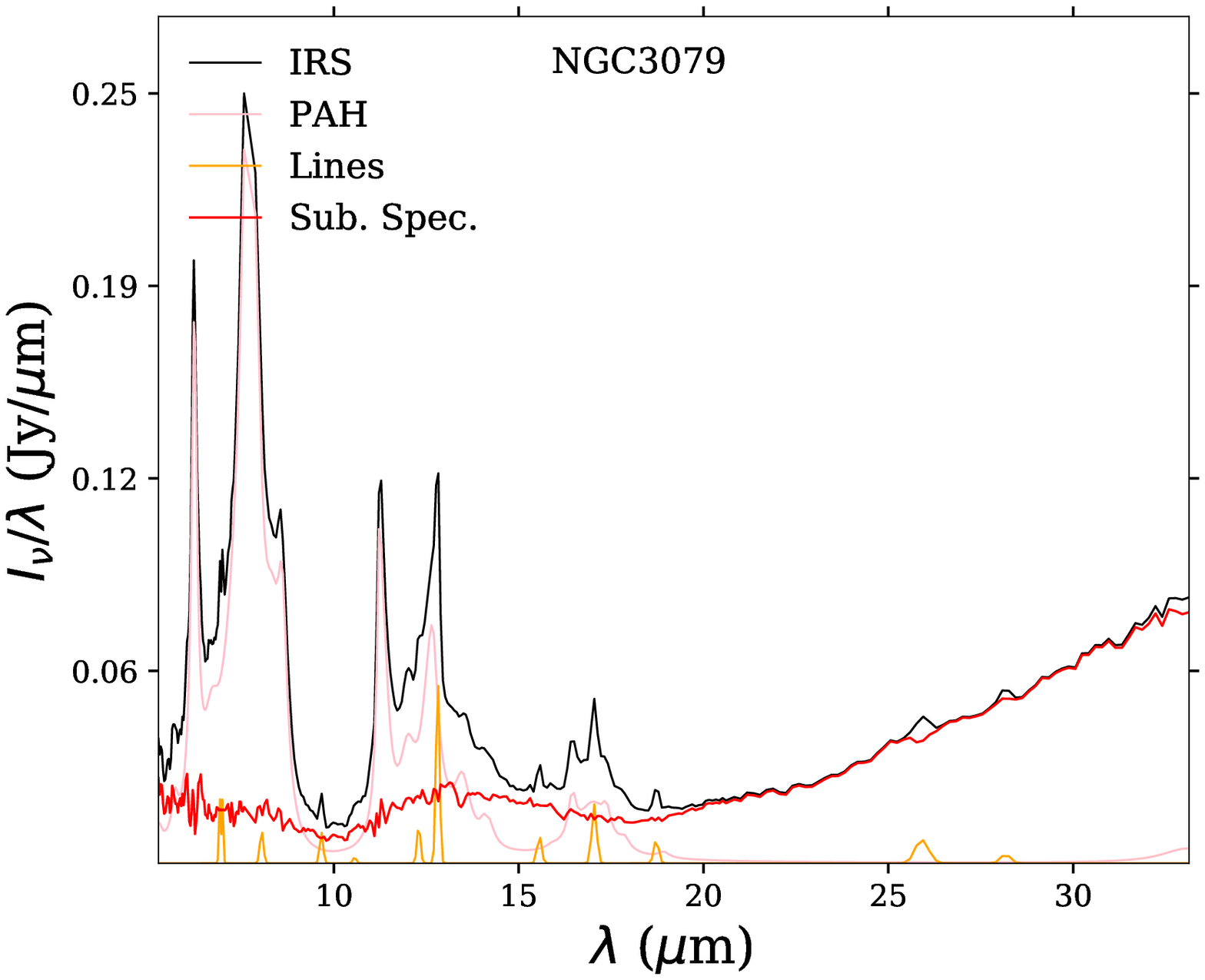}
\end{minipage} \hfill
\begin{minipage}[b]{0.325\linewidth}
\includegraphics[width=\textwidth]{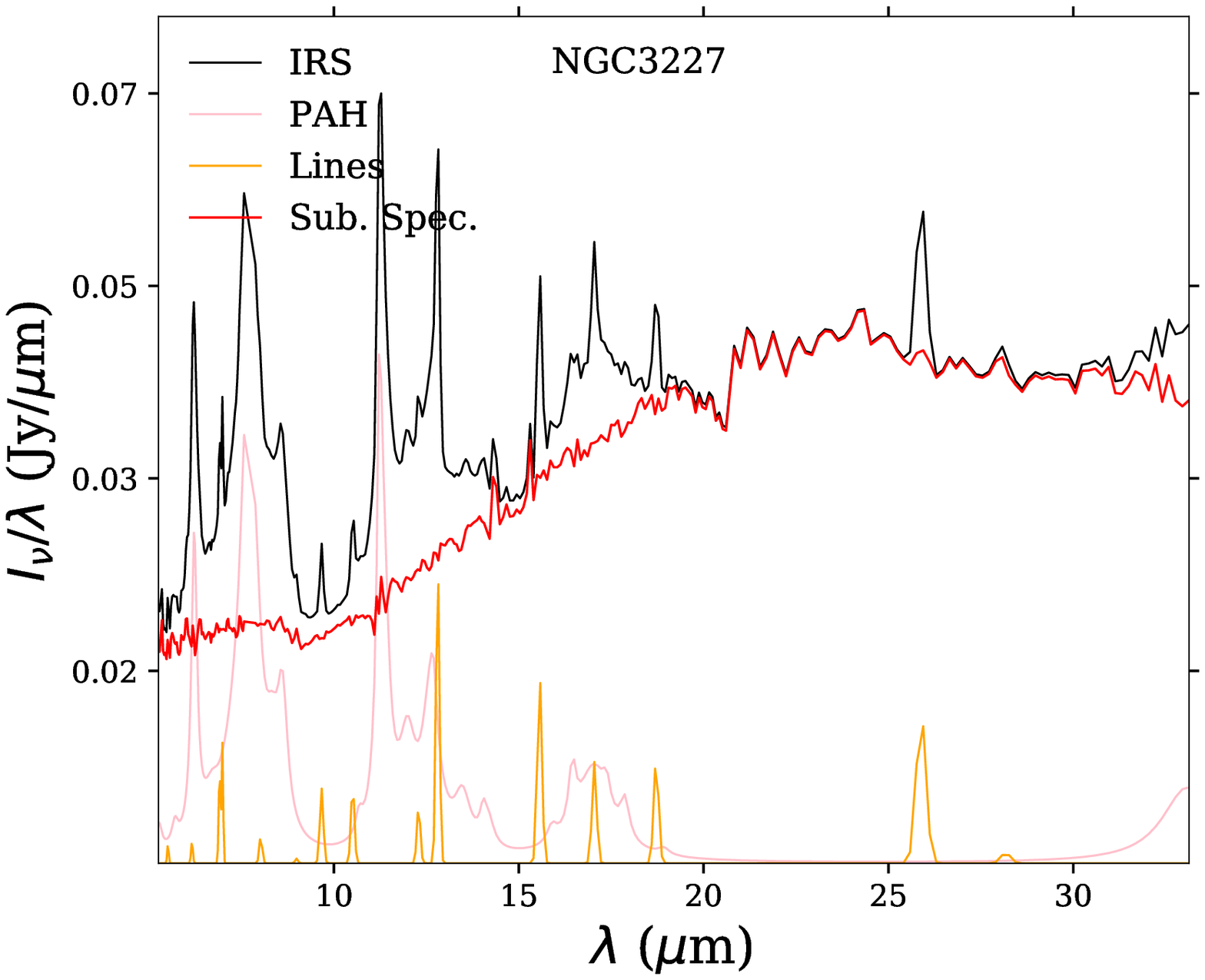}
\end{minipage} \hfill
\begin{minipage}[b]{0.325\linewidth}
\includegraphics[width=\textwidth]{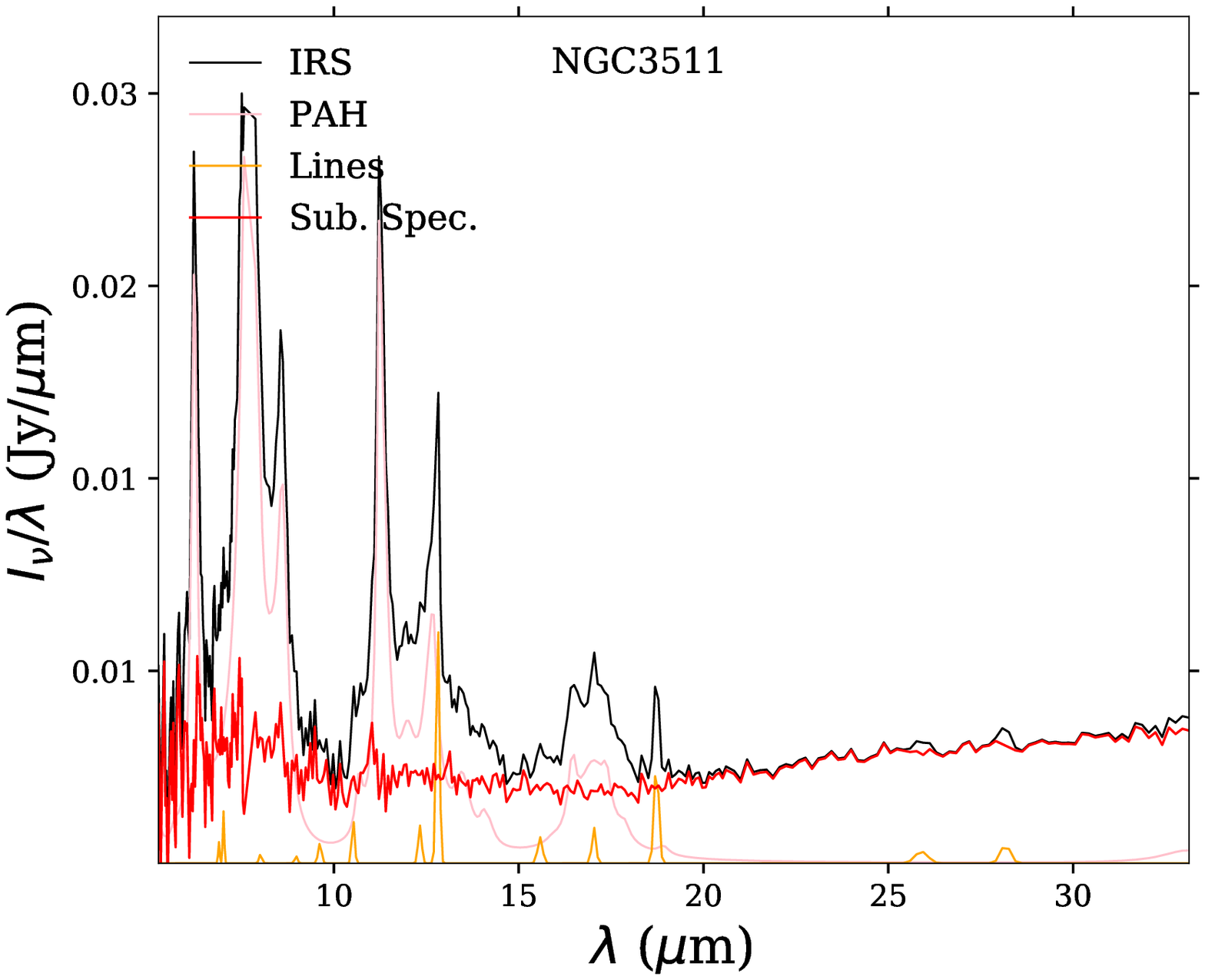}
\end{minipage} \hfill
\begin{minipage}[b]{0.325\linewidth}
\includegraphics[width=\textwidth]{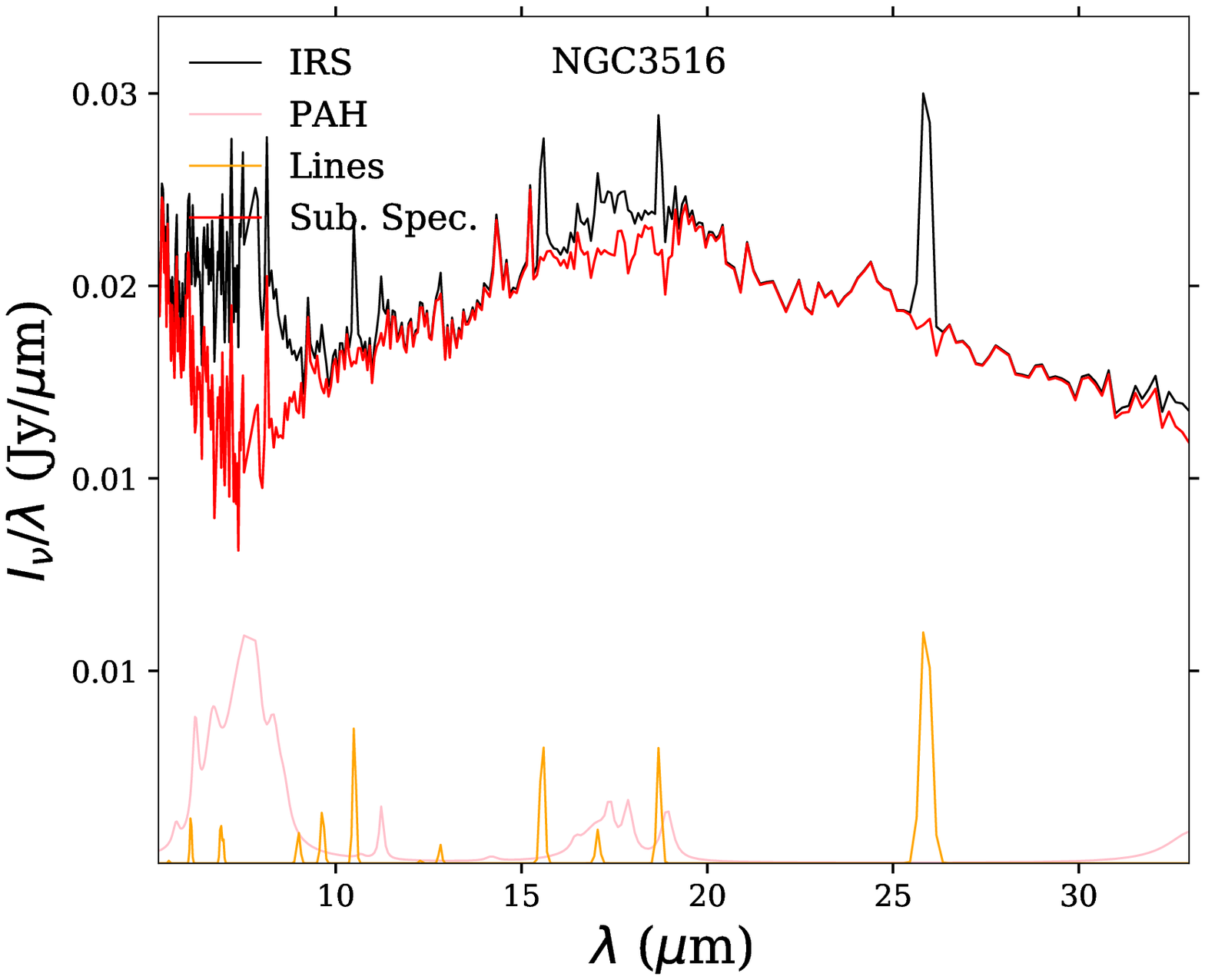}
\end{minipage} \hfill
\begin{minipage}[b]{0.325\linewidth}
\includegraphics[width=\textwidth]{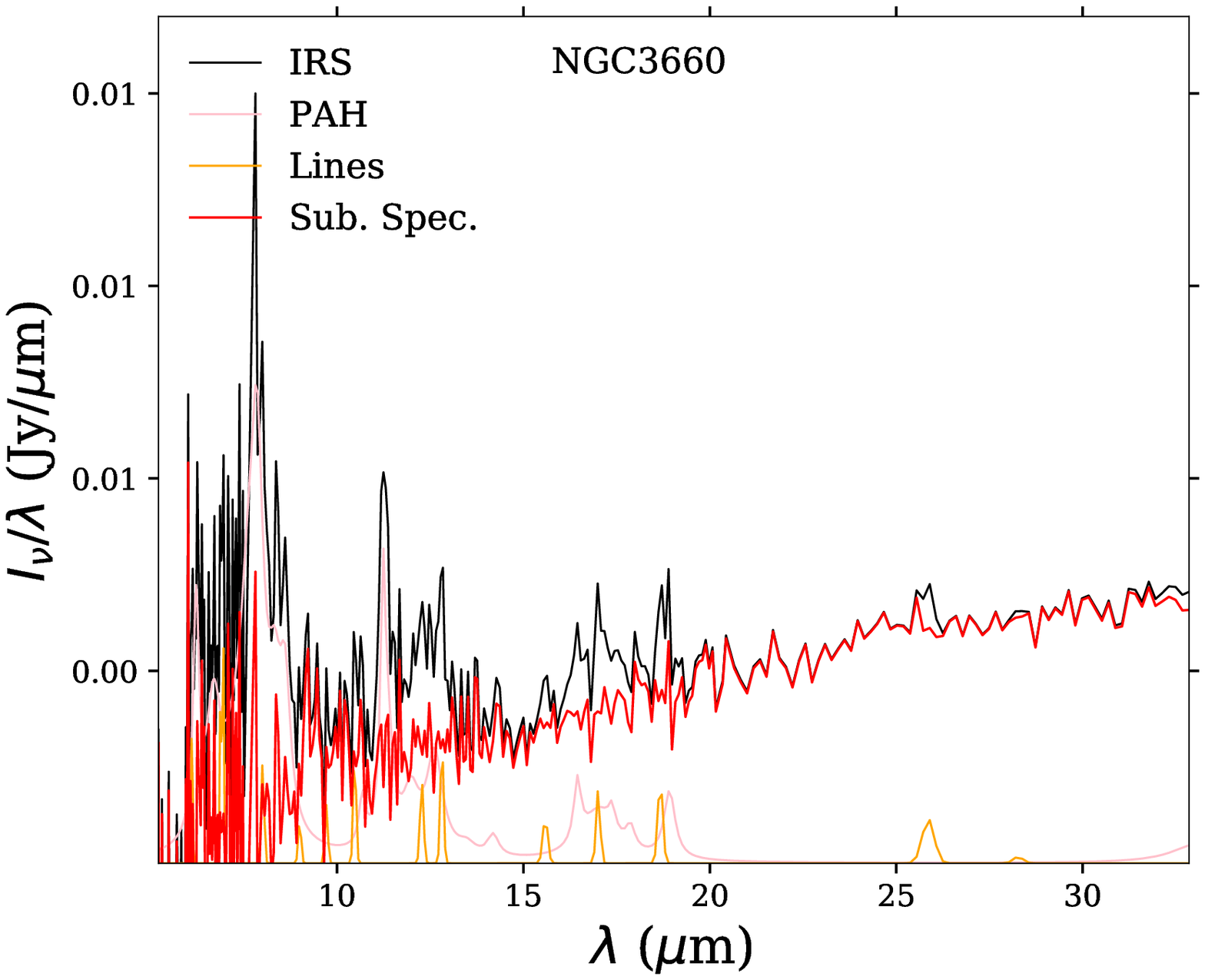}
\end{minipage} \hfill
\begin{minipage}[b]{0.325\linewidth}
\includegraphics[width=\textwidth]{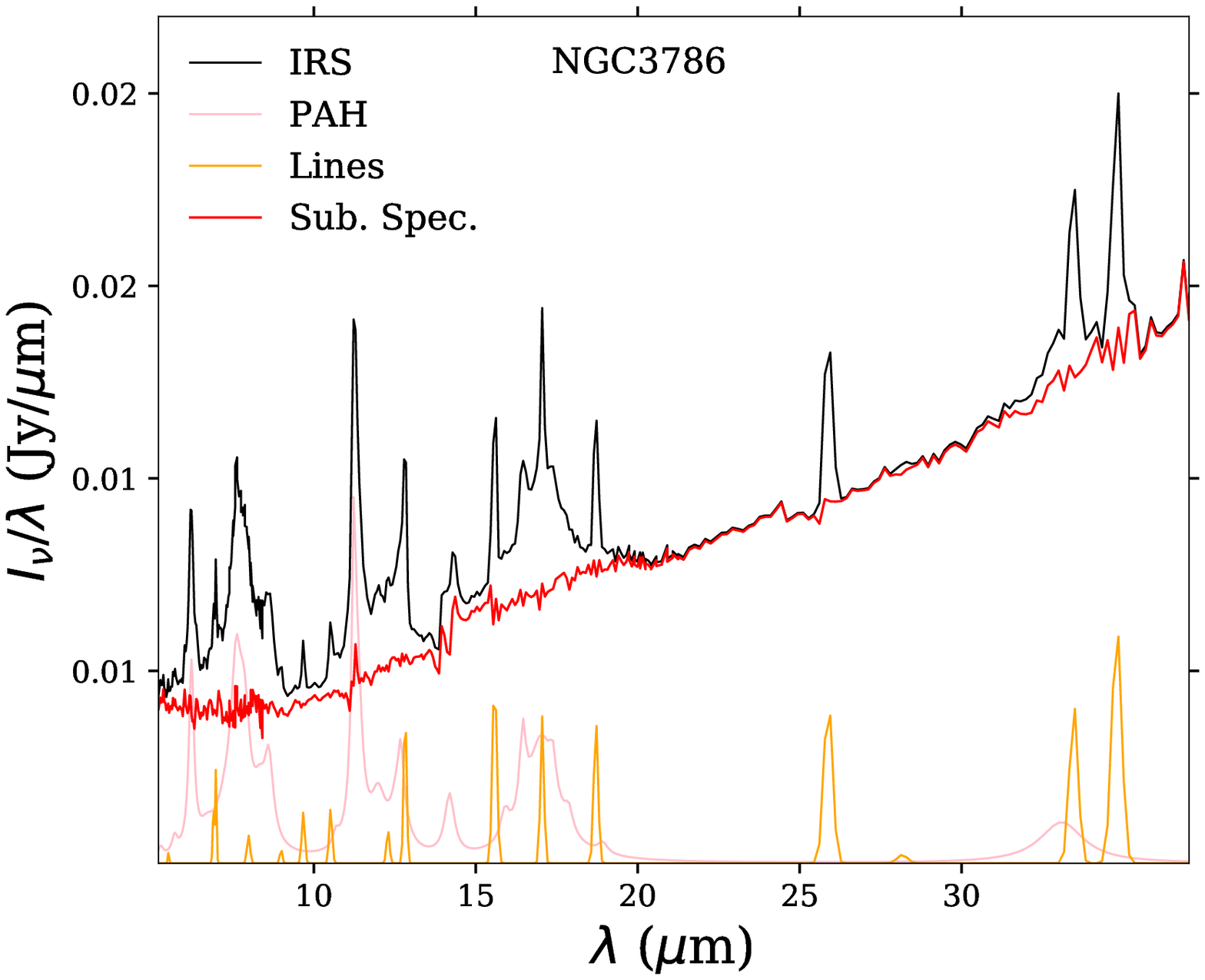}
\end{minipage} \hfill
\begin{minipage}[b]{0.325\linewidth}
\includegraphics[width=\textwidth]{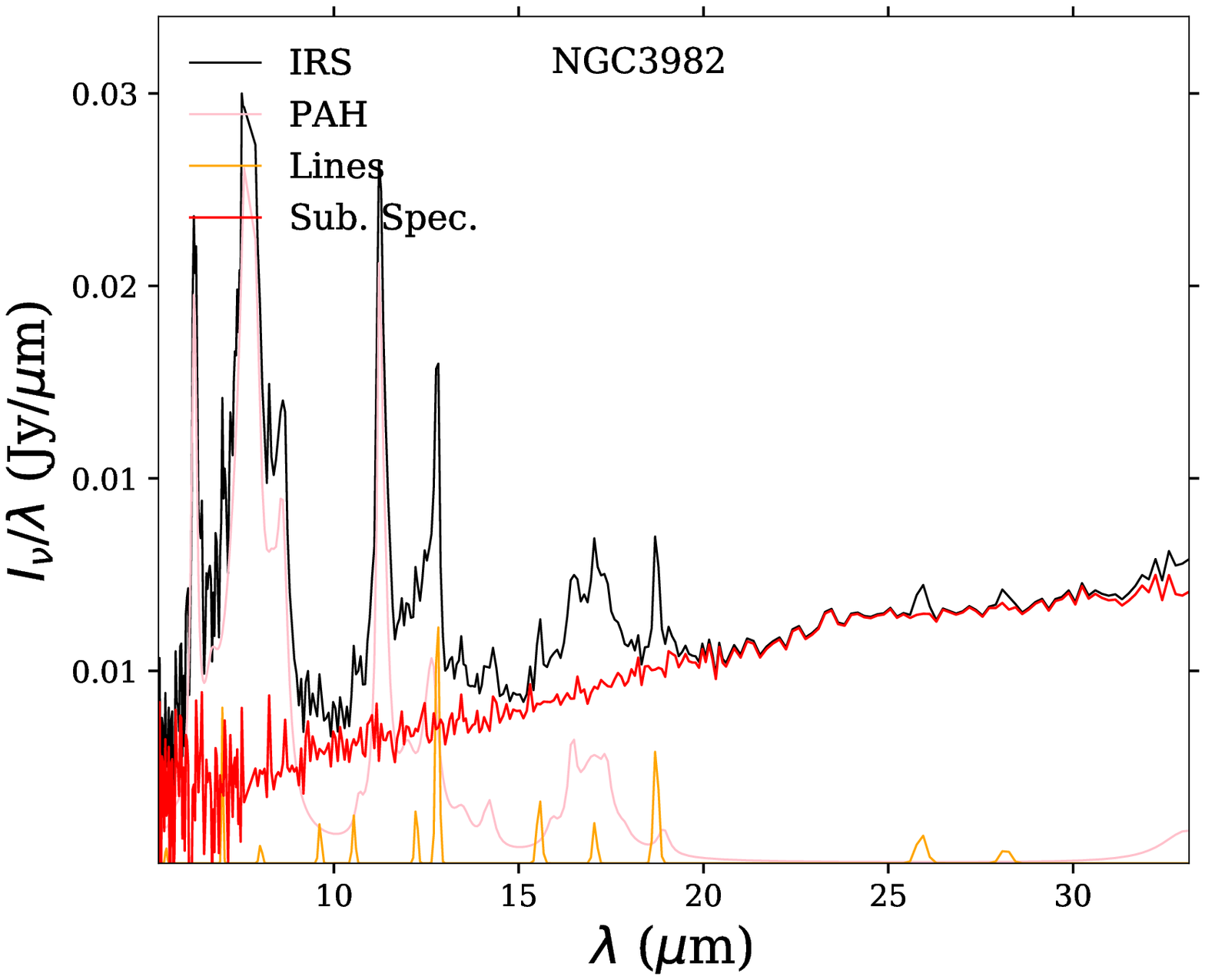}
\end{minipage} \hfill
\begin{minipage}[b]{0.325\linewidth}
\includegraphics[width=\textwidth]{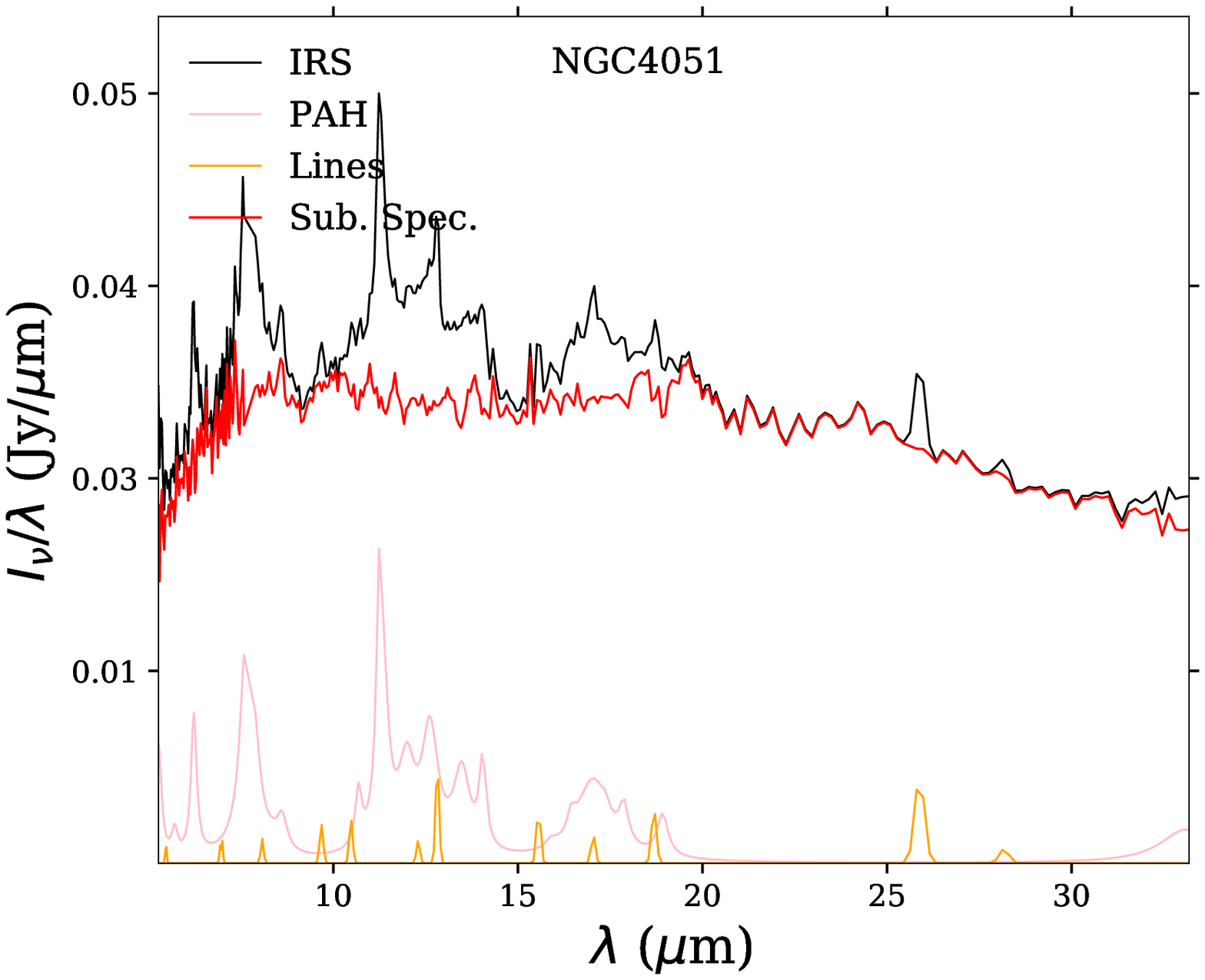}
\end{minipage} \hfill
\begin{minipage}[b]{0.325\linewidth}
\includegraphics[width=\textwidth]{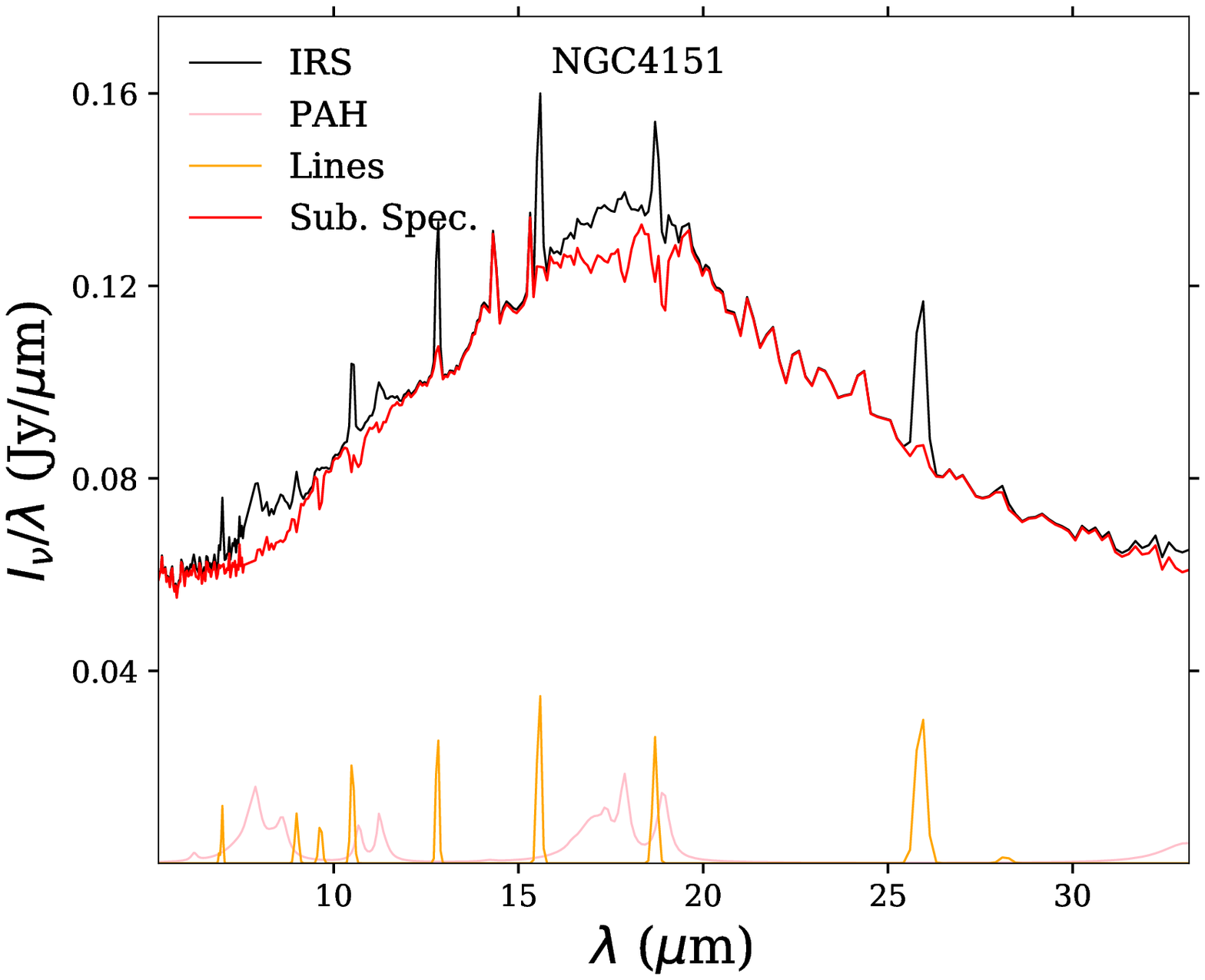}
\end{minipage} \hfill
\begin{minipage}[b]{0.325\linewidth}
\includegraphics[width=\textwidth]{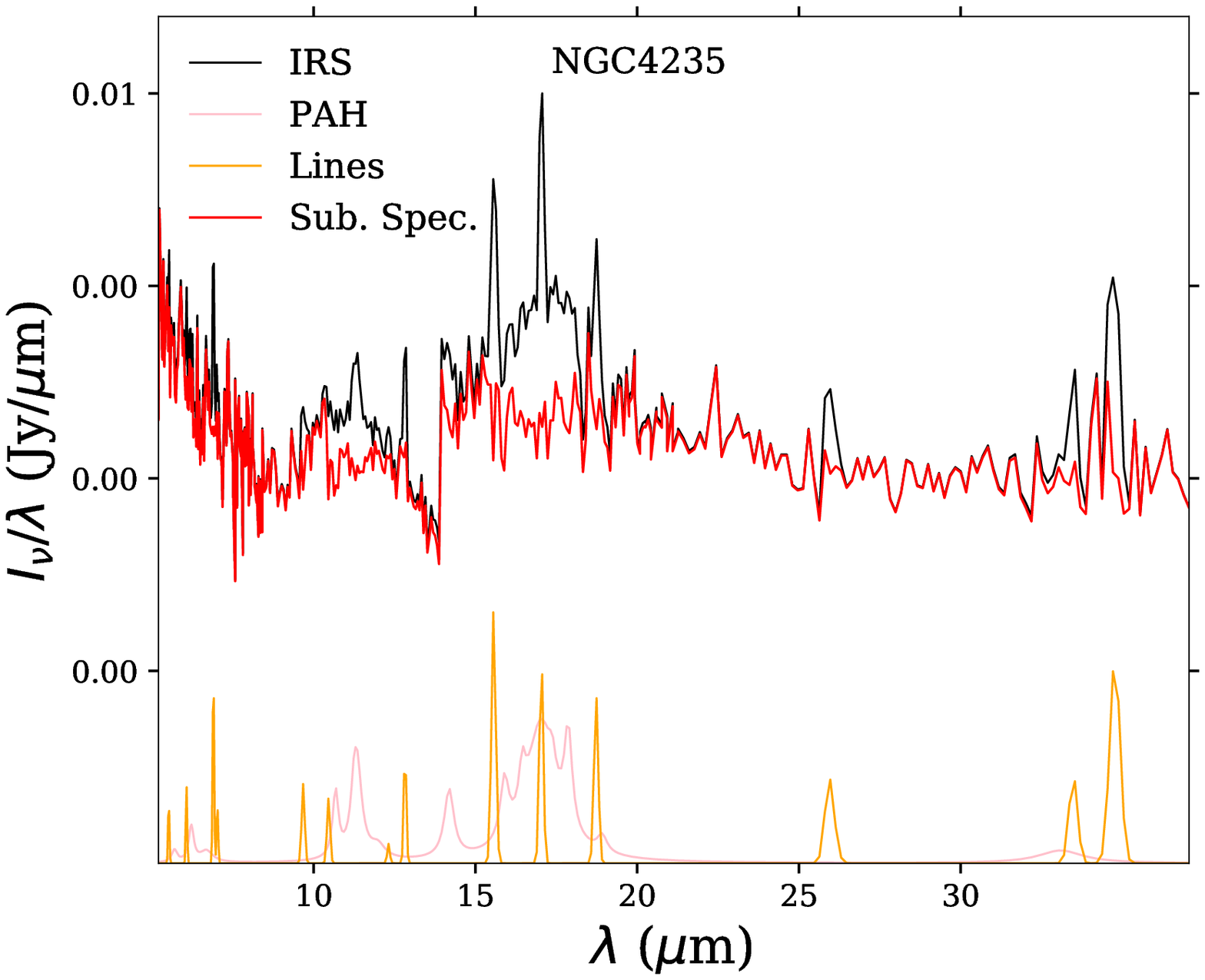}
\end{minipage} \hfill
\caption{continued from previous page.}
\setcounter{figure}{0}
\end{figure}

\begin{figure}

\begin{minipage}[b]{0.325\linewidth}
\includegraphics[width=\textwidth]{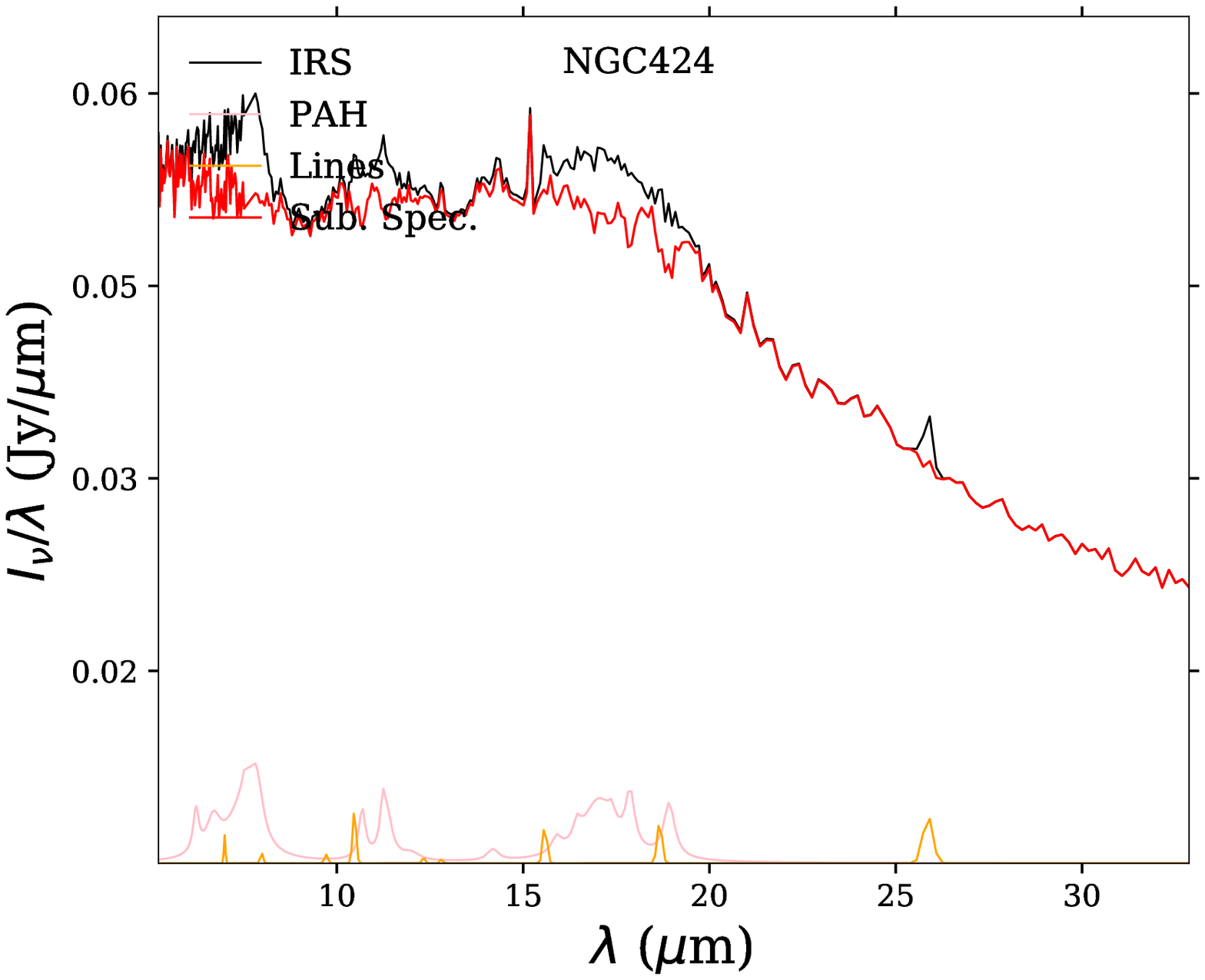}
\end{minipage} \hfill
\begin{minipage}[b]{0.325\linewidth}
\includegraphics[width=\textwidth]{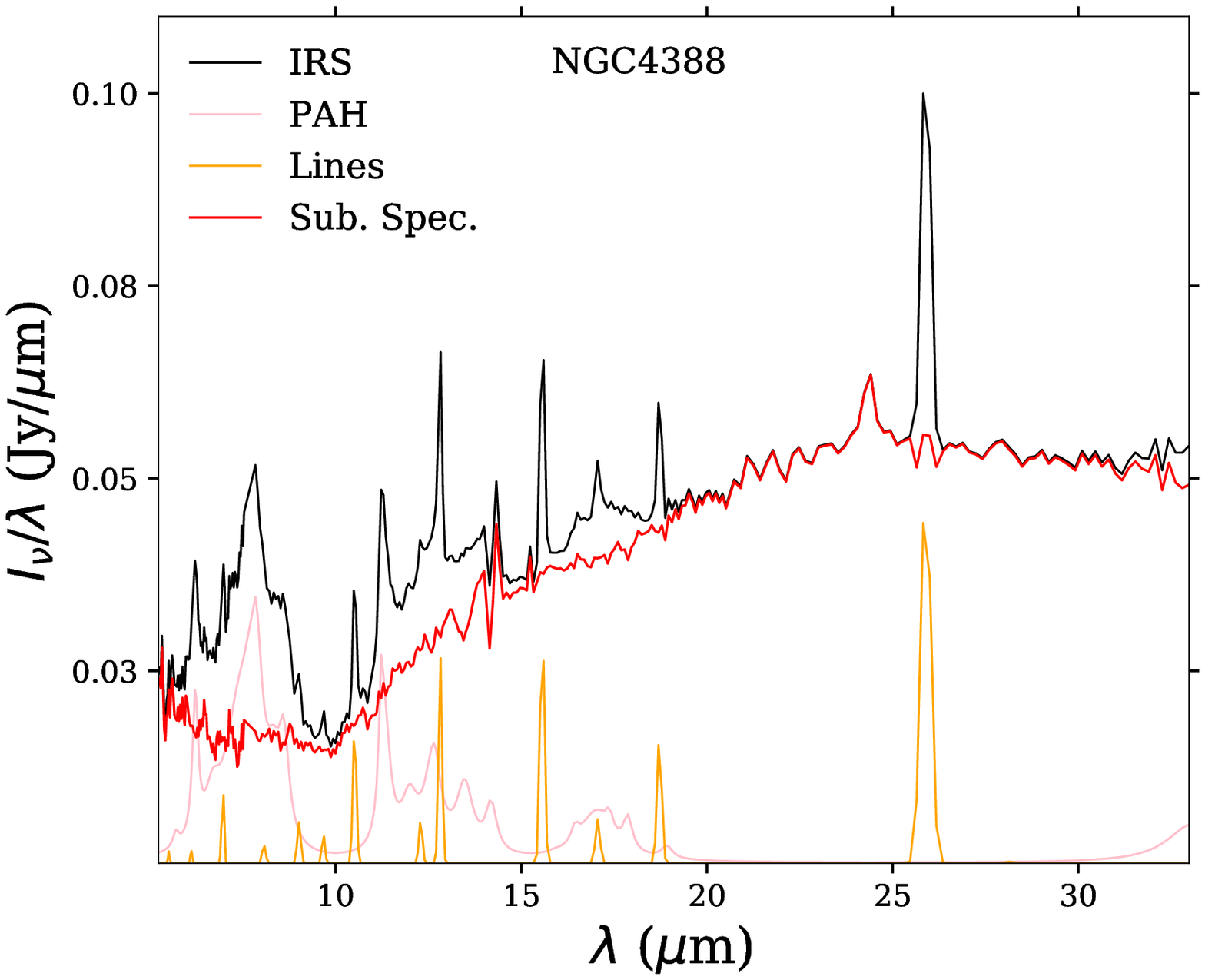}
\end{minipage} \hfill
\begin{minipage}[b]{0.325\linewidth}
\includegraphics[width=\textwidth]{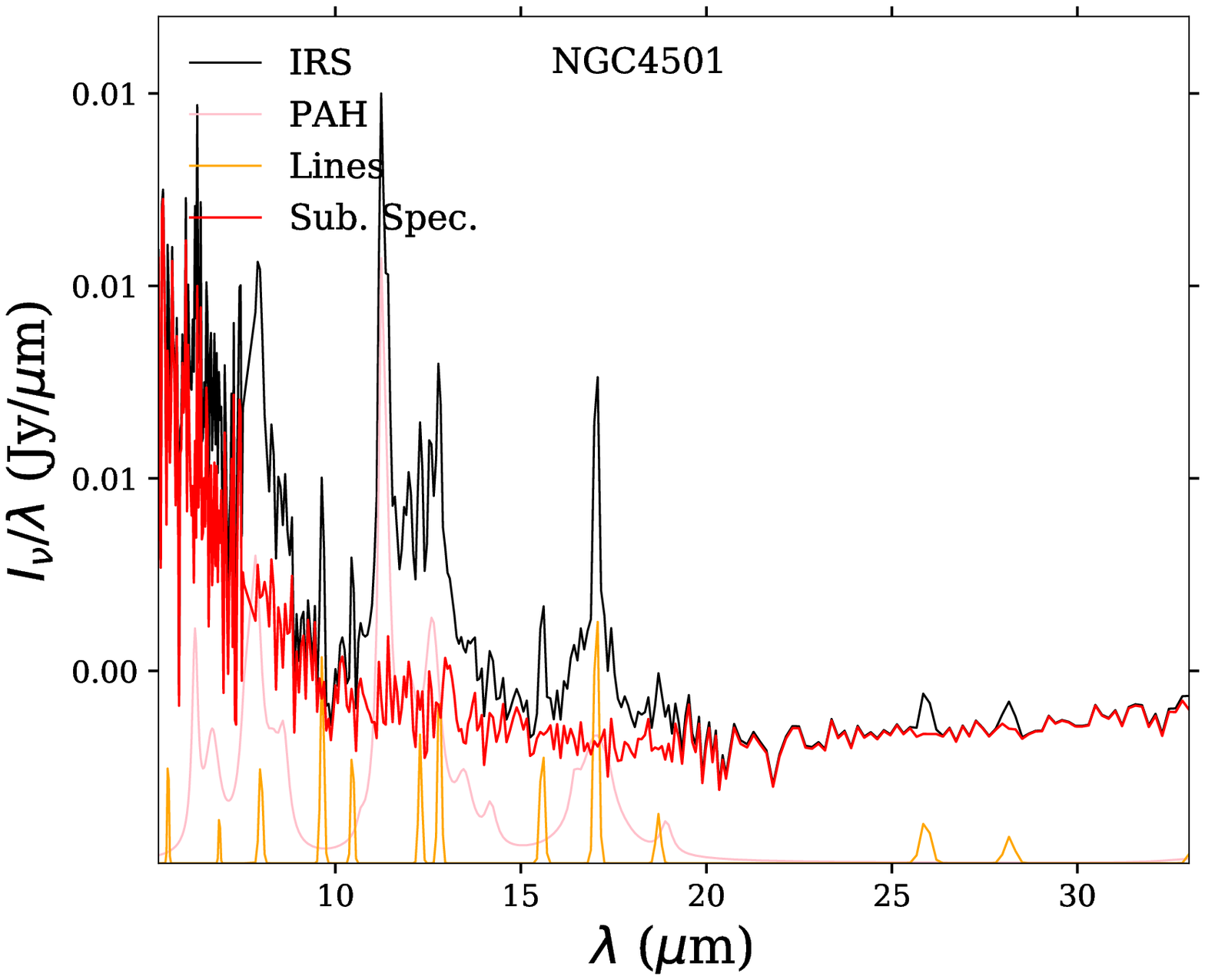}
\end{minipage} \hfill
\begin{minipage}[b]{0.325\linewidth}
\includegraphics[width=\textwidth]{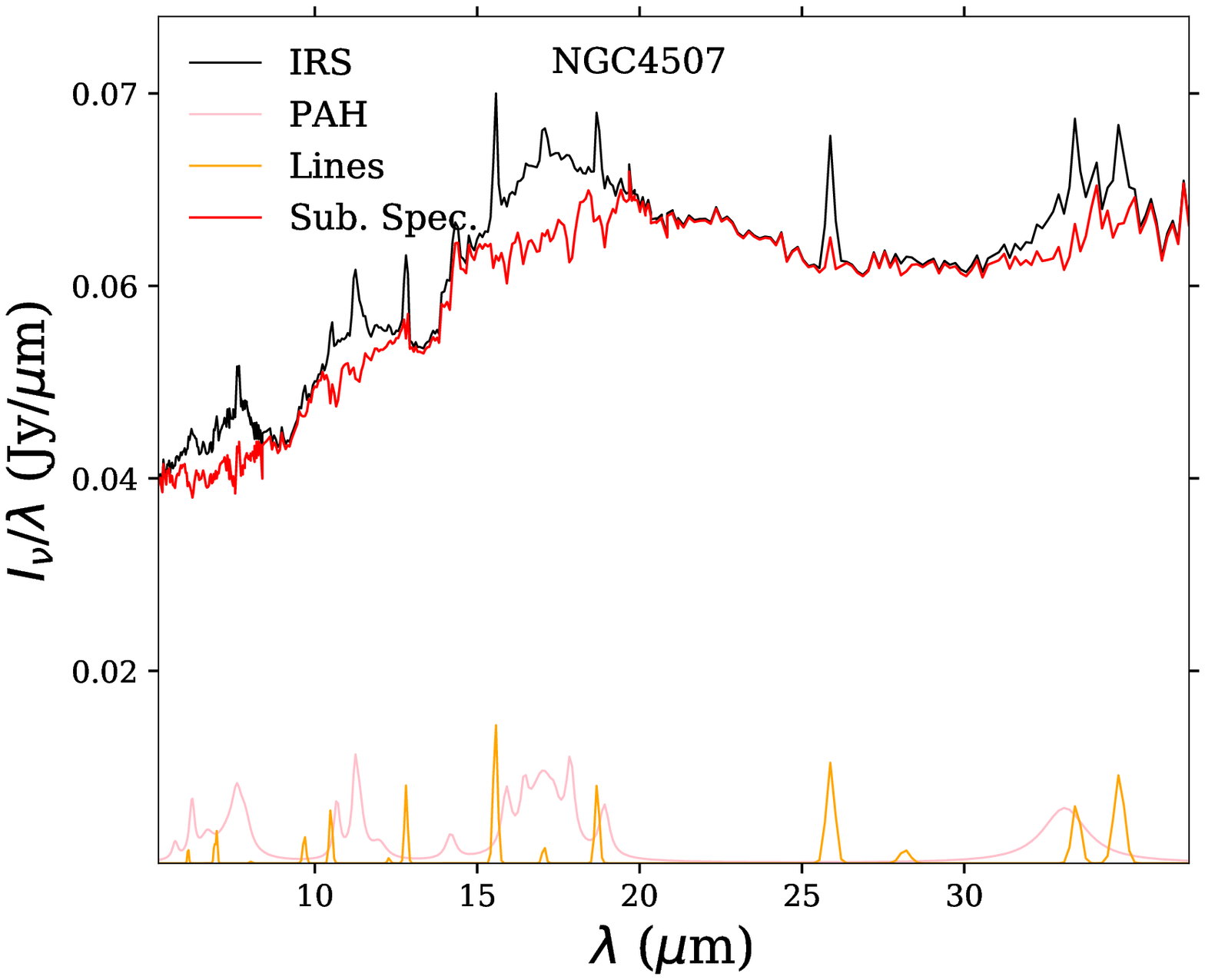}
\end{minipage} \hfill
\begin{minipage}[b]{0.325\linewidth}
\includegraphics[width=\textwidth]{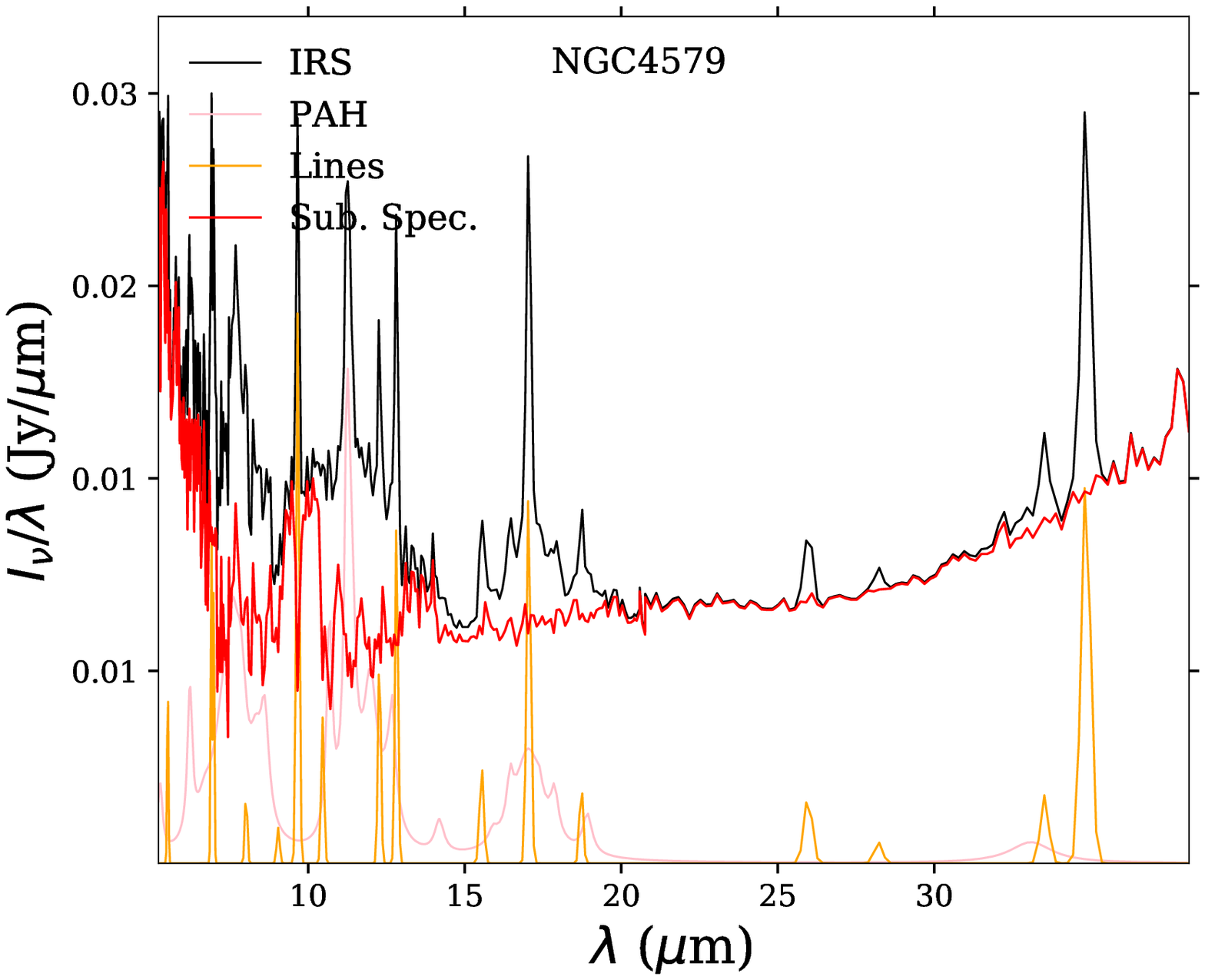}
\end{minipage} \hfill
\begin{minipage}[b]{0.325\linewidth}
\includegraphics[width=\textwidth]{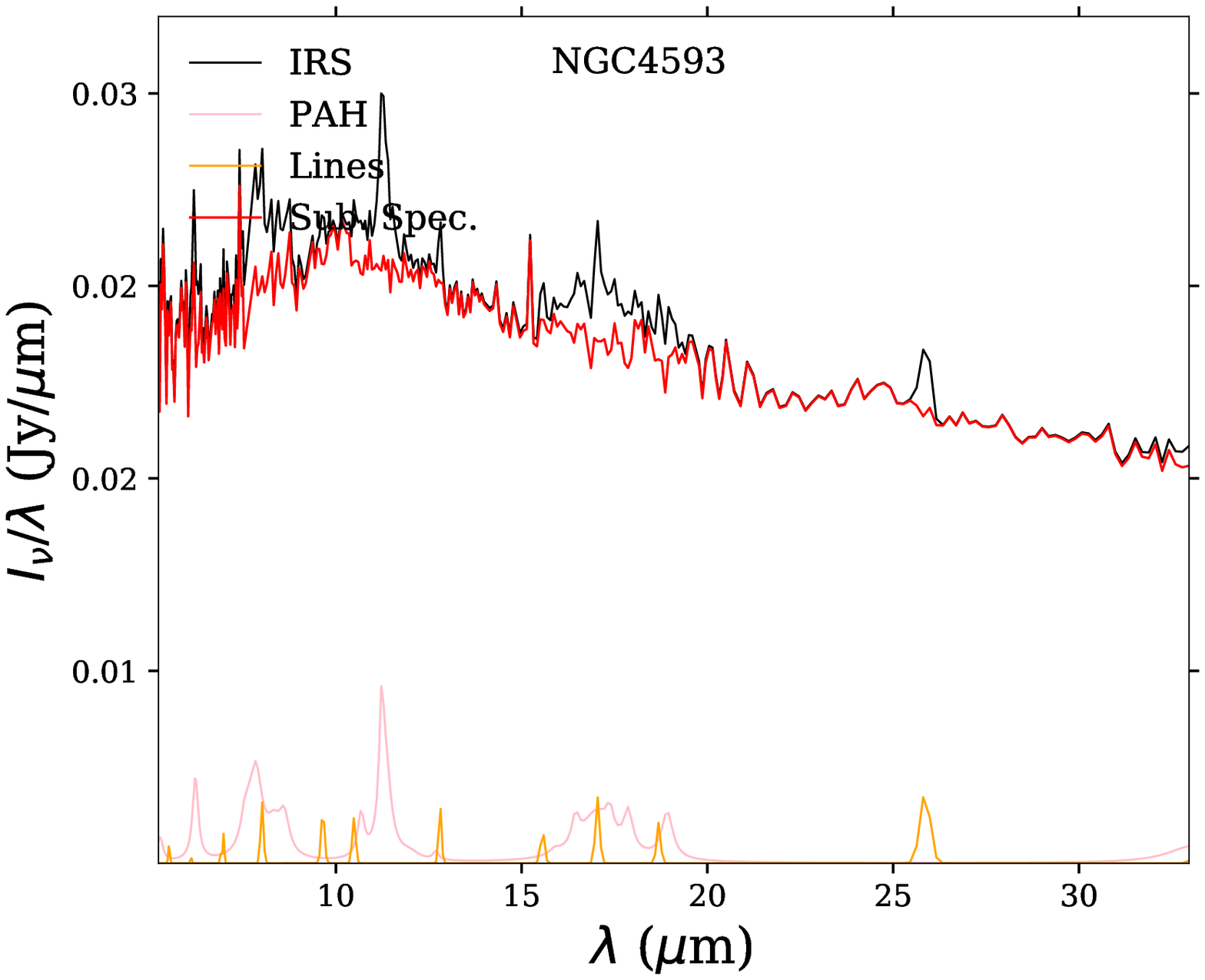}
\end{minipage} \hfill
\begin{minipage}[b]{0.325\linewidth}
\includegraphics[width=\textwidth]{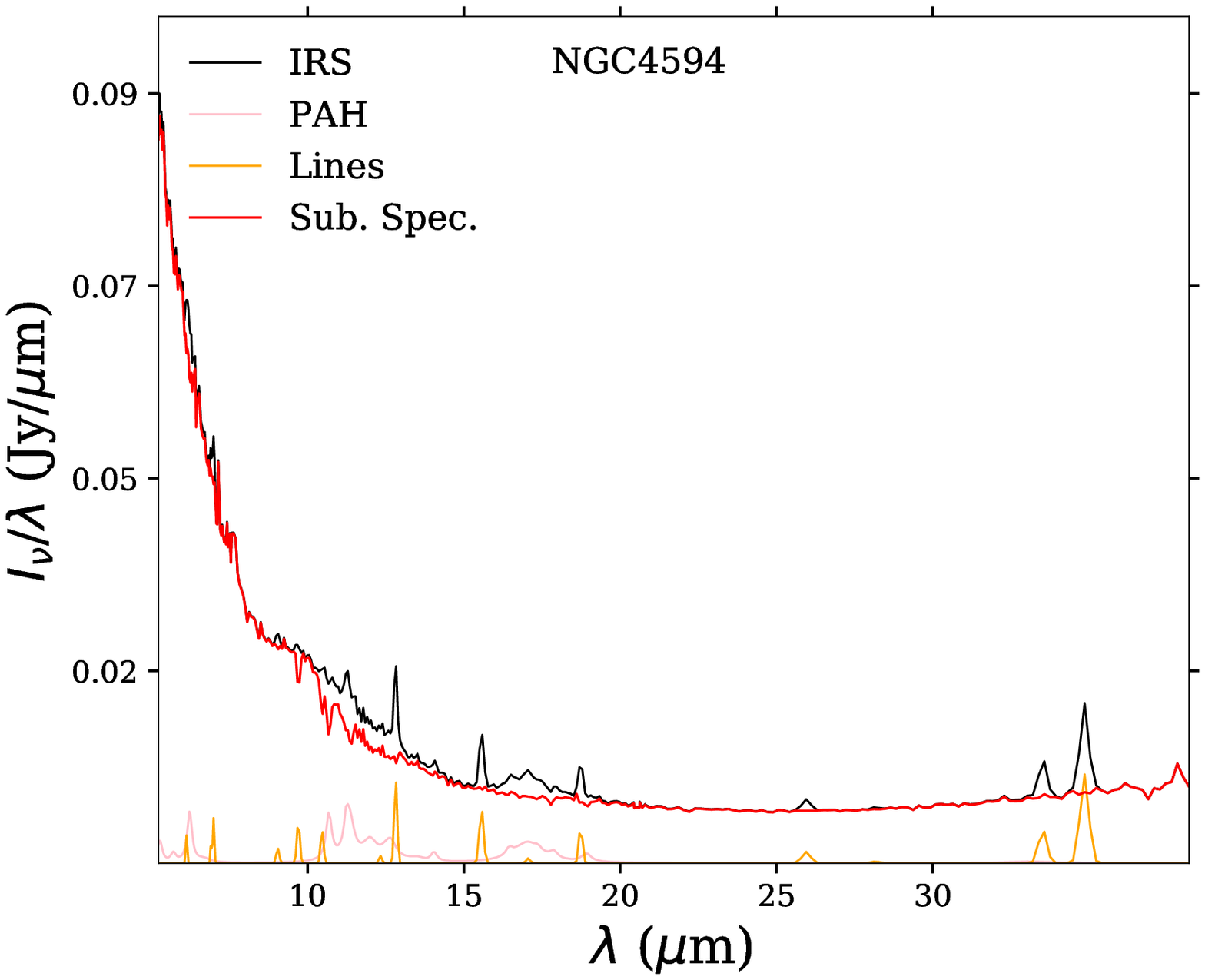}
\end{minipage} \hfill
\begin{minipage}[b]{0.325\linewidth}
\includegraphics[width=\textwidth]{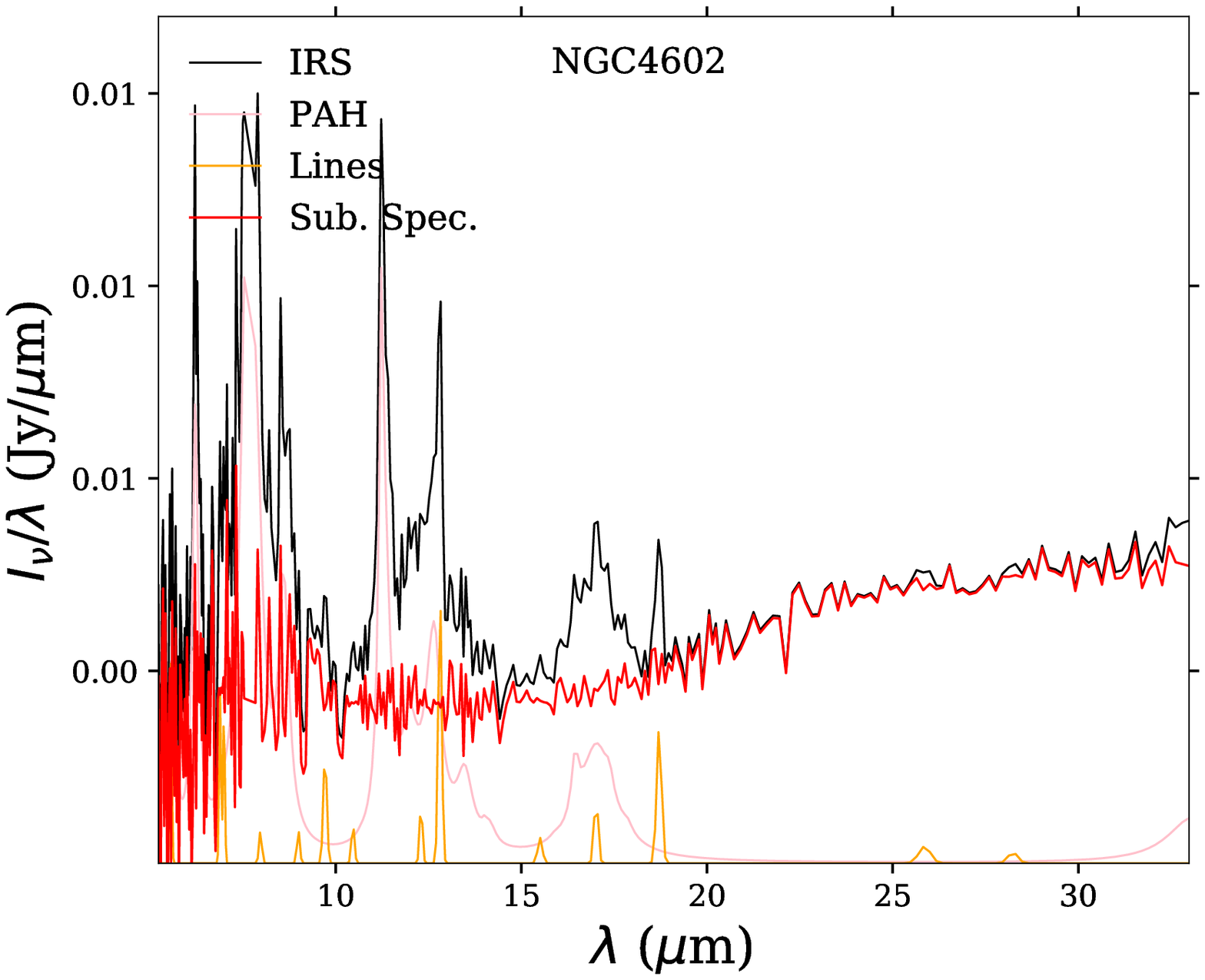}
\end{minipage} \hfill
\begin{minipage}[b]{0.325\linewidth}
\includegraphics[width=\textwidth]{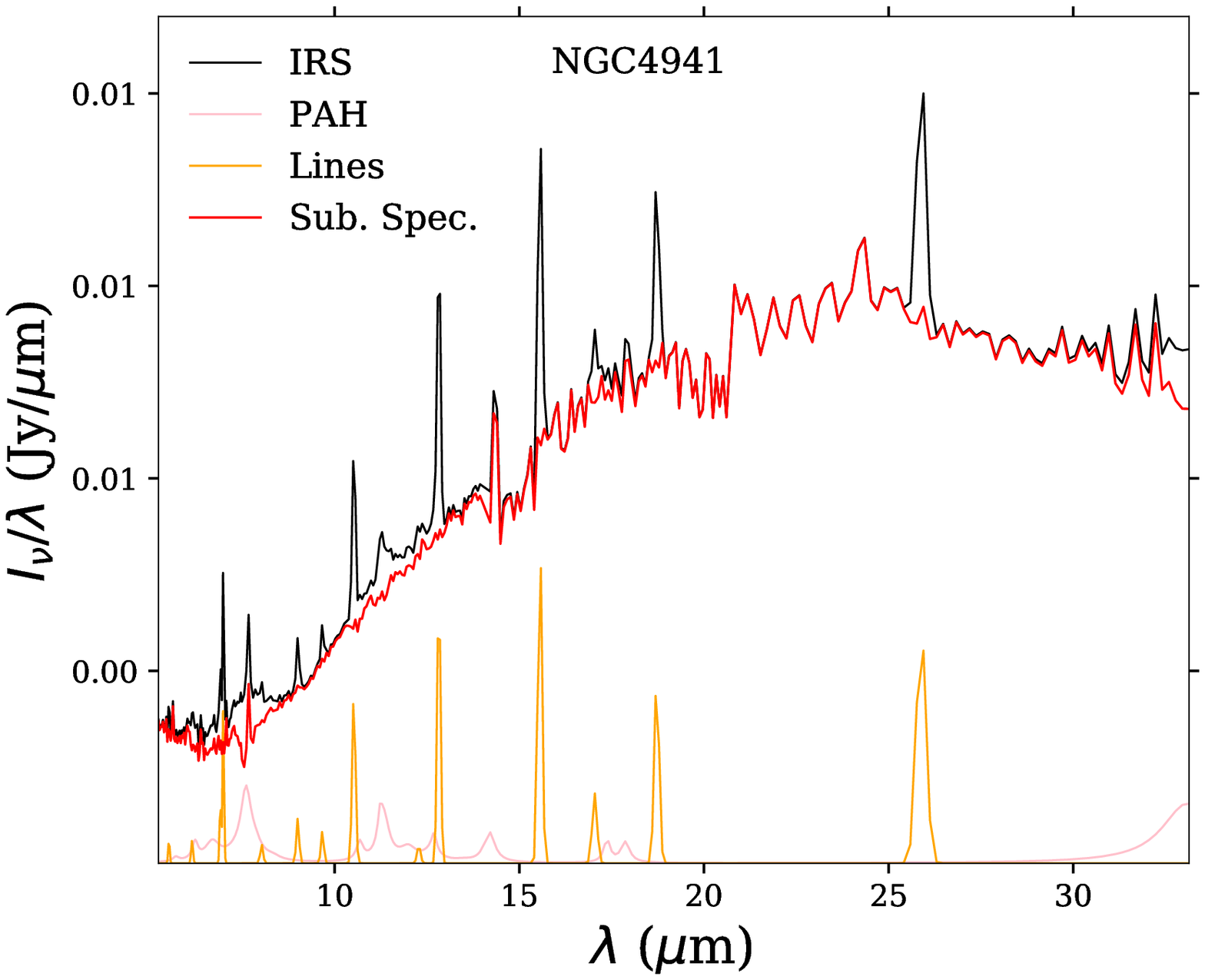}
\end{minipage} \hfill
\begin{minipage}[b]{0.325\linewidth}
\includegraphics[width=\textwidth]{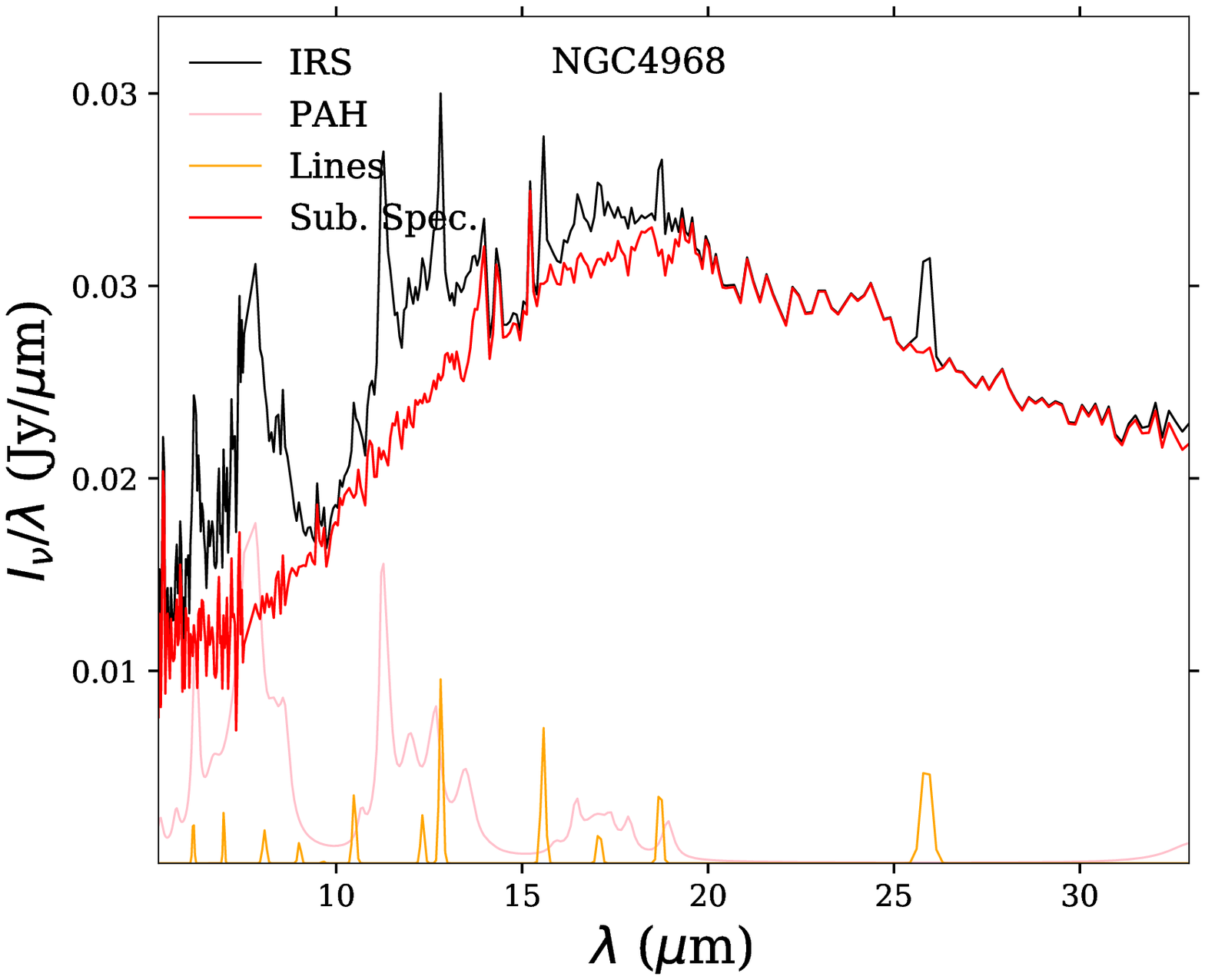}
\end{minipage} \hfill
\begin{minipage}[b]{0.325\linewidth}
\includegraphics[width=\textwidth]{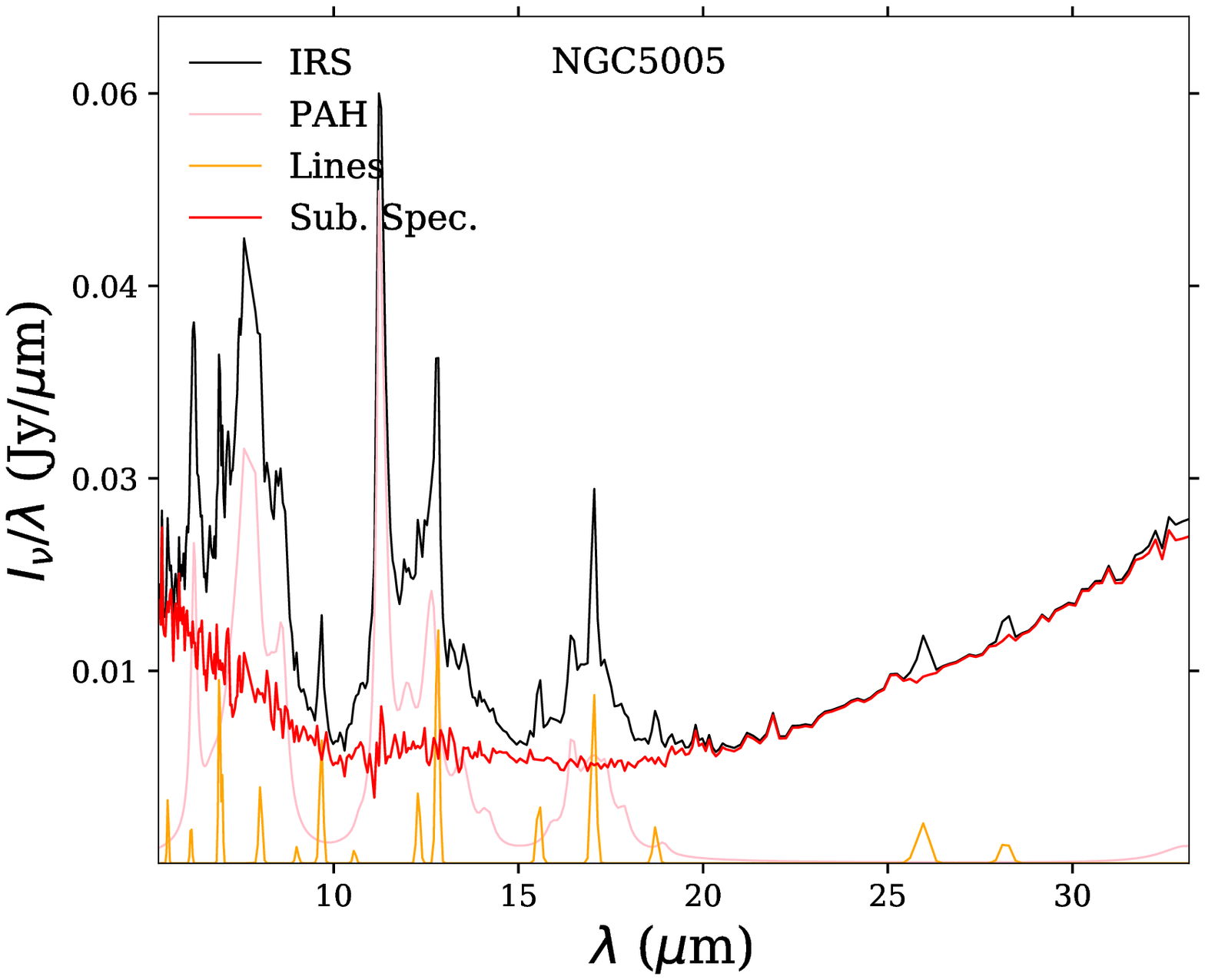}
\end{minipage} \hfill
\begin{minipage}[b]{0.325\linewidth}
\includegraphics[width=\textwidth]{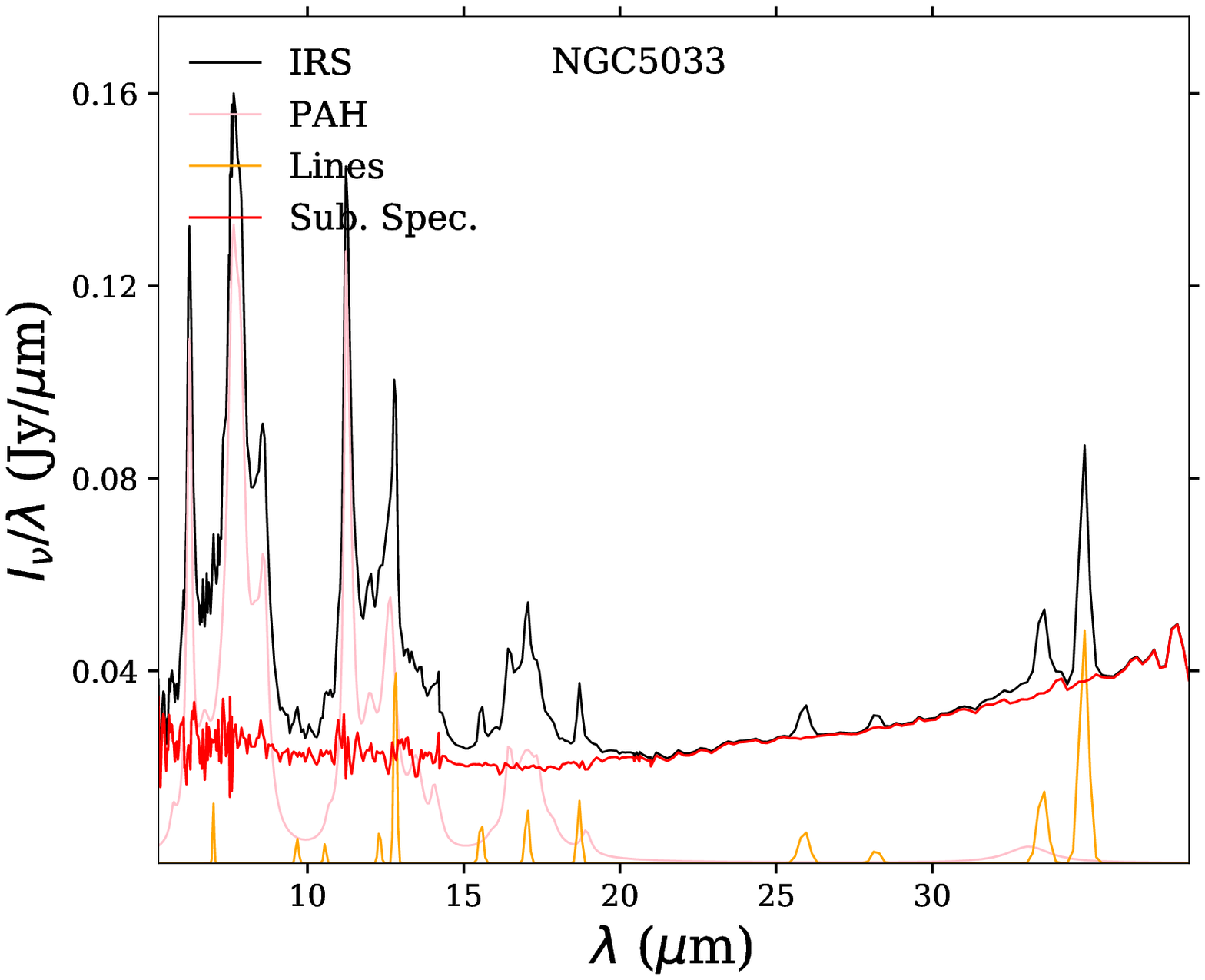}
\end{minipage} \hfill
\begin{minipage}[b]{0.325\linewidth}
\includegraphics[width=\textwidth]{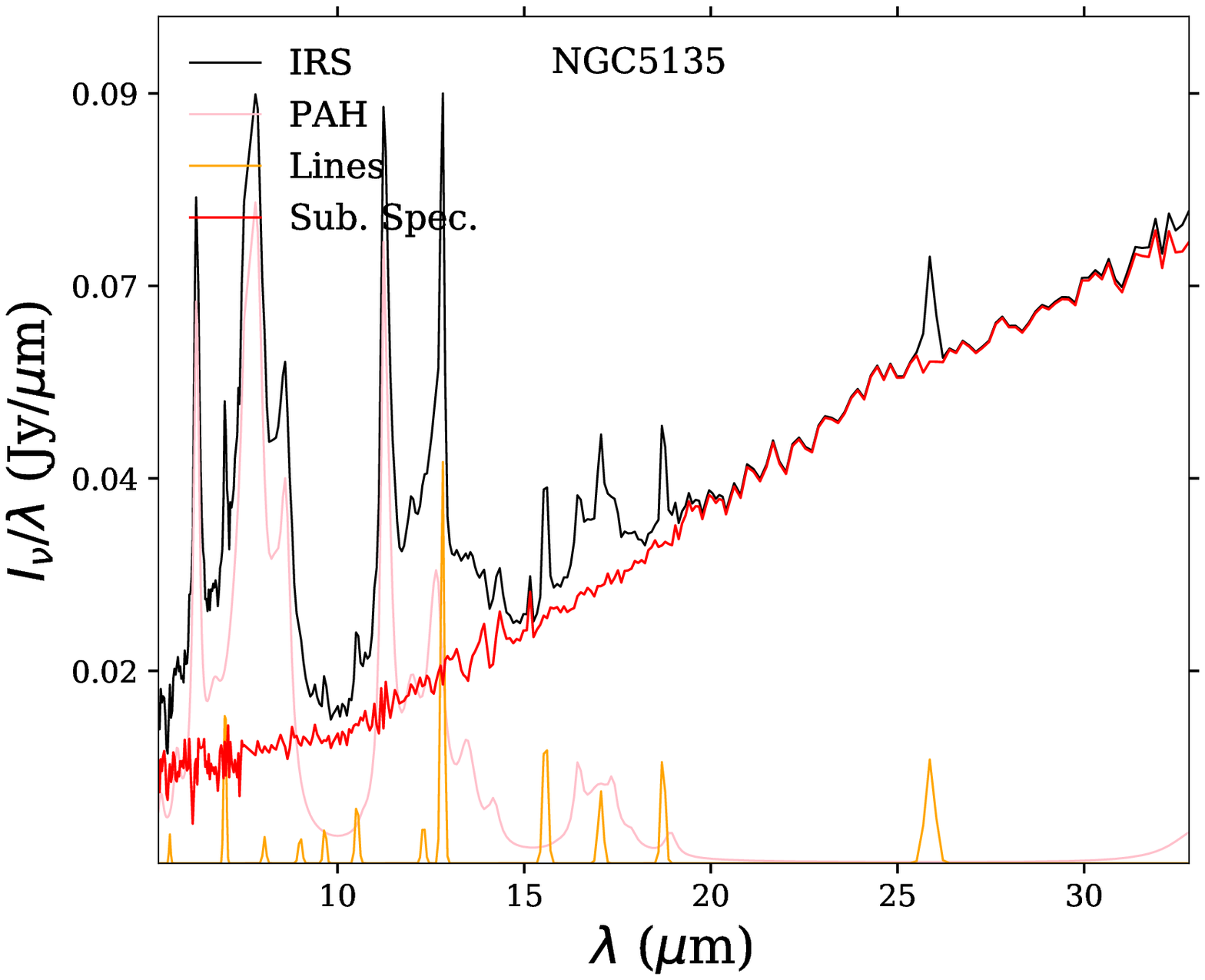}
\end{minipage} \hfill
\begin{minipage}[b]{0.325\linewidth}
\includegraphics[width=\textwidth]{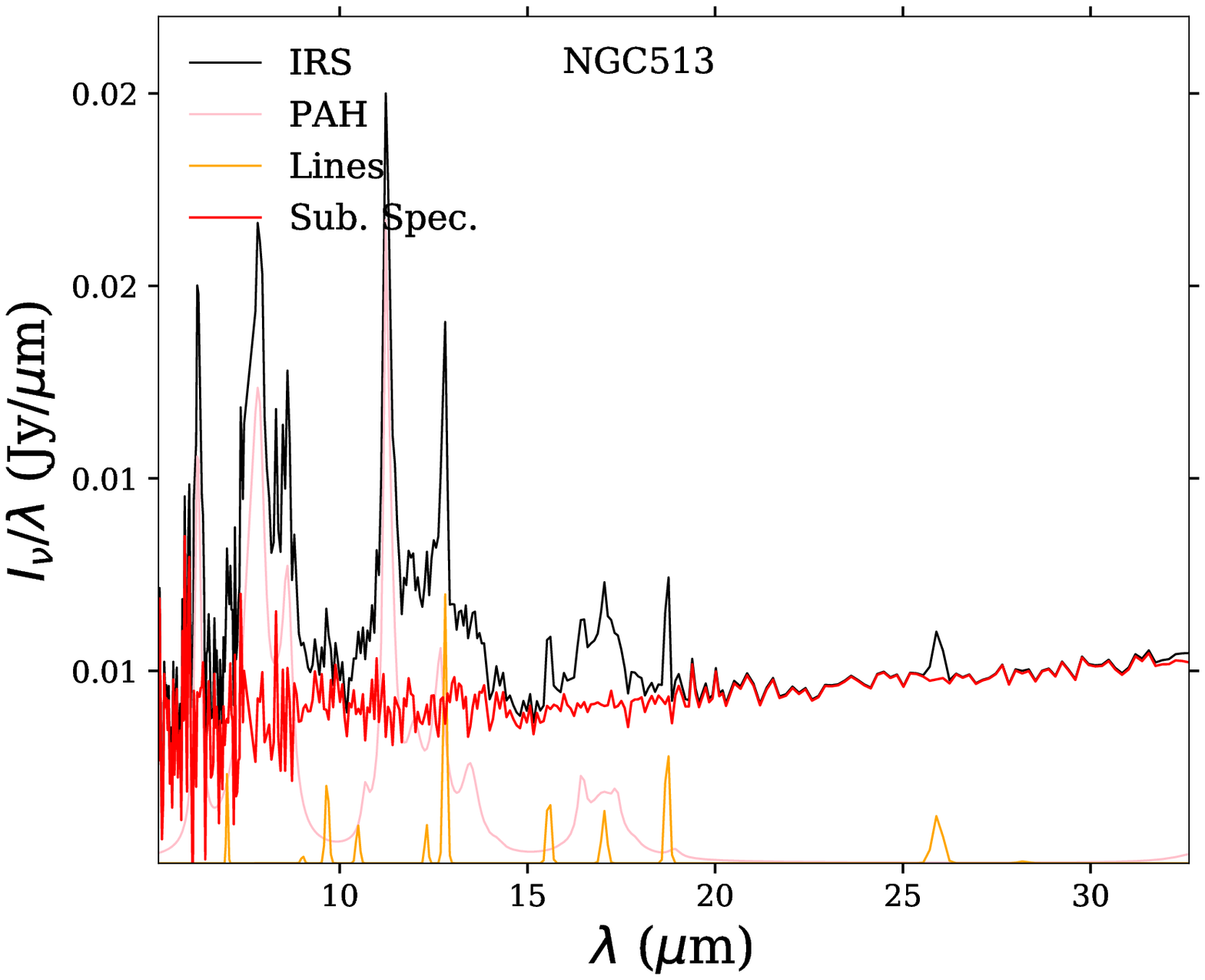}
\end{minipage} \hfill
\begin{minipage}[b]{0.325\linewidth}
\includegraphics[width=\textwidth]{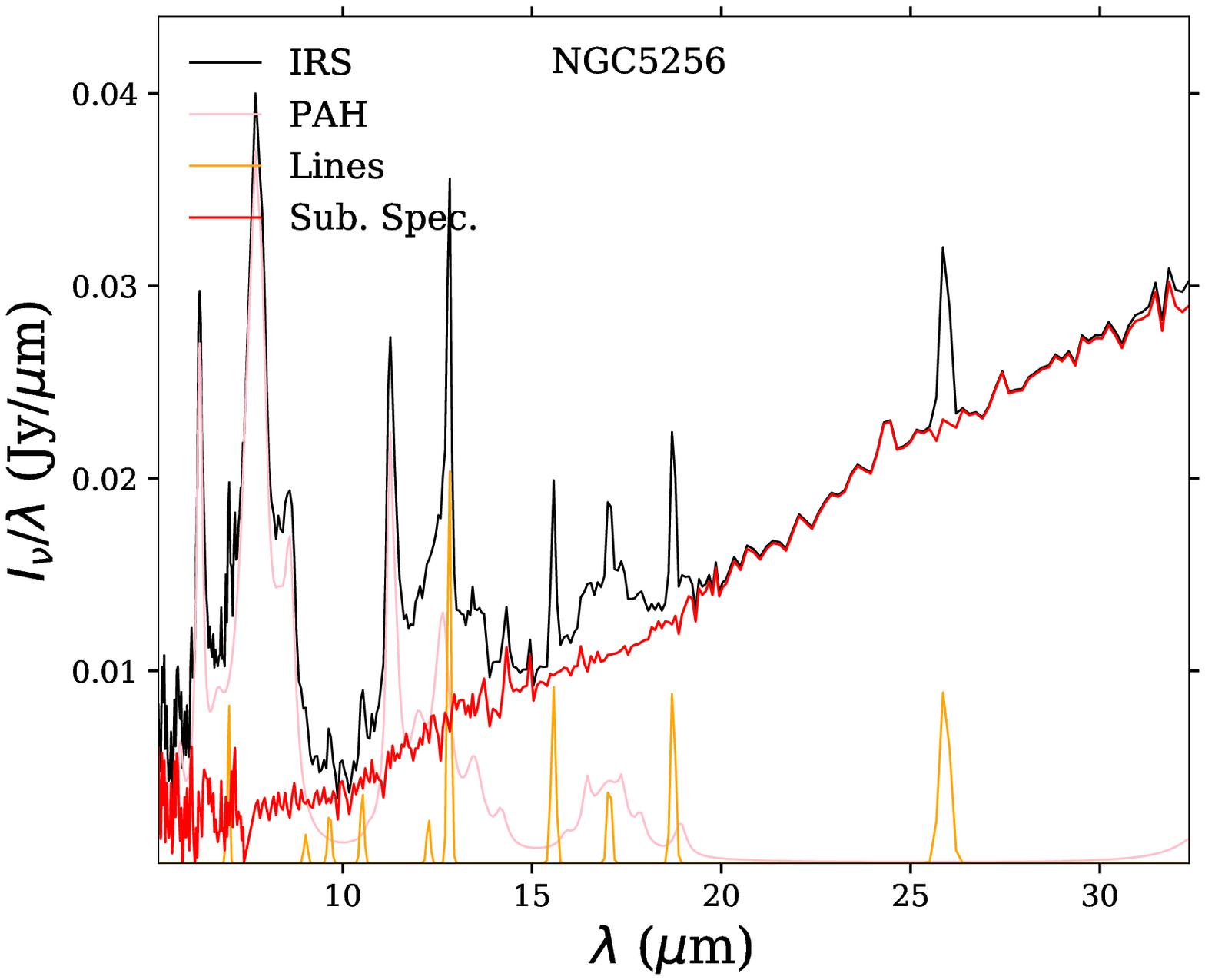}
\end{minipage} \hfill
\caption{continued from previous page.}
\setcounter{figure}{0}
\end{figure}

\begin{figure}

\begin{minipage}[b]{0.325\linewidth}
\includegraphics[width=\textwidth]{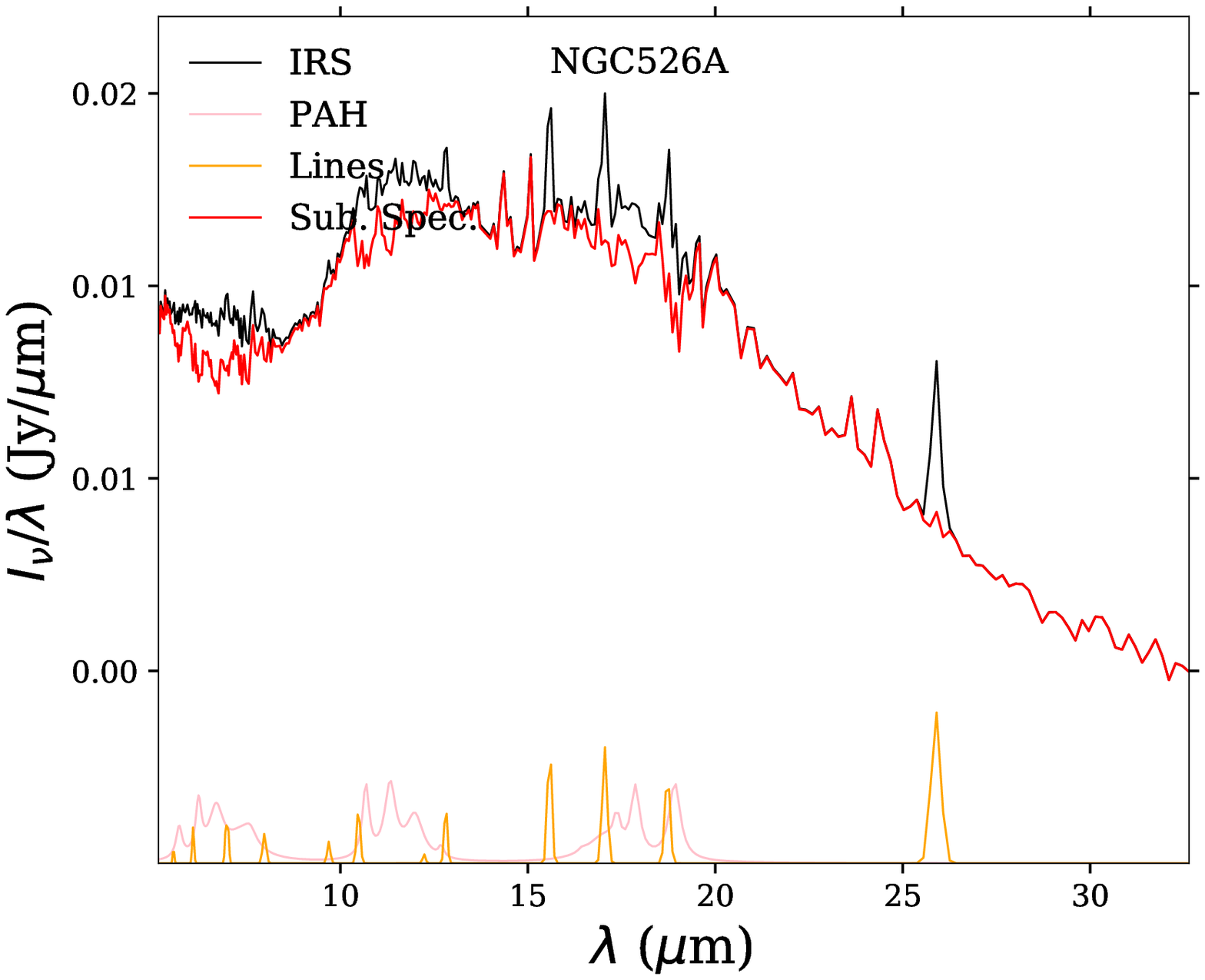}
\end{minipage} \hfill
\begin{minipage}[b]{0.325\linewidth}
\includegraphics[width=\textwidth]{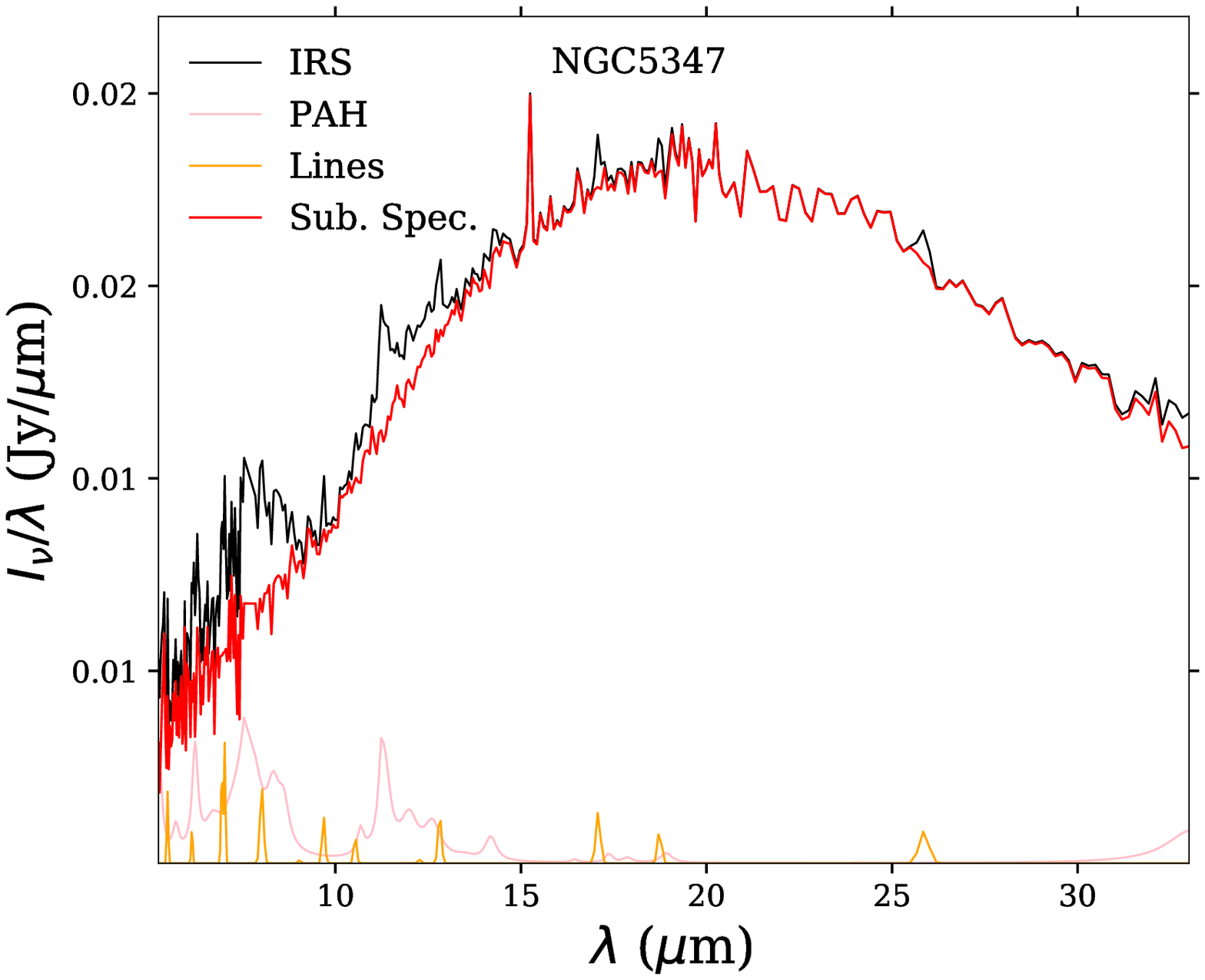}
\end{minipage} \hfill
\begin{minipage}[b]{0.325\linewidth}
\includegraphics[width=\textwidth]{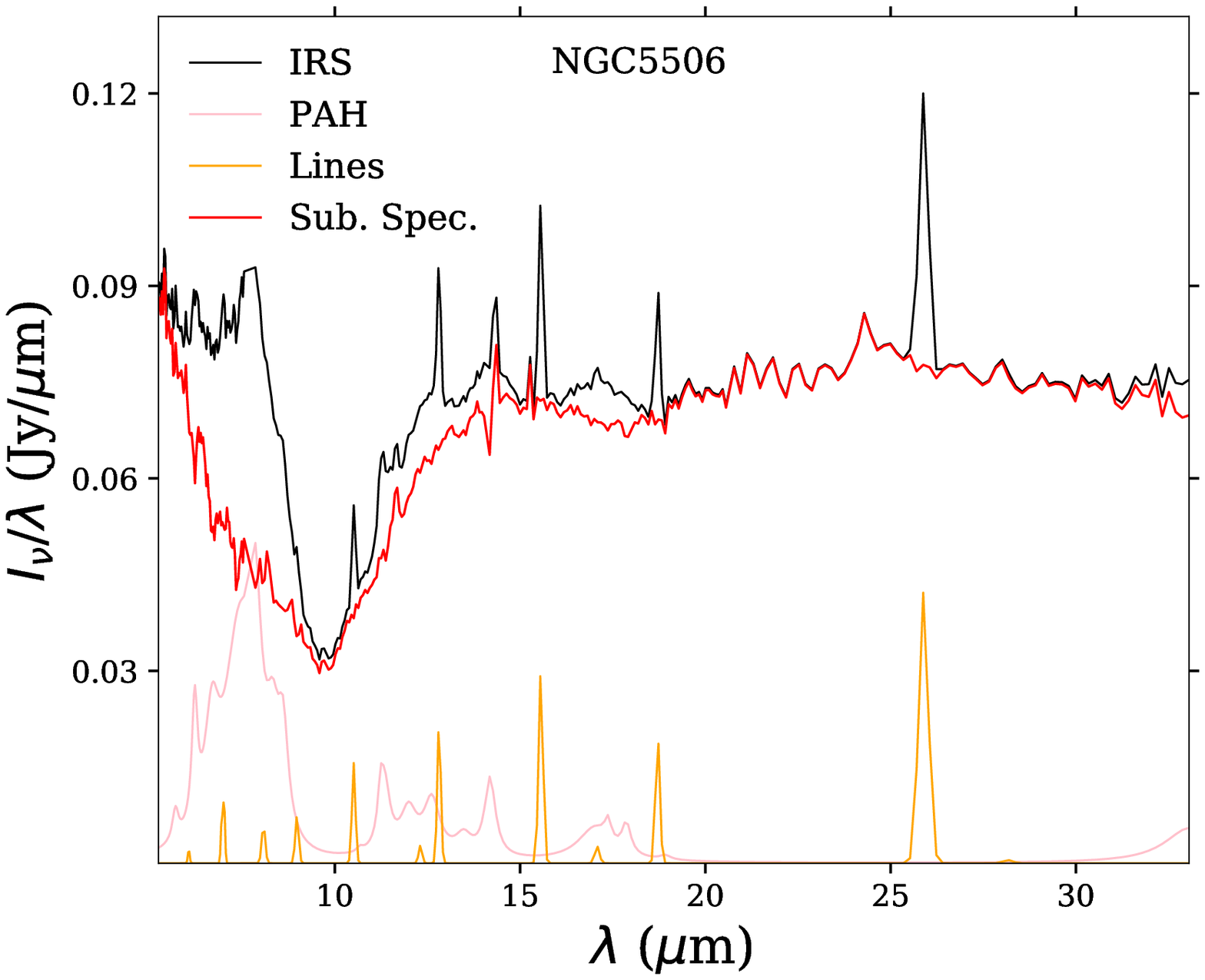}
\end{minipage} \hfill
\begin{minipage}[b]{0.325\linewidth}
\includegraphics[width=\textwidth]{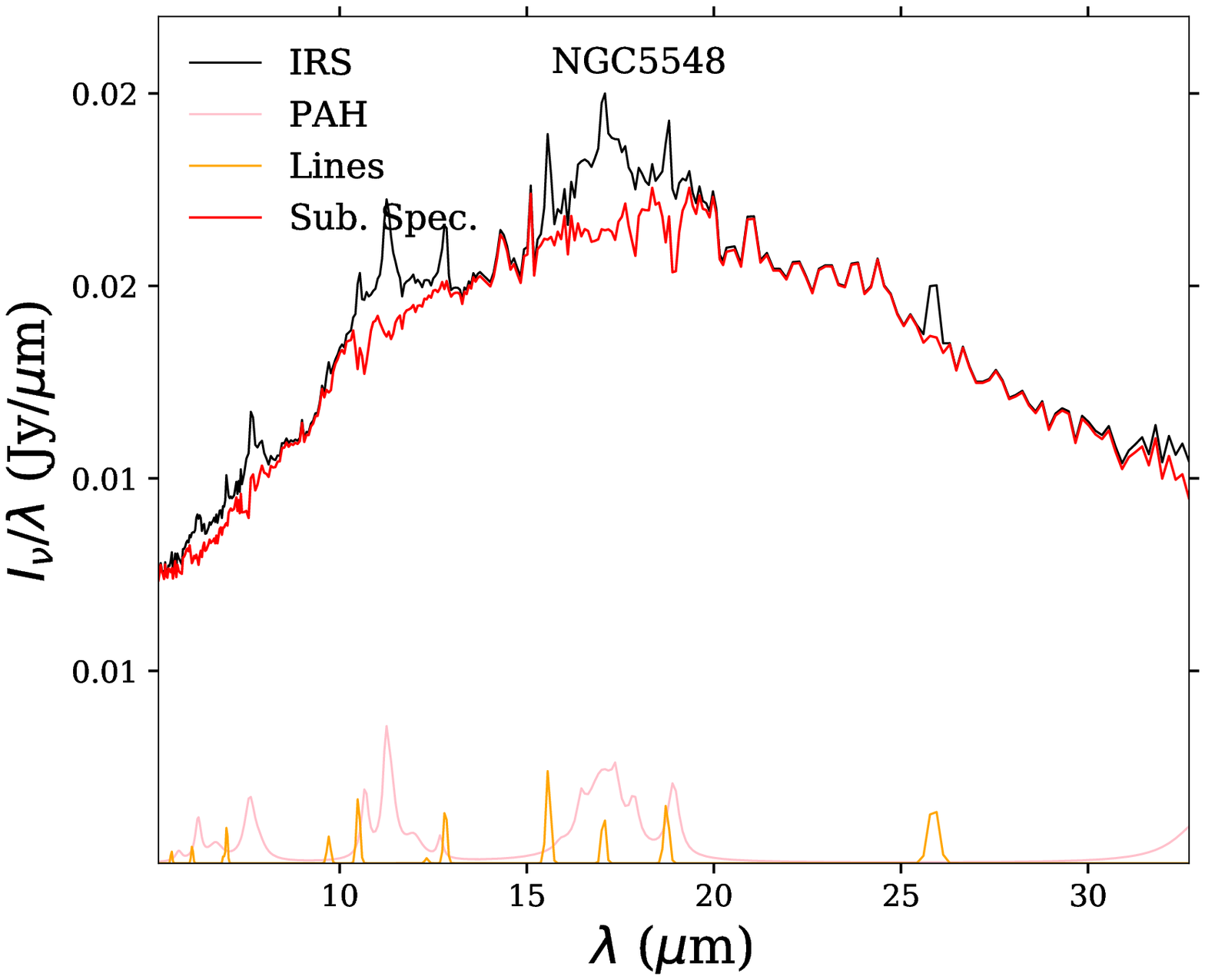}
\end{minipage} \hfill
\begin{minipage}[b]{0.325\linewidth}
\includegraphics[width=\textwidth]{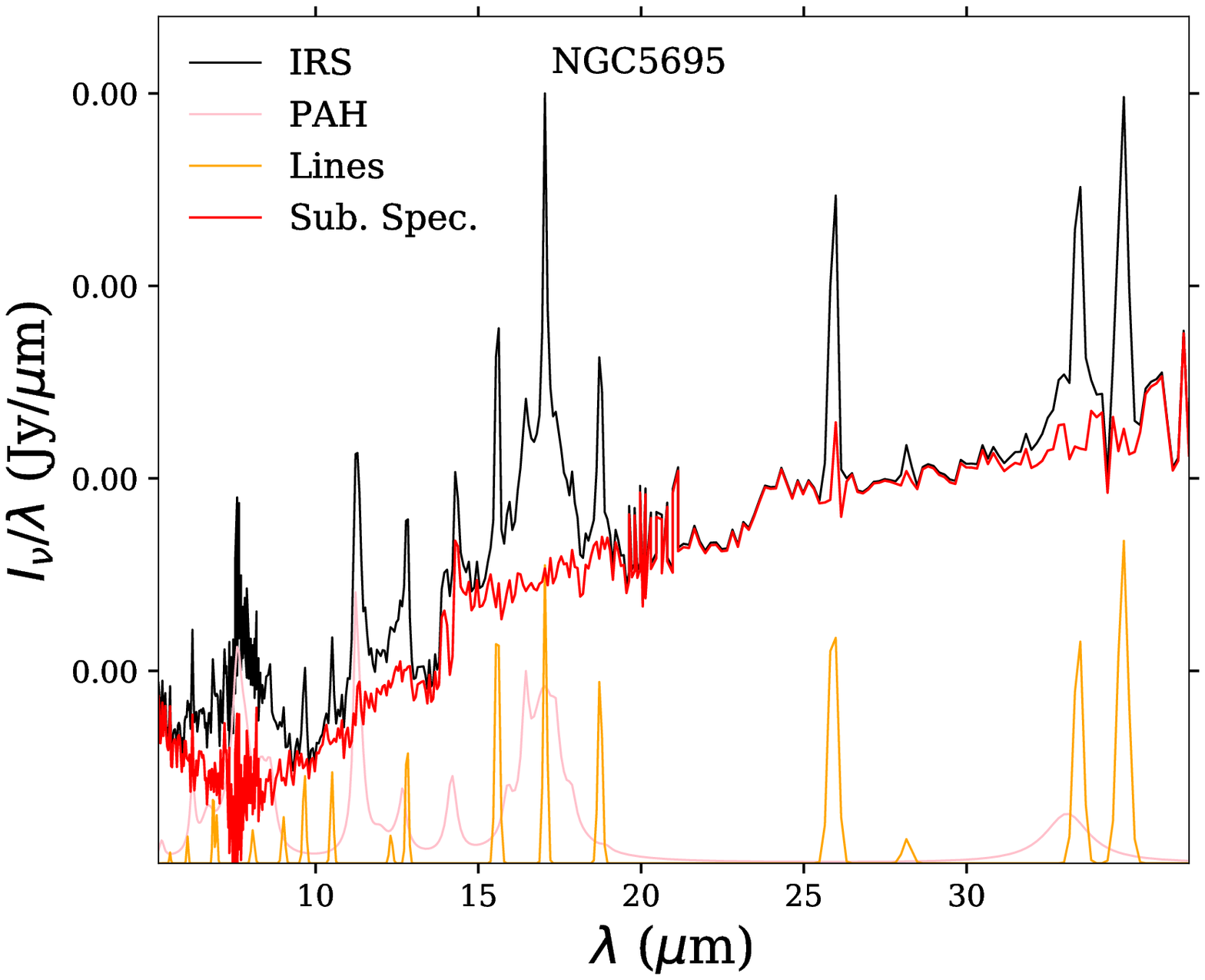}
\end{minipage} \hfill
\begin{minipage}[b]{0.325\linewidth}
\includegraphics[width=\textwidth]{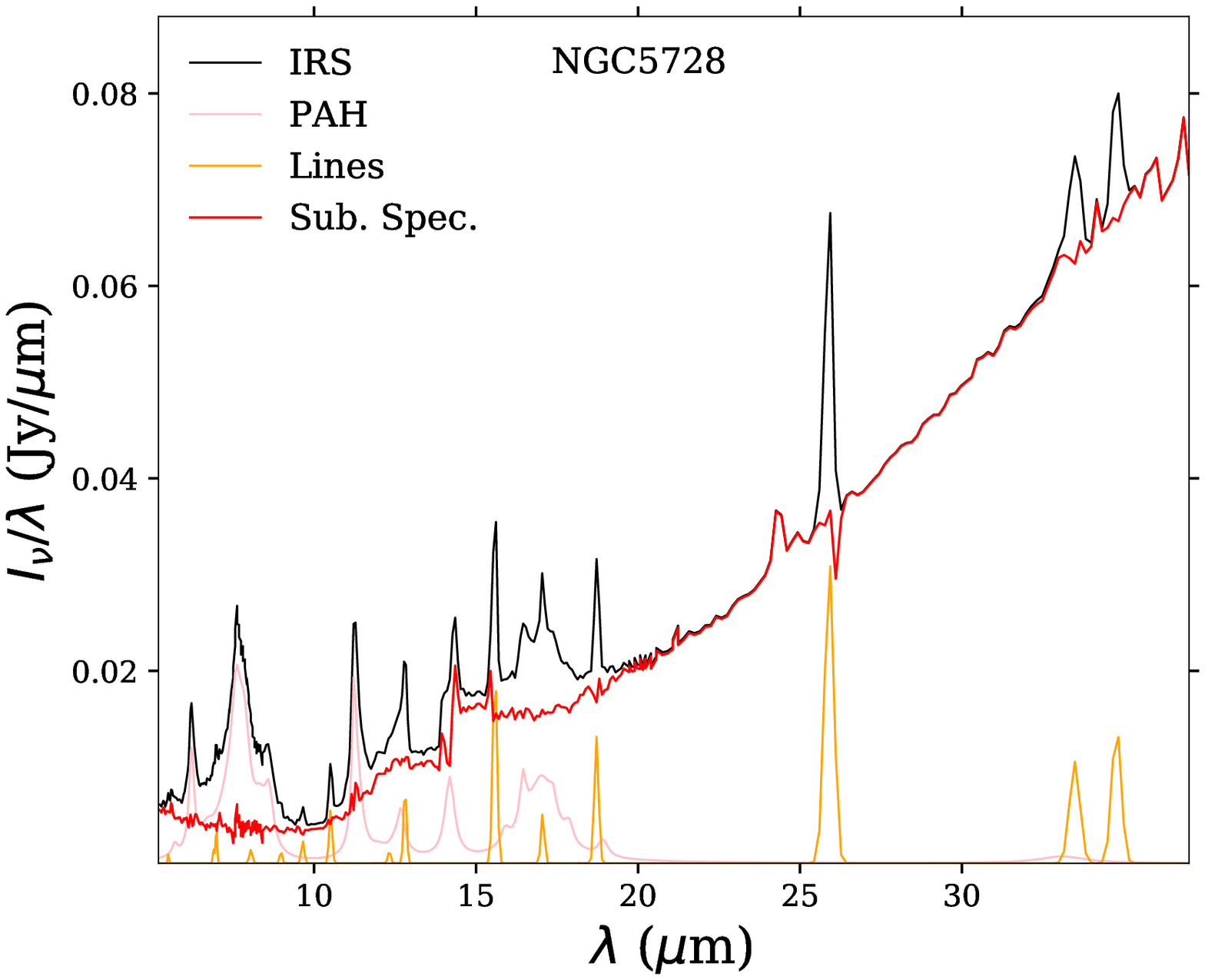}
\end{minipage} \hfill
\begin{minipage}[b]{0.325\linewidth}
\includegraphics[width=\textwidth]{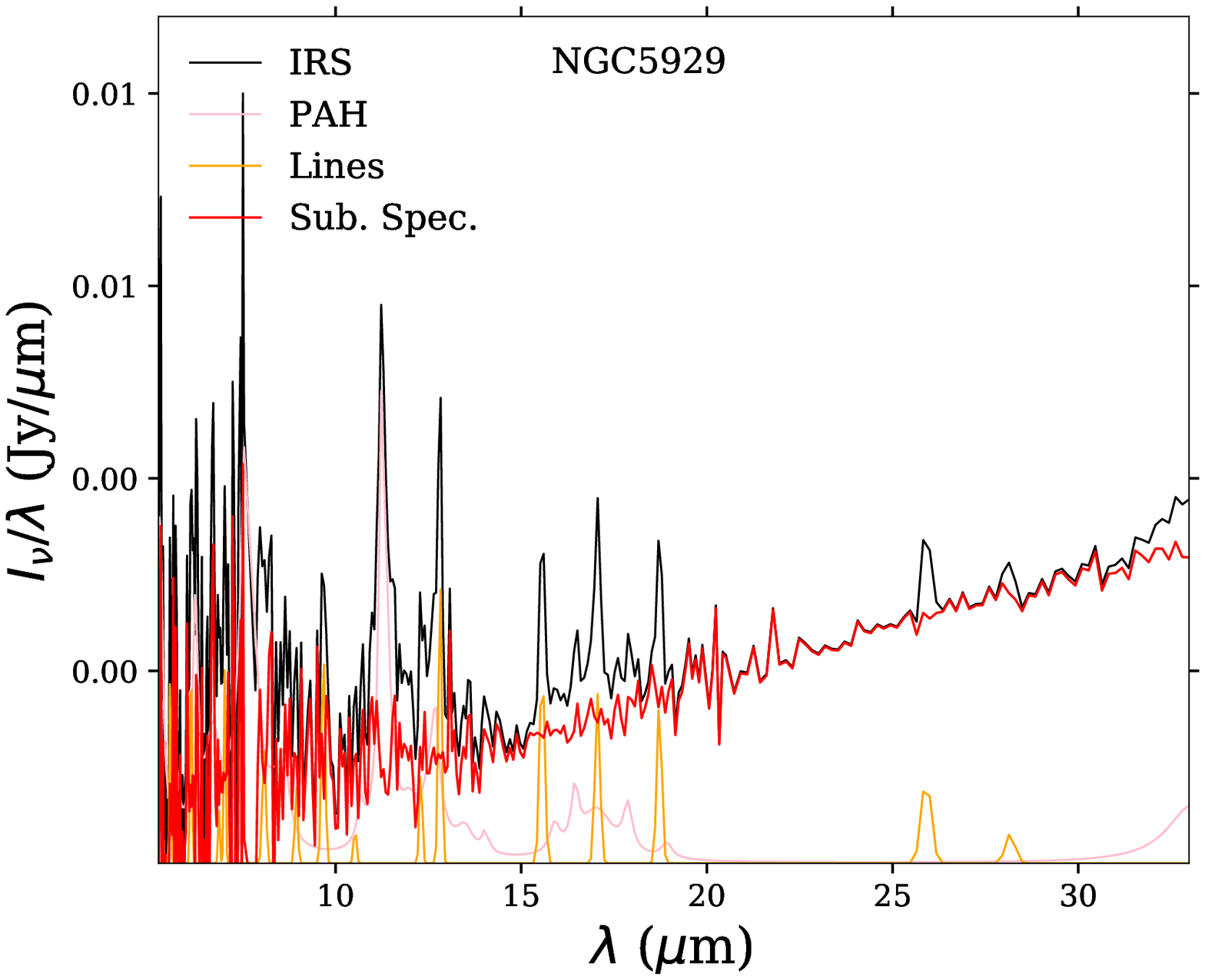}
\end{minipage} \hfill
\begin{minipage}[b]{0.325\linewidth}
\includegraphics[width=\textwidth]{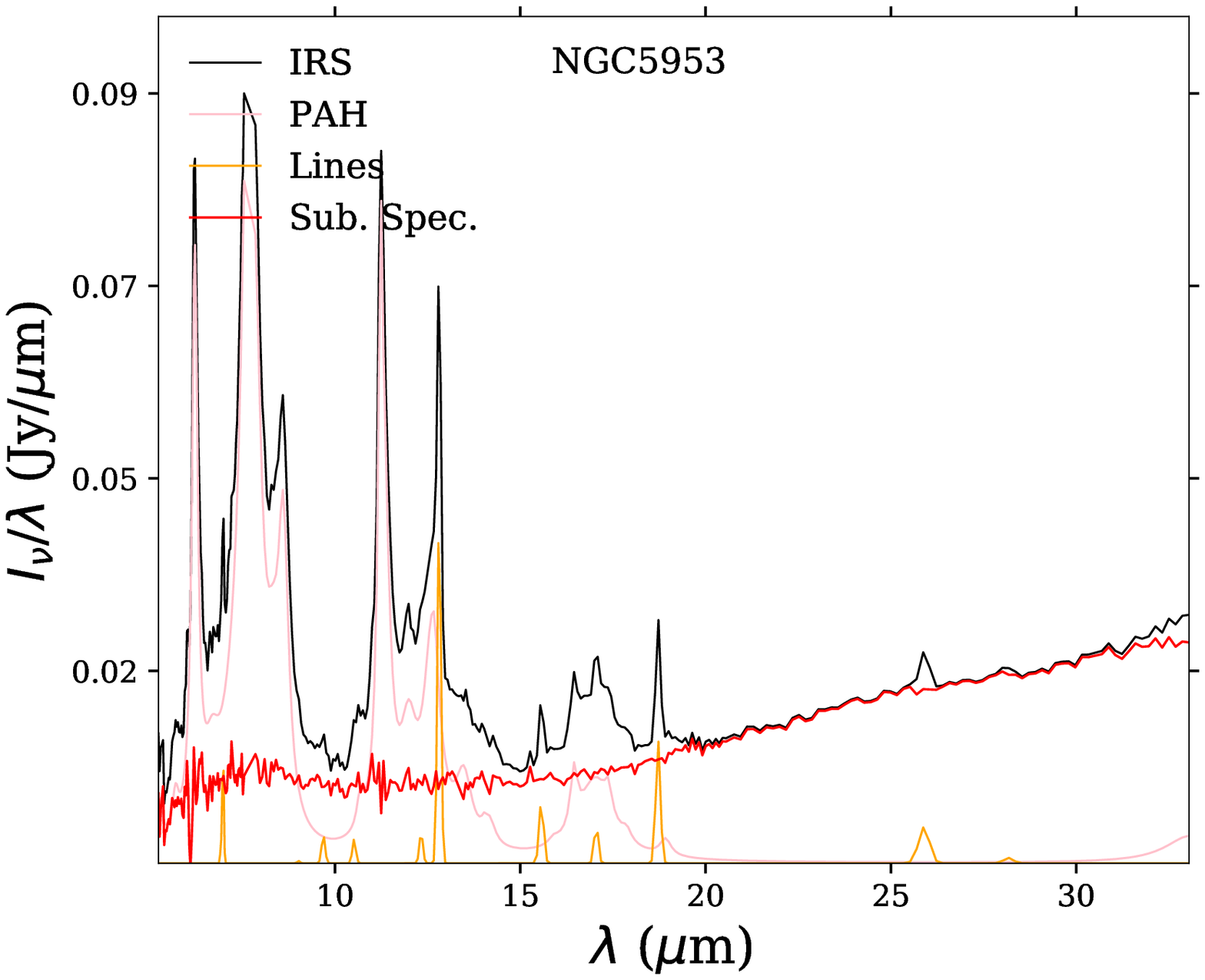}
\end{minipage} \hfill
\begin{minipage}[b]{0.325\linewidth}
\includegraphics[width=\textwidth]{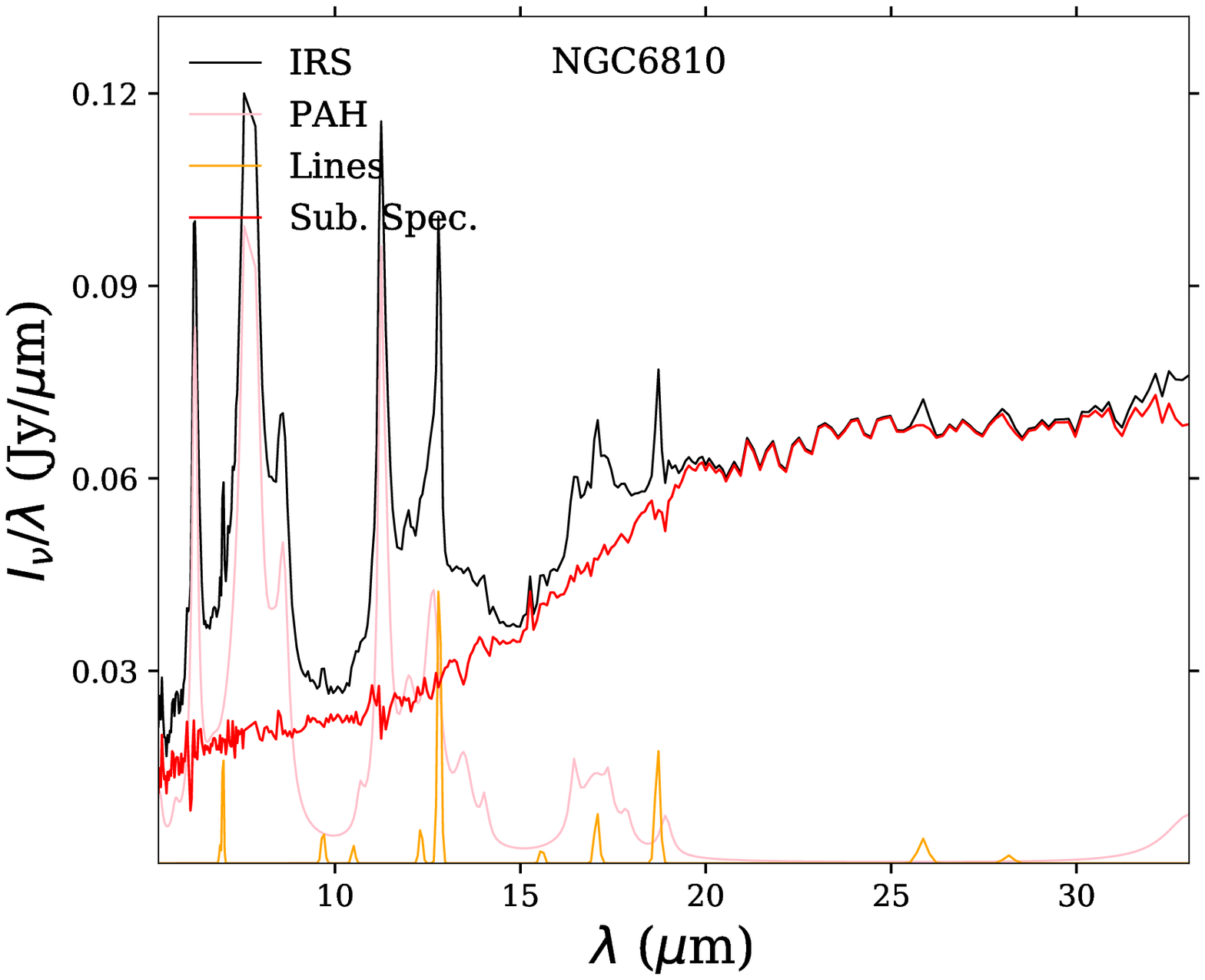}
\end{minipage} \hfill
\begin{minipage}[b]{0.325\linewidth}
\includegraphics[width=\textwidth]{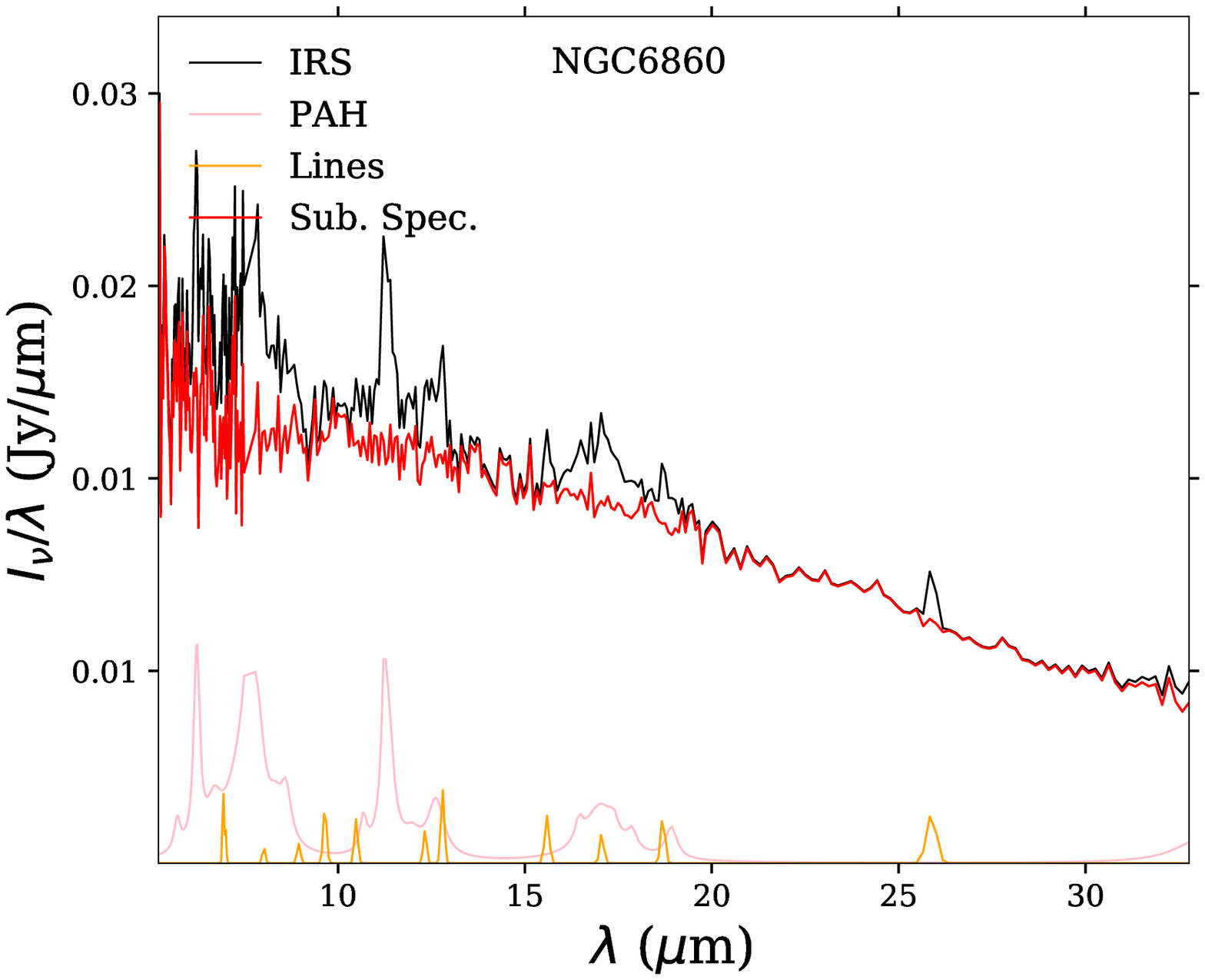}
\end{minipage} \hfill
\begin{minipage}[b]{0.325\linewidth}
\includegraphics[width=\textwidth]{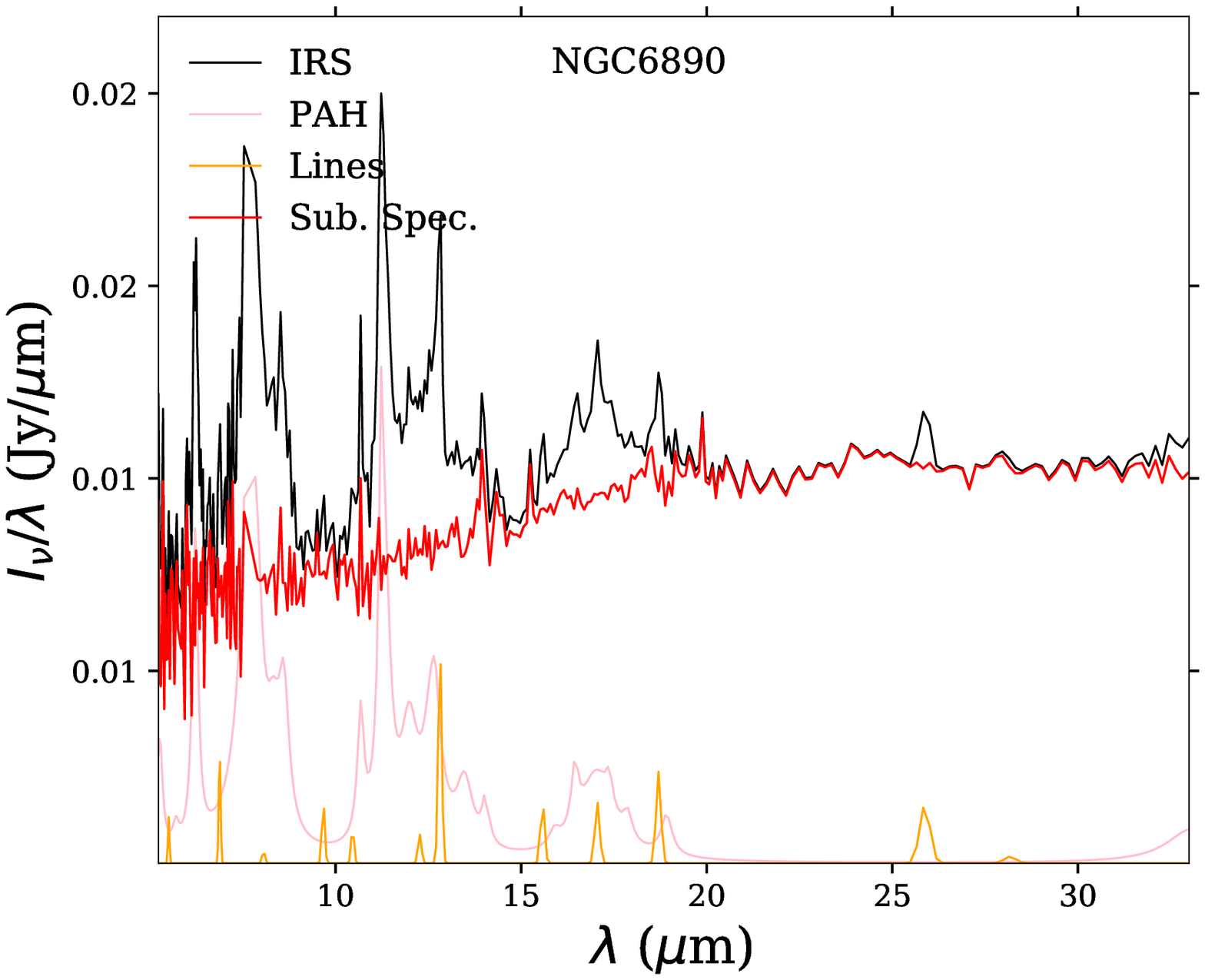}
\end{minipage} \hfill
\begin{minipage}[b]{0.325\linewidth}
\includegraphics[width=\textwidth]{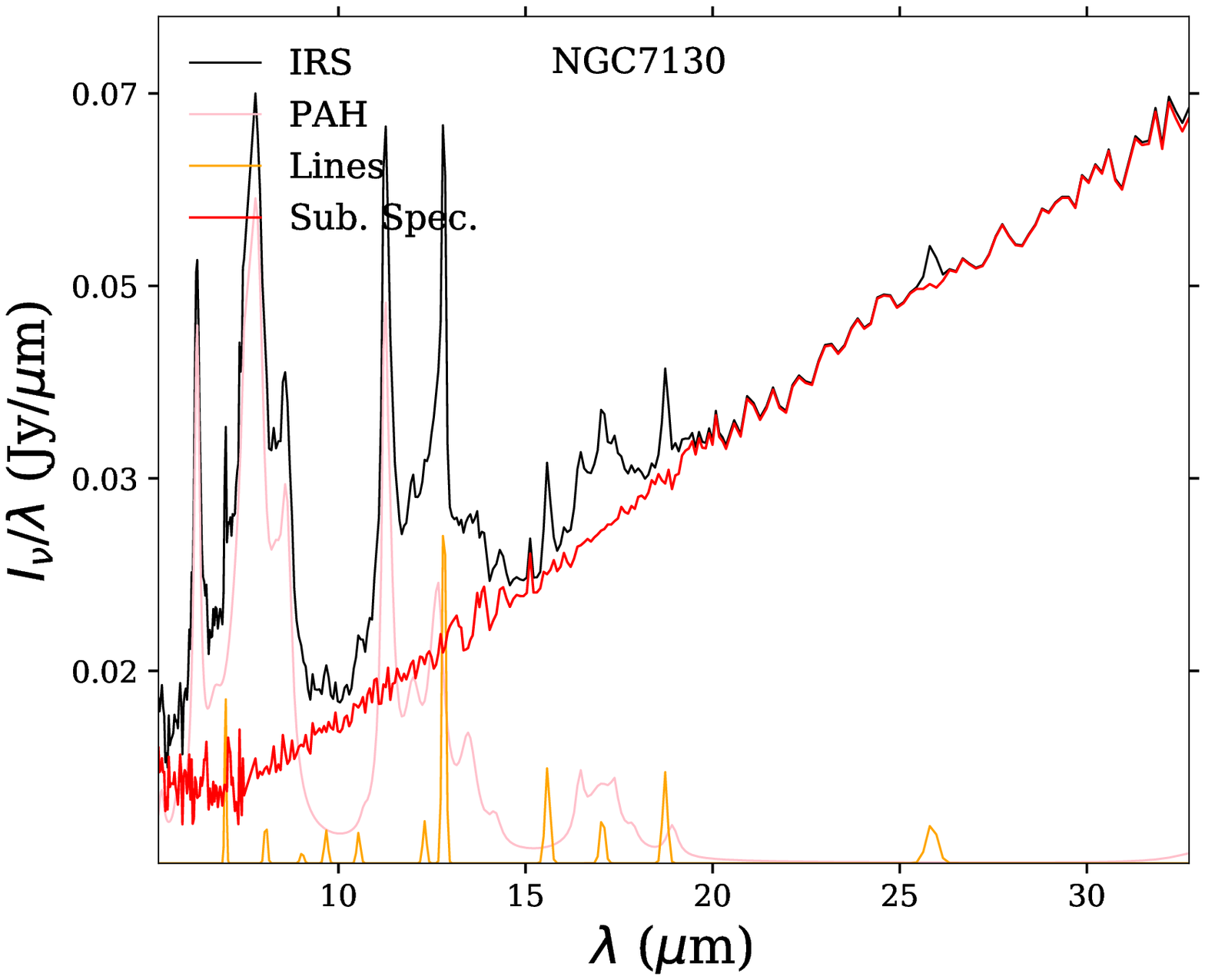}
\end{minipage} \hfill
\begin{minipage}[b]{0.325\linewidth}
\includegraphics[width=\textwidth]{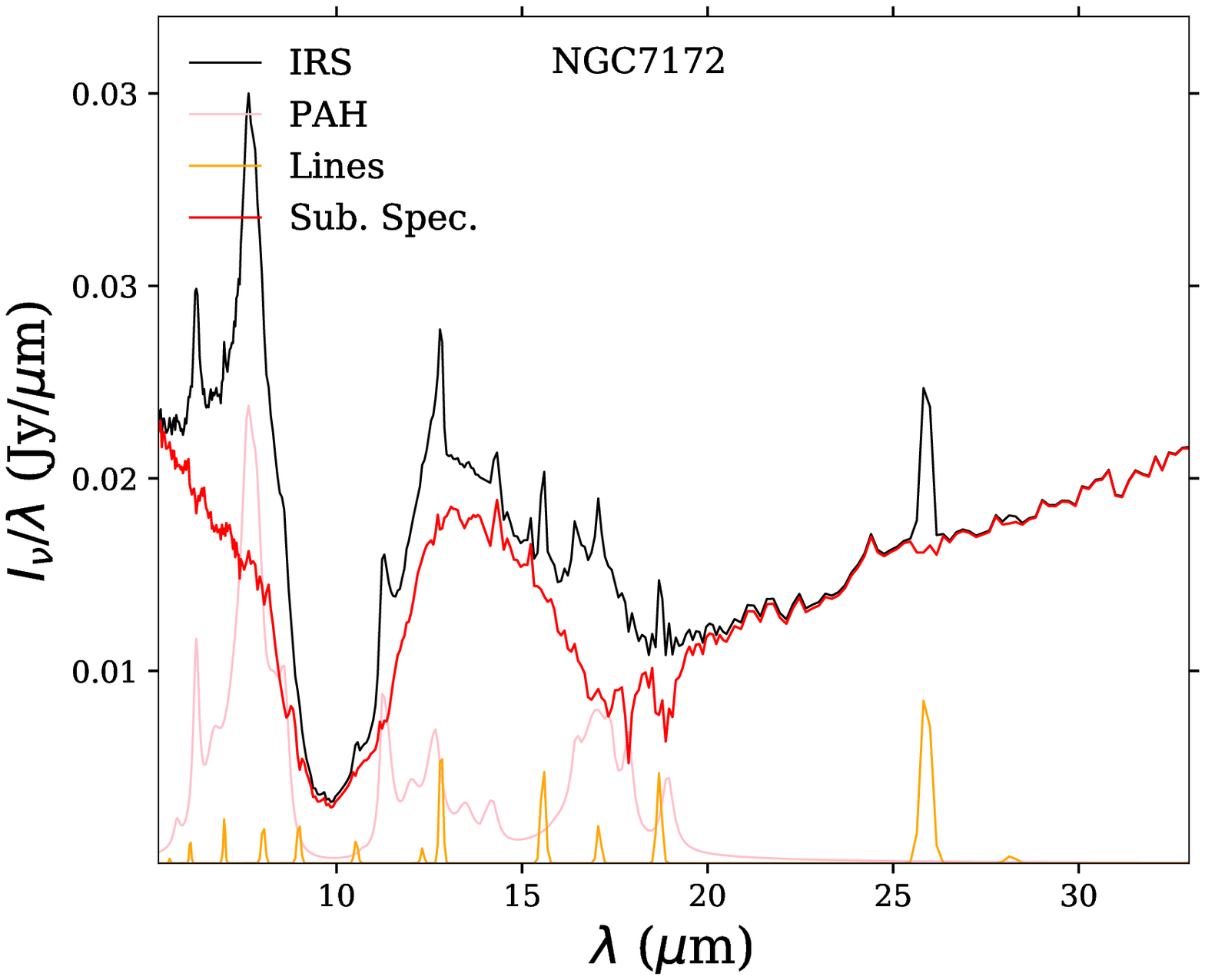}
\end{minipage} \hfill
\begin{minipage}[b]{0.325\linewidth}
\includegraphics[width=\textwidth]{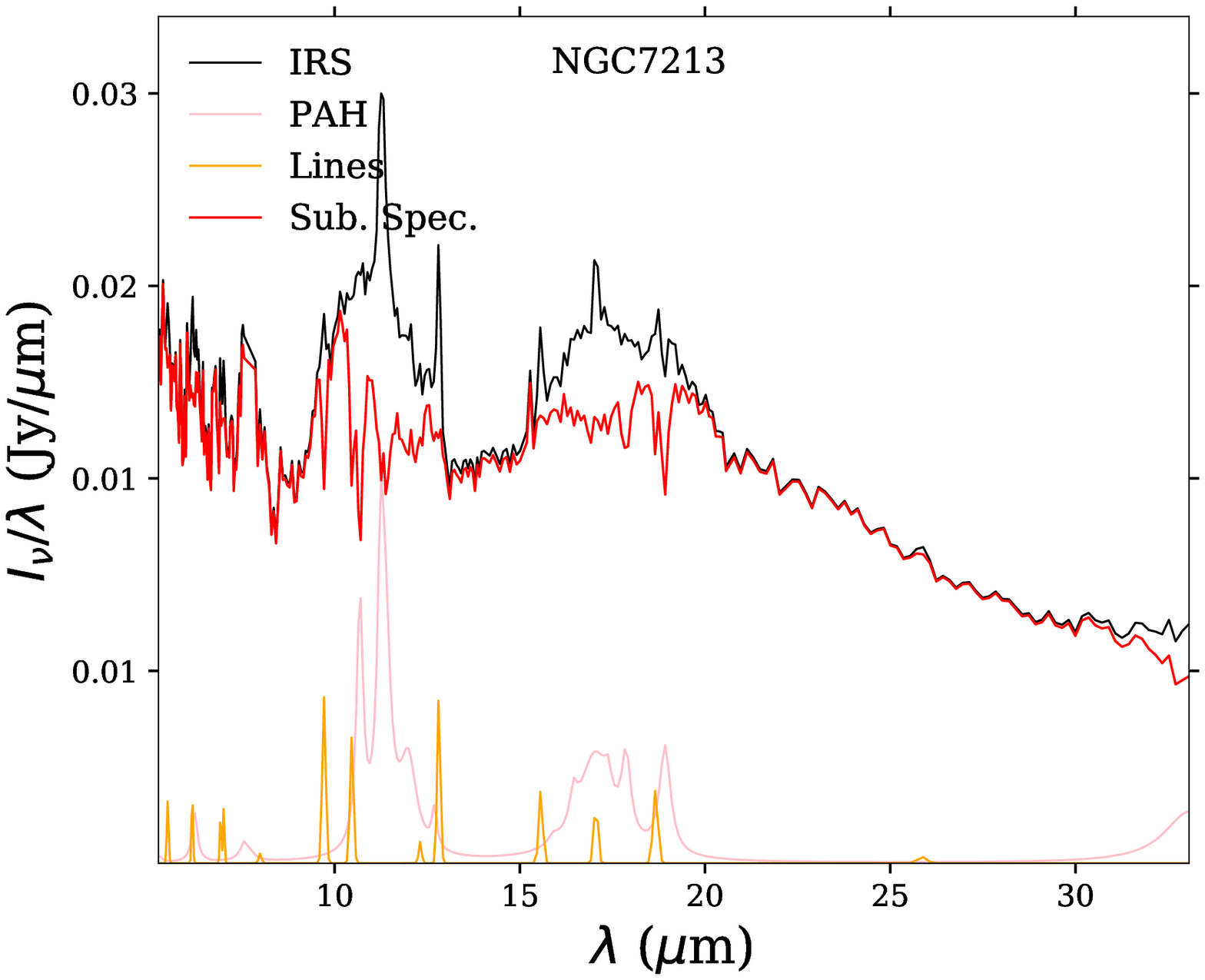}
\end{minipage} \hfill
\begin{minipage}[b]{0.325\linewidth}
\includegraphics[width=\textwidth]{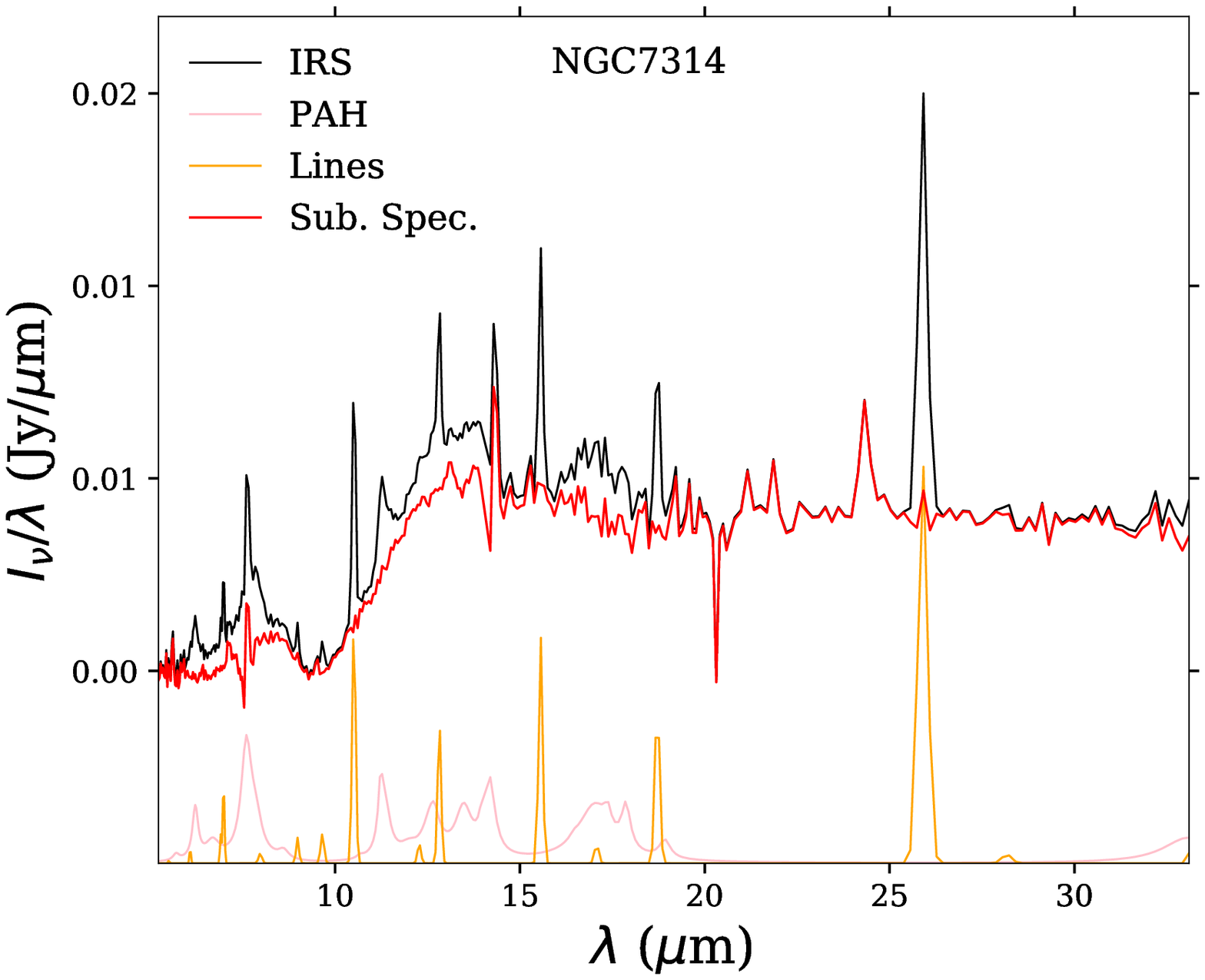}
\end{minipage} \hfill
\caption{continued from previous page.}
\setcounter{figure}{0}
\end{figure}

\begin{figure}

\begin{minipage}[b]{0.325\linewidth}
\includegraphics[width=\textwidth]{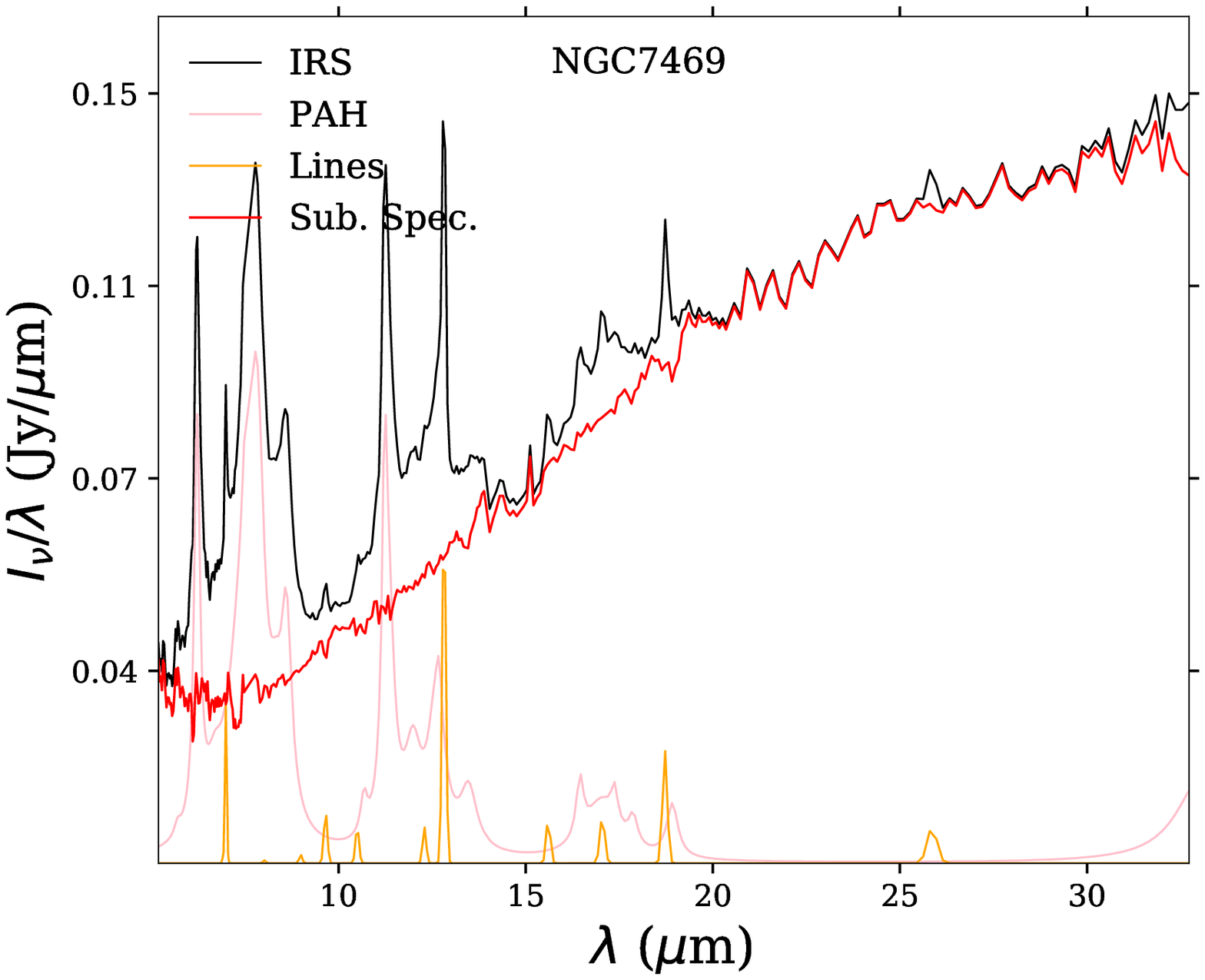}
\end{minipage} \hfill
\begin{minipage}[b]{0.325\linewidth}
\includegraphics[width=\textwidth]{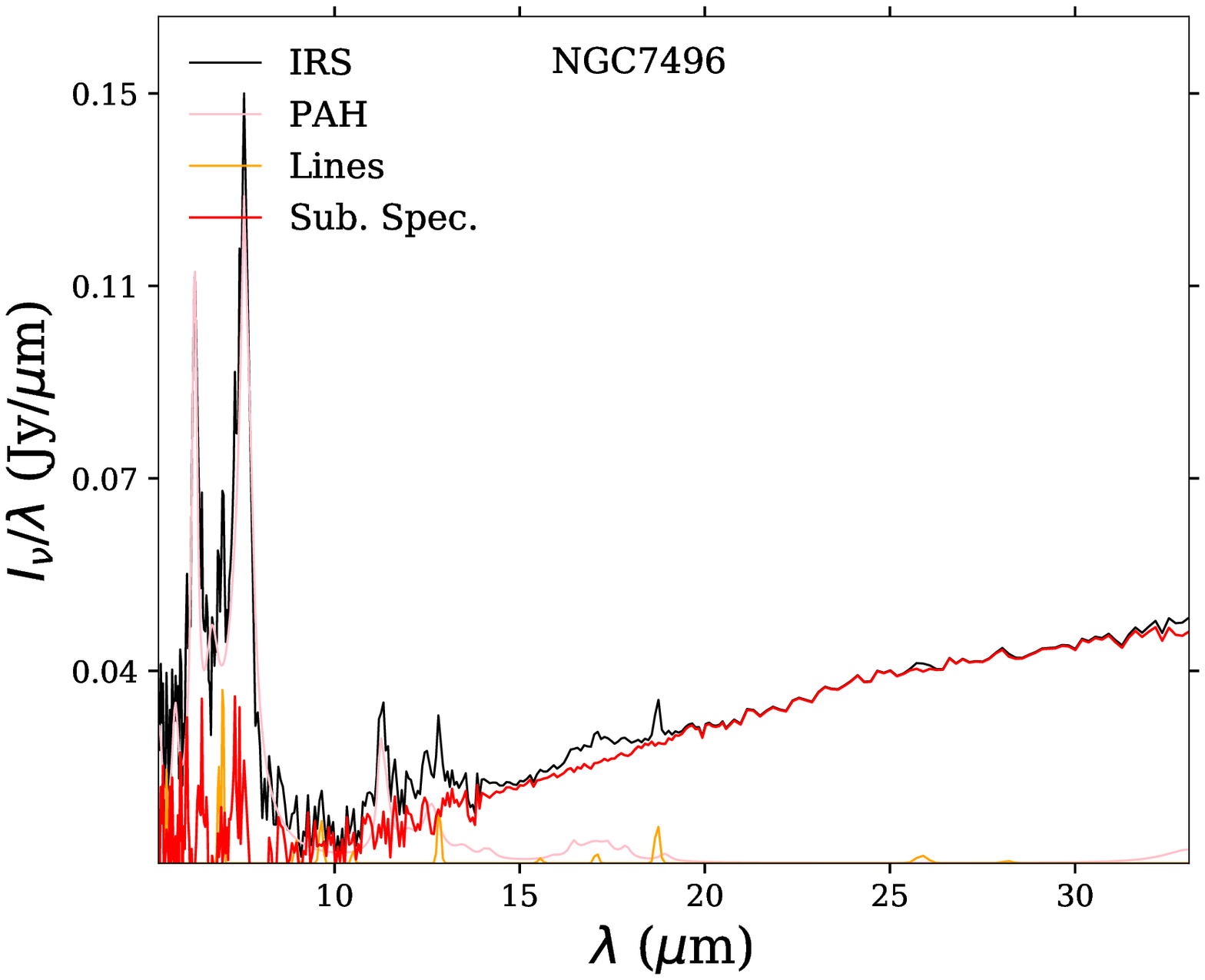}
\end{minipage} \hfill
\begin{minipage}[b]{0.325\linewidth}
\includegraphics[width=\textwidth]{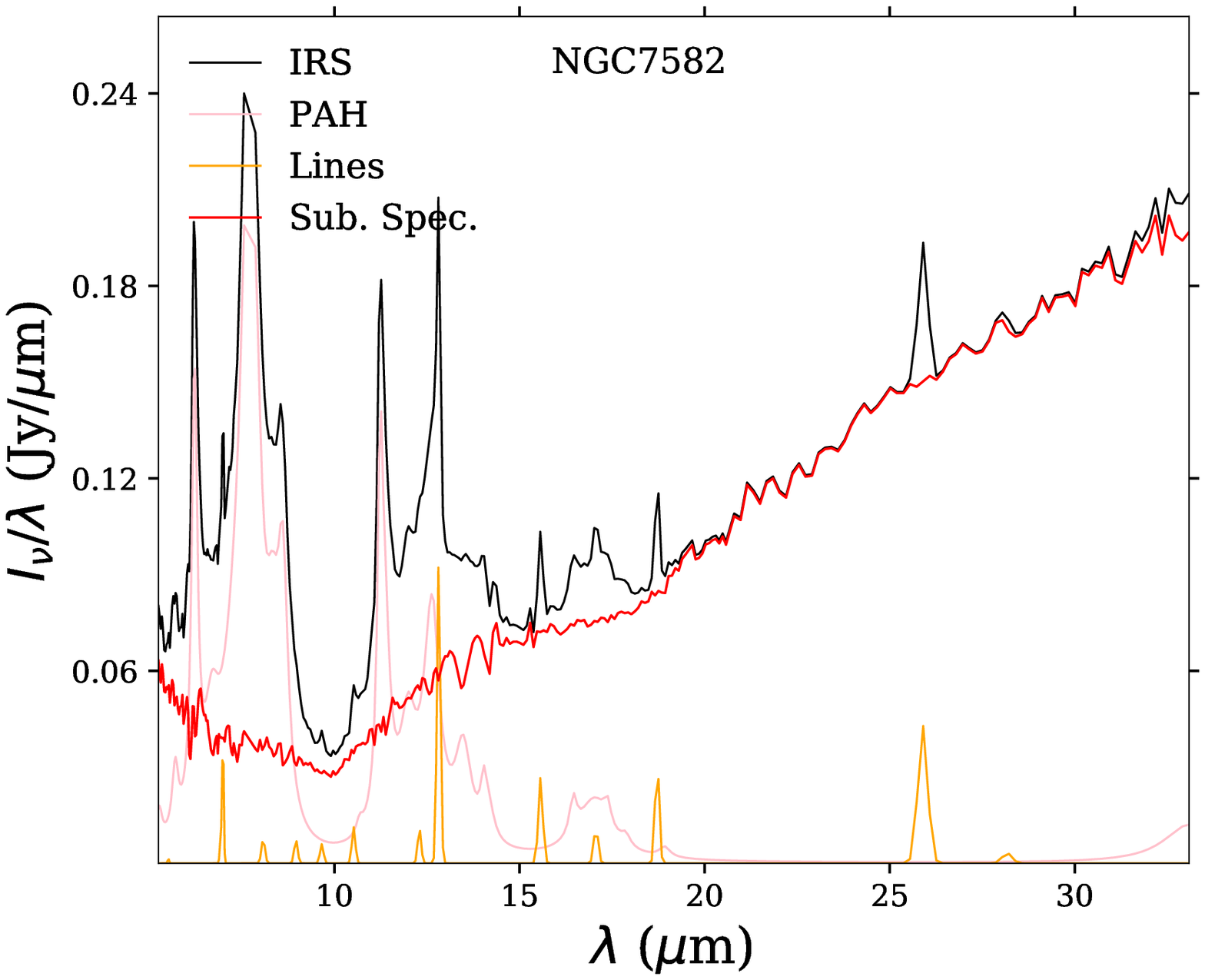}
\end{minipage} \hfill
\begin{minipage}[b]{0.325\linewidth}
\includegraphics[width=\textwidth]{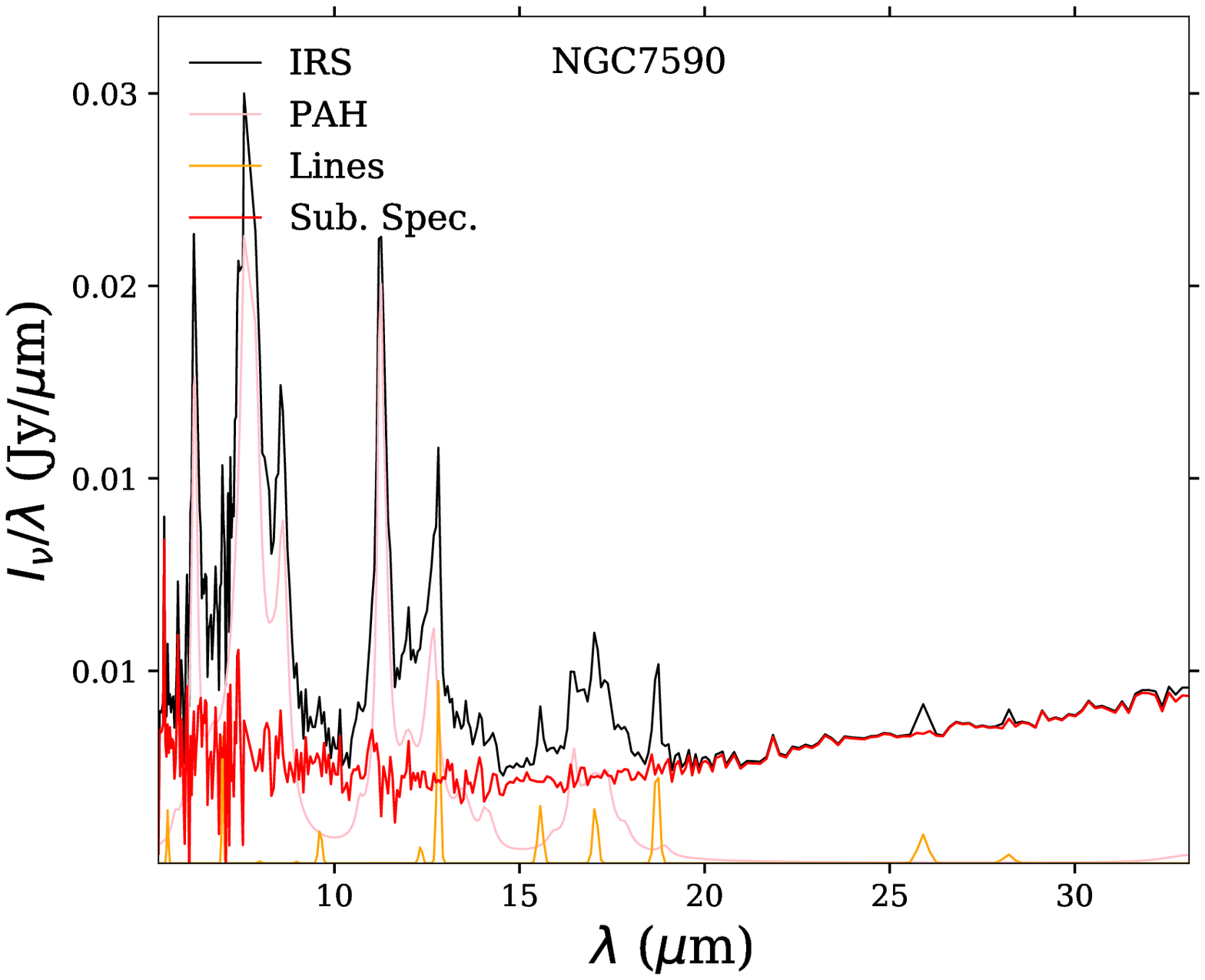}
\end{minipage} \hfill
\begin{minipage}[b]{0.325\linewidth}
\includegraphics[width=\textwidth]{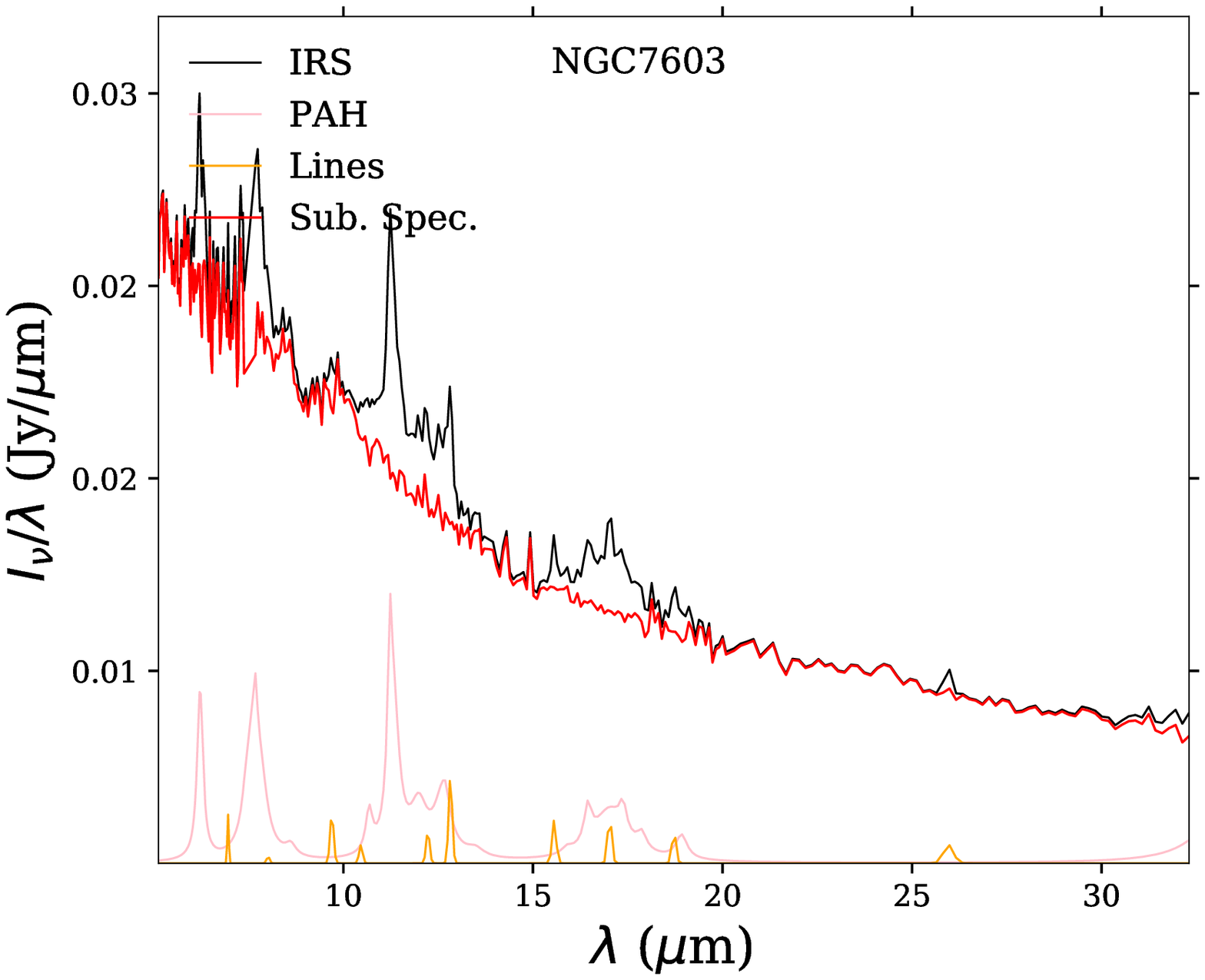}
\end{minipage} \hfill
\begin{minipage}[b]{0.325\linewidth}
\includegraphics[width=\textwidth]{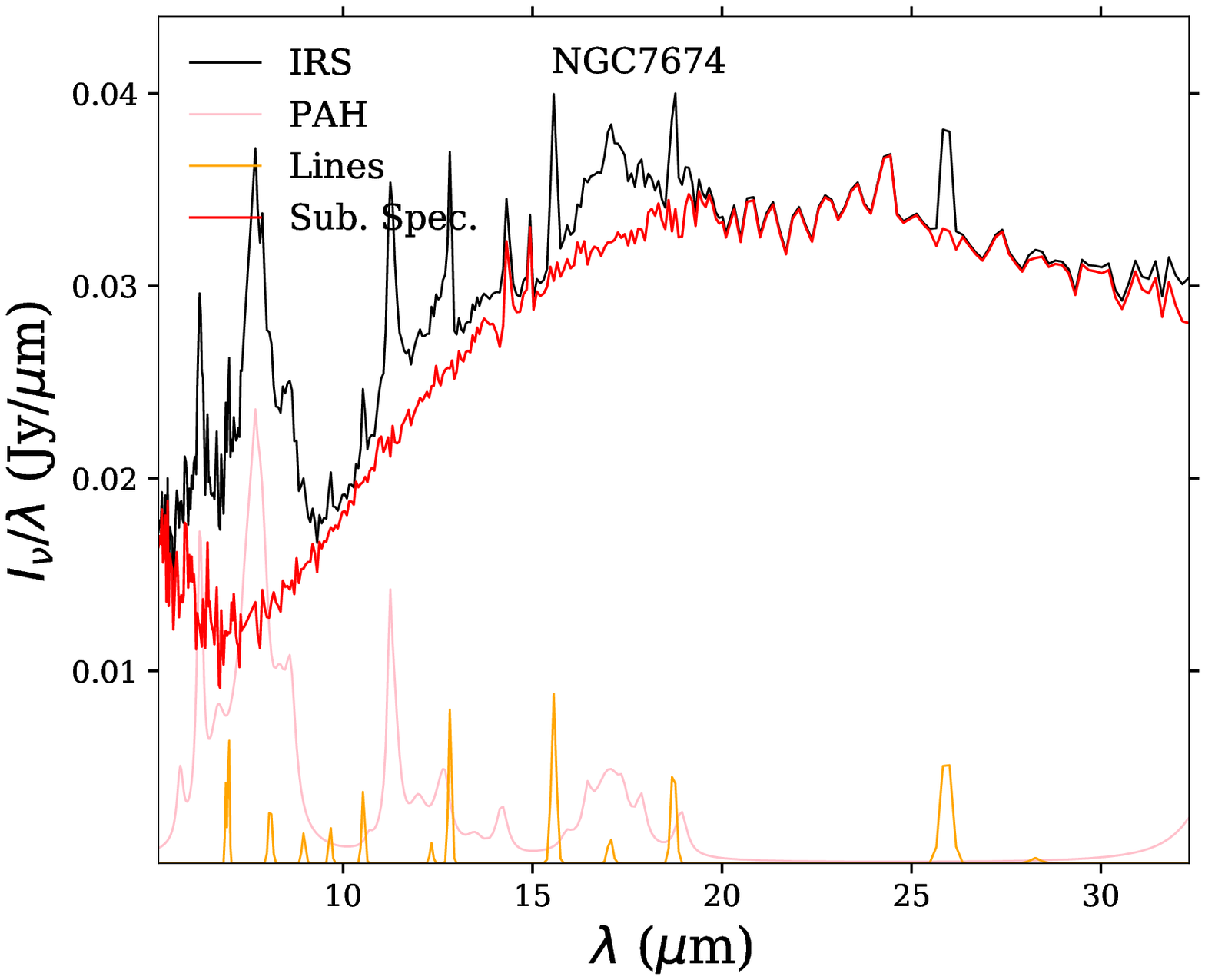}
\end{minipage} \hfill
\begin{minipage}[b]{0.325\linewidth}
\includegraphics[width=\textwidth]{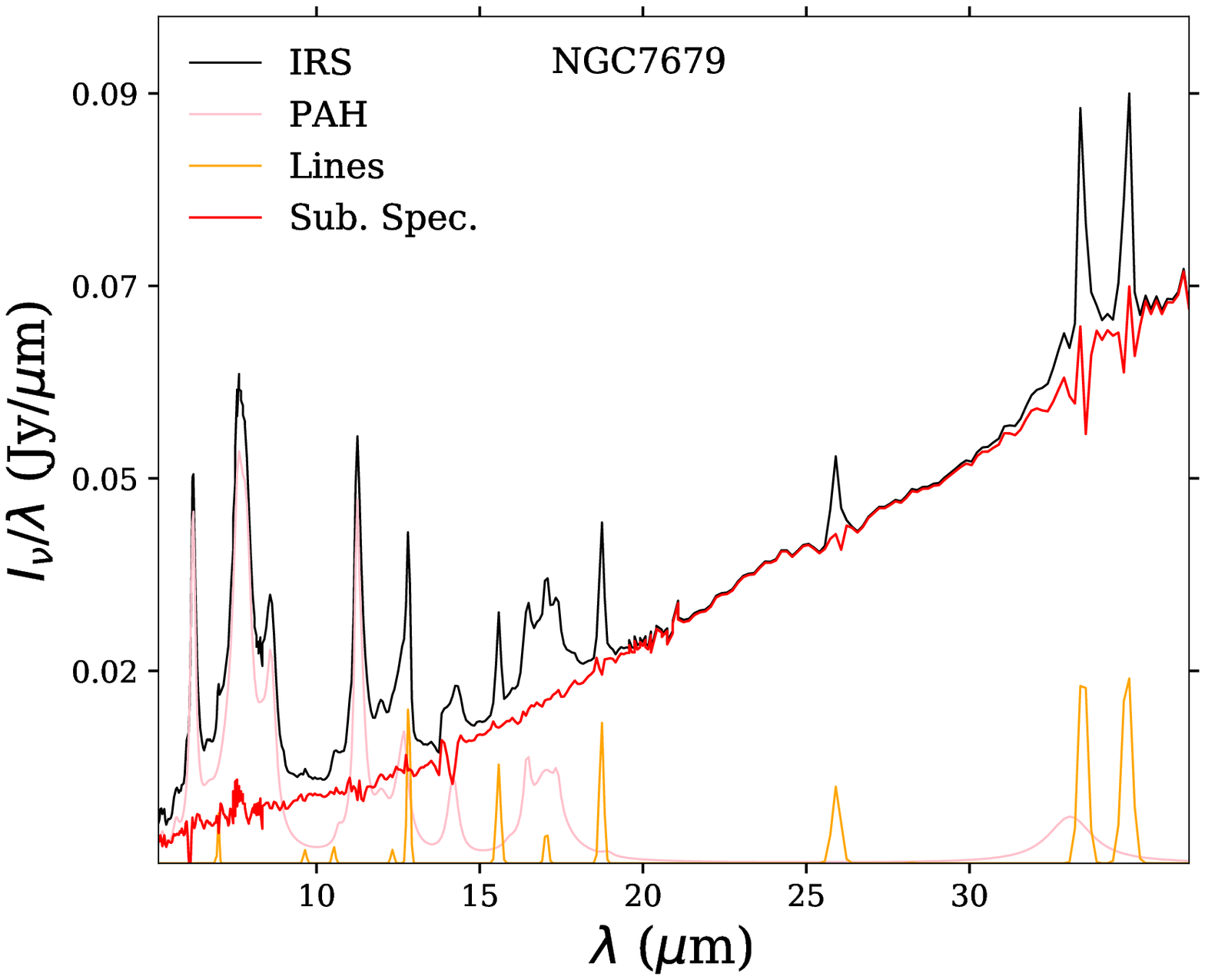}
\end{minipage} \hfill
\begin{minipage}[b]{0.325\linewidth}
\includegraphics[width=\textwidth]{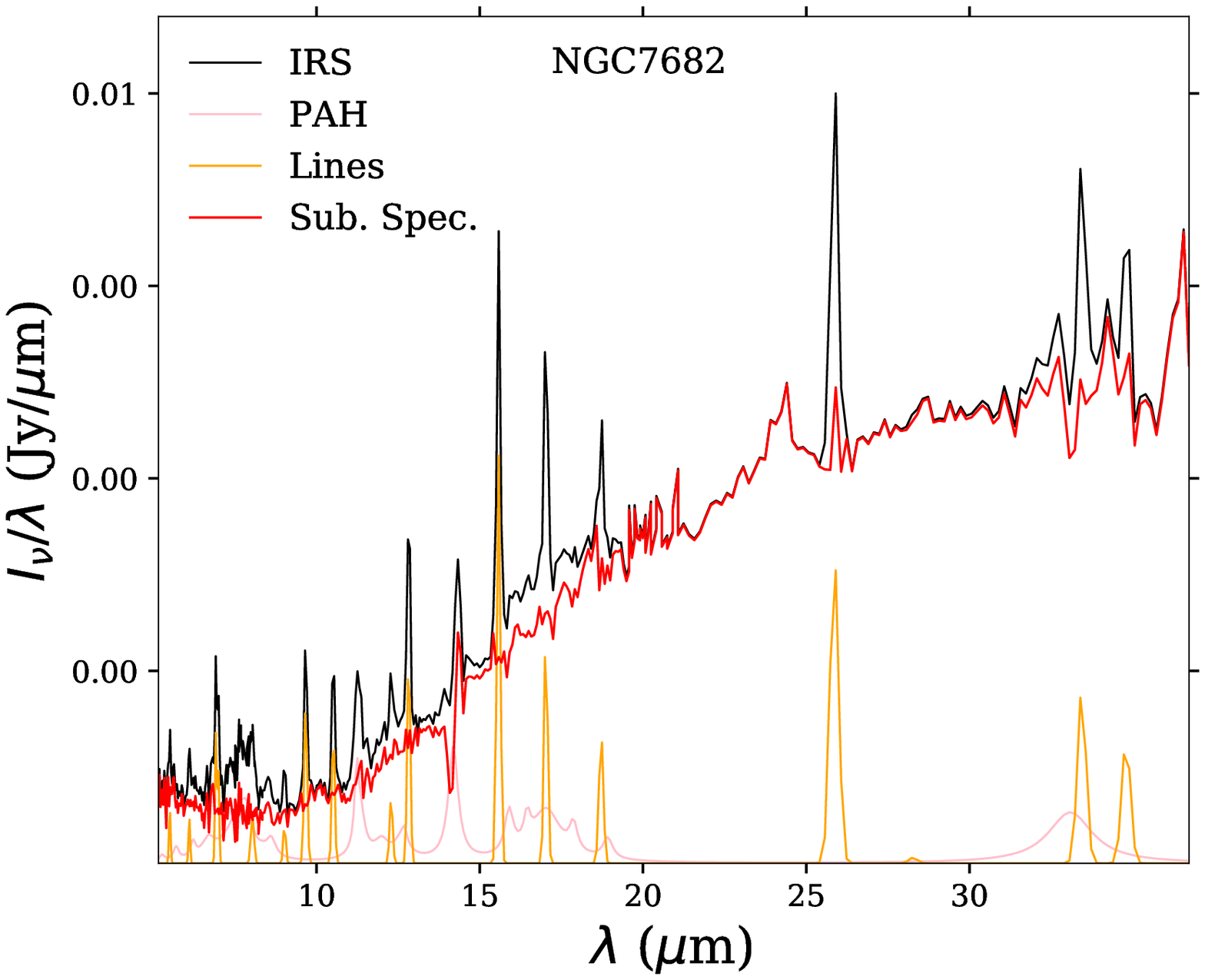}
\end{minipage} \hfill
\begin{minipage}[b]{0.325\linewidth}
\includegraphics[width=\textwidth]{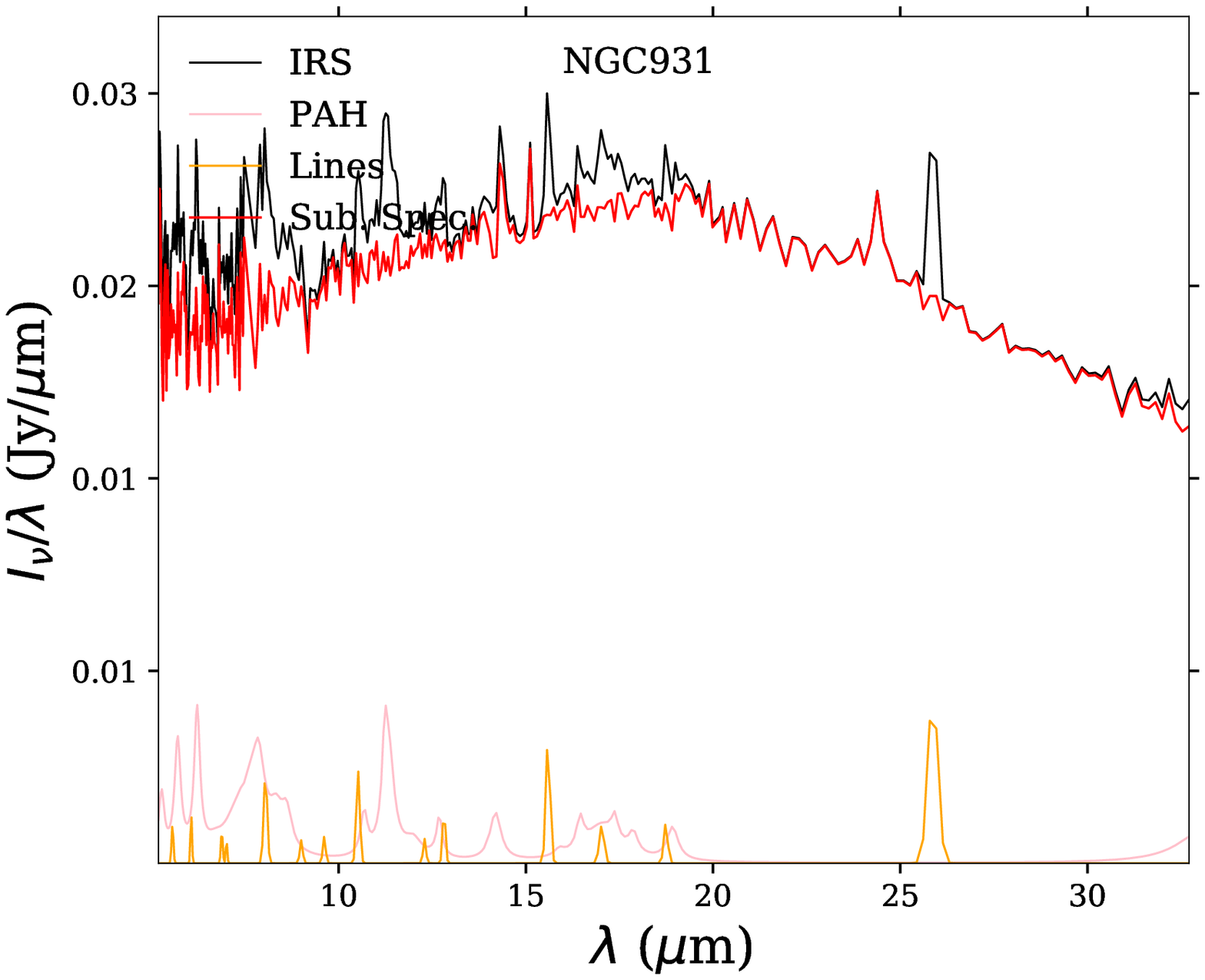}
\end{minipage} \hfill
\begin{minipage}[b]{0.325\linewidth}
\includegraphics[width=\textwidth]{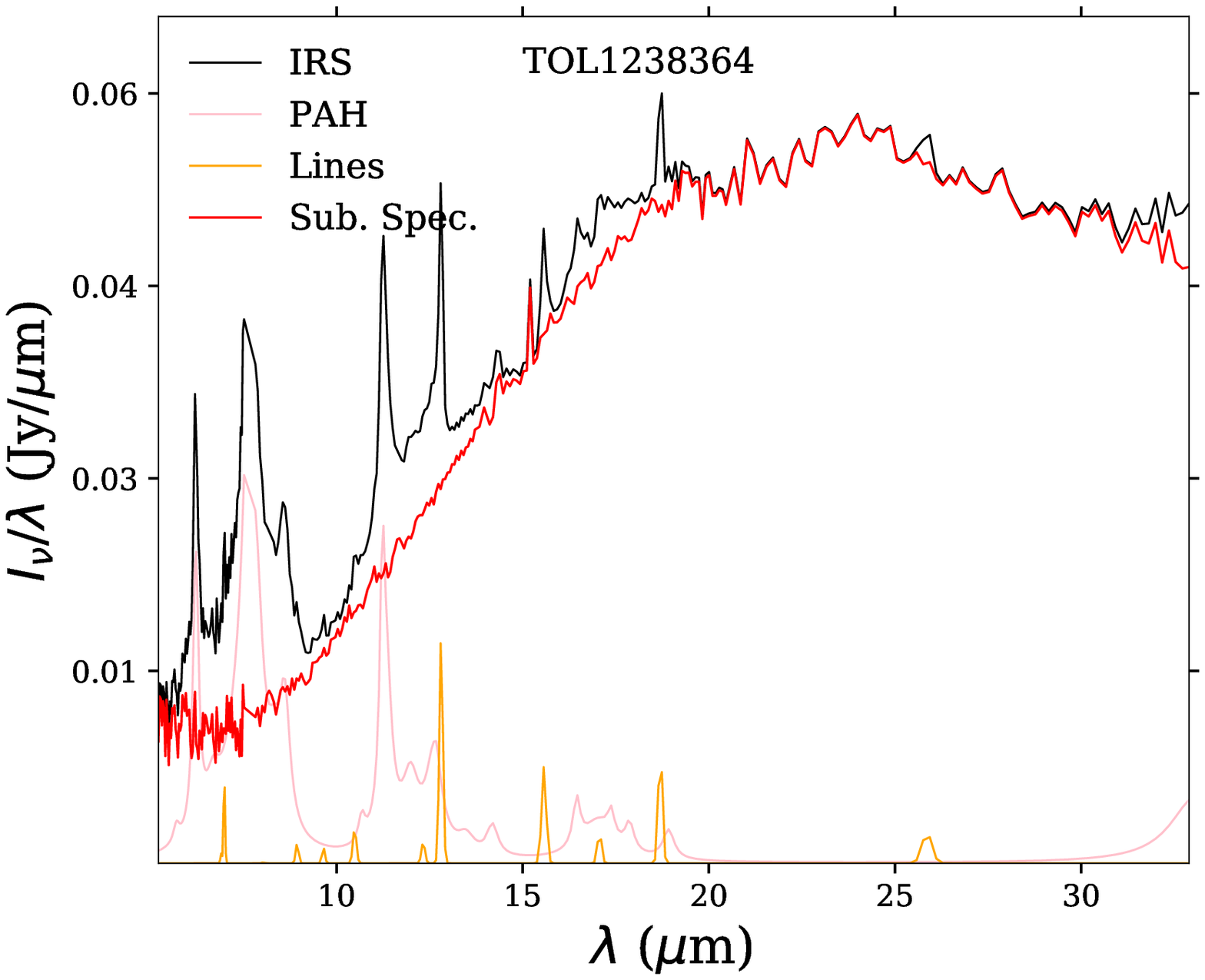}
\end{minipage} \hfill
\begin{minipage}[b]{0.325\linewidth}
\includegraphics[width=\textwidth]{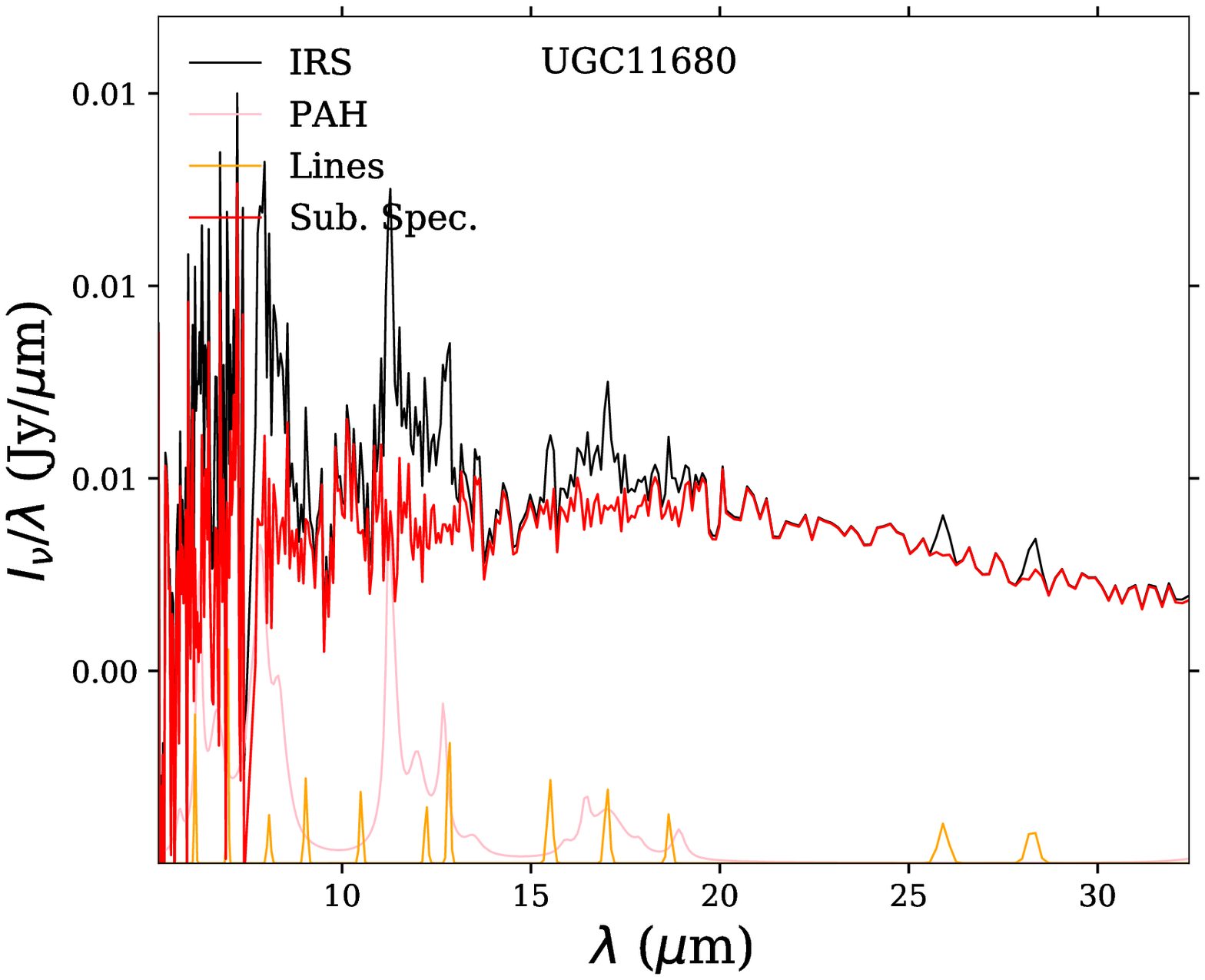}
\end{minipage} \hfill
\begin{minipage}[b]{0.325\linewidth}
\includegraphics[width=\textwidth]{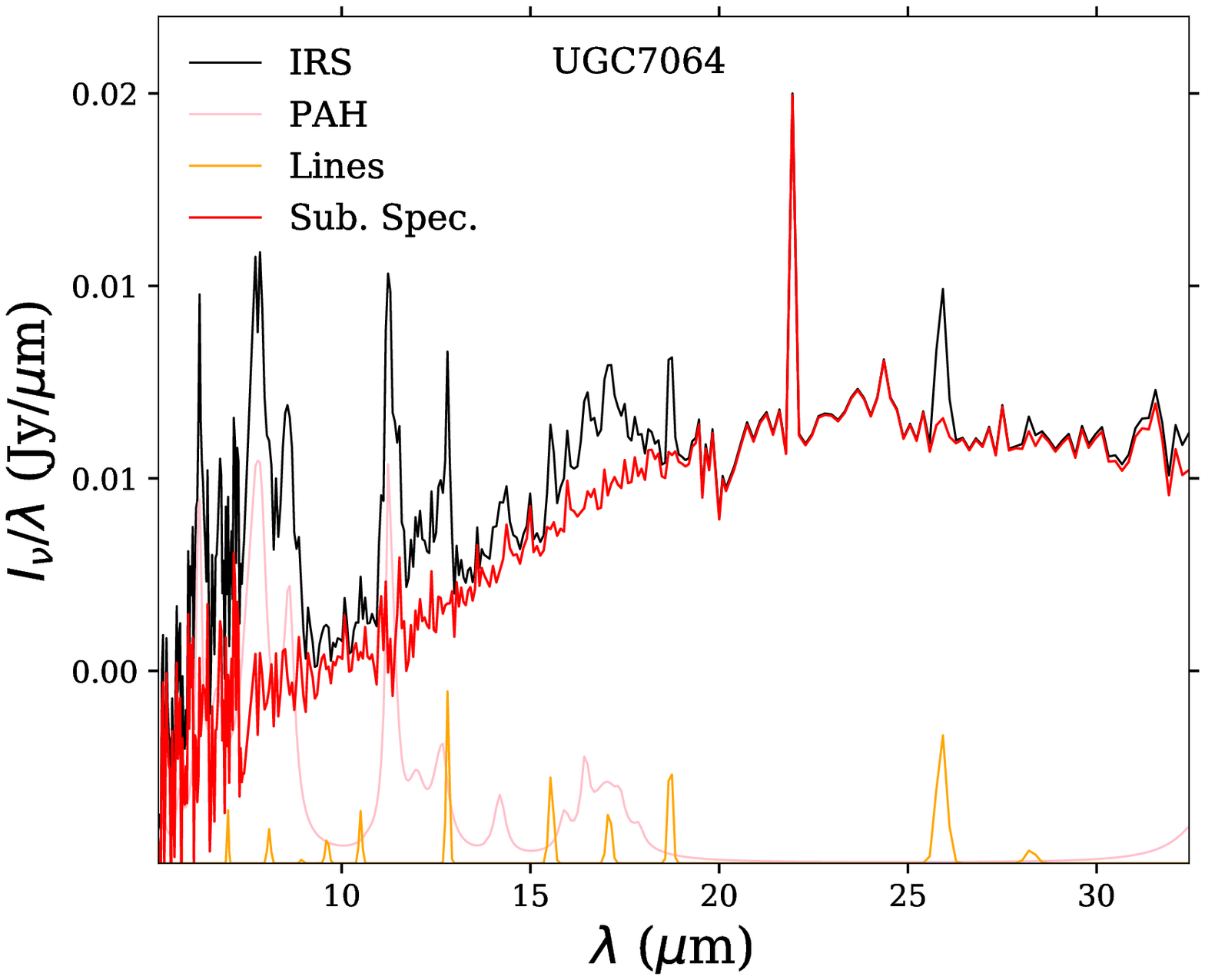}
\end{minipage} \hfill
\label{fig:cont}
\caption{continued from previous page.}
\setcounter{figure}{0}
\end{figure}

\newpage

\section[]{Individual Adjusts}

\renewcommand{\thefigure}{C\arabic{figure}}
\setcounter{figure}{0}

\begin{figure}
\begin{minipage}[b]{0.325\linewidth}
\includegraphics[width=\textwidth]{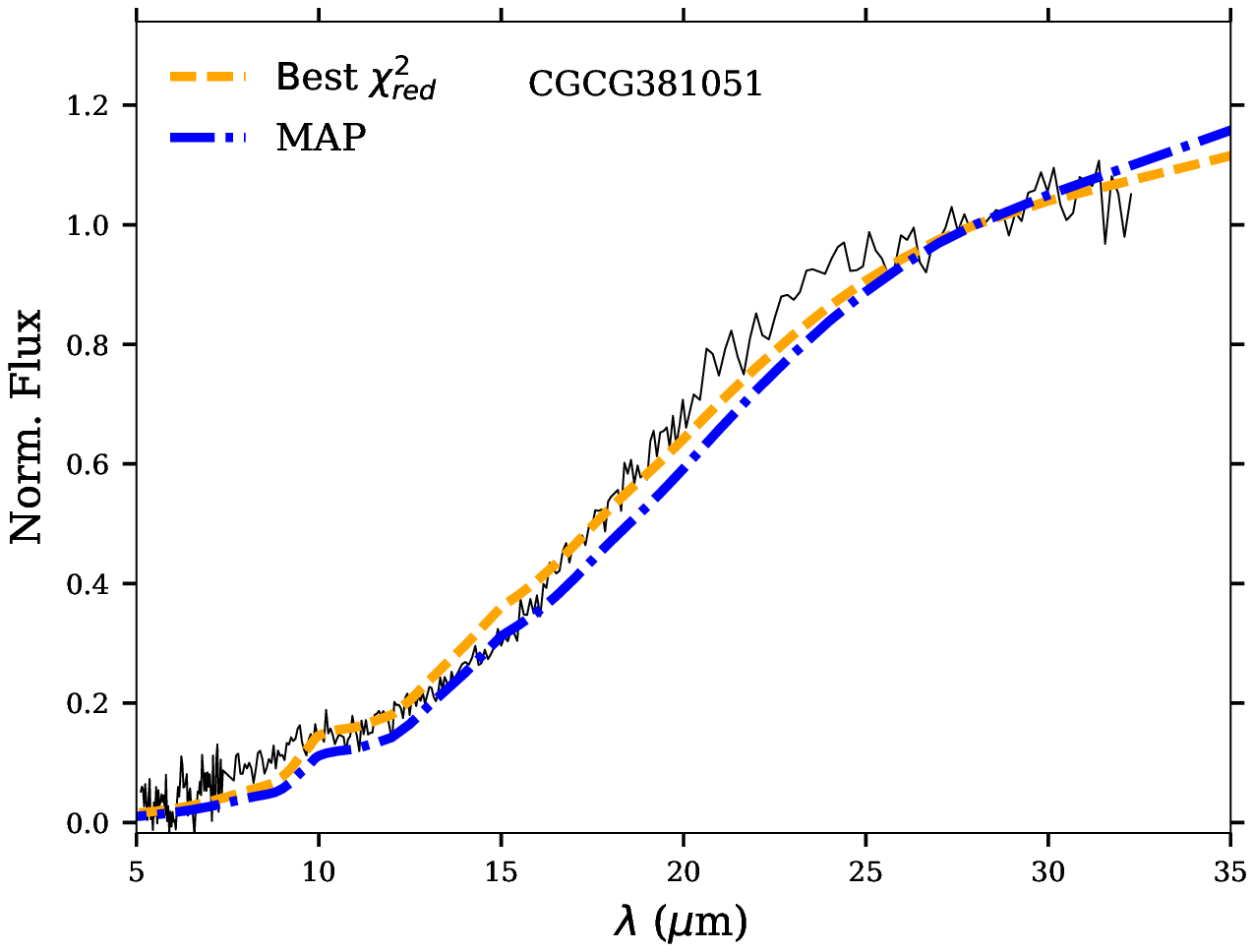}
\end{minipage} \hfill
\begin{minipage}[b]{0.325\linewidth}
\includegraphics[width=\textwidth]{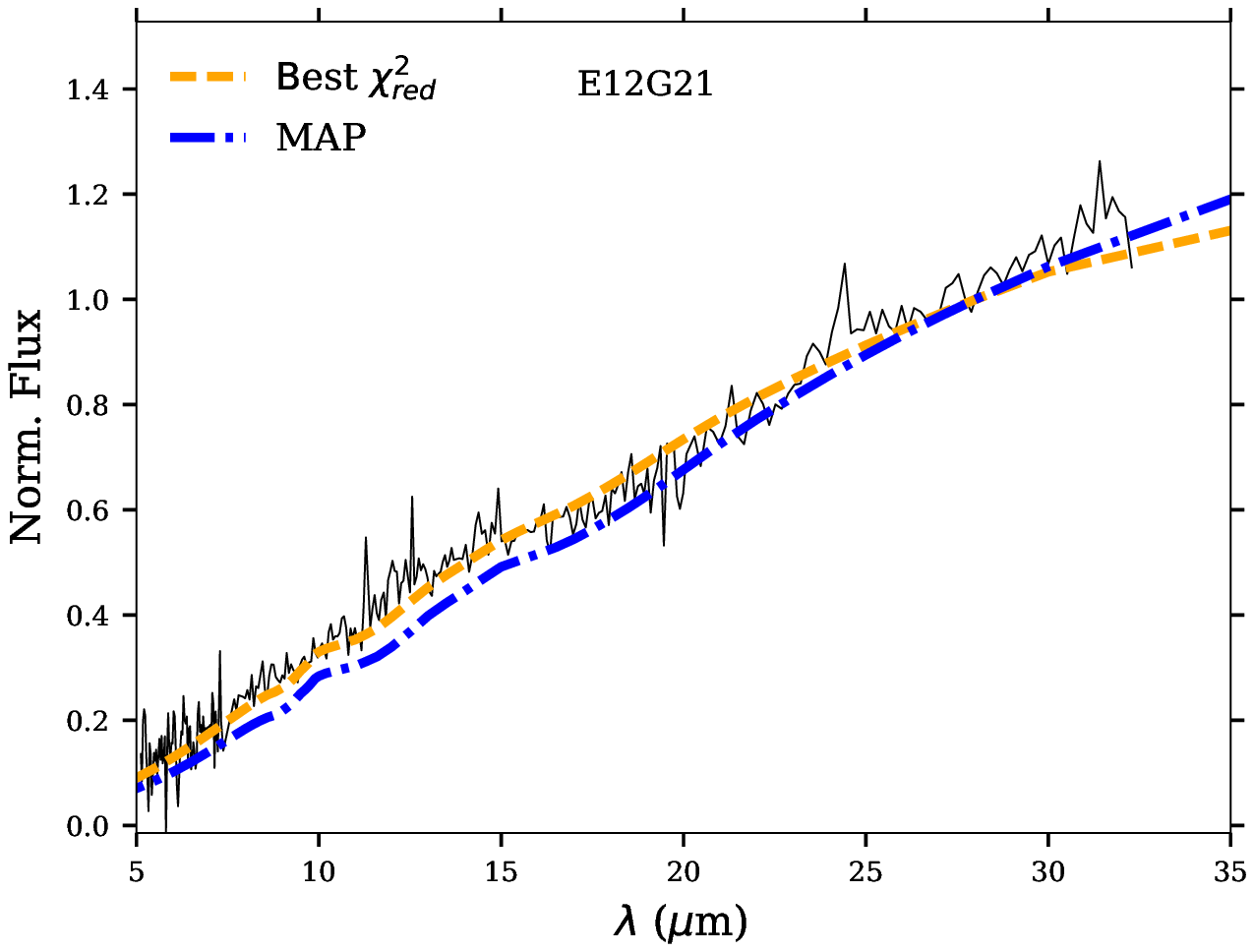}
\end{minipage} \hfill
\begin{minipage}[b]{0.325\linewidth}
\includegraphics[width=\textwidth]{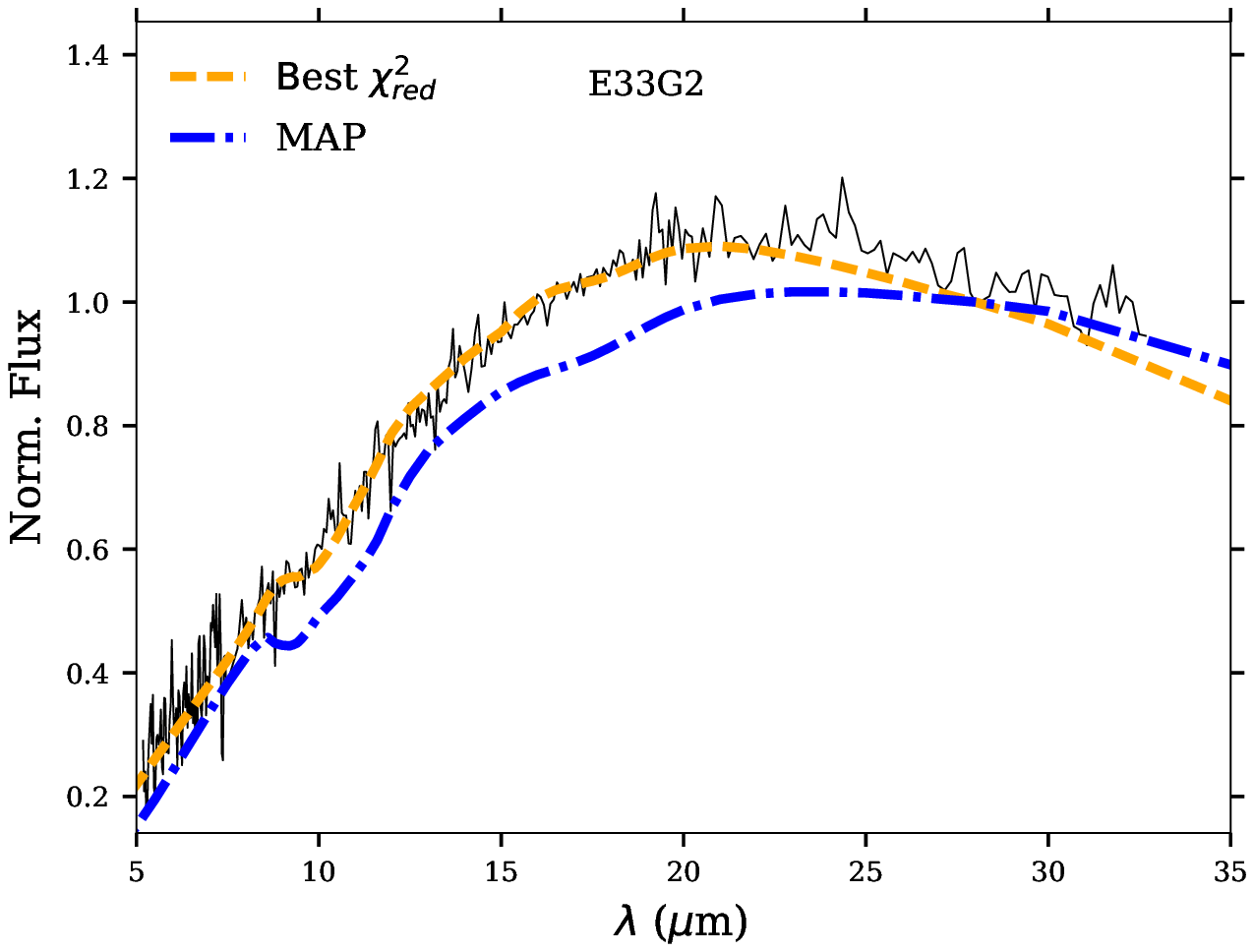}
\end{minipage} \hfill
\begin{minipage}[b]{0.325\linewidth}
\includegraphics[width=\textwidth]{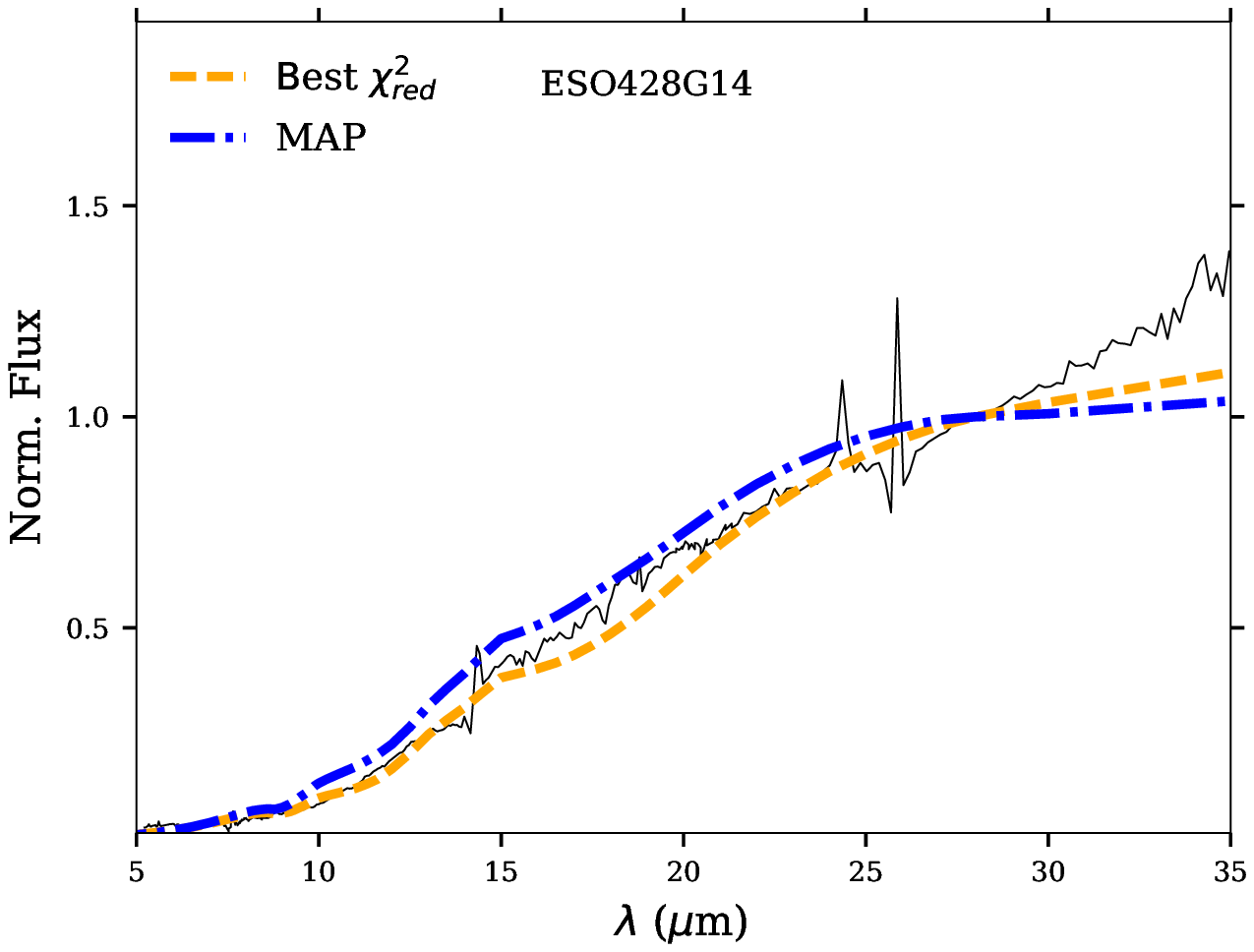}
\end{minipage} \hfill
\begin{minipage}[b]{0.325\linewidth}
\includegraphics[width=\textwidth]{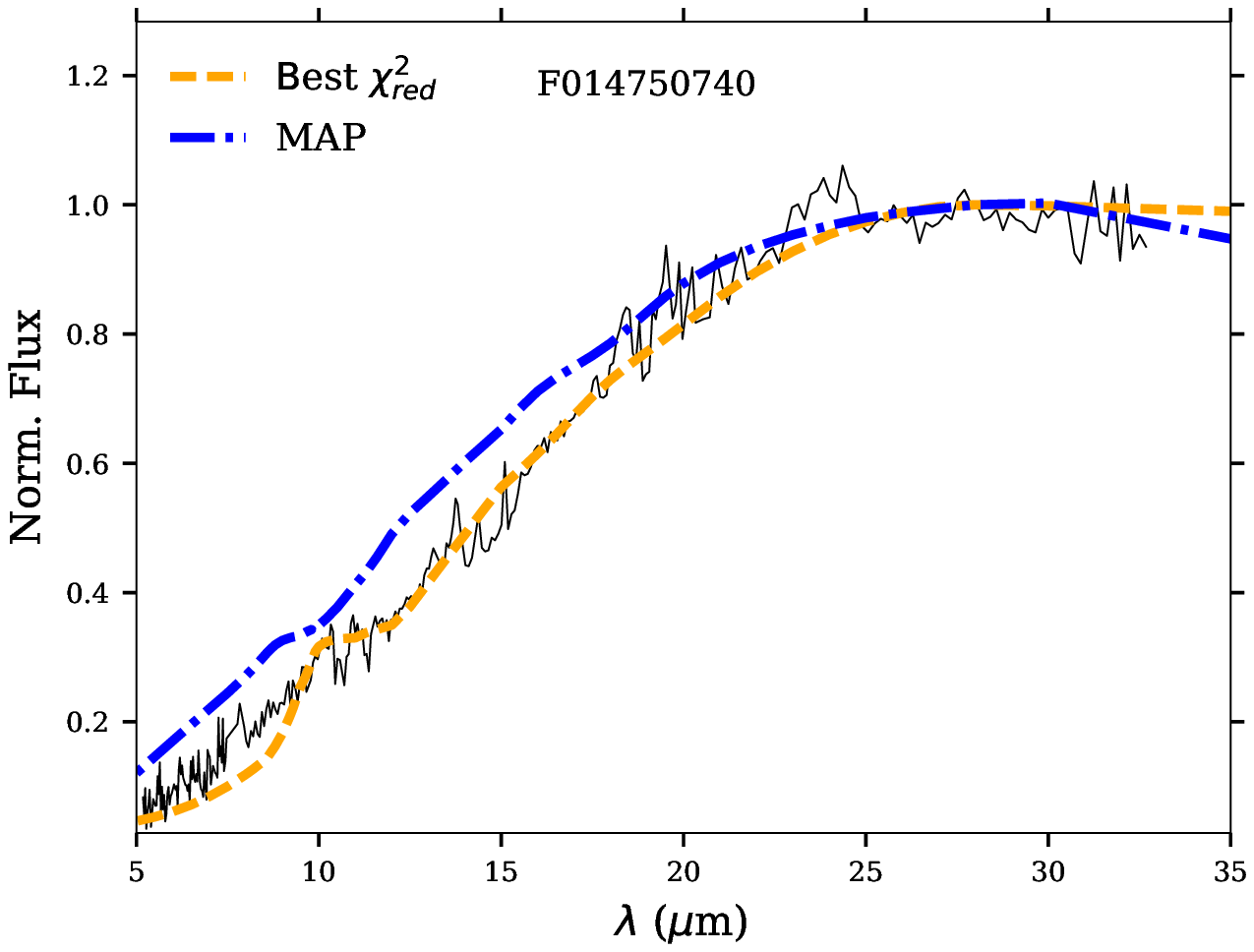}
\end{minipage} \hfill
\begin{minipage}[b]{0.325\linewidth}
\includegraphics[width=\textwidth]{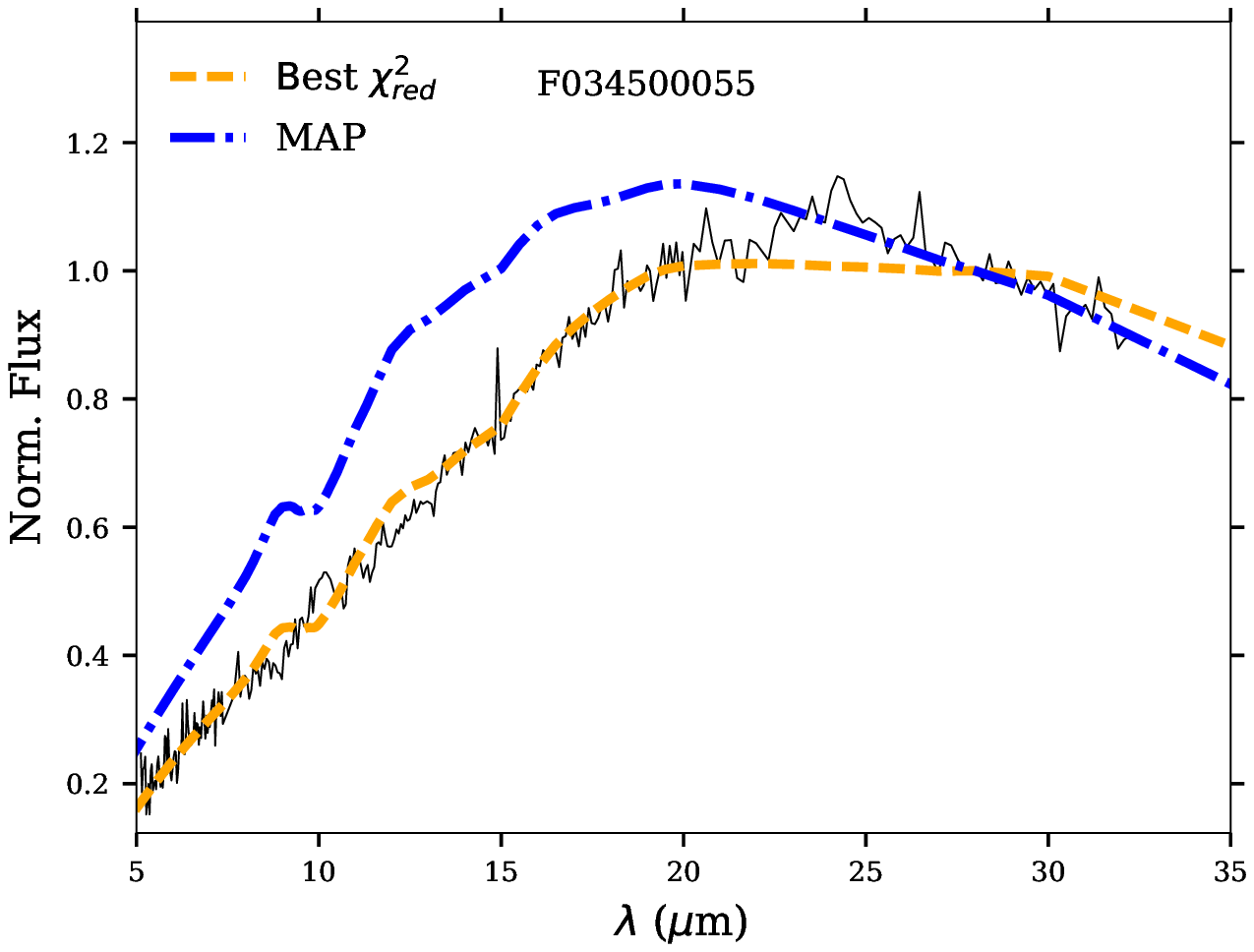}
\end{minipage} \hfill
\begin{minipage}[b]{0.325\linewidth}
\includegraphics[width=\textwidth]{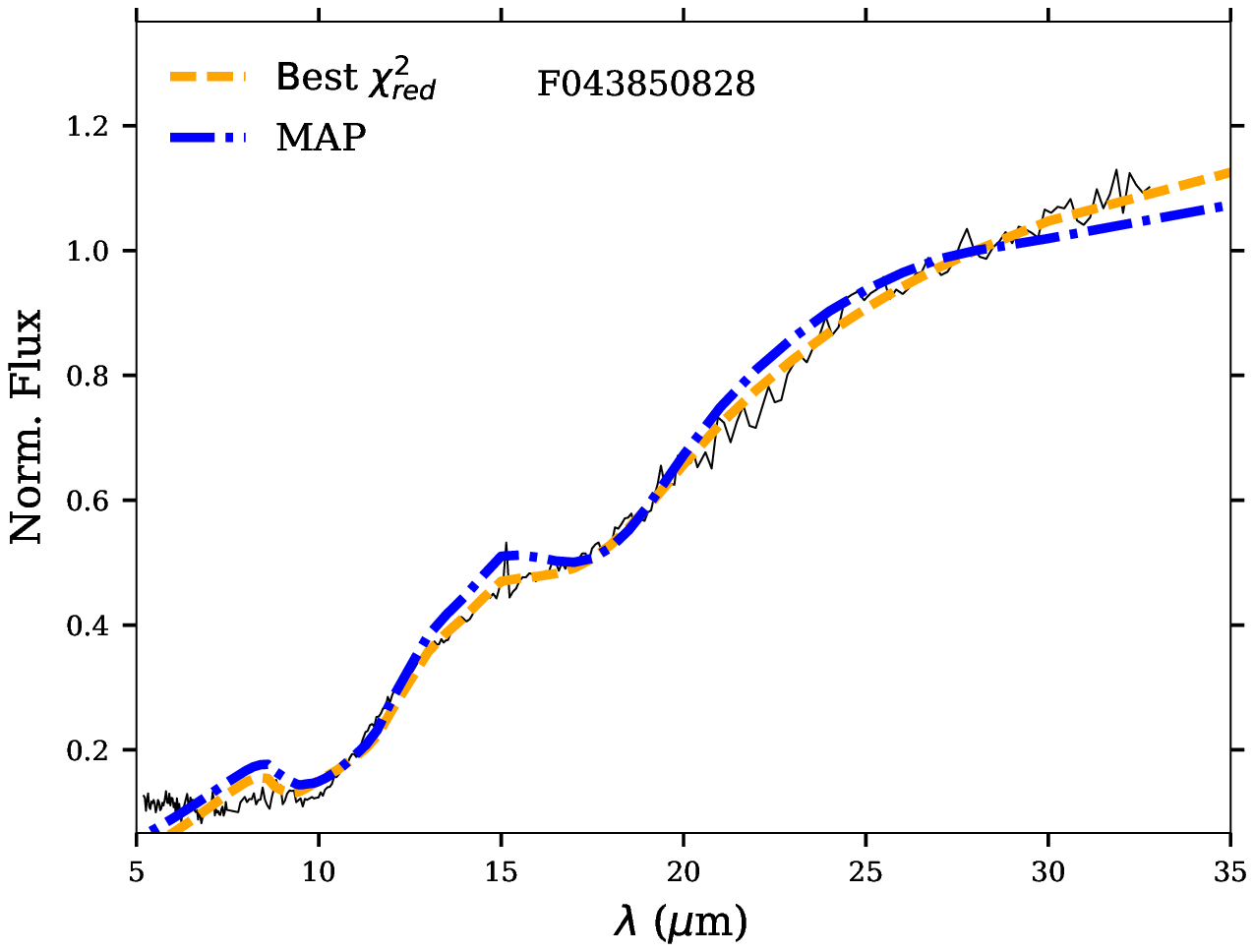}
\end{minipage} \hfill
\begin{minipage}[b]{0.325\linewidth}
\includegraphics[width=\textwidth]{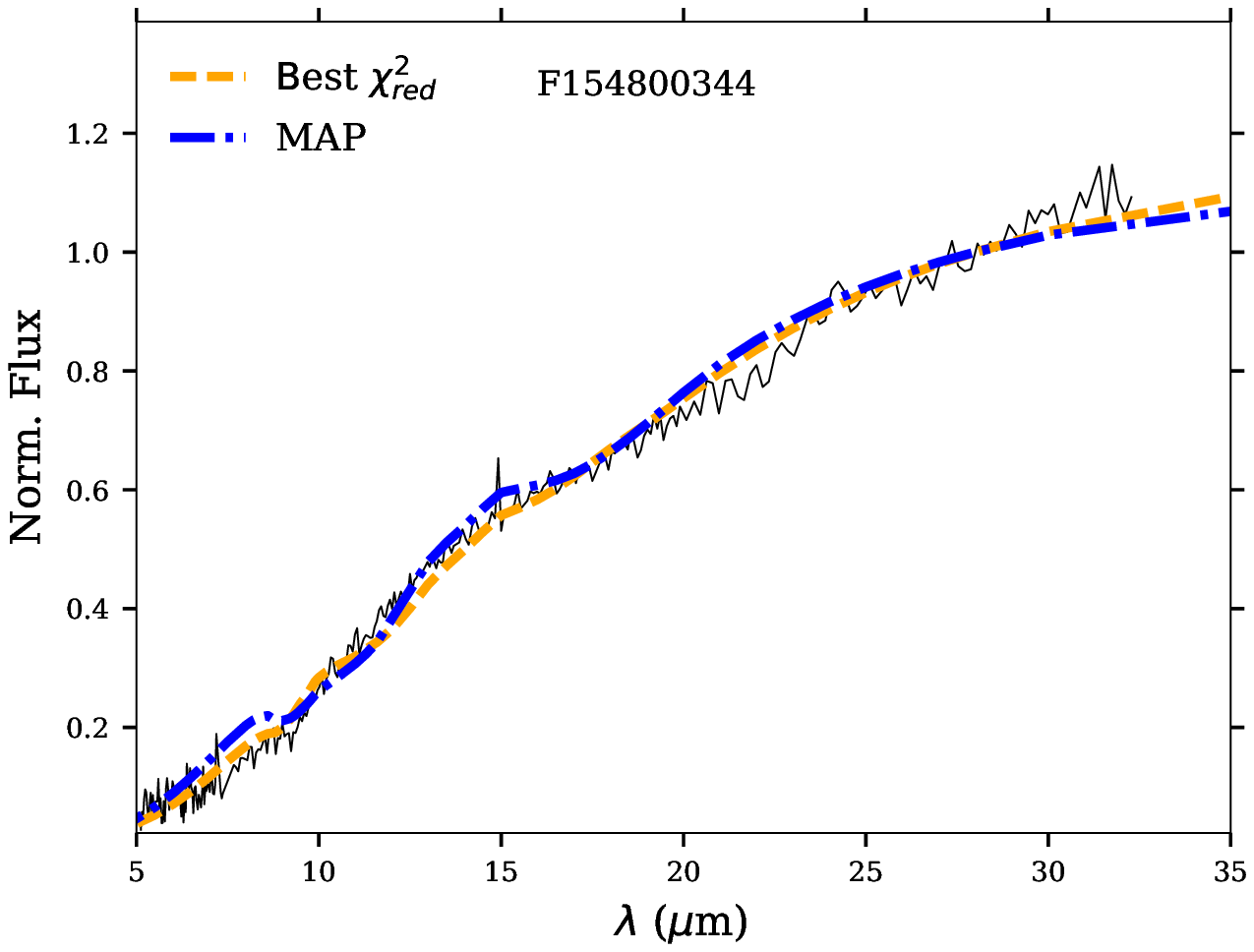}
\end{minipage} \hfill
\begin{minipage}[b]{0.325\linewidth}
\includegraphics[width=\textwidth]{IC4329A_bestMAP_JY.eps}
\end{minipage} \hfill
\caption{Individual fitting results. We present the adjusts for the best fit using ${\chi_{red}}^2$ (yellow dashed line) and the MAP (blue dotted-dashed line) distribution from BayesCLUMPY. The observed spectra and the SED models are normalized at 28$\mu$m.}
\setcounter{figure}{0}
\end{figure}

\begin{figure}

\begin{minipage}[b]{0.325\linewidth}
\includegraphics[width=\textwidth]{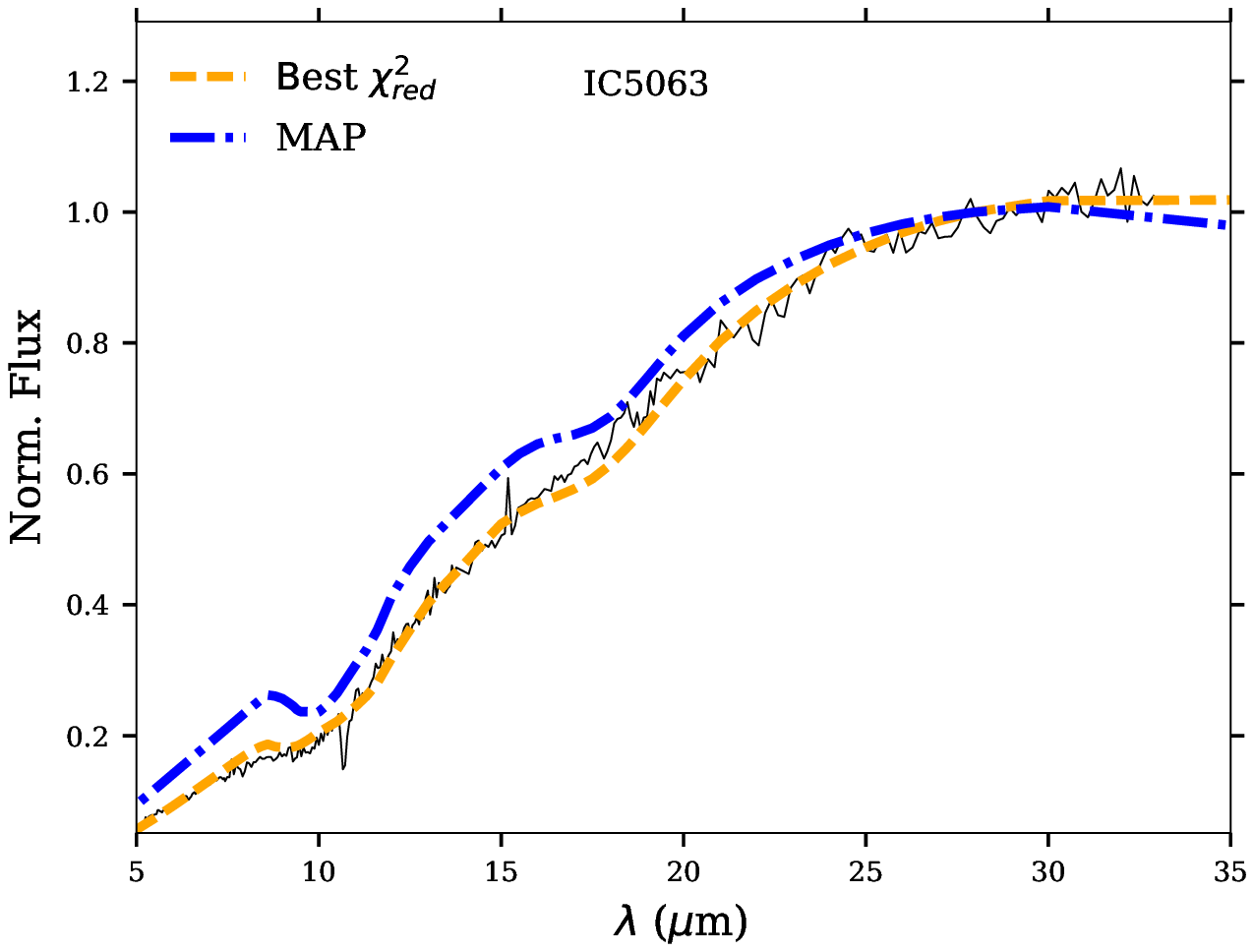}
\end{minipage} \hfill
\begin{minipage}[b]{0.325\linewidth}
\includegraphics[width=\textwidth]{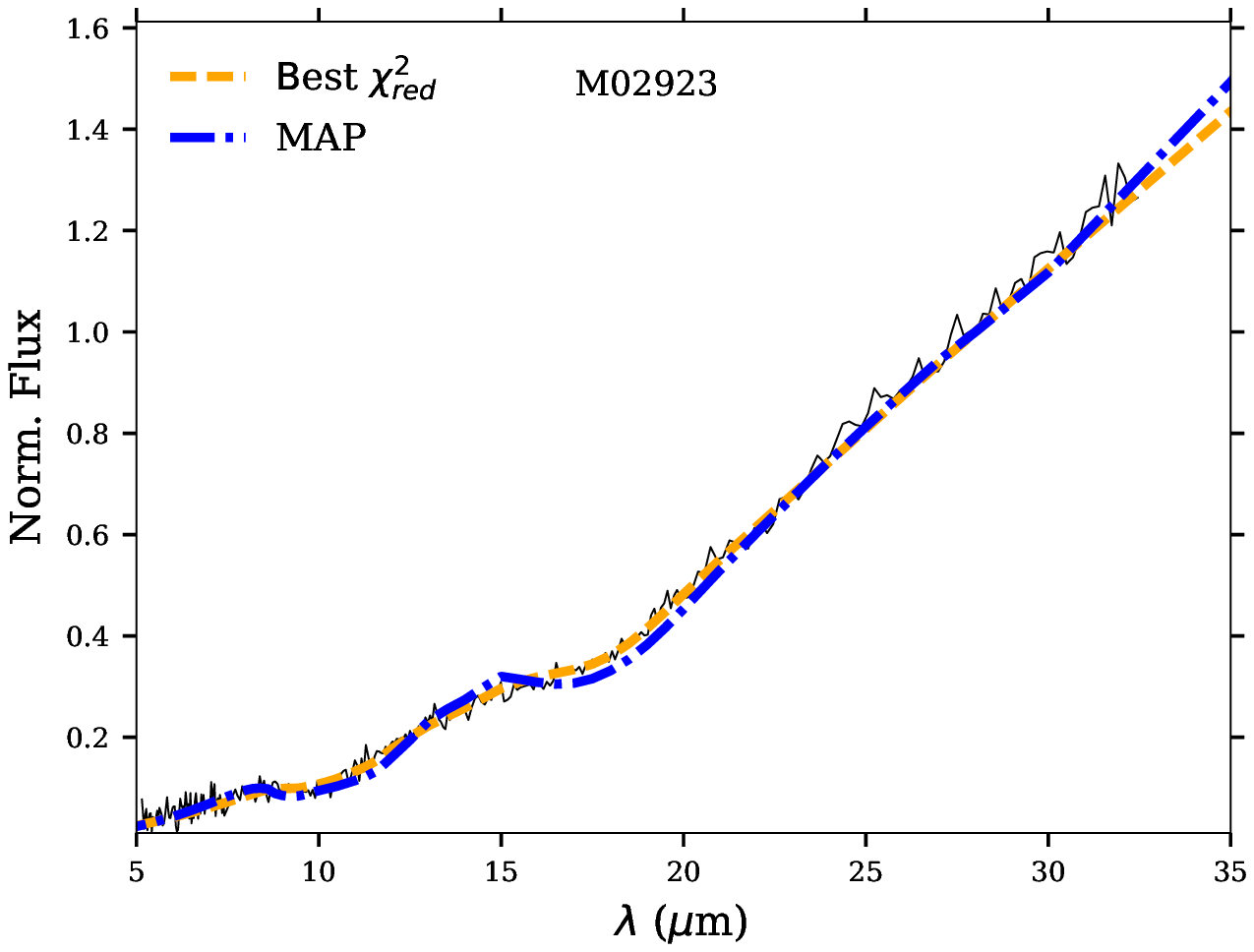}
\end{minipage} \hfill
\begin{minipage}[b]{0.325\linewidth}
\includegraphics[width=\textwidth]{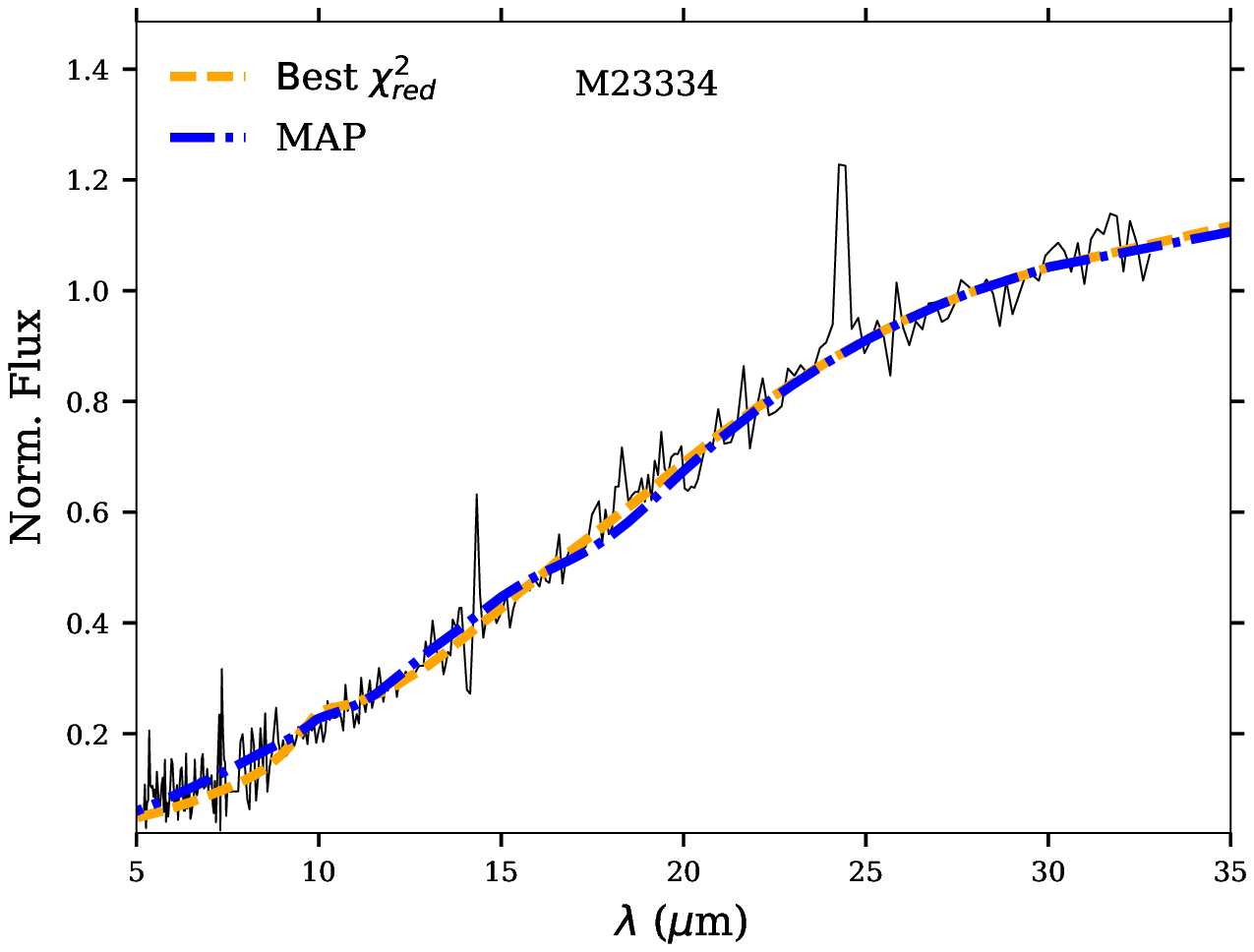}
\end{minipage} \hfill
\begin{minipage}[b]{0.325\linewidth}
\includegraphics[width=\textwidth]{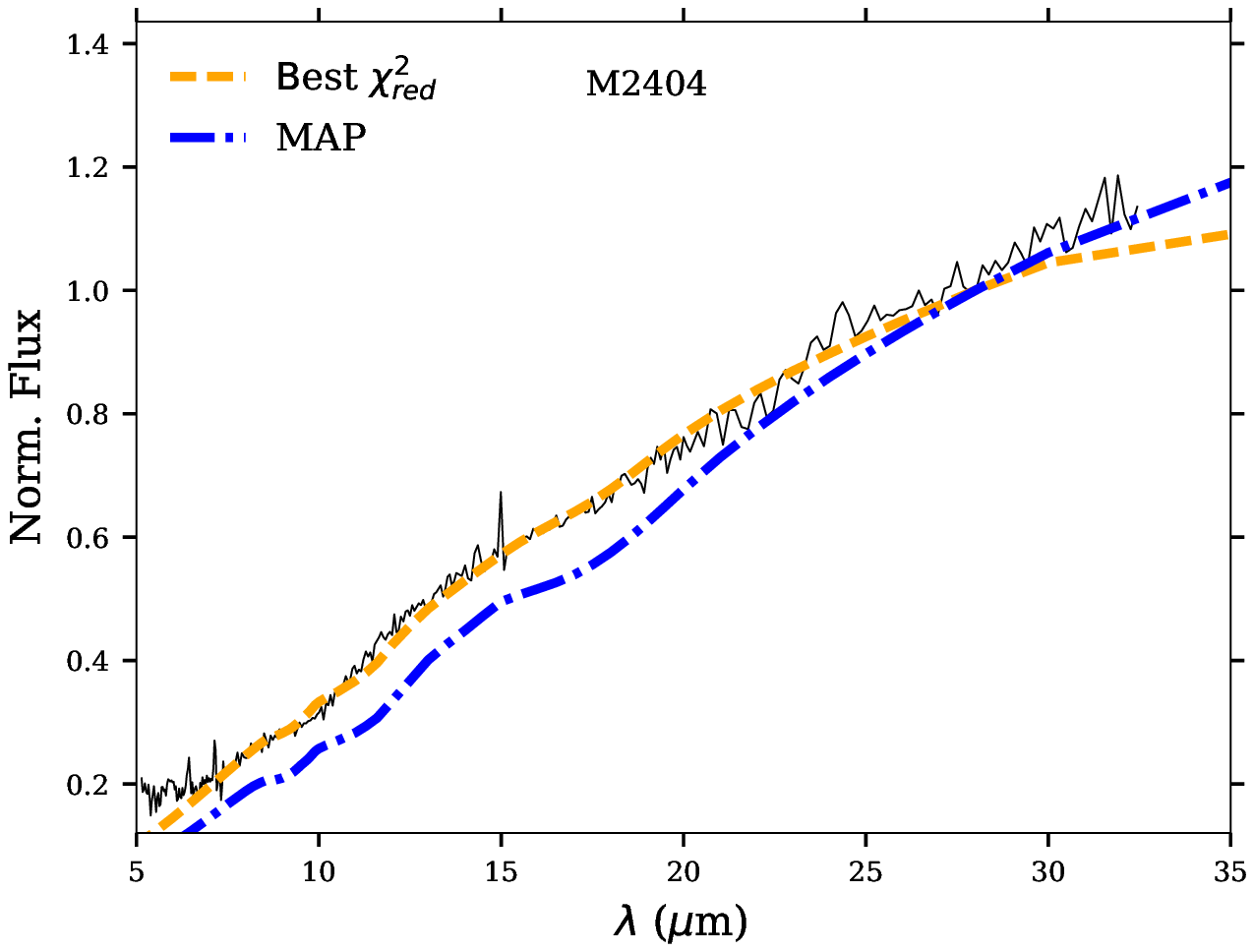}
\end{minipage} \hfill
\begin{minipage}[b]{0.325\linewidth}
\includegraphics[width=\textwidth]{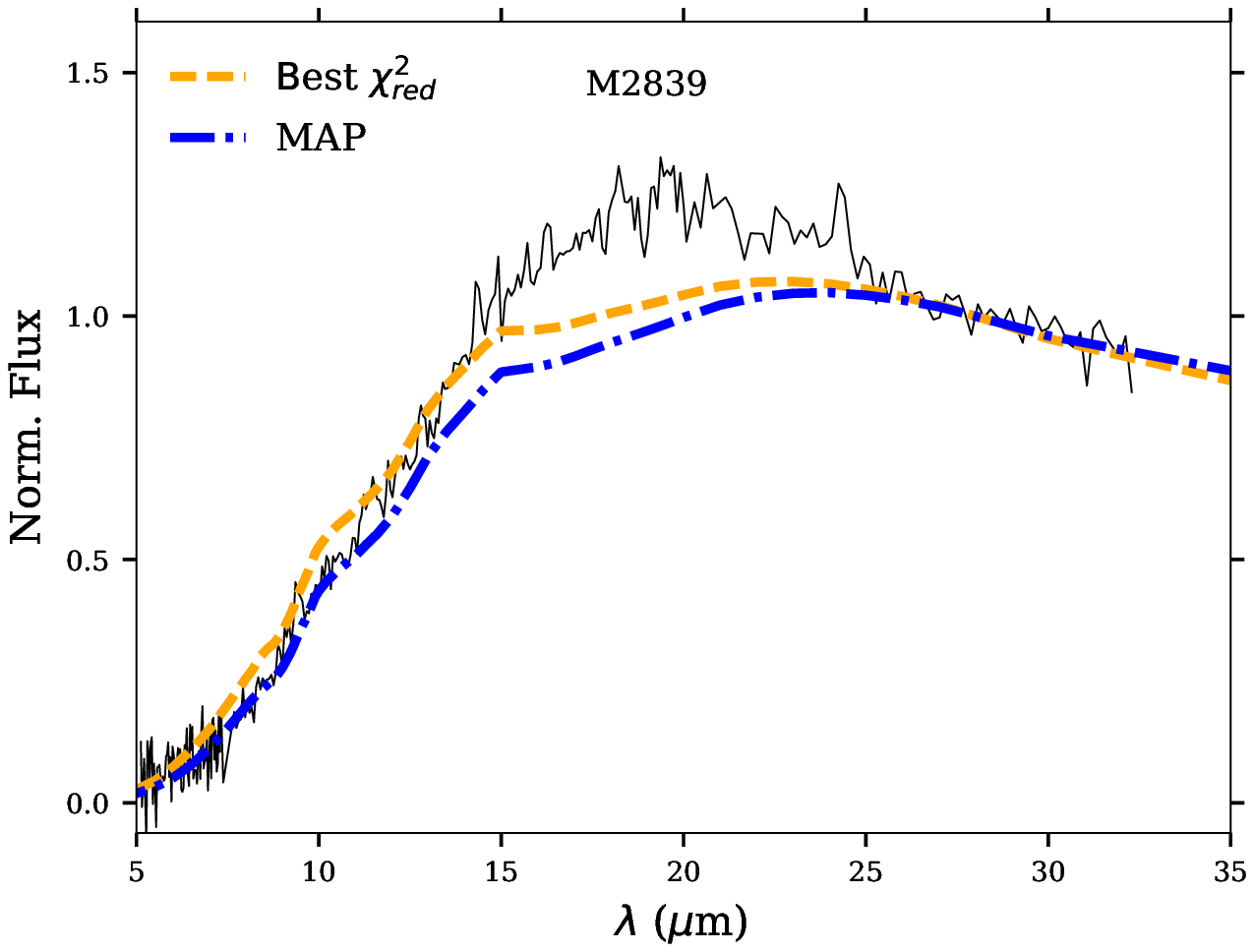}
\end{minipage} \hfill
\begin{minipage}[b]{0.325\linewidth}
\includegraphics[width=\textwidth]{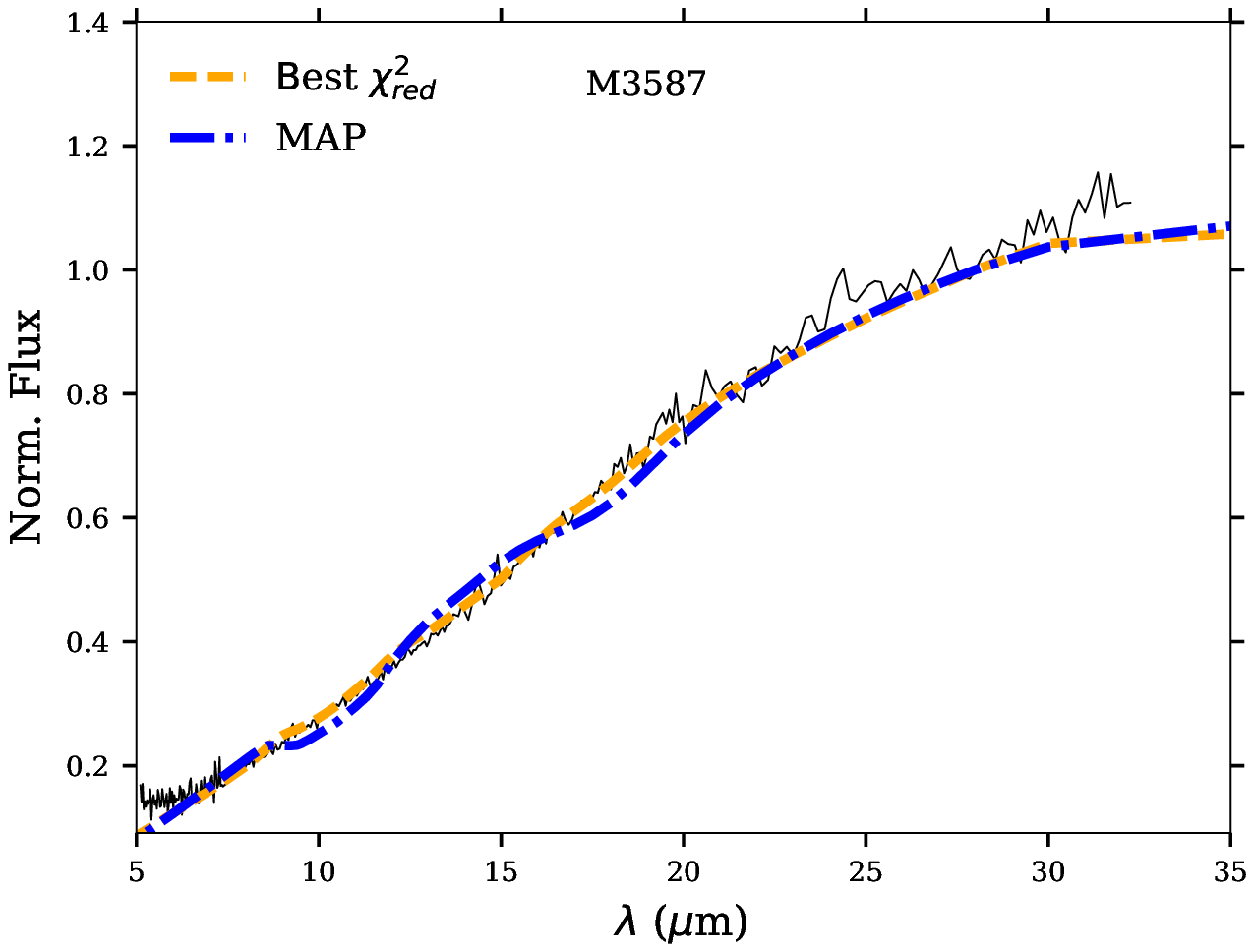}
\end{minipage} \hfill
\begin{minipage}[b]{0.325\linewidth}
\includegraphics[width=\textwidth]{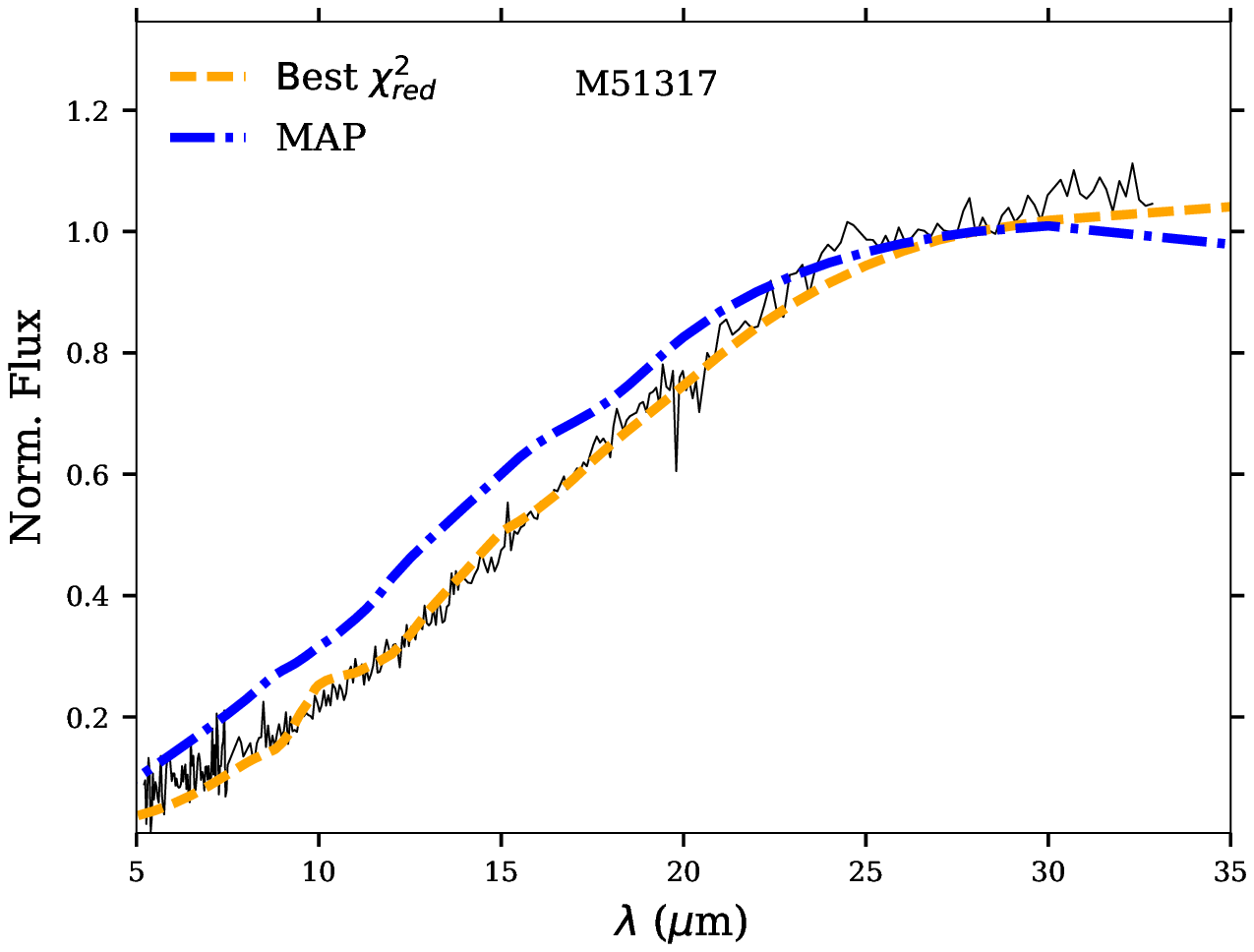}
\end{minipage} \hfill
\begin{minipage}[b]{0.325\linewidth}
\includegraphics[width=\textwidth]{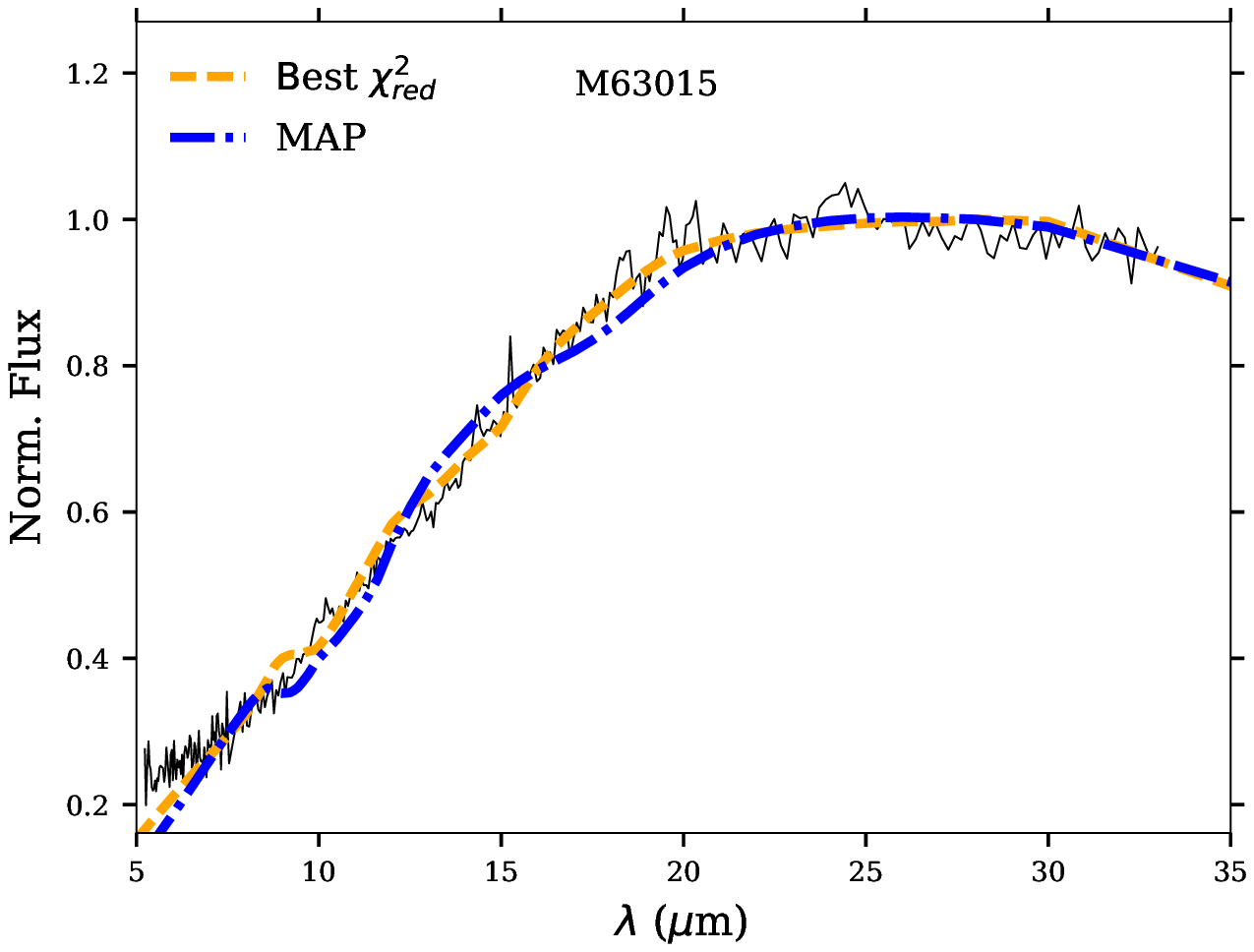}
\end{minipage} \hfill
\begin{minipage}[b]{0.325\linewidth}
\includegraphics[width=\textwidth]{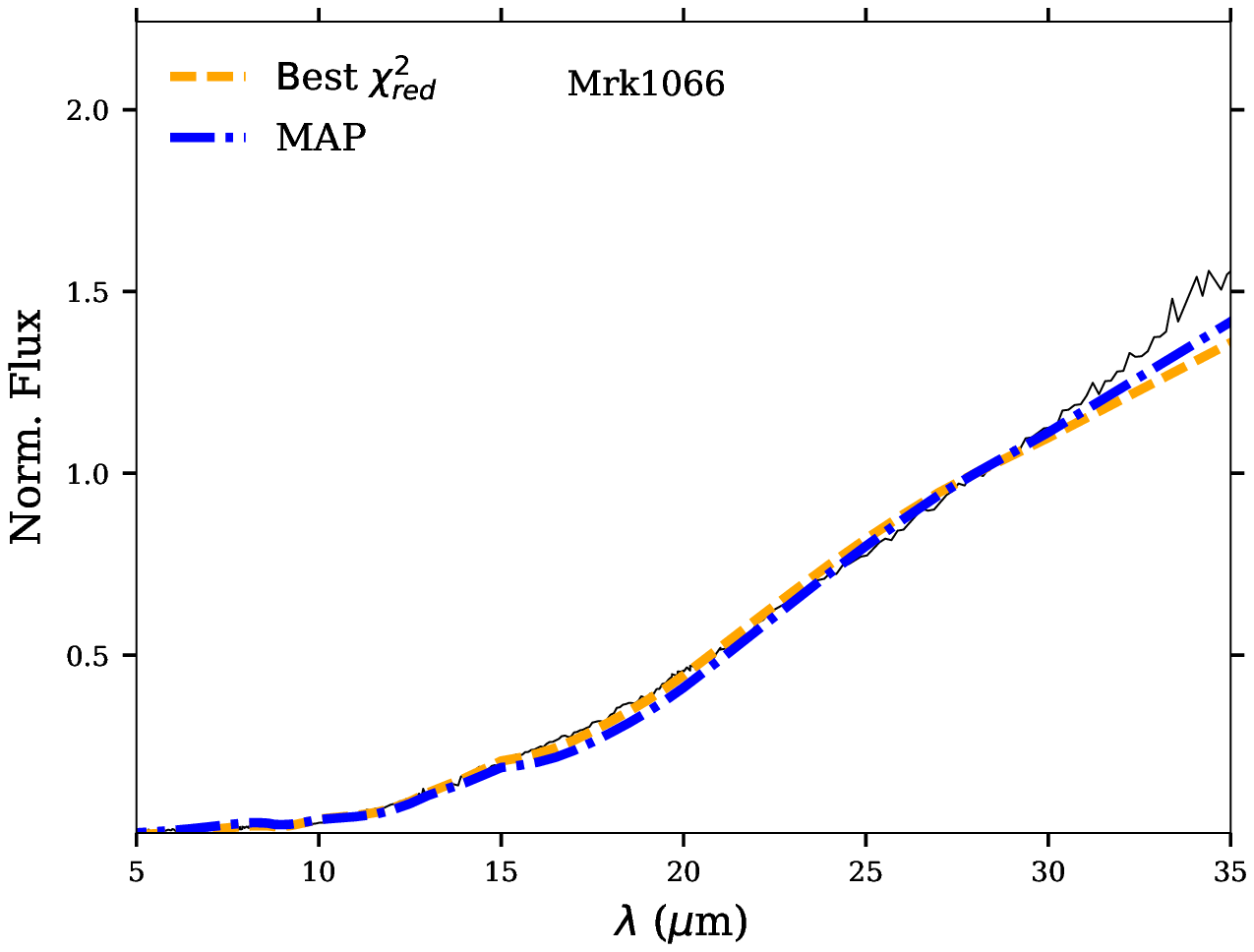}
\end{minipage} \hfill
\begin{minipage}[b]{0.325\linewidth}
\includegraphics[width=\textwidth]{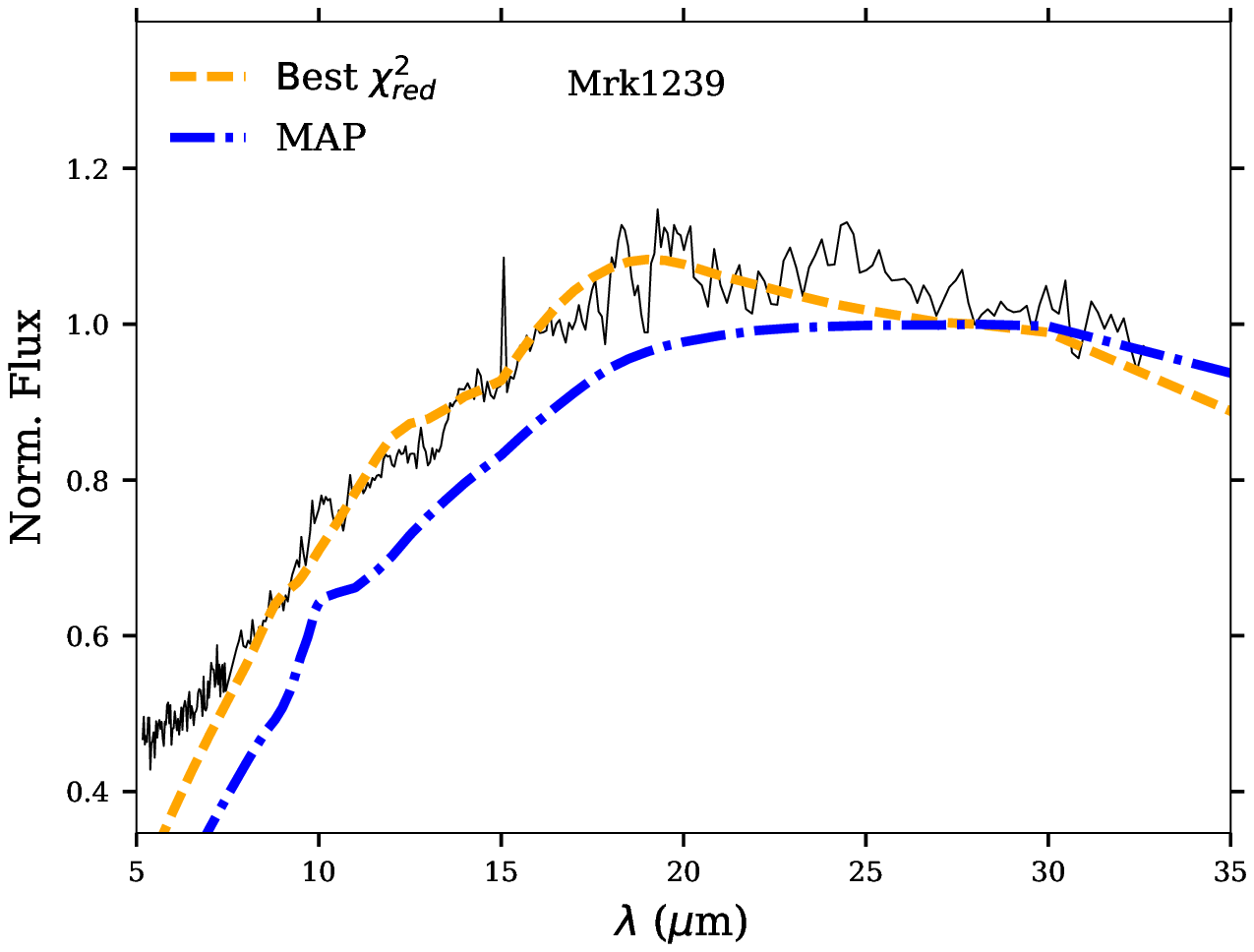}
\end{minipage} \hfill
\begin{minipage}[b]{0.325\linewidth}
\includegraphics[width=\textwidth]{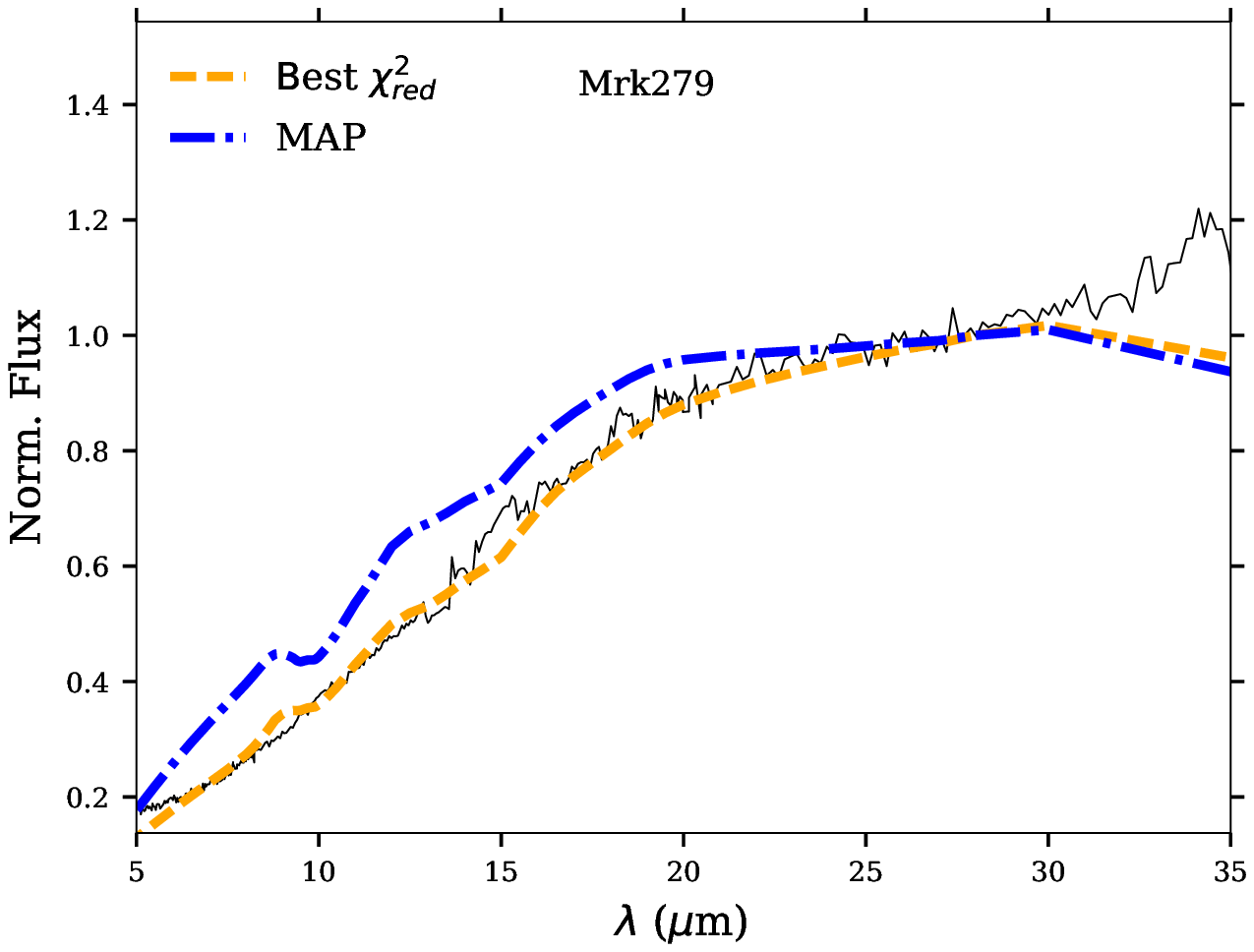}
\end{minipage} \hfill
\begin{minipage}[b]{0.325\linewidth}
\includegraphics[width=\textwidth]{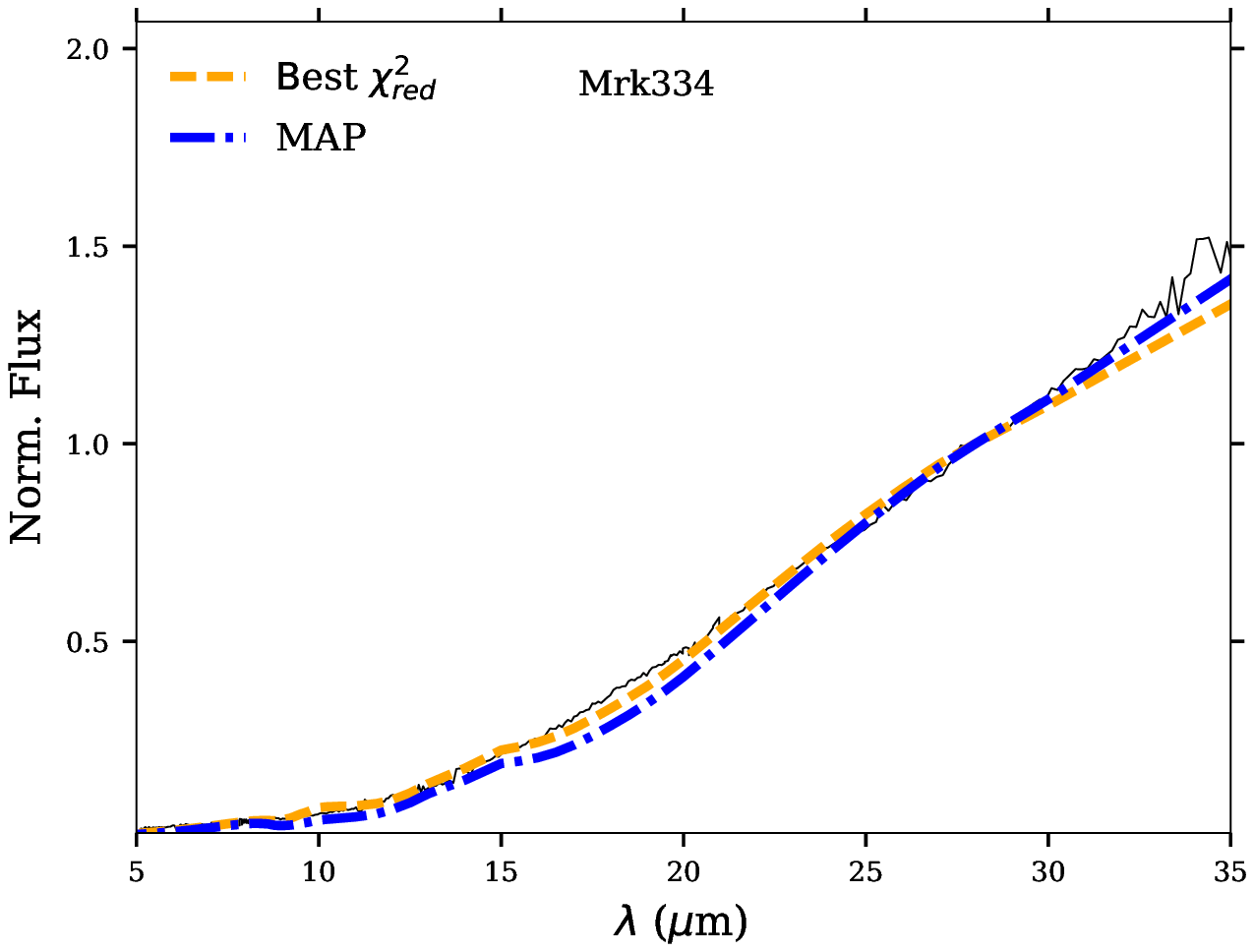}
\end{minipage} \hfill
\begin{minipage}[b]{0.325\linewidth}
\includegraphics[width=\textwidth]{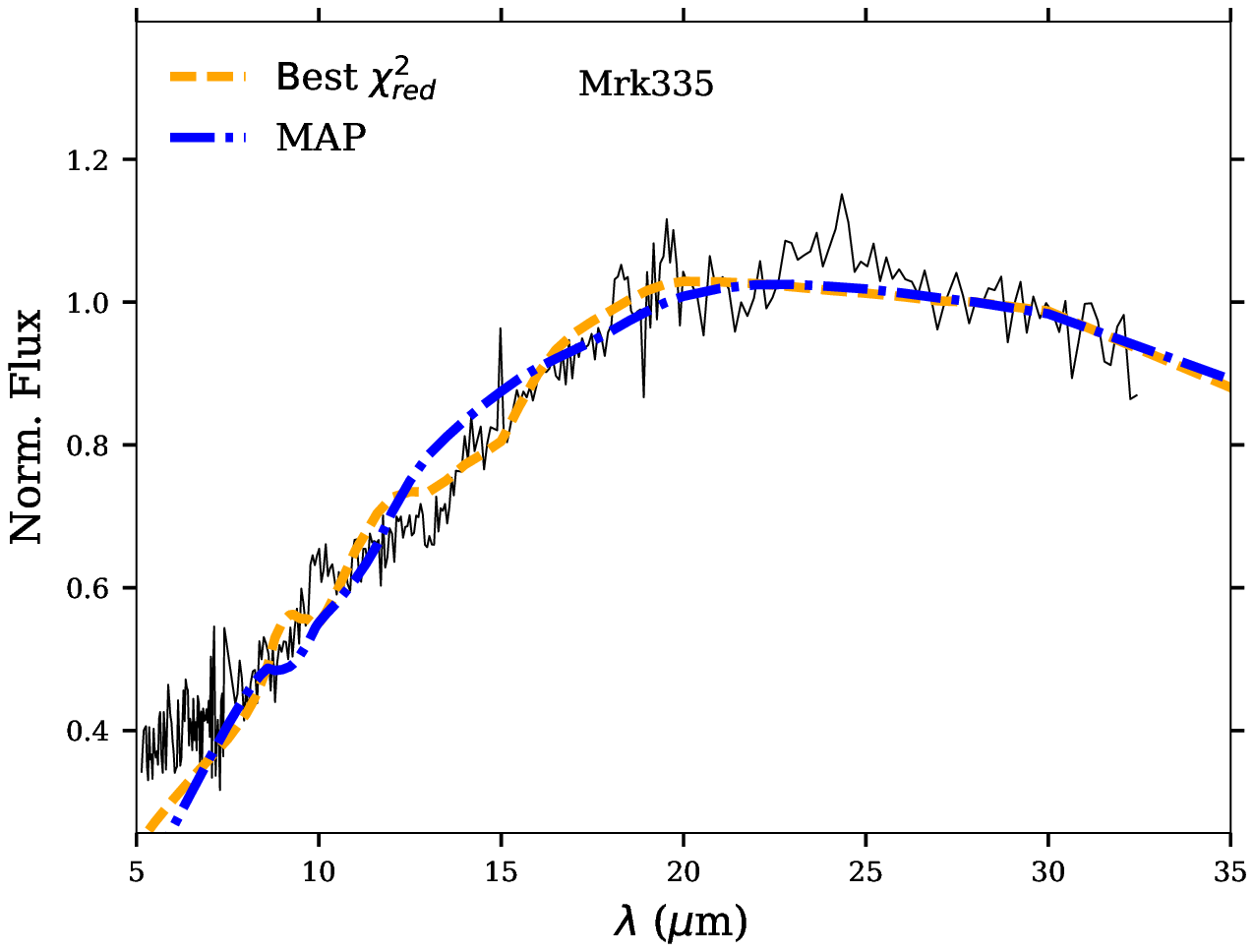}
\end{minipage} \hfill
\begin{minipage}[b]{0.325\linewidth}
\includegraphics[width=\textwidth]{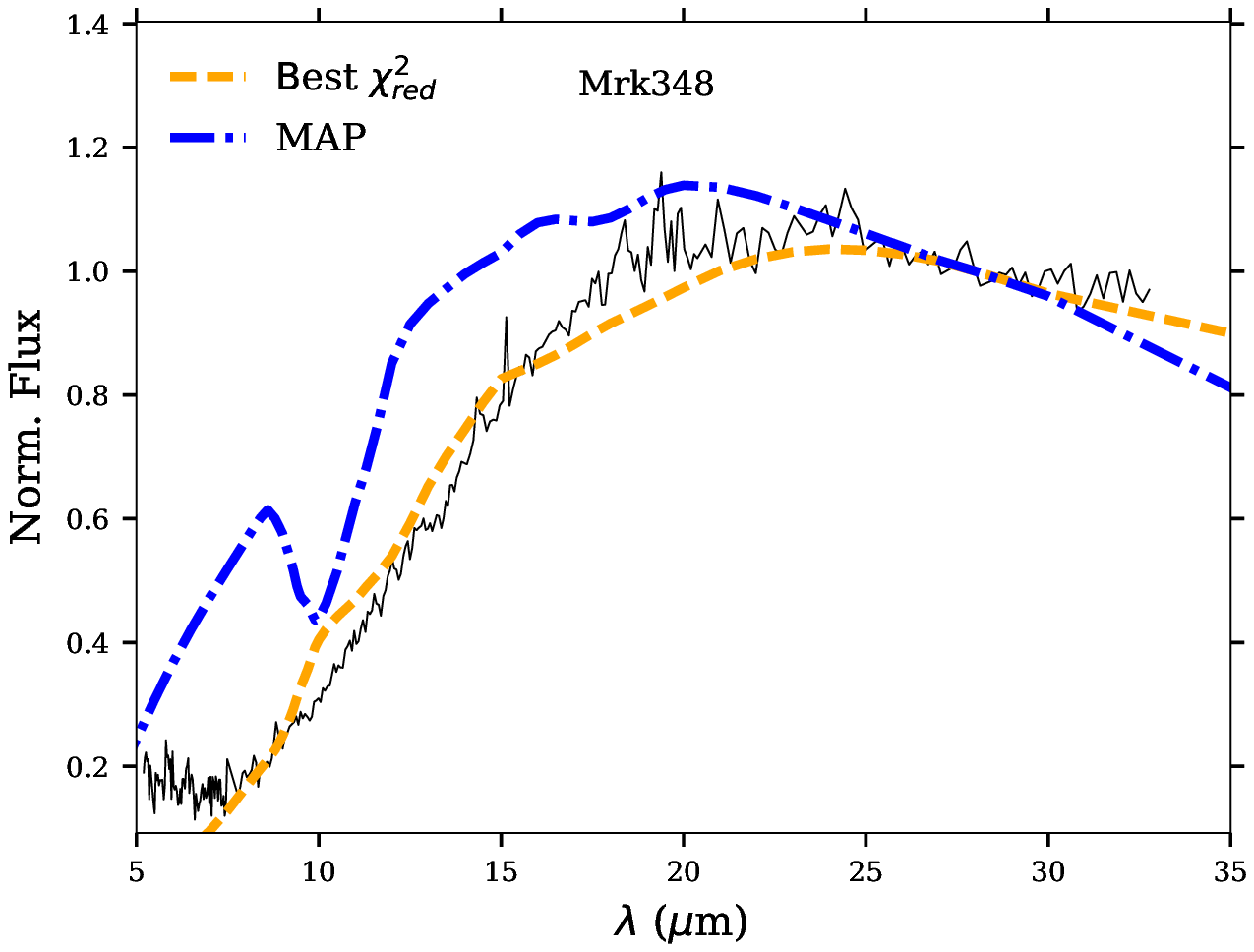}
\end{minipage} \hfill
\begin{minipage}[b]{0.325\linewidth}
\includegraphics[width=\textwidth]{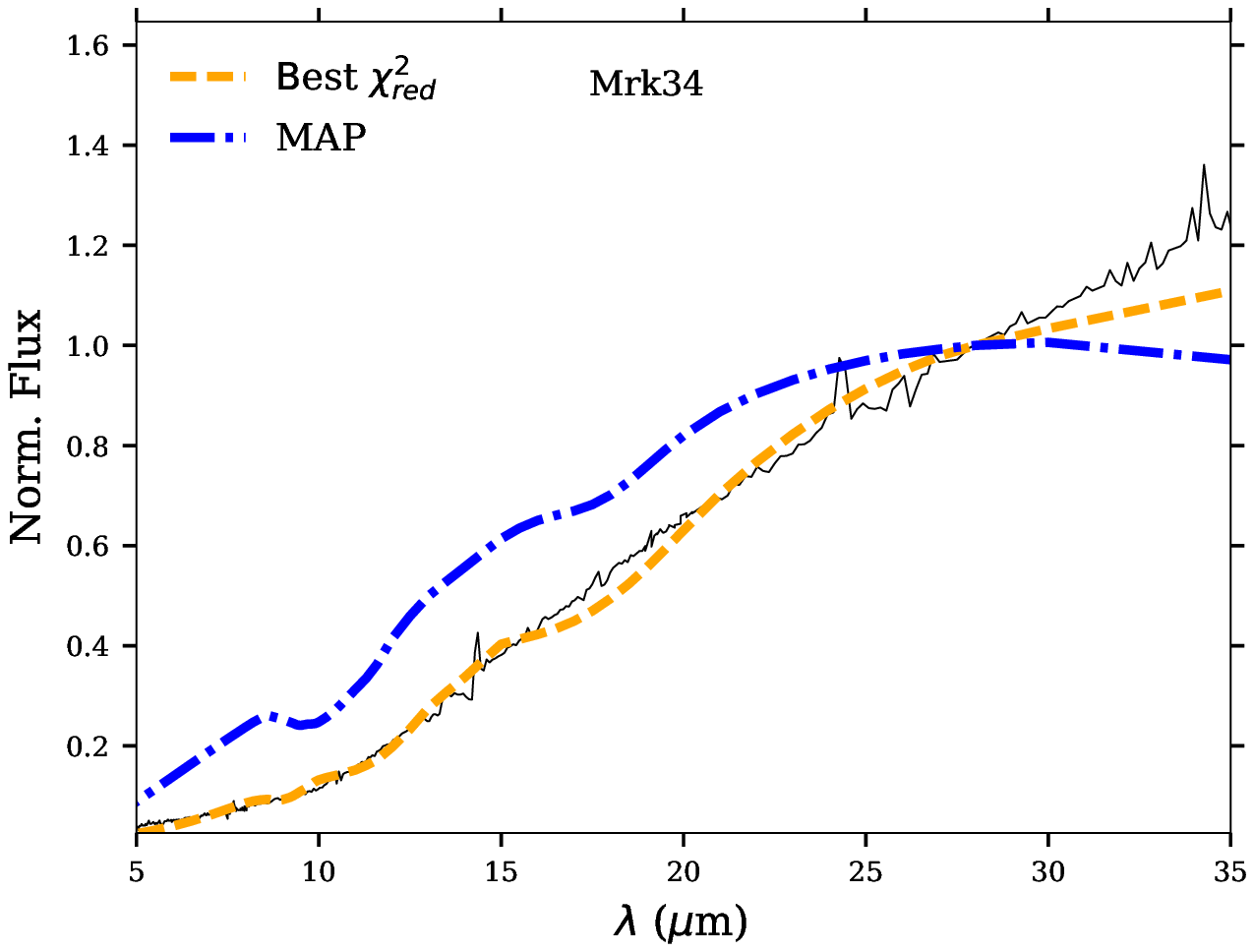}
\end{minipage} \hfill
\caption{continued from previous page.}
\setcounter{figure}{0}
\end{figure}

\begin{figure}

\begin{minipage}[b]{0.325\linewidth}
\includegraphics[width=\textwidth]{Mrk3_bestMAP_JY.eps}
\end{minipage} \hfill
\begin{minipage}[b]{0.325\linewidth}
\includegraphics[width=\textwidth]{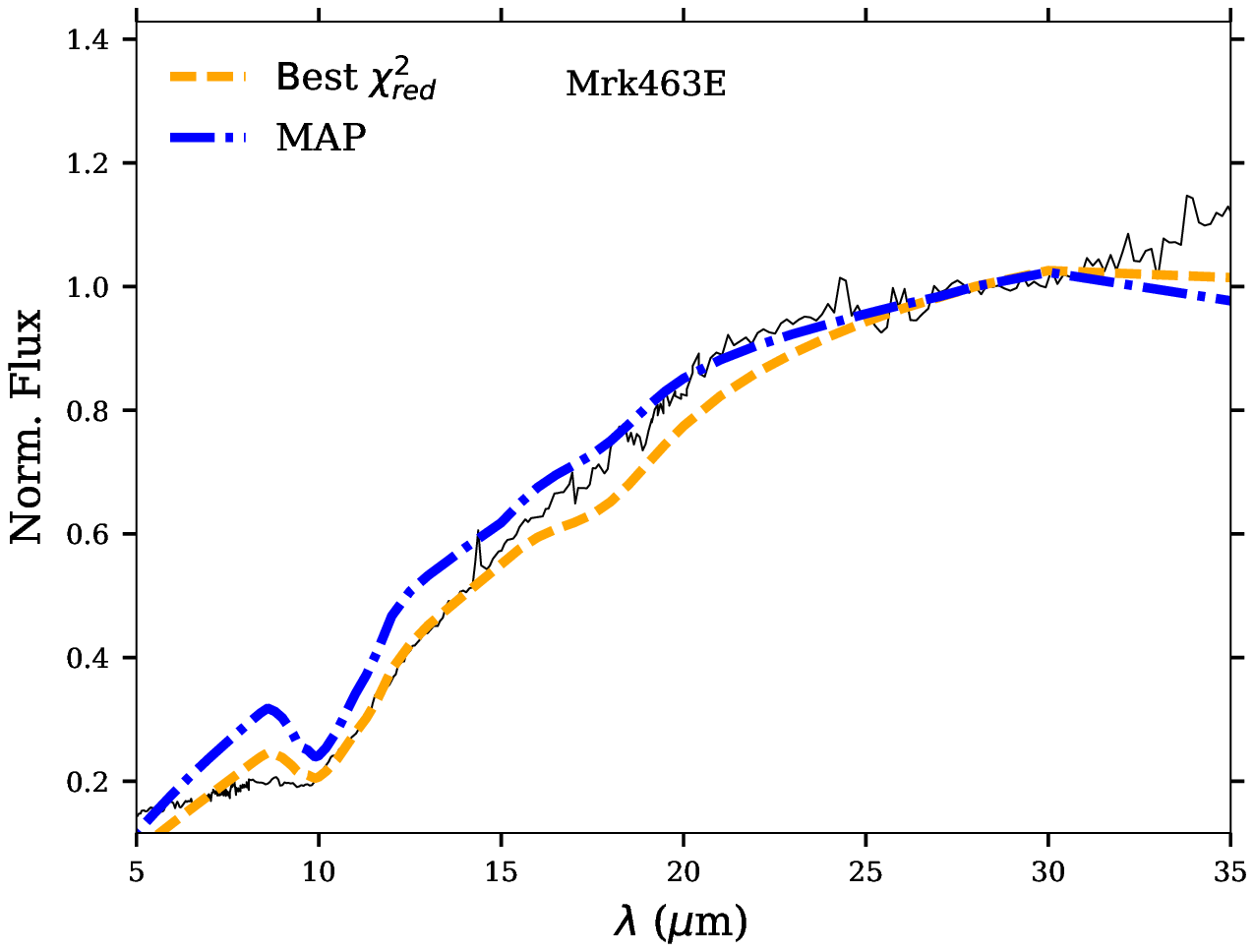}
\end{minipage} \hfill
\begin{minipage}[b]{0.325\linewidth}
\includegraphics[width=\textwidth]{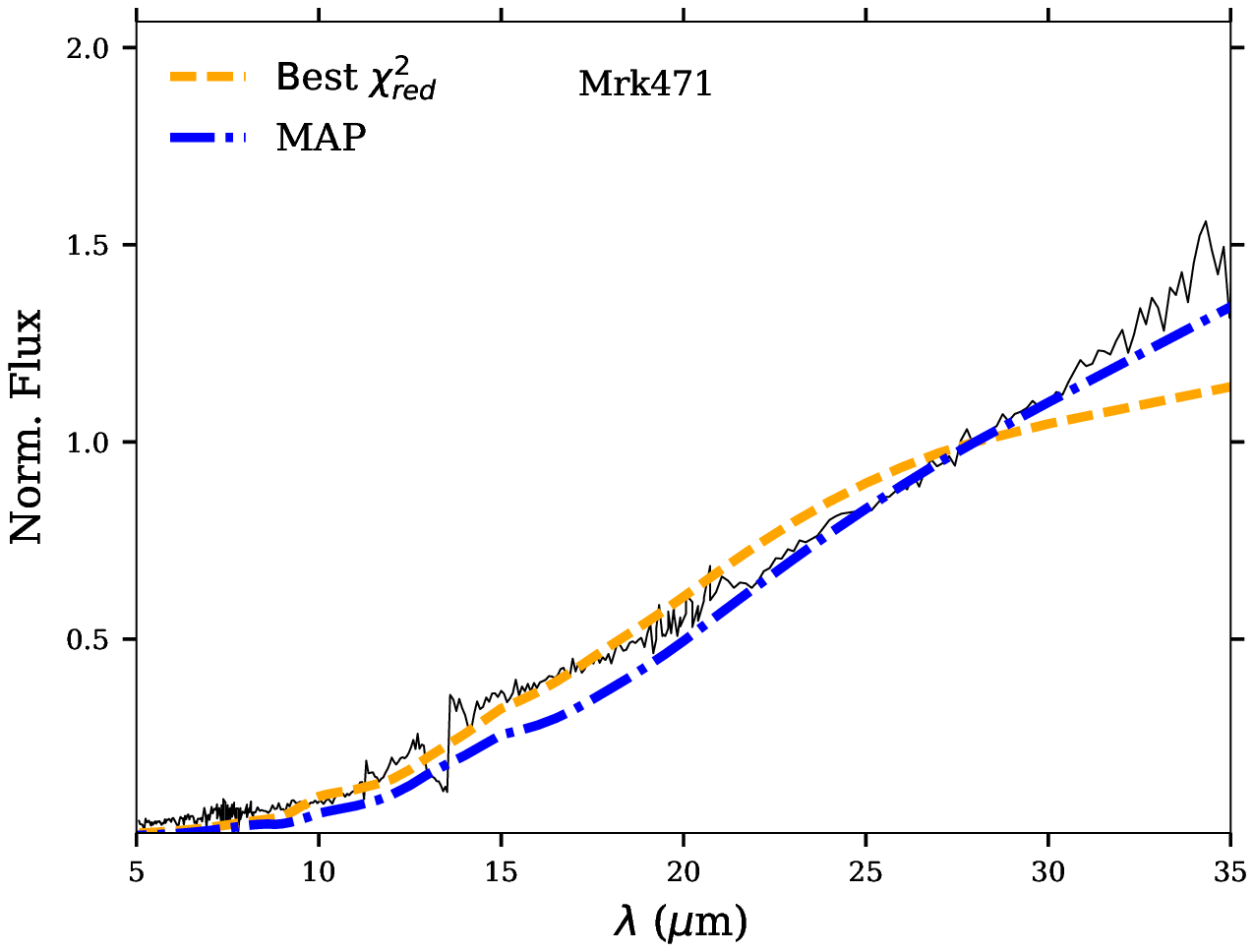}
\end{minipage} \hfill
\begin{minipage}[b]{0.325\linewidth}
\includegraphics[width=\textwidth]{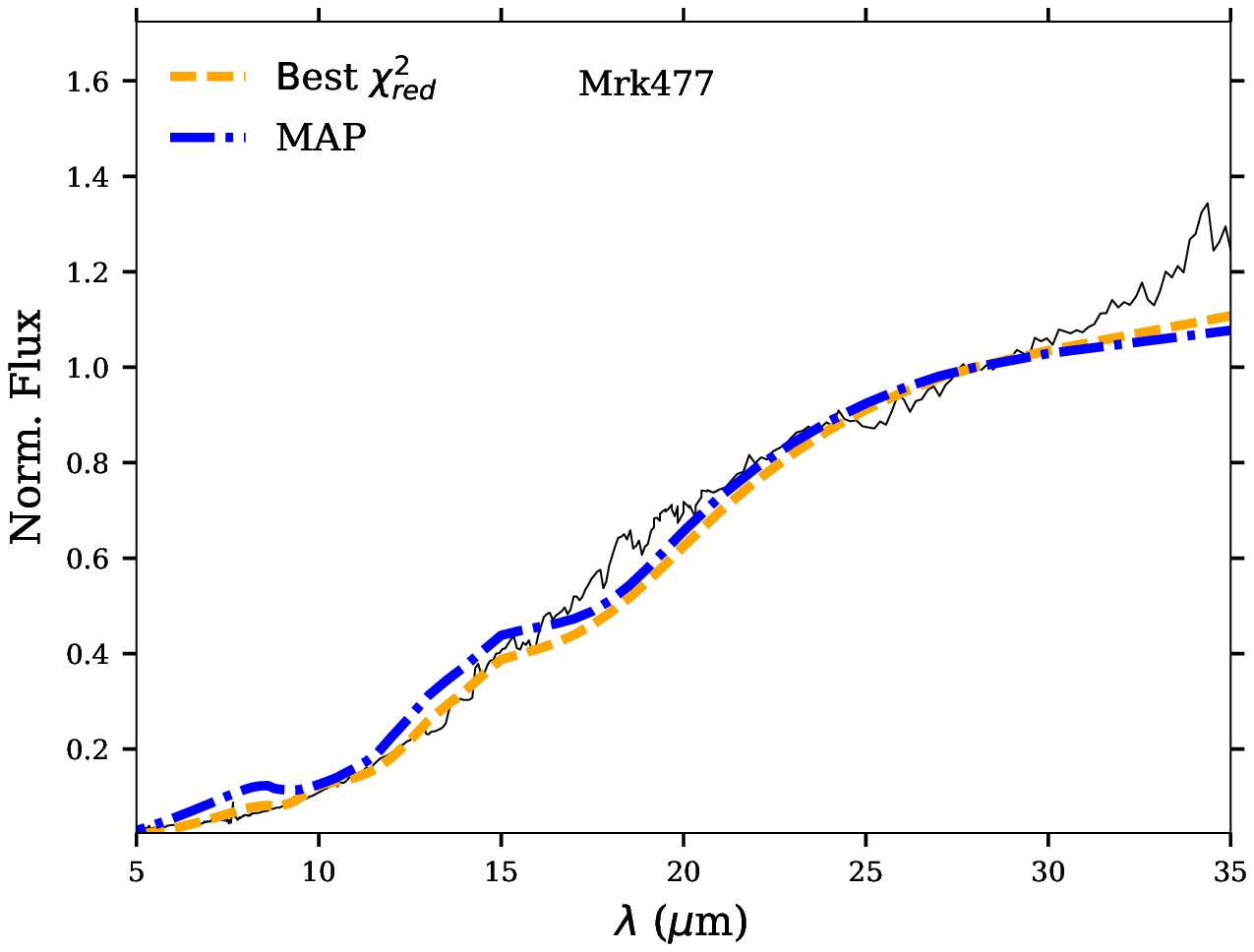}
\end{minipage} \hfill
\begin{minipage}[b]{0.325\linewidth}
\includegraphics[width=\textwidth]{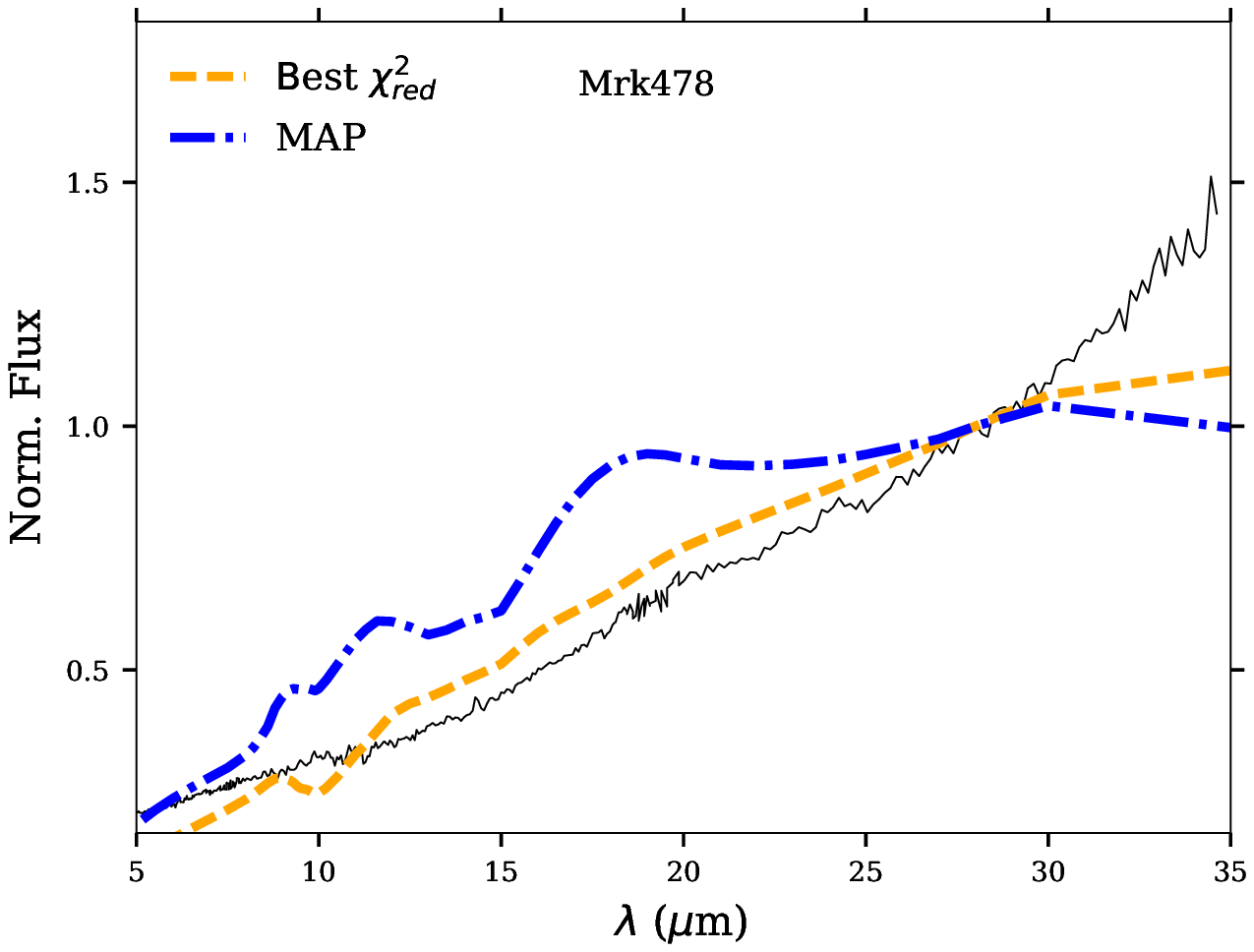}
\end{minipage} \hfill
\begin{minipage}[b]{0.325\linewidth}
\includegraphics[width=\textwidth]{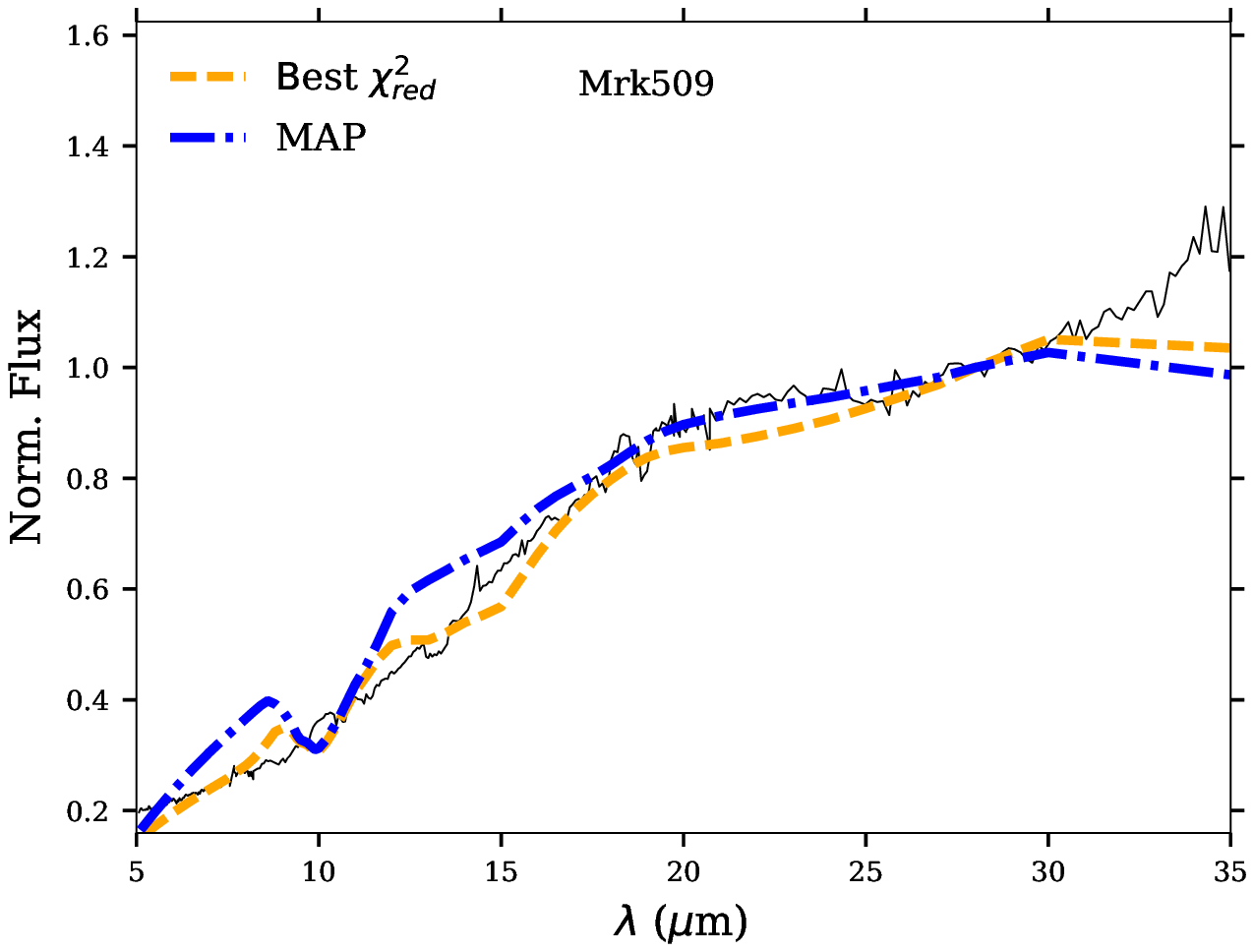}
\end{minipage} \hfill
\begin{minipage}[b]{0.325\linewidth}
\includegraphics[width=\textwidth]{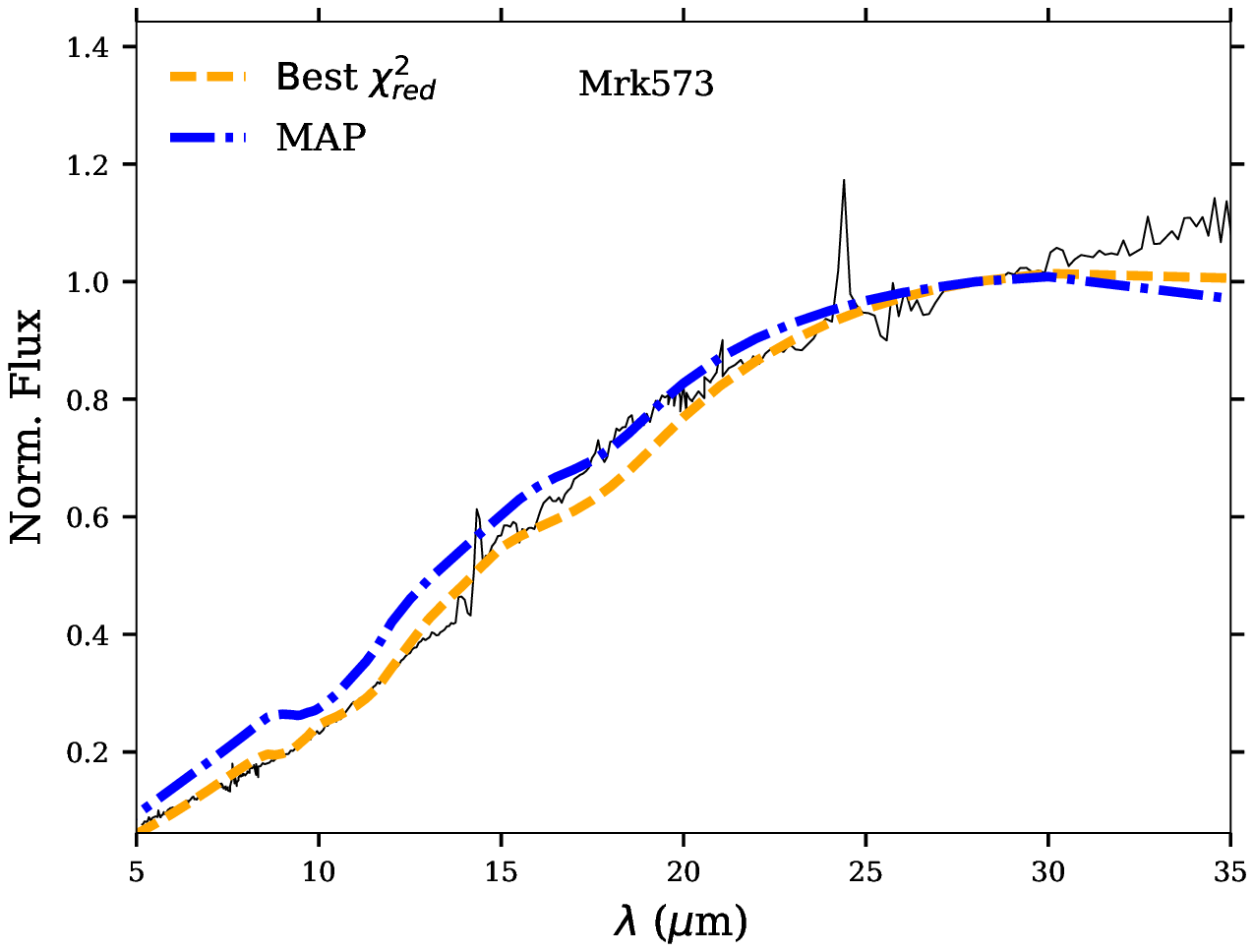}
\end{minipage} \hfill
\begin{minipage}[b]{0.325\linewidth}
\includegraphics[width=\textwidth]{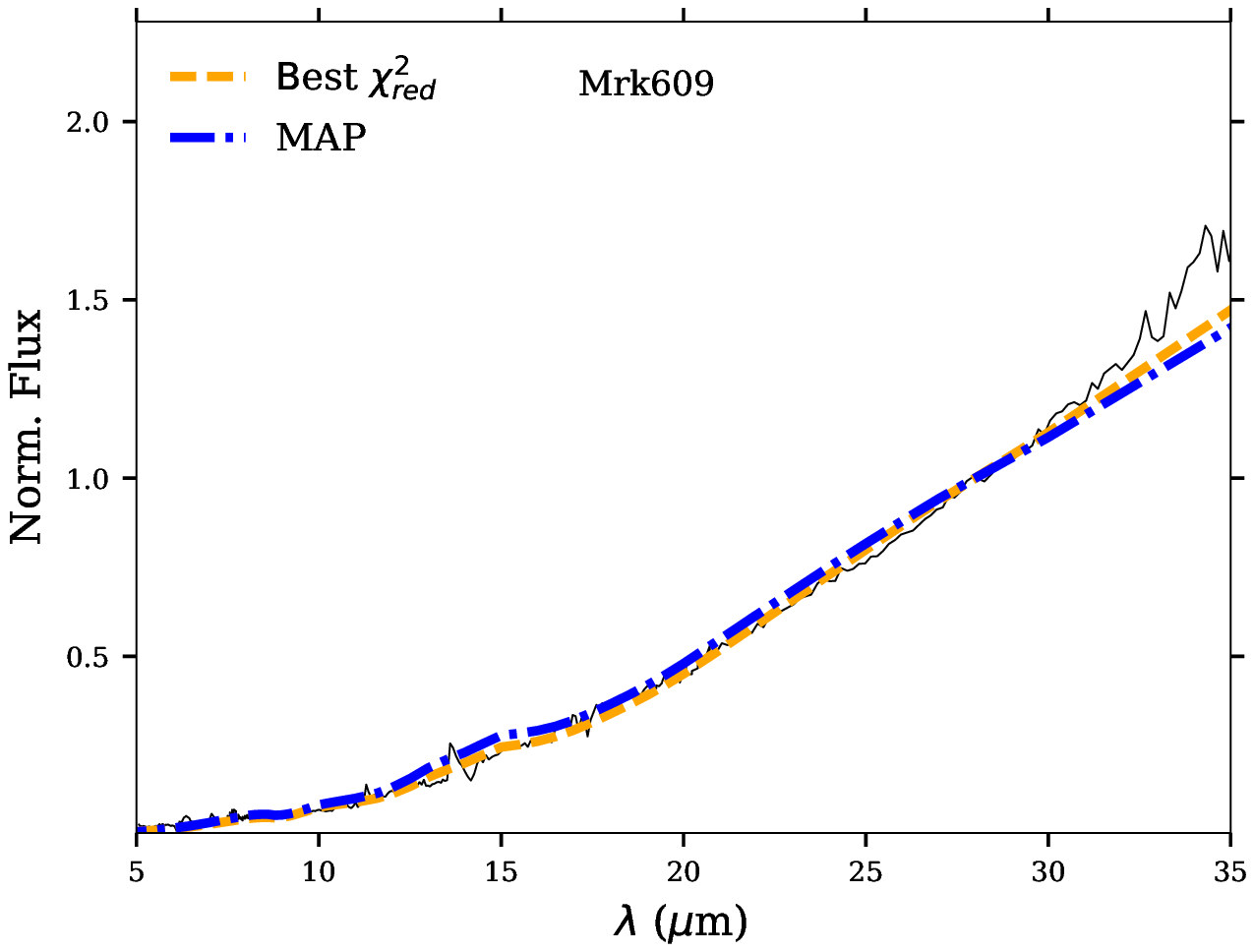}
\end{minipage} \hfill
\begin{minipage}[b]{0.325\linewidth}
\includegraphics[width=\textwidth]{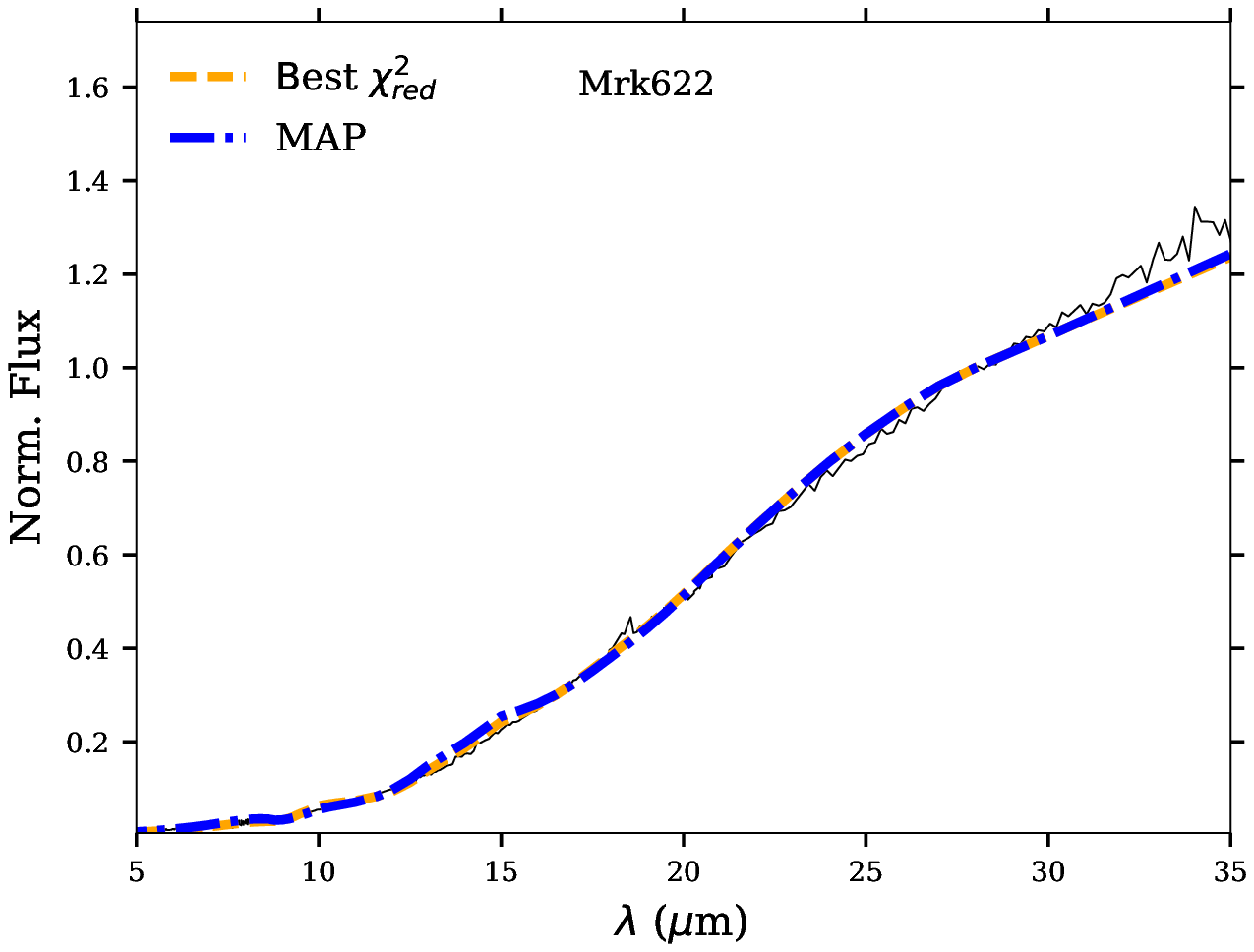}
\end{minipage} \hfill
\begin{minipage}[b]{0.325\linewidth}
\includegraphics[width=\textwidth]{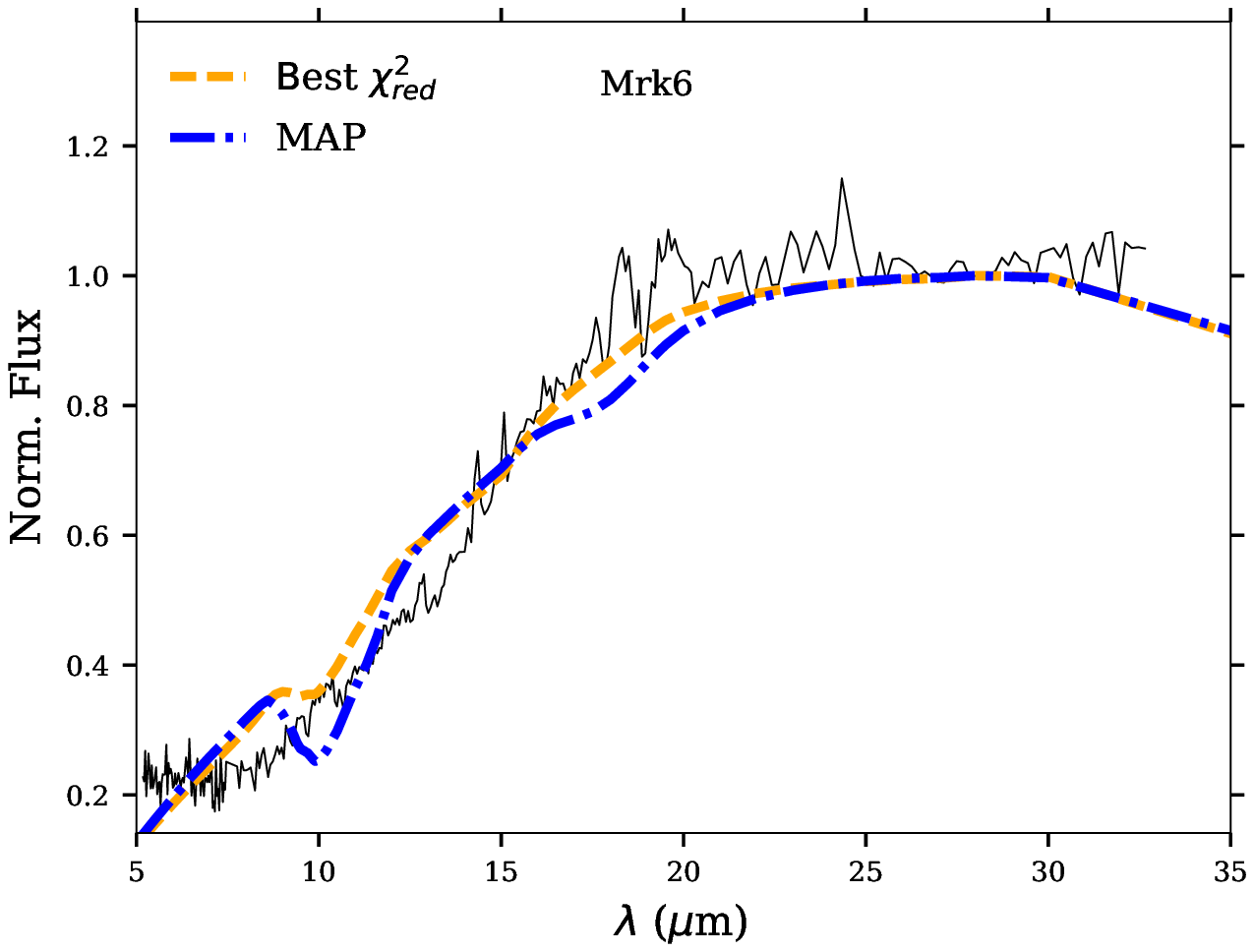}
\end{minipage} \hfill
\begin{minipage}[b]{0.325\linewidth}
\includegraphics[width=\textwidth]{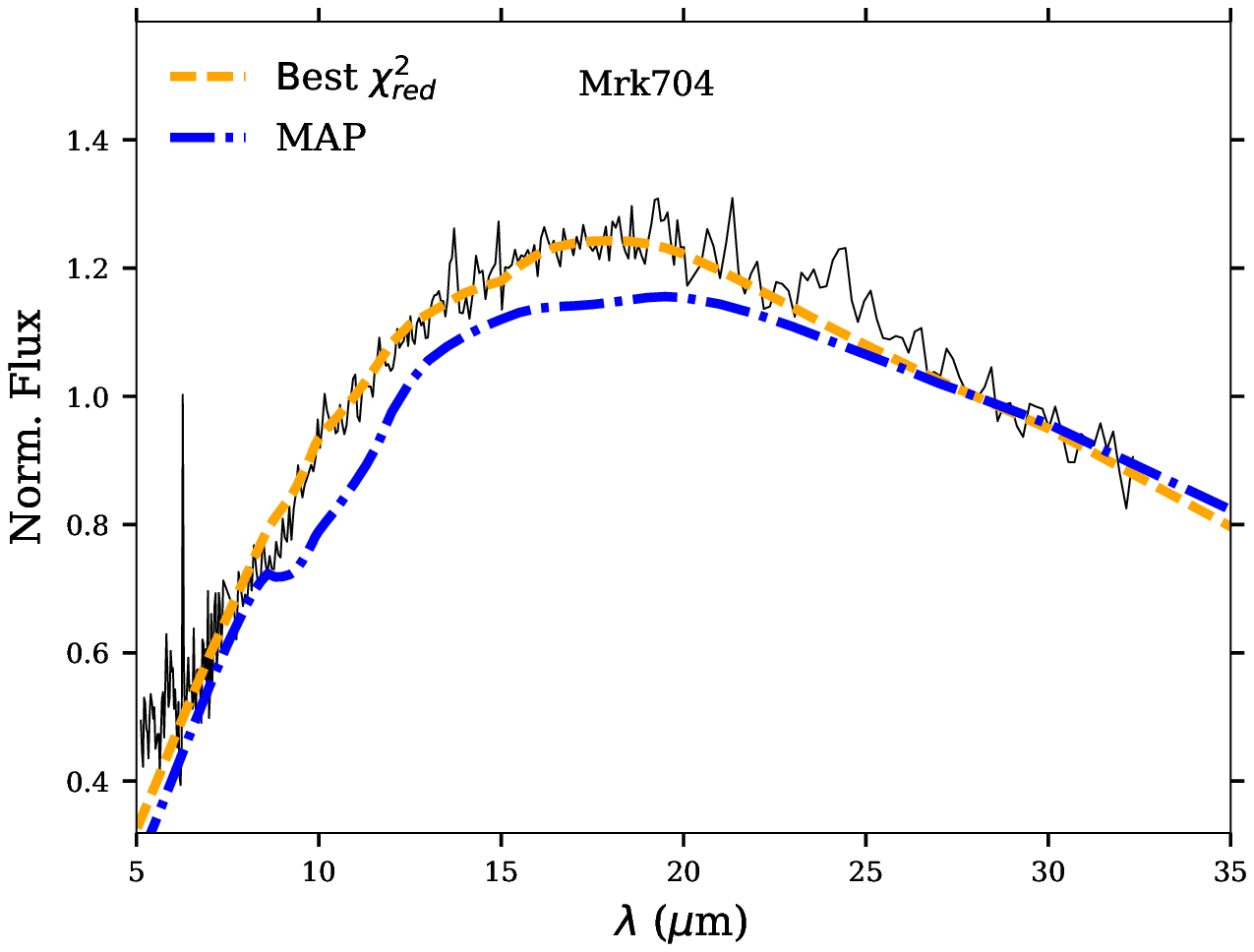}
\end{minipage} \hfill
\begin{minipage}[b]{0.325\linewidth}
\includegraphics[width=\textwidth]{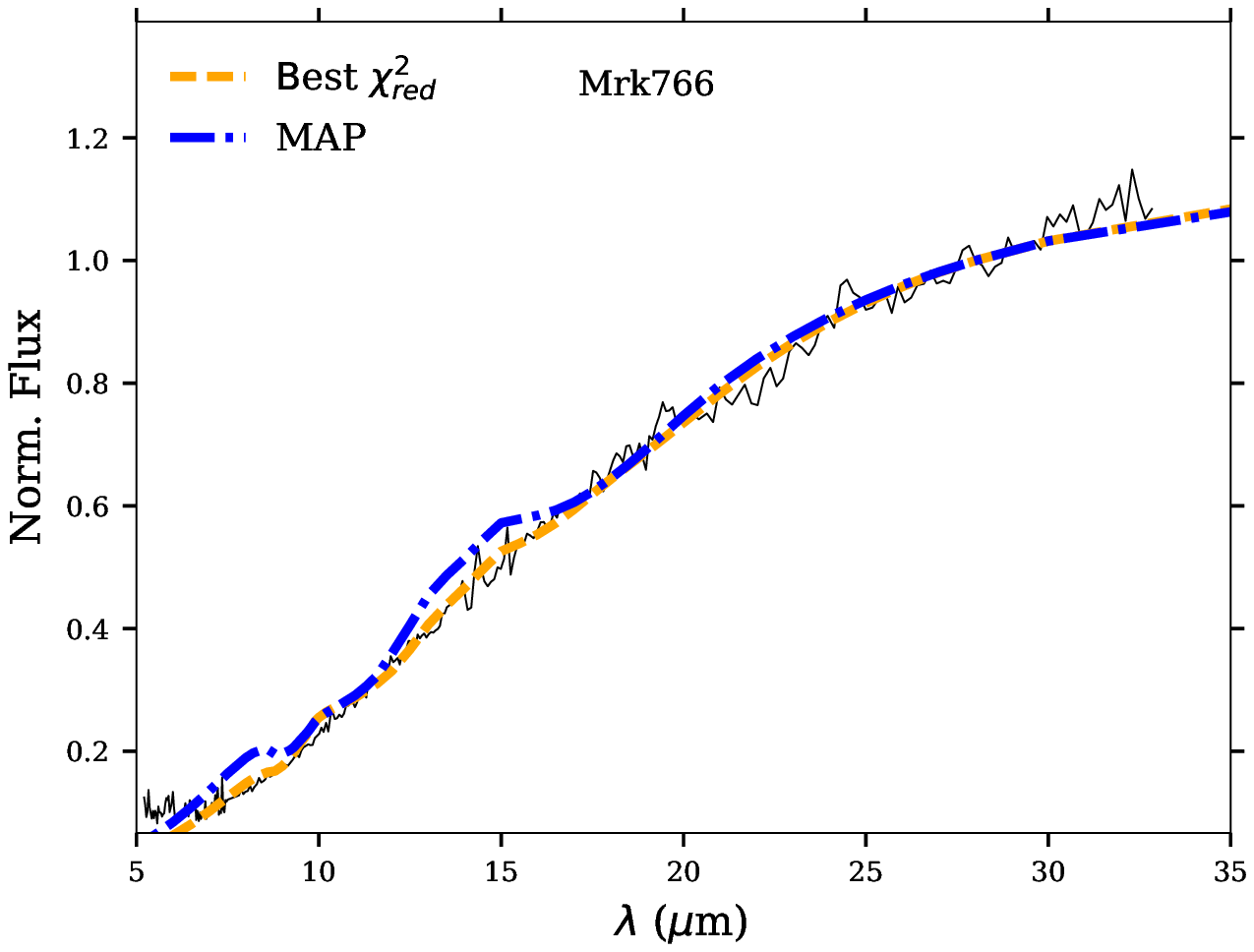}
\end{minipage} \hfill
\begin{minipage}[b]{0.325\linewidth}
\includegraphics[width=\textwidth]{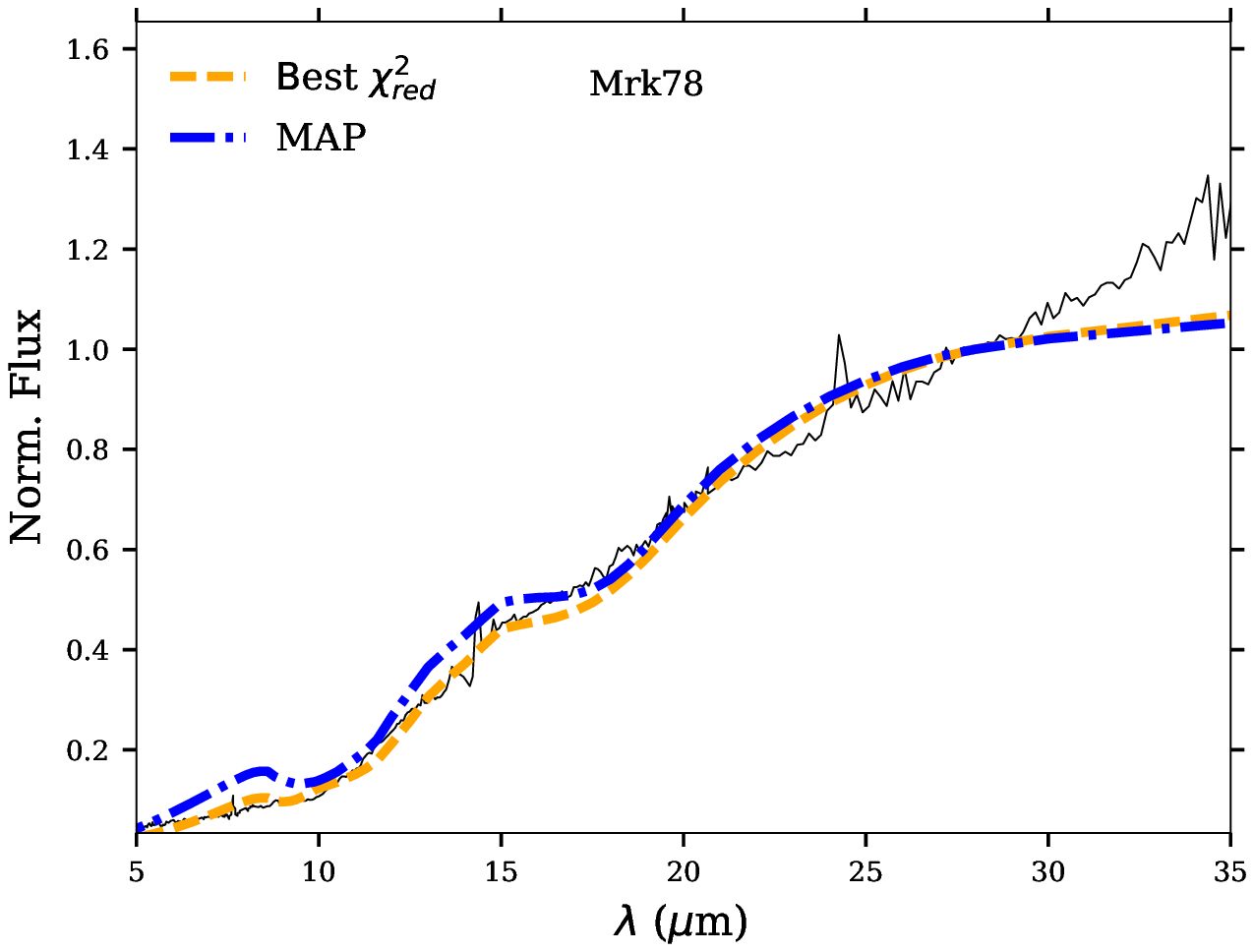}
\end{minipage} \hfill
\begin{minipage}[b]{0.325\linewidth}
\includegraphics[width=\textwidth]{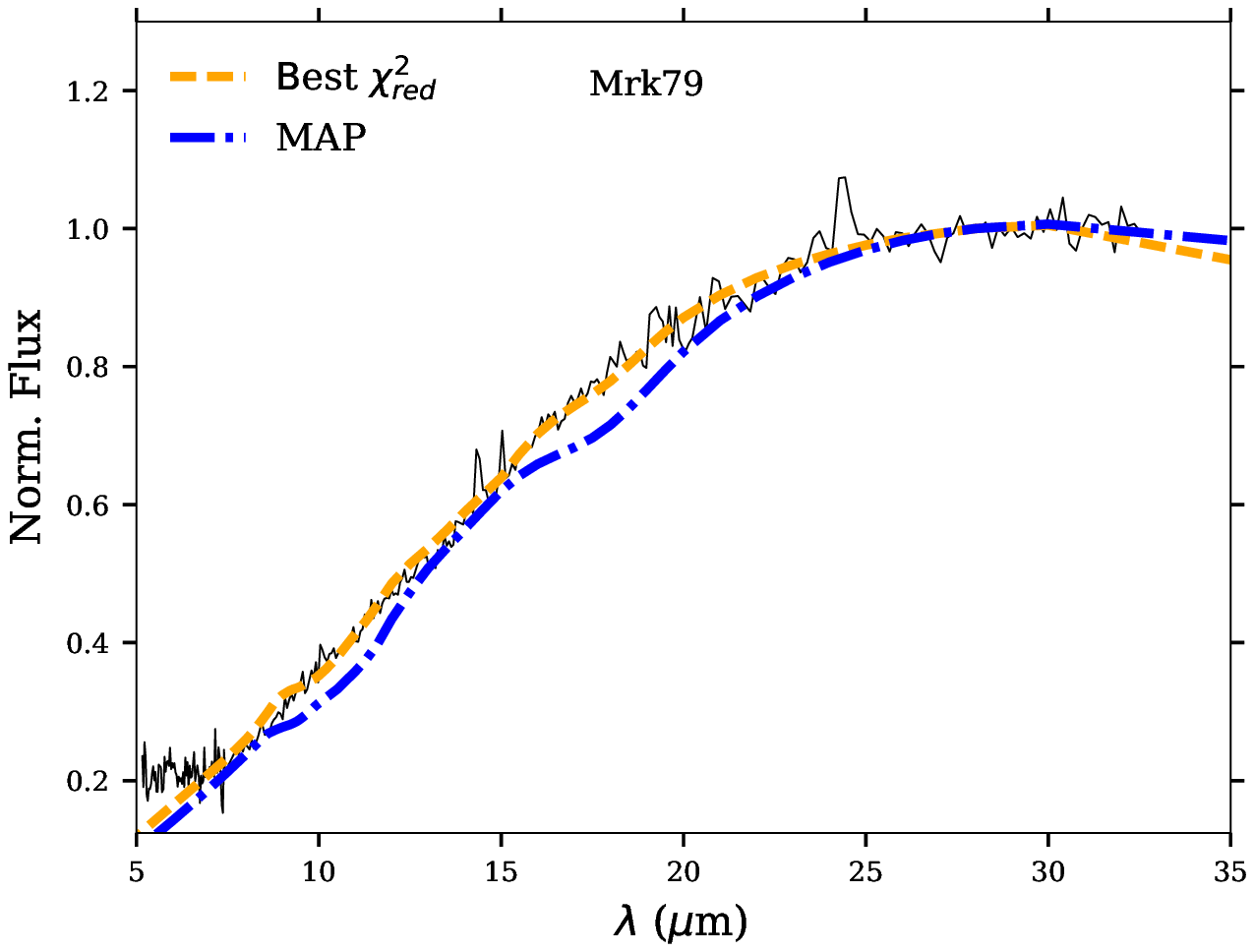}
\end{minipage} \hfill
\begin{minipage}[b]{0.325\linewidth}
\includegraphics[width=\textwidth]{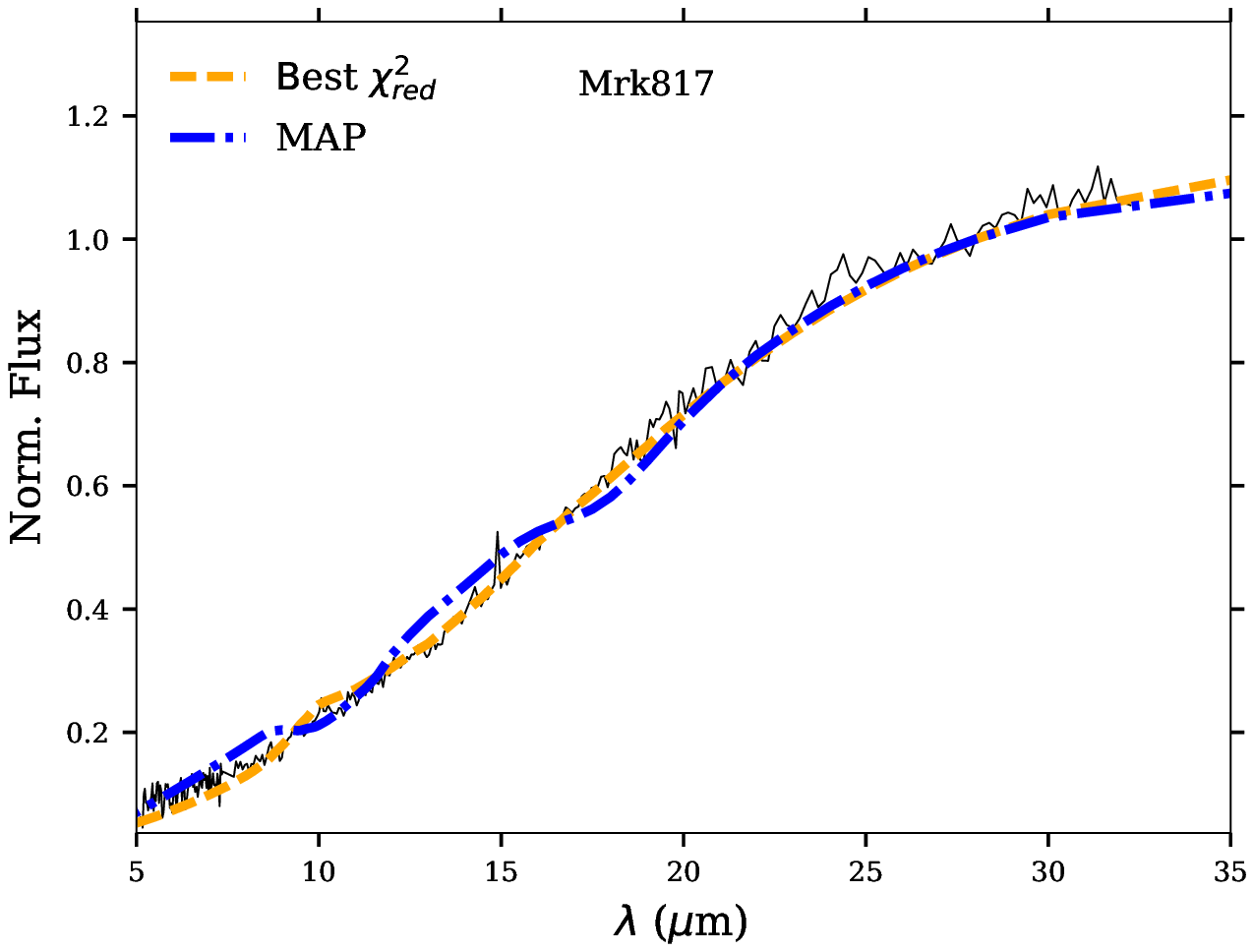}
\end{minipage} \hfill
\caption{continued from previous page.}
\setcounter{figure}{0}
\end{figure}

\begin{figure}

\begin{minipage}[b]{0.325\linewidth}
\includegraphics[width=\textwidth]{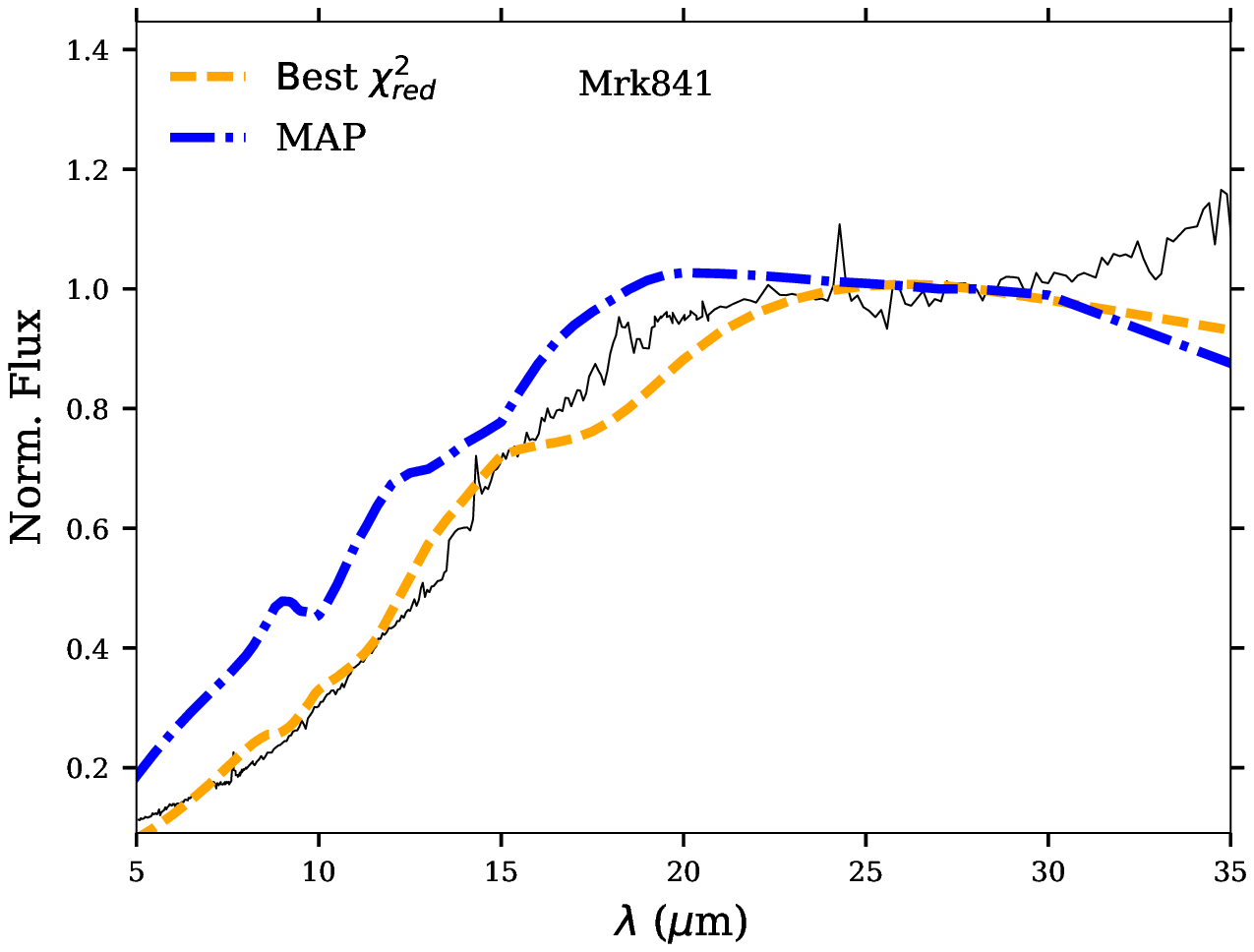}
\end{minipage} \hfill
\begin{minipage}[b]{0.325\linewidth}
\includegraphics[width=\textwidth]{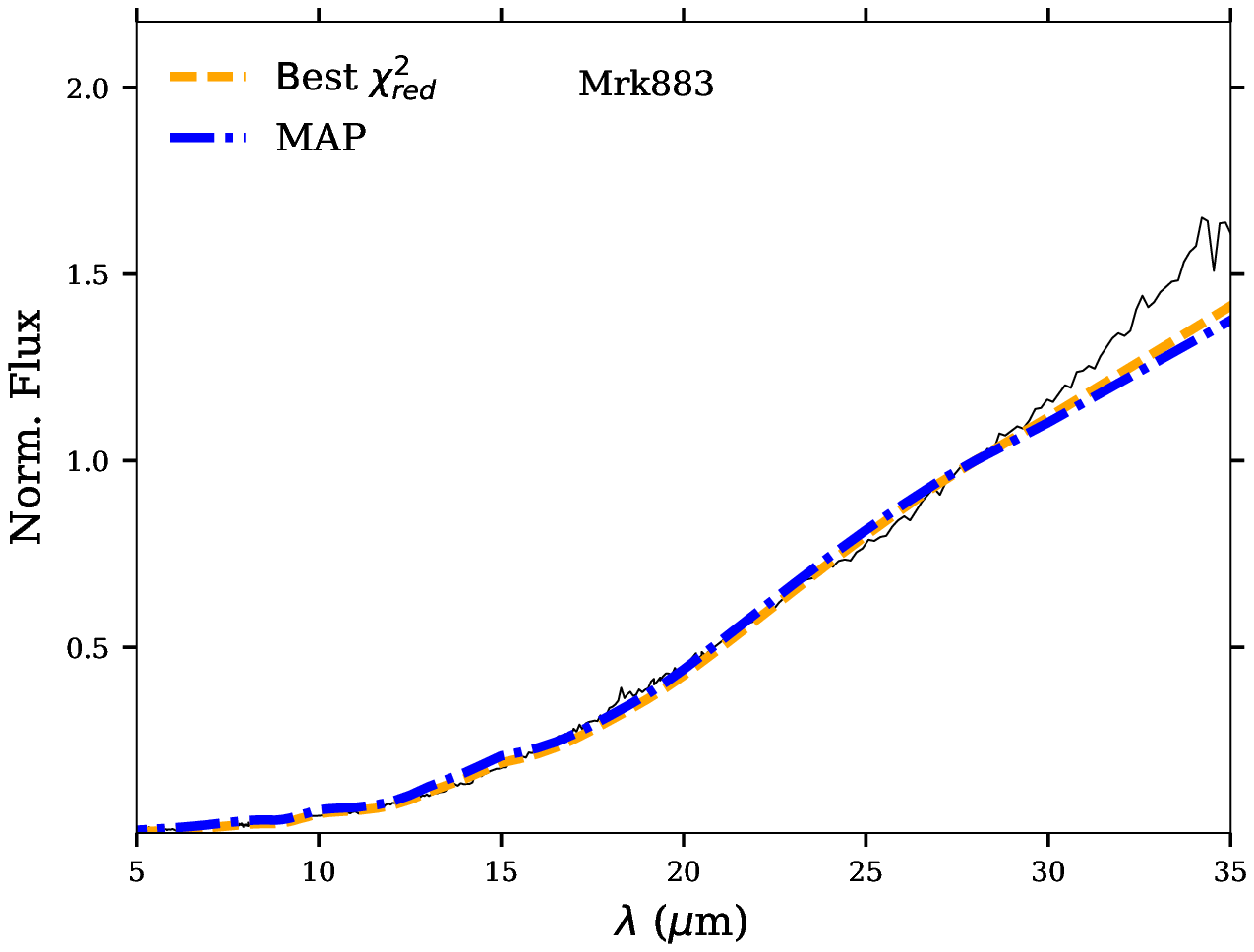}
\end{minipage} \hfill
\begin{minipage}[b]{0.325\linewidth}
\includegraphics[width=\textwidth]{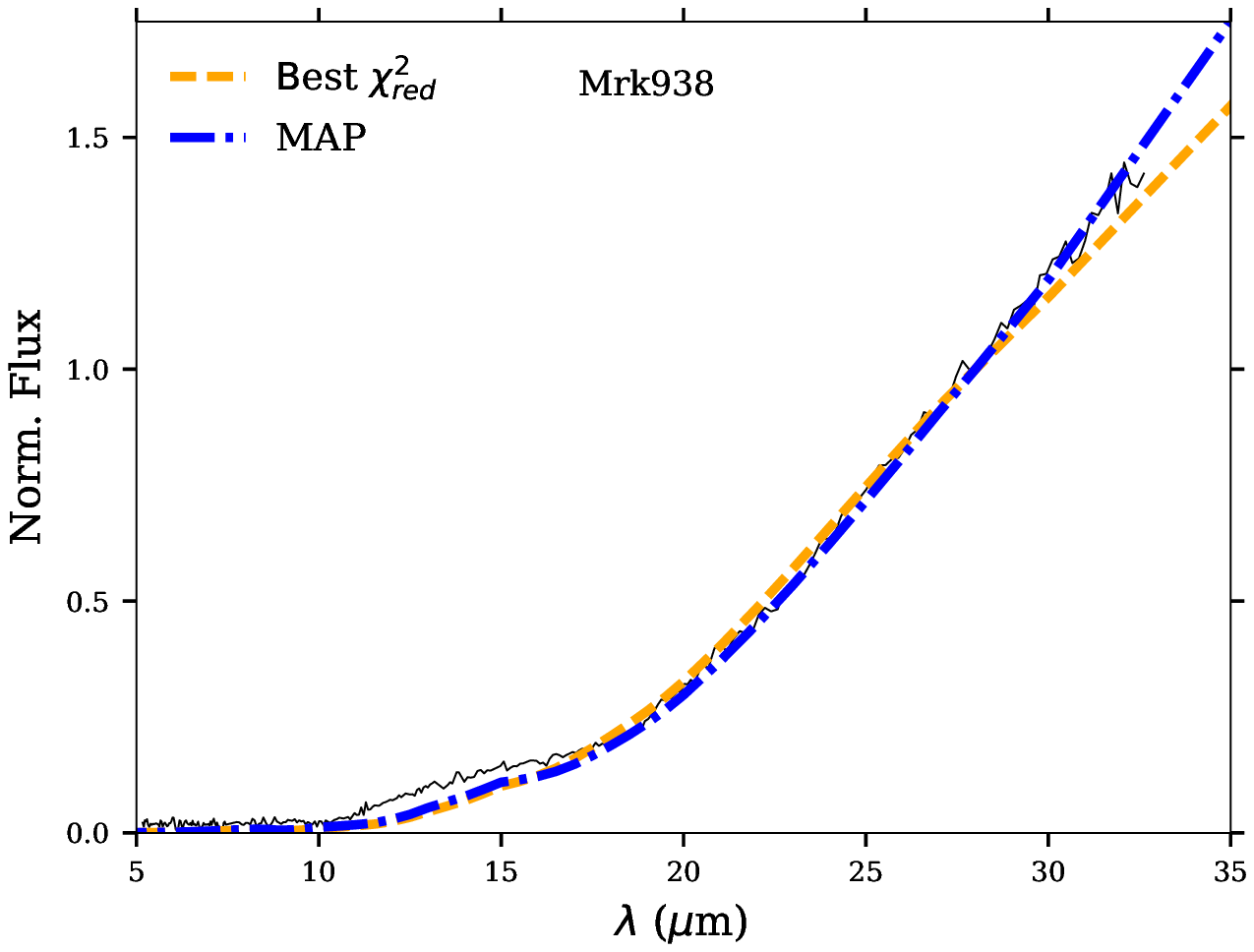}
\end{minipage} \hfill
\begin{minipage}[b]{0.325\linewidth}
\includegraphics[width=\textwidth]{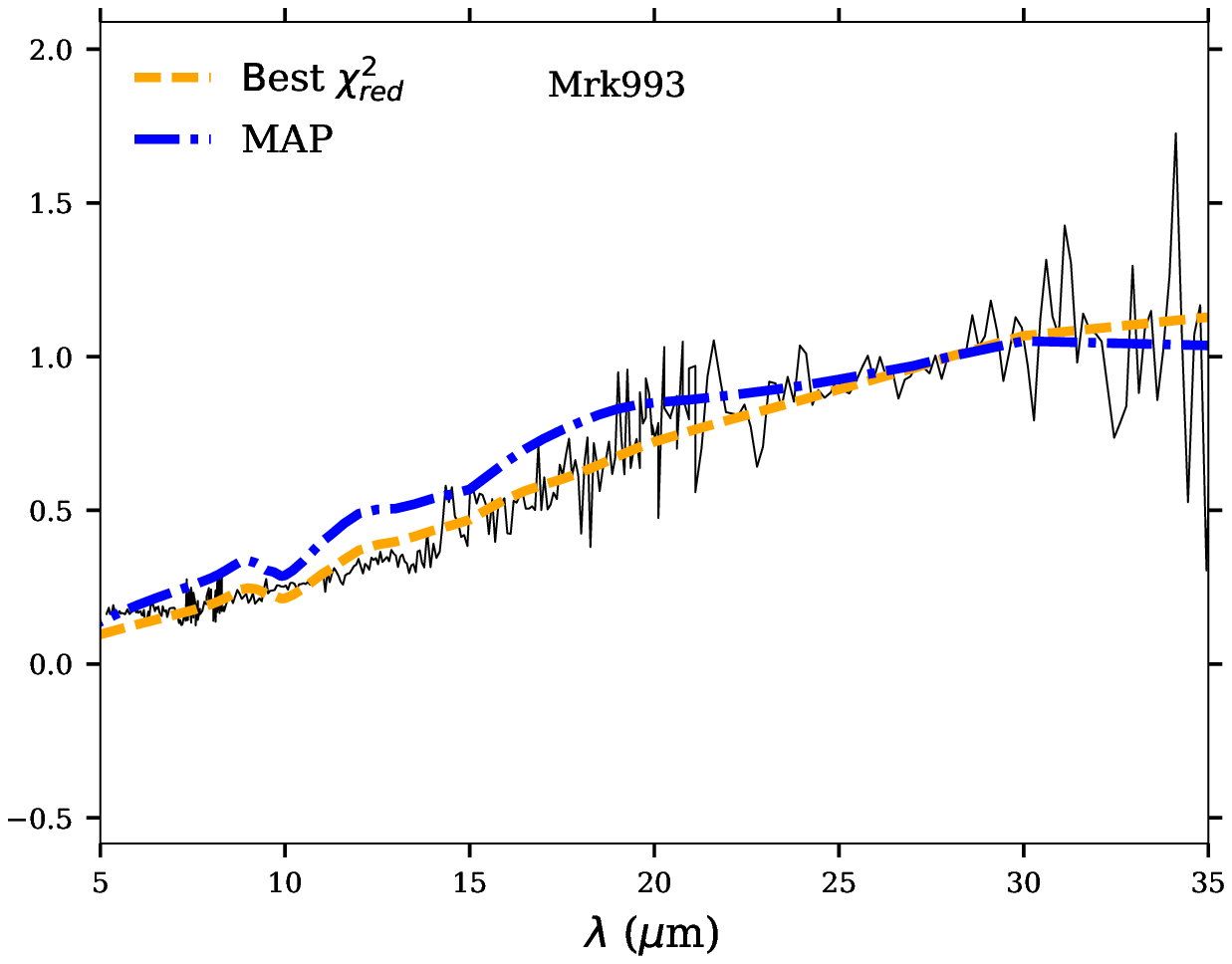}
\end{minipage} \hfill
\begin{minipage}[b]{0.325\linewidth}
\includegraphics[width=\textwidth]{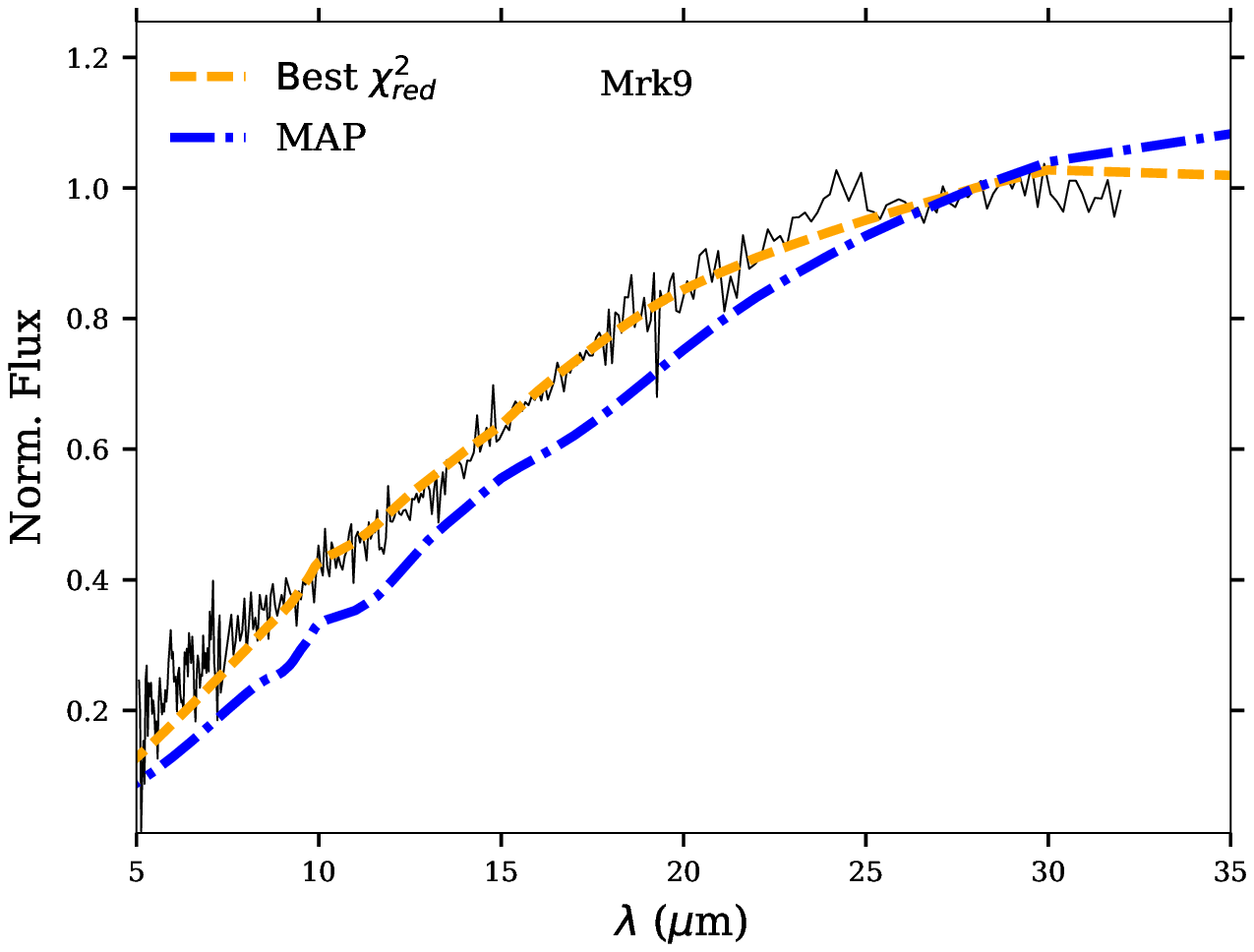}
\end{minipage} \hfill
\begin{minipage}[b]{0.325\linewidth}
\includegraphics[width=\textwidth]{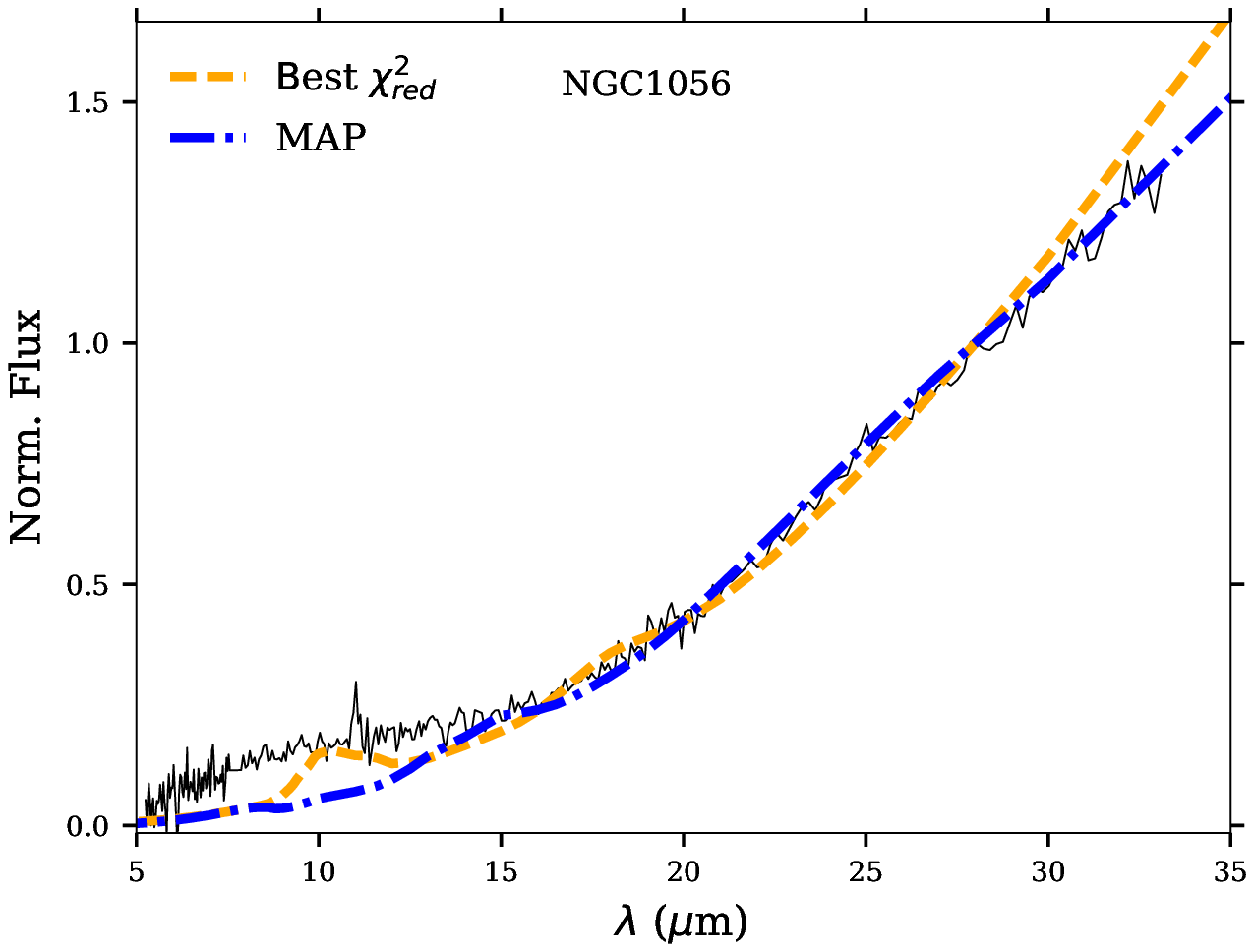}
\end{minipage} \hfill
\begin{minipage}[b]{0.325\linewidth}
\includegraphics[width=\textwidth]{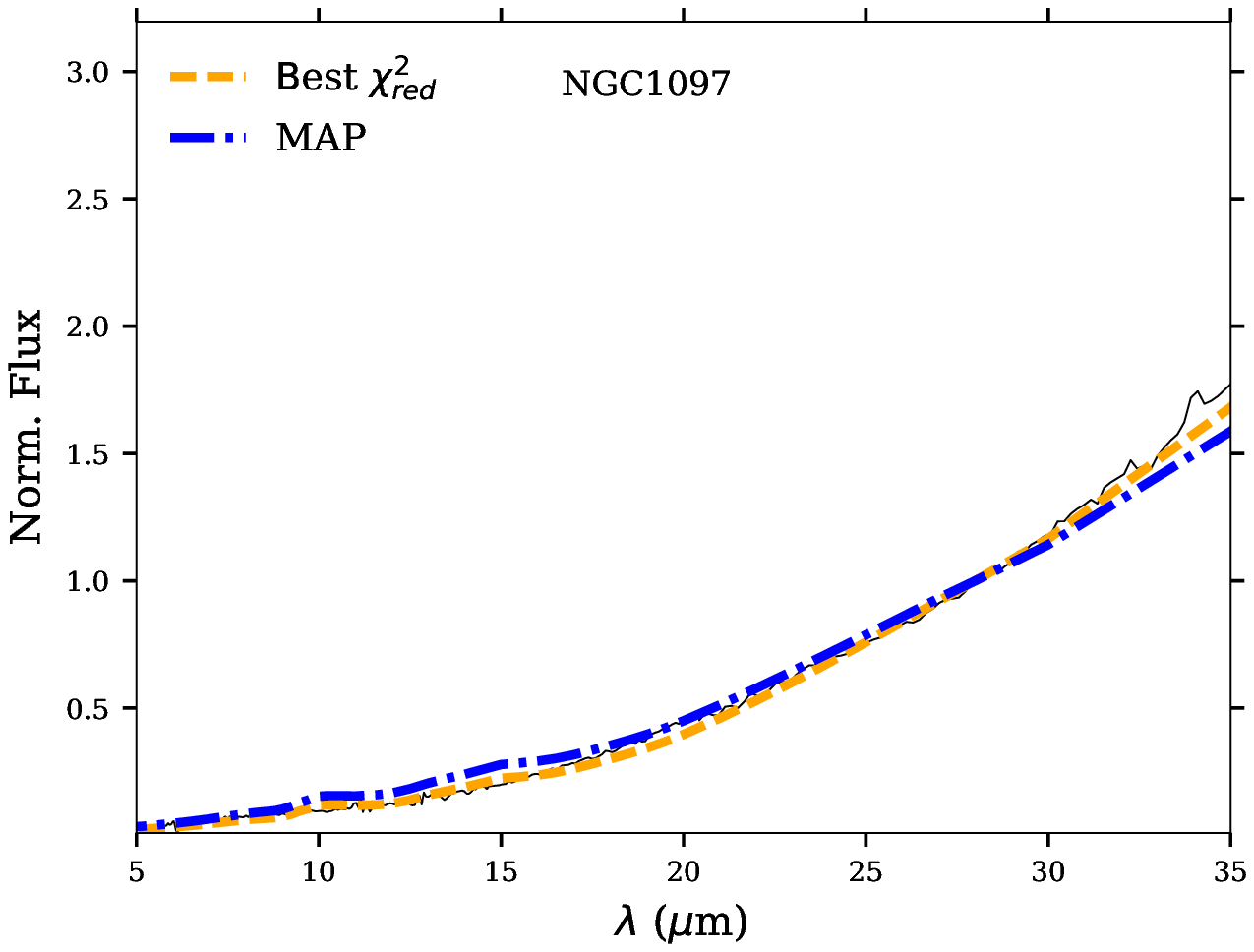}
\end{minipage} \hfill
\begin{minipage}[b]{0.325\linewidth}
\includegraphics[width=\textwidth]{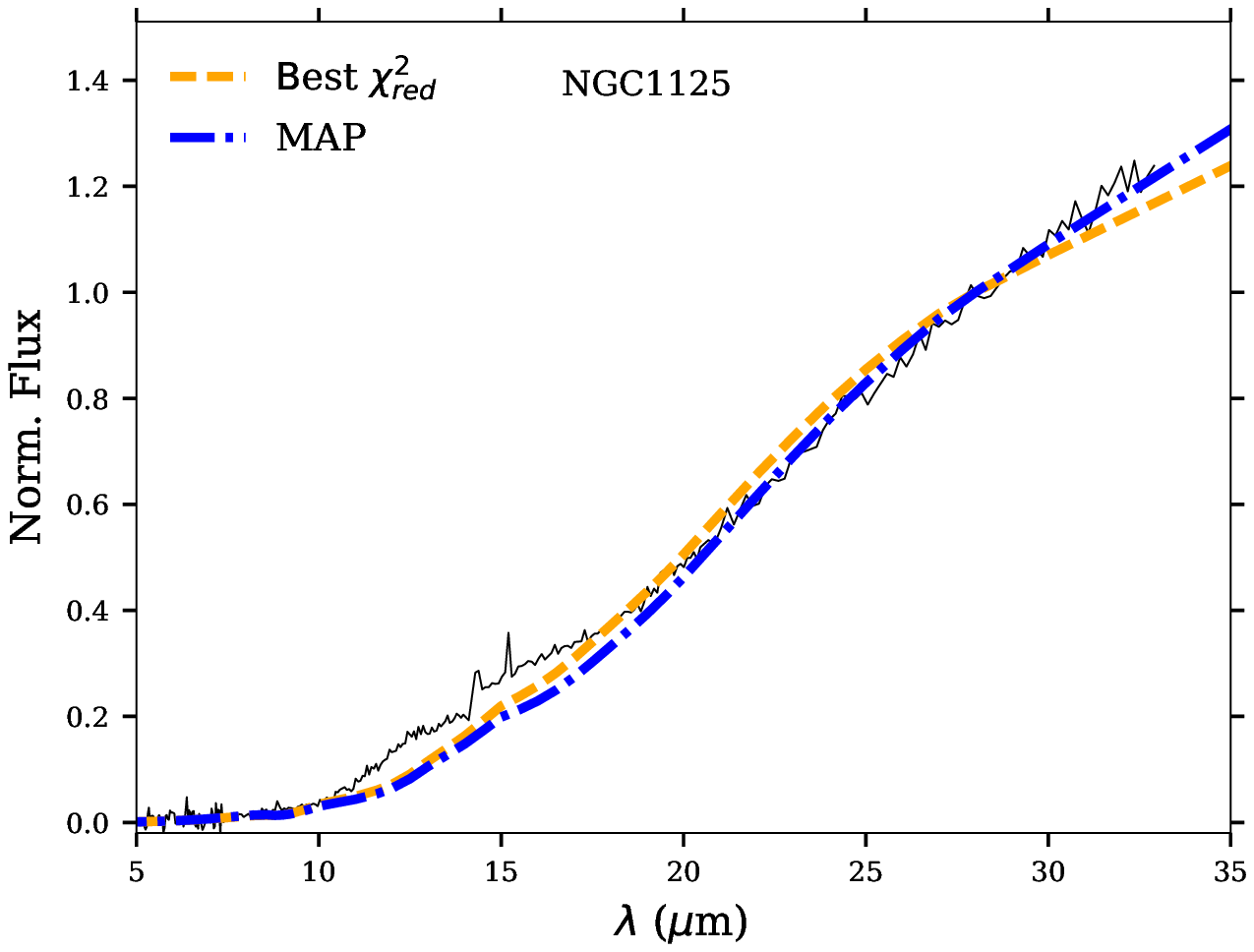}
\end{minipage} \hfill
\begin{minipage}[b]{0.325\linewidth}
\includegraphics[width=\textwidth]{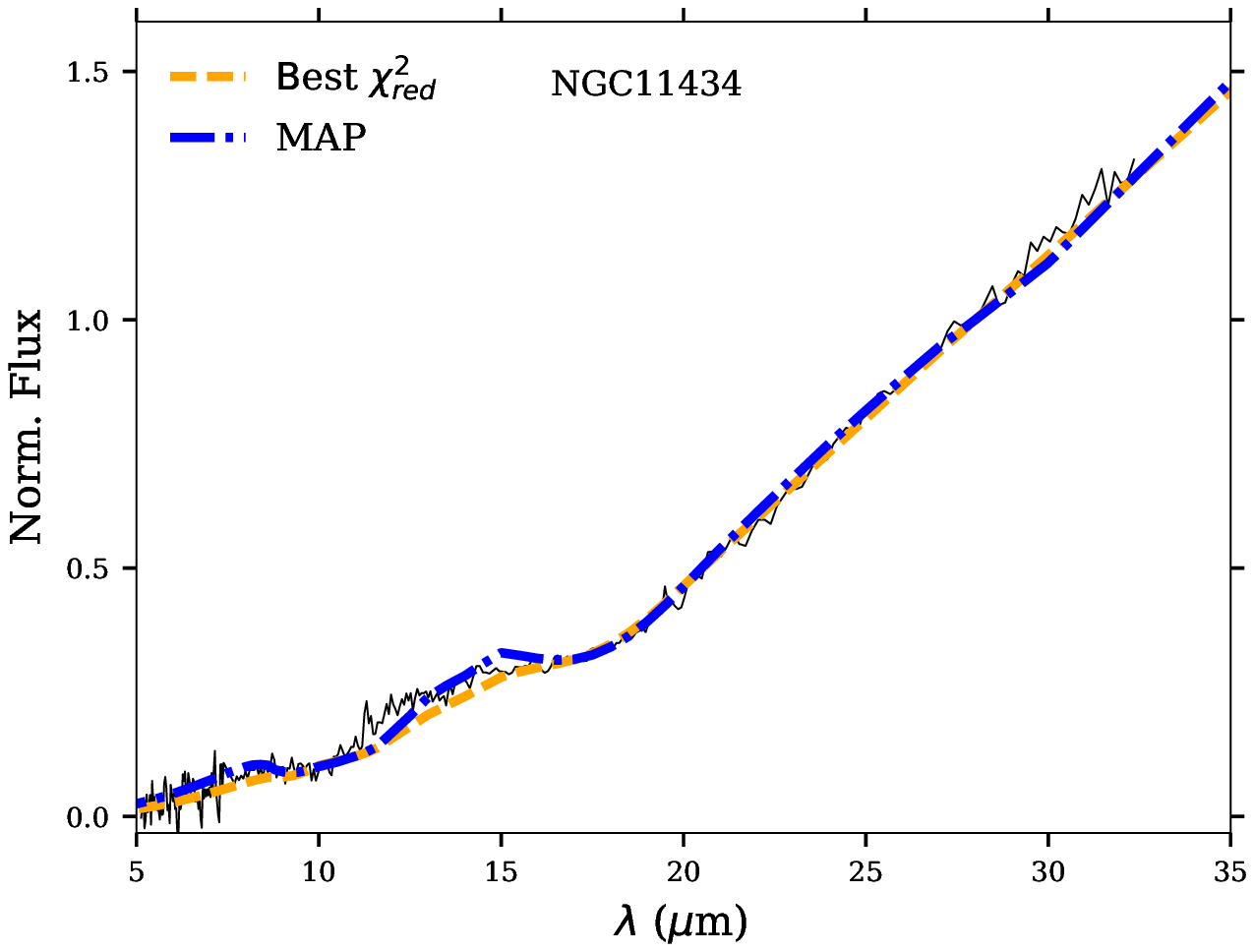}
\end{minipage} \hfill
\begin{minipage}[b]{0.325\linewidth}
\includegraphics[width=\textwidth]{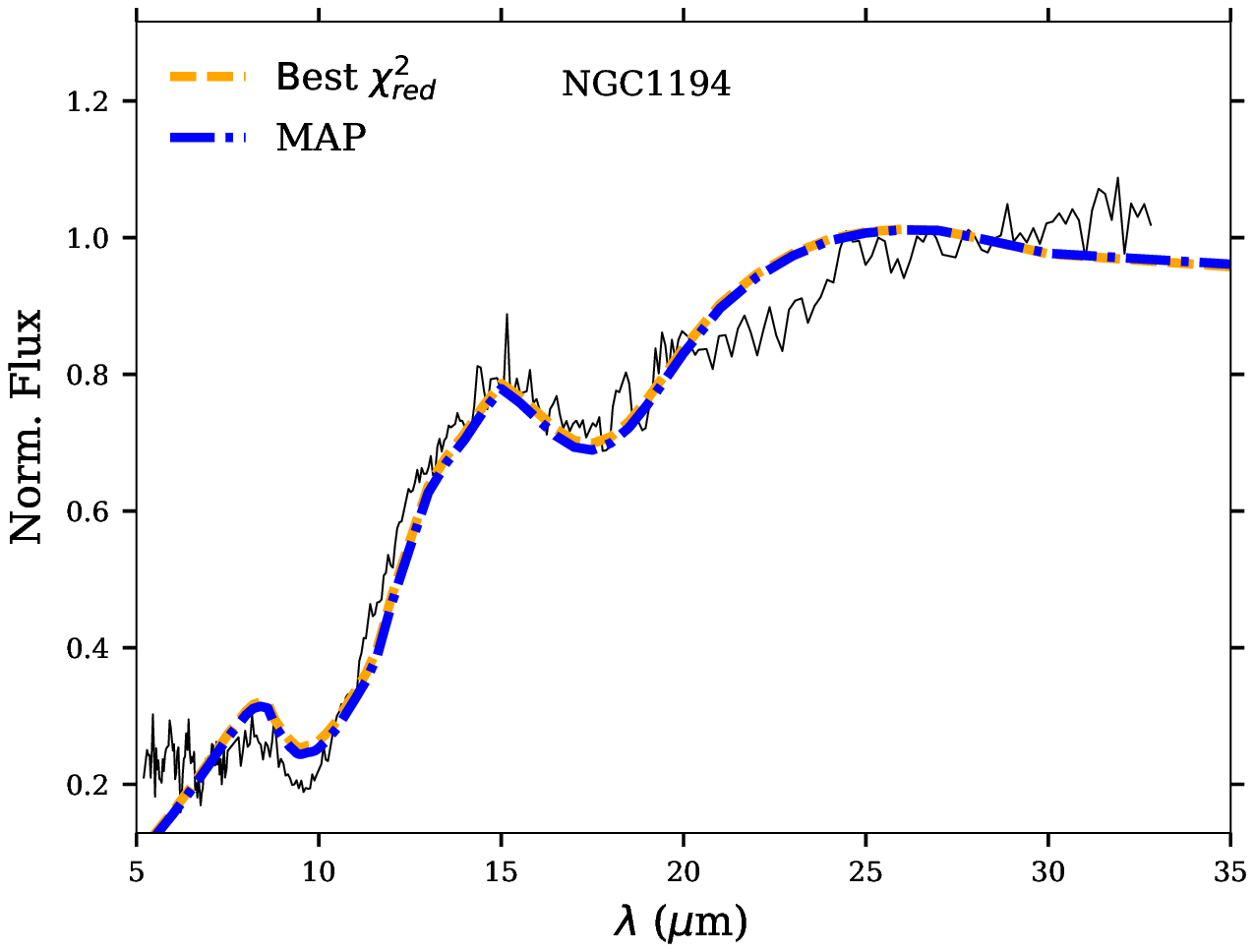}
\end{minipage} \hfill
\begin{minipage}[b]{0.325\linewidth}
\includegraphics[width=\textwidth]{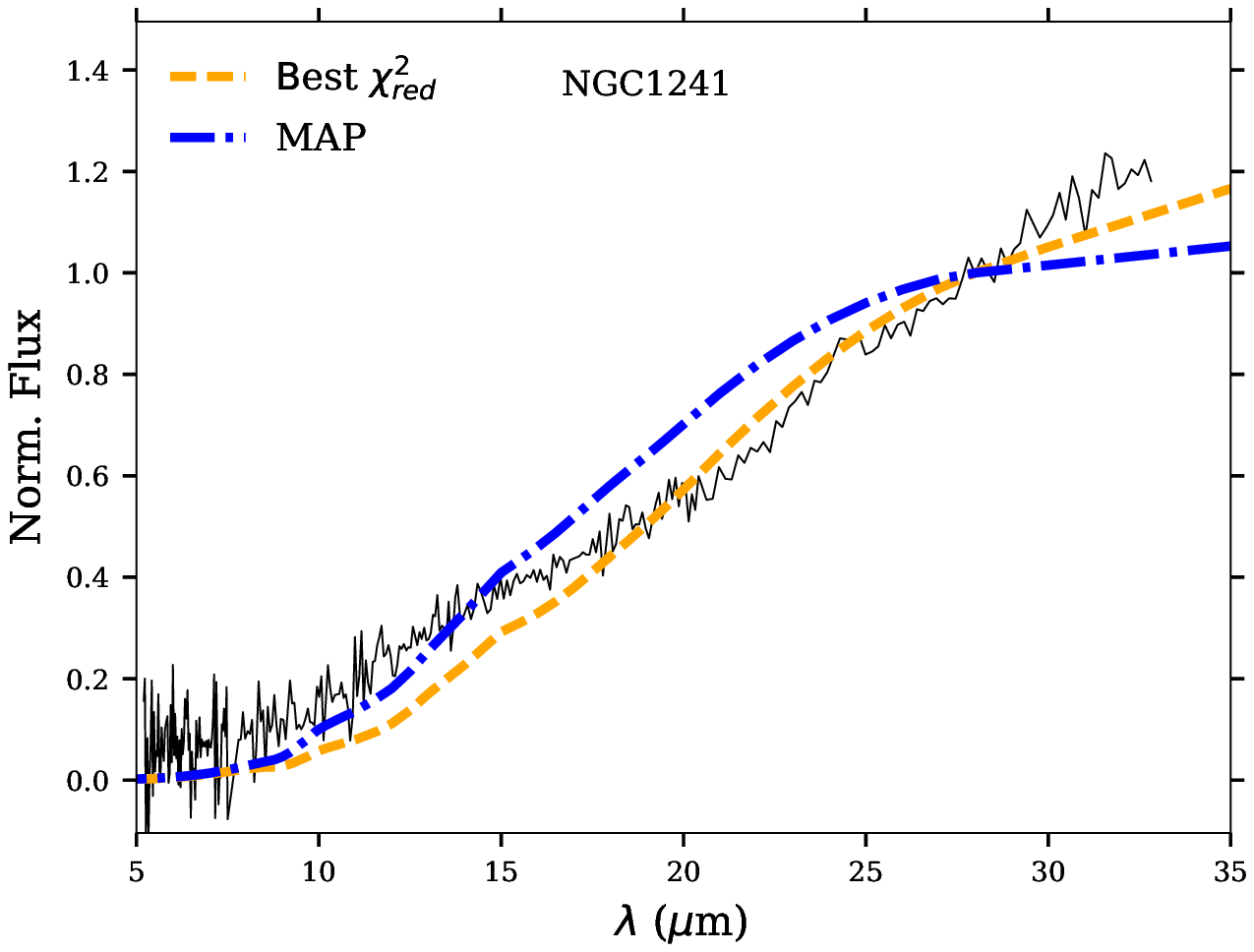}
\end{minipage} \hfill
\begin{minipage}[b]{0.325\linewidth}
\includegraphics[width=\textwidth]{NGC1275_bestMAP_JY.eps}
\end{minipage} \hfill
\begin{minipage}[b]{0.325\linewidth}
\includegraphics[width=\textwidth]{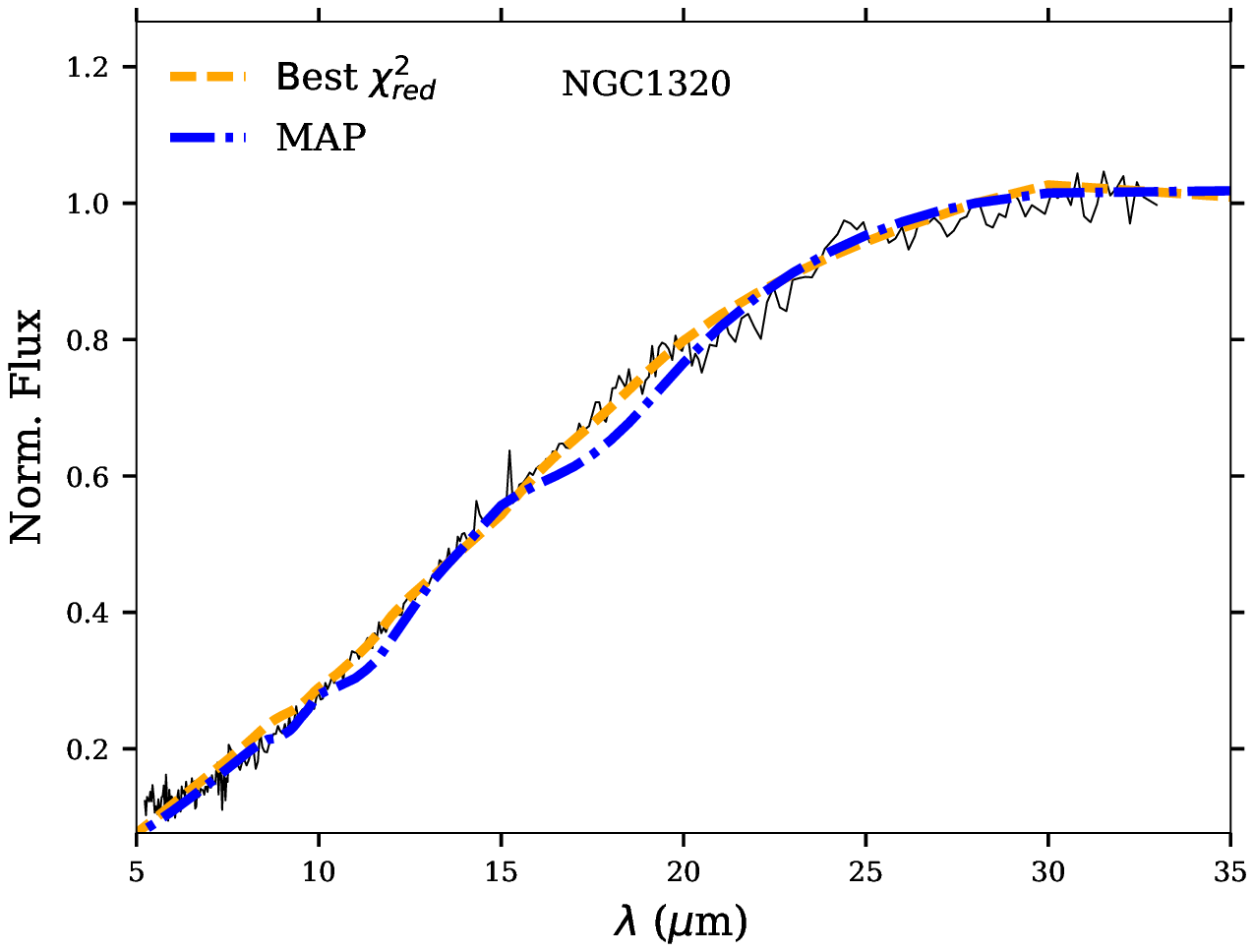}
\end{minipage} \hfill
\begin{minipage}[b]{0.325\linewidth}
\includegraphics[width=\textwidth]{NGC1365_bestMAP_JY.eps}
\end{minipage} \hfill
\begin{minipage}[b]{0.325\linewidth}
\includegraphics[width=\textwidth]{NGC1386_bestMAP_JY.eps}
\end{minipage} \hfill
\caption{continued from previous page.}
\setcounter{figure}{0}
\end{figure}

\begin{figure}

\begin{minipage}[b]{0.325\linewidth}
\includegraphics[width=\textwidth]{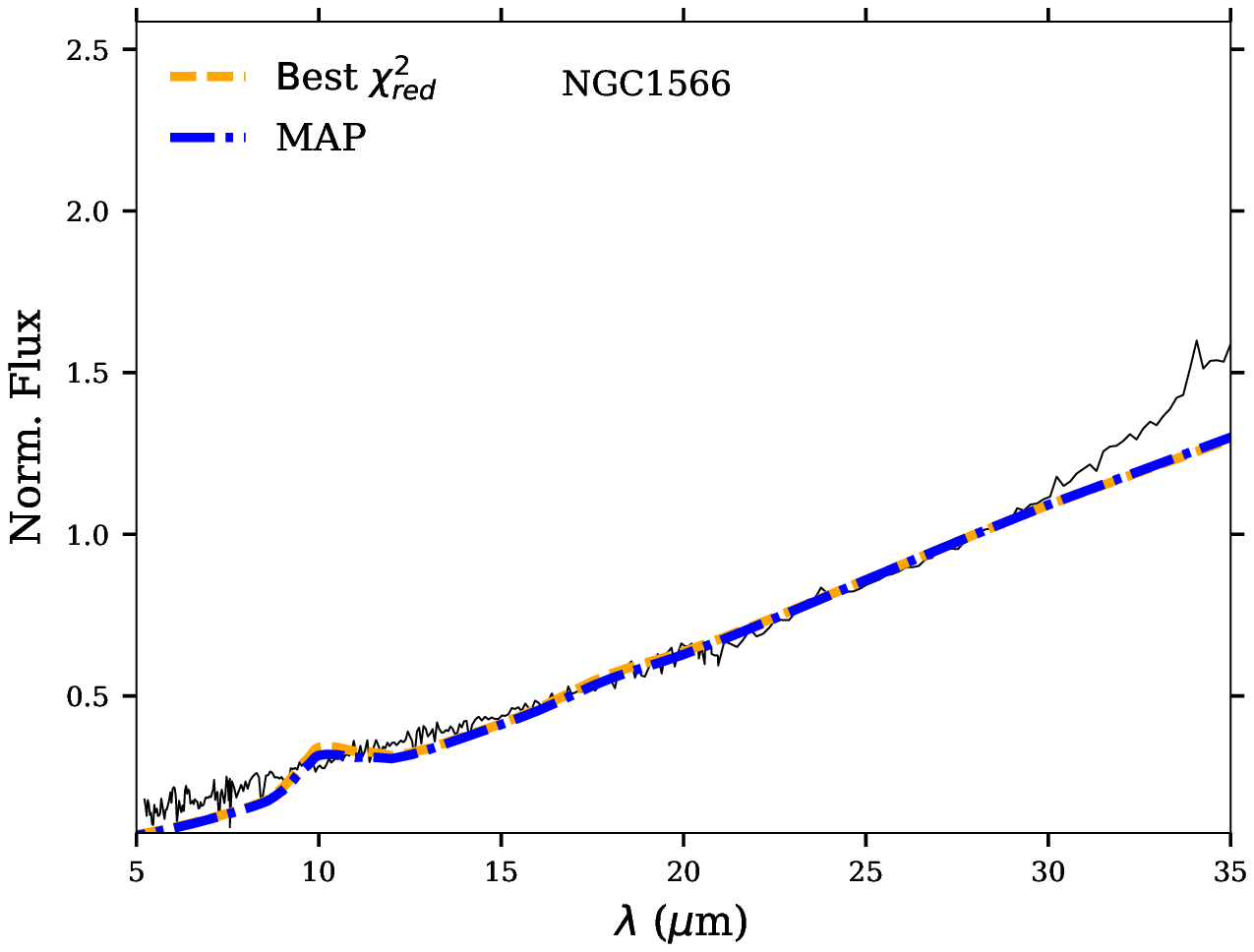}
\end{minipage} \hfill
\begin{minipage}[b]{0.325\linewidth}
\includegraphics[width=\textwidth]{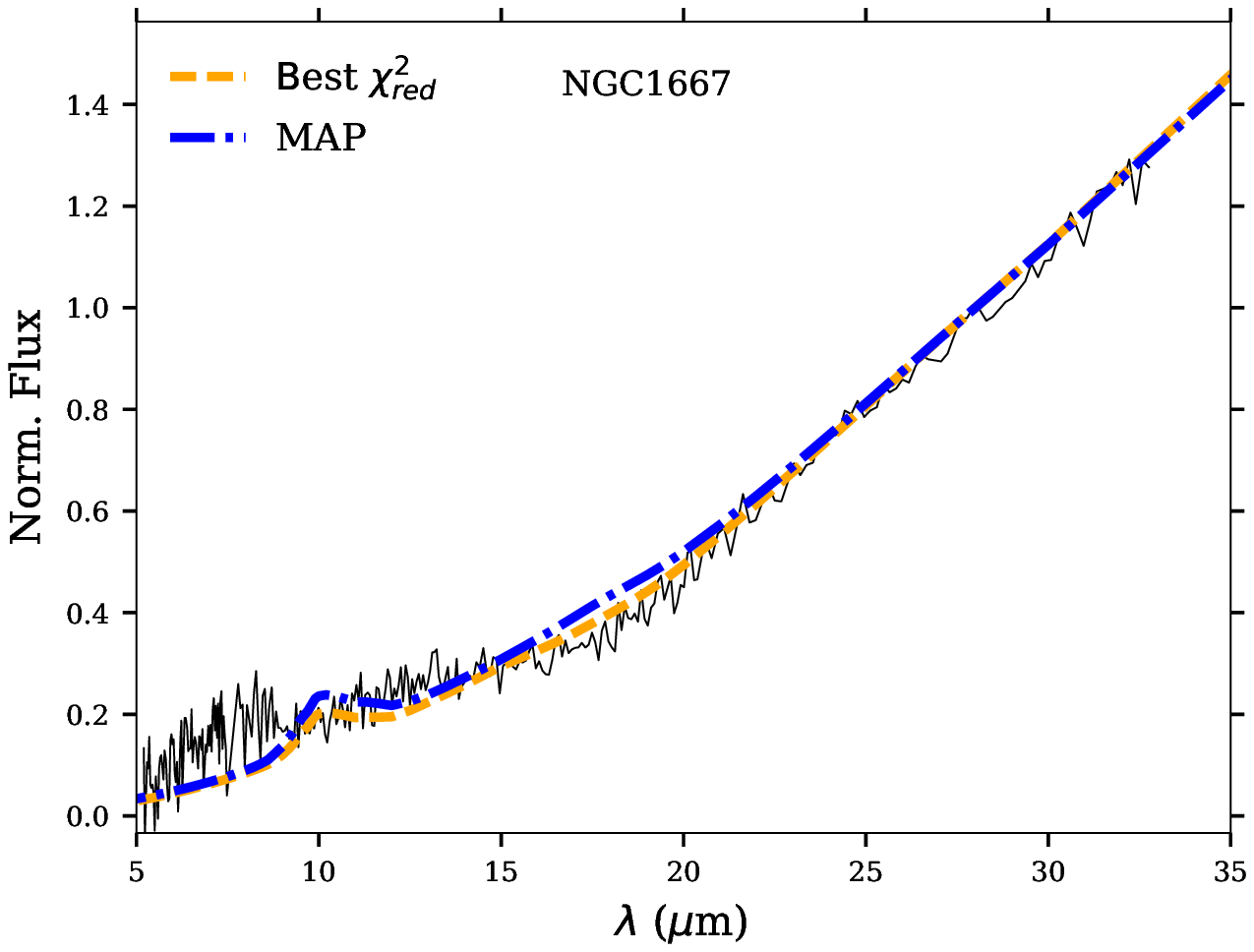}
\end{minipage} \hfill
\begin{minipage}[b]{0.325\linewidth}
\includegraphics[width=\textwidth]{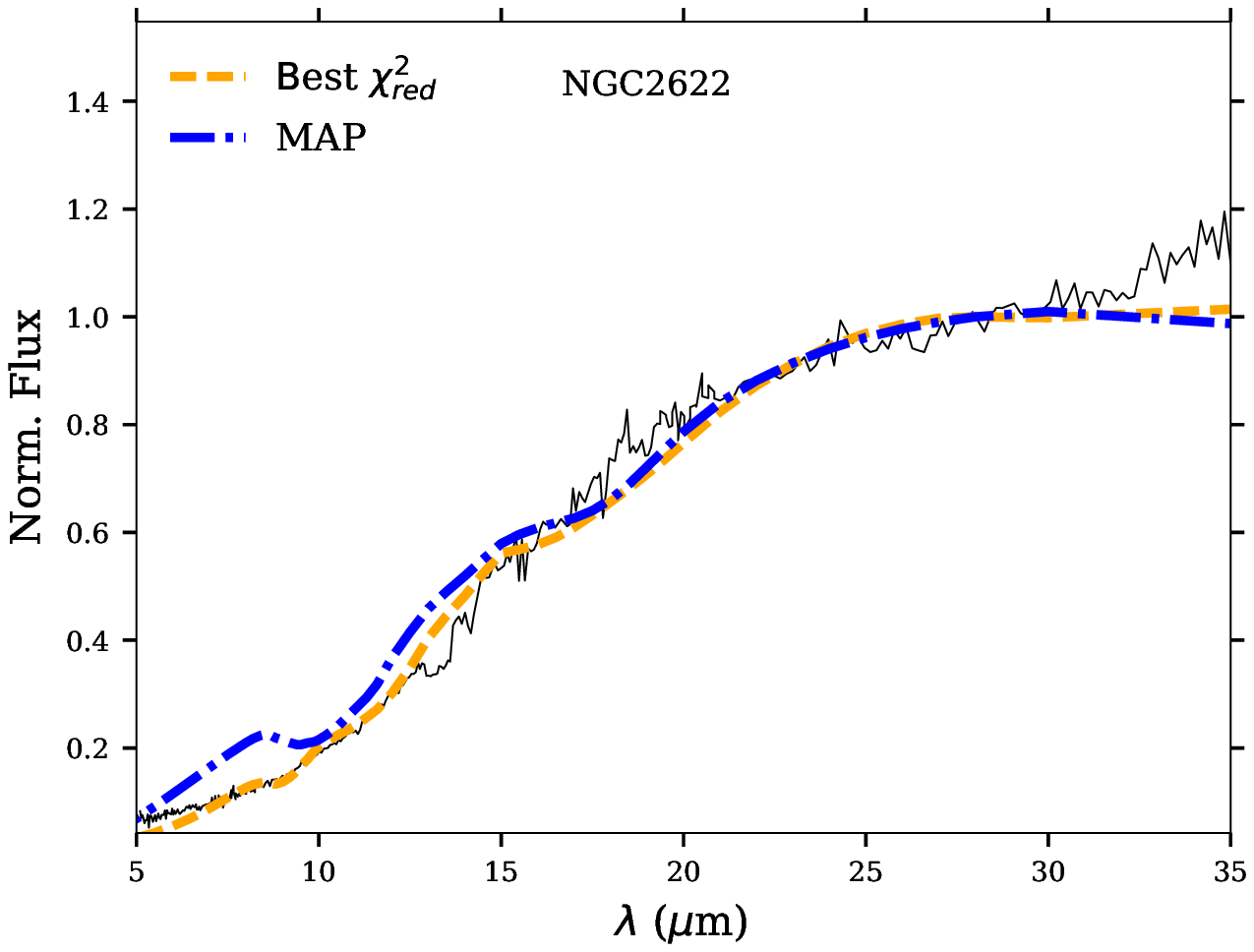}
\end{minipage} \hfill
\begin{minipage}[b]{0.325\linewidth}
\includegraphics[width=\textwidth]{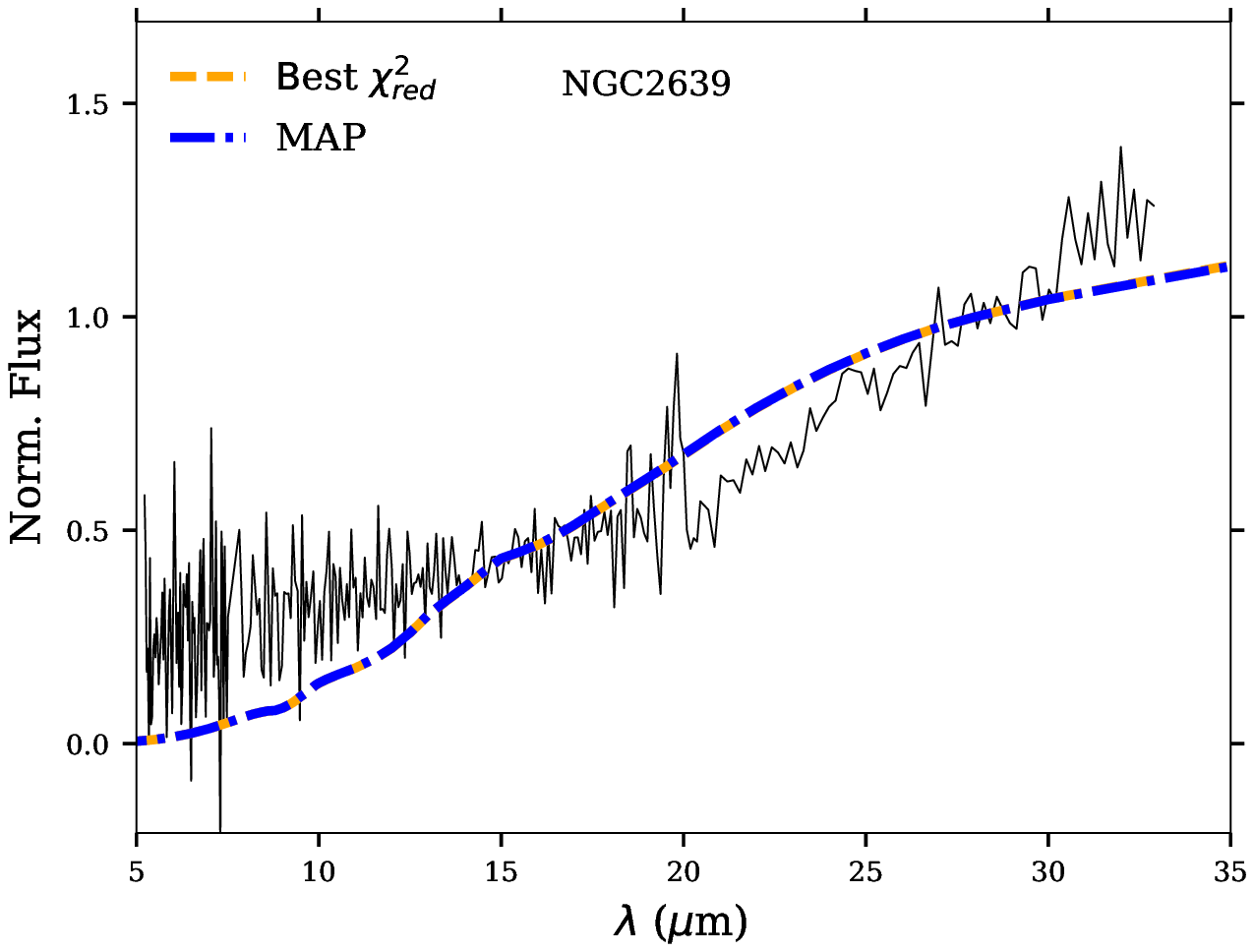}
\end{minipage} \hfill
\begin{minipage}[b]{0.325\linewidth}
\includegraphics[width=\textwidth]{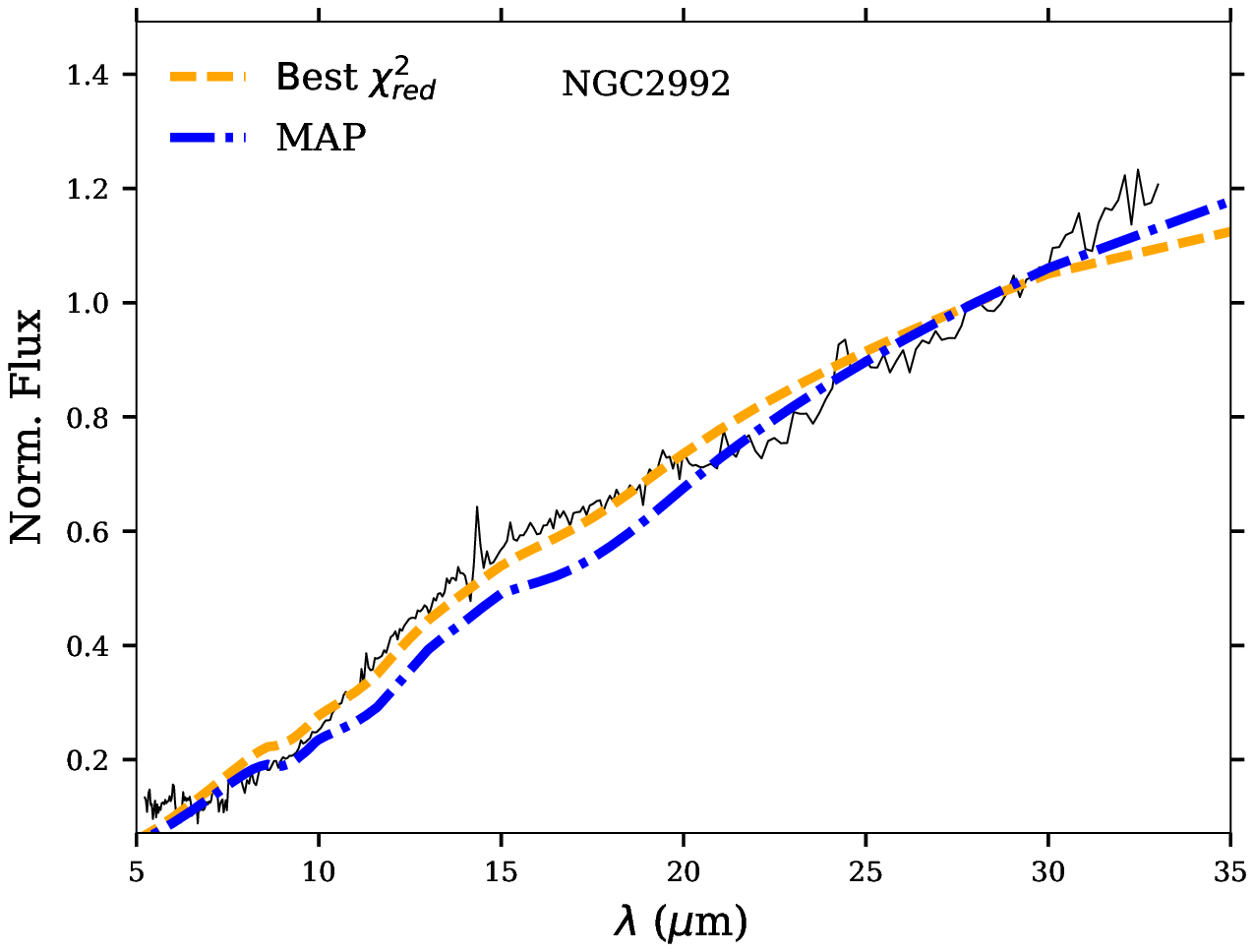}
\end{minipage} \hfill
\begin{minipage}[b]{0.325\linewidth}
\includegraphics[width=\textwidth]{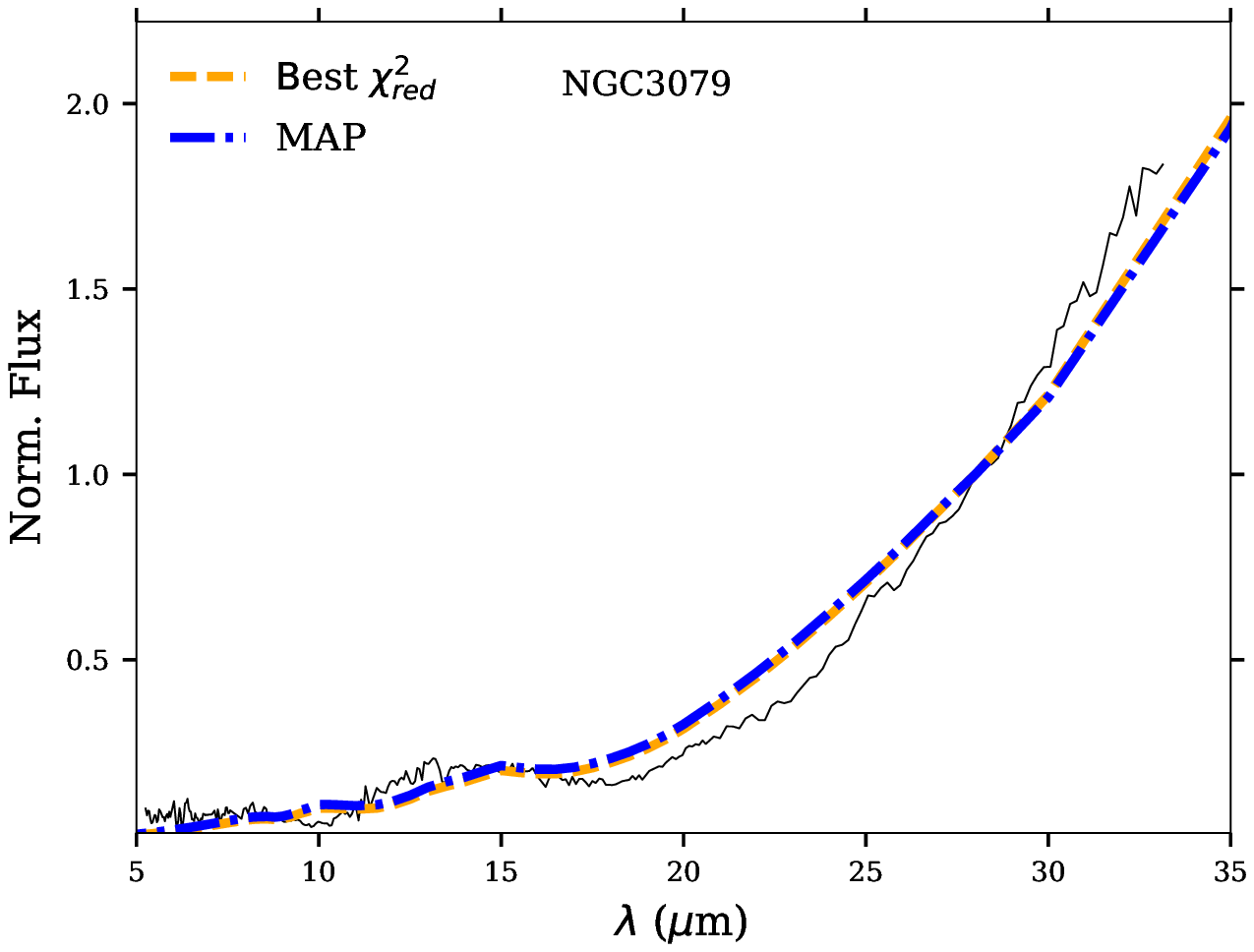}
\end{minipage} \hfill
\begin{minipage}[b]{0.325\linewidth}
\includegraphics[width=\textwidth]{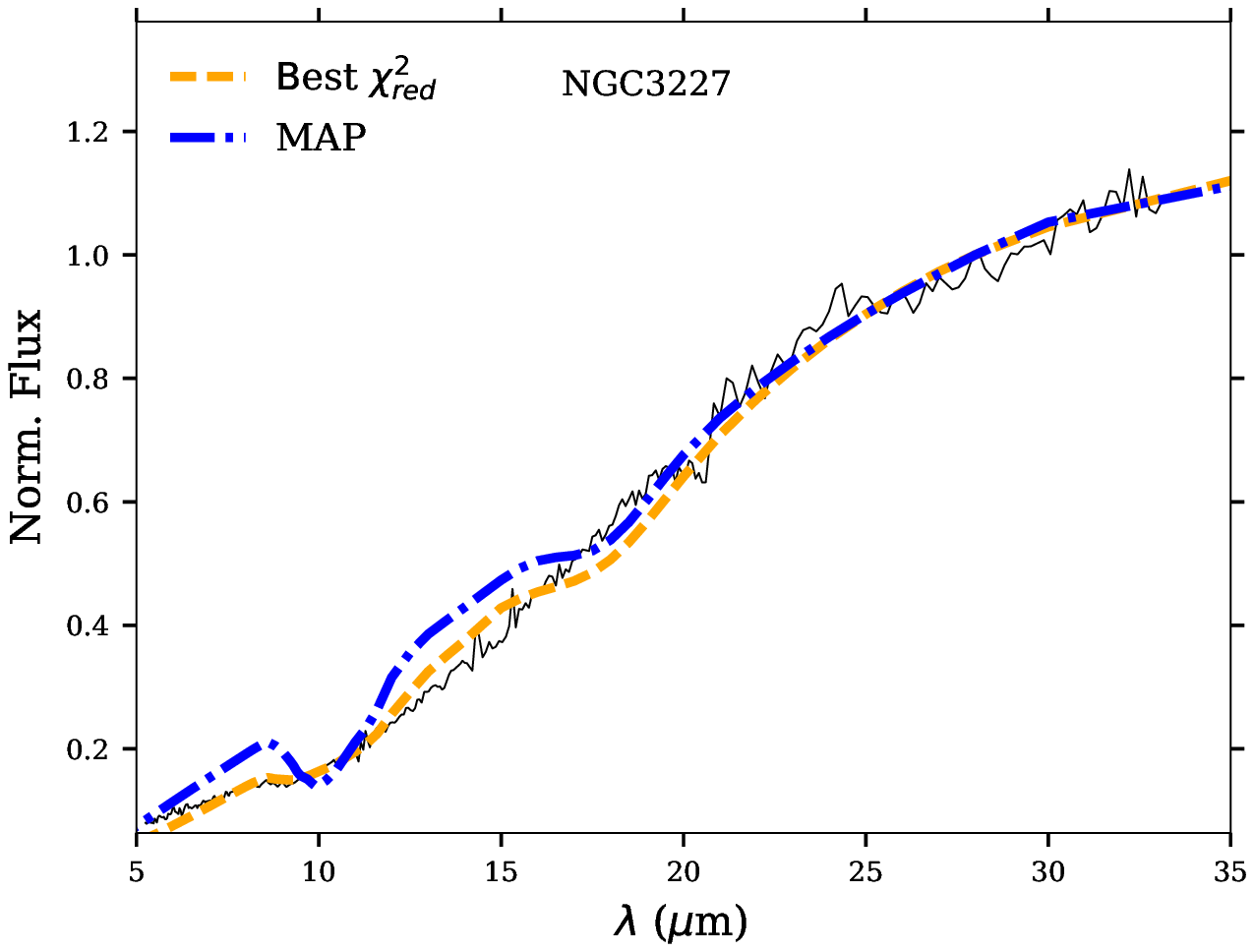}
\end{minipage} \hfill
\begin{minipage}[b]{0.325\linewidth}
\includegraphics[width=\textwidth]{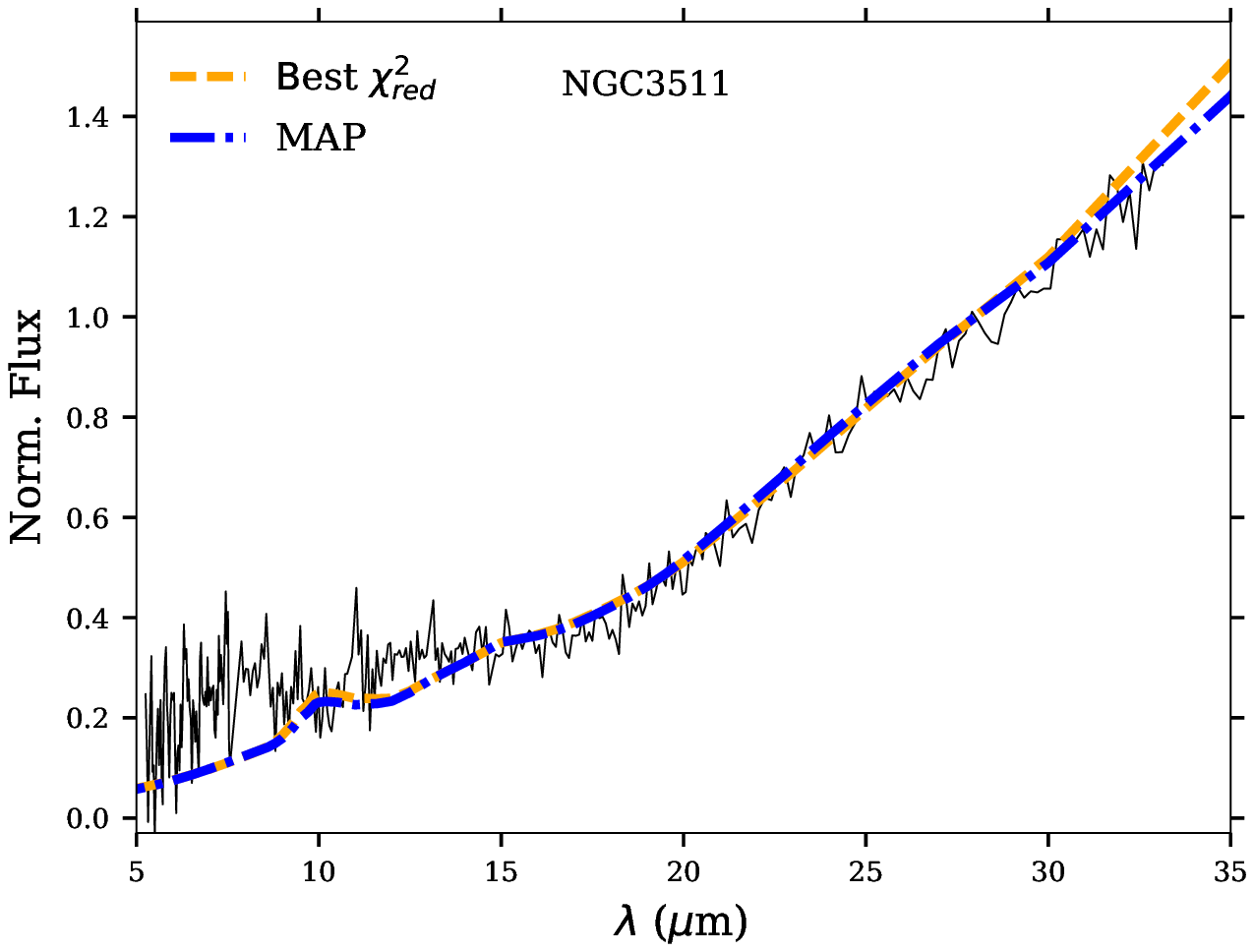}
\end{minipage} \hfill
\begin{minipage}[b]{0.325\linewidth}
\includegraphics[width=\textwidth]{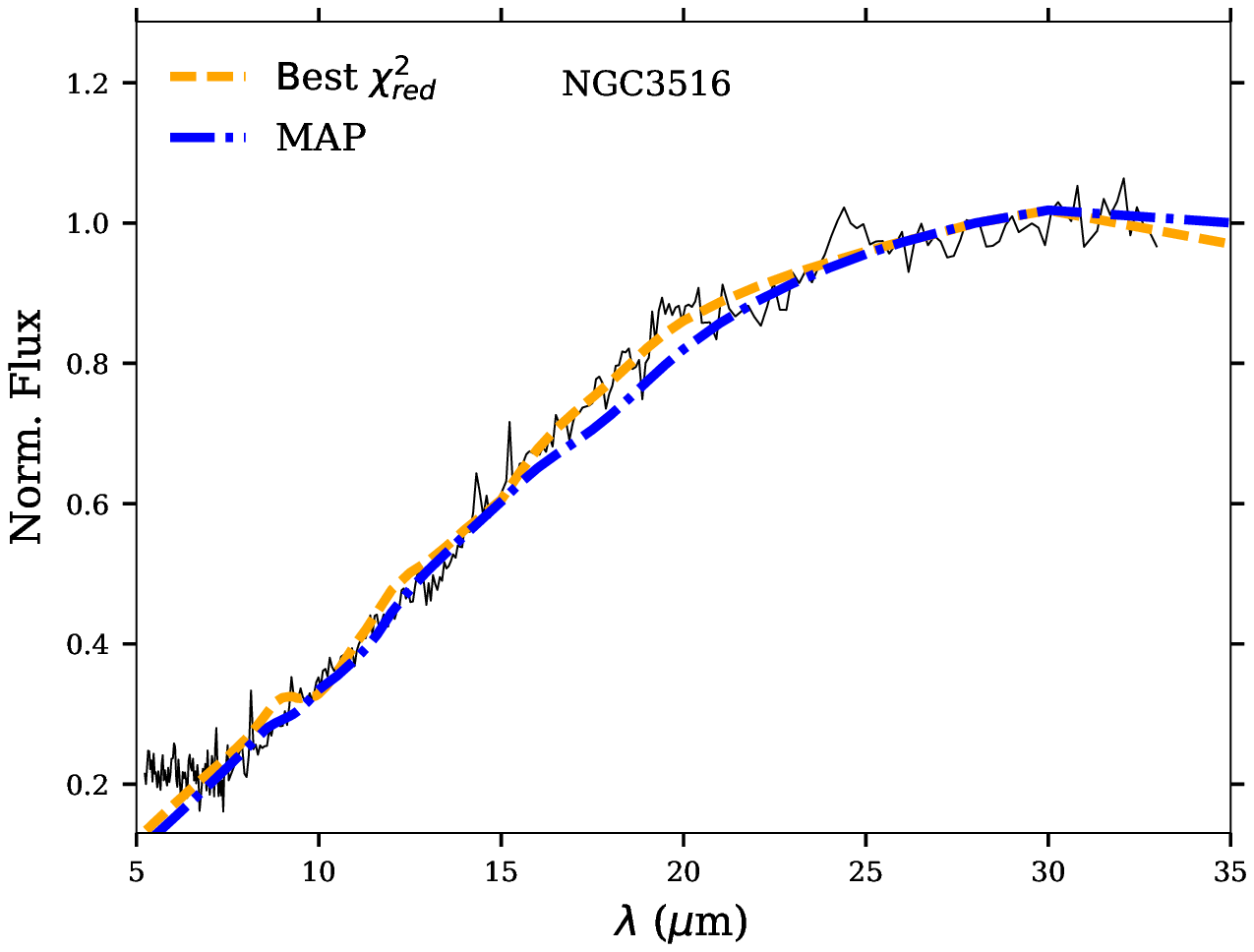}
\end{minipage} \hfill
\begin{minipage}[b]{0.325\linewidth}
\includegraphics[width=\textwidth]{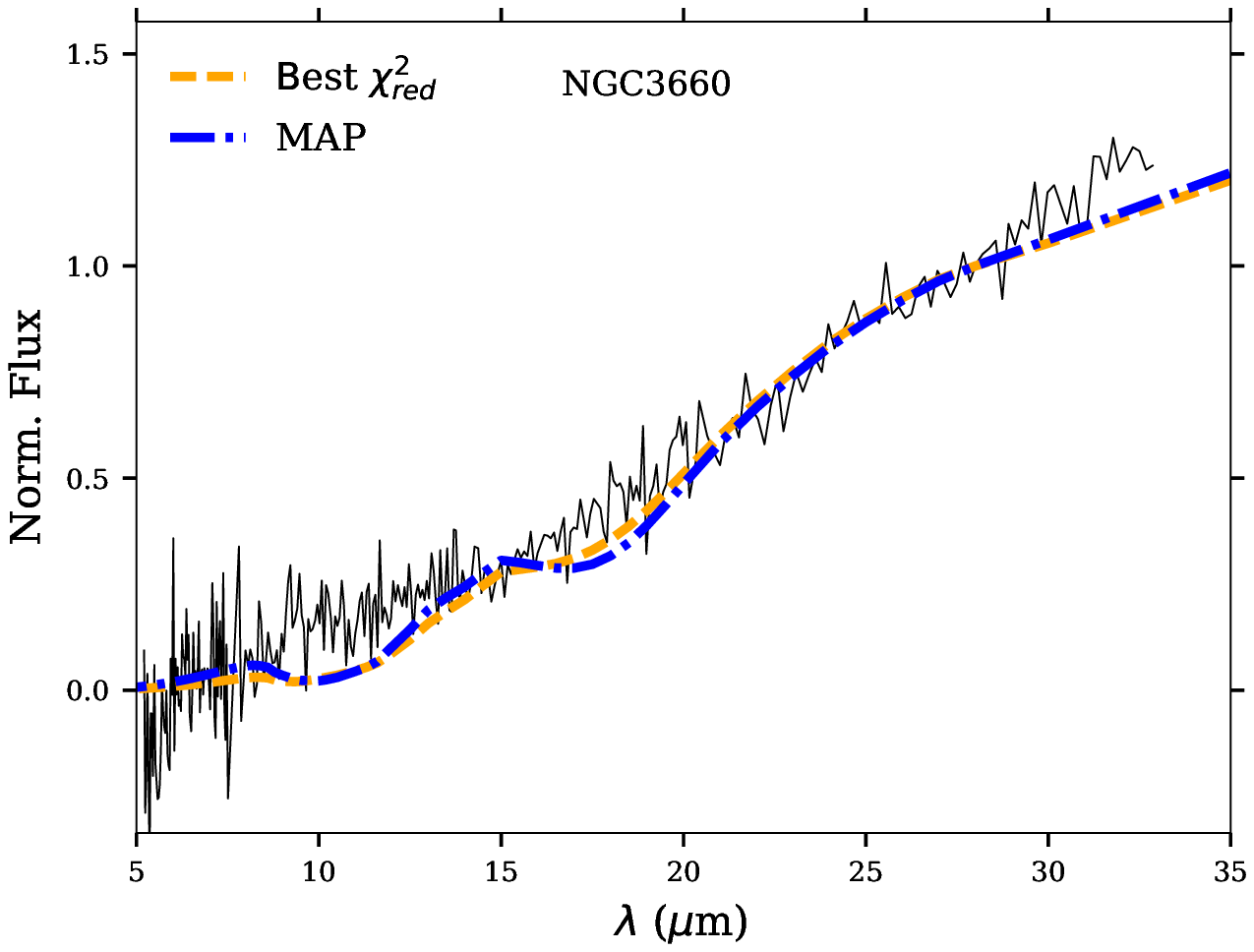}
\end{minipage} \hfill
\begin{minipage}[b]{0.325\linewidth}
\includegraphics[width=\textwidth]{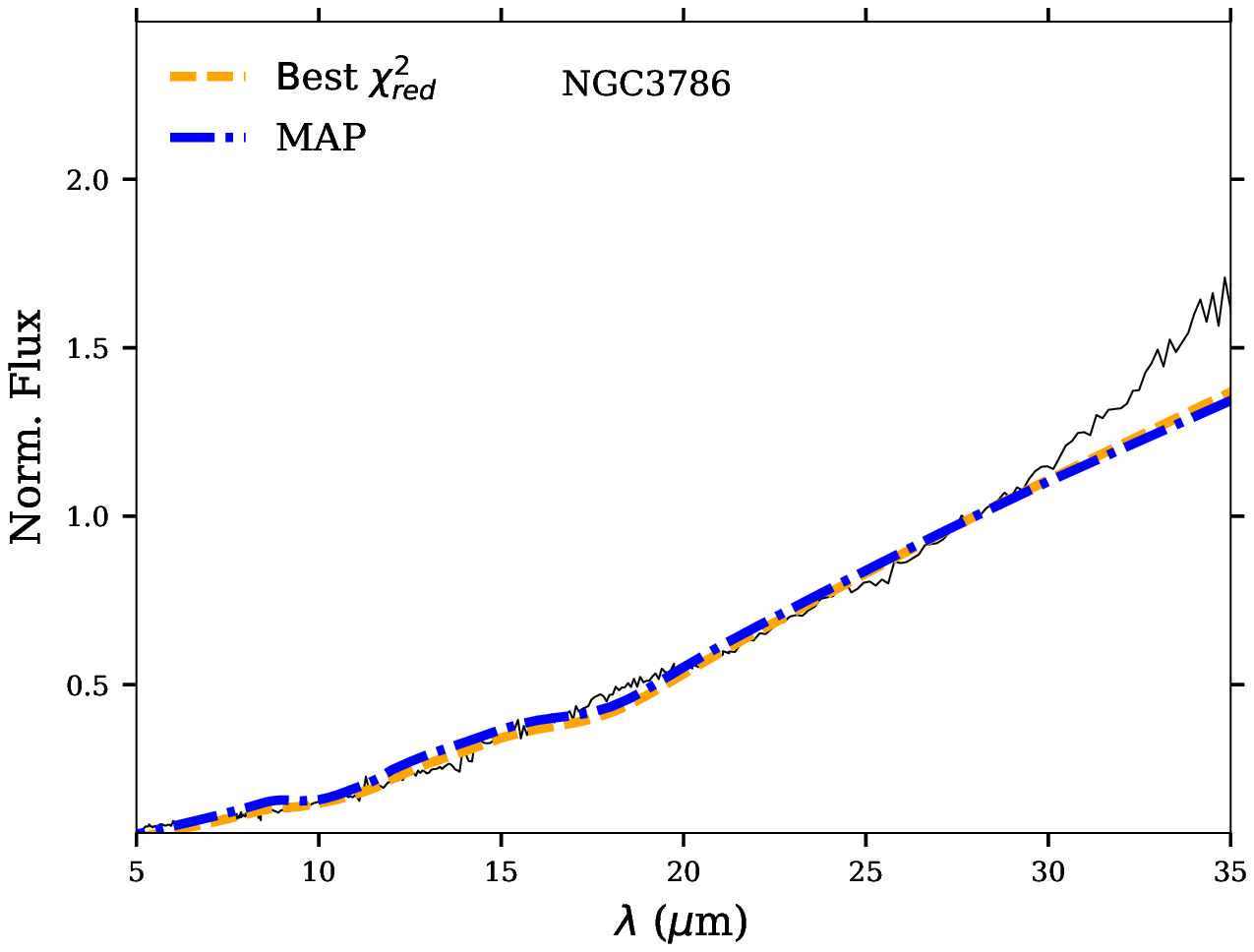}
\end{minipage} \hfill
\begin{minipage}[b]{0.325\linewidth}
\includegraphics[width=\textwidth]{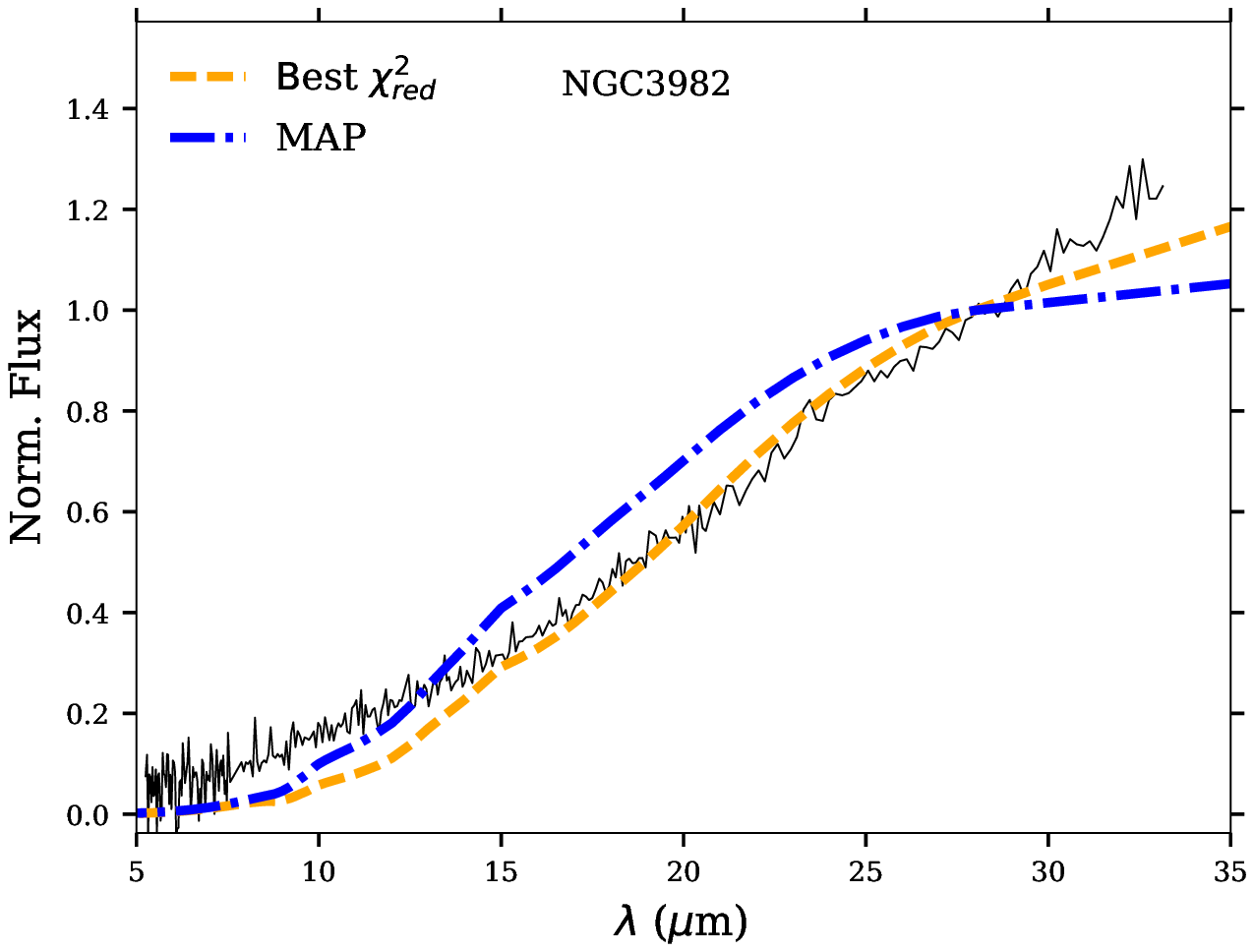}
\end{minipage} \hfill
\begin{minipage}[b]{0.325\linewidth}
\includegraphics[width=\textwidth]{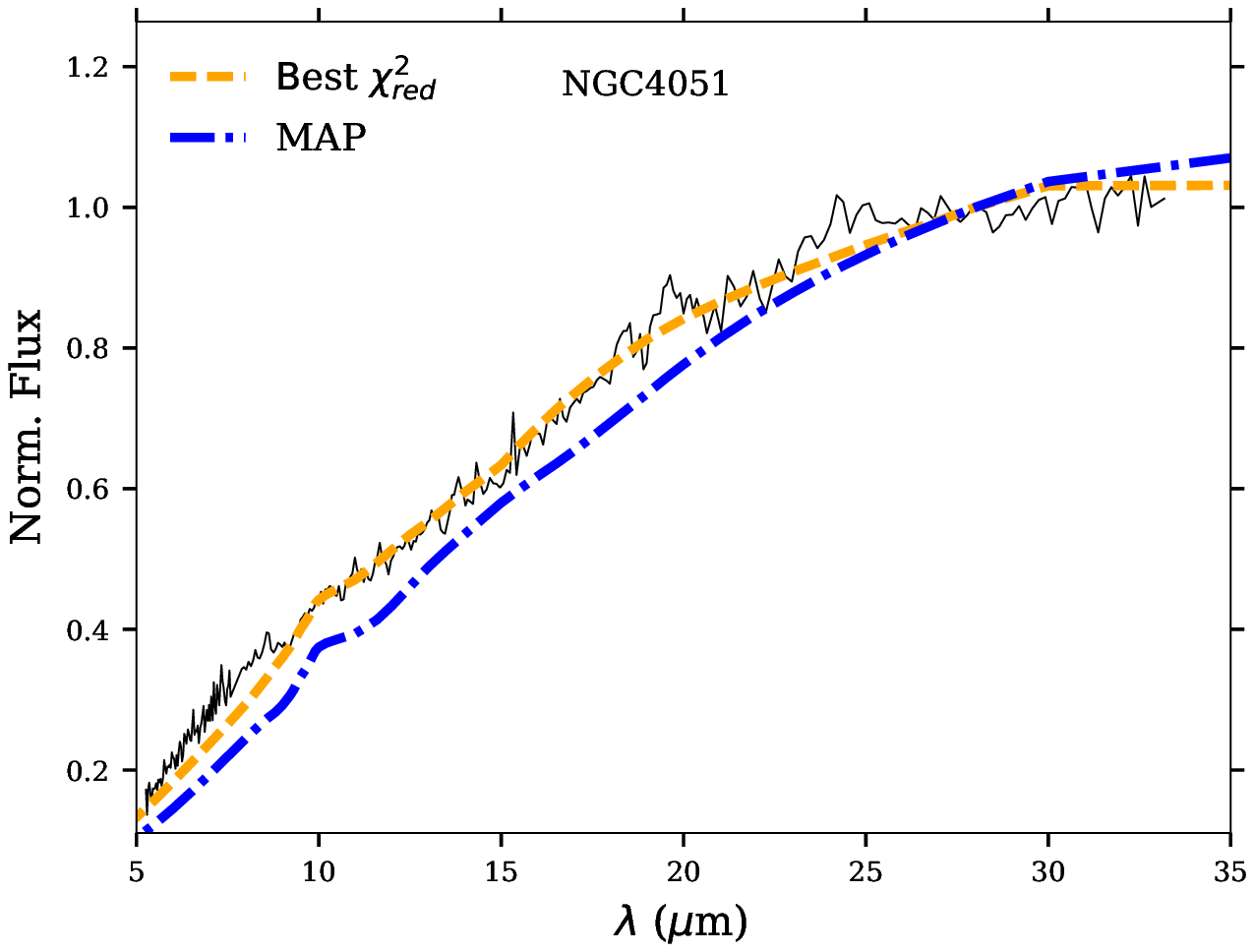}
\end{minipage} \hfill
\begin{minipage}[b]{0.325\linewidth}
\includegraphics[width=\textwidth]{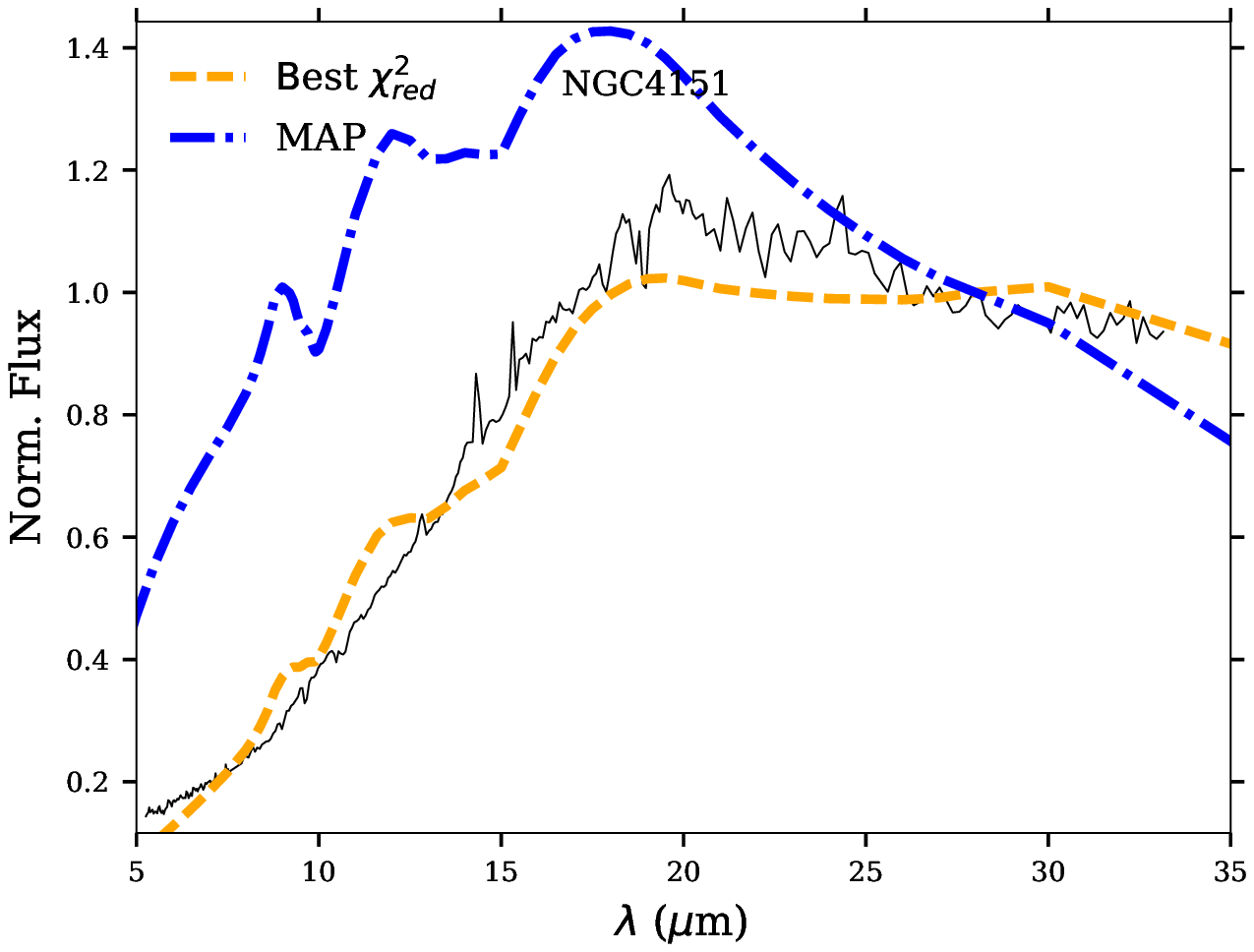}
\end{minipage} \hfill
\begin{minipage}[b]{0.325\linewidth}
\includegraphics[width=\textwidth]{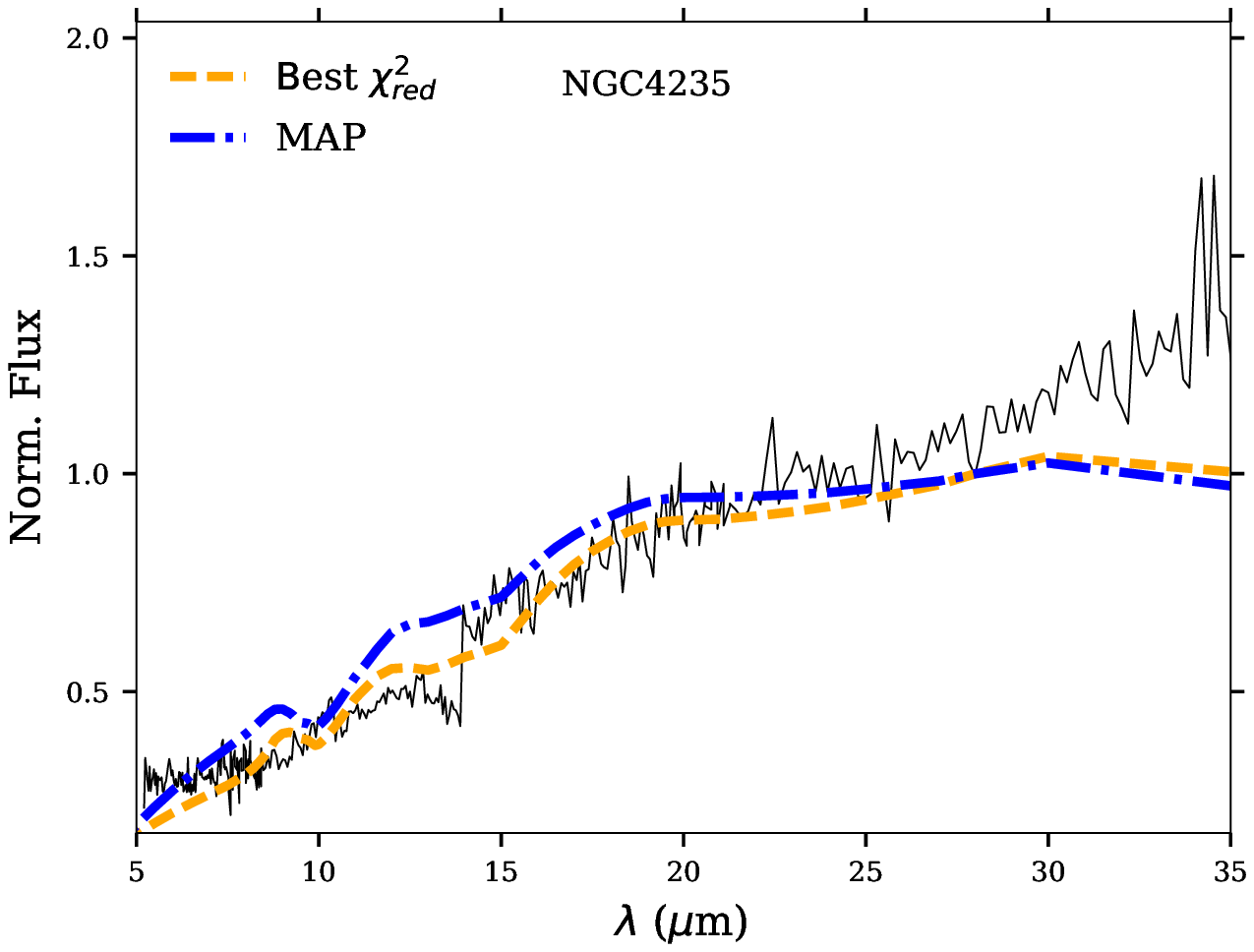}
\end{minipage} \hfill
\caption{continued from previous page.}
\setcounter{figure}{0}
\end{figure}

\begin{figure}

\begin{minipage}[b]{0.325\linewidth}
\includegraphics[width=\textwidth]{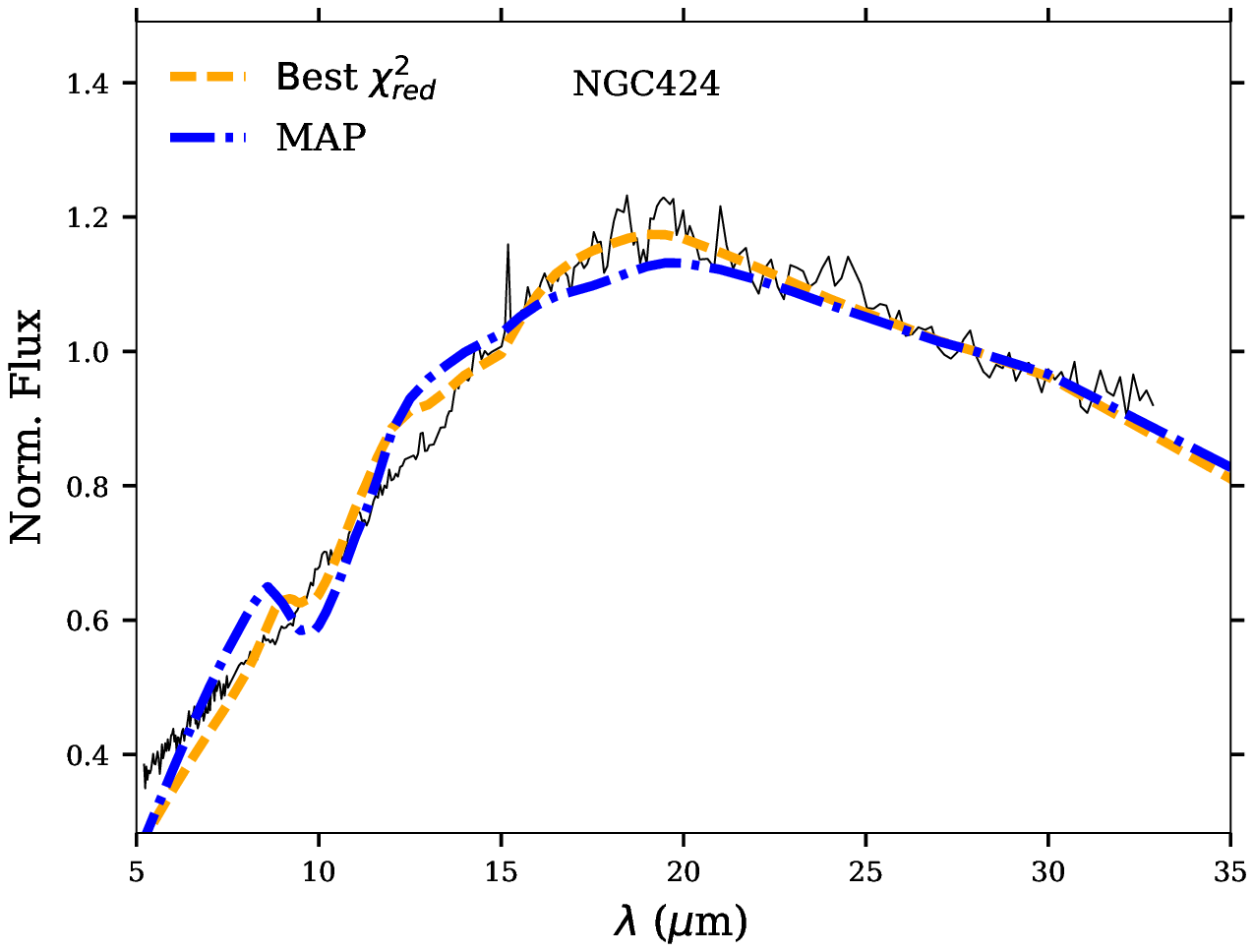}
\end{minipage} \hfill
\begin{minipage}[b]{0.325\linewidth}
\includegraphics[width=\textwidth]{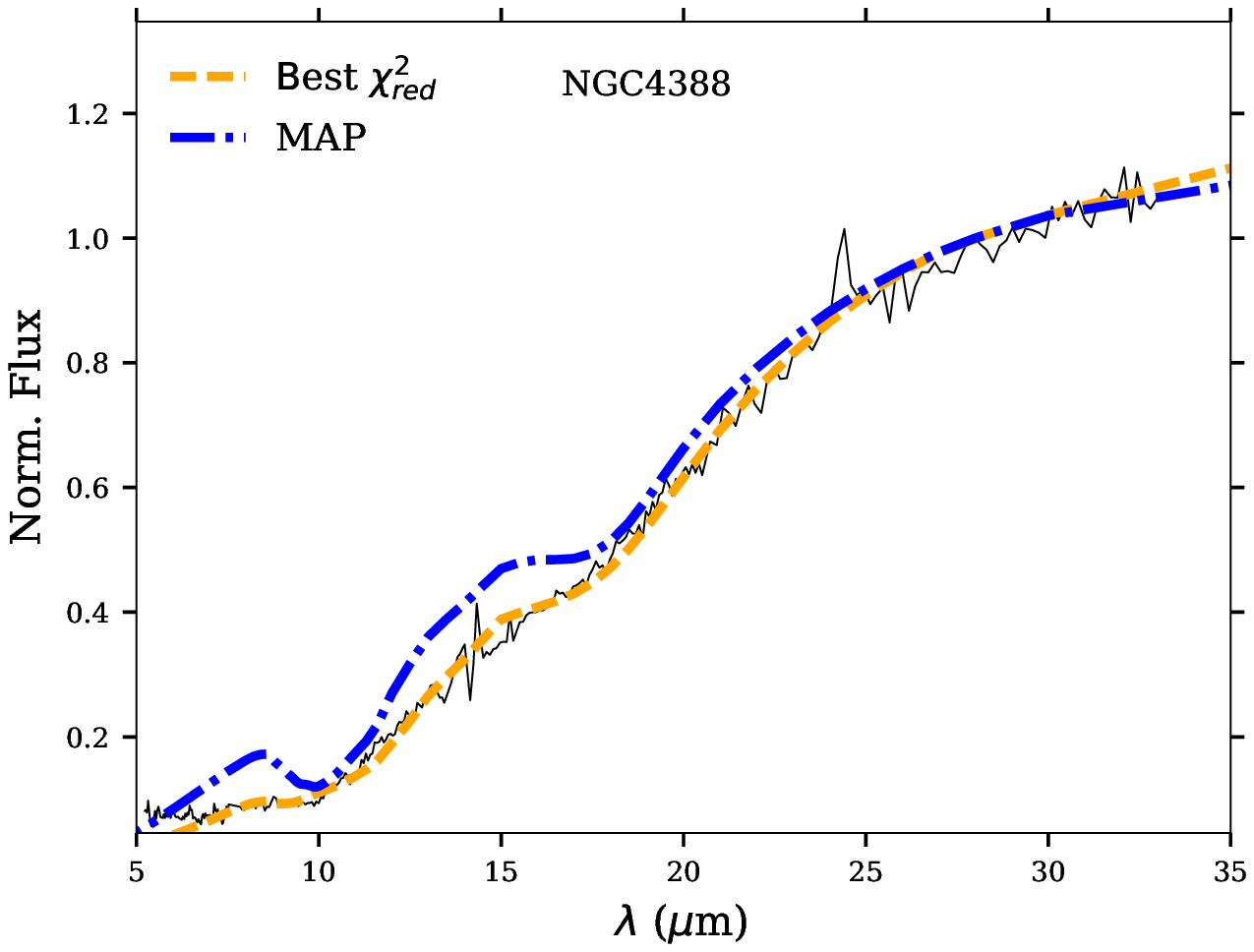}
\end{minipage} \hfill
\begin{minipage}[b]{0.325\linewidth}
\includegraphics[width=\textwidth]{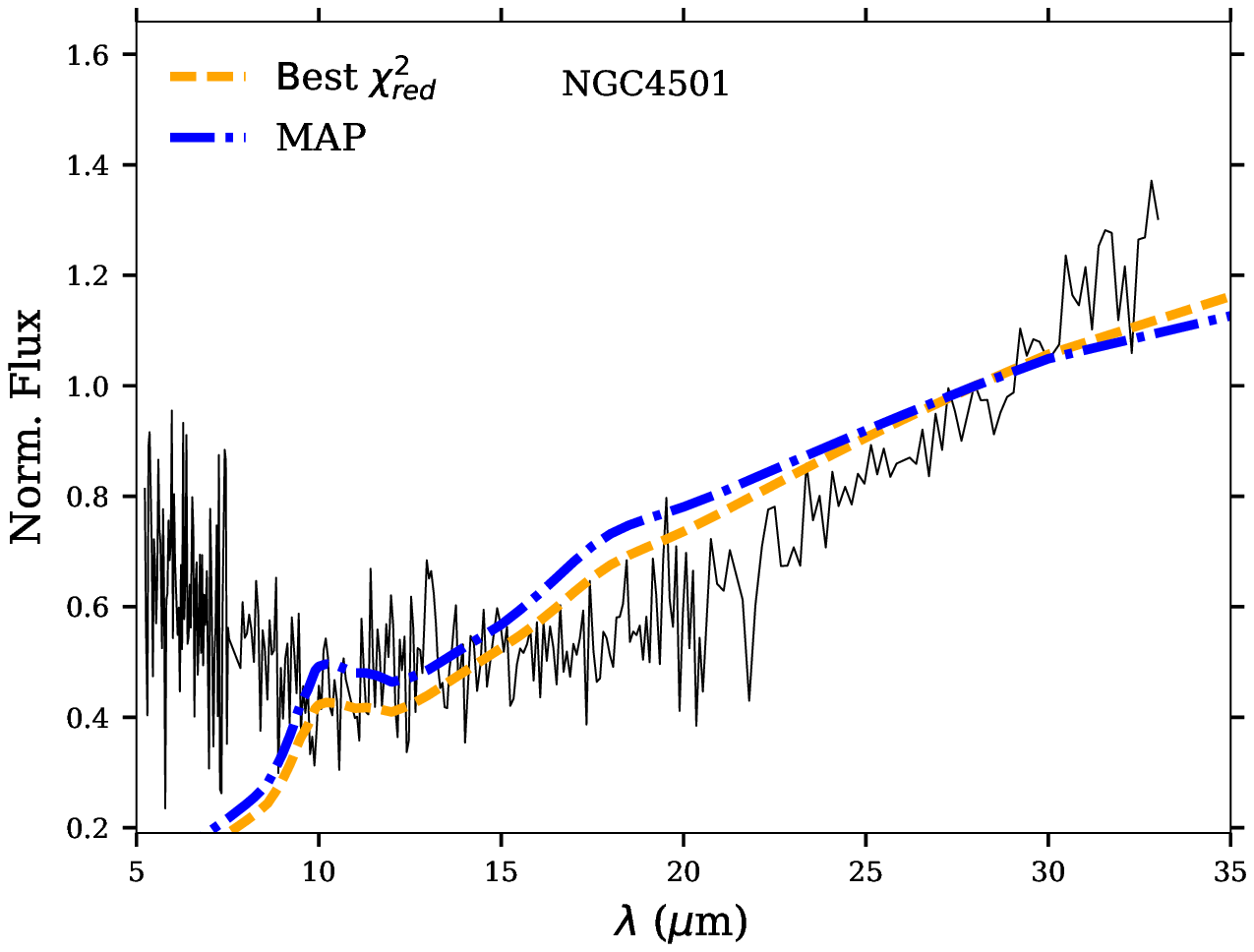}
\end{minipage} \hfill
\begin{minipage}[b]{0.325\linewidth}
\includegraphics[width=\textwidth]{NGC4507_bestMAP_JY.eps}
\end{minipage} \hfill
\begin{minipage}[b]{0.325\linewidth}
\includegraphics[width=\textwidth]{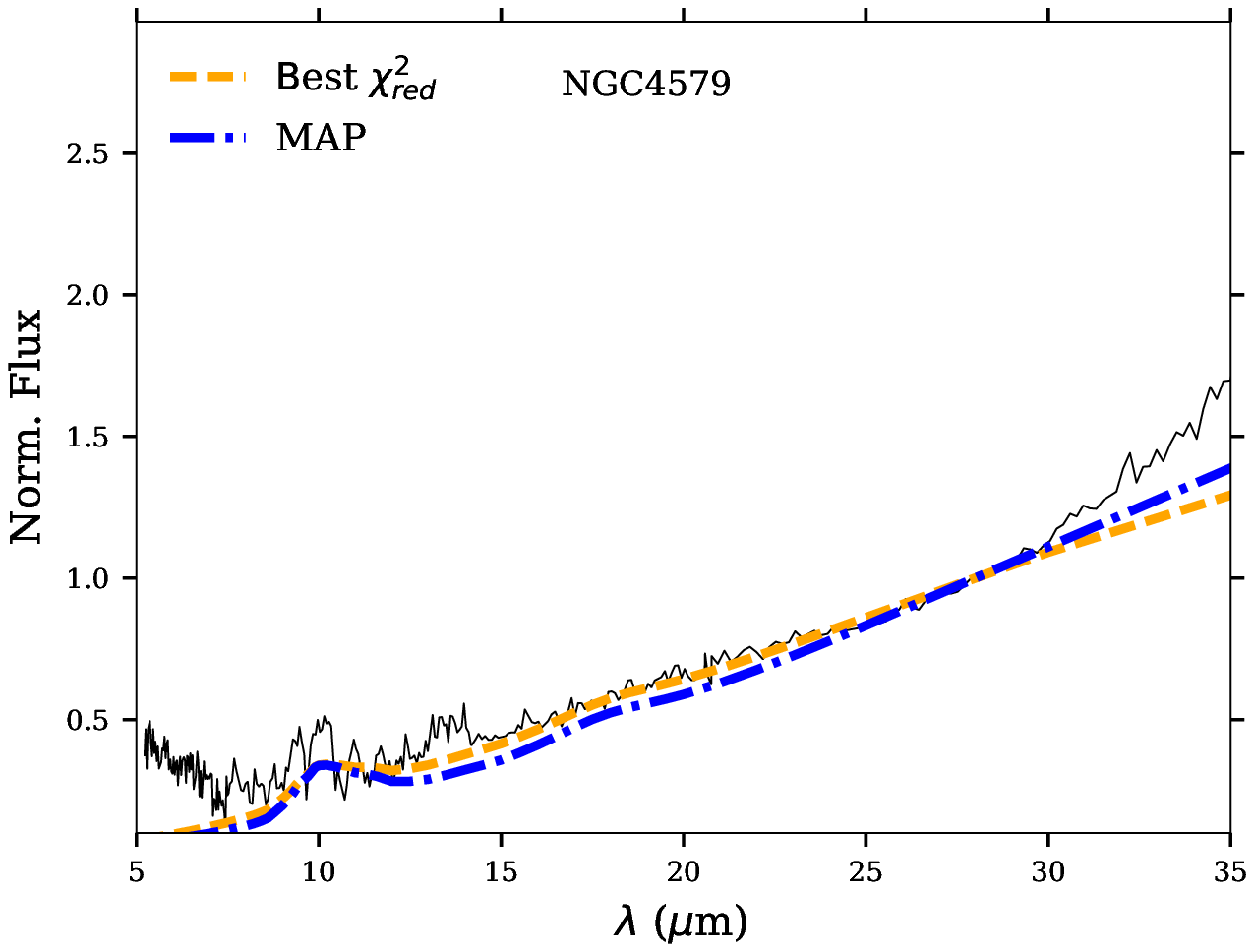}
\end{minipage} \hfill
\begin{minipage}[b]{0.325\linewidth}
\includegraphics[width=\textwidth]{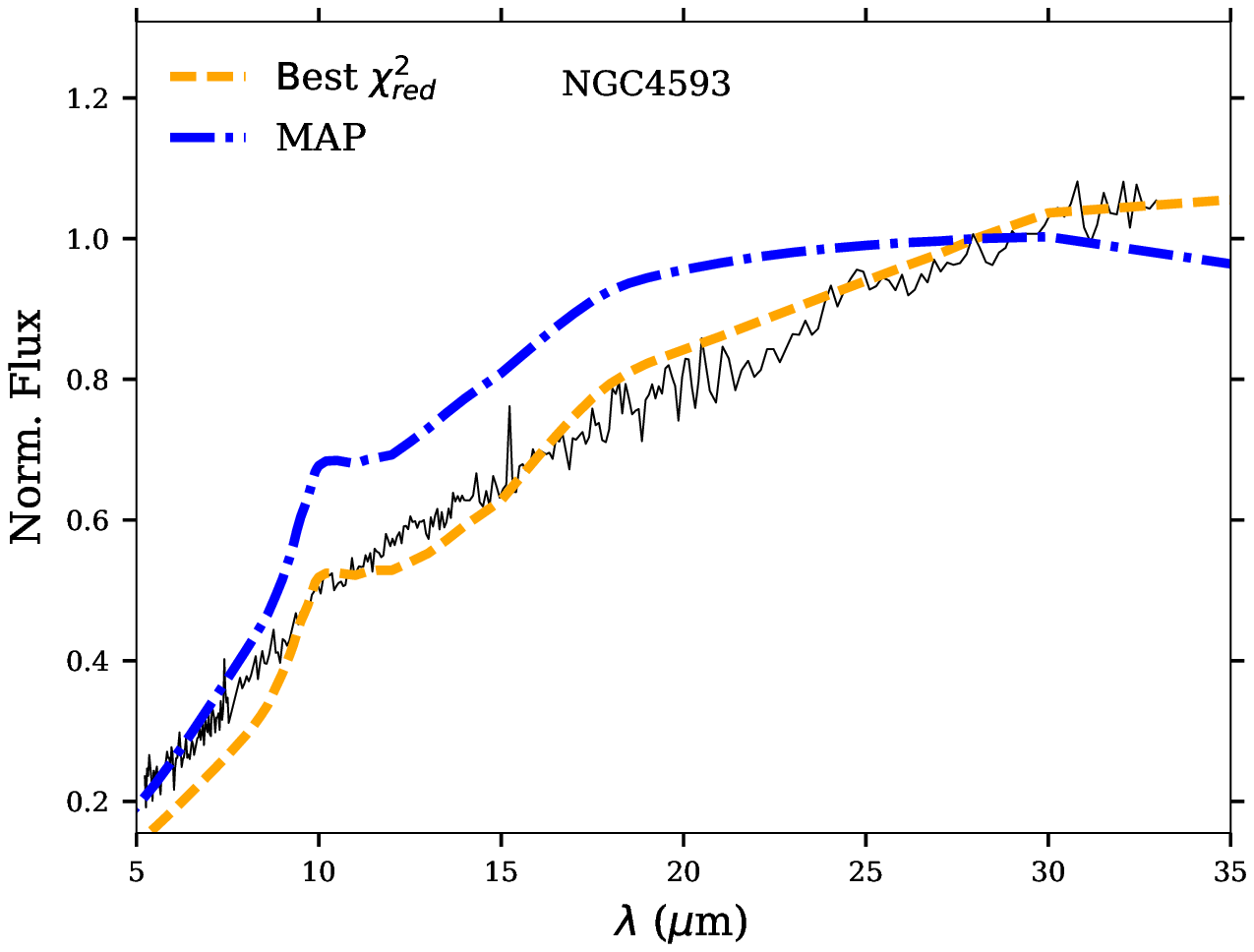}
\end{minipage} \hfill
\begin{minipage}[b]{0.325\linewidth}
\includegraphics[width=\textwidth]{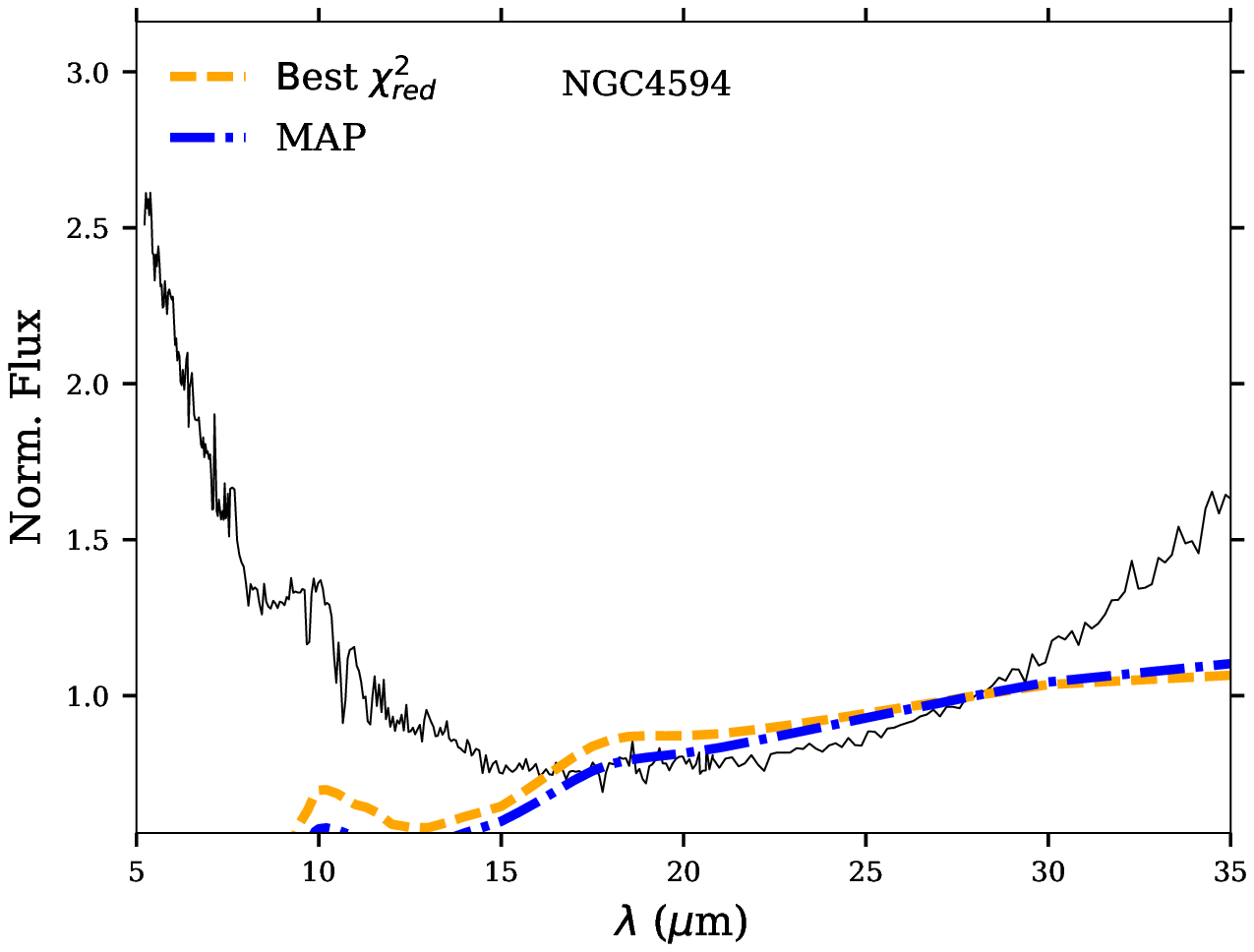}
\end{minipage} \hfill
\begin{minipage}[b]{0.325\linewidth}
\includegraphics[width=\textwidth]{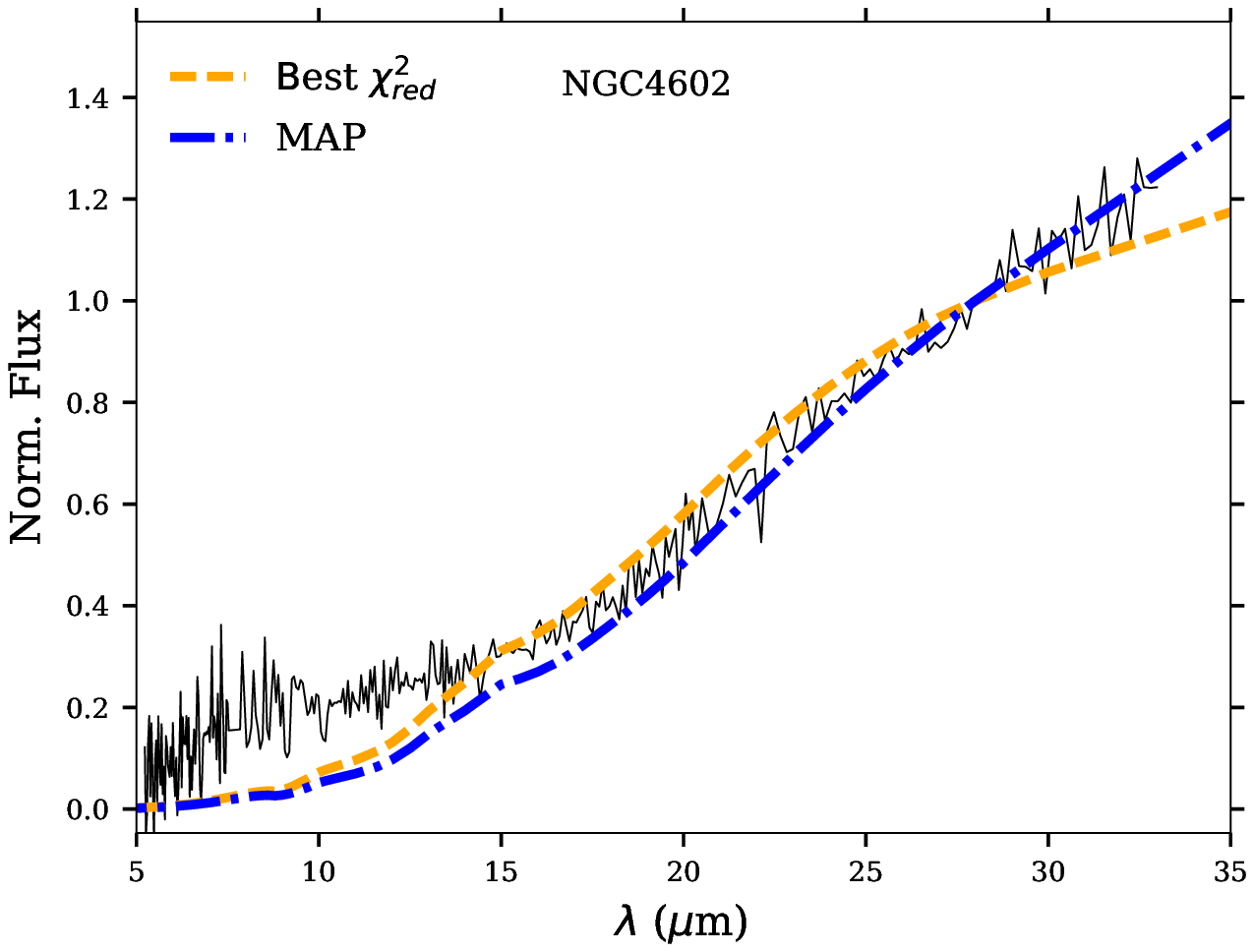}
\end{minipage} \hfill
\begin{minipage}[b]{0.325\linewidth}
\includegraphics[width=\textwidth]{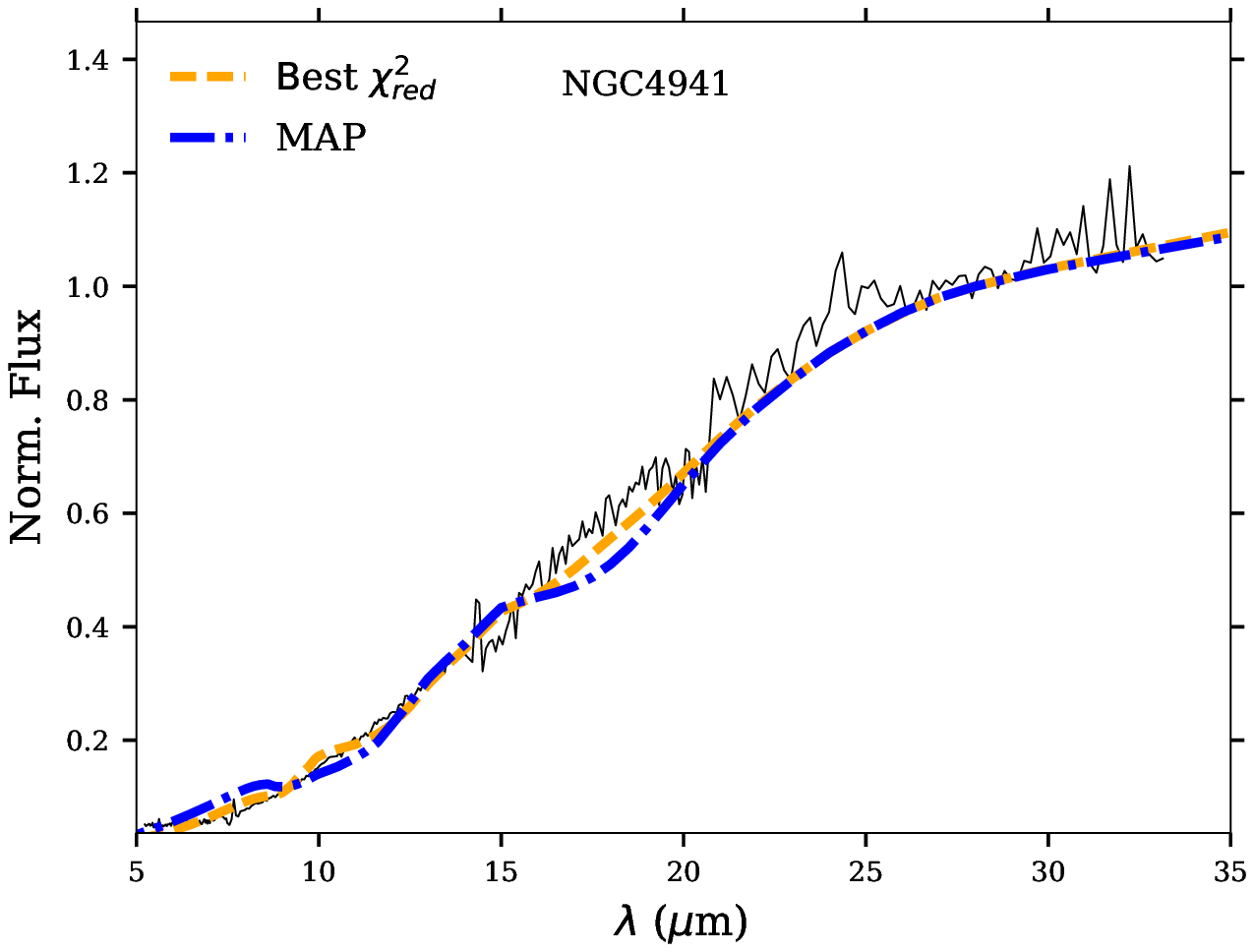}
\end{minipage} \hfill
\begin{minipage}[b]{0.325\linewidth}
\includegraphics[width=\textwidth]{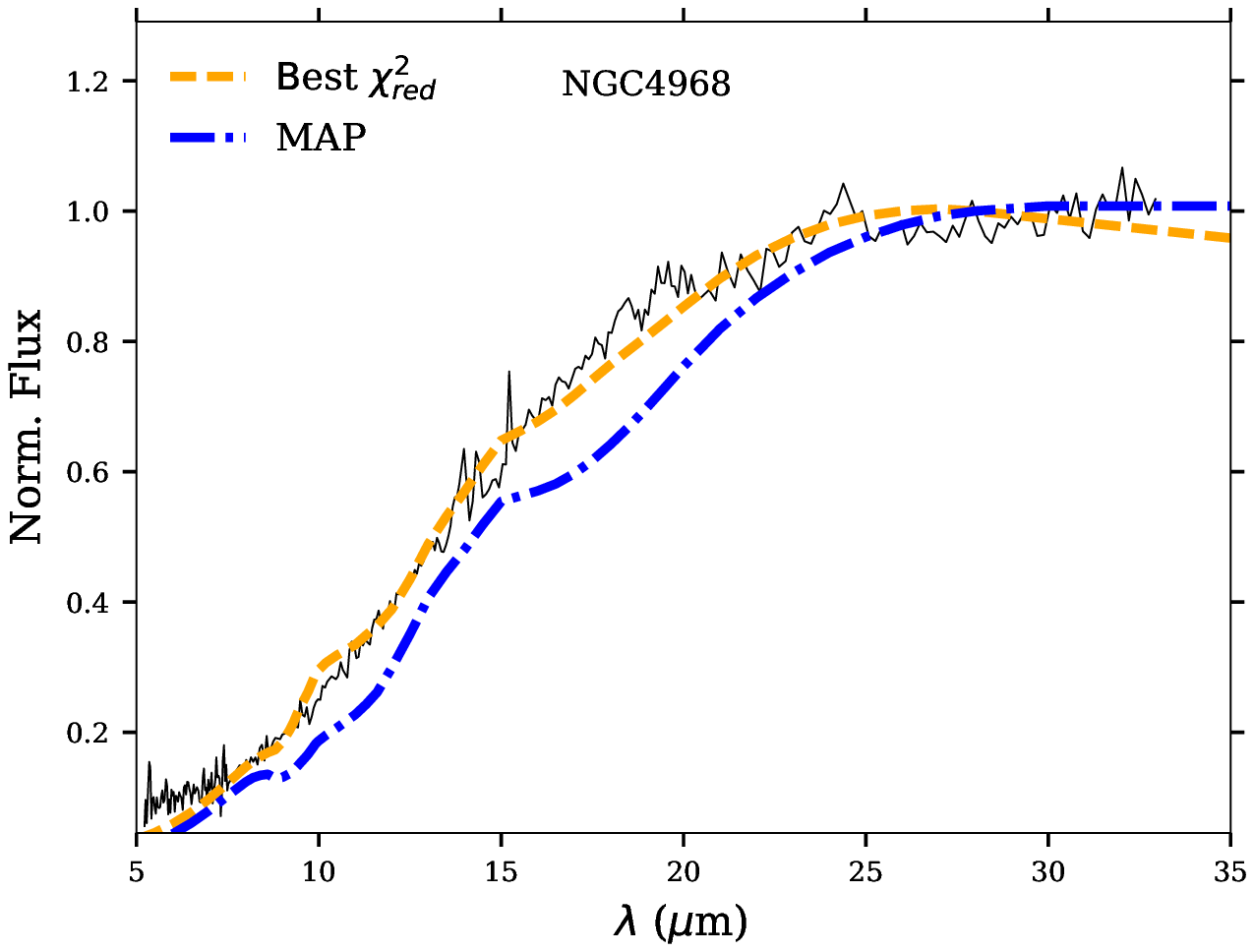}
\end{minipage} \hfill
\begin{minipage}[b]{0.325\linewidth}
\includegraphics[width=\textwidth]{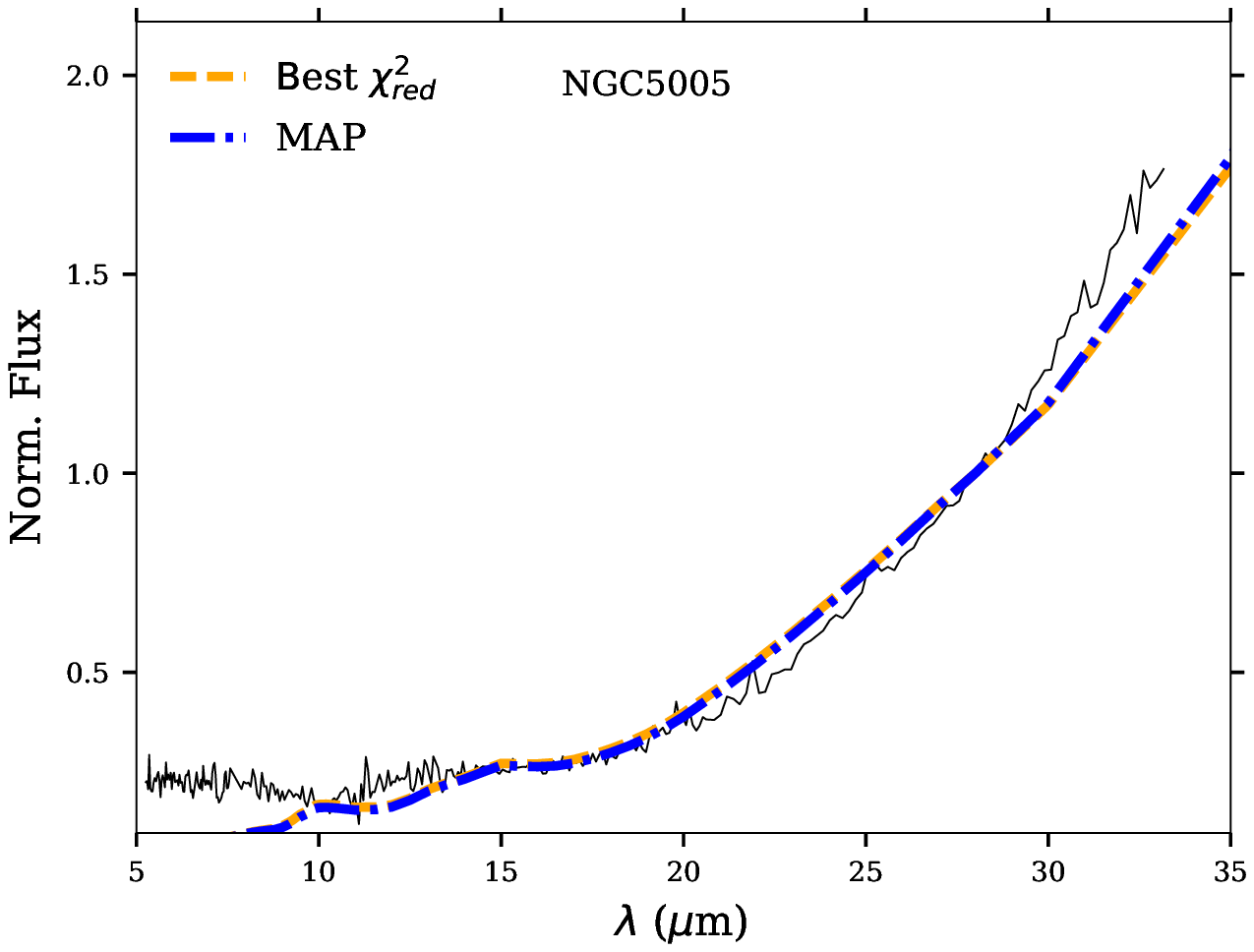}
\end{minipage} \hfill
\begin{minipage}[b]{0.325\linewidth}
\includegraphics[width=\textwidth]{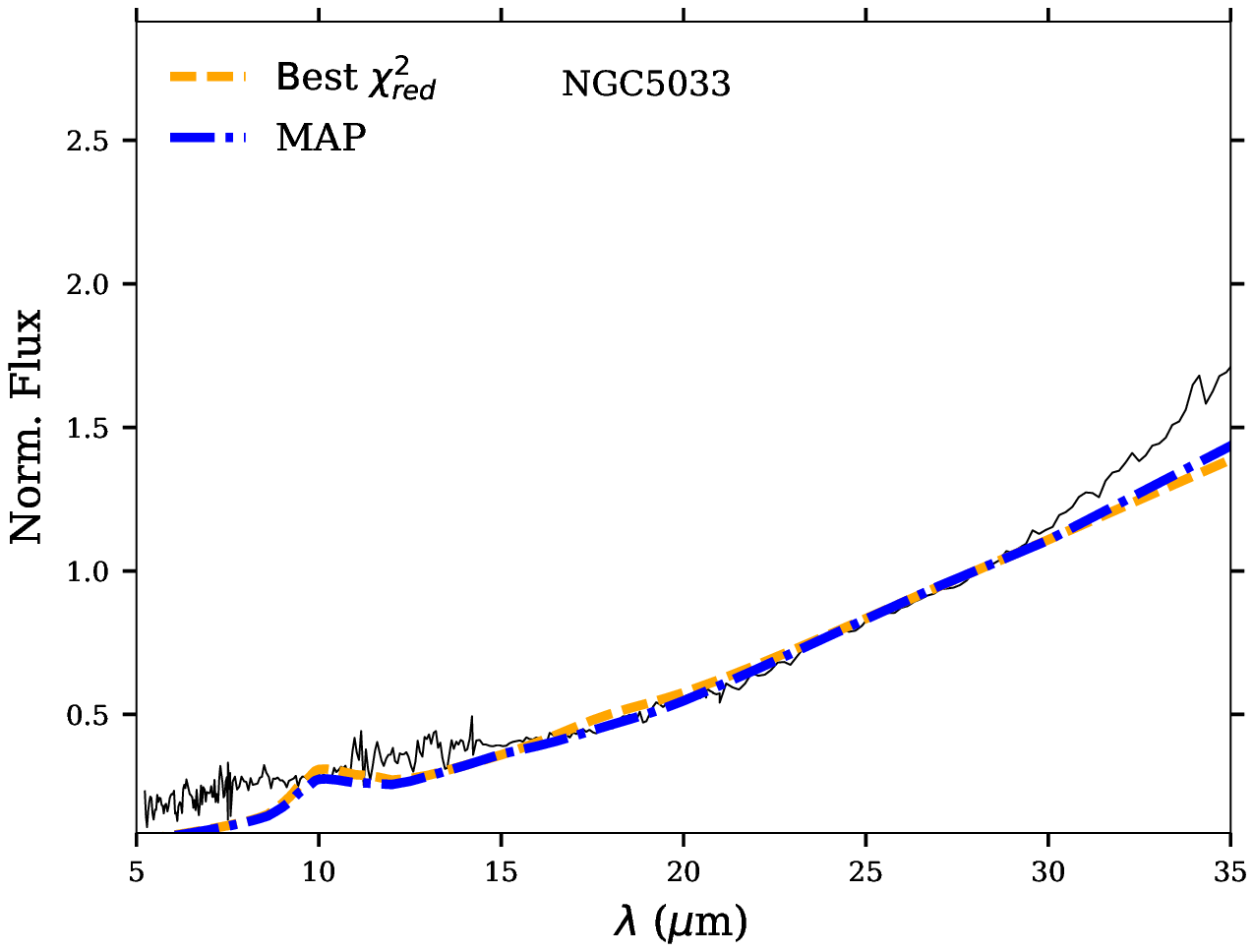}
\end{minipage} \hfill
\begin{minipage}[b]{0.325\linewidth}
\includegraphics[width=\textwidth]{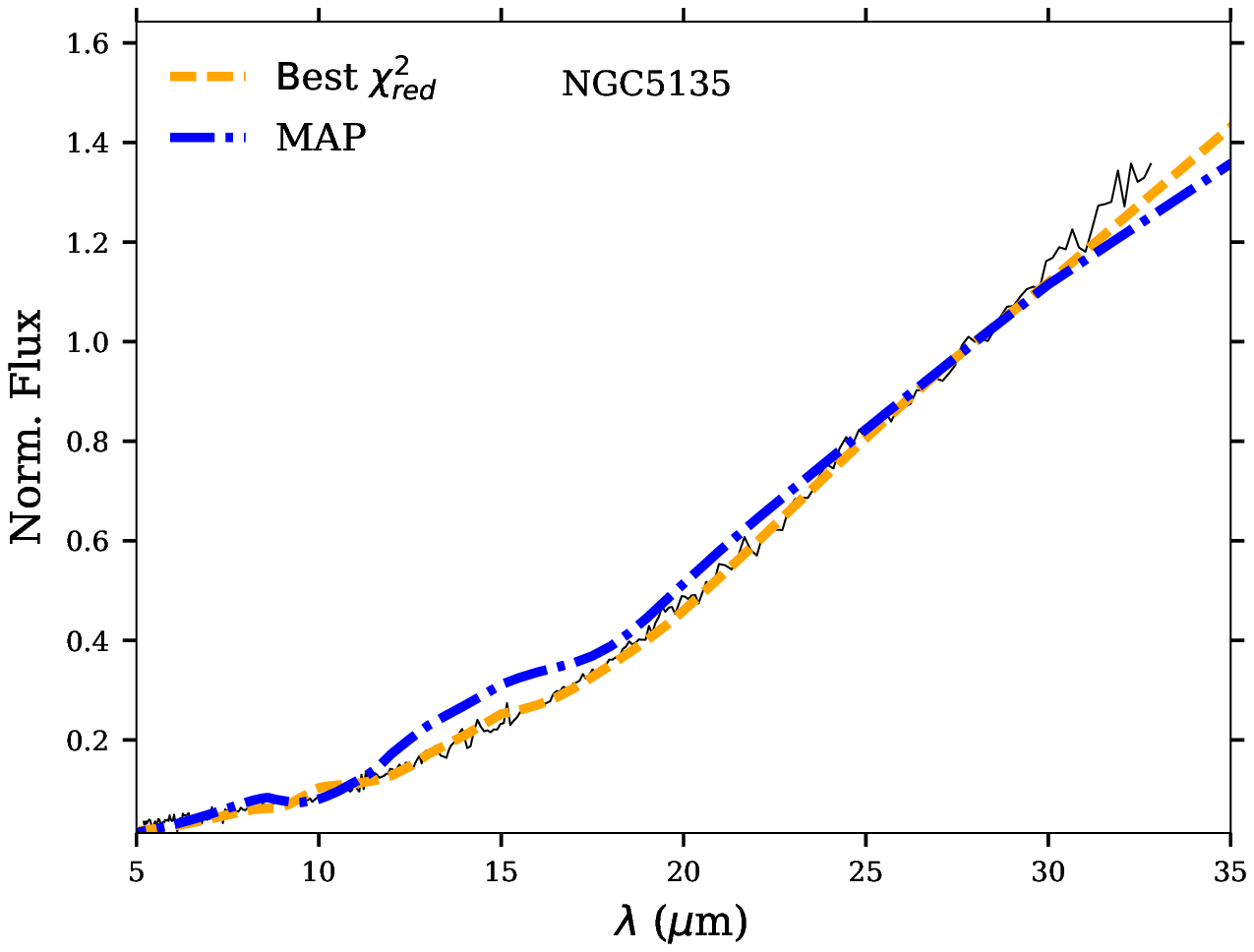}
\end{minipage} \hfill
\begin{minipage}[b]{0.325\linewidth}
\includegraphics[width=\textwidth]{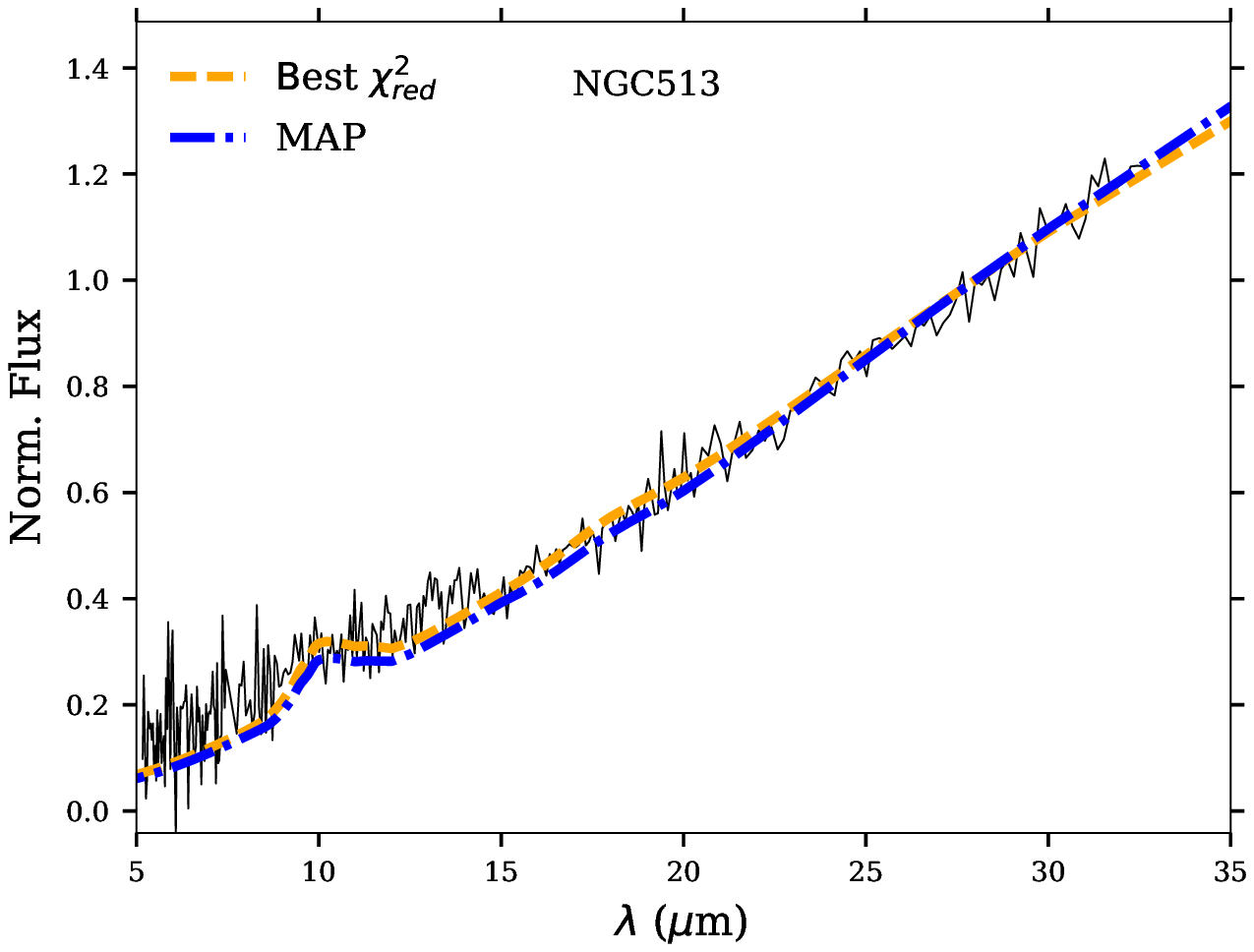}
\end{minipage} \hfill
\begin{minipage}[b]{0.325\linewidth}
\includegraphics[width=\textwidth]{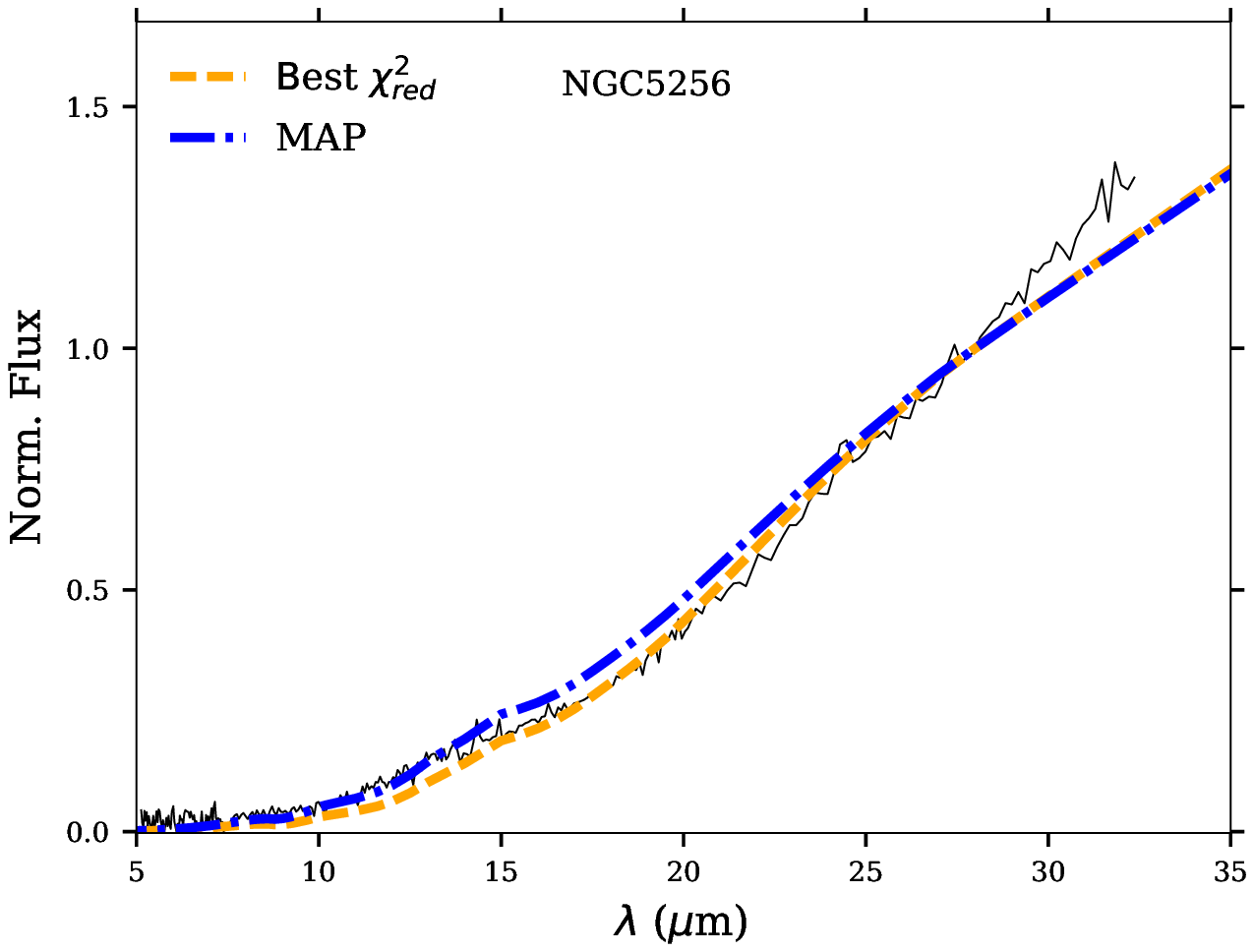}
\end{minipage} \hfill
\caption{continued from previous page.}
\setcounter{figure}{0}
\end{figure}

\begin{figure}

\begin{minipage}[b]{0.325\linewidth}
\includegraphics[width=\textwidth]{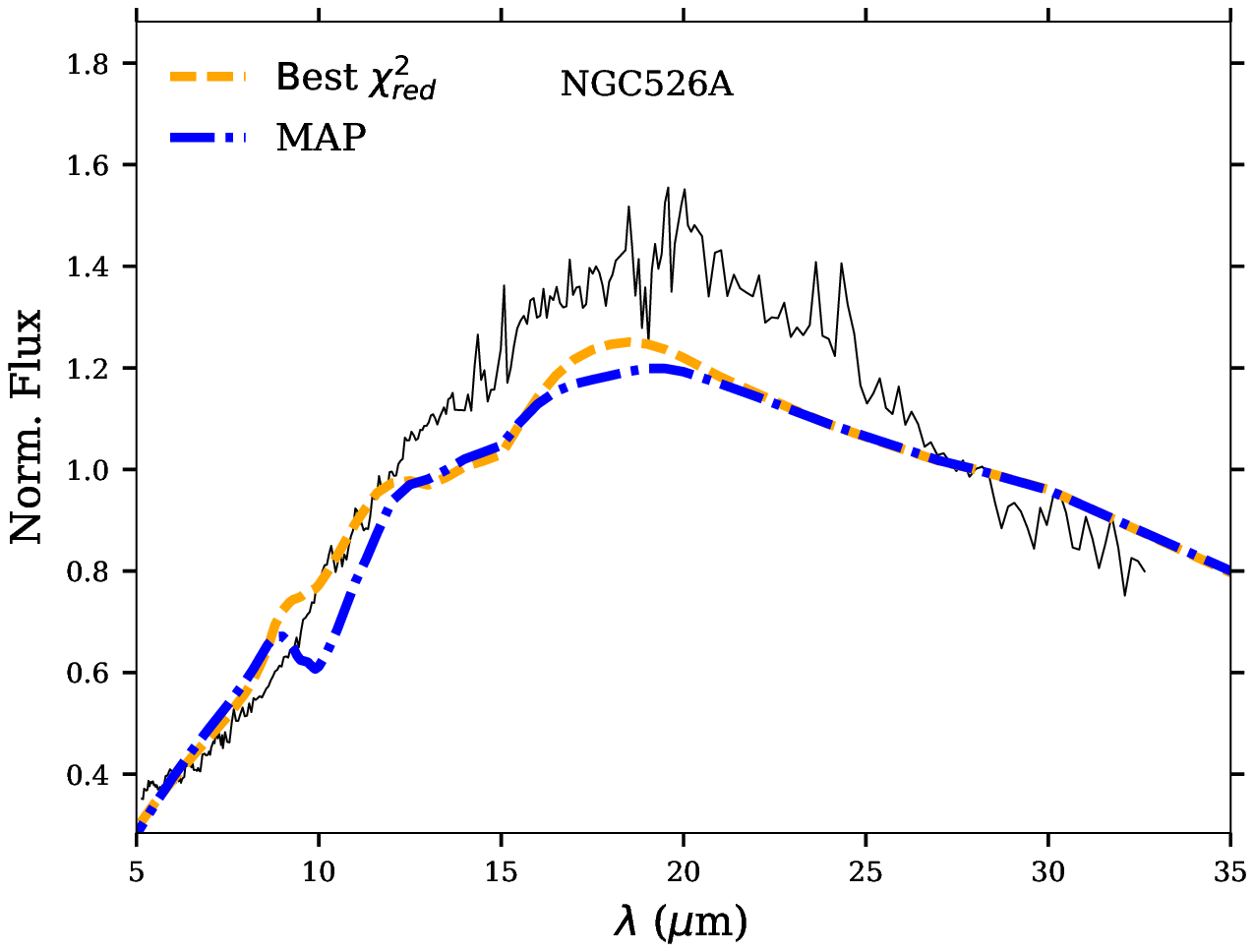}
\end{minipage} \hfill
\begin{minipage}[b]{0.325\linewidth}
\includegraphics[width=\textwidth]{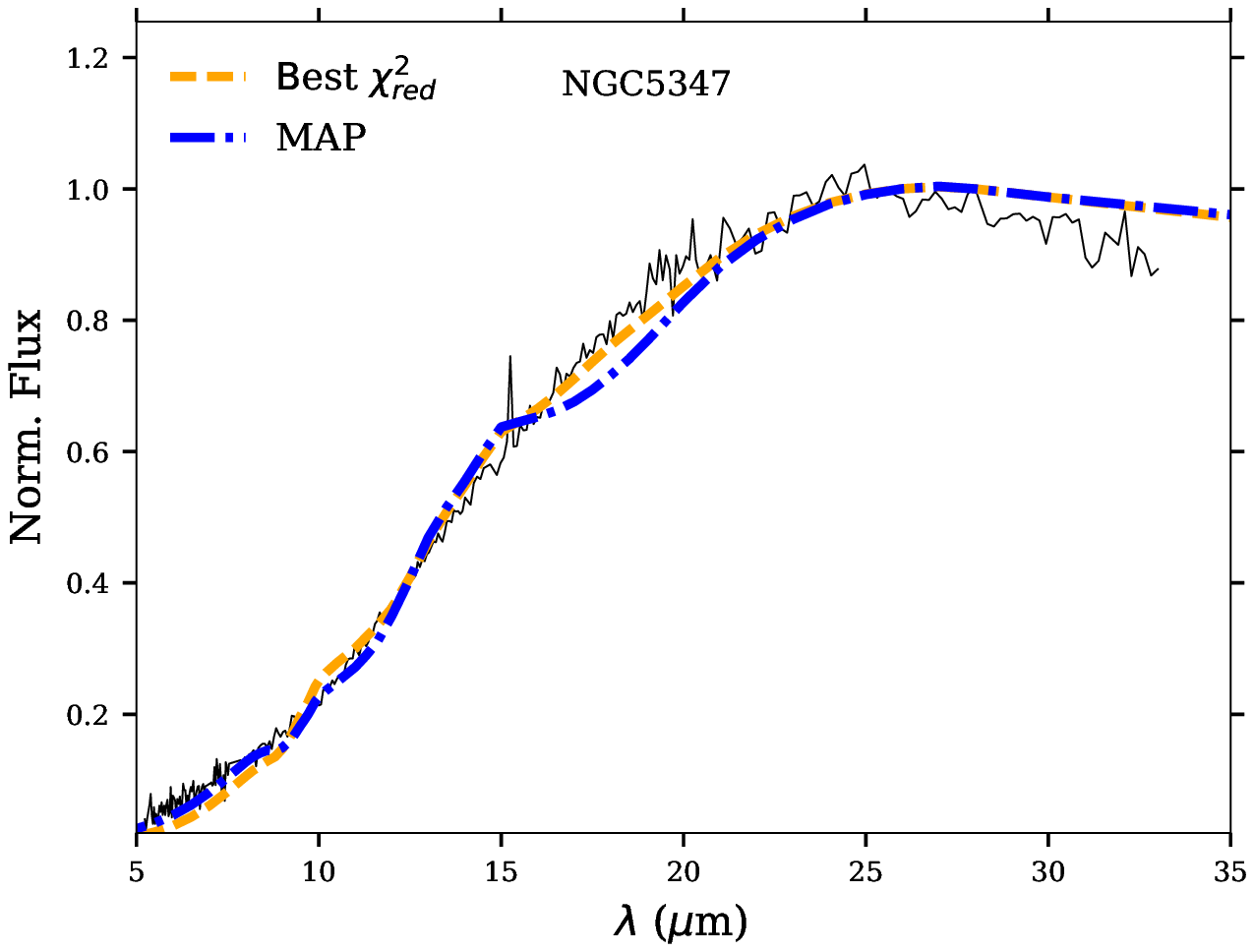}
\end{minipage} \hfill
\begin{minipage}[b]{0.325\linewidth}
\includegraphics[width=\textwidth]{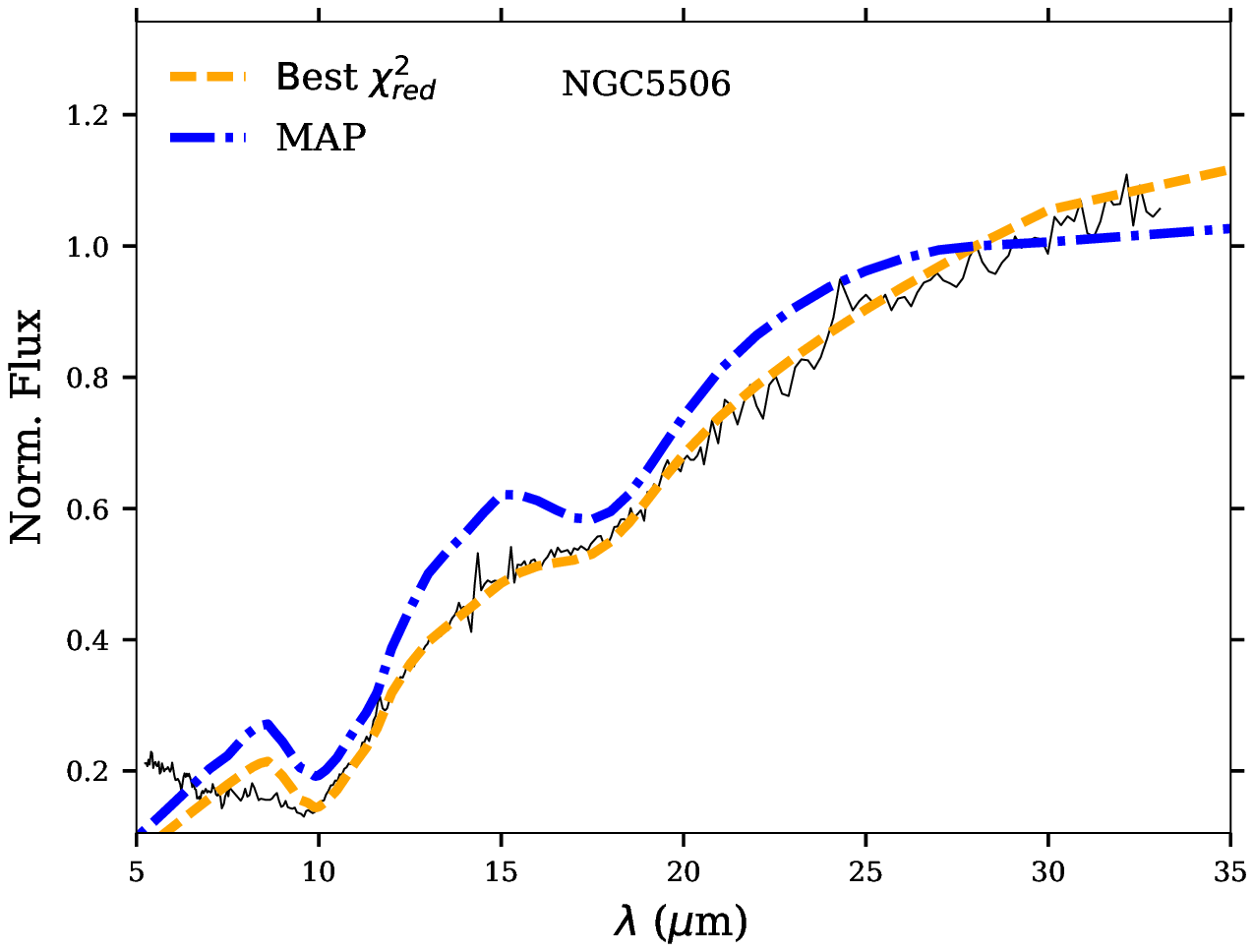}
\end{minipage} \hfill
\begin{minipage}[b]{0.325\linewidth}
\includegraphics[width=\textwidth]{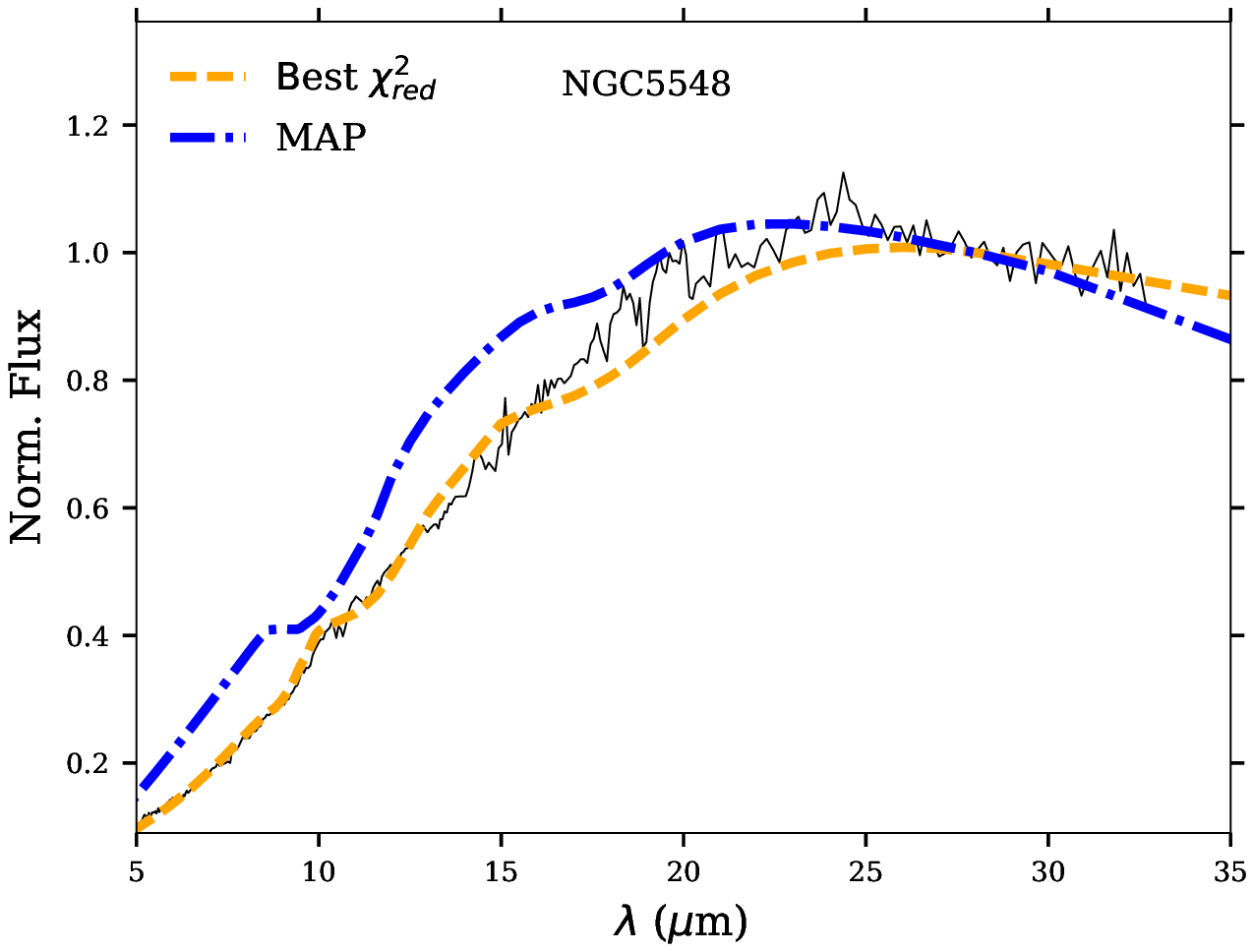}
\end{minipage} \hfill
\begin{minipage}[b]{0.325\linewidth}
\includegraphics[width=\textwidth]{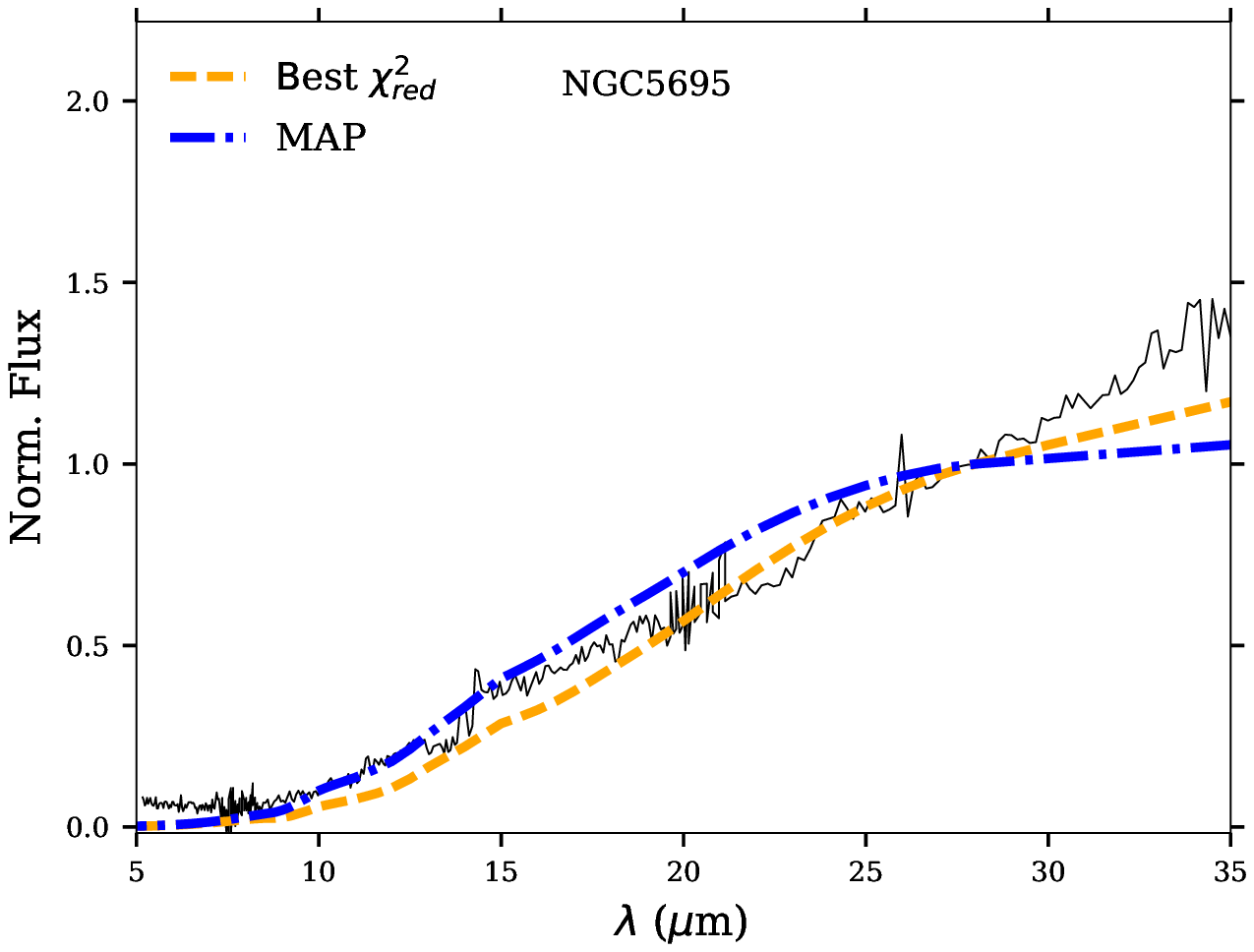}
\end{minipage} \hfill
\begin{minipage}[b]{0.325\linewidth}
\includegraphics[width=\textwidth]{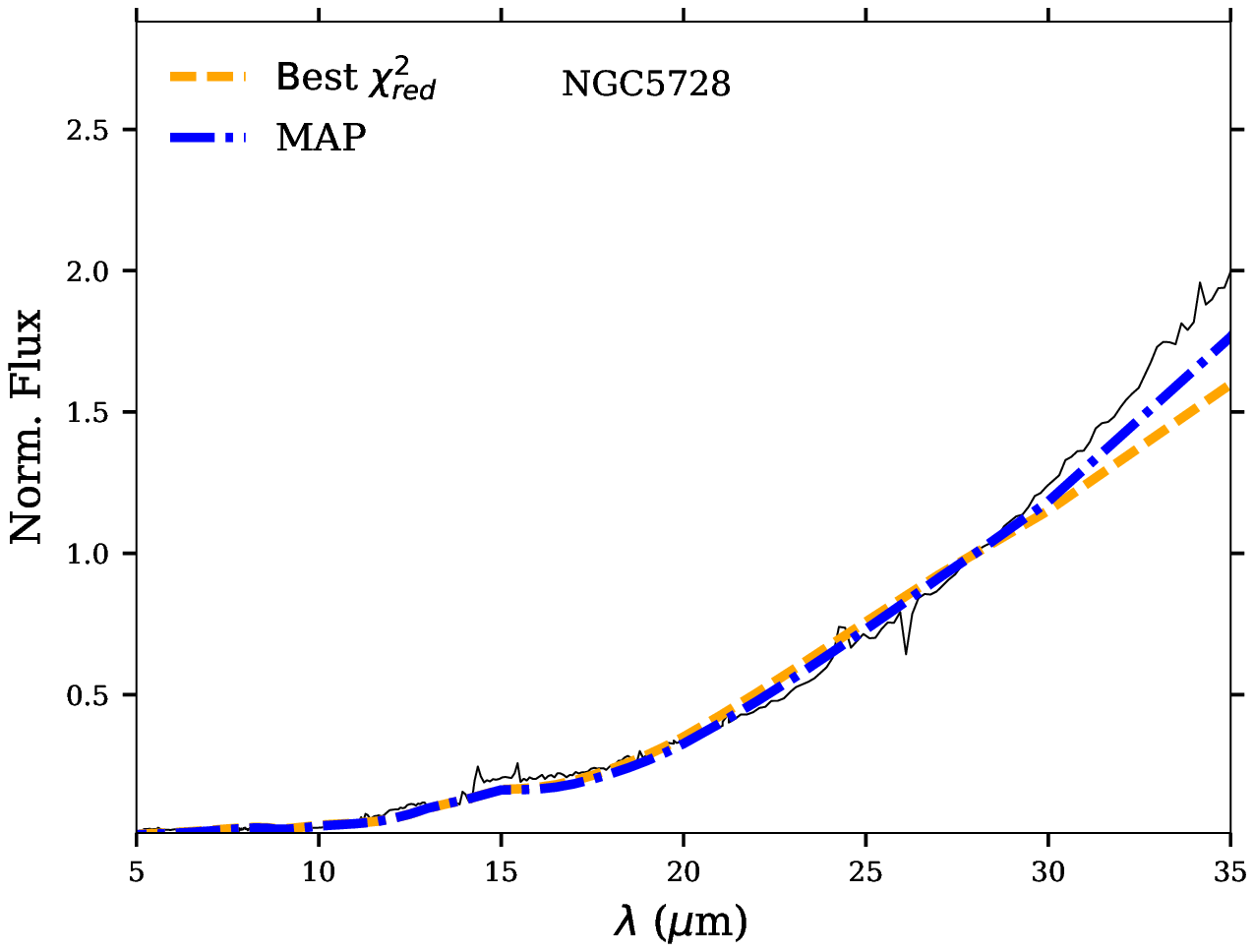}
\end{minipage} \hfill
\begin{minipage}[b]{0.325\linewidth}
\includegraphics[width=\textwidth]{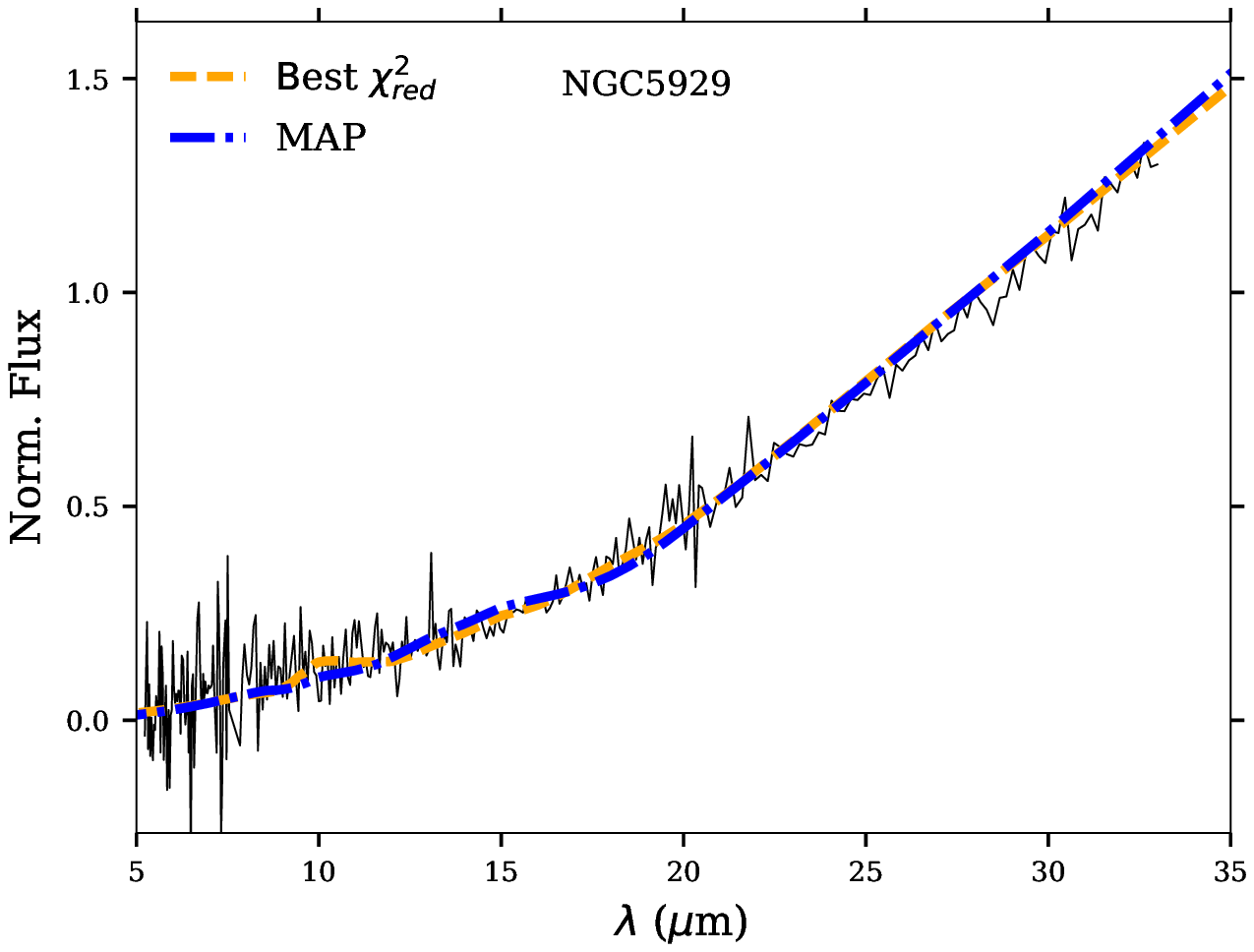}
\end{minipage} \hfill
\begin{minipage}[b]{0.325\linewidth}
\includegraphics[width=\textwidth]{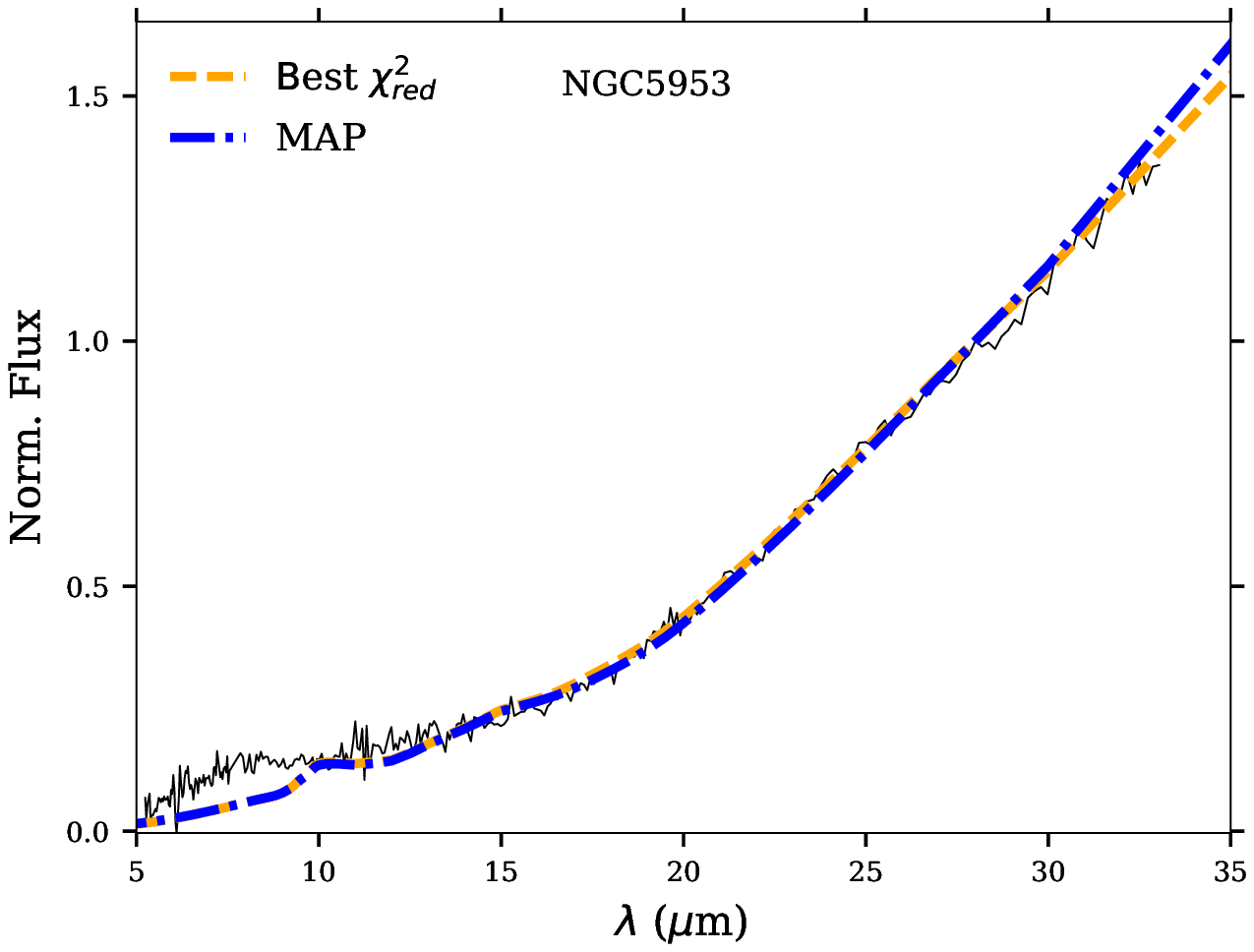}
\end{minipage} \hfill
\begin{minipage}[b]{0.325\linewidth}
\includegraphics[width=\textwidth]{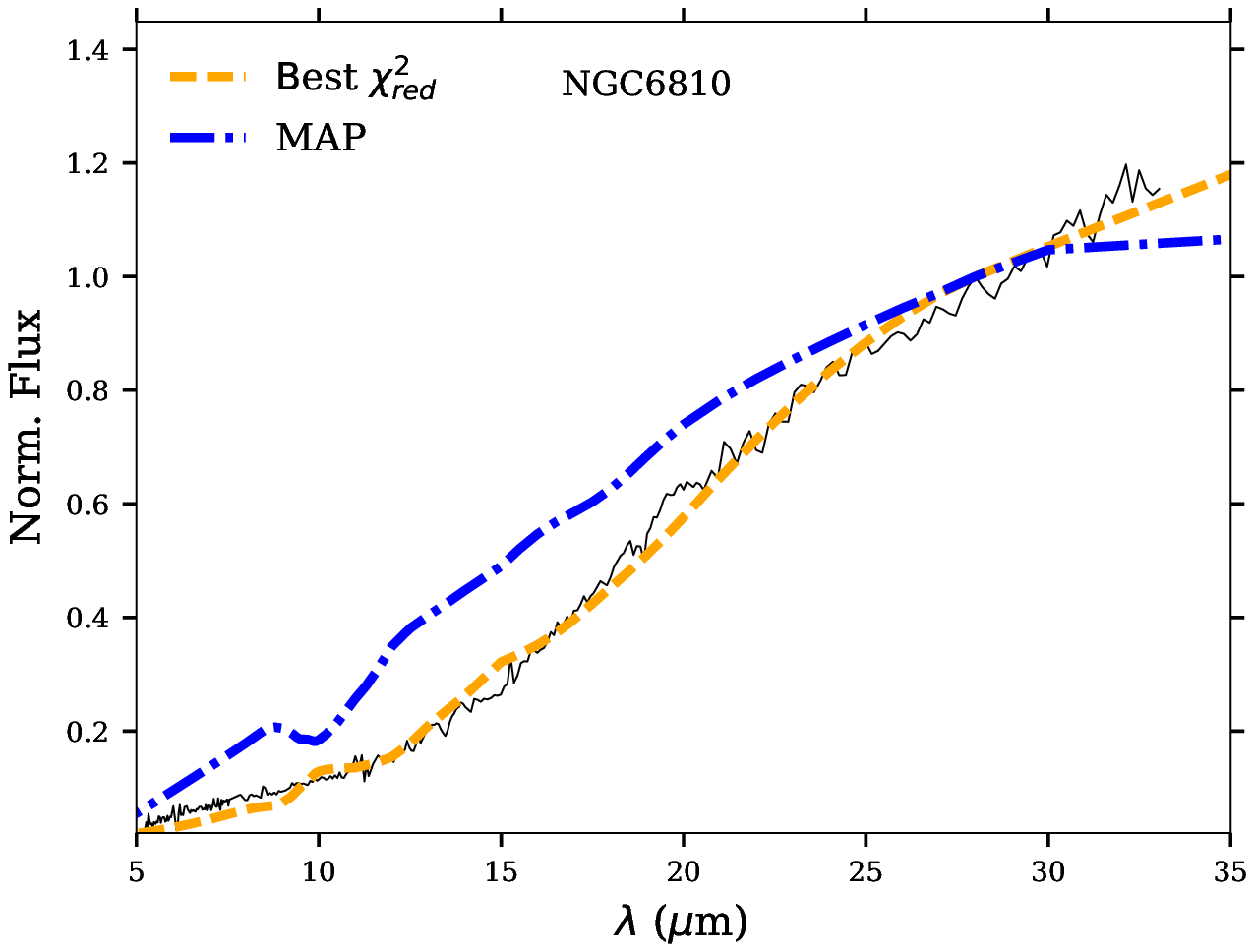}
\end{minipage} \hfill
\begin{minipage}[b]{0.325\linewidth}
\includegraphics[width=\textwidth]{NGC6860_bestMAP_JY.eps}
\end{minipage} \hfill
\begin{minipage}[b]{0.325\linewidth}
\includegraphics[width=\textwidth]{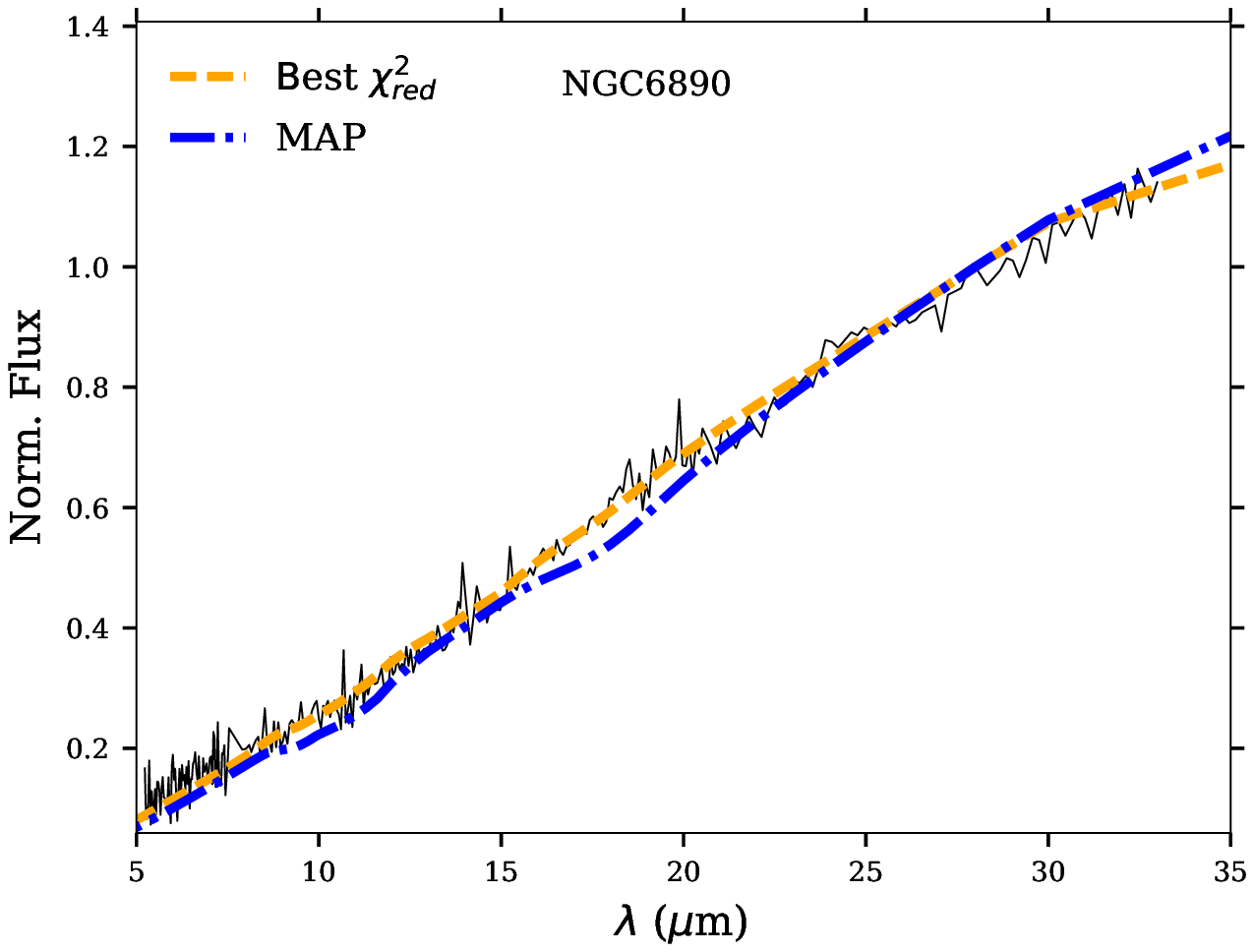}
\end{minipage} \hfill
\begin{minipage}[b]{0.325\linewidth}
\includegraphics[width=\textwidth]{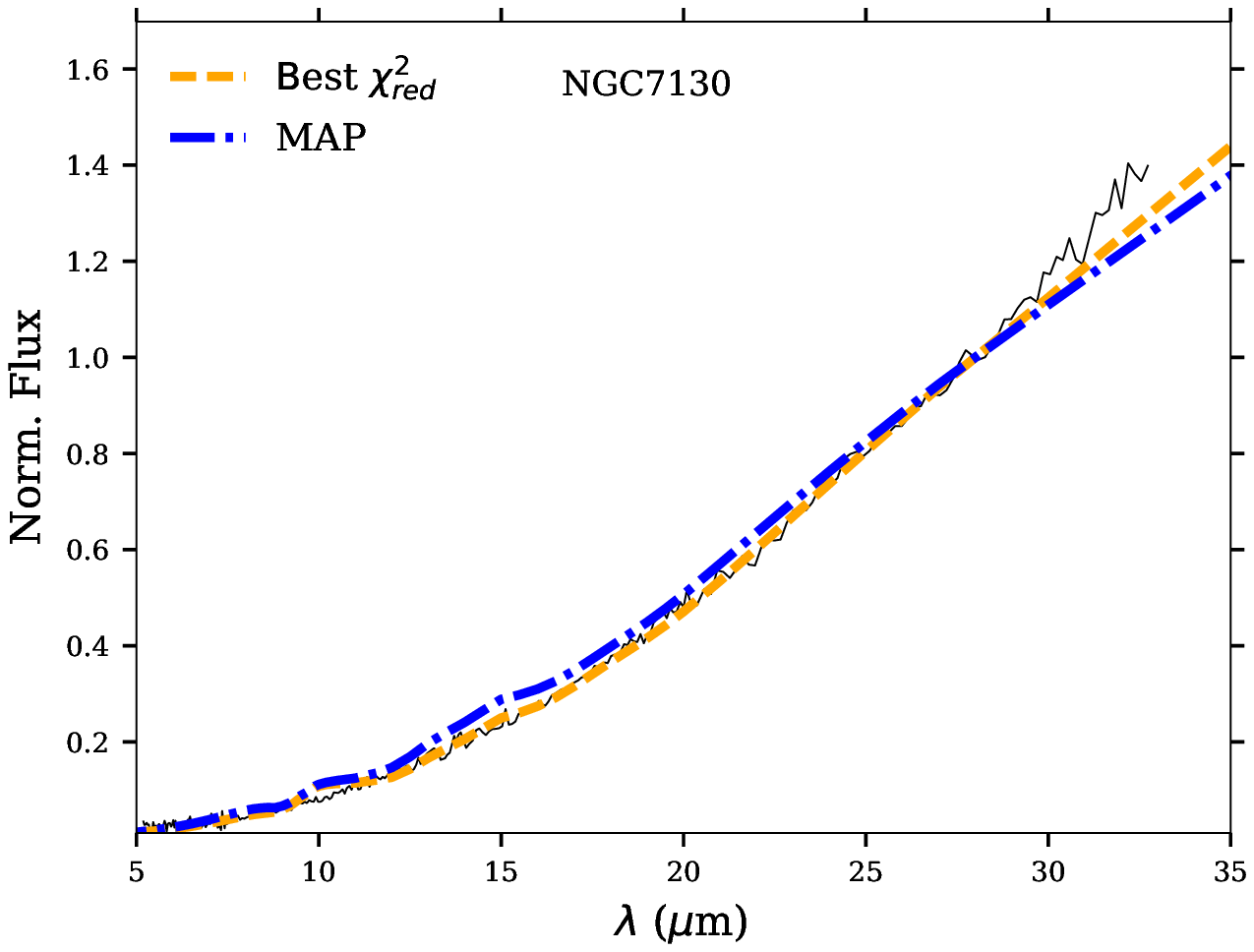}
\end{minipage} \hfill
\begin{minipage}[b]{0.325\linewidth}
\includegraphics[width=\textwidth]{NGC7172_bestMAP_JY.eps}
\end{minipage} \hfill
\begin{minipage}[b]{0.325\linewidth}
\includegraphics[width=\textwidth]{NGC7213_bestMAP_JY.eps}
\end{minipage} \hfill
\begin{minipage}[b]{0.325\linewidth}
\includegraphics[width=\textwidth]{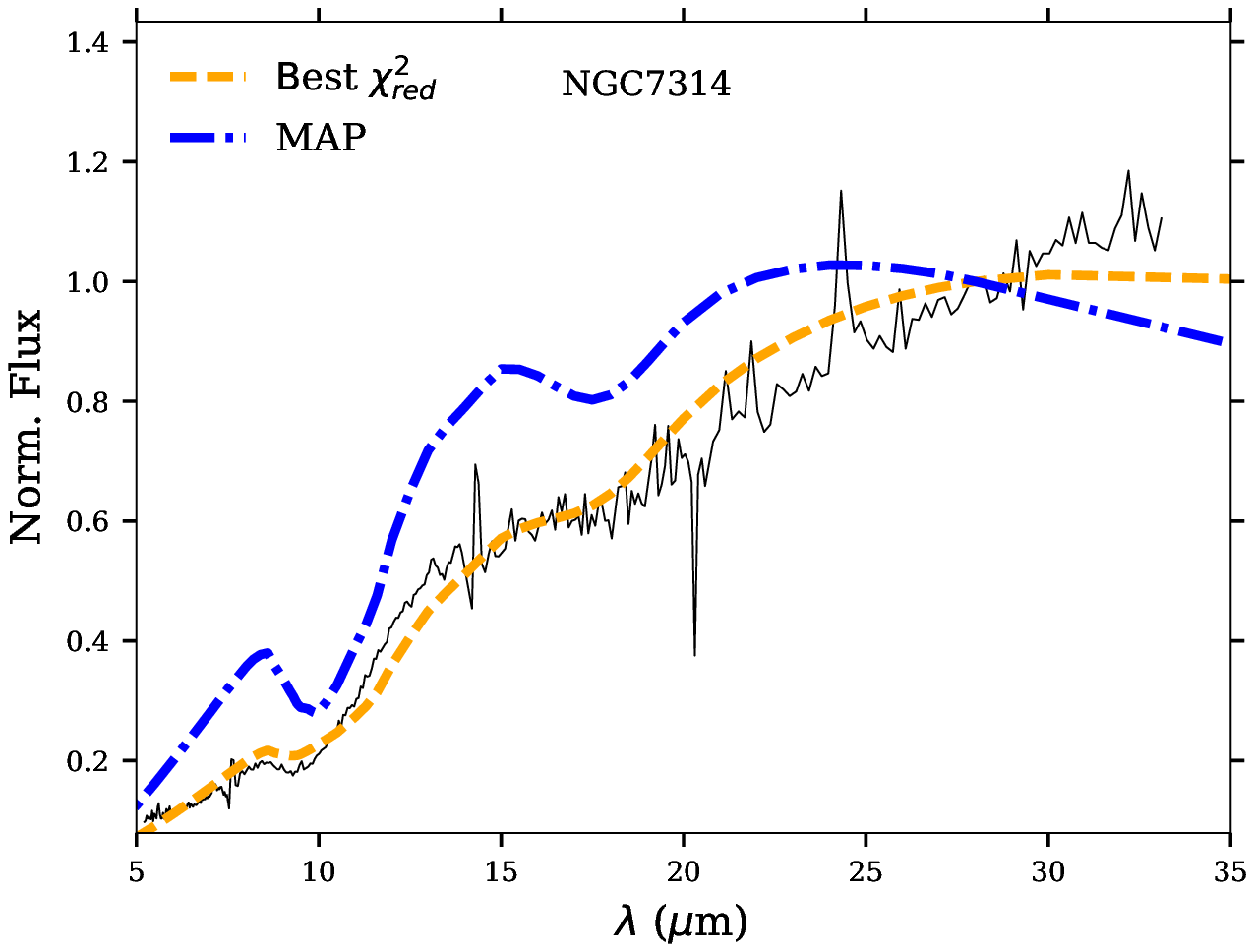}
\end{minipage} \hfill
\caption{continued from previous page.}
\setcounter{figure}{0}
\end{figure}

\begin{figure}

\begin{minipage}[b]{0.325\linewidth}
\includegraphics[width=\textwidth]{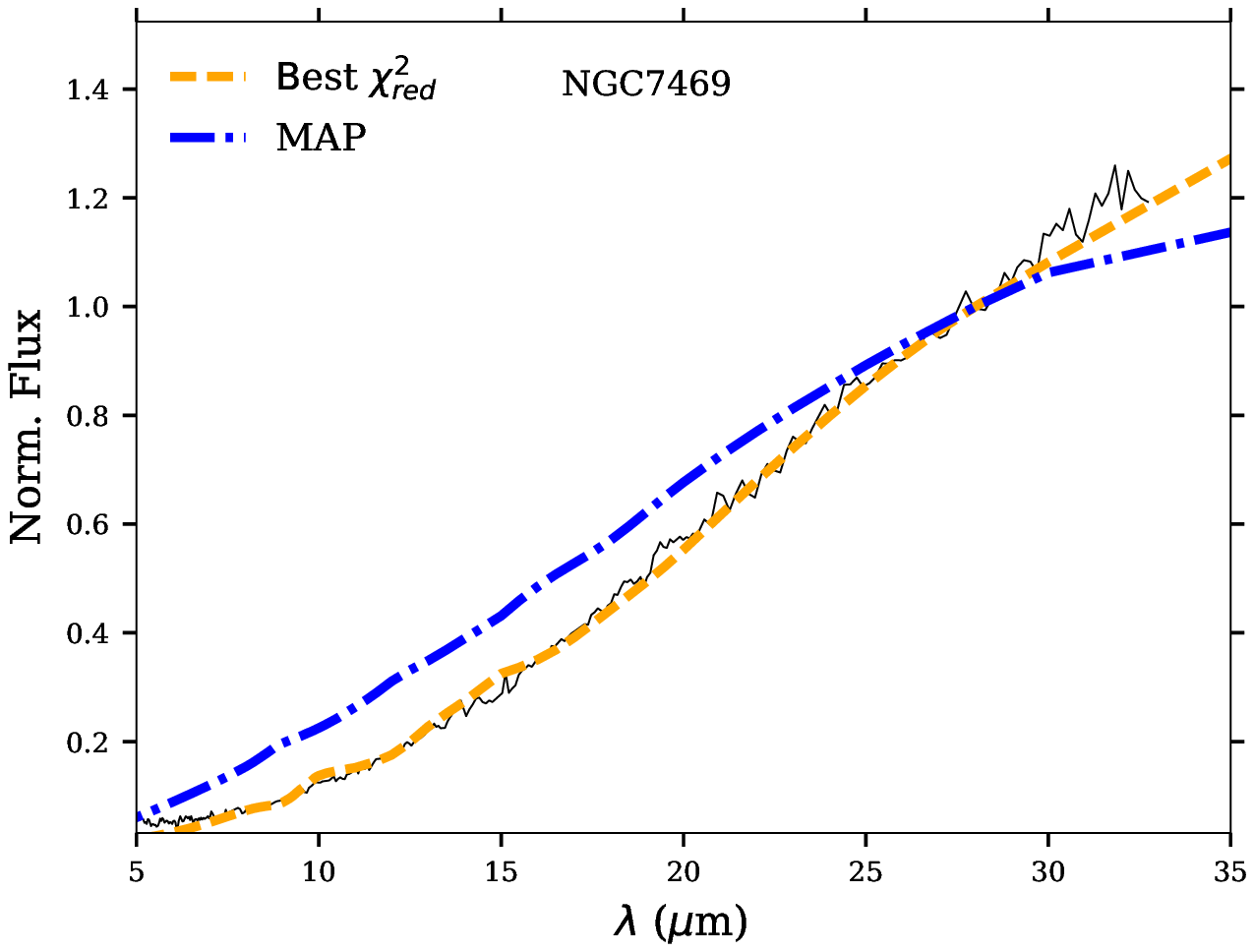}
\end{minipage} \hfill
\begin{minipage}[b]{0.325\linewidth}
\includegraphics[width=\textwidth]{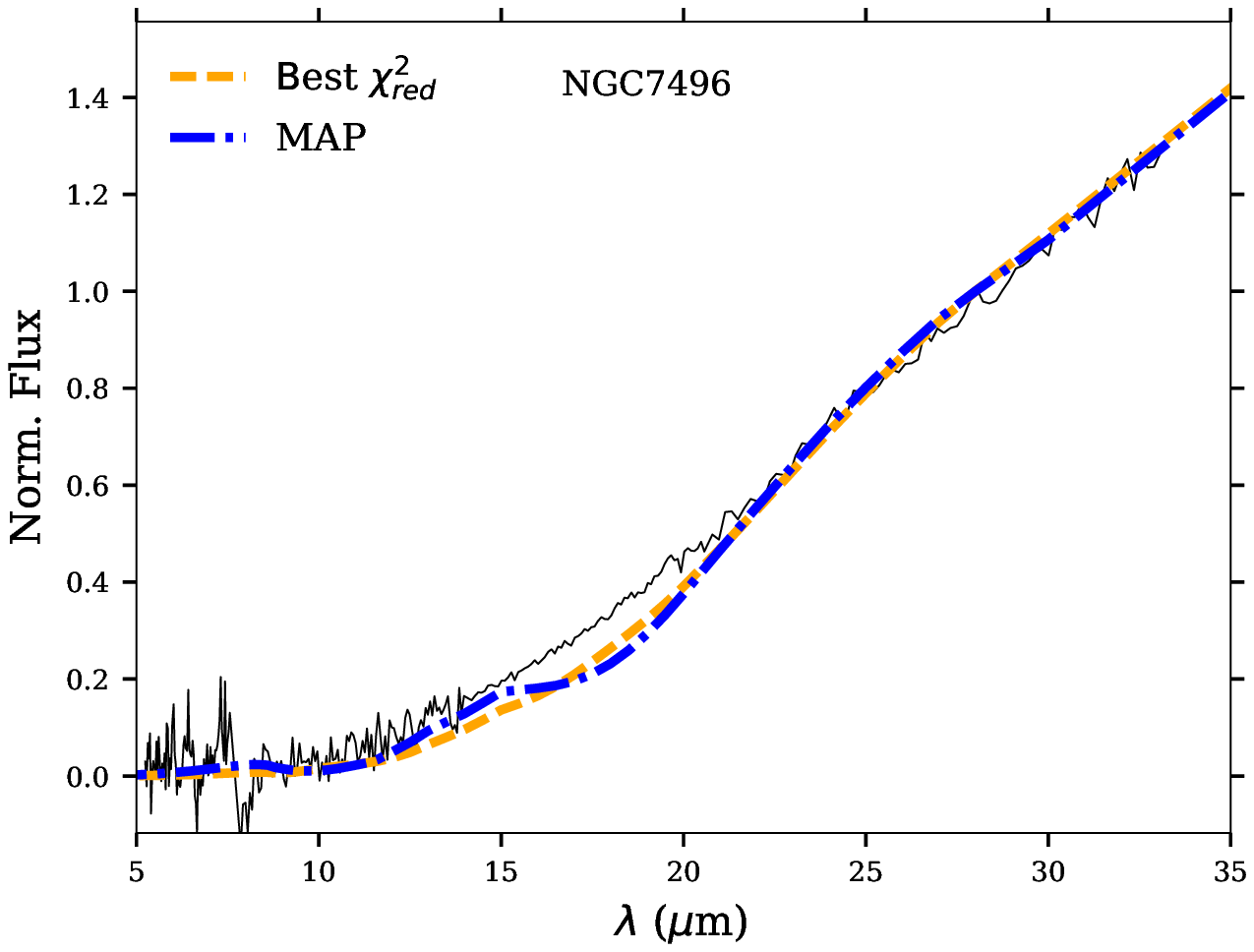}
\end{minipage} \hfill
\begin{minipage}[b]{0.325\linewidth}
\includegraphics[width=\textwidth]{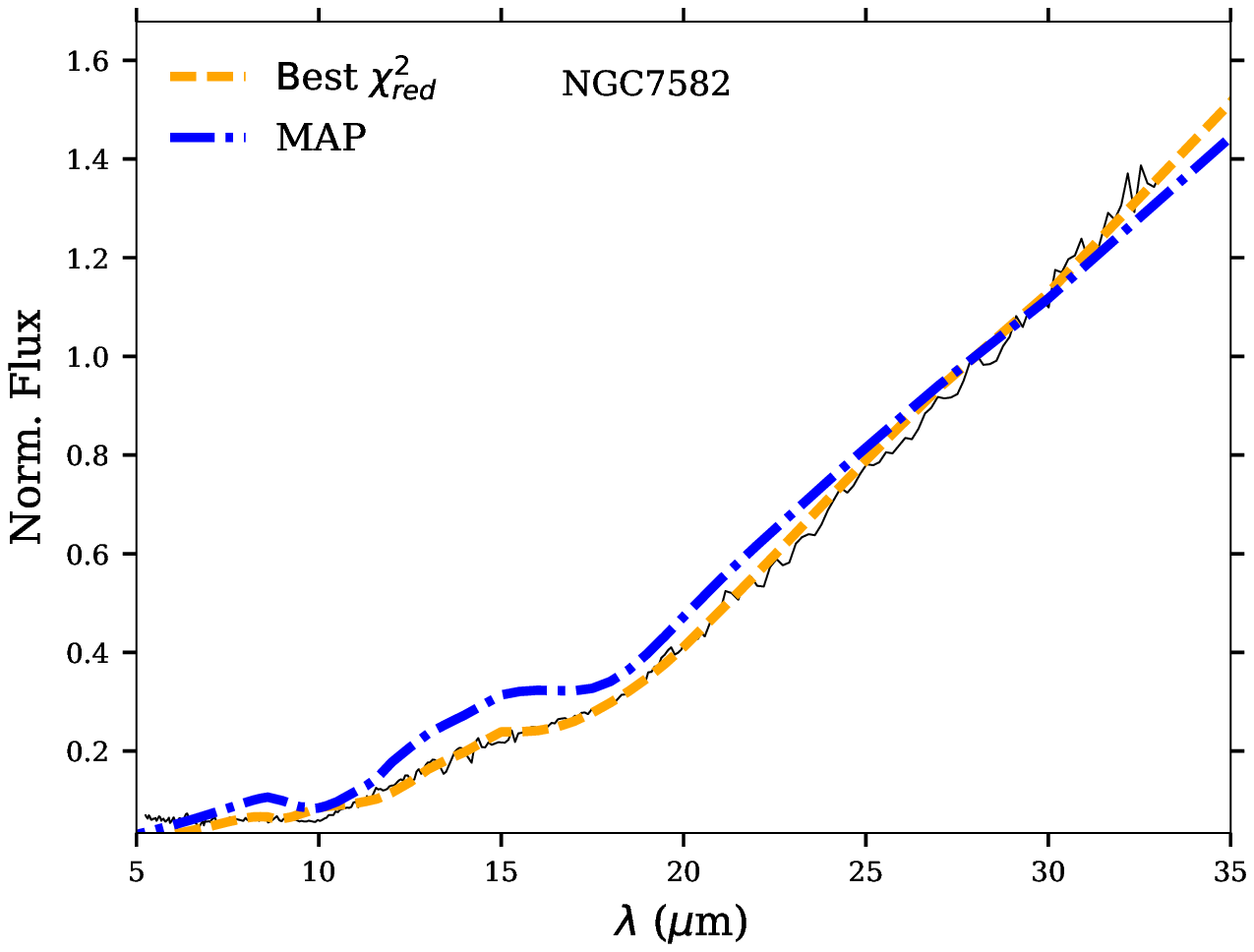}
\end{minipage} \hfill
\begin{minipage}[b]{0.325\linewidth}
\includegraphics[width=\textwidth]{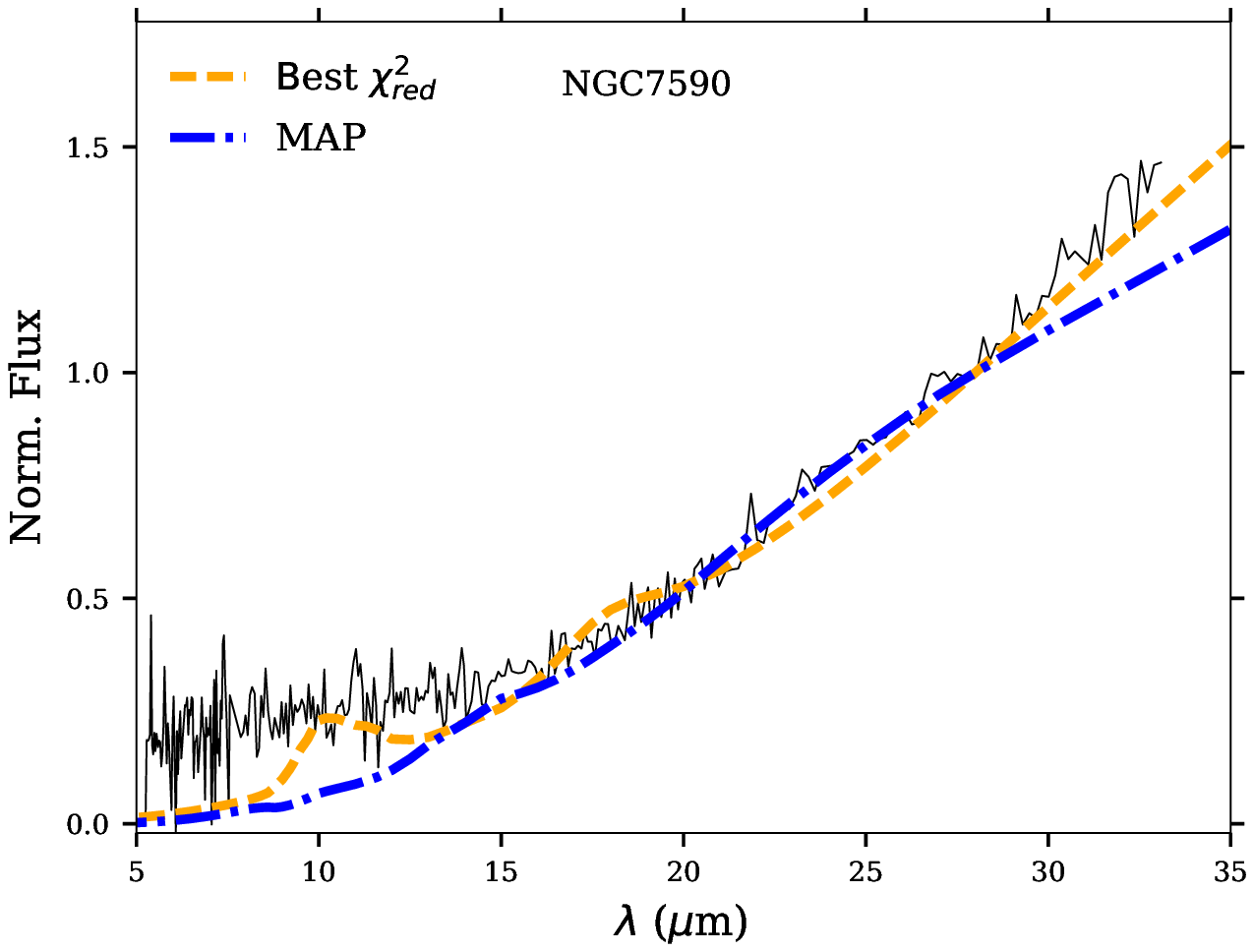}
\end{minipage} \hfill
\begin{minipage}[b]{0.325\linewidth}
\includegraphics[width=\textwidth]{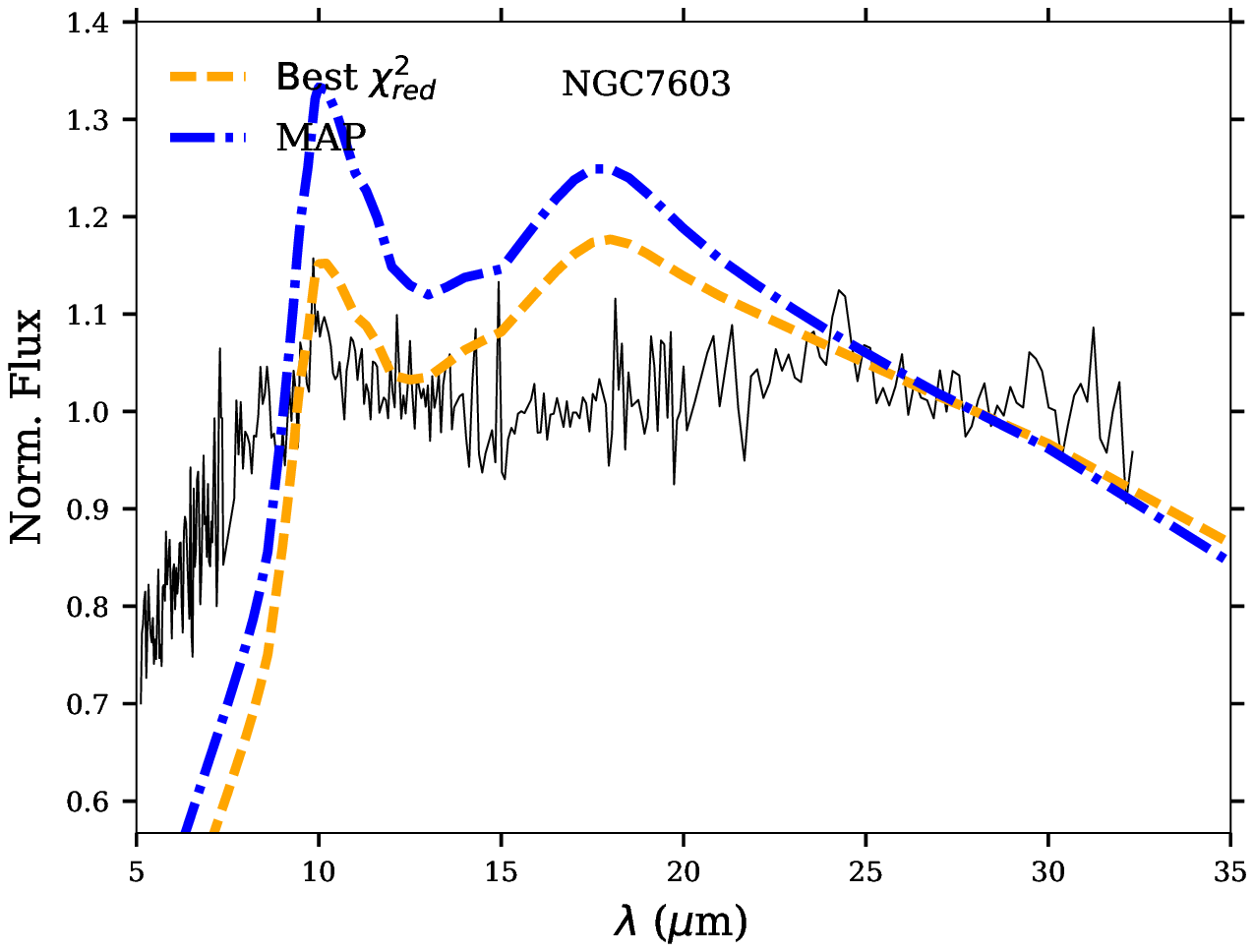}
\end{minipage} \hfill
\begin{minipage}[b]{0.325\linewidth}
\includegraphics[width=\textwidth]{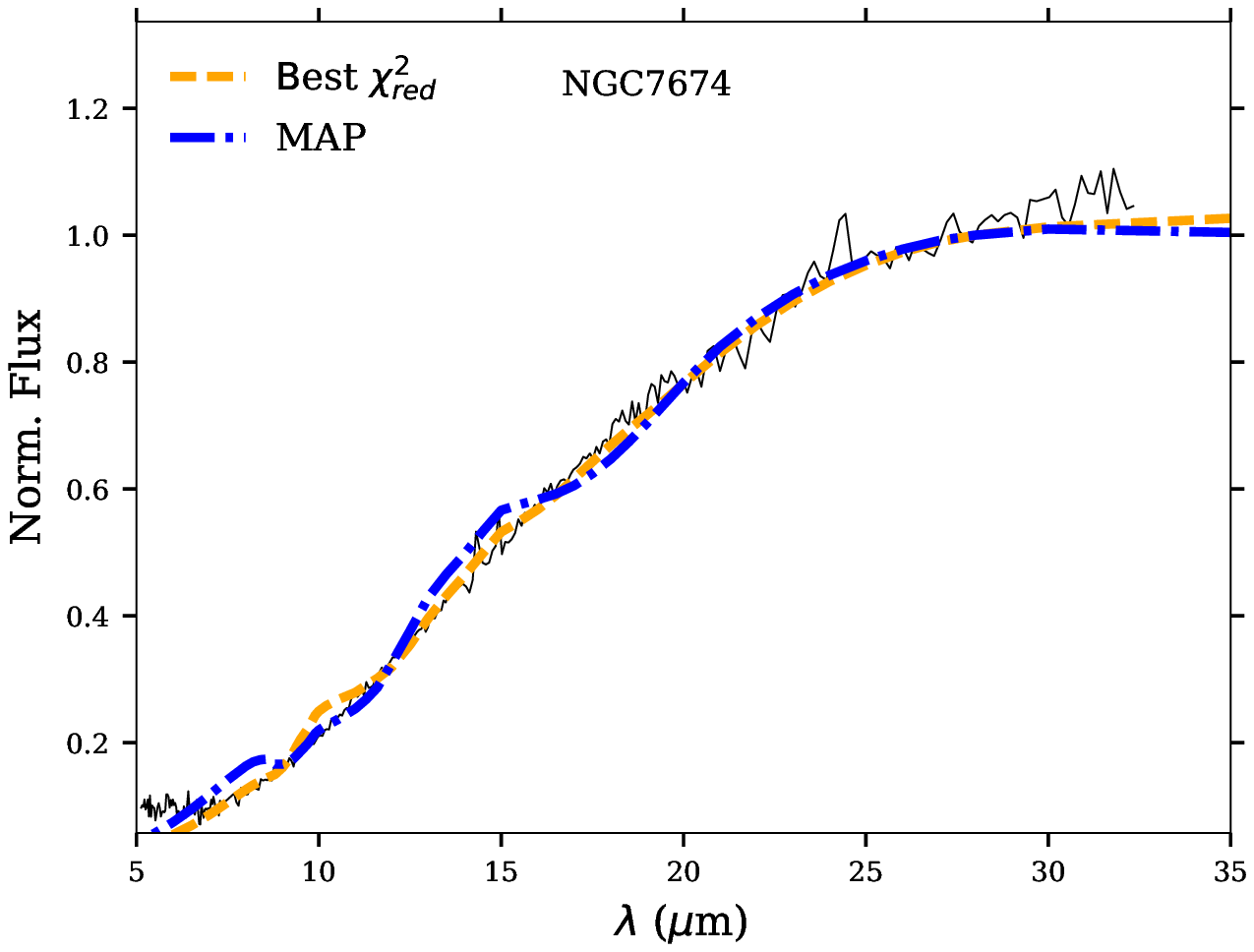}
\end{minipage} \hfill
\begin{minipage}[b]{0.325\linewidth}
\includegraphics[width=\textwidth]{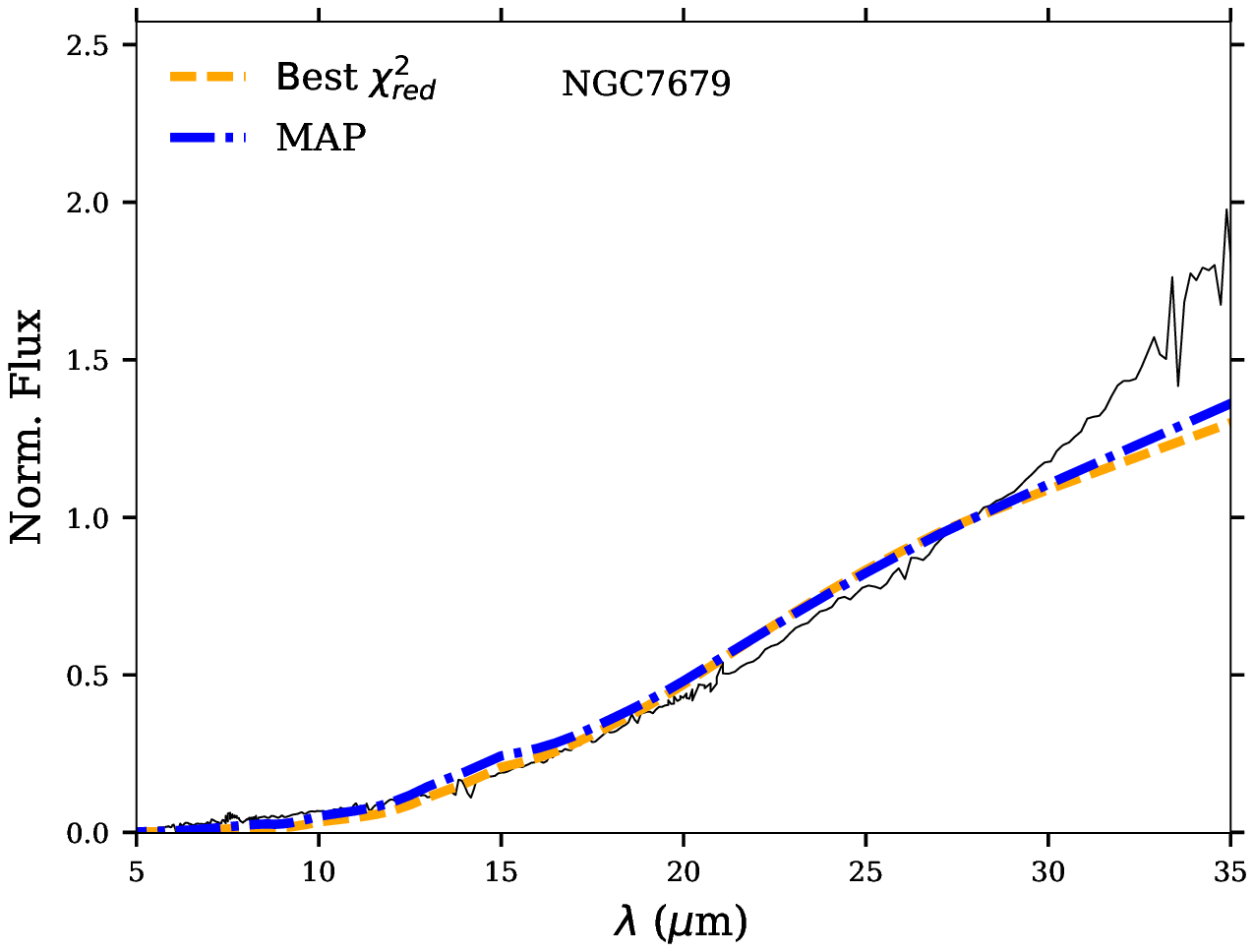}
\end{minipage} \hfill
\begin{minipage}[b]{0.325\linewidth}
\includegraphics[width=\textwidth]{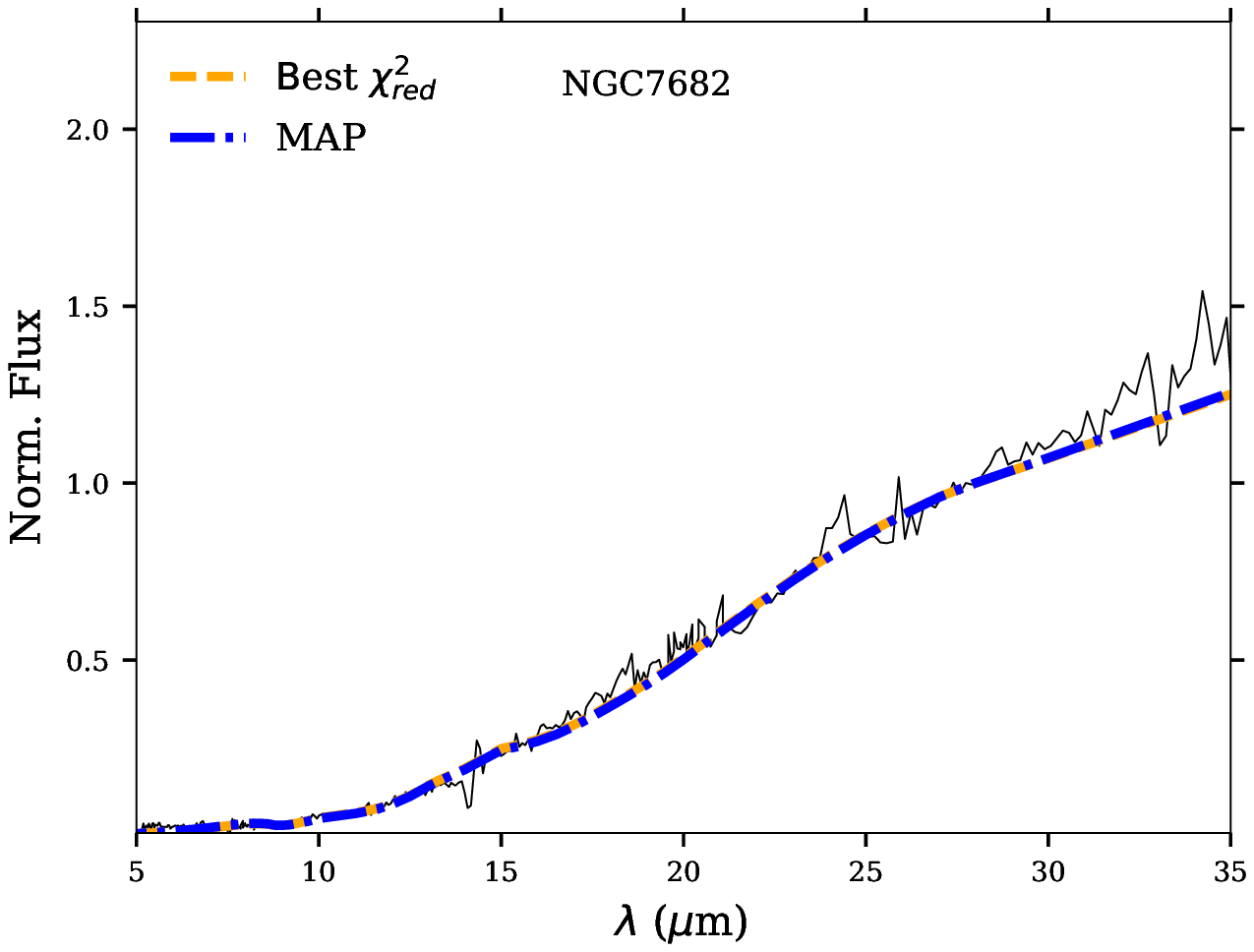}
\end{minipage} \hfill
\begin{minipage}[b]{0.325\linewidth}
\includegraphics[width=\textwidth]{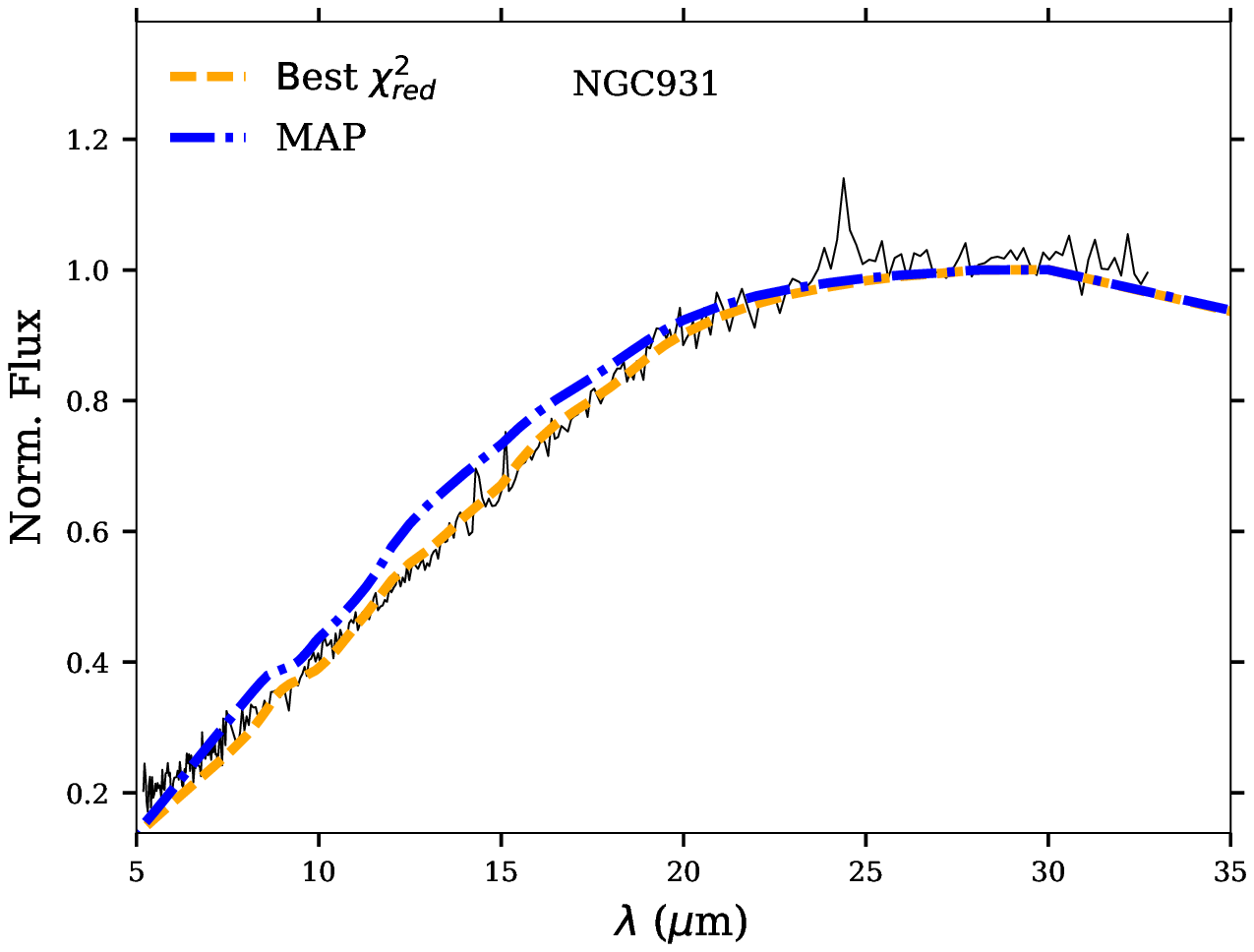}
\end{minipage} \hfill
\begin{minipage}[b]{0.325\linewidth}
\includegraphics[width=\textwidth]{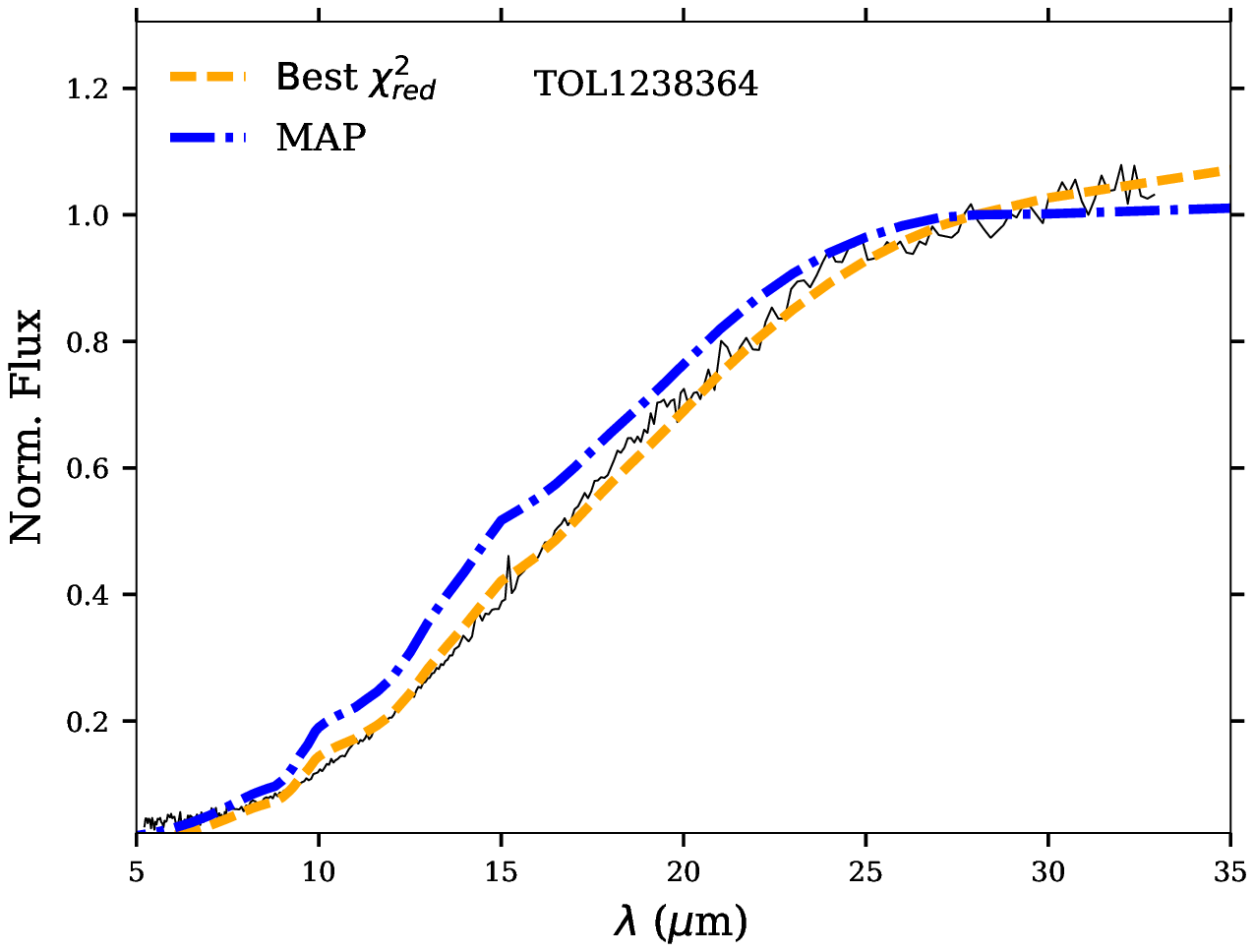}
\end{minipage} \hfill
\begin{minipage}[b]{0.325\linewidth}
\includegraphics[width=\textwidth]{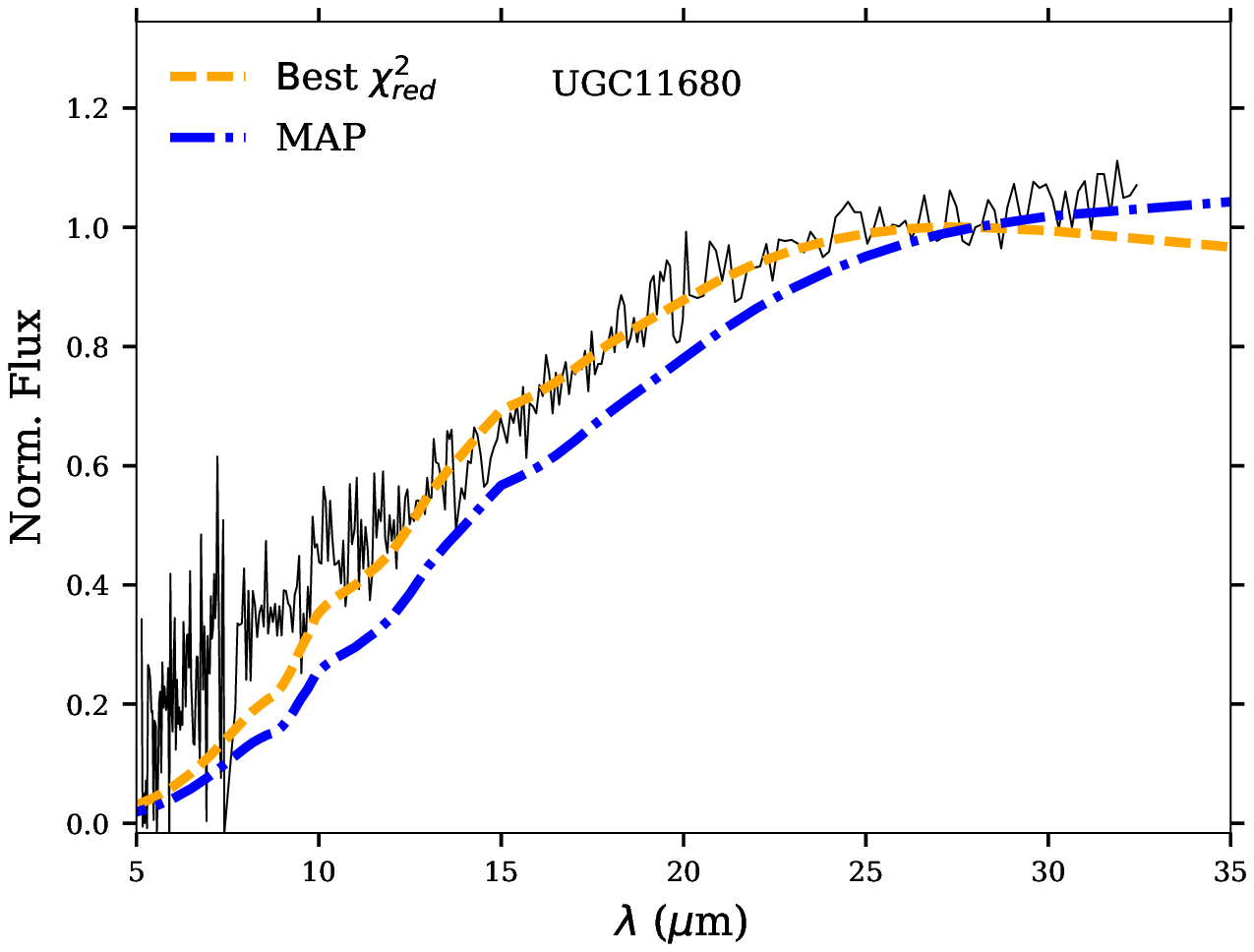}
\end{minipage} \hfill
\begin{minipage}[b]{0.325\linewidth}
\includegraphics[width=\textwidth]{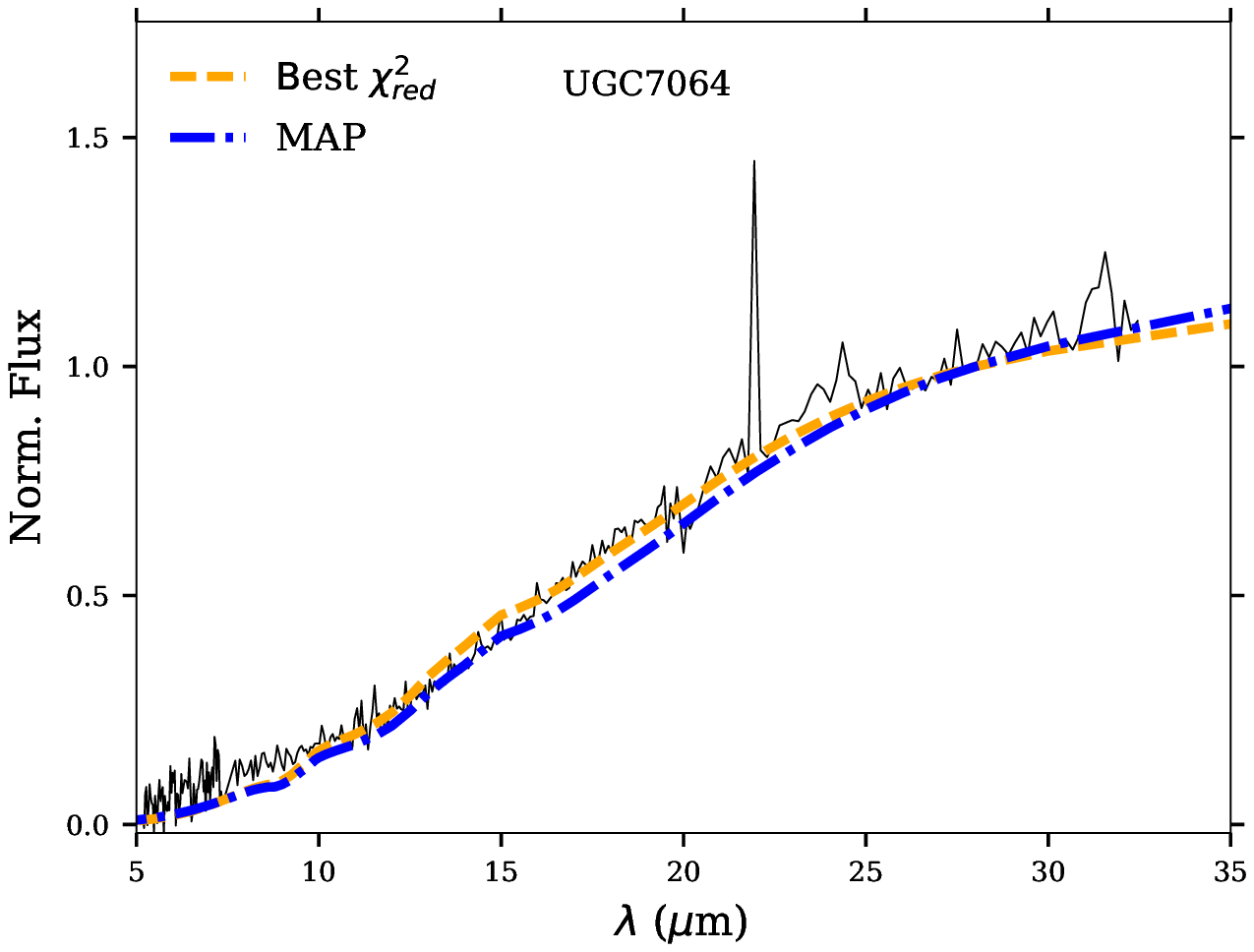}
\end{minipage} \hfill
\caption{continued from previous page.}
\setcounter{figure}{0}
\end{figure}

\label{lastpage}

\end{document}